\documentclass[]{aa}
\usepackage[varg]{txfonts}
\usepackage{graphicx}
\usepackage{natbib}
\usepackage{mathrsfs}
\usepackage{xcolor}
\usepackage{lscape}
\bibpunct{(}{)}{;}{a}{}{,}
\graphicspath{{figures/}}
\usepackage{adjustbox}
\usepackage{longtable}
\usepackage{caption}

\def\deg{\ensuremath{\,{\rm deg}}\xspace}

\def\fh{\ensuremath{^{\mathrm h}}}
\def\fm{\ensuremath{^{\mathrm m}}}
\def\fs{\ensuremath{^{\mathrm s}}}
\def\fdg{\ensuremath{^\circ}}
\def\fmin{\ensuremath{^\prime}}
\def\fsec{\ensuremath{^{\prime\prime}}}
\newcommand\gdr[1]{\gaia~DR#1}
\newcommand\egdr[1]{\gaia~EDR#1}
\newcommand{\gaia}{Gaia\xspace}
\newcommand{\hip}{Hipparcos\xspace}

\newcommand{\tyctwo}{Tycho-2\xspace}
\newcommand{\orvara}{\texttt{orvara}\xspace}
\newcommand{\exofast}{\texttt{EXOFASTv2}\xspace}
\def\teff{\ensuremath{T_{\rm eff}}\xspace}
\def\logg{\ensuremath{\log g}\xspace}
\def\gmag{\ensuremath{G}\xspace}
\def\gbp{\ensuremath{G_{\rm BP}}\xspace}
\def\grp{\ensuremath{G_{\rm RP}}\xspace}
\def\parallax{\ensuremath{\varpi}\xspace}
\def\feh{\ensuremath{[\rm Fe/H]}\xspace}
\providecommand{\yr}{\ensuremath{\,\rm yr}\xspace}
\providecommand{\pc}{\ensuremath{\,\rm pc}\xspace}
\providecommand{\Lsun}{\ensuremath{\,{\rm L}_{\odot}}\xspace}
\providecommand{\Msun}{\ensuremath{\,{\rm M}_{\odot}}\xspace}
\providecommand{\Rsun}{\ensuremath{\,{\rm R}_{\odot}}\xspace}
\providecommand{\Mjup}{\ensuremath{\,{\rm M}_{\rm Jup}}\xspace}
\providecommand{\Mearth}{\ensuremath{\,{\rm M}_\oplus}\xspace}
\providecommand{\mps}{\ensuremath{\,{\rm ms}^{-1}}\xspace}
\providecommand{\kmps}{\ensuremath{\,{\rm kms}^{-1}}\xspace}
\providecommand{\days}{\ensuremath{\,\rm d}\xspace}
\providecommand{\au}{\ensuremath{\,\rm au}\xspace}
\providecommand{\deg}{\ensuremath{\,\rm deg}\xspace}
\def\msini{{M$\sin{i}$}\xspace}
\def\mas{{mas}\xspace}
\def\muasyr{{$\rm{\mu}$as yr$^{-2}$}\xspace}

\def\nbinary{218\xspace}    % number of stars in the binary sample ----------------------
\def\nnew{130\xspace}       % number of unpublished binary systems ----------------------
\def\npub{88\xspace}        % number of published binary systems ------------------------

\def\nohgca{40\xspace}      % number of targets in the sample not in HGCA ---------------
\def\inhgca{178\xspace}     % number of targets in the sample in HGCA -------------------
\def\goodorvara{132\xspace} % number of targets with good RV+astrometry solutions -------
\def\badorvara{44\xspace}   % number of targets with bad RV+astrometry solutions --------
\def\rvonly{86\xspace}      % number of targets with only RV solutions ------------------

\def\browndwarfs{28\xspace} % number of RV brown dwarfs in the sample -------------------
\def\bdorvara{18\xspace}    % number of BD in the sample with RV+PMA solution -----------
\def\staysbd{7\xspace}      % number of RV brown dwarfs confirmed as such by orvara -----
\def\bdtostar{11\xspace}    % number of RV brown dwarfs becoming stars with orvara ------

\def\nostype{4\xspace}      % number of systems with no s-type stable orbit -------------

\title{The CORALIE survey for southern extrasolar planets}
\subtitle{XIX. Brown dwarfs and stellar companions unveiled by radial velocity and astrometry
            \thanks{The radial velocity measurements and additional data products discussed in this paper are available on the DACE web platform at \url{https://dace.unige.ch/radialVelocities}. A copy of the data is also available at the CDS via anonymous ftp to \url{cdsarc.u-strasbg.fr} (\url{130.79.128.5}) or via \url{http://cdsarc.u-strasbg.fr/viz-bin/qcat?J/A+A/}. Appendices \ref{app:rvplots} and \ref{app:detection-limits}, containing orbital solution and detection limit plots, are available in the online version of this paper.}\fnmsep
            \thanks{Based on observations collected with the CORALIE spectrograph mounted on the 1.2 m Swiss telescope at La Silla Observatory}
            }

\author{D.~Barbato\inst{\ref{obsge},\ref{oato}}
        \and D.~S\'{e}gransan\inst{\ref{obsge}}
        \and S.~Udry\inst{\ref{obsge}}
        \and N.~Unger\inst{\ref{obsge}}
        \and F.~Bouchy\inst{\ref{obsge}}
        \and C.~Lovis\inst{\ref{obsge}}
        \and M.~Mayor\inst{\ref{obsge}}
        \and F.~Pepe\inst{\ref{obsge}}
        \and D.~Queloz\inst{\ref{eth},\ref{cavlab}}
        \and N.C.~Santos\inst{\ref{iace},\ref{uniporto}}
        \and J.B.~Delisle\inst{\ref{obsge}}
        \and P.~Figueira\inst{\ref{obsge}}
        \and M.~Marmier\inst{\ref{obsge}}
        \and E.~C.~Matthews\inst{\ref{mpia},\ref{obsge}}
        \and G.~Lo Curto\inst{\ref{eso}}
        \and J.~Venturini\inst{\ref{obsge}}
        \and G.~Chaverot\inst{\ref{obsge}}
        \and M. Cretignier\inst{\ref{obsge}}
        \and J.F. Otegi\inst{\ref{obsge}}
        \and M.~Stalport\inst{\ref{obsge}}
        }
\institute{
            Department of Astronomy, University of Geneva, Chemin Pegasi 51, CH-1290 Versoix, Switzerland\\\email{domenico.barbato@unige.ch}\label{obsge}
            \and INAF – Osservatorio Astrofisico di Torino, Via Osservatorio 20, I-10025 Pino Torinese, Italy \label{oato}
            \and Max-Planck-Institut für Astronomie, Königstuhl 17, D-69117 Heidelberg, Germany \label{mpia}
            \and Department of Astronomy, University of Geneva, Ch. d'Ecogia 16, CH-1290 Versoix, Switzerland\label{ecogia}
            \and European Southern Observatory, Casilla 19001, Santiago, Chile \label{eso}
            \and ETH Zurich, Department of Physics, Wolfgang-Pauli-Strasse 2, CH-8093 Zurich, Switzerland \label{eth}
            \and Astrophysics Group, Cavendish Laboratory, JJ Thomson Avenue, CB3 0HE Cambridge, UK \label{cavlab}
            \and Instituto de Astrof\'isica e Ci\^encias do Espa\c{c}o, Universidade do Porto, CAUP, Rua das Estrelas, 4150-762 Porto, Portugal \label{iace}
            \and Departamento de F\'isica e Astronomia, Faculdade de Ci\^encias, Universidade do Porto, Rua do Campo Alegre, 4169-007 Porto, Portugal \label{uniporto}
            }
            
\date{Received <date> / Accepted <date>}
\abstract
        {A historical planet-search on a sample of 1647 nearby southern main sequence stars has been ongoing since 1998 with the CORALIE spectrograph at La Silla Observatory, with a backup subprogram dedicated to the monitoring of binary stars.}
        {We review 25 years of CORALIE measurements and search for Doppler signals consistent with stellar or brown dwarf companions to produce an updated catalog of both known and previously unpublished binary stars in the planet-search sample, assessing the binarity fraction of the stellar population and providing perspective for more precise planet-search in the binary sample.}
        {We perform new analysis on the CORALIE planet-search sample radial velocity measurements, searching for stellar companions and obtaining orbital solutions for both known and new binary systems. We perform simultaneous radial velocity and proper motion anomaly fits on the subset of these systems for which Hipparcos and Gaia astrometry measurements are available, obtaining accurate estimates of true mass for the companions.}
        {We find \nbinary stars in the CORALIE sample to have at least one stellar companion, \nnew of which are not yet published in the literature and for which we present orbital solutions. The use of proper motion anomaly allow us to derive true masses for the stellar companions in \goodorvara systems, which we additionally use to estimate stability regions for possible planetary companions on circumprimary or circumbinary orbits. Finally, we produce detection limit maps for each star in the sample and obtain occurrence rates of $0.43^{+0.23}_{-0.11}\%$ and $12.69^{+0.87}_{-0.77}\%$ for brown dwarf and stellar companions respectively in the CORALIE sample.}
        {}
\keywords{astrometry -- proper motions -- stars: fundamental parameters -- binaries: general -- techniques: radial velocities -- planets and satellites: dynamical evolution and stability}

\begin{document}
    \titlerunning{Brown dwarfs and stellar companions unveiled by radial velocity and astrometry}
    \maketitle
  
    \section{Introduction} \label{sec:introduction}
        Since June 1998, the historical CORALIE exoplanet-search survey has been continuously monitoring a southern hemisphere volume-limited sample composed of 1647 main sequence stars located within 50\pc from the Sun and having spectral types ranging from F8 to K0 \citep{queloz2000,udry2000}. As of the time of writing, the survey has collected more than 60000 radial velocity measurements using the CORALIE Echelle spectrograph mounted on the Euler Telescope at La Silla Observatory, with average measurement precision of $\sim$5\mps and an average timespan of $\sim$7600 \days.
        \par This uniquely long and continuous survey is especially suited for the detection of giant planets with semimajor axes as large as 10\au \citep{tamuz2008,segransan2010,marmier2013,rickman2019} and brown dwarfs \citep{udry2002,santos2002,rickman2019}, as well as contributing to statistical studies of the frequency of planetary companions and its dependence on stellar properties \citep[see e.g.][]{santos2001,udry2007,mayor2011}, making the almost 25-year long CORALIE survey an invaluable asset to the field of exoplanetology. Finally, it is worth remarking that the continuous monitoring of the less-active stars in the volume-limited sample also makes the CORALIE survey a fertile ground for the search of low-mass exoplanetary candidates suitable for follow-up study with higher-precision instruments such as HARPS \citep{pepe2000,mayor2003} and ESPRESSO \citep{pepe2014}, expanding its contributions to the search of exoplanetary bodies toward the realm of terrestrial companions.
        \par During the sample selection process \cite[see][]{udry2000}, known large amplitude binary stars were collected in a low-priority subprogram within the planet-search survey, due to the disruptive influence that a close-in stellar companion would have on the stability of the inner region of a planetary system \citep{holman1999,musielak2005,turrini2005,marzari2016} and in order to limit both the blending effect produced by double-lined spectroscopic binaries (SB2) and the observational effort necessary to disentangle the planetary signal from the higher-amplitude stellar contribution. On the other hand, longer-period binary companions producing only linear trends in radial velocity were instead considered still to be good candidates for the planet search survey, both due to the weak gravitational effect that the distant stellar companion would produce on the inner regions of planetary systems and the fact that such linear trends can easily be corrected for \cite[such as Gl86 and HD41004AB, see][]{queloz2000,santos2002}. This selection strategy also reflected the initial bias against low-separation binaries in favour of stellar environments similar to that of the Solar System \citep{eggenberger2010,quarles2020}. Still, the large number of measurements collected during almost 25 years of observations with CORALIE have unveiled the binary nature of a non-negligible portion of the stars selected over a wide range of orbital periods, and an updated assessment of the binary population in the sample is a necessary step for advancing the analysis of the CORALIE exoplanet-search survey.
        \par The long-term search for stellar companions within the CORALIE survey represents a key contribution to the current endeavours to further our understanding of stellar formation. Both observational and theoretical studies have now shown that most stars form with at least one stellar companions, and more specifically that about half of FGK stars are th part of a binary system \citep[see e.g.][]{moe2017,halbwachs2018,offner2022} in which the main-sequence companions appear to follow a lognormal separation distribution peaking around 40\au and roughly uniform mass-ratio distributions \citep{duquennoy1991,melo2003,raghavan2010,tokovinin2014}. It has also been shown that tighter solar-type binaries seem to favor larger mass ratios \citep{lucy1979,tokovinin2000,moe2017}, suggesting a common formation and evolution history in a shared circumbinary disc; the fact that wider ($a>200$\au) binaries also feature a small but significant fraction of high mass ratio systems \citep{elbadry2019} is similarly an indication that at least some wide stellar companion form at intermediate separation and undergo outward migration in later stages of their dynamical evolution. Many different theoretical models have been proposed to explain the formation and observed characteristics of binary systems, such as the fragmentation of filaments and cores in star-forming regions \citep{konyves2015,pineda2015,guszejnov2015,guszenjov2017} and of massive accretion discs around individual forming stars \citep{bonnell1994,gammie2001,kratter2010,harsono2011}, and continuous study of the statistics of binary systems is essential in deepening our understanding of stellar formation.
        \par Considering instead the brown dwarf companion population around solar-type stars, robust study of its demographics is hindered by the currently low number of detection, as less than 100 brown dwarf companions are currently known to orbit such stars \citep[see e.g.][]{ma2014,grieves2017}, but recent work suggest that only $\sim4\%$ of solar-type stars have brown dwarf companions \citep{offner2022}. However, a notable characteristic is a clear paucity of brown dwarf companions around solar-type primary stars on close-in orbits in what is commonly referred to as the brown dwarf desert and that could be explained by post-formation migration processes \citep[see e.g.][]{grether2006,sahlmann2011a,shahaf2019,kiefer2019}.
        \par The importance of the characterization of binary stellar systems within the scope of a radial velocity survey aimed at the detection of planetary companions is clear by virtue of the effect that the presence or absence of an additional stellar companion has on the formation and stability of planetary bodies is a fundamental theme in exoplanetary science. While works such as \cite{roell2012} found the binarity rate among planet-hosting stars to be about four times smaller than for single solar-type stars, the high sample heterogeneity and observational bias are still impediments toward a full understanding of exoplanet demographics in the binary environment \citep[see e.g.][and references therein for reviews on the subject]{thebault2015,quarles2020}. More recently, \cite{ngo2017} found no evidence that host binarity alters the distribution of planet properties in systems characterized by radial velocity observations, while \cite{su2021} reports a positive correlation between planetary multiplicity and stellar orbital separations in circumprimary planetary systems.
        \par In this paper we present the results of a new analysis of the CORALIE measurements of the 1647 stars in the sample, specifically aimed at the search of radial velocity signals comparable with stellar or brown dwarf companions of the target stars. More precisely, in this study we focus on a specific region of the binary companion parameter space, namely a region limited both in orbital separation as a result of the 25\yr duration of the CORALIE survey and in mass regimes, as we focus on companion having minimum mass higher then 40\Mjup. Companions populating the rest of the parameter space will be the main focus of future papers in this series. We find a total of \nbinary stars in the sample to have at least one such companion, among which \nnew are previously unpublished ones and \npub are instead already known and for which we present updated orbital solutions. Additionally, we present further refined orbital solution for a subset of \goodorvara binary stars in the sample using astrometry constraints provided by \hip \citep{perryman1997} and \gaia Early Data Release 3 (\egdr{3}, \citealt{gaia2021}) proper motion measurements.
        \par Our paper is organised as follows: in Sect.~\ref{sec:hosts} we describe the physical characteristics of the stars in host stars in our sample. In Sect.~\ref{sec:rv} we present an overview of the CORALIE observational campaign and of the search for radial velocity signals compatible with brown dwarf and stellar companions, while in Sect.~\ref{sec:astrometry} we obtain estimates of dynamical masses for a subset of the presented companions using \hip and \gaia proper motion measurements. In Sect.~\ref{sec:notable-cases} and \ref{sec:exoplanet-search} we respectively discuss a few systems especially worthy of note and the prospects for follow-up search for exoplanets in the systems comprising our sample, while in Sect.~\ref{sec:occurrence-rates} we derive occurrence rate values for brown dwarfs and stellar companion in the CORALIE exoplanetary search sample, before concluding and discussing the results of this work in Sect.~\ref{sec:conclusions}.
    
    \section{Host stars characteristics} \label{sec:hosts}
        \begin{figure}[t]
            \centering
            \includegraphics[width=\linewidth]{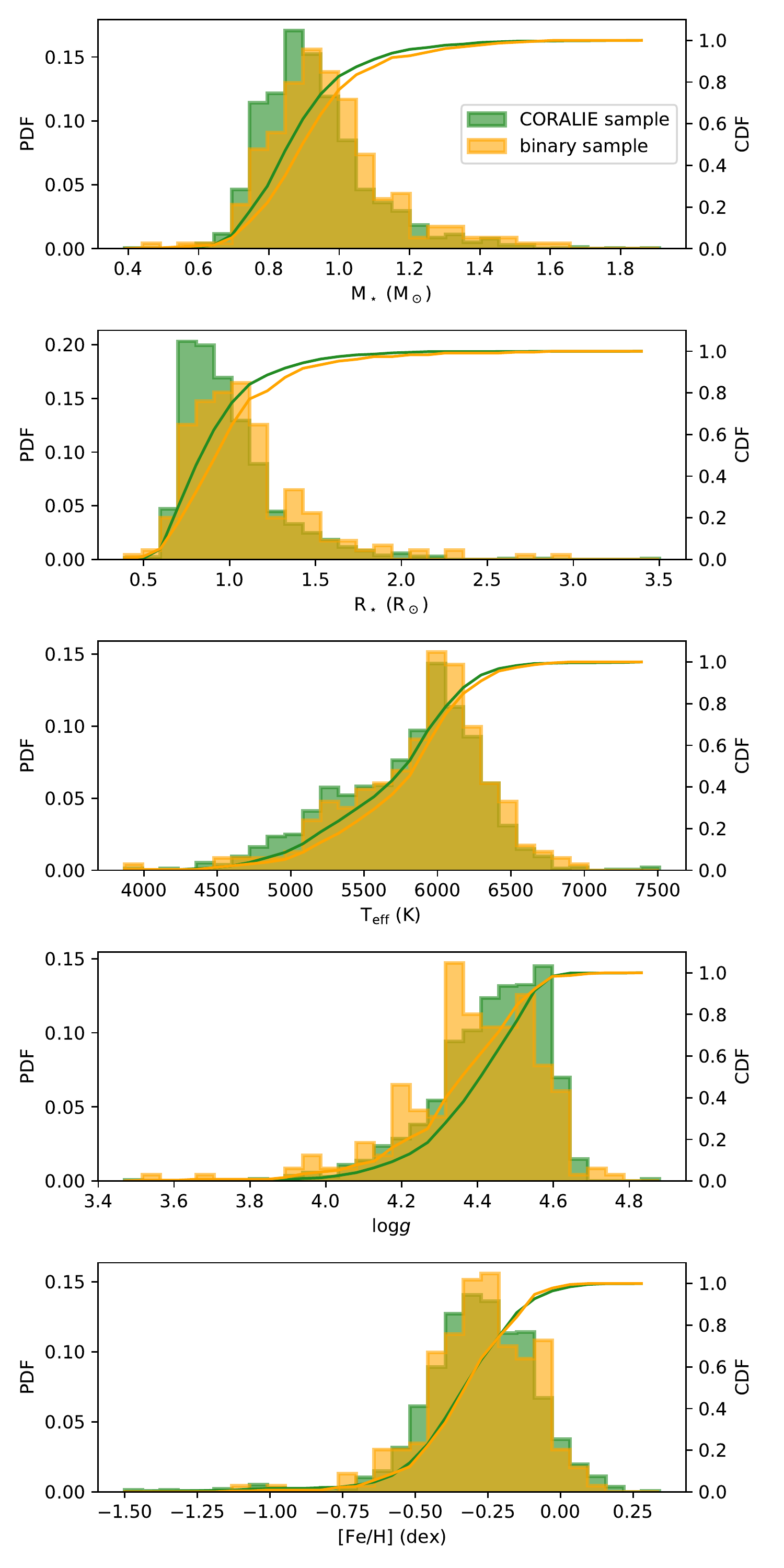}
            \caption{Distributions of stellar mass, radius, effective temperature, surface gravity and metallicity of the \nbinary stars composing the sample discussed in this work (orange) and for the whole CORALIE exoplanetary search sample (green).}
            \label{fig:sample-hosts}
        \end{figure}
        Out of the 1647 stars composing the CORALIE exoplanet-search sample, we first exclude known SB2 identified as such either by archive query or by identifying double peaks in the cross-correlation function (CCF) of the radial velocity spectra. From this we further identify a subset of \nbinary stars for which we detect a robust radial velocity signal hinting at the presence of a massive companion having minimum mass \msini$\gtrapprox$40 \Mjup; the radial velocity analysis that led to this selection is fully detailed in Sect.~\ref{sec:rv}. For clarity and simplicity, we'll refer to this subsample of \nbinary stars hosting stellar companions that is the main focus of this work simply as "the binary sample" throughout this paper, while the larger CORALIE exoplanet-search sample will be referred to as "the CORALIE sample".
        \par In order to have an updated and homogeneous characterization of the physical properties of every star in the sample, we fit the stellar Spectral Energy Distribution (SED) of each star, using the MESA Isochrones and Stellar Tracks (MIST) \citep{dotter2016,choi2016} via the \texttt{IDL} suite \exofast \citep{eastman2019}. With this method, the stellar parameters are simultaneously constrained by the SED and the MIST isochrones, since the SED primarily constrains the stellar radius $R_\star$ and effective temperature \teff, while a penalty for straying from the MIST evolutionary tracks ensures that the resulting star is physical in nature \citep[see][for more details on the method]{eastman2019}. For each star, we fitted all available archival magnitudes from Tycho $B_T$ and $V_T$ bands \citep{hog2000}, Johnson’s $B$, $V$ and 2MASS $J$, $H$, $K$ bands from the UCAC4 catalog \citep{zacharias2012}, WISE bands \citep{cutri2014}, and Gaia \gmag, \gbp and \grp bands \citep{gaia2016}, imposing Gaussian priors on each star's effective temperature \teff and metallicity \feh based on their respective values in the \cite{anders2019} catalog, as well as on the stellar parallax \parallax based on Gaia~EDR3 astrometric measurement \citep{gaia2021}.
        \par The stellar parameters derived from the SED fitting for the binary sample are listed in Table~\ref{table:star-parameters-short}, along with each star's archival spectral type, while the distribution of a few selected parameters for both the binary and CORALIE sample are plotted in Fig.~\ref{fig:sample-hosts}. The median value of host star mass in our sample is of 0.94\Msun, and the average relative error on this parameter is 10\%; median values and average errors for other stellar parameters of interest are 1.02\Rsun and 4\% for stellar radii, 5985~K and 3.72\% for effective temperature, 4.41 and 1.41\% for surface gravity \logg, -0.27~dex and 95\% for metallicity. In order to compare the distributions of the binary sample with those of the larger CORALIE sample we perform a Kolmogorov-Smirnov test for each stellar parameter derived from the SED fits, finding p-values $<0.05$ for M$_\star$ ($p=0.008$) R$_\star$ ($p=0.009$) and $\log{g}$ ($p=0.014$), suggesting that the underlying population of the binary sample is not the same as the overall CORALIE sample. Indeed, we find the median values of stellar mass, radius and surface gravity in the CORALIE sample to be 0.91\Msun, 0.95\Rsun and 4.45, suggesting therefore that the underlying population of our binary sample consists of slightly more massive and larger stars than the underlying population of the overall exoplanetary search sample.
        
        \begin{sidewaystable*}
            \caption{Stellar parameters for the sample discussed in this work, derived by the SED fits described in Sect.~\ref{sec:hosts} except spectral types, retrieved from Simbad.}\label{table:star-parameters-short}
            \centering
            % \begin{adjustbox}{scale=0.62,center}
            \begin{tabular}{l c c c c c c c c c c}
                \hline\hline
        			Name & $\alpha$(J2000) & $\delta$(J2000) & SpType & M$_\star$ & R$_\star$ & L$_\star$ & $\rho_\star$ & $\log{g}$ & T$_{\rm eff}$ & [$\rm Fe\ H$] \\
        			 & & & & [\Msun] & [\Rsun] & [\Lsun] & [cgs] & [cgs] & [K] & [dex] \\
                \hline			
                    HD225155&00\fh03\fm53.37\fs&-28\fdg23\fmin37.70\fsec&G5IV&${1.15}^{+0.15}_{-0.16}$&${1.41}\pm{0.10}$&${2.78}^{+0.75}_{-0.59}$&${0.59}^{+0.16}_{-0.13}$&${4.21}^{+0.08}_{-0.09}$&${6290}^{+370}_{-340}$&${-0.22}^{+0.27}_{-0.28}$\\
        			HD1815&00\fh22\fm23.56\fs&-27\fdg01\fmin57.05\fsec&K2V&${0.71}\pm{0.04}$&${0.67}\pm{0.03}$&${0.24}^{+0.04}_{-0.03}$&${3.30}^{+0.41}_{-0.36}$&${4.63}\pm{0.03}$&${4930}^{+160}_{-130}$&${-0.30}^{+0.20}_{-0.21}$\\
        			HD1926&00\fh23\fm04.73\fs&-65\fdg07\fmin16.11\fsec&F8/G0V&${1.00}^{+0.14}_{-0.12}$&${1.15}^{+0.14}_{-0.15}$&${1.90}^{+0.64}_{-0.56}$&${0.95}^{+0.41}_{-0.28}$&${4.33}\pm{0.10}$&${6320}\pm{260}$&${-0.49}^{+0.29}_{-0.28}$\\
        			HD2070&00\fh24\fm44.81\fs&-51\fdg02\fmin37.90\fsec&G0V&${1.15}^{+0.14}_{-0.15}$&${1.36}\pm{0.04}$&${2.66}^{+0.51}_{-0.37}$&${0.65}^{+0.12}_{-0.11}$&${4.23}^{+0.06}_{-0.07}$&${6320}^{+330}_{-270}$&${-0.23}^{+0.24}_{-0.28}$\\
        			HD2098&00\fh25\fm01.41\fs&-30\fdg41\fmin51.41\fsec&G2V&${1.01}^{+0.15}_{-0.12}$&${1.16}\pm{0.17}$&${1.68}^{+0.67}_{-0.56}$&${0.94}^{+0.47}_{-0.31}$&${4.33}^{+0.11}_{-0.12}$&${6080}^{+310}_{-300}$&${-0.22}^{+0.32}_{-0.33}$\\
        			HD3222&00\fh35\fm02.81\fs&-63\fdg41\fmin42.64\fsec&K2V&${0.78}^{+0.05}_{-0.04}$&${0.76}\pm{0.02}$&${0.46}\pm{0.07}$&${2.54}^{+0.25}_{-0.23}$&${4.57}\pm{0.03}$&${5450}^{+200}_{-190}$&${-0.41}\pm{0.21}$\\
        			HD3277&00\fh35\fm34.25\fs&-39\fdg44\fmin46.65\fsec&G8V&${0.94}^{+0.11}_{-0.10}$&${1.02}\pm{0.14}$&${1.22}^{+0.44}_{-0.36}$&${1.28}^{+0.55}_{-0.38}$&${4.40}\pm{0.10}$&${5990}\pm{230}$&${-0.30}^{+0.27}_{-0.28}$\\
        			HD3359&00\fh36\fm04.40\fs&-49\fdg07\fmin41.28\fsec&G8V&${0.94}^{+0.09}_{-0.08}$&${0.98}\pm{0.07}$&${1.00}^{+0.23}_{-0.20}$&${1.42}^{+0.31}_{-0.26}$&${4.43}\pm{0.06}$&${5830}\pm{230}$&${-0.17}\pm{0.28}$\\
        			HD3795&00\fh40\fm32.79\fs&-23\fdg48\fmin17.72\fsec&K0V&${1.05}^{+0.34}_{-0.17}$&${1.25}^{+0.80}_{-0.34}$&${1.76}^{+3.90}_{-0.96}$&${0.79}^{+0.94}_{-0.57}$&${4.28}^{+0.20}_{-0.36}$&${5800}^{+500}_{-260}$&${-0.04}^{+0.28}_{-0.35}$\\
        			HD4392&00\fh45\fm41.87\fs&-48\fdg18\fmin04.56\fsec&G4V&${0.92}^{+0.11}_{-0.09}$&${1.02}^{+0.14}_{-0.13}$&${1.30}^{+0.44}_{-0.37}$&${1.26}^{+0.54}_{-0.38}$&${4.40}\pm{0.10}$&${6100}^{+210}_{-200}$&${-0.47}\pm{0.24}$\\
        			HD4747&00\fh49\fm26.76\fs&-23\fdg12\fmin44.86\fsec&G8V&${1.02}\pm{0.09}$&${1.70}\pm{0.04}$&${2.45}^{+0.26}_{-0.16}$&${0.29}\pm{0.04}$&${3.98}\pm{0.05}$&${5540}^{+160}_{-120}$&${-0.03}^{+0.22}_{-0.24}$\\
        			HD5562&00\fh56\fm21.26\fs&-63\fdg57\fmin30.03\fsec&G8IV&${1.19}^{+0.22}_{-0.17}$&${1.89}\pm{0.07}$&${4.22}^{+1.10}_{-0.86}$&${0.25}^{+0.07}_{-0.05}$&${3.96}\pm{0.09}$&${6030}^{+430}_{-400}$&${-0.14}^{+0.25}_{-0.29}$\\
        			HD7320&01\fh13\fm18.82\fs&-01\fdg51\fmin43.72\fsec&G5V&${0.88}\pm{0.06}$&${0.87}\pm{0.02}$&${0.75}^{+0.09}_{-0.07}$&${1.87}^{+0.19}_{-0.18}$&${4.50}\pm{0.04}$&${5760}^{+180}_{-150}$&${-0.30}^{+0.19}_{-0.20}$\\
        			HD8129&01\fh20\fm30.01\fs&-19\fdg56\fmin56.73\fsec&G7V&${0.94}^{+0.11}_{-0.10}$&${1.00}\pm{0.13}$&${1.06}^{+0.41}_{-0.33}$&${1.36}^{+0.55}_{-0.39}$&${4.42}^{+0.09}_{-0.10}$&${5850}\pm{270}$&${-0.20}^{+0.31}_{-0.32}$\\
        			HD9770&01\fh35\fm01.00\fs&-29\fdg54\fmin37.34\fsec&K1V&${1.12}^{+0.23}_{-0.19}$&${1.40}^{+0.26}_{-0.29}$&${2.80}^{+1.90}_{-1.30}$&${0.61}^{+0.46}_{-0.23}$&${4.22}^{+0.15}_{-0.14}$&${6290}^{+570}_{-460}$&${-0.27}^{+0.32}_{-0.33}$\\
        			HD9905&01\fh36\fm10.09\fs&-29\fdg23\fmin32.47\fsec&K1V&${0.95}^{+0.21}_{-0.13}$&${1.11}^{+0.75}_{-0.29}$&${1.38}^{+2.40}_{-0.73}$&${0.99}^{+1.10}_{-0.75}$&${4.33}^{+0.20}_{-0.39}$&${5830}\pm{270}$&${-0.28}^{+0.28}_{-0.29}$\\
        			HD10519&01\fh42\fm14.91\fs&-17\fdg53\fmin19.47\fsec&G2V&${1.08}^{+0.19}_{-0.15}$&${1.46}^{+0.21}_{-0.24}$&${2.77}^{+1.10}_{-0.87}$&${0.50}^{+0.31}_{-0.17}$&${4.16}^{+0.14}_{-0.13}$&${6180}^{+330}_{-310}$&${-0.29}\pm{0.27}$\\
        			HD11131&01\fh49\fm23.34\fs&-10\fdg42\fmin13.08\fsec&G3V&${0.96}^{+0.17}_{-0.12}$&${1.06}^{+0.30}_{-0.20}$&${1.30}^{+1.00}_{-0.54}$&${1.16}^{+0.82}_{-0.57}$&${4.38}^{+0.14}_{-0.19}$&${5960}\pm{260}$&${-0.27}^{+0.32}_{-0.33}$\\
        			HD11264&01\fh49\fm35.56\fs&-46\fdg46\fmin07.19\fsec&G5V&${1.02}^{+0.14}_{-0.13}$&${1.17}\pm{0.14}$&${2.10}^{+0.73}_{-0.59}$&${0.92}^{+0.37}_{-0.25}$&${4.32}^{+0.09}_{-0.10}$&${6430}^{+310}_{-270}$&${-0.57}^{+0.30}_{-0.31}$\\
        			HD11352&01\fh51\fm31.19\fs&-07\fdg44\fmin23.57\fsec&G5V&${0.89}\pm{0.07}$&${0.89}^{+0.06}_{-0.05}$&${0.91}^{+0.18}_{-0.16}$&${1.79}^{+0.32}_{-0.29}$&${4.49}^{+0.05}_{-0.06}$&${5990}\pm{200}$&${-0.45}^{+0.23}_{-0.24}$\\
        			HD13945&02\fh15\fm16.20\fs&-23\fdg16\fmin52.93\fsec&G6IV&${1.03}^{+0.10}_{-0.11}$&${1.12}\pm{0.03}$&${1.63}^{+0.23}_{-0.20}$&${1.04}\pm{0.14}$&${4.35}^{+0.05}_{-0.06}$&${6170}^{+230}_{-220}$&${-0.25}^{+0.24}_{-0.27}$\\
        			HD14629&02\fh20\fm42.92\fs&-39\fdg02\fmin01.44\fsec&K3V&${0.74}\pm{0.04}$&${0.71}\pm{0.02}$&${0.36}^{+0.05}_{-0.04}$&${2.88}^{+0.23}_{-0.21}$&${4.60}\pm{0.03}$&${5270}^{+180}_{-170}$&${-0.40}^{+0.19}_{-0.20}$\\
        			HD14802&02\fh22\fm32.59\fs&-23\fdg49\fmin00.47\fsec&G0V&${1.46}^{+0.20}_{-0.23}$&${1.84}\pm{0.07}$&${6.20}^{+2.70}_{-1.40}$&${0.33}^{+0.09}_{-0.07}$&${4.07}\pm{0.09}$&${6730}^{+750}_{-510}$&${-0.12}^{+0.25}_{-0.28}$\\
        			HD15064&02\fh24\fm33.88\fs&-40\fdg50\fmin25.64\fsec&G1V&${1.29}^{+0.15}_{-0.20}$&${1.66}\pm{0.05}$&${3.70}^{+0.84}_{-0.54}$&${0.40}\pm{0.08}$&${4.11}^{+0.07}_{-0.09}$&${6210}^{+390}_{-290}$&${-0.03}^{+0.23}_{-0.27}$\\
        			HD16287&02\fh36\fm41.76\fs&-03\fdg09\fmin22.09\fsec&K1V&${0.81}\pm{0.05}$&${0.78}\pm{0.04}$&${0.41}^{+0.07}_{-0.06}$&${2.38}^{+0.29}_{-0.28}$&${4.56}^{+0.03}_{-0.04}$&${5220}^{+170}_{-160}$&${-0.09}\pm{0.20}$\\
        			HD17155&02\fh43\fm34.21\fs&-46\fdg27\fmin17.51\fsec&K4V&${0.75}\pm{0.04}$&${0.71}\pm{0.02}$&${0.27}^{+0.04}_{-0.03}$&${2.96}^{+0.23}_{-0.22}$&${4.61}\pm{0.03}$&${4960}^{+150}_{-140}$&${-0.14}^{+0.15}_{-0.16}$\\
        			HD17289&02\fh43\fm35.47\fs&-62\fdg55\fmin09.10\fsec&G0V&${1.10}^{+0.18}_{-0.16}$&${1.37}^{+0.21}_{-0.24}$&${2.65}^{+1.30}_{-1.00}$&${0.62}^{+0.40}_{-0.21}$&${4.22}\pm{0.13}$&${6290}^{+410}_{-380}$&${-0.29}^{+0.28}_{-0.27}$\\
        			HD17152&02\fh44\fm28.95\fs&-24\fdg24\fmin56.33\fsec&G8V&${0.92}\pm{0.08}$&${0.96}\pm{0.06}$&${0.98}^{+0.19}_{-0.16}$&${1.49}^{+0.28}_{-0.24}$&${4.44}^{+0.05}_{-0.06}$&${5870}^{+210}_{-200}$&${-0.26}^{+0.24}_{-0.25}$\\
        			HD18168&02\fh54\fm02.78\fs&-35\fdg54\fmin16.87\fsec&K3V&${0.91}^{+0.08}_{-0.07}$&${0.94}\pm{0.06}$&${0.91}^{+0.17}_{-0.16}$&${1.56}^{+0.29}_{-0.26}$&${4.46}^{+0.05}_{-0.06}$&${5790}\pm{210}$&${-0.20}^{+0.25}_{-0.27}$\\
        			HD18809&03\fh00\fm19.71\fs&-37\fdg27\fmin16.16\fsec&G4V&${0.93}\pm{0.08}$&${0.96}\pm{0.03}$&${1.03}^{+0.15}_{-0.13}$&${1.50}\pm{0.19}$&${4.45}^{+0.04}_{-0.05}$&${5940}^{+200}_{-190}$&${-0.31}^{+0.24}_{-0.25}$\\
        			HD18907&03\fh01\fm37.62\fs&-28\fdg05\fmin29.37\fsec&G9V&${0.48}^{+0.52}_{-0.33}$&${0.48}^{+0.66}_{-0.30}$&${0.05}^{+1.40}_{-0.04}$&${6.30}^{+32.00}_{-5.40}$&${4.77}^{+0.35}_{-0.46}$&${3990}^{+2000}_{-650}$&${-0.23}^{+0.32}_{-0.30}$\\
        			HD20916&03\fh20\fm11.75\fs&-52\fdg01\fmin54.67\fsec&K0V&${0.81}\pm{0.05}$&${0.80}\pm{0.02}$&${0.59}^{+0.08}_{-0.07}$&${2.25}^{+0.21}_{-0.20}$&${4.54}\pm{0.03}$&${5670}\pm{190}$&${-0.44}^{+0.21}_{-0.22}$\\
        			HD22705&03\fh36\fm53.40\fs&-49\fdg57\fmin28.87\fsec&G2V&${1.01}^{+0.11}_{-0.12}$&${1.11}\pm{0.06}$&${1.94}^{+0.43}_{-0.35}$&${1.05}^{+0.21}_{-0.20}$&${4.36}^{+0.06}_{-0.07}$&${6470}^{+310}_{-300}$&${-0.61}^{+0.28}_{-0.29}$\\
        			HD24492&03\fh40\fm48.90\fs&-81\fdg47\fmin20.65\fsec&G6V&${0.95}^{+0.11}_{-0.10}$&${1.02}\pm{0.12}$&${1.23}^{+0.41}_{-0.34}$&${1.28}^{+0.47}_{-0.35}$&${4.41}^{+0.08}_{-0.09}$&${6010}\pm{250}$&${-0.30}^{+0.29}_{-0.30}$\\
        			HD23308&03\fh42\fm09.85\fs&-45\fdg57\fmin28.39\fsec&F7V&${1.16}^{+0.33}_{-0.25}$&${1.35}^{+0.50}_{-0.39}$&${2.70}^{+3.80}_{-1.70}$&${0.67}^{+0.87}_{-0.36}$&${4.25}\pm{0.21}$&${6340}^{+530}_{-500}$&${-0.23}\pm{0.29}$\\
        			HD23576&03\fh44\fm45.42\fs&-38\fdg49\fmin05.05\fsec&G1V&${1.03}^{+0.10}_{-0.11}$&${1.12}\pm{0.03}$&${1.65}^{+0.21}_{-0.16}$&${1.03}\pm{0.14}$&${4.35}^{+0.05}_{-0.06}$&${6170}^{+220}_{-180}$&${-0.26}^{+0.23}_{-0.25}$\\
        			HD25874&04\fh02\fm26.97\fs&-61\fdg21\fmin25.16\fsec&G2V&${1.00}^{+0.12}_{-0.11}$&${1.11}\pm{0.11}$&${1.50}^{+0.41}_{-0.35}$&${1.05}^{+0.34}_{-0.26}$&${4.36}^{+0.08}_{-0.09}$&${6060}^{+240}_{-230}$&${-0.22}\pm{0.27}$\\
                    ...&...&...&...&...&...&...&...&...&...&...\\
                \hline
            \end{tabular}
            % \end{adjustbox}
            \tablefoot{Full table is available at the CDS. A portion is shown here for guidance regarding its form and content.}
        \end{sidewaystable*}
        
    \section{Radial velocity observations and analysis} \label{sec:rv}
        \begin{figure}[t]
            \centering
            \includegraphics[width=\linewidth]{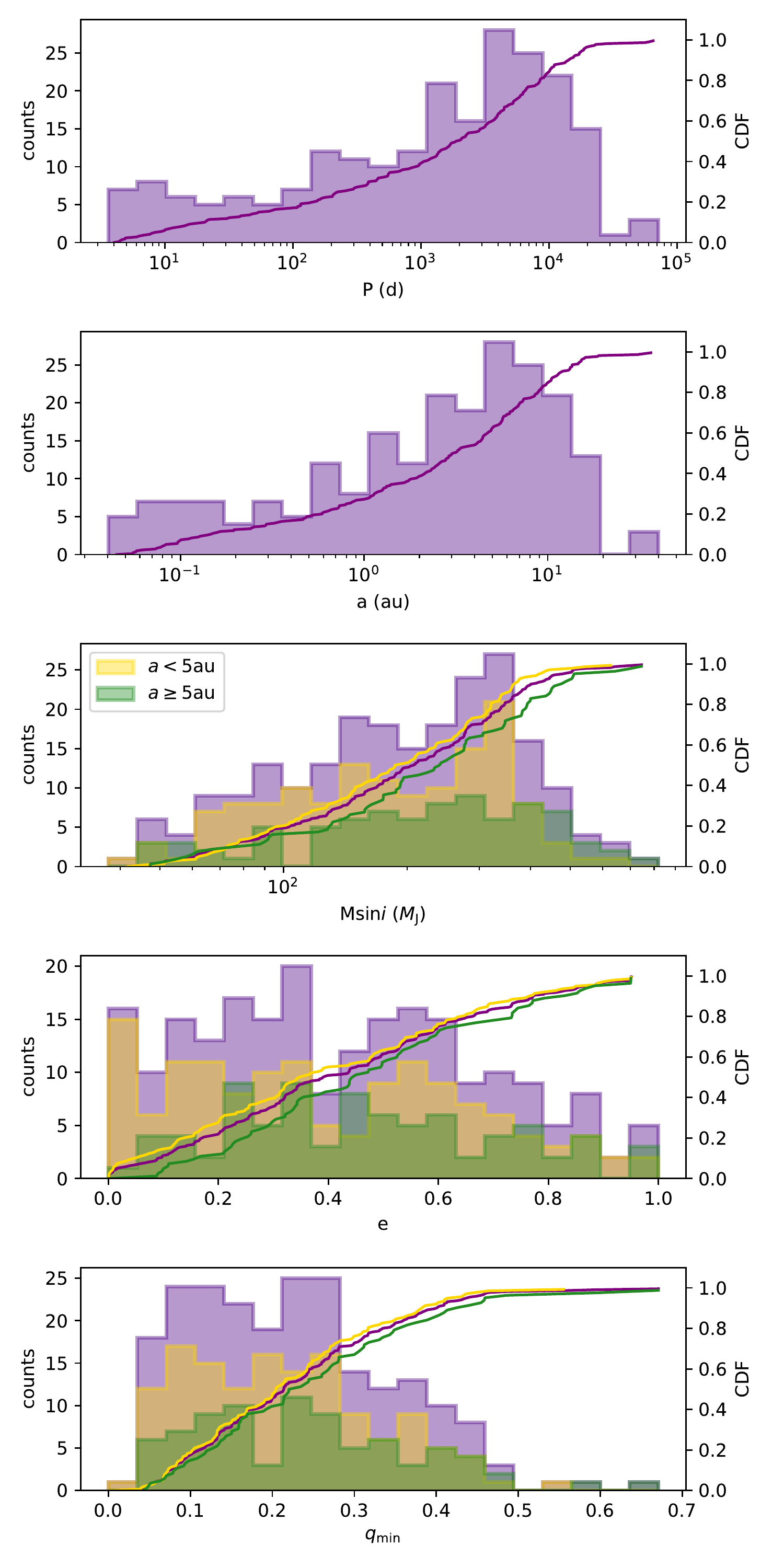}
            \caption{Distributions of orbital period, semimajor axis, minimum mass, eccentricity and mass ratio $q$ of the \msini>40\Mjup companions identified in the sample and characterized via radial velocity analysis. The minimum masses, eccentricity and $q$ distributions for inner ($a<5$\au) and outer ($a\geq5$\au) companions found in the sample are shown in yellow and green respectively.}
            \label{fig:rv-solution-hist}
        \end{figure}
        \begin{figure*}[t]
            \centering
            \includegraphics[width=\linewidth]{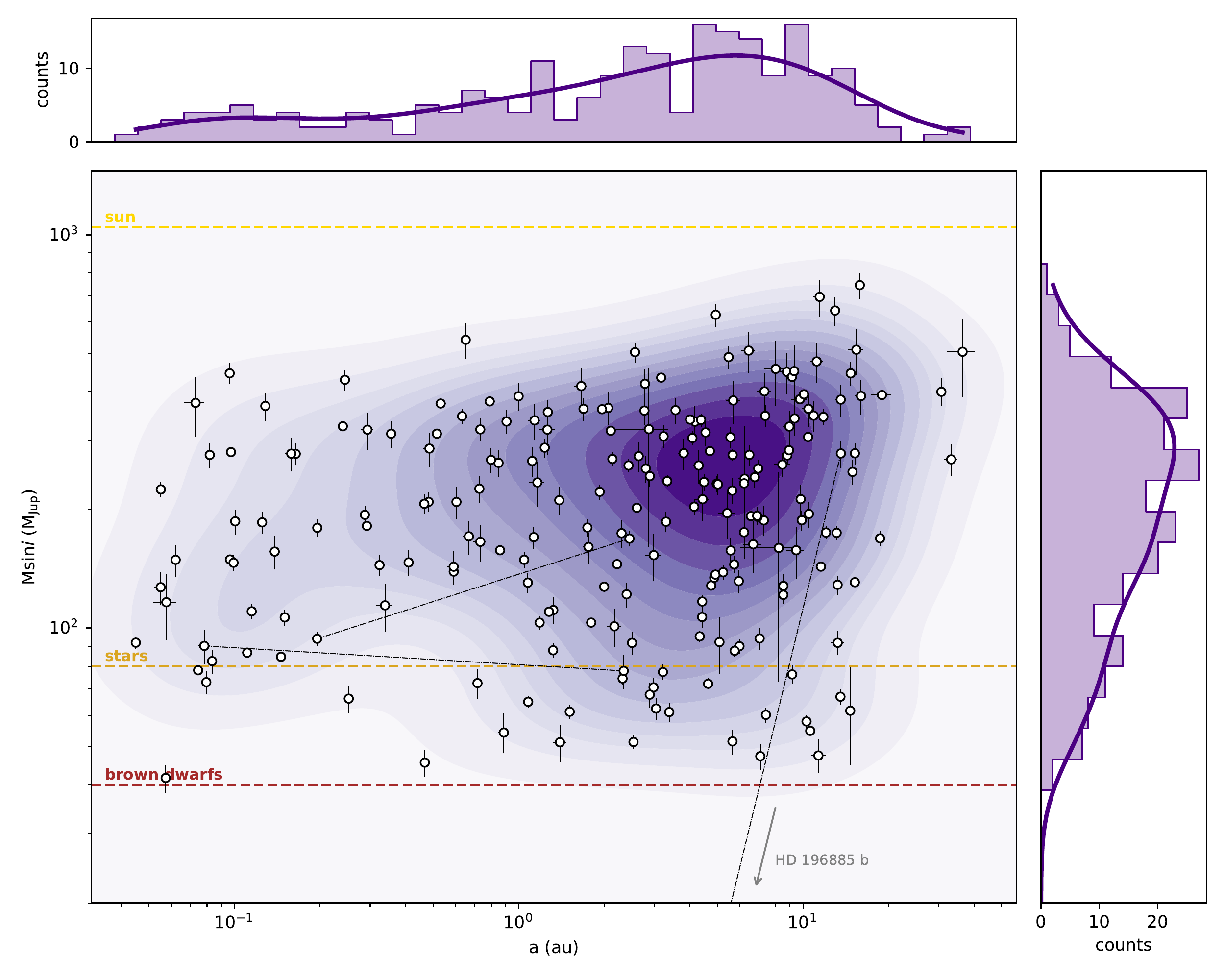}
            \caption{Distribution in the \msini-$a$ parameter space of the \msini$>$40\Mjup companions identified in the sample and characterized via radial velocity analysis. In the main plot, the kernel density estimation of the population is plotted as contour levels, the components of multiple systems are connected by black dash-dotted lines, while the horizontal dashed brown, orange and yellow lines respectively indicate the brown dwarf (40\Mjup), dwarf star (80\Mjup) and solar-mass thresholds. The top-left and right-hand histograms show the distribution and kernel density estimation of the semimajor axes and minimum mass of the companions, respectively.}
            \label{fig:rv-population}
        \end{figure*}
        Since its first observations in June 1998, the CORALIE spectrograph went through two significant upgrades in June 2007 and in November 2014 in order to increase the overall efficiency and accuracy of the instrument. Specifically, the 2007 upgrade consisted in the replacement of CORALIE's fiber link and cross-disperser optics \citep{segransan2010}, while the 2014 upgrade consisted in replacing CORALIE's fiber link with octagonal fibers \citep{chazelas2012} and adding a Fabry-P\'{e}rot calibration unit \citep{cersullo2017}. Both interventions on the instrument introduced small offsets between the radial velocity measurements collected before and after each upgrade, depending on such parameters as the spectral type and systemic velocity of the observed star; during the course of the timeseries analysis we therefore consider CORALIE as three different instruments, each marked by the different upgrades. During the course of this work we'll refer to the original CORALIE dataset as CORALIE-98 (C98), to the dataset collected after the first upgrade as CORALIE-07 (C07) and to the one collected after the most recent upgrade as CORALIE-14 (C14). For selected stars in our sample, we additionally include in the analysis the measurements collected at lower precision ($\sim$300\mps) with the  CORrelation-RAdial-VELocities (CORAVEL) spectrometer \citep{baranne1979} between 1981 and 1998, especially when the CORAVEL data are numerous enough to help identify long-period signals and constraining the orbital parameters of the companions found.
        \par As of the time of writing, over the course of almost 25 years of observations on the CORALIE sample we collected a total of 62600 radial velocity measurements for the 1647 stars in the sample, averaging 38 datapoints per star, with median photon-noise uncertainty and timespan of 5.21\mps and 7698\days. In order to search for Doppler signals consistent with the presence of stellar companions in the CORALIE radial velocity timeseries, we follow an iterative process of investigation of successive dominant peaks in the radial velocity periodogram, such as described in \cite{delisle2016}. As mentioned in Sect.~\ref{sec:hosts}, we once again note that stars found to be SB2s in the CORALIE sample are excluded from the following analysis.
        \par First of all, we consider in our analysis only the 1497 stars for which a total of at least 10 CORALIE measurements have been collected over the years, to ensure robust identification of significant signals. We model instrumental offsets, noise and stellar jitter for each star in the CORALIE sample following the formalism detailed in \cite{diaz2016} and \cite{delisle2018}, computing false alarm probabilities (FAPs) on the periodogram of the residuals as described in \cite{baluev2008}. The periodogram's main peak is considered significant if characterized by a FAP lower than 0.1\% and is modeled as a Keplerian, and the thusly obtained new radial velocity residuals is again investigated for significant signals, re-adjusting the jitter, noise and offsets at each step of the iterative process. This method is however valid only when enough measurements are available to compute a value of FAP; for those cases in which no robust value of FAP is obtained but a clear variation having a scatter in excess of the observation formal errors is present in the radial velocity measurements, we still model the data with a Keplerian model, assessing its significance using the difference between the Bayesian Information Criterion ($\Delta\rm{BIC}$) of the Keplerian and flat models, computed as:
        \begin{equation}    \label{eq:bic}
            \rm{BIC}=k\log{n}-2\log{\mathscr{L}},
        \end{equation}
        with $k$ as the number of model parameters, $n$ the number of datapoints and $\log\mathscr{L}$ the maximised log-likelihood of the model evaluated following \cite{delisle2020}. Additionally, the longer-period signals found during this search are also modelled as linear or quadratic trends instead, and are included in the final sample that is the main focus of this work only if $\Delta\rm{BIC}>10$ in favour of the Keplerian solution. Overall, the signal search process is similar to that undertaken in parallel in Unger et al., in prep., focused instead on identifying new giant planets and brown dwarf companions in the CORALIE sample.
        \par At the end of this analysis, we identified a total of \nbinary stars featuring at least one signal compatible with a companion minimum mass \msini$\gtrapprox$40\Mjup within 1$\sigma$, a threshold between giant planets and brown dwarfs we select following the findings reported in \cite{sahlmann2011b}, composing the binary sample representing the main focus of this work. The CORALIE radial velocity dataset for this sample is comprised of a total of 7226 CORALIE measurements, averaging 33 datapoints per star and featuring a median radial velocity uncertainty of 5.31\mps and an observational timespan of 7581\days; all data products are publicly available at the Data and Analysis Center for Exoplanets (DACE)\footnote{\url{https://dace.unige.ch}}. We run a Markov chain Monte Carlo (MCMC) analysis for each star in our sample based on the algorithm described in \cite{diaz2016} and \cite{delisle2016,delisle2018} in order to obtain the posterior distribution of the model parameters, using initial conditions drawn from the orbital solutions we obtained during our preliminary iterative signal search and computing each parameter's confidence intervals for a 68.27\% confidence level.
        \par A summary of the bestfit radial velocity orbital solutions for all companions found in the sample is listed in the left portion of Table~\ref{table:full-solutions}, the distributions of selected orbital parameters and mass ratio are plotted in the histograms shown in Fig~\ref{fig:rv-solution-hist}, while the companions distribution on the \msini-$a$ parameter space is shown in Fig~\ref{fig:rv-population}, and finally the phase folded radial velocity curves for every companion in the sample are collected in Appendix~\ref{app:rvplots}.
        \par It can be seen that the population of the companions having \msini$>$40\Mjup identified by radial velocity in the sample peaks at $\sim$5.92\au and $\sim$290\Mjup (around 0.27 \Msun) and that the orbital elements cover a large variety, with periods ranging from 4\days (for HD196998B) to 65213\days (HD3795B), semimajor axes from 0.045\au (HD196998B) to 36.40\au (HD3795B), minimum masses from 41.60 \Mjup ($\sim$0.04\Msun, HD30774B) to $\sim$0.71\Msun (HD181199B) and eccentricities values from fully circular (HD207450B) to 0.95 (HD137763B). It can be also seen that all the companions in the sample have a minimum mass below the solar mass.
        \par Another point of interest is the distinction between single–lined binaries (SB1) and double–lined spectroscopic binaries (SB2) in the sample. Following \cite{halbwachs2003} we can use the minimum mass ratio parameter $q_{\rm min}$ between secondary and primary component to identify possible SB2s as those having $q>0.8$. By doing so we find no binary systems in the sample with such high value of $q$ as characterized by our radial velocity solutions.
        \par To search for differences in the properties of inner ($a<$5\au) and outer ($a\geq$5\au) companions we again perform a Kolmogorov-Smirnov test of the orbital elements of the detected companions, finding p-values only marginally lower than 0.05 for minimum mass ($p=0.03$) and eccentricity ($p=0.04$) suggesting a possible difference in the respective distributions for inner and outer companions. Indeed, as shown in Fig.~\ref{fig:rv-solution-hist}, more companions are found on low-eccentricity inner orbits than outer ones, likely as a result of orbit circularization effects.
        %there seem to be a surplus of 65$\lesssim$\msini$\lesssim$150\Mjup companion on closer orbits than those with similar minimum masses found on outer orbits, and more companions are found on low-eccentricity inner orbits than outer ones, likely as a result of orbit circularization effects.
        \par Finally, we find three stars in the sample to host more than one companion. As described by our radial velocity solution, HD94340 is a triple star system in which the 1.28\Msun primary is orbited by a \msini$\sim$0.08\Msun companion on a 6.84\days orbit and by a \msini$\sim$0.07\Msun body with an orbital period of $\sim$1123\days, while HD206276 hosts a \msini$\sim$0.09\Msun companion on a 32\days orbit and a \msini$\sim$0.16\Msun at 1374\days, while HD196885 hosts both an inner giant planet with minimum mass of 1.95\Mjup and orbital period of 1330\days and an outer stellar companion with \msini$\sim$0.26\Msun on a 14912\days orbit.
        Two of these systems are already known in the literature (see \citealt{tokovinin2006,tokovinin2012} for HD94340 and \citealt{correia2008} for HD196885) but in the present work we provide updated orbital parameters for all components, especially with the use of astrometry constrains detailed in Sect.~\ref{sec:astrometry}, while for HD206276 we provide an updated solution for the outer stellar companion, whose presence was already hinted at by astrometric observations, and present a new inner stellar companion (see Sections~\ref{subsec:HD206276}-\ref{subsec:HD196885} for more details on the specific systems).
        
    \section{Astrometric constraints} \label{sec:astrometry}
        \begin{figure}[t]
            \centering
            \includegraphics[width=\linewidth]{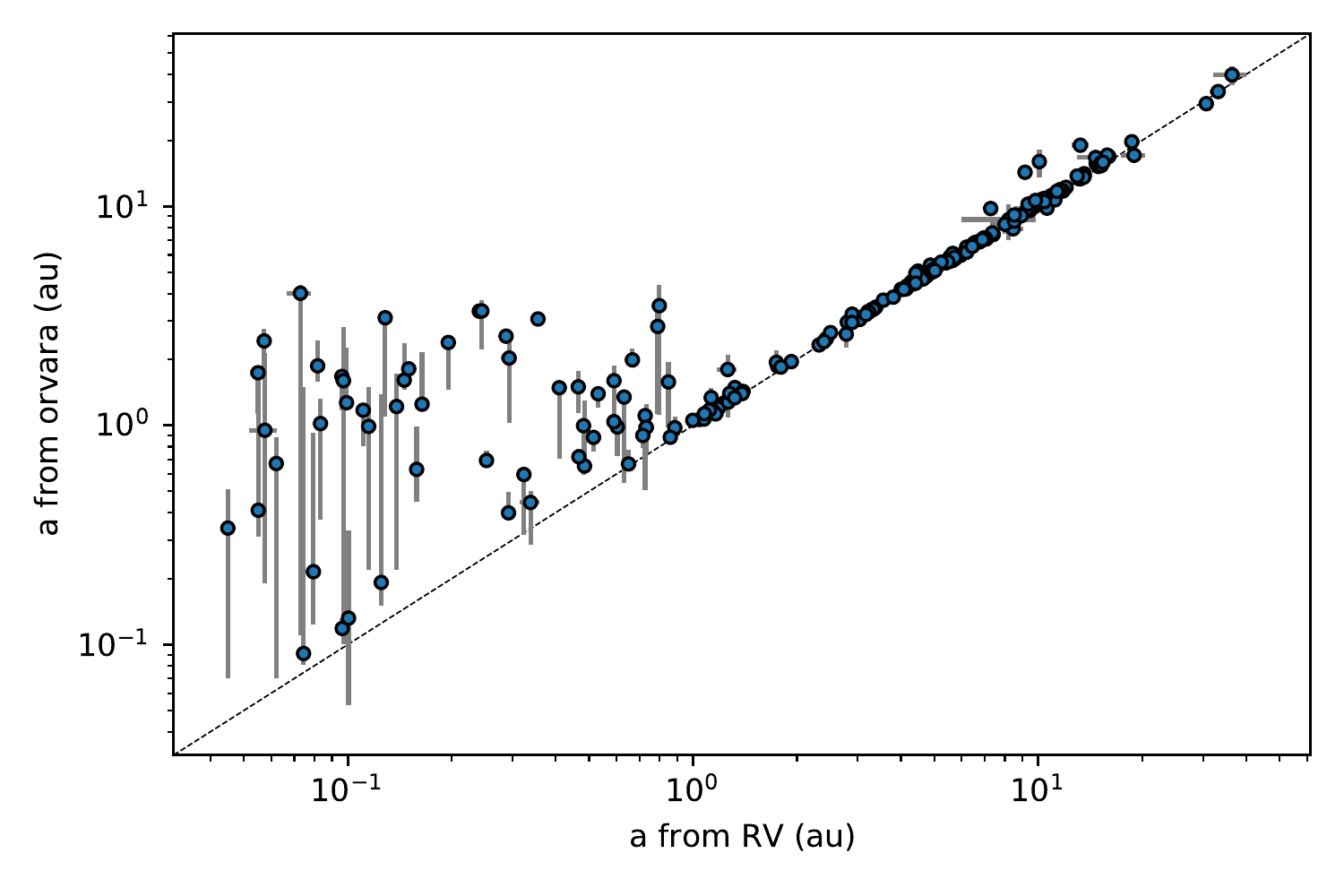}
            \caption{Comparison between the best-fit semimajor axes obtained by the simultaneous fitting of radial velocity timeseries and proper motions variations and those obtained by the fitting of radial velocities alone.}
            \label{fig:rv-dpm-diagonal-sma}
        \end{figure}
        \begin{figure*}[t]
            \centering
            \includegraphics[width=\linewidth]{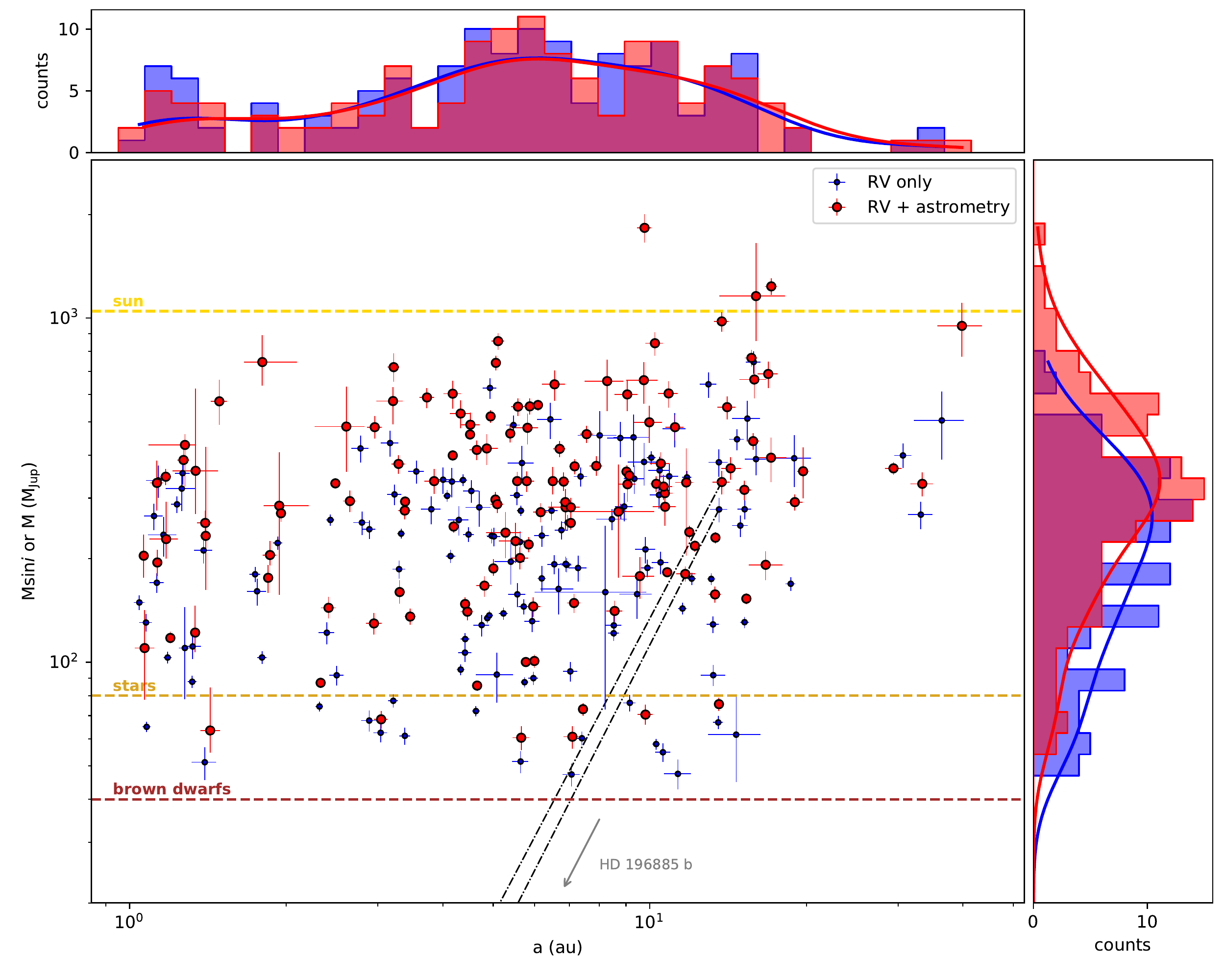}
            \caption{Comparison between the distributions in the \msini-$a$ parameter space of the \msini$>40$\Mjup companions orbiting the \goodorvara stars in the sample for which we performed simultaneous radial velocity and proper motion anomaly fits. The blue dots and histogram respectively show the parameter space position and distribution of semimajor axes and minimum mass as retrieved by the radial velocity analysis alone, while the respective red plot elements refer instead to the results of the simultaneous astrometry and radial velocity fits, and therefore true dynamical mass instead of minimum mass. The components of multiple systems are connected by gray dash-dotted lines, while the horizontal dashed brown, orange and yellow lines respectively indicate the brown dwarf (40\Mjup), dwarf star (80\Mjup) and solar-mass (1047.58\Mjup) thresholds.}
            \label{fig:rv-dpm-distributions}
        \end{figure*}
        In order to achieve a more complete characterization of the orbital parameters of the stellar companions of the CORALIE sample identified in our radial velocity analysis we use the variations in proper motion measurements $\delta\mu$ in conjunction with the radial velocity data already analysed in Sect~\ref{sec:rv} to derive precise dynamical mass and orbits for the massive companions in the sample. Specifically, we use the proper motion anomalies between the \hip \citep{perryman1997} and \gaia \citep{gaia2016} epochs, which have been shown to be able to detect accelerations as small as a few \muasyr \citep{sahlmann2016,brandt2018}. Such an approach has been successfully used more and more commonly in recent years to characterize both stellar and substellar companions \citep[see e.g.][]{calissendorff2018,snellen2018,kervella2019anomaly,kervella2019rrlyrae2,kervella2019rrlyrae1,kervella2020,kervella2022,brandt2019,brandt2020,brandt2021,damasso2020proxima,damasso2020pimen,makarov2021dpav,makarov2021eeri,venner2021,feng2021,llop2021}, especially with the advent of more and more precise proper motions measurements provided by the different Gaia Data Releases.
        \par In this work we use the proper motion measurements provided by \egdr{3}, which are on average about 3-4 times more precise than those provided by the previous \gdr{2} \citep{gaia2021,lindegren2021,brandt2021hgca}. More specifically, we use the variations $\Delta\mu_{\alpha, \delta}$ from a purely linear motion as reported in the \hip-\gaia Catalog of Accelerations \citep[HGCA,][]{brandt2018,brandt2021hgca}, in which the \hip and \egdr{3} catalogues have been cross-calibrated to account for systematics and shift all proper motions in the \egdr{3}. For this purpose, we use the open source code \orvara \citep{brandt2021orvara}, an MCMC orbit fitting code that is able to fit Keplerian orbits to any combination of HCGA proper motion variations, absolute astrometry, relative astrometry, and radial velocities to obtain precise dynamical masses and orbital elements. We use a version of \orvara that has been especially modified to accept priors on orbital periods and semimajor axes\footnote{\url{https://github.com/nicochunger/orvara/tree/period-prior}}, in addition to the priors on primary mass and stellar jitter already included in the original version of the code, to help in achieving better convergence and more constrained orbital elements especially for intermediate-separation binaries.
        \par We find \nohgca stars in our binary sample not to be included in the HGCA, leaving us with \inhgca stars for which the simultaneous use of proper motion anomalies and radial velocity is in principle viable. For each of them, we impose priors on the primary mass equal to the stellar masses obtained with the SED fits (see Sect.~\ref{sec:hosts}), and on orbital periods equal to the periods retrieved from the radial velocity analysis (see Sect.~\ref{sec:rv}).
        \begin{figure}[t]
            \centering
            \includegraphics[width=\linewidth]{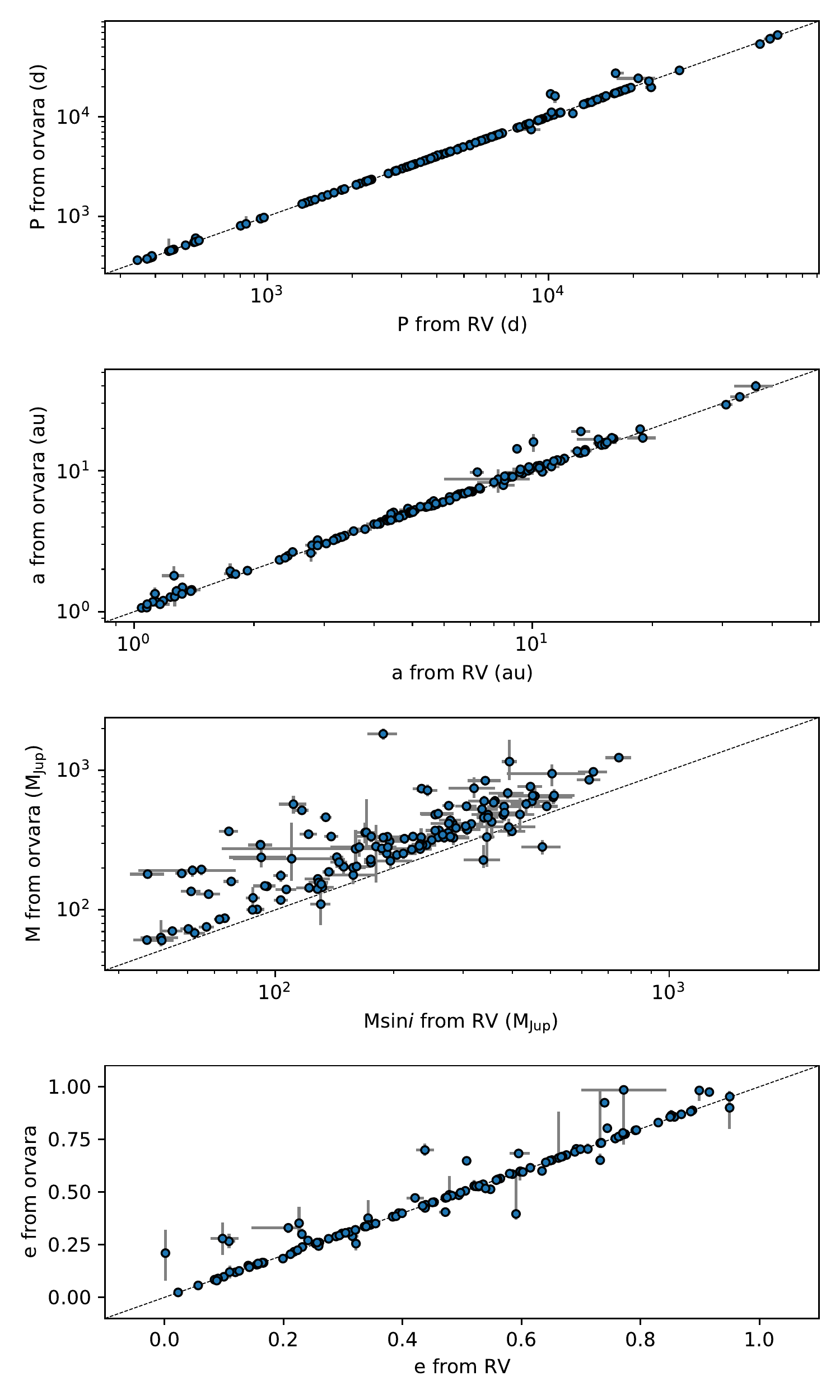}
            \caption{Comparison between the best-fit orbital period, semimajor axis, minimum (or true) mass and eccentricity values of \msini$>$40\Mjup companions as obtained by the simultaneous fitting of radial velocity timeseries and proper motions variations and those obtained by the fitting of radial velocities alone for the \goodorvara stars considered in Sect.~\ref{sec:astrometry}.}
            \label{fig:rv-dpm-diagonal}
        \end{figure}
        \begin{figure}[t]
            \centering
            \includegraphics[width=\linewidth]{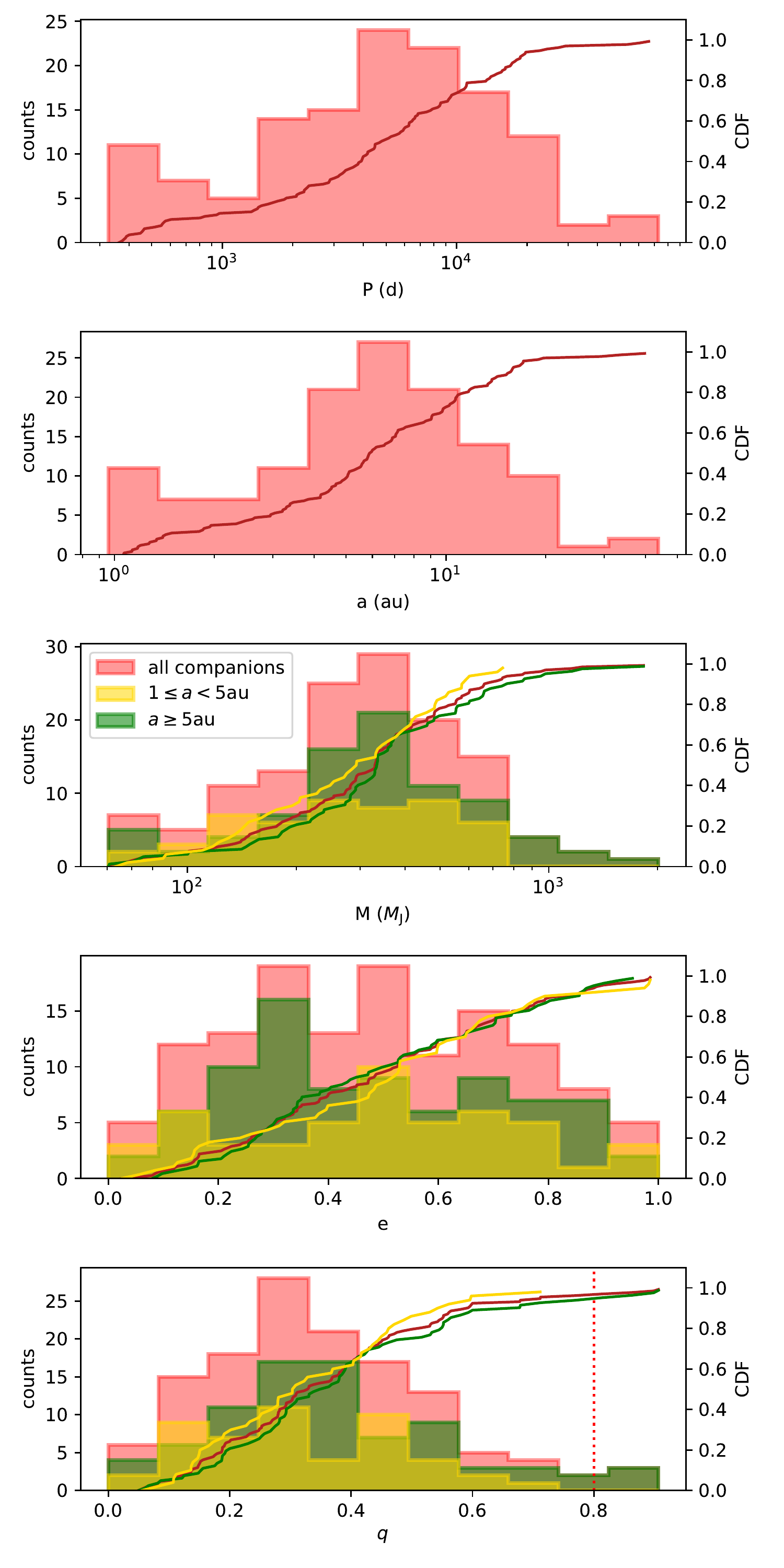}
            \caption{Distributions of orbital period, semimajor axis, minimum mass, eccentricity and mass ratio $q$ of \msini$>$40\Mjup companions as obtained by the simultaneous radial velocity and proper motion anomaly fits for the \goodorvara stars considered in Sect.~\ref{sec:astrometry}. The minimum masses, eccentricity and $q$ distributions for inner ($1\leq a<5$\au) and outer ($a\geq5$\au) companions found in the sample are shown in yellow and green respectively.}
            \label{fig:rv-dpm-hist}
        \end{figure}
        \par As the use of proper motion anomaly is based on two measurements separated by almost 25\yr, the method is clearly more efficient for long orbital periods \citep{kervella2019anomaly}; when analysing in such a manner the entirety of our sample, which as discussed in Sect.~\ref{sec:rv} features a large variety of orbital periods, special cautions must be taken. This is particularly evident when comparing the semimajor axes for the companions in the sample as characterized by radial velocities alone with those resulting from the \orvara simultaneous fit of proper motion variations and radial velocities. Such a comparison, shown in Fig.~\ref{fig:rv-dpm-diagonal-sma}, highlights a large discrepancy between the values obtained by the two methods for the companions that radial velocity characterized as having shorter separations, and that \orvara typically overestimates the orbital periods of such close-in stellar companions. For the purposes of this work, we identify $a_{\rm RV}\sim$1\au as a reasonable threshold above which astrometry constraints are indeed helpful in constraining the orbital elements of the stellar companions in our sample, and we decide to focus our following analysis only on the \goodorvara stars in our sample for which our radial velocity analysis identified a companion with semimajor axis greater than 1\au. In the cases of HD206276 and HD94340, being both triple star systems hosting one companion below 1\au, we perform the \orvara analysis using as radial velocity timeseries the residuals obtained after subtracting the Keplerian signal of the inner stellar companion at 0.19 and 0.08\au respectively, so that the timeseries used for the simultaneous fit contain in principle only the contribution of the outer companions in the systems. Lastly, we must note that binary systems with mass ratios $q>0.6$ could, in principle, be affected by the luminosity of the secondary companion shifting the photocenter orbit from the primary orbit. To account for this, we analysed all the CCFs for the systems having such mass ratios, failing to detect any secondary peak; this corresponds to a lower limit on magnitude ratio of 2.5-3.0, which would lead at most to a barycentric semimajor axis underestimation of of 5-9\%.
        \par A summary of the best-fit solution obtained for these \goodorvara systems using \orvara is listed in Table~\ref{table:full-solutions}, while Fig~\ref{fig:rv-dpm-distributions} compares the distributions of the stellar companions as characterized by radial velocity alone and the simultaneous use of radial velocity and proper motion anomaly. Generally speaking, this plot shows again the good agreement in orbital separations for the companions in the sample between the two different solutions, with the semimajor axis distribution from the radial velocity fits alone and the one obtained with the addition of proper motion anomalies peaking both at $\sim$6.30\au. Of clear interest is also the comparison between the mass distributions obtained by the two solutions, as the determination of orbital inclinations derived from the use of astrometry permits the derivation of the true dynamical mass of the stellar companions, shifting the peak of the mass distribution upward to $\sim$343\Mjup ($\sim$0.33\Msun) from the peak value of $\sim$262\Mjup ($\sim$0.25\Msun) for the \msini distribution derived from radial velocity alone. Fig~\ref{fig:rv-dpm-diagonal} shows a comparison between the values of selected orbital parameters of \msini$>$40\Mjup companions as obtained by radial velocities and those obtained using also astrometry constraints, while Fig~\ref{fig:rv-dpm-hist} shows the distribution of the same parameters as obtained by the simultaneous radial velocity and proper motion anomaly fits. Once again, we generally find good agreement between the orbital separations found from the two solutions and between the eccentricities as well. As done for the orbital parameters obtained from the radial velocity solutions (see Sect.~\ref{sec:rv}) we again perform Kolmogorov–Smirnov tests to search for significant differences in the distributions of the properties of inner (1$\leq a<$5\au) and outer ($a\geq$5au) companions, finding this time no significant difference between the respective distributions of true mass ($p=0.39$) and eccentricities ($p=0.66$).
        \par The true masses derived from the simultaneous fit once again show the majority of the companions found in the sample to lie below the solar mass value within the respective uncertainties, although the precise estimate of orbital inclination provided by the astrometry measurements allow some companions to reveal themselves to have a true mass greater than 1\Msun, namely the ones found orbiting HD27019 (M$=1.74\pm0.17$\Msun), HD223084 (M$=0.93\pm0.06$\Msun), HD173872 (M$={1.10}_{-0.29}^{+0.47}$\Msun), HD181199A (M$=1.18\pm0.06$\Msun) and HD3795 (M$={0.90}_{-0.17}^{+0.15}$\Msun).
        \begin{sidewaystable*}
            \caption{Best-fit orbital solutions for the binary systems identified in the sample, left side reporting the results from the radial velocity fits (see Sect.~\ref{sec:rv} and right side referring to the simultaneous radial velocities and proper motions fits (see Sect.~\ref{sec:astrometry}). We note that systems with mass ratio $q>0.6$ could in principle be affected by photocenter bias leading to underestimating the barycentric semimajor axis at most by 5-9\%}\label{table:full-solutions}
            \begin{adjustbox}{scale=0.56,center}
            \centering
            \begin{tabular}{l |c c c c c c c c c c |c c c c c c c c |c}
    			\hline\hline
    			    & \multicolumn{10}{c|}{RV only solution} &\multicolumn{8}{c|}{RV+$\Delta\mu$ solution} &\\
    			\hline
        			Name & P & K & e & $\lambda_{\mathrm{0}}$ & $\omega$ & \multicolumn{2}{c}{\msini} & $a$ & $t_{\rm span}/P$ & $q_{\rm min}$ & \multicolumn{2}{c}{M} & a & P & e & i & $\omega$ & $q$ & Previous Publication\\
        			& [$\mathrm{d}$] & [$\mathrm{kms^{-1}}$] & & [$\mathrm{deg}$] & [$\mathrm{deg}$] & [\Mjup] & [\Msun] & [$\mathrm{au}$] & & & [\Mjup] & [\Msun] & [\au] & [d] & [$\mathrm{deg}$] & [$\mathrm{deg}$] & & \\
    			\hline		
            			HD225155&${2276.36}^{+0.30}_{-0.23}$&${4.137}\pm{0.003}$&${0.321}\pm{0.001}$&${79.91}\pm{0.05}$&${295.24}^{+0.13}_{-0.15}$&${278.39}^{+25.14}_{-26.40}$&${0.27}^{+0.02}_{-0.03}$&${3.80}^{+0.16}_{-0.17}$&${3.44}$&${0.23}$& ${336}_{-28}^{+28}$&${0.32}\pm{0.03}$& ${3.85}_{-0.16}^{+0.15}$&${2276.23}\pm{0.24}$& ${0.32004}_{-0.00068}^{+0.00066}$& ${77.7}_{-3.9}^{+5.0}$& ${295.37}_{-0.15}^{+0.15}$&${0.28}$&---\\
            			HD1815&${1160.40}\pm{0.05}$&${6.624}^{+0.002}_{-0.003}$&${0.130}\pm{0.001}$&${146.15}\pm{0.02}$&${81.03}^{+0.19}_{-0.22}$&${269.22}^{+10.84}_{-11.00}$&${0.26}\pm{0.01}$&${2.13}\pm{0.04}$&${10.62}$&${0.36}$& --- & --- & --- & --- & --- & --- & --- & ---&---\\
            			HD1926&${156.33}\pm{0.01}$&${7.904}^{+0.001}_{-0.002}$&${0.039}\pm{0.001}$&${-297.16}\pm{0.01}$&${64.91}^{+0.35}_{-0.33}$&${209.34}^{+19.07}_{-19.98}$&${0.20}\pm{0.02}$&${0.60}\pm{0.03}$&${69.73}$&${0.20}$& --- & --- & --- & --- & --- & --- & --- & ---&\cite{tokovinin2014}\\
            			HD2070&${115.20}\pm{0.01}$&${15.110}\pm{0.001}$&${0.347}\pm{0.001}$&${191.37}\pm{0.01}$&${355.98}^{+0.02}_{-0.01}$&${372.55}^{+31.61}_{-33.19}$&${0.36}\pm{0.03}$&${0.53}\pm{0.02}$&${118.79}$&${0.31}$& --- & --- & --- & --- & --- & --- & --- & ---&\cite{tokovinin2014}\\
            			HD2098&${2693.21}^{+0.47}_{-0.50}$&${4.909}\pm{0.002}$&${0.141}\pm{0.001}$&${-319.46}^{+0.07}_{-0.10}$&${317.56}^{+0.39}_{-0.44}$&${334.97}^{+32.36}_{-34.00}$&${0.32}\pm{0.03}$&${4.17}^{+0.18}_{-0.20}$&${5.20}$&${0.32}$& ${528}_{-49}^{+48}$&${0.50}\pm{0.05}$& ${4.33}_{-0.20}^{+0.18}$&${2690.29}^{+1.57}_{-2.37}$& ${0.1509}_{-0.0057}^{+0.0087}$& ${55.7}_{-1.6}^{+1.7}$& ${315.5}_{-1.7}^{+1.3}$&${0.50}$&\cite{tokovinin2014}\\
            			HD3222&${15676.93}^{+2.18}_{-2.37}$&${1.771}\pm{0.002}$&${0.321}\pm{0.001}$&${222.25}^{+0.10}_{-0.07}$&${162.19}^{+0.38}_{-0.27}$&${175.00}^{+6.95}_{-7.08}$&${0.17}\pm{0.01}$&${12.04}^{+0.22}_{-0.23}$&${0.74}$&${0.21}$& ${218.0}_{-8.2}^{+8.0}$&${0.21}\pm{0.01}$& ${12.21}_{-0.23}^{+0.22}$&${15675.43}\pm{9.86}$& ${0.3202}_{-0.0021}^{+0.0022}$& ${109.57}_{-0.83}^{+0.80}$& ${162.43}_{-0.61}^{+0.62}$&${0.27}$&---\\
            			HD3277&${46.15}\pm{0.01}$&${4.071}\pm{0.002}$&${0.285}\pm{0.001}$&${132.80}\pm{0.02}$&${320.50}\pm{0.07}$&${66.17}^{+5.04}_{-5.25}$&${0.06}\pm{0.01}$&${0.25}\pm{0.01}$&${231.47}$&${0.07}$& --- & --- & --- & --- & --- & --- & --- & ---&\cite{sahlmann2011a}\\
            			HD3359&${22.14}\pm{0.01}$&${22.295}^{+0.016}_{-0.006}$&${0.342}\pm{0.001}$&${102.46}^{+0.09}_{-0.11}$&${5.87}^{+0.16}_{-0.30}$&${277.39}^{+17.27}_{-17.81}$&${0.26}\pm{0.02}$&${0.16}\pm{0.01}$&${529.05}$&${0.28}$& --- & --- & --- & --- & --- & --- & --- & ---&---\\
            			HD3795&${65213.79}^{+1409.42}_{-4386.49}$&${2.844}^{+0.061}_{-0.103}$&${0.495}^{+0.009}_{-0.017}$&${20.15}^{+0.53}_{-0.55}$&${32.80}^{+0.83}_{-0.76}$&${504.25}^{+107.02}_{-116.63}$&${0.48}^{+0.10}_{-0.11}$&${36.41}^{+3.74}_{-4.27}$&${0.12}$&${0.46}$& ${949}_{-177}^{+159}$&${0.91}^{+0.15}_{-0.17}$& ${39.8}_{-4.1}^{+3.7}$&${66110.25}\pm{4017.75}$& ${0.485}_{-0.020}^{+0.019}$& ${118.4}_{-2.3}^{+2.1}$& ${34.4}_{-1.5}^{+1.6}$&${0.86}$&---\\
            			HD4392&${548.84}^{+0.06}_{-0.08}$&${3.706}^{+0.004}_{-0.008}$&${0.615}\pm{0.001}$&${148.17}^{+0.05}_{-0.06}$&${226.43}^{+0.17}_{-0.18}$&${111.20}^{+8.69}_{-9.06}$&${0.11}\pm{0.01}$&${1.32}\pm{0.05}$&${22.16}$&${0.12}$& ${573}_{-83}^{+88}$&${0.55}\pm{0.08}$& ${1.488}_{-0.057}^{+0.051}$&${548.73}^{+0.05}_{-0.07}$& ${0.61518}_{-0.00088}^{+0.0010}$& ${15.2}_{-1.4}^{+1.8}$& ${226.21}_{-0.12}^{+0.11}$&${0.60}$&\cite{tokovinin2014}\\
            			HD4747&${12196.71}^{+295.98}_{-231.17}$&${0.703}\pm{0.005}$&${0.732}\pm{0.001}$&${55.33}\pm{1.10}$&${266.93}\pm{0.25}$&${54.81}^{+3.28}_{-3.36}$&${0.05}\pm{0.01}$&${10.60}^{+0.35}_{-0.36}$&${0.69}$&${0.05}$& ${70.5}_{-4.6}^{+4.8}$&${0.07}\pm{0.01}$& ${9.81}_{-0.33}^{+0.32}$&${10782.18}^{+255.67}_{-237.41}$& ${0.651}_{-0.023}^{+0.032}$& ${41.2}_{-2.4}^{+2.1}$& ${282.3}_{-4.6}^{+2.9}$&${0.07}$&\cite{peretti2019}\\
            			HD5562&${3968.53}^{+0.80}_{-0.67}$&${4.605}\pm{0.004}$&${0.336}\pm{0.001}$&${266.55}\pm{0.03}$&${354.46}^{+0.15}_{-0.12}$&${379.31}^{+45.32}_{-48.29}$&${0.36}^{+0.04}_{-0.05}$&${5.68}^{+0.30}_{-0.34}$&${2.03}$&${0.30}$& ${480}_{-50}^{+48}$&${0.46}\pm{0.05}$& ${5.82}_{-0.31}^{+0.28}$&${3968.37}^{+0.62}_{-0.58}$& ${0.33599}_{-0.00056}^{+0.00052}$& ${97.5}_{-1.8}^{+1.7}$& ${354.43}_{-0.14}^{+0.14}$&${0.39}$&---\\
            			HD7320&${13476.31}^{+4.81}_{-1.13}$&${10.277}\pm{0.001}$&${0.950}\pm{0.001}$&${265.41}^{+0.37}_{-0.41}$&${91.74}^{+0.54}_{-0.61}$&${344.50}^{+15.56}_{-15.93}$&${0.33}^{+0.01}_{-0.02}$&${11.79}^{+0.24}_{-0.25}$&${0.97}$&${0.38}$& ${333}_{-129}^{+86}$&${0.32}^{+0.08}_{-0.12}$& ${11.75}_{-0.57}^{+0.37}$&${13475.94}\pm{2.05}$& ${0.90}_{-0.10}^{+0.08}$& ${90.28}_{-1.1}^{+0.79}$& ${100.5}_{-16}^{+1.7}$&${0.36}$&\cite{latham2002}\\
            			HD8129&${5269.10}^{+145.40}_{-5.90}$&${9.054}^{+0.708}_{-1.216}$&${0.949}^{+0.007}_{-0.043}$&${83.40}^{+3.17}_{-5.96}$&${281.18}^{+2.99}_{-4.77}$&${239.88}^{+86.78}_{-89.67}$&${0.23}^{+0.08}_{-0.09}$&${6.23}^{+0.29}_{-0.30}$&${1.33}$&${0.24}$& --- & --- & --- & --- & --- & --- & --- & ---&\cite{tokovinin2014}\\
            			HD9770&${1673.32}^{+0.08}_{-0.09}$&${2.562}\pm{0.001}$&${0.312}\pm{0.001}$&${351.33}^{+0.03}_{-0.04}$&${292.65}^{+0.10}_{-0.09}$&${153.31}^{+20.34}_{-21.74}$&${0.15}\pm{0.02}$&${2.98}^{+0.18}_{-0.21}$&${3.97}$&${0.13}$& --- & --- & --- & --- & --- & --- & --- & ---&\cite{watson2001}\\
            			HD9905&${10126.26}^{+122.66}_{-156.35}$&${1.608}^{+0.021}_{-0.018}$&${0.298}^{+0.007}_{-0.008}$&${272.30}^{+0.44}_{-0.58}$&${355.47}^{+1.18}_{-1.42}$&${157.80}^{+22.61}_{-24.24}$&${0.15}\pm{0.02}$&${9.46}^{+0.63}_{-0.73}$&${0.96}$&${0.16}$& ${178}_{-25}^{+24}$&${0.17}\pm{0.02}$& ${9.57}_{-0.71}^{+0.62}$&${10212.39}^{+80.36}_{-76.70}$& ${0.3032}_{-0.0042}^{+0.0043}$& ${86.8}_{-2.0}^{+2.3}$& ${354.35}_{-0.88}^{+0.88}$&${0.18}$&---\\
            			HD10519&${7731.73}^{+0.98}_{-1.24}$&${6.793}\pm{0.001}$&${0.765}\pm{0.001}$&${172.69}\pm{0.04}$&${155.68}\pm{0.04}$&${448.35}^{+51.11}_{-54.22}$&${0.43}\pm{0.05}$&${8.78}^{+0.44}_{-0.50}$&${1.90}$&${0.40}$& ${600}_{-63}^{+58}$&${0.57}\pm{0.06}$& ${9.05}_{-0.49}^{+0.42}$&${7737.46}\pm{9.50}$& ${0.76442}_{-0.00081}^{+0.00081}$& ${95.4}_{-4.6}^{+3.3}$& ${155.76}_{-0.19}^{+0.19}$&${0.53}$&\cite{tokovinin2014}\\
            			HD11131&${3358.89}^{+1.48}_{-1.45}$&${5.333}^{+0.010}_{-0.009}$&${0.676}\pm{0.001}$&${92.10}^{+0.05}_{-0.06}$&${52.60}\pm{0.11}$&${281.96}^{+32.38}_{-34.27}$&${0.27}\pm{0.03}$&${4.70}^{+0.24}_{-0.27}$&${3.49}$&${0.28}$& ${418}_{-44}^{+42}$&${0.40}\pm{0.04}$& ${4.86}_{-0.26}^{+0.24}$&${3358.80}^{+2.30}_{-2.34}$& ${0.6756}_{-0.0010}^{+0.0010}$& ${58.2}_{-1.1}^{+1.1}$& ${52.59}_{-0.12}^{+0.12}$&${0.42}$&---\\
            			HD11264&${15530.07}^{+7.65}_{-7.74}$&${3.200}^{+0.007}_{-0.005}$&${0.288}\pm{0.001}$&${297.94}^{+0.08}_{-0.10}$&${237.76}^{+0.32}_{-0.33}$&${381.16}^{+34.04}_{-35.71}$&${0.36}\pm{0.03}$&${13.58}^{+0.54}_{-0.59}$&${0.80}$&${0.36}$& ${551}_{-44}^{+42}$&${0.53}\pm{0.04}$& ${14.08}_{-0.58}^{+0.53}$&${15529.70}\pm{9.86}$& ${0.2889}_{-0.0022}^{+0.0022}$& ${65.69}_{-0.27}^{+0.27}$& ${237.65}_{-0.45}^{+0.44}$&${0.52}$&\cite{boffin2003}\\
            			HD11352&${464.75}^{+0.14}_{-0.09}$&${8.412}\pm{0.001}$&${0.232}\pm{0.001}$&${123.01}^{+0.11}_{-0.19}$&${178.44}^{+0.22}_{-0.34}$&${287.70}^{+15.59}_{-16.01}$&${0.27}^{+0.01}_{-0.02}$&${1.23}\pm{0.03}$&${23.78}$&${0.31}$& ${387}_{-28}^{+50}$&${0.37}^{+0.05}_{-0.03}$& ${1.270}_{-0.033}^{+0.031}$&${464.80}\pm{0.16}$& ${0.239}_{-0.013}^{+0.014}$& ${107}_{-14}^{+14}$& ${171.4}_{-4.2}^{+4.6}$&${0.42}$&---\\
            			HD13945&${8.36}\pm{0.01}$&${8.147}\pm{0.003}$&${0.097}\pm{0.001}$&${349.19}\pm{0.07}$&${281.55}\pm{0.43}$&${82.43}^{+5.77}_{-5.98}$&${0.08}\pm{0.01}$&${0.08}\pm{0.01}$&${967.35}$&${0.08}$& --- & --- & --- & --- & --- & --- & --- & ---&\cite{tokovinin2014}\\
            			HD14629&${512.16}\pm{0.02}$&${6.923}^{+0.134}_{-0.103}$&${0.887}^{+0.002}_{-0.001}$&${110.94}^{+0.27}_{-0.34}$&${29.86}^{+0.18}_{-0.23}$&${103.27}^{+4.20}_{-4.16}$&${0.10}\pm{0.01}$&${1.18}\pm{0.02}$&${14.81}$&${0.13}$& ${117.7}_{-4.0}^{+5.3}$&${0.11}\pm{0.01}$& ${1.198}_{-0.016}^{+0.017}$&${512.15}\pm{0.02}$& ${0.8875}_{-0.0015}^{+0.0020}$& ${88}_{-13}^{+12}$& ${29.83}_{-0.27}^{+0.20}$&${0.15}$&---\\
            			HD14802&${9699.95}^{+0.46}_{-0.36}$&${5.457}\pm{0.001}$&${0.331}\pm{0.001}$&${211.55}^{+0.01}_{-0.02}$&${261.94}^{+0.02}_{-0.03}$&${695.52}^{+70.96}_{-75.03}$&${0.66}\pm{0.07}$&${11.44}^{+0.51}_{-0.57}$&${1.43}$&${0.45}$& --- & --- & --- & --- & --- & --- & --- & ---&---\\
            			HD15064&${142.56}\pm{0.01}$&${18.327}^{+0.014}_{-0.010}$&${0.247}\pm{0.001}$&${74.14}^{+0.04}_{-0.03}$&${195.43}^{+0.08}_{-0.09}$&${540.92}^{+54.46}_{-57.40}$&${0.52}\pm{0.05}$&${0.65}\pm{0.03}$&${80.07}$&${0.40}$& --- & --- & --- & --- & --- & --- & --- & ---&\cite{tokovinin2014}\\
            			HD16287&${14.84}\pm{0.01}$&${10.699}\pm{0.003}$&${0.208}\pm{0.001}$&${328.08}\pm{0.13}$&${10.61}\pm{0.11}$&${110.32}^{+4.65}_{-4.74}$&${0.11}\pm{0.01}$&${0.11}\pm{0.01}$&${1077.29}$&${0.13}$& --- & --- & --- & --- & --- & --- & --- & ---&---\\
            			HD17155&${1426.20}^{+0.05}_{-0.04}$&${2.581}^{+0.004}_{-0.001}$&${0.774}\pm{0.001}$&${50.71}^{+0.04}_{-0.07}$&${276.62}\pm{0.04}$&${74.44}^{+2.43}_{-2.49}$&${0.07}\pm{0.01}$&${2.32}\pm{0.04}$&${8.18}$&${0.10}$& ${87.2}_{-3.0}^{+3.0}$&${0.08}\pm{0.01}$& ${2.330}_{-0.038}^{+0.036}$&${1426.01}\pm{0.69}$& ${0.7751}_{-0.0056}^{+0.0065}$& ${66.2}_{-1.3}^{+1.3}$& ${275.9}_{-1.1}^{+1.0}$&${0.11}$&---\\
            			HD17289&${561.68}^{+0.09}_{-0.07}$&${1.395}\pm{0.005}$&${0.528}^{+0.002}_{-0.003}$&${86.54}\pm{0.23}$&${51.66}^{+0.39}_{-0.43}$&${51.26}^{+5.45}_{-5.74}$&${0.05}\pm{0.01}$&${1.40}^{+0.07}_{-0.08}$&${17.08}$&${0.04}$& ${63.4}_{-8.7}^{+21}$&${0.06}^{+0.02}_{-0.01}$& ${1.429}_{-0.059}^{+0.063}$&${561.77}\pm{0.07}$& ${0.5261}_{-0.0025}^{+0.0025}$& ${96}_{-41}^{+34}$& ${51.61}_{-0.57}^{+0.58}$&${0.06}$&\cite{sahlmann2011a}\\
            			HD17152&${65.51}\pm{0.01}$&${9.768}\pm{0.060}$&${0.611}\pm{0.002}$&${299.43}\pm{0.24}$&${239.61}\pm{0.18}$&${144.62}^{+8.78}_{-9.01}$&${0.14}\pm{0.01}$&${0.32}\pm{0.01}$&${174.21}$&${0.15}$& --- & --- & --- & --- & --- & --- & --- & ---&---\\
            			HD18168&${9.41}\pm{0.01}$&${45.740}^{+0.014}_{-0.013}$&${0.114}\pm{0.001}$&${331.70}\pm{0.02}$&${268.78}^{+0.15}_{-0.17}$&${444.66}^{+26.76}_{-27.67}$&${0.42}\pm{0.03}$&${0.10}\pm{0.01}$&${710.19}$&${0.46}$& --- & --- & --- & --- & --- & --- & --- & ---&---\\
            			HD18809&${115.17}\pm{0.01}$&${9.283}\pm{0.001}$&${0.121}\pm{0.001}$&${506.30}\pm{0.07}$&${243.00}^{+0.73}_{-0.68}$&${209.41}^{+11.75}_{-12.08}$&${0.20}\pm{0.01}$&${0.48}\pm{0.01}$&${104.50}$&${0.22}$& --- & --- & --- & --- & --- & --- & --- & ---&---\\
            			HD18907&${9860.58}^{+68.11}_{-38.37}$&${2.414}^{+0.015}_{-0.014}$&${0.218}^{+0.003}_{-0.004}$&${-246.47}^{+0.36}_{-0.27}$&${306.98}^{+0.94}_{-0.90}$&${159.91}^{+89.41}_{-86.87}$&${0.15}^{+0.09}_{-0.08}$&${8.21}^{+1.63}_{-2.21}$&${1.27}$&${0.32}$& ${274}_{-98}^{+101}$&${0.26}^{+0.10}_{-0.09}$& ${8.7}_{-1.7}^{+1.5}$&${9880.01}^{+40.18}_{-36.52}$& ${0.2165}_{-0.0029}^{+0.0029}$& ${122.44}_{-0.32}^{+0.33}$& ${307.28}_{-0.87}^{+0.85}$&${0.54}$&\cite{jenkins2015}\\
            			HD20916&${2234.53}^{+0.93}_{-0.96}$&${1.616}^{+0.007}_{-0.003}$&${0.520}^{+0.002}_{-0.003}$&${210.60}^{+0.28}_{-0.41}$&${350.60}^{+0.35}_{-0.32}$&${77.33}^{+3.41}_{-3.47}$&${0.07}\pm{0.01}$&${3.21}\pm{0.07}$&${3.75}$&${0.09}$& ${160}_{-12}^{+13}$&${0.15}\pm{0.01}$& ${3.305}_{-0.071}^{+0.068}$&${2233.72}^{+2.96}_{-2.81}$& ${0.529}_{-0.025}^{+0.029}$& ${140.8}_{-111}^{+7.7}$& ${350.0}_{-1.6}^{+1.7}$&${0.19}$&---\\
            			HD22705&${204.29}\pm{0.01}$&${7.885}^{+0.026}_{-0.022}$&${0.182}\pm{0.002}$&${74.91}\pm{0.26}$&${358.68}^{+0.95}_{-0.91}$&${226.19}^{+17.56}_{-18.28}$&${0.22}\pm{0.02}$&${0.73}\pm{0.03}$&${54.00}$&${0.21}$& --- & --- & --- & --- & --- & --- & --- & ---&\cite{makarov2007}\\
            			HD24492&${1566.84}^{+0.16}_{-0.17}$&${4.680}^{+0.007}_{-0.006}$&${0.157}\pm{0.001}$&${230.82}^{+0.10}_{-0.09}$&${250.96}^{+0.21}_{-0.20}$&${254.68}^{+19.35}_{-20.17}$&${0.24}\pm{0.02}$&${2.80}^{+0.10}_{-0.11}$&${5.18}$&${0.26}$& ${482}_{-36}^{+37}$&${0.46}^{+0.04}_{-0.03}$& ${2.960}_{-0.10}^{+0.093}$&${1566.78}^{+0.44}_{-0.47}$& ${0.1568}_{-0.0015}^{+0.0014}$& ${136.5}_{-1.7}^{+1.6}$& ${250.91}_{-0.42}^{+0.40}$&${0.49}$&---\\
            			HD23308&${4.47}\pm{0.01}$&${13.003}\pm{0.002}$&${0.001}\pm{0.001}$&${237.89}\pm{0.01}$&${24.01}\pm{14.35}$&${116.40}^{+21.12}_{-23.26}$&${0.11}\pm{0.02}$&${0.06}\pm{0.01}$&${1436.19}$&${0.10}$& --- & --- & --- & --- & --- & --- & --- & ---&---\\
            			HD23576&${17109.11}^{+3.05}_{-2.97}$&${3.834}\pm{0.002}$&${0.439}\pm{0.001}$&${322.37}^{+0.12}_{-0.15}$&${44.24}^{+0.18}_{-0.19}$&${444.53}^{+31.23}_{-32.30}$&${0.42}\pm{0.03}$&${14.71}^{+0.46}_{-0.49}$&${0.78}$&${0.41}$& ${766}_{-44}^{+43}$&${0.73}\pm{0.04}$& ${15.68}_{-0.46}^{+0.44}$&${17109.77}\pm{9.86}$& ${0.4395}_{-0.0014}^{+0.0015}$& ${123.89}_{-0.19}^{+0.20}$& ${44.19}_{-0.46}^{+0.48}$&${0.71}$&---\\
            			HD25874&${4469.97}^{+4.72}_{-3.31}$&${2.207}^{+0.005}_{-0.006}$&${0.473}\pm{0.002}$&${230.86}^{+0.16}_{-0.18}$&${95.86}^{+0.21}_{-0.20}$&${157.59}^{+12.33}_{-12.89}$&${0.15}\pm{0.01}$&${5.56}^{+0.20}_{-0.22}$&${2.77}$&${0.15}$& ${201}_{-15}^{+15}$&${0.19}\pm{0.01}$& ${5.63}_{-0.22}^{+0.21}$&${4475.04}\pm{5.84}$& ${0.4730}_{-0.0023}^{+0.0023}$& ${61.82}_{-0.54}^{+0.55}$& ${96.14}_{-0.20}^{+0.21}$&${0.19}$&---\\
            			HD26491&${9517.81}^{+4.90}_{-4.09}$&${4.439}\pm{0.003}$&${0.564}\pm{0.001}$&${58.41}\pm{0.12}$&${221.36}^{+0.08}_{-0.09}$&${382.04}^{+51.71}_{-55.59}$&${0.36}\pm{0.05}$&${9.75}^{+0.58}_{-0.67}$&${1.32}$&${0.36}$& ${498}_{-64}^{+59}$&${0.48}\pm{0.06}$& ${9.98}_{-0.66}^{+0.57}$&${9521.70}^{+8.40}_{-8.04}$& ${0.5642}_{-0.0011}^{+0.0011}$& ${97.7}_{-15}^{+1.4}$& ${221.34}_{-0.14}^{+0.14}$&${0.48}$&---\\
            			HD27019&${6810.51}^{+4.14}_{-2.80}$&${3.144}^{+0.039}_{-0.012}$&${0.740}^{+0.004}_{-0.002}$&${310.68}^{+0.25}_{-0.20}$&${157.82}^{+0.30}_{-0.26}$&${187.96}^{+16.12}_{-16.67}$&${0.18}\pm{0.02}$&${7.28}^{+0.28}_{-0.31}$&${1.73}$&${0.19}$& ${1828}_{-172}^{+177}$&${1.74}^{+0.17}_{-0.16}$& ${9.77}_{-0.32}^{+0.30}$&${6817.76}\pm{9.86}$& ${0.924}_{-0.015}^{+0.012}$& ${29.9}_{-3.0}^{+3.3}$& ${170.3}_{-1.4}^{+1.2}$&${0.53}$&\cite{gomez2016}\\
            			HD28454&${6.29}\pm{0.01}$&${7.162}^{+0.002}_{-0.001}$&${0.001}\pm{0.001}$&${142.75}\pm{0.03}$&${213.27}^{+44.24}_{-37.04}$&${78.13}^{+4.68}_{-4.82}$&${0.07}\pm{0.01}$&${0.07}\pm{0.01}$&${1105.73}$&${0.06}$& --- & --- & --- & --- & --- & --- & --- & ---&---\\
            			HD29813&${4488.14}^{+2.66}_{-1.06}$&${2.908}^{+0.003}_{-0.004}$&${0.254}\pm{0.001}$&${179.29}^{+0.11}_{-0.05}$&${24.90}^{+0.27}_{-0.11}$&${223.72}^{+16.61}_{-17.23}$&${0.21}\pm{0.02}$&${5.63}^{+0.19}_{-0.21}$&${2.61}$&${0.22}$& ${336}_{-23}^{+23}$&${0.32}\pm{0.02}$& ${5.80}_{-0.21}^{+0.19}$&${4488.41}^{+1.94}_{-2.01}$& ${0.2542}_{-0.0015}^{+0.0015}$& ${53.71}_{-0.66}^{+0.67}$& ${25.08}_{-0.27}^{+0.27}$&${0.33}$&\cite{tokovinin2014}\\
            			HIP22059&${13855.08}^{+16.54}_{-15.13}$&${0.624}\pm{0.004}$&${0.100}^{+0.006}_{-0.007}$&${350.51}^{+0.37}_{-0.47}$&${303.34}^{+3.36}_{-3.83}$&${57.90}^{+2.05}_{-2.08}$&${0.06}\pm{0.01}$&${10.29}^{+0.17}_{-0.18}$&${0.50}$&${0.08}$& ${182.7}_{-6.2}^{+6.1}$&${0.17}\pm{0.01}$& ${10.80}_{-0.18}^{+0.17}$&${13855.76}\pm{14.98}$& ${0.0977}_{-0.0086}^{+0.0088}$& ${158.42}_{-0.28}^{+0.27}$& ${303.1}_{-5.6}^{+4.7}$&${0.25}$&\cite{rickman2022}\\
            			HD30501&${2076.27}^{+2.81}_{-2.97}$&${1.756}^{+0.011}_{-0.010}$&${0.758}^{+0.004}_{-0.005}$&${358.19}^{+1.46}_{-1.65}$&${71.79}^{+0.55}_{-0.65}$&${62.44}^{+3.82}_{-3.92}$&${0.06}\pm{0.01}$&${3.04}\pm{0.09}$&${3.18}$&${0.07}$& ${68.3}_{-4.2}^{+4.0}$&${0.07}\pm{0.01}$& ${3.045}_{-0.093}^{+0.087}$&${2075.72}^{+3.14}_{-3.10}$& ${0.7539}_{-0.0036}^{+0.0041}$& ${104.5}_{-3.3}^{+2.5}$& ${71.11}_{-0.77}^{+0.68}$&${0.08}$&\cite{sahlmann2011a}\\
            			HD30517&${29184.35}^{+6.42}_{-5.07}$&${1.238}\pm{0.004}$&${0.001}\pm{0.001}$&${295.44}^{+0.21}_{-0.16}$&${33.19}^{+277.23}_{-342.19}$&${169.12}^{+7.71}_{-7.87}$&${0.16}\pm{0.01}$&${18.66}^{+0.40}_{-0.42}$&${0.26}$&${0.19}$& ${359}_{-60}^{+62}$&${0.34}\pm{0.06}$& ${19.71}_{-0.49}^{+0.48}$&${29183.48}\pm{9.86}$& ${0.21}_{-0.13}^{+0.11}$& ${120.6}_{-18}^{+8.8}$& ${88}_{-17}^{+12}$&${0.40}$&---\\
            			HD30774&${4.93}\pm{0.01}$&${5.000}\pm{0.002}$&${0.003}\pm{0.001}$&${217.67}\pm{0.04}$&${310.85}\pm{6.78}$&${41.59}^{+3.30}_{-3.42}$&${0.04}\pm{0.01}$&${0.06}\pm{0.01}$&${1349.68}$&${0.04}$& --- & --- & --- & --- & --- & --- & --- & ---&\cite{tokovinin2014}\\
            			HD31143&${8669.86}^{+662.20}_{-755.93}$&${3.319}^{+0.041}_{-0.035}$&${0.472}^{+0.009}_{-0.011}$&${149.51}^{+5.41}_{-6.02}$&${125.46}^{+5.34}_{-5.47}$&${260.50}^{+18.58}_{-18.63}$&${0.25}\pm{0.02}$&${8.46}\pm{0.57}$&${0.97}$&${0.30}$& ${372}_{-24}^{+25}$&${0.36}\pm{0.02}$& ${7.89}_{-0.23}^{+0.21}$&${7436.49}^{+58.44}_{-47.48}$& ${0.4047}_{-0.0091}^{+0.0093}$& ${95.6}_{-2.8}^{+2.9}$& ${148.0}_{-2.5}^{+2.3}$&${0.43}$&---\\
            			HD32778&${5248.65}^{+1.18}_{-2.37}$&${1.774}^{+0.003}_{-0.002}$&${0.346}\pm{0.001}$&${426.36}^{+0.14}_{-0.09}$&${168.75}^{+0.28}_{-0.21}$&${131.69}^{+8.92}_{-9.24}$&${0.13}\pm{0.01}$&${5.94}^{+0.19}_{-0.20}$&${2.28}$&${0.14}$& ${145.3}_{-10}^{+9.3}$&${0.14}\pm{0.01}$& ${5.97}_{-0.20}^{+0.19}$&${5247.91}^{+4.02}_{-4.38}$& ${0.3440}_{-0.0027}^{+0.0026}$& ${92.2}_{-1.5}^{+1.5}$& ${169.07}_{-0.52}^{+0.53}$&${0.16}$&---\\
            			HD33487&${1366.25}^{+0.11}_{-0.09}$&${5.661}^{+0.001}_{-0.002}$&${0.155}\pm{0.001}$&${234.97}\pm{0.05}$&${161.92}^{+0.17}_{-0.20}$&${259.17}^{+9.60}_{-9.80}$&${0.25}\pm{0.01}$&${2.43}\pm{0.04}$&${4.77}$&${0.32}$& ${331}_{-11}^{+11}$&${0.32}\pm{0.01}$& ${2.487}_{-0.041}^{+0.040}$&${1366.25}\pm{0.15}$& ${0.15496}_{-0.00075}^{+0.00072}$& ${100.4}_{-2.0}^{+2.0}$& ${162.02}_{-0.23}^{+0.25}$&${0.40}$&---\\
            			HD34101&${803.63}\pm{0.01}$&${3.601}\pm{0.001}$&${0.084}\pm{0.001}$&${269.82}^{+0.03}_{-0.02}$&${273.99}^{+0.20}_{-0.29}$&${160.85}^{+14.01}_{-14.68}$&${0.15}\pm{0.01}$&${1.76}^{+0.07}_{-0.08}$&${14.58}$&${0.16}$& ${205}_{-14}^{+20}$&${0.20}^{+0.02}_{-0.01}$& ${1.862}_{-0.061}^{+0.053}$&${803.62}\pm{0.02}$& ${0.08388}_{-0.00049}^{+0.00050}$& ${79}_{-13}^{+13}$& ${274.32}_{-0.48}^{+0.48}$&${0.20}$&\cite{nidever2002}\\
            			HD34540&${696.52}^{+0.07}_{-0.09}$&${8.391}^{+0.020}_{-0.013}$&${0.033}\pm{0.002}$&${3430.25}\pm{0.19}$&${149.29}^{+2.98}_{-1.48}$&${361.11}^{+23.40}_{-24.27}$&${0.34}\pm{0.02}$&${1.69}\pm{0.05}$&${15.71}$&${0.35}$& --- & --- & --- & --- & --- & --- & --- & ---&---\\
            			HD39012&${3313.56}^{+0.28}_{-0.54}$&${3.565}\pm{0.001}$&${0.166}\pm{0.001}$&${470.13}^{+0.02}_{-0.03}$&${264.19}^{+0.19}_{-0.22}$&${235.18}^{+11.59}_{-11.82}$&${0.22}\pm{0.01}$&${4.48}^{+0.10}_{-0.11}$&${2.86}$&${0.26}$& ${741}_{-37}^{+37}$&${0.71}\pm{0.04}$& ${5.06}_{-0.10}^{+0.10}$&${3311.61}\pm{1.79}$& ${0.1634}_{-0.0014}^{+0.0015}$& ${151.86}_{-0.67}^{+0.65}$& ${263.21}_{-0.55}^{+0.55}$&${0.81}$&---\\
            			HD40107&${4166.79}^{+1.30}_{-1.81}$&${2.824}^{+0.002}_{-0.003}$&${0.692}^{+0.001}_{-0.002}$&${-251.00}^{+0.20}_{-0.13}$&${93.02}^{+0.14}_{-0.11}$&${134.37}^{+4.56}_{-4.62}$&${0.13}\pm{0.01}$&${4.87}\pm{0.08}$&${1.76}$&${0.17}$& ${462}_{-27}^{+29}$&${0.44}\pm{0.03}$& ${5.393}_{-0.087}^{+0.086}$&${4173.35}^{+9.86}_{-10.59}$& ${0.705}_{-0.016}^{+0.017}$& ${24.1}_{-1.2}^{+132}$& ${94.6}_{-1.9}^{+2.0}$&${0.58}$&---\\
            			HD42286&${9449.67}^{+4.28}_{-2.97}$&${3.754}\pm{0.003}$&${0.531}\pm{0.001}$&${67.87}^{+0.04}_{-0.05}$&${17.08}\pm{0.04}$&${274.86}^{+13.87}_{-14.23}$&${0.26}\pm{0.01}$&${8.80}^{+0.20}_{-0.21}$&${1.24}$&${0.35}$& ${358}_{-16}^{+16}$&${0.34}\pm{0.02}$& ${9.02}_{-0.20}^{+0.20}$&${9447.92}\pm{9.50}$& ${0.53078}_{-0.00056}^{+0.00056}$& ${99.9}_{-1.1}^{+1.0}$& ${17.08}_{-0.12}^{+0.12}$&${0.45}$&---\\
            			HD43848&${2345.72}^{+2.00}_{-3.78}$&${2.537}^{+0.003}_{-0.065}$&${0.916}^{+0.001}_{-0.002}$&${-178.73}^{+0.41}_{-0.29}$&${203.12}^{+0.35}_{-0.25}$&${61.17}^{+3.52}_{-3.51}$&${0.06}\pm{0.01}$&${3.38}\pm{0.08}$&${3.11}$&${0.07}$& ${135.8}_{-7.2}^{+7.4}$&${0.13}\pm{0.01}$& ${3.463}_{-0.084}^{+0.081}$&${2343.63}^{+3.21}_{-3.36}$& ${0.9746}_{-0.0041}^{+0.0020}$& ${82}_{-17}^{+18}$& ${192.27}_{-0.45}^{+1.0}$&${0.15}$&\cite{sahlmann2011a}\\
            			HD45701&${18709.40}^{+12.78}_{-13.20}$&${2.691}\pm{0.006}$&${0.166}\pm{0.002}$&${63.47}^{+0.21}_{-0.26}$&${118.33}^{+0.46}_{-0.41}$&${389.21}^{+38.22}_{-40.29}$&${0.37}\pm{0.04}$&${16.00}^{+0.71}_{-0.78}$&${0.80}$&${0.31}$& ${688}_{-62}^{+59}$&${0.66}\pm{0.06}$& ${16.89}_{-0.78}^{+0.71}$&${18708.84}\pm{13.15}$& ${0.1646}_{-0.0047}^{+0.0048}$& ${49.15}_{-0.27}^{+0.27}$& ${118.29}_{-0.59}^{+0.59}$&${0.55}$&---\\
            			HD46569&${9335.06}^{+2.66}_{-2.51}$&${1.712}^{+0.001}_{-0.002}$&${0.585}\pm{0.001}$&${58.31}^{+0.05}_{-0.07}$&${278.71}^{+0.07}_{-0.08}$&${195.13}^{+14.52}_{-15.08}$&${0.19}\pm{0.01}$&${10.49}^{+0.37}_{-0.40}$&${1.21}$&${0.12}$& ${310}_{-23}^{+22}$&${0.30}\pm{0.02}$& ${10.69}_{-0.40}^{+0.37}$&${9334.33}\pm{9.50}$& ${0.5848}_{-0.0010}^{+0.0010}$& ${44.74}_{-0.62}^{+0.66}$& ${278.55}_{-0.23}^{+0.22}$&${0.19}$&---\\
                        HD48189&${254.33}\pm{0.01}$&${10.383}\pm{0.024}$&${0.590}\pm{0.002}$&${12.25}\pm{0.26}$&${196.92}\pm{0.21}$&${263.17}^{+20.44}_{-21.24}$&${0.25}\pm{0.02}$&${0.85}\pm{0.03}$&${56.17}$&${0.25}$& --- & --- & --- & --- & --- & --- & --- & ---&\cite{zunigafernandez2021}\\
            			HD52698&${3013.66}^{+0.46}_{-0.43}$&${4.947}^{+0.007}_{-0.005}$&${0.763}\pm{0.001}$&${116.65}^{+0.03}_{-0.04}$&${181.31}^{+0.06}_{-0.07}$&${203.58}^{+8.84}_{-9.05}$&${0.19}\pm{0.01}$&${4.14}^{+0.08}_{-0.09}$&${3.91}$&${0.23}$& ${248}_{-10}^{+10}$&${0.24}\pm{0.01}$& ${4.197}_{-0.086}^{+0.082}$&${3013.75}\pm{0.66}$& ${0.7630}_{-0.0010}^{+0.0010}$& ${104.1}_{-1.2}^{+1.1}$& ${181.339}_{-0.085}^{+0.085}$&${0.28}$&---\\
            			HD53680&${1685.20}^{+0.37}_{-0.38}$&${1.252}^{+0.001}_{-0.002}$&${0.481}\pm{0.001}$&${147.86}^{+0.06}_{-0.08}$&${226.17}^{+0.13}_{-0.17}$&${51.33}^{+1.95}_{-1.99}$&${0.05}\pm{0.01}$&${2.53}\pm{0.05}$&${3.46}$&${0.07}$& --- & --- & --- & --- & --- & --- & --- & ---&\cite{sahlmann2011a}\\
            			HD58696&${345.28}\pm{0.01}$&${18.545}^{+0.010}_{-0.007}$&${0.899}\pm{0.001}$&${231.34}\pm{0.02}$&${336.05}^{+0.03}_{-0.02}$&${337.58}^{+34.91}_{-36.75}$&${0.32}^{+0.03}_{-0.04}$&${1.14}^{+0.05}_{-0.06}$&${35.17}$&${0.24}$& ${228}_{-28}^{+65}$&${0.22}^{+0.06}_{-0.03}$& ${1.176}_{-0.083}^{+0.062}$&${362.59}^{+0.04}_{-17.17}$& ${0.9823}_{-0.051}^{+0.0077}$& ${88}_{-21}^{+30}$& ${310}_{-36}^{+26}$&${0.16}$&\cite{tokovinin2014}\\
            			HD59099&${2312.88}\pm{0.54}$&${5.515}^{+0.003}_{-0.005}$&${0.598}\pm{0.001}$&${50.04}^{+0.13}_{-0.10}$&${64.21}^{+0.09}_{-0.10}$&${339.17}^{+29.36}_{-30.83}$&${0.32}\pm{0.03}$&${4.01}^{+0.16}_{-0.17}$&${3.30}$&${0.25}$& ${603}_{-57}^{+54}$&${0.58}\pm{0.05}$& ${4.18}_{-0.17}^{+0.16}$&${2311.52}^{+2.89}_{-43.83}$& ${0.5968}_{-0.041}^{+0.0036}$& ${134.3}_{-86}^{+1.2}$& ${64.02}_{-6.0}^{+0.35}$&${0.45}$&\cite{tokovinin2014}\\
            			HD59380&${998.45}\pm{0.06}$&${2.966}\pm{0.003}$&${0.560}^{+0.002}_{-0.001}$&${150.62}^{+0.13}_{-0.34}$&${358.22}^{+0.12}_{-0.28}$&${145.43}^{+10.79}_{-11.23}$&${0.14}\pm{0.01}$&${2.22}\pm{0.08}$&${8.05}$&${0.11}$& --- & --- & --- & --- & --- & --- & --- & ---&\cite{fekel2018}\\
            			% HD61033&${278.06}^{+0.04}_{-0.01}$&${11.030}^{+0.005}_{-0.002}$&${0.233}^{+0.007}_{-0.005}$&${78.38}^{+0.40}_{-1.40}$&${165.87}^{+0.79}_{-1.52}$&${335.28}^{+22.78}_{-23.70}$&${0.32}\pm{0.02}$&${0.91}\pm{0.03}$&${30.04}$&${0.33}$& --- & --- & --- & --- & --- & --- & --- & ---&\cite{tokovinin2014}\\
            			% HD62644&${380.41}\pm{0.01}$&${9.873}\pm{0.002}$&${0.732}\pm{0.001}$&${170.12}\pm{0.02}$&${62.55}\pm{0.02}$&${319.67}^{+41.46}_{-44.43}$&${0.31}\pm{0.04}$&${1.26}^{+0.07}_{-0.09}$&${18.09}$&${0.20}$& ${745}_{-108}^{+148}$&${0.71}^{+0.14}_{-0.10}$& ${1.80}_{-0.14}^{+0.30}$&${380.41}^{+23.01}_{-0.01}$& ${0.73223}_{-0.00091}^{+0.25}$& ${89}_{-10}^{+12}$& ${62.579}_{-0.064}^{+63}$&${0.46}$&\cite{setiawan2004}\\
            			% HD62848&${5916.22}^{+12.08}_{-11.56}$&${2.026}^{+0.021}_{-0.024}$&${0.394}^{+0.012}_{-0.011}$&${98.86}^{+1.65}_{-1.74}$&${218.08}^{+2.38}_{-2.65}$&${163.39}^{+23.76}_{-25.48}$&${0.16}\pm{0.02}$&${6.68}^{+0.45}_{-0.52}$&${1.42}$&${0.16}$& ${282}_{-41}^{+37}$&${0.27}\pm{0.04}$& ${6.89}_{-0.51}^{+0.44}$&${5913.03}\pm{9.13}$& ${0.400}_{-0.011}^{+0.012}$& ${42.93}_{-0.82}^{+0.84}$& ${215.7}_{-2.6}^{+2.6}$&${0.27}$&---\\
            			% HD63077&${8405.69}^{+28.89}_{-7.16}$&${4.762}\pm{0.013}$&${0.443}^{+0.002}_{-0.001}$&${174.42}\pm{0.08}$&${184.38}^{+0.16}_{-0.17}$&${435.55}^{+33.26}_{-34.38}$&${0.42}\pm{0.03}$&${9.15}^{+0.31}_{-0.33}$&${1.74}$&${0.40}$& --- & --- & --- & --- & --- & --- & --- & ---&---\\
               
                        ...&...&...&...&...&...&...&...&...&...&...&...&...&...&...&...&...&...&...&...\\
                \hline
            \end{tabular}
            \end{adjustbox}
            \tablefoot{Full table is available at the CDS. A portion is shown here for guidance regarding its form and content.}
        \end{sidewaystable*}
        \par As previously done in Sect.~\ref{sec:rv}, we follow \cite{halbwachs2003} and select a threshold of mass ratio $q>0.8$ for SB2s in the sample, using this time the companion true mass values obtained by the joint radial velocity and proper motion anomalies analysis. By virtue of the true masses derived, this time we find four systems to have $q>0.8$, namely HD3795 ($q=0.86$), HD39012 ($q=0.81$), HD181199A ($q=0.90$) and HD223084 ($q=0.90$) which are therefore possible undetected blended binaries.
        \par In addition to the companions detected in this sample we also report some results from Unger et al., in prep., in which a similar joint analysis of radial velocity and astrometric measurements is performed on the planetary companions detected in the CORALIE sample. Specifically, the possible brown dwarf companions found to be orbiting stars HD162020 and HD112758, having respective minimum masses of $14.96\pm0.53$\Mjup and $32.7\pm1.9$\Mjup are found to have true masses of $0.392\pm0.005$\Msun and $0.245\pm0.001$\Msun. While we shall not go into the details of the orbital solutions of these two companions in the present work as they are thoroughly analysed in Unger et al., in prep., these additional stellar objects in the CORALIE sample will be here considered in the occurrence rates analysis detailed in Sect.~\ref{fig:sample-detection}.
    
    \section{A few notable cases} \label{sec:notable-cases}
        
        \subsection{Brown dwarfs in the sample} \label{subsec:brown-dwarfs}
            Following the radial velocity analysis detailed in Sect.~\ref{sec:rv}, we find \browndwarfs companions in the sample to have a minimum mass between 40 and 80 \Mjup and that are therefore describable as possible brown dwarfs. While the low number of such companions found in the sample do not allow for robust statistical analysis, it is possible to note that most of them are found at orbital separations larger than 0.5\au from the primary star, in a further example of the brown dwarfs desert observed around solar-type stars \citep{grether2006,shahaf2019,kiefer2019}. As the simultaneous fit of radial velocities and proper motion anomalies performed and discussed in Sect.~\ref{sec:astrometry} allow for the determination of orbital inclination and therefore true mass of the companions, it is of clear interest to discuss how many and which of these possible brown dwarfs are confirmed to be as such by this analysis and which are instead revealed to be stellar companions instead.
            \par It is first of all important to note that the primary stars hosting three of these possible brown dwarf companions (namely those found orbiting HD53680, HD153284, HD184860A) are not included in the HGCA, and therefore no astrometric analysis is possible for these stars, and that seven additional possible brown dwarf companions (HD3277, HD28454, HD30774, HD89707, HD151528, HD164427A and HD219709) are found to be on orbits tighter than the 1\au threshold that we have set for reliable analysis with \orvara; this therefore leaves us with \bdorvara such companions for which we are able to determine true masses and confirm or reject their nature as brown dwarfs.
            \par Of these \bdorvara objects, our use of proper motion anomalies confirm \staysbd companions (those orbiting HD4747, HD17289, HD30501, HD74014, HD112863, HD167665 and HD206505) to remain in the brown dwarf mass range, while the remaining \bdtostar (orbiting HIP22059, HD17155, HD20916, HD43848, HD78746, HD87359, HD94340, HD119559, HD154697, HD195010 and HD217580) are found to have a true mass above 80\Mjup and are therefore revealed to be stellar companions. Of the latter, the companion characterized by the larger difference in minimum and true mass is the one found orbiting star HD119559, which starting from a minimum mass of $76.267^{+4.149}_{-4.274}$\Mjup is revealed by astrometric constraints to have a true mass of $0.35\pm0.02$\Msun.
            \par Finally, we again report some results from Unger et al., in prep. like done in Sect.~\ref{sec:astrometry}, in this work the usage of Gaia DR3 astrometric measurements allow for four of the aforementioned eleven possible brown dwarf companions to be revealed as stellar companions having a true mass above 80\Mjup, namely the ones found around HD3277 ($0.468^{+0.023}_{-0.005}$\Msun), HD89707 ($0.100\pm0.001$\Msun), HD151528 ($0.1403^{+0.0001}_{-0.0030}$\Msun), and HD164427A ($0.3369^{+0.003}_{-0.002}$\Mjup). As mentioned before, we refer to Unger et al., in prep. for further details on the respective orbital solution, but they will considered in the present work as stellar companions in the occurrence rates analysis detailed in Sect.~\ref{subsec:detection-limits}.
        
        \subsection{A new triple star system: HD206276 (HIP107143)}\label{subsec:HD206276}
            \begin{figure}[t!]
                \includegraphics[width=\linewidth]{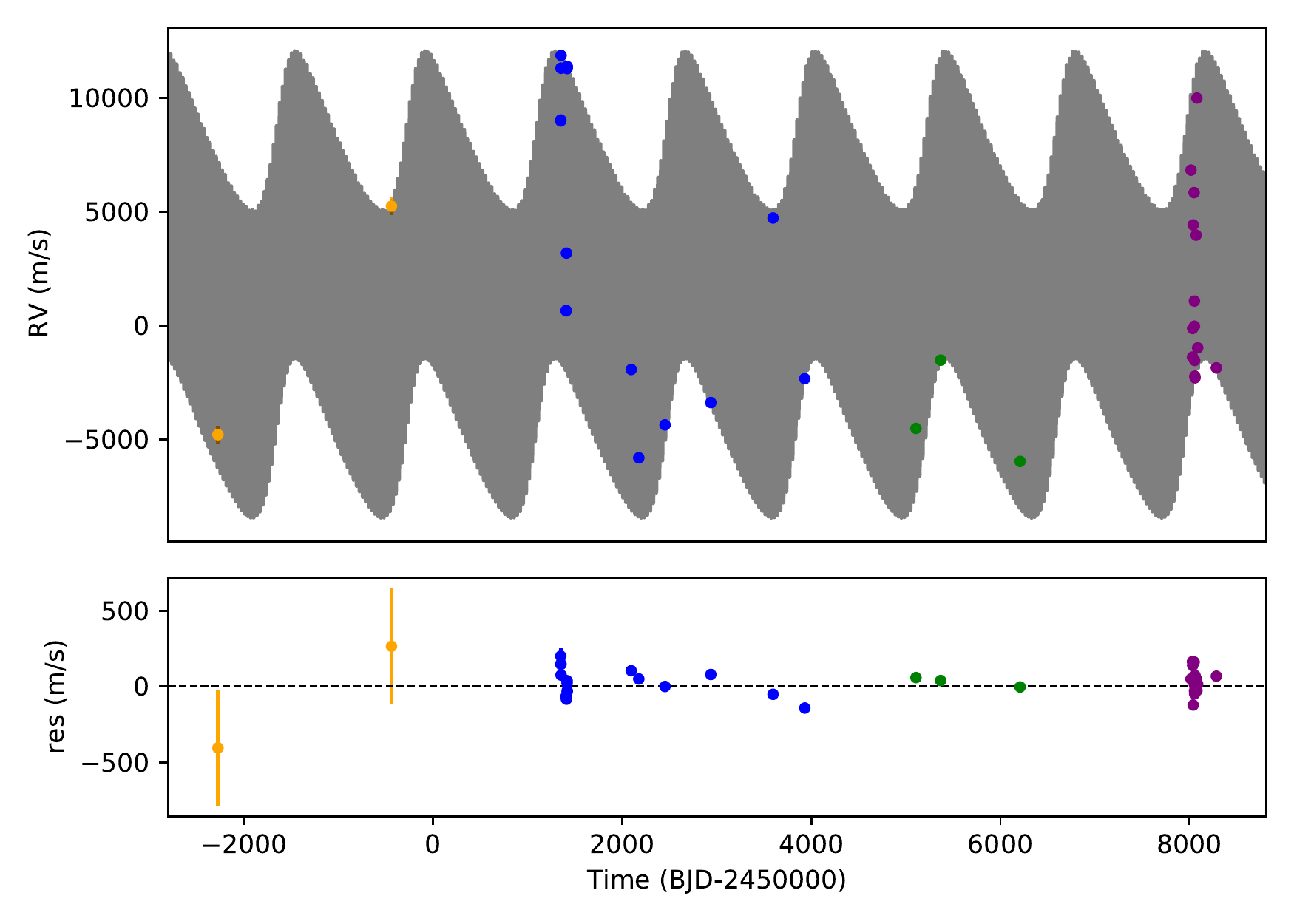}
                \\
                \includegraphics[width=\linewidth]{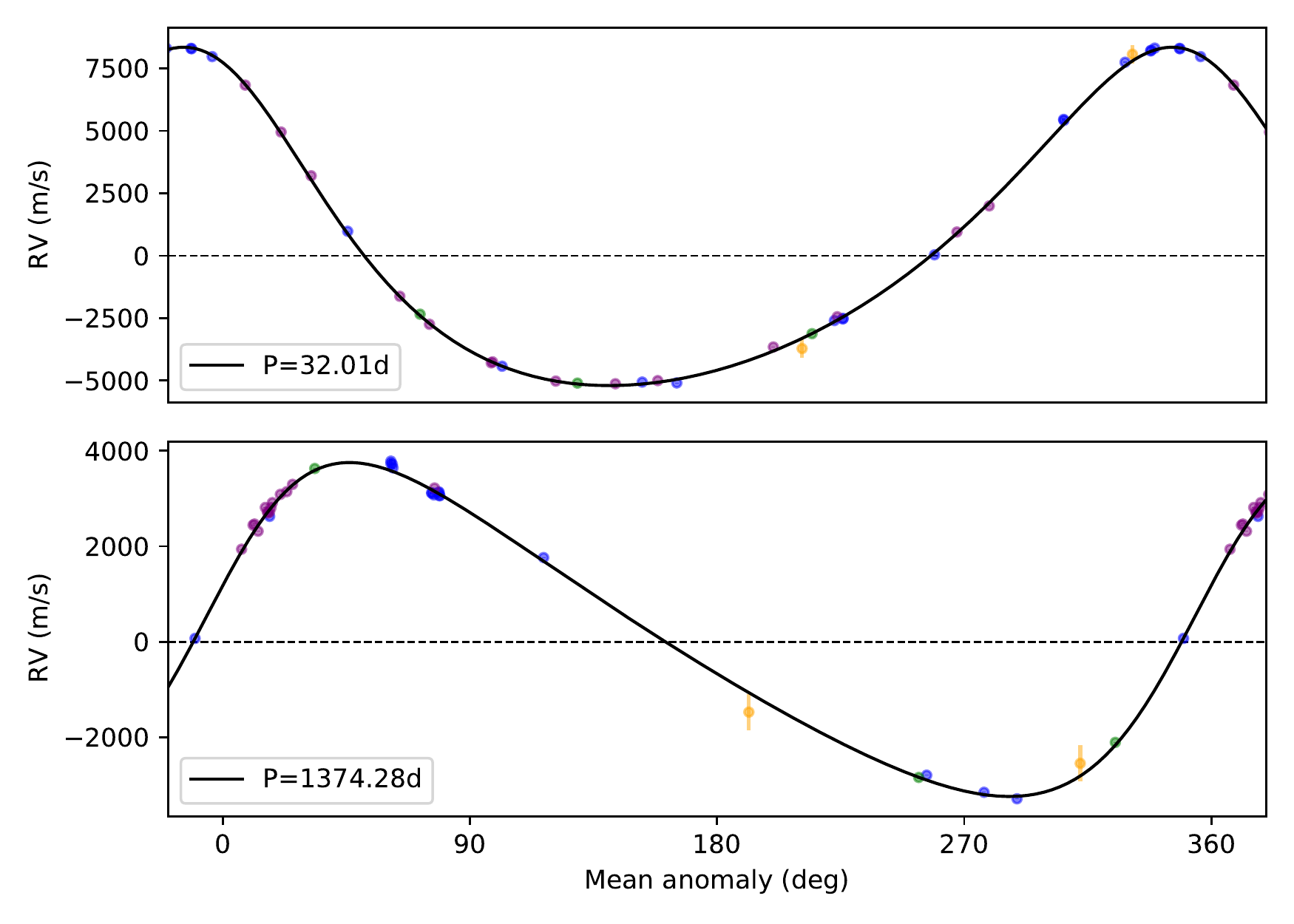}
                \caption{Radial velocity best-fit solution and phase folded model curves for triple stellar system HD206276, with CORAVEL, CORALIE98, CORALIE07 and CORALIE14 measurements shown in orange, blue, green and purple respectively.}
                \label{fig:HD206276-rvfit}
            \end{figure}
            The possible presence of a long-period massive companion around K3~V star HD206276 was first noted in \cite{goldin2007}, in which the use of a genetic optimization-based algorithm allowed for obtain additional orbital solutions for a subsample of \hip stars with previous stochastic solutions. In the cited work, HD206276 was reported as possibly hosting a massive companion on a $1338^{+171}_{-73}$\days orbit with a $0.14^{+0.21}_{-0.11}$ eccentricity and an orbital inclination of $40\pm6$\deg. However, no radial velocity measurements have been published ever since or used to confirm thet companion or to provide a better orbital solution for the system.
            \par Within the scope of the CORALIE exoplanetary search, we have observed HD206276 over a total of 6932\days collecting 35 radial velocity measurements (divided as 18 C98, 3 C07 and 14 C14); the additional usage of two CORAVEL measurements brings the total of datapoints available for our analysis to  37 over the course of 10584\days. We identify in the timeseries periodogram a highly significant peak (FAP=$4.3\cdot10^{-6}$) at 32\days, with an one-Keplerian residual peak present at 1363\days with FAP=$3.1\cdot10^{-4}$ corresponding to the astrometric signal reported in \cite{goldin2007}, with no further significant signals present in the two-Keplerian residuals. We therefore present our two-Keplerian bestfit model (shown in Fig.~\ref{fig:HD206276-rvfit}) with which we confirm the presence of the outer companions having an orbital period $P_{\rm C}=1374.27\pm0.97$\days, semiamplitude $K_{\rm C}=3.50\pm0.03$\kmps and eccentricity $e_{\rm C}=0.275\pm0.008$, while we also report the detection of an inner companion with $P_{\rm B}=32.005\pm0.0002$\days, $K_{\rm B}=6.8\pm0.03$\kmps and eccentricity $e_{\rm B}=0.255\pm0.003$. By virtue of the primary star having a mass of $0.88^{+0.06}_{-0.05}$\Msun (see Sect.~\ref{sec:hosts}), we derive values of minimum masses and semimajor axes of $168.89^{+7.59}_{-7.66}$\Mjup (corresponding to $0.161\pm0.007$\Msun) and $2.45\pm0.05$\au for the outer companion and of $93.96^{+4.09}_{-4.19}$\Mjup (corresponding to $0.089\pm0.004$\Msun) and $0.195\pm0.004$\au for the inner one.
            \par As mentioned in Sect.~\ref{sec:astrometry}, when performing the simultaneous fit of radial velocities and proper motion anomaly with \orvara we first subtract the Keplerian signal of the inner companion, since its short period of 32\days is not detectable though the use of proper motion measurements at the two \hip and \gaia epochs, therefore using for the joint analysis a radial velocity timeseries containing in principle only the contribution of the outer companion. However, we find the joint \orvara to be unable to converge, failing to contraint the orbital elements of the outer stellar companions, likely due to the 32\days stellar companion still producing a significant astrometric contribution to the proper motion anomalies. The same failure to converge is obtained when trying to model the triple system by jointly fitting our radial velocity with the Hipparcos epoch astrometric time series via the kepmodel python package \citep[see][]{delisle2016,delisle2022} and the samsam MCMC sampler \citep[e.g.][]{delisle2018}. We additionally note that this triple system is not listed amongst the \gdr{3} astrometric orbital solutions validated in \cite{holl2022}, further highlighting the difficulties in disentagling the astrometric signature of the inner stellar companion. For this system we are therefore able to provide only the orbital solutions resulting from the radial velocity fit alone.

        \subsection{An updated triple star system: HD94340 (HIP53217)}\label{subsec:HD94340}
            \begin{figure}[t!]
                \includegraphics[width=\linewidth]{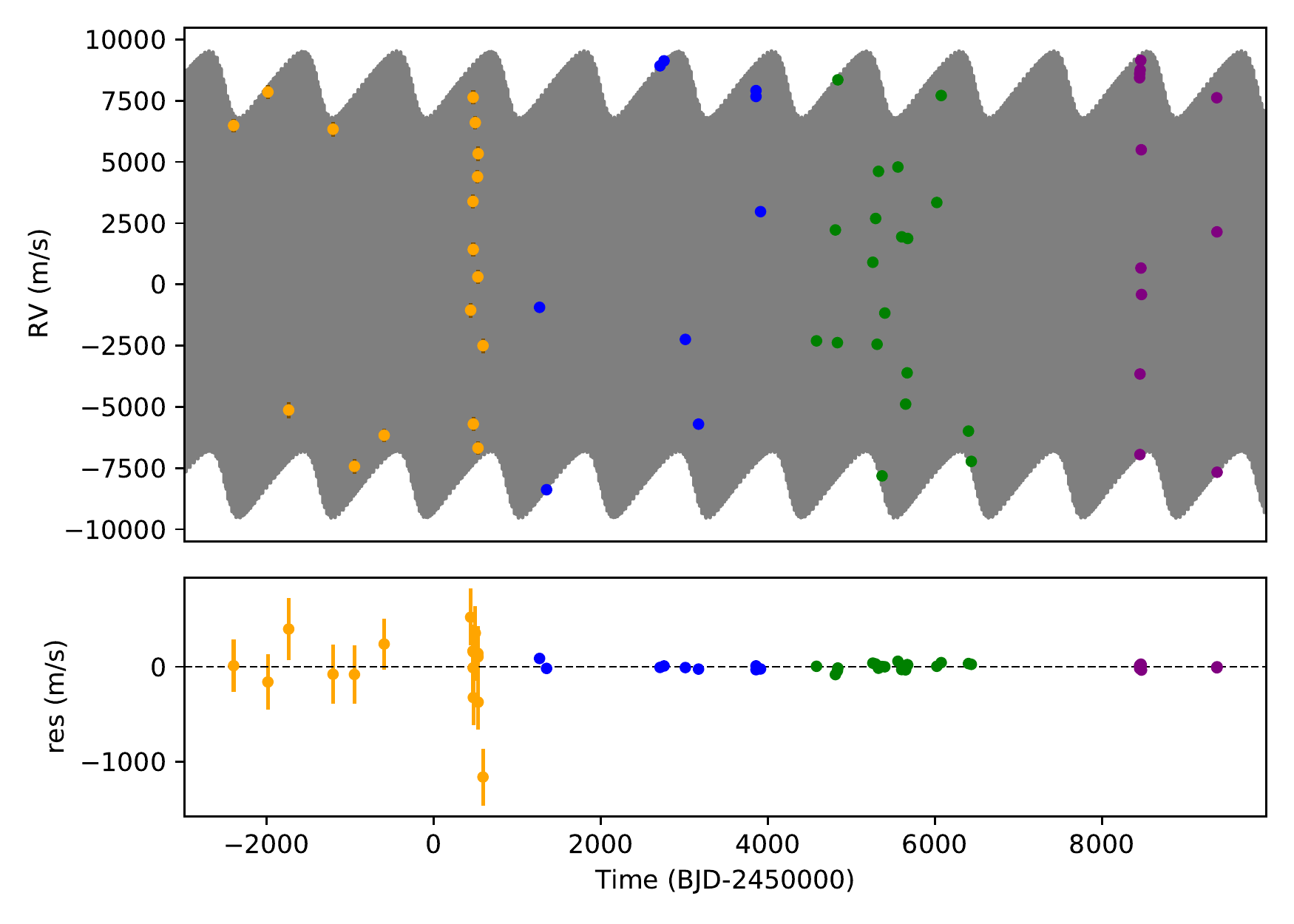}
                \\
                \includegraphics[width=\linewidth]{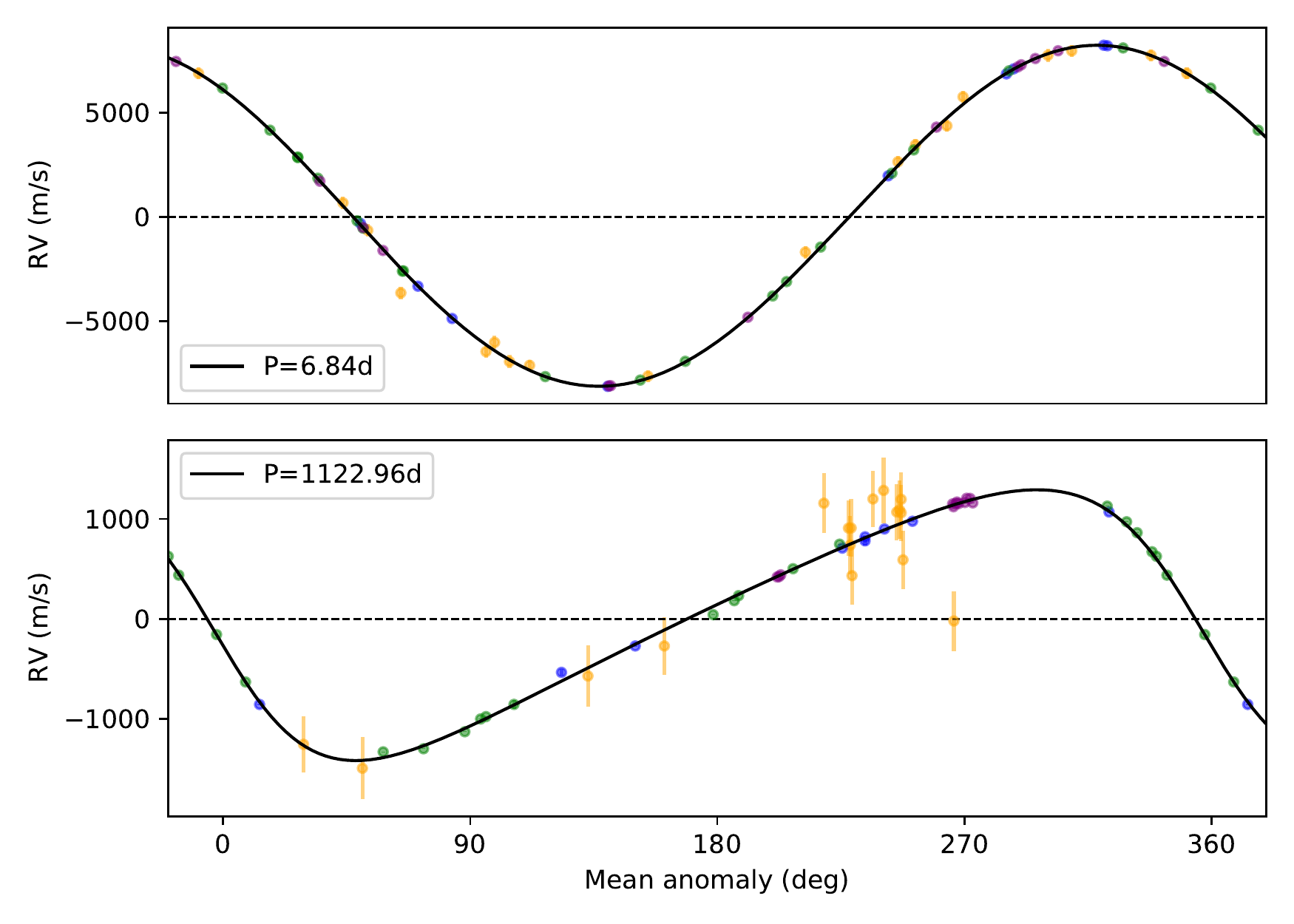}
                \\
                \centering
                \includegraphics[width=0.8\linewidth]{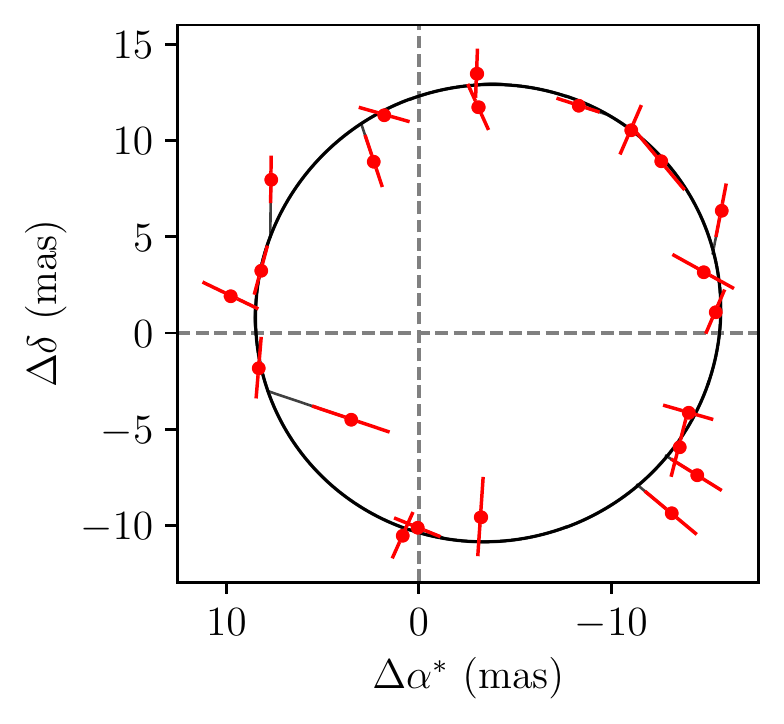}
                \caption{Top two rows: same as Fig.~\ref{fig:HD206276-rvfit} but for HD94340. Bottom panel: best-fit solution astrometric orbit induced by HD94340C, Hipparcos intermediate astrometric data shown in red.}
                \label{fig:HD94340-rvfit}
            \end{figure}
            First hints on the multiple nature of the G4~V star HD94340 were reported in \cite{makarov2005}, in which the presence of a stellar companion to the primary star is suggested by the detection of a large proper motion acceleration between \hip and \tyctwo measurements. More specifically, using preliminary results from the CORALIE survey, \cite{tokovinin2006} reports the presence of a stellar companion with an orbital period of 6.8\days and hints of a possible second massive companion at 1200\days; a joint analysis of \hip proper motion anomaly, adaptive optics and speckle interferometry presented in \cite{tokovinin2012} confirm the triple nature of the system, win an inner companion of 6.8\days and an outer one with an estimated orbital period of 3.3\yr period and 40\mas axis unresolved by speckle interferometry.
            \par Over the course of our survey, we collected a total of 63 radial velocity measurements (17 CORAVEL, 10 C98, 20 C07 and 16 C14) with an observational timespan of 11772\days. The timeseries periodogram features a high significance (FAP=$1.8\cdot10^{-45}$) peak at 6.84\days and the one-Keplerian solution residuals show an additionally highly significant (FAP=$9.6\cdot10^{-19}$) signal at 1120\days, both signals clearly consistent with literature values, with no further residual significant peak and no correspondence between the identified significant peaks and stellar activity signals among the activity indicators analysed. According to our two-Keplerian bestfit model (shown in the top two rows of Fig.~\ref{fig:HD94340-rvfit}), we find the inner companion to have an orbital period $P_{\rm B}=6.84\pm0.01$\days, semiamplitude $K_{\rm B}={8.182}\pm{0.003}$\kmps and eccentricity $e_{\rm B}={0.009}\pm{0.001}$, while for the outer companion we find $P_{\rm C}={1122.96}^{+0.60}_{-0.48}$\days, $K_{\rm C}={1.356}^{+0.008}_{-0.010}$\kmps and $e_{\rm C}={0.305}\pm{0.004}$; having obtained from the SED fit of the primary star (see Sect.~\ref{sec:hosts}) a stellar mass of $1.28^{+0.15}_{-0.19}$\Msun we derive values of minimum masses and semimajor axes of ${90.08}^{+8.71}_{-9.13}$\Mjup and ${0.08}\pm{0.01}$\au for the inner companion and of ${77.85}^{+7.54}_{-7.93}$\Mjup and ${2.34}^{+0.11}_{-0.12}$\au for the outer companion.
            \par As mentioned in Sect.~\ref{sec:astrometry} and already done for HD206276, we subtract from the radial velocity timeseries the Keplerian signal of the 6.84\days inner companion, performing the \orvara fit using the thusly obtained residuals. From the results of this simultaneous radial velocity and proper motion anomaly fit we find for the outer companion a true mass of M$_{C}={269}_{-45}^{+51}$\Mjup (corresponding to ${0.26}_{-0.04}^{+0.05}$\Msun), inclination $i_{\rm C}={18.2}_{-2.3}^{+2.8}$\deg, with values of semimajor axis ($a_{\rm C}={2.44}_{-0.12}^{+0.11}$\au) and eccentricity ($e_{\rm C}={0.332}_{-0.022}^{+0.023}$) in good agreement with those derived from fitting the radial velocity measurements alone. %The bestfit solutions for the proper motions anomaly curves are shown in the bottom row of Fig.~\ref{fig:HD94340-rvfit}.
            % Finally, if we assume coplanarity between the stellar components of the system we can estimate the inner companion to have a true mass of $M_{\rm B}=288.45^{+29.23}_{-27.82}$\Mjup ($0.27^{+0.03}_{-0.02}$\Msun).
            \par We additionaly note that the outer stellar companion in this system is part of the \gdr{3} astrometric orbital solutions validated in \cite{holl2022}, in which it is characterized by having an orbital period of $1213.8\pm22.0$\days and eccentricity $0.30\pm0.01$. The inclination corresponding to this solution is $i_{\rm C}=20.9^{+1.1}_{-1.3}$\deg, the companion true mass is M$_{\rm C}=0.37\pm0.03$\Msun, the relative semi-major axis is $a_{\rm C}=2.63^{+0.12}_{-0.13}$\au. This astrometric-only solution would correspond to a minimum-mass of $M_{\rm C}\sin{i_{\rm C}}=137\pm15$\Mjup or $0.131\pm0.015$\Msun, which we note is larger by a factor 1.75 than the radial-velocity solution. The parallax is $22.508 \pm 0.036$ instead of $19.72 \pm 0.56$ for the \gdr{3} single-star solution.
            \par Moreover, we additionally analyzed jointly the radial velocity and Hipparcos epoch astrometric time series using the kepmodel python package \citep[see][]{delisle2016,delisle2022} and the samsam MCMC sampler \citep[e.g.][]{delisle2018}. While the inclination of the inner 6.8~d companion remains, as expected, unconstrained, the outer 1100~d companion is detected by Hipparcos and its inclination and true mass are constrained. We find an inclination of $i_{\rm C}=13.2\pm0.8$\deg, corrsponding to a true mass of M$_{\rm C}=0.39\pm0.04$\Msun. The relative semi-major axis is $a_{\rm C}=2.5\pm0.1$\au ($53.8 \pm 2.7$\mas), and we find an orbital period for the outer companion of $1122.9\pm0.4$\days, an eccentricity of $0.305^{+0.004}_{-0.003}$, $\omega_{\rm C}=98.9^{+0.7}_{-0.8}$\deg, $\Omega = 8.6\pm4.0$\deg and a revised parallax of $21.4^{+0.7}_{-0.8}$\mas  instead of the $19.58 \pm 1.46$\mas reported for the Hipparcos single-star solution.
            \par We find some discrepancy between the joint Hipparcos and radial velocity solution, the \orvara solution, and the \gdr{3} astrometric orbit solution, in particular between the respective outer companion orbital inclination values. We note that the orbital period is slightly longer than the \gdr{3} timespan, which makes the \gdr{3} solution typically less accurate. Moreover, the errorbars of the \gdr{3} solution are probably underestimated, which is confirmed by the poor matching between the \gdr{3} period and the well determined RV period (about 4$\sigma$). Finally, in this regime of period, the \orvara solution is expected to be more sensitive to the Gaia scanning-law, which could lead to inaccurate results. We therefore adopt the joint Hipparcos and radial velocity solution for this system.

        \subsection{A planet-hosting system: HD196885 (HIP 101966)}\label{subsec:HD196885}
            \begin{figure*}[t!]
                \includegraphics[width=0.5\linewidth]{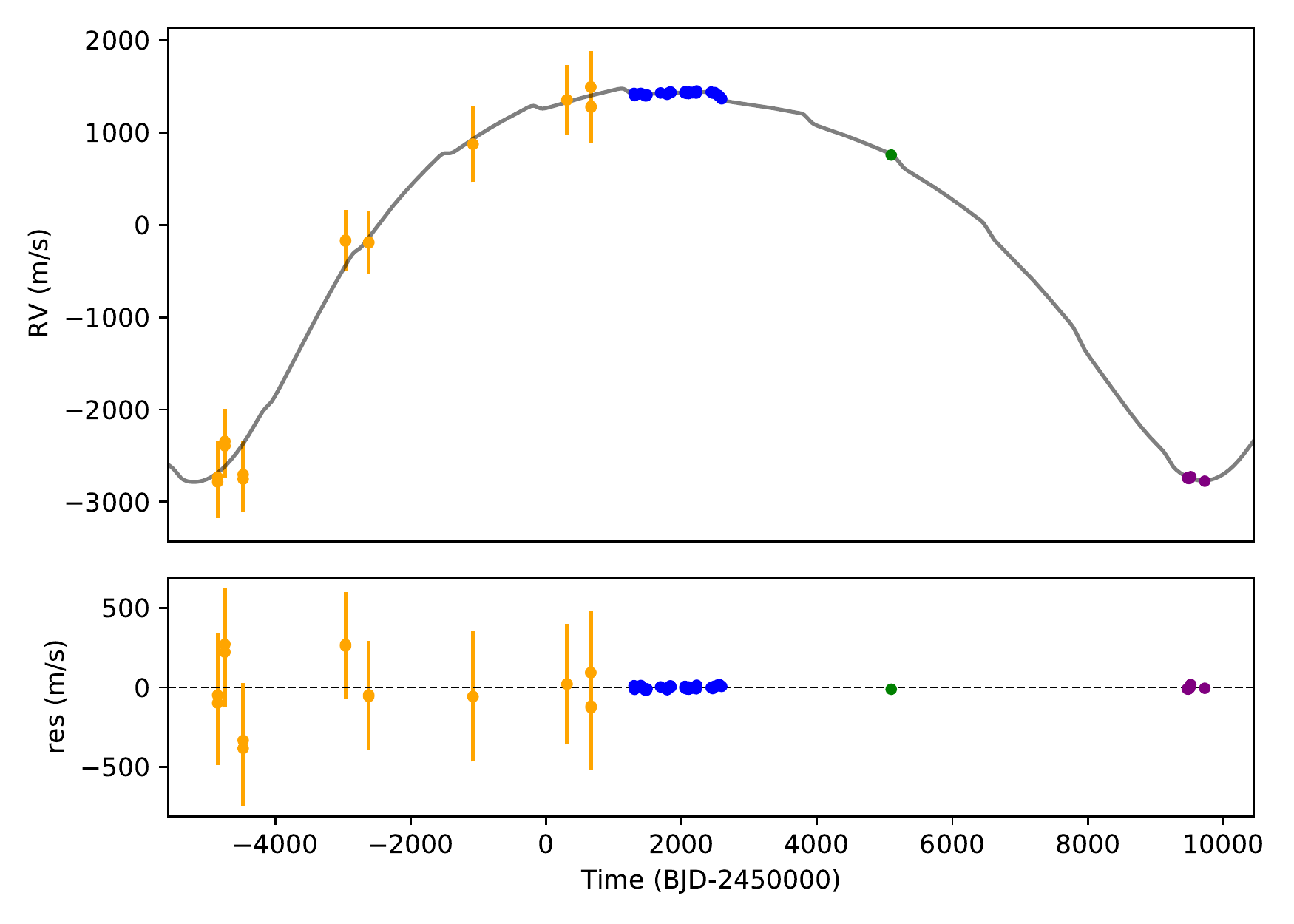}
                \includegraphics[width=0.5\linewidth]{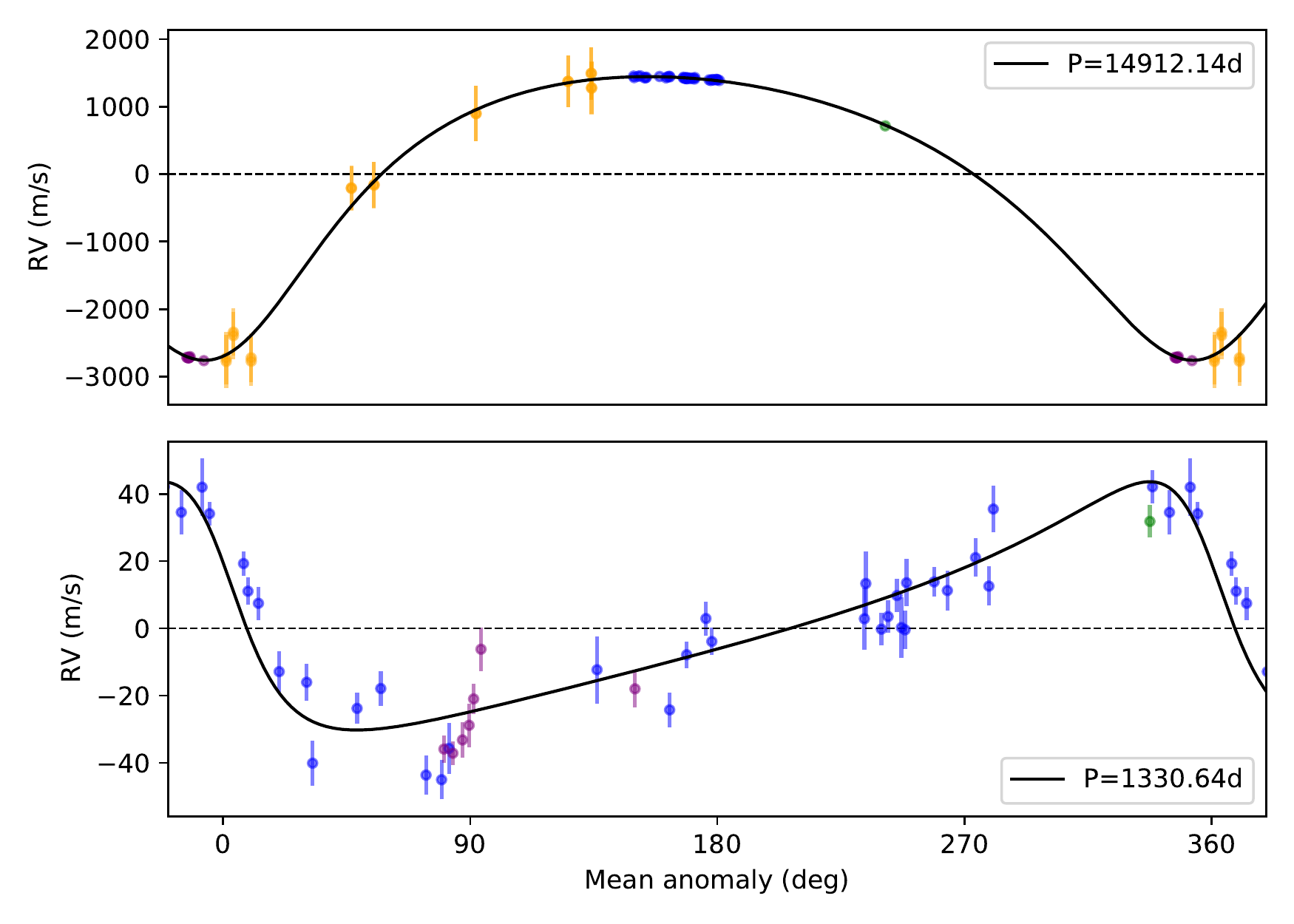}
                \\
                \includegraphics[width=\linewidth]{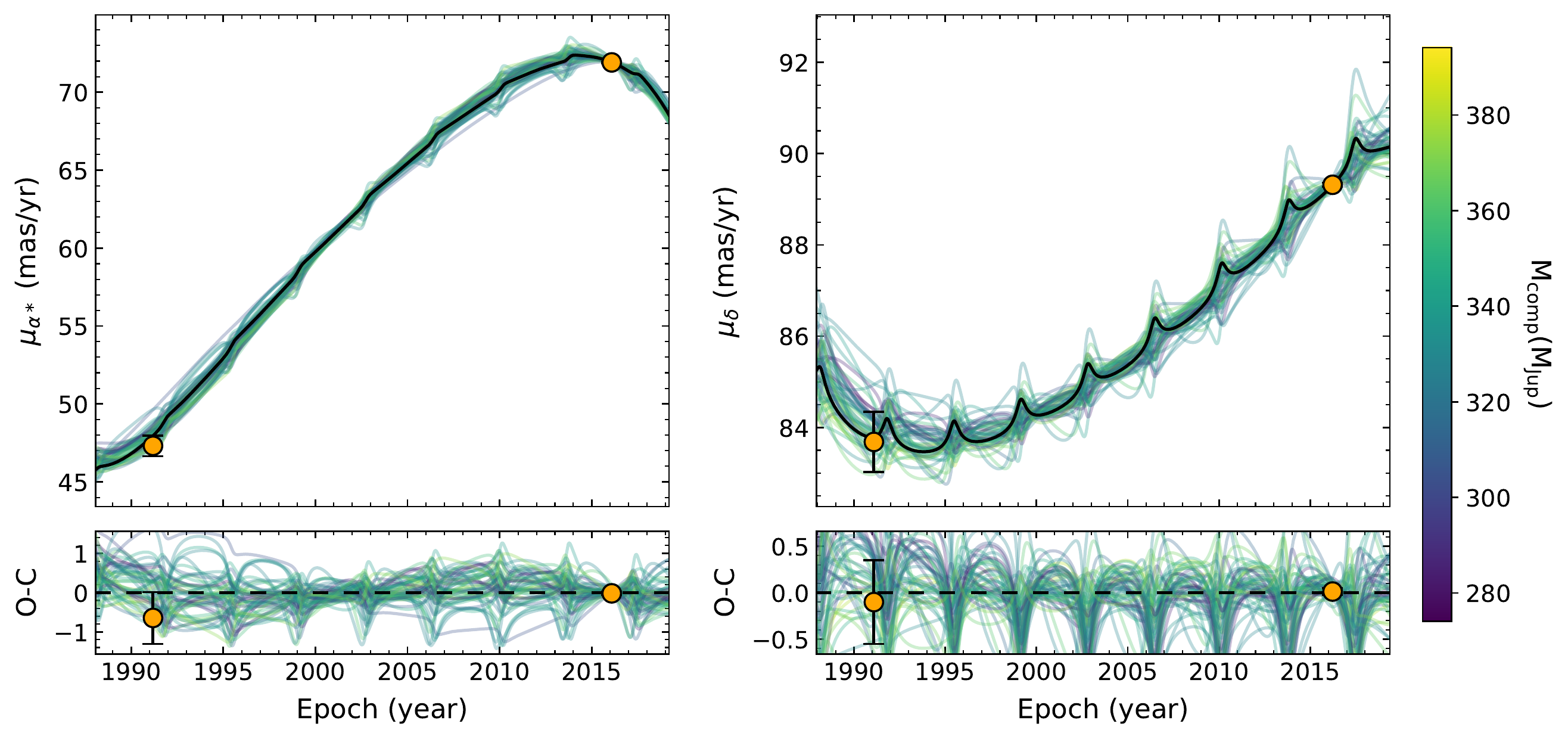}
                \caption{Best-fit orbital solutions for HD196885. Top row: Radial velocity solution and phase folded model curves, with CORAVEL, CORALIE98, CORALIE07 and CORALIE14 measurements shown in orange, blue, green and purple respectively. Bottom row: Observed and fitted proper motions in right ascension and declination. The thick black line is the best-fit orbit obtained by the simultaneous fit of radial velocities and proper motion anomalies, with 50 orbits randomly drawn from the posterior distributions and color-coded according to the mass of the outer massive companion, and the proper motion measurements from \hip and \egdr{3} are shown in orange. Note that in the radial velocity phase folded plot for HD196885~b the low-precision CORAVEL data are not shown in order to better show the radial velocity curve of the low-amplitude inner companion.}
                \label{fig:HD196885-fullfit}
            \end{figure*}
            The presence of a possible stellar companion to the F8~V star HD196885 was first reported in \cite{chauvin2006,chauvin2007} as a result of a near-infrared adaptive optics survey, while \cite{correia2008} additionally detected a planetary companion of having minimum mass of 3\Mjup and orbital period of 1349\days using CORAVEL, CORALIE and ELODIE radial velocity measurements. Follow-up relative astrometry and radial velocity observations \citep{fischer2009,chauvin2011} confirmed the binary nature of the system as well as the orbital parameters of the planetary body found around the primary star.
            \par In the scope of this work, we analyse 18 CORAVEL and 41 CORALIE radial velocity measurements (divided as 33 C98, 1 C07 and 7 C14), over a total of 14574\days of and we are able to provide yet another update on the HD196885 multiple system. The timeseries periodogram features a highly significant region (FAP=$1.08\cdot10^{-8}$) at periodicity higher than 16800\days, indicating either a Keplerian signal longer that our observational timespan or a long-term drift trend, and a residual peak at 1323\days with a FAP level of $1\cdot10^{-3}$ after subtracting a one-Keplerian solution, with no further significant residual signals. In our two-Keplerian solution, shown in the top row of Fig.~\ref{fig:HD196885-fullfit}, we find the outer stellar companion to have orbital period $P_{\rm B}=14912.14^{+1.13}_{-1.16}$\days, semiamplitude $K_{\rm B}={2.102}^{+0.001}_{-0.001}$\kmps and eccentricity $e_{\rm B}={0.322}^{+0.001}_{-0.003}$, for which we then derive using a primary mass of ${1.24}^{+0.12}_{-0.15}$\Msun (see Sect.~\ref{sec:hosts}) values of minimum mass M$_{\rm B}\sin{i_{\rm B}}$=${278.19}^{+21.98}_{-22.84}$\Mjup and semimajor axis $a_{\rm B}$=${13.59}^{+0.50}_{-0.54}$\au. Similarly, we characterize the inner planetary companion as having orbital period $P_{\rm b}={1330.64}^{+0.07}_{-0.43}$\days, semiamplitude $K_{\rm c}=36.94^{+1.24}_{-1.67}$\mps and eccentricity $e_{\rm b}={0.521}^{+0.325}_{-0.085}$, and we derive a minimum mass M$_{\rm b}\sin{i_{\rm b}}$=${1.96}^{+0.34}_{-0.53}$\Mjup and semimajor axis $a_{\rm b}$=${2.54}^{+0.10}_{-0.11}$\au.
            \par As both companions of the primary star orbit beyond our 1\au threshold, we perform the \orvara fit on the original radial velocity timeseries instead of removing one of the companion. As a result, we find for the outer companion a true mass of M$_{B}={334}_{-27}^{+26}$\Mjup (corresponding to $0.32\pm0.02$\Msun) and inclination $i_{\rm B}={101.9}_{-1.5}^{+1.6}$\deg, and for the inner one M$_{b}={2.67}_{-0.63}^{+1.4}$\Mjup and inclination $i_{\rm C}={89}_{-44}^{+42}$\deg; the bestfit solutions for the proper motions anomaly curves are shown in the bottom row of Fig.~\ref{fig:HD196885-fullfit}.
            \par The simultaneous usage of radial velocity and proper motion measurements therefore allow us to confirm the planetary and stellar nature of the inner and outer companion, respectively. Following \cite{tokovinin1993}, we additionally compute the relative orientation of angular momenta $\varphi$ to try and provide further information on the system's configuration. However, due to the highly unconstrained value of the longitude of the ascending node of the inner planetary companion ($\Omega_{\rm b}=169_{-121}^{+142}$\deg), we obtain a relative orientation of $\varphi=87$\deg, a value too close to the 90\deg threshold proposed in \cite{tokovinin1993} to provide any further robust statement on the configuration of the HD196855 system components.
    
    \section{Prospects for exoplanetary search} \label{sec:exoplanet-search}
        As discussed in Sect.~\ref{sec:introduction}, the binarity of a stellar system has a deep influence on the formation and evolution of planetary companions, especially when the orbital separations between the stellar components of the systems is below a few hundreds \au, in which case most studies in the literature suggest that the formation and survival of more massive and close-in exoplanets than those found around single stars \citep[see e.g.][]{fontanive2019,fontanive2021,cadman2022}. Additionally and more closely related to the present work, different studies have highlighted how planetary formation and evolution is especially affected by binary separations less than $~$100\au \citep[see e.g.][]{mayer2005,moe2021}. \cite{fontanive2019} derives a binary fraction of $79.0^{+13.2}_{-14.7}\%$ for stars hosting giant planets and brown dwarfs on orbits shorter than 1 au, again supporting the critical influence that the presence of stellar companions have on the formation and evolution of such planetary systems. Similarly, based on a literature search for binary companions to exoplanet-hosting stars within 200 \pc, \cite{fontanive2021} finds that while exoplanets found on circumprimary orbits in very wide binary systems show similar physical properties than those around single stars, tighter binary systems with separations up to a few hundred astronomical units tend to promote instead the formation and survival of more massive giant planets and brown dwarfs with shorter orbital periods and typically in single-planet configurations. The same work also suggests that the properties of close-in exoplanets in wide binary systems are consistent with them being the result of formation via fragmentation in a gravitationally unstable disc. This result is further supported by the simulations detailed in \cite{cadman2022}, which find that intermediate separations between the component of a binary stellar system promoting fragmentation is consistent with those featured in the systems displaying an excess of close-in giant planets and brown dwarf \citep{wang2014,kraus2016,ngo2016}
        \par While we have detected no new robust exoplanetary signal in the radial velocity analysis of our binary sample the low-to-intermediate orbital separations we derived for the stellar companions discussed in this work, ranging from $\sim$0.045\au to $\sim$36.40\au, makes our sample a significant opportunity for the search of exoplanetary companions in binary systems and a suitable testing field for planetary formation theoretical models, provided a larger number of radial velocity measurements with high enough density and precision are collected to successfully disentangle the stellar companion's radial velocity signal from that of any lower-mass body that can orbit the system.
    
    \subsection{Planetary stability in the binary sample} \label{subsec:dynamics}
        In order to provide a first assessment of the regions in which exoplanets could be found on stable orbits in the systems of our sample, we use the analytical stability criteria provided in \cite{ballantyne2021} and based on the numerical simulations performed in \cite{holman1999}.
        \par Considering a planet on a circumprimary (or S-type) orbit in a binary system, \cite{ballantyne2021} defines the critical semimajor axis $a_{\rm cS}$ as the maximum stable distance from the primary star:
        \begin{equation} \label{eq:acs}
            \begin{split}
                a_{\rm cS} = a_{\rm bin} \Bigl( & 0.464 - 0.38\mu - 0.631e + 0.586\mu e +\\
                            & 0.15e^2 - 0.198\mu e^2 \Bigr)
            \end{split}
        \end{equation}
        being $a_{\rm bin}$ and $e$ the semimajor axis and eccentricity of the binary, and where:
        \begin{equation} \label{eq:mu}
            \mu=\frac{m_s}{m_p+m_s}
        \end{equation}
        with $m_p$ and $m_s$ as the masses of the primary and secondary stellar components of the binary. Similarly, for a circumbinary (or P-type) orbit the critical semimajor axis $a_{\rm cP}$, being the minimum stable distance from the binary system, is given by:
        \begin{equation} \label{eq:acp}
            \begin{split}
                a_{cP} = a_{\rm bin} \Bigl( & 1.6 + 5.1e - 2.22e^2 + 4.12\mu - 4.27\mu e -\\ 
                        & 5.09\mu^2 + 4.61\mu^2 e^2 \Bigr)
            \end{split}
        \end{equation}
        While finally considering a planet on a circumsecondaty orbit, the maximum stable distance from the secondary star $a_{\rm cS, sec}$ is given by using again Eq.~\ref{eq:acs} with instead:
        \begin{equation} \label{eq:mu_sec}
            \mu=\frac{m_p}{m_p+m_s}
        \end{equation}
        \par As noted in \cite{ballantyne2021}, these stability criteria are to be taken as a first-order indication of the circumprimary or circumbinary stability of a planet, since there is a variety of mechanisms that can further enhance or disrupt the stability of such orbits \citep[see e.g.][and references therein]{pilat2002,pilat2003,parker2013,lam2018,quarles2018,quarles2020,kong2021}. However, since a full dynamical characterization of the binary systems in our sample is beyond the scope of the present work, we assume the validity of the defined stability criteria for the purpose of producing a first estimate of the stability regions of the considered systems.
        \par Additionally, we consider as the minimum stable distance from the primary star the Roche limit of the host star, defined as:
        \begin{equation} \label{eq:roche}
            d_{\rm Roche} =  2.423~R_\star \sqrt[3]{\frac{\rho_\star}{\rho_{\rm pl}}}
        \end{equation}
        where $R_\star$ and $\rho_\star$ are the radius and density of the primary star and $\rho_{\rm pl}$ is the density of the orbiting planet. We use for each primary component the stellar parameters derived from the SED fits detailed in Sect~\ref{sec:hosts}, while for the planetary density used to compute the Roche limits we use the Earth density $\rho_\oplus$ as a lower limit scenario. The values of stability limits $d_{\rm Roche}$, $a_{\rm cS}$ and $a_{\rm cP}$ computed for the systems in the sample are listed in Table~\ref{table:stable-regions-short}
        \begin{table}[t]
            \caption{Boundaries of the circumprimary, circumbinary and circumsecondary} stability regions of the systems in the binary sample as described in Sect.~\ref{subsec:dynamics}. The Note column indicates with value of the detected companion mass is used for the stability estimation.\label{table:stable-regions-short}
            \centering
            \begin{tabular}{l c c c c c}
    			\hline\hline
        			Name & $d_{\rm Roche}$ & $a_{\rm cS}$ & $a_{\rm cP}$ & $a_{\rm cS, sec}$ & Note\\
			        & [\au] & [\au] & [\au] & [\au] & \\
    			\hline
        			HD225155&0.008&0.889&13.035&0.428&M$_{\rm true}$\\
        			HD1815&0.006&0.646&6.016&0.340&\msini\\
        			HD1926&0.007&0.229&1.396&0.086&\msini\\
        			HD2070&0.008&0.115&1.843&0.058&\msini\\
        			HD2098&0.007&1.185&12.659&0.756&M$_{\rm true}$\\
        			HD3222&0.007&2.839&41.274&1.336&M$_{\rm true}$\\
        			HD3277&0.007&0.071&0.766&0.021&\msini\\
        			HD3359&0.007&0.036&0.566&0.018&\msini\\
        			HD3795&0.007&5.067&144.914&4.650&M$_{\rm true}$\\
        			HD4392&0.007&0.145&5.935&0.109&M$_{\rm true}$\\
        			HD4747&0.007&1.096&39.734&0.387&M$_{\rm true}$\\
        			HD5562&0.008&1.233&20.116&0.702&M$_{\rm true}$\\
        			HD7320&0.007&0.166&51.340&0.095&M$_{\rm true}$\\
        			HD8129&0.007&-&27.509&-&\msini\\
        			HD9770&0.008&0.766&9.656&0.269&\msini\\
        			HD9905&0.007&2.421&31.232&0.951&M$_{\rm true}$\\
        			HD10519&0.007&0.477&38.102&0.345&M$_{\rm true}$\\
        			HD11131&0.007&0.410&20.108&0.260&M$_{\rm true}$\\
        			HD11264&0.007&3.054&46.823&2.032&M$_{\rm true}$\\
        			HD11352&0.007&0.315&4.046&0.183&M$_{\rm true}$\\
        			...&...&...&...&...\\
                \hline
            \end{tabular}
            \tablefoot{Full table is available at the CDS. A portion is shown here for guidance regarding its form and content.}
        \end{table}
        \par
\begin{figure*}[t!]
            \centering
            \includegraphics[width=0.82\linewidth]{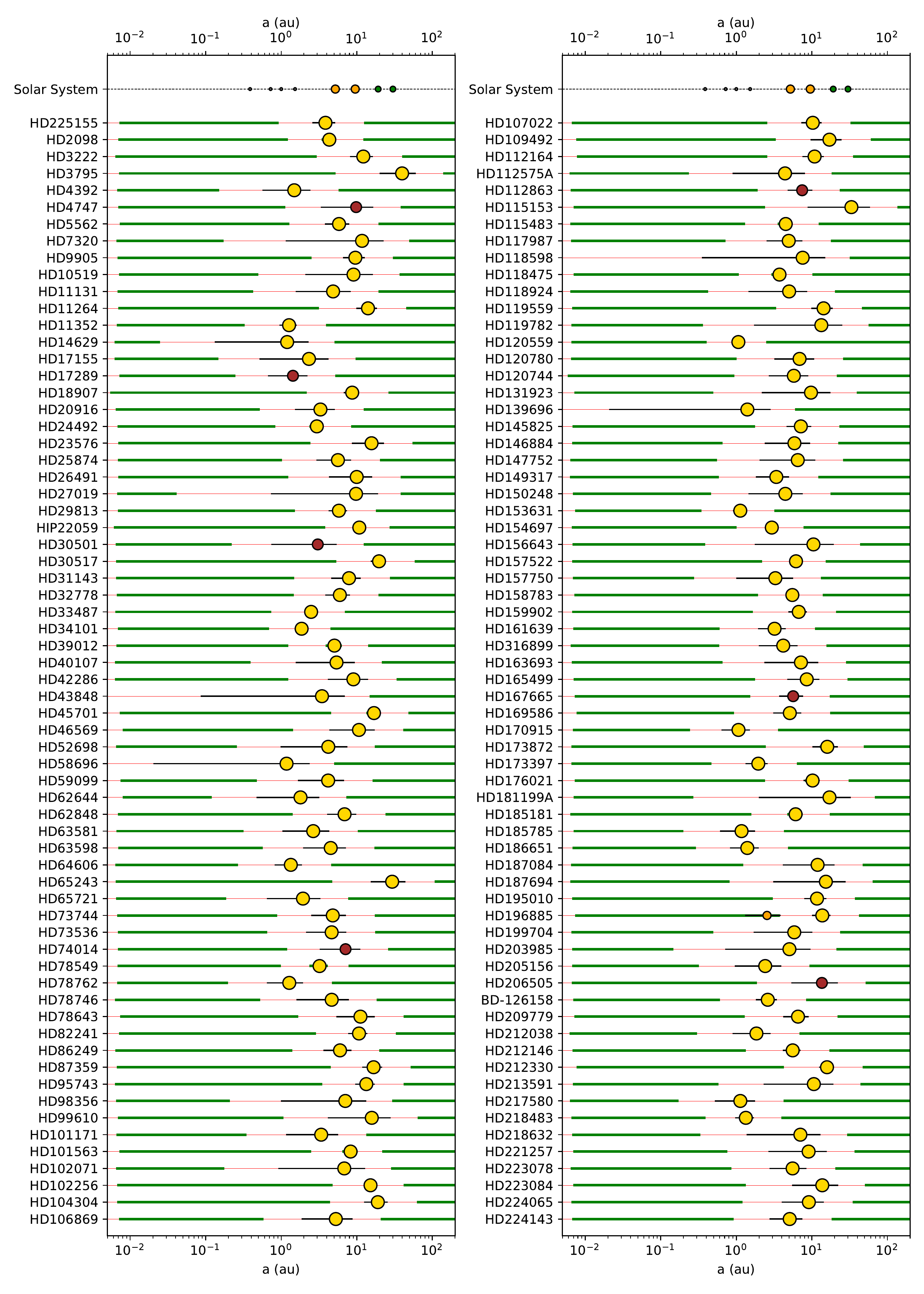}
            \caption{Overview of the architectures of the \goodorvara binary systems for which true masses were derived using simultaneous radial velocity and proper motion anomaly fits, compared to that of the Solar System. For each system, the companion periastron and apoastron are connected by a thin black line, while the thick green and thin red lines highlight the stable and unstable regions for additional planetary companions respectively as detailed in Sect.~\ref{subsec:dynamics}; we note that circumbinary stable regions are typically smaller than the companion marker size due to the logarithmic axis used. The mass of each companion is represented by the marker growing size and different color, namely grey for terrestrial (M$<$2\Mearth), green for Neptune-mass (10\Mearth$<$M$<$30\Mearth), orange for giant planets (30\Mearth$<$M$<$40\Mjup), brown for brown dwarfs (40\Mjup$<$M$<$80\Mjup) and yellow for stellar companions (M$>$80\Mjup).}
            \label{fig:binary-architecture-mtrue}
        \end{figure*}

 We first focus our attention on the binary systems for which we obtained true mass values from the simultaneous fit of radial velocities and proper motion anomalies performed with \orvara and detailed in Sect.~\ref{sec:astrometry}, ignoring for the moment the \nohgca systems absent from the HGCA and the \badorvara systems with orbital period too short to be detected by proper motion anomalies. We therefore consider in the following analysisc only the \goodorvara systems for which we have derived values of companion true dynamical mass. Additionally we note that the dynamical stability of a planet in the triple systems HD206276 and HD94340 is likely beyond the validity of the \cite{ballantyne2021} criteria and therefore warrants further study and focused analysis such as numerical simulations, and that the stability assessments presented here for these two systems should then be interpreted as first-order estimates.
        % assuming coplanarity for the inner stellar companions of HD206276 and HD94340 (see Sects.~\ref{subsec:HD206276} and \ref{subsec:HD94340}), for which we additionally note that the dynamical stability of a planet in a triple stellar system is likely beyond the validity of the \cite{ballantyne2021} criteria and therefore warrants further study and focused analysis such as numerical simulations. 
        Fig.~\ref{fig:binary-architecture-mtrue} shows the system architectures compared with that of the Solar System, the stability regions represented by green bands.
        \begin{figure*}[t!]
            \centering
            \includegraphics[width=0.82\linewidth]{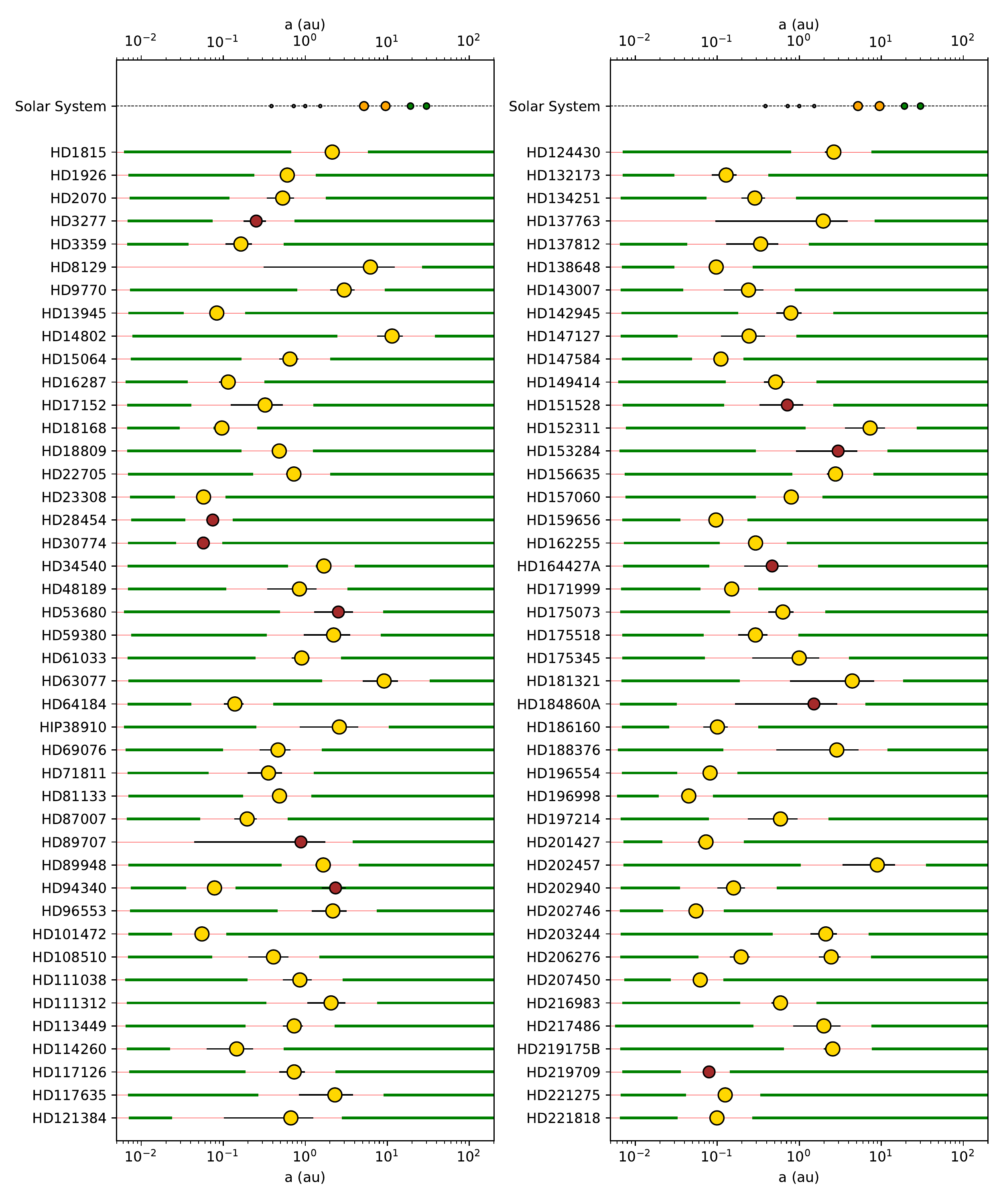}
            \caption{Same as Fig.~\ref{fig:binary-architecture-mtrue} but for the \rvonly systems with only radial velocity solutions available, for which the dynamically stable and unstable regions for additional planetary companions are computed using the \msini value of the detected companions.}
            \label{fig:binary-architecture-msini}
        \end{figure*}
        \par From our stability estimates, we note that only \nostype systems in our sample do not allow for stable S-type orbits, namely the HD58696, HD43848, HD139696 and HD118598 systems, due to the high eccentricities (0.98, 0.97, 0.98 and 0.95 respectively) of their secondary components as derived by the simultaneous radial velocity and proper motion analysis. Additionally, we find both triple systems HD206276 and HD94340 to have a very narrow stable region for S-type orbits, spanning below 0.05\au and therefore unlikely to host circumprimary planetary companions, especially by virtue of the presence of the two stellar companions discussed in Sects.~\ref{subsec:HD206276}-\ref{subsec:HD94340}. The system featuring the widest S-type stability region in the sample is HD30517, with said region spanning 5.16\au, while the system for which the secondary companion has the larger impact on planetary stability is instead HD3795, in which no planetary orbit appear to be stable from an orbital separation of 5.07\au to 144.91\au from the primary star. The latter system is also the one in the sample characterized by the wider minimum P-type stable orbit in the sample, while the system with the tighter circumbinary stable orbit is HD120559, having a $a_{\rm cP}$ of 2.58\au. Finally focusing on planetary companions orbiting the secondary component of our binary systems, the largest stable circumsecondary region in our sample is found in the HD3795 system with a $a_{\rm cS, sec}$ of 4.65\au.
        \par Considering instead the \rvonly binary systems in the sample for which we have only radial velocity orbital solution and therefore only minimum mass values for the companions detected in these systems, we apply the same stability criterion using the value of \msini as the $m_s$ in Eqs.~\ref{eq:acs}-\ref{eq:acp}; the stable regions thus obtained are then to be considered estimates of the maximum range of semimajor axes in which stable orbits for planetary companions are possible. The architecture of these systems is then shown in Fig.~\ref{fig:binary-architecture-msini}, from which it can be noted that 3 systems (HD8129, HD89707 and HD137763) do not allow for stable S-type orbits, again by virtue of the large eccentricities (0.95 for all of them) of the detected companions.
    
    \subsection{Detection limits} \label{subsec:detection-limits}
        \begin{figure}[t]
            \centering
            \includegraphics[width=\linewidth]{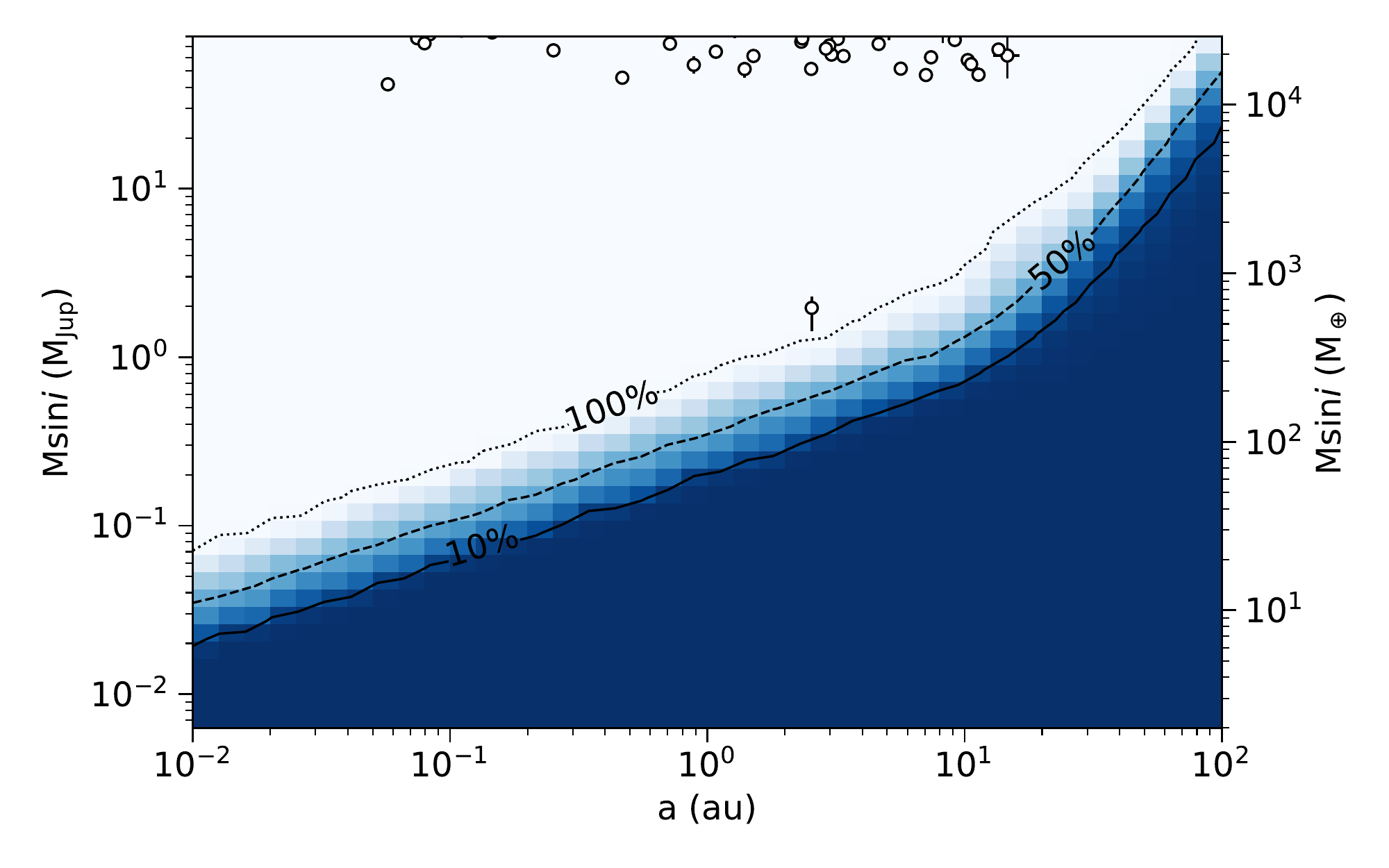}
            \caption{Completeness map of the binary sample, focused on the substellar (2\Mearth$<$\msini$<$80\Mjup) companion regime. Detection frequency contour levels of 10, 50 and 100\% are respectively shown as solid, dashed and dotted curves, while the companions detected in our search are shown as white circles.}
            \label{fig:sample-detection}
        \end{figure}
        In order to investigate the detection capabilities of the CORAVEL and CORALIE data analysed so far we compute the detection limits for the binary sample object of the present work, especially focusing on the substellar (\msini$<$80\Mjup) regime.
        \par To this end we follow a injection and retrieval scheme similar to the one pursued in \cite{barbato2018}, injecting synthetic companion signals into the radial velocity residuals timeseries of each star in the sample, obtained by subtracting from the original radial velocity data the contribution of the companions detected and characterized as described in Sect.~\ref{sec:rv}. The synthetic signals are generated over a 40x40 grid of semimajor axes $a_{\rm inj}$ evenly spaced in logarithm from 0.01 to 100\au and minimum masses M$_{\rm inj}$ similarly spaced from 2\Mearth to 80\Mjup. For each of the 1600 ($a_{\rm inj}$, M$_{inj}$) realizations we generate and inject into the residuals 500 synthetic radial velocity curves with randomly drawn values of mean longitude $\lambda_{0, \rm{inj}}$, eccentricity $e_{\rm inj}$ and argument of periastron $\omega_{\rm inj}$. Lastly, we add to each synthetic data thus generated a random Gaussian noise with an amplitude equal to the standard deviation of the instrumental uncertainties of the original timeseries. Each of the resulting $8\cdot10^5$ synthetic radial velocity timeseries is then fitted with a flat model and a Keplerian one, and the injected signal is considered as detected only if the $\Delta$BIC between the two models is at least 10 points in favour of the Keplerian model. The resulting detection limit maps obtained for each stars in our binary sample are collected in Appendix~\ref{app:detection-limits}, highlighting both the dynamically unstable regions estimated in Sect.~\ref{subsec:dynamics} and the parameter space region of additional companions that we can exclude based on the available radial velocity measurements.
        \par We show in Fig.~\ref{fig:sample-detection} the averaged completeness map for the whole binary sample; while the CORALIE and CORAVEL measurements collected for the sample do not have the precision and sampling necessary to ensure detection completeness for planetary companion below 10\Mearth and we have only partially completeness for giant companions below the mass of Jupiter in the explored range of orbital separation, we are instead complete for companions above 1\Mjup within 1.80\au.
        
    \section{Occurrence rates for stellar and brown dwarfs companions in the sample} \label{sec:occurrence-rates}
        From the detection limit map produced in Sect.~\ref{subsec:detection-limits} and shown in Fig~\ref{fig:sample-detection} it is possible to notice that the radial velocity measurements collected for the sample analysed in the present paper allow us to reach full detection completeness for both brown dwarf companions (40$<$\msini$<$80\Mjup) with semimajor axis below $\sim$62\au and stellar (\msini$>$80\Mjup) companions within 100\au from the primary stars. The thorough analysis of the detection limits map produced instead for the whole CORALIE sample will be the focus of a future paper in the series, but for the purposes of the present work it is possible to report that the detection completeness for the aforementioned brown dwarf and stellar companion parameter space is confirmed for the overall CORALIE sample. We exclude from the following analysis the five companions with $q>0.8$ identified in Sect.~\ref{sec:rv} and Sect.~\ref{sec:astrometry}.
        \par We can use this information to provide an assessment of the occurrence rate $f$ for brown dwarfs and stellar companions using the binomial distribution:
            \begin{equation}  \label{eq:binomial}
                p(m;N,f)=\frac{N!}{m!(N-m)!}\ f^m (1-f)^{N-m}
            \end{equation}
        being $N$ the size of the CORALIE search sample and $m$ the number of detections within the parameter space here taken into account. In order to derive $f$ we follow the approach described in \cite{burgasser2003}, \cite{mccarthy2004} and previously applied in different occurrence rate sudies \citep[such as][to name a few]{sozzetti2009,faria2016,barbato2018,santos2011}, where the inverse binomial function $p^\prime(f;m,N)\propto p(m;N,f)$ is normalized to 1 over a range of $f$ values bound between 0 and 1. This yield the result:
            \begin{equation}    \label{eq:inverse-binomial}
                \int_{0}^{1} p^\prime(f;m,N)\,df=1 \rightarrow p^\prime = (N+1)p
            \end{equation}
        and the occurrence rate $f$ is then found as the value corresponding to the mode of $p^\prime$. Lastly, upper and lower uncertainty limits $f_{\rm U}$, $f_{\rm L}$ corresponding to 1$\sigma$ confidence limits are computed by finding the range covering 68\% of the $p^\prime$ distribution by numerically solving the relation:
            \begin{equation}    \label{eq:binomial-uncertainties}
                \sum_{i=0}^{m} \frac{(N+1)!}{i!(N+1-i)!}x^i (1-x)^{N+i-1} = 
                \begin{cases}
                    0.84, &~x=f_{\rm L}\\
                    0.16, &~x=f_{\rm U}
                \end{cases}
            \end{equation}
        \par Firstly, considering the 209 stellar companions detected (203 from this work and 6 from Unger et al., in prep.) we find an occurrence rate for stellar companions of $f_\star=12.69^{+0.87}_{-0.77}\%$. If again we distinguish between the 127 inner and 82 outer stellar companions setting the threshold at 5\au, we find occurrence rate values of $7.71^{+0.71}_{-0.60}\%$ and $4.98^{+0.59}_{-0.48}\%$ respectively. It can be noted that these occurrence rates are much lower than the $\sim$50\% computed for the CORAVEL survey presented in \cite{duquennoy1991}, but it is important to underline not only the fact that the sample analysed in the cited work was composed of 164 stars and therefore much smaller in size than the whole CORALIE sample we instead considered in the present study, but also that the CORAVEL sample also included a number of SB2 that we instead excluded from our analysis, making a direct comparison between the two studies non-trivial.
        \par Considering the 13 detections with 40$<$\msini$<$80\Mjup for which the astrometric analysis either confirm the brown dwarf nature or is not possible (see Sect.~\ref{subsec:brown-dwarfs} and Unger et al., in prep.) we obtain an occurrence rate for brown dwarf companions in the CORALIE sample of $f_{\rm BD}=0.79^{+0.29}_{-0.16}\%$; considering only the 8 such companions found within 5\au from the primary star we obtain an occurrence rate of close-in brown dwarfs of $0.49^{+0.24}_{-0.12}\%$, while for the 5 wider brown dwarf companions we obtain an occurrence rate of $0.30^{+0.21}_{-0.07}\%$, values that we note to be compatible within 1$\sigma$. While this apparent surplus of brown dwarfs on closer orbits might be interpreted as in opposition with the known brown dwarf desert, it is important to remember that, while the CORALIE survey is certainly able to detect the large amplitude signals of brown dwarf companions on wide orbits as evident from Fig.~\ref{fig:sample-detection}, its 25\yr timespan allows only for the robust identification of Keplerian signals corresponding to an orbital separation up to $\sim$7\au while wider companions would be instead detected as radial velocity trends which have not been considered for analysis in the present work (see Sect.~\ref{sec:rv}): a number of possible long-period brown dwarf companions in the CORALIE sample are therefore possibly still to be detected, leading to an larger occurrence rate difference between the two population, and the same applies to the stellar companions in the sample.
        \par By instead considering only the 7 brown dwarfs in the sample confirmed as such by the joint radial velocity and proper motion analysis we obtain an occurrence rate of $f_{\rm BD}=0.43^{+0.23}_{-0.11}\%$, which we therefore propose as a lower limit on the occurrence rate of brown dwarfs in the sample. Selecting again a threshold of 5\au between inner and outer brown dwarfs, we find occurrence rate values of $0.12^{+0.17}_{-0.03}\%$ and $0.30^{+0.21}_{-0.07}\%$ respectively. While we now find a lower occurrence rate of inner brown dwarfs, we must again note that these two values are compatible within 1$\sigma$.
        
    \section{Discussion and conclusions} \label{sec:conclusions}
        We present in this paper the results of the homogeneous analysis performed in search for brown dwarf and stellar companions to the 1647 stars comprising the long-term CORALIE exoplanetary survey, in order to produce an updated catalog of binary objects in the sample using a combination of radial velocity and astrometry measurements. As a result, we find \nbinary stars in the CORALIE sample to host at least one stellar or brown dwarf companion, \npub of which are already known in the literature and for which we present updated orbital solutions, and \nnew of which are instead not known so far and for which we therefore provide first assessment of the orbital parameters.
        \par Furthermore, by combining radial velocity measurements and astrometric accelerations as computed between the \hip and \egdr{3} we are able to derive precise dynamical masses of \goodorvara stellar and brown dwarf companions with an orbital separation down to 1\au. Notably, we are also able to confirm the planetary nature of HD196885~b as well as the brown dwarf nature of \staysbd companions with 40$\leq$\msini$\leq$80\Mjup, while we find \bdtostar companions with minimum masses within this range to be revealed as stellar-mass companions, again stressing the power of joint usage of radial velocity and astrometric measurements in painting a full picture of system characterization.
        \par The detection completeness analysis we perform on the sample also allow us to derive occurrence rates $f_\star=12.69^{+0.87}_{-0.77}\%$ and $f_{\rm BD}=0.43^{+0.23}_{-0.11}\%$ for stellar and brown dwarf companions respectively. While our occurrence rates also show an apparent overabundance of stellar and brown dwarf companions below 5\au compared to those found on wider orbits, it is imperative to stress that in the present work we have considered only those companions that are found to be best characterized by Keplerian models instead of linear or parabolic trends, and therefore a possibly large number of wide massive companions are still to be found and fully characterized by continuous observations by spectrographs and especially direct imaging instruments, and will be the subject of future papers in this series.
        \par The binary sample presented and characterized in this work can not only represent an important element for follow-up studies on binary statistics and comparison with formation and evolution theoretical models for both stellar and brown dwarf companions, but also an unparalleled opportunity for the search of exoplanetary bodies in binary systems, as theoretical models have shown that planetary formation and survival is deeply influenced by stellar companions within $\sim$100\au such as those detailed in the present study. Both the dynamical stability assessment and detection limit maps we have produced show that there is still significant space for the discovery of exoplanets on circumprimary and circumbinary orbits around the stars here analysed, and continued follow-up observations will allow in the near future to deeply probe the exoplanetary discovery space in the sample, allowing this catalog to be used as testing field for models of planetary formation in binary systems.

    \begin{acknowledgements}
        % Thanking the referee ----------------------------------------------------------
        The authors wish to thank the referee, Dr. F.~Kiefer, for the thorough and useful comments which significantly improved the quality of the manuscript.
        % General PlanetS acknowledgments -----------------------------------------------
        This work has been carried out within the framework of the National Centre of Competence in Research PlanetS supported by the Swiss National Science Foundation under grants 51NF40\_182901 and 51NF40\_205606. The authors acknowledge the financial support of the SNSF.
        % EULER and CORALIE acknowledgements --------------------------------------------
        The 120 cm EULER telescope and the CORALIE spectrograph were funded by the SNSF and the University of Geneva.
        % General DACE acknowledgments --------------------------------------------------
        This publication makes use of the Data \& Analysis Center for Exoplanets (DACE), which is a facility based at the University of Geneva (CH) dedicated to extrasolar planets data visualisation, exchange and analysis. DACE is a platform of the Swiss National Centre of Competence in Research (NCCR) PlanetS, federating the Swiss expertise in Exoplanet research. The DACE platform is available at \url{https://dace.unige.ch}.
        % Author acknowledgements -------------------------------------------------------
        NCS acknowledges support from the European Research Council through the grant agreement 101052347 (FIERCE) and by FCT - Funda\c{c}\~{a}o para a Ci\^{e}ncia e a Tecnologia through national funds and by FEDER through COMPETE2020 - Programa Operacional Competitividade e Internacionaliza\c{c}\~{a}o by these grants: UIDB/04434/2020; UIDP/04434/2020.
        % Gaia acknowledgment -----------------------------------------------------------
        This work has made use of data from the European Space Agency (ESA) mission {\it Gaia} (\url{https://www.cosmos.esa.int/gaia}), processed by the {\it Gaia} Data Processing and Analysis Consortium (DPAC, \url{https://www.cosmos.esa.int/web/gaia/dpac/consortium}). Funding for the DPAC has been provided by national institutions, in particular the institutions participating in the {\it Gaia} Multilateral Agreement.
        % SIMBAD acknowledgment ---------------------------------------------------------
        This research has made use of the SIMBAD database, operated at CDS, Strasbourg, France.
        % Software acknowledgments ------------------------------------------------------
        The authors made use of 
        \textsc{ASTROPY} \citep[a community-developed core Python package for Astronomy][]{software:astropy2013,software:astropy2018}, 
        \textsc{MATPLOTLIB} \citep{software:matplotlib}, 
        \textsc{NUMPY} \citep{software:numpy}, 
        \textsc{SCIPY} \citep{software:scipy} 
        and 
        \textsc{SEABORN} \citep{software:seaborn}.
        % Silly, personal acknowledgment? -----------------------------------------------
        DB also wishes to thank N.~Gaiman for his inspiring words about the illusion of permanence and stellar transience.
    \end{acknowledgements}
  
  \bibliographystyle{aa}
  % \bibliography{binaries-ref}
  \bibliography{coralie-binaries}
  
  \begin{appendix}
    
    \clearpage
    \onecolumn
    \section{RV orbital solutions phase-folded plots} \label{app:rvplots}
    	{
    	\centering
    	\begin{longtable}{c c c c}
		\includegraphics[width=0.22\linewidth]{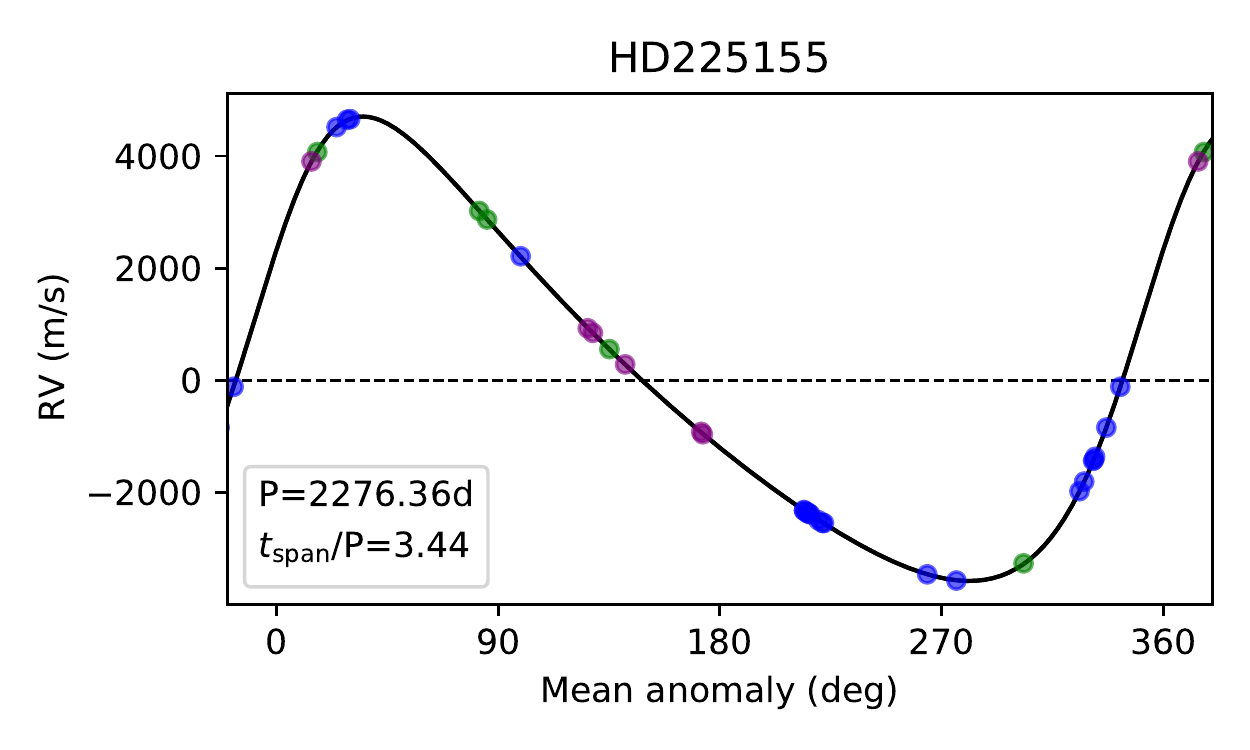}&
		\includegraphics[width=0.22\linewidth]{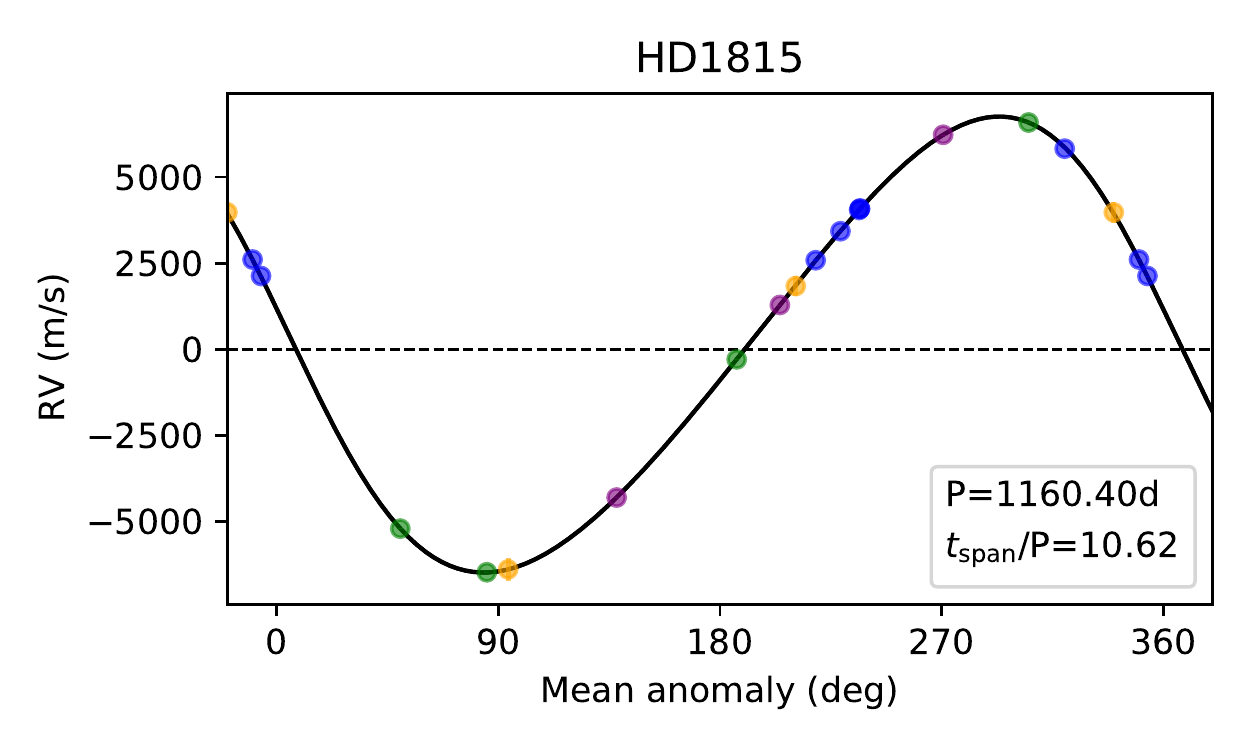}&
		\includegraphics[width=0.22\linewidth]{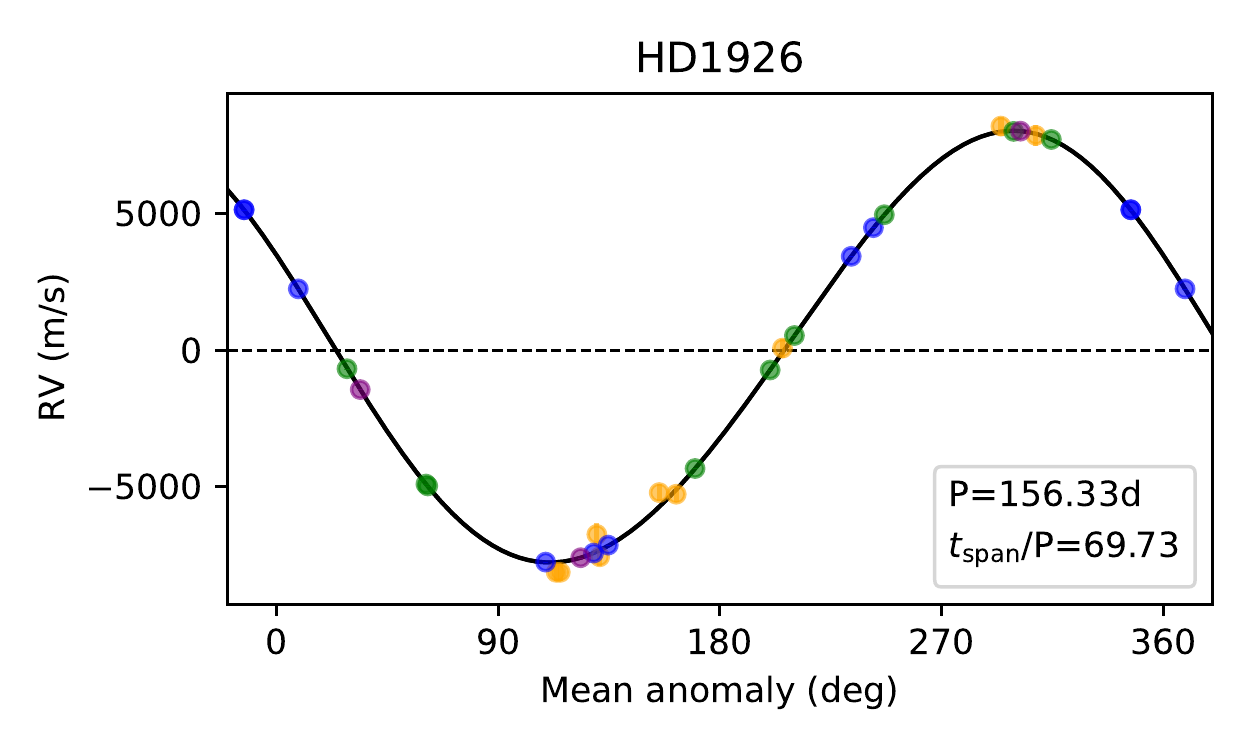}&
		\includegraphics[width=0.22\linewidth]{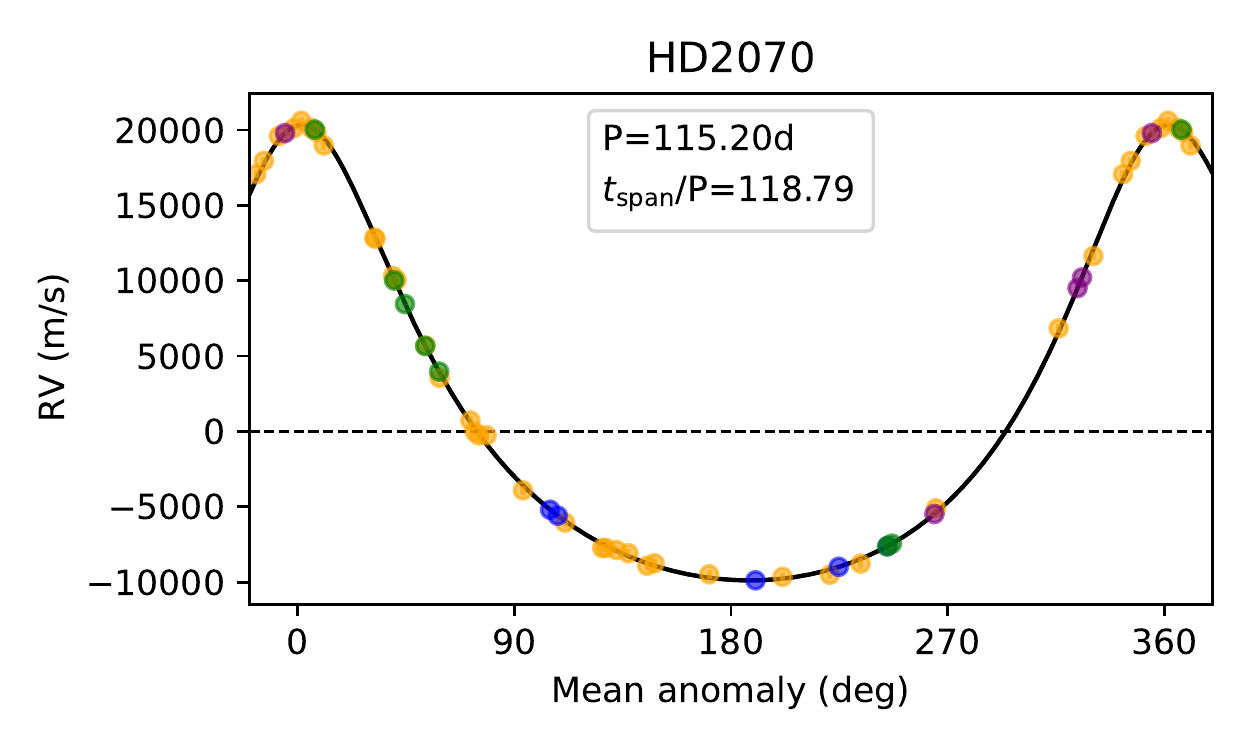}\\

		\includegraphics[width=0.22\linewidth]{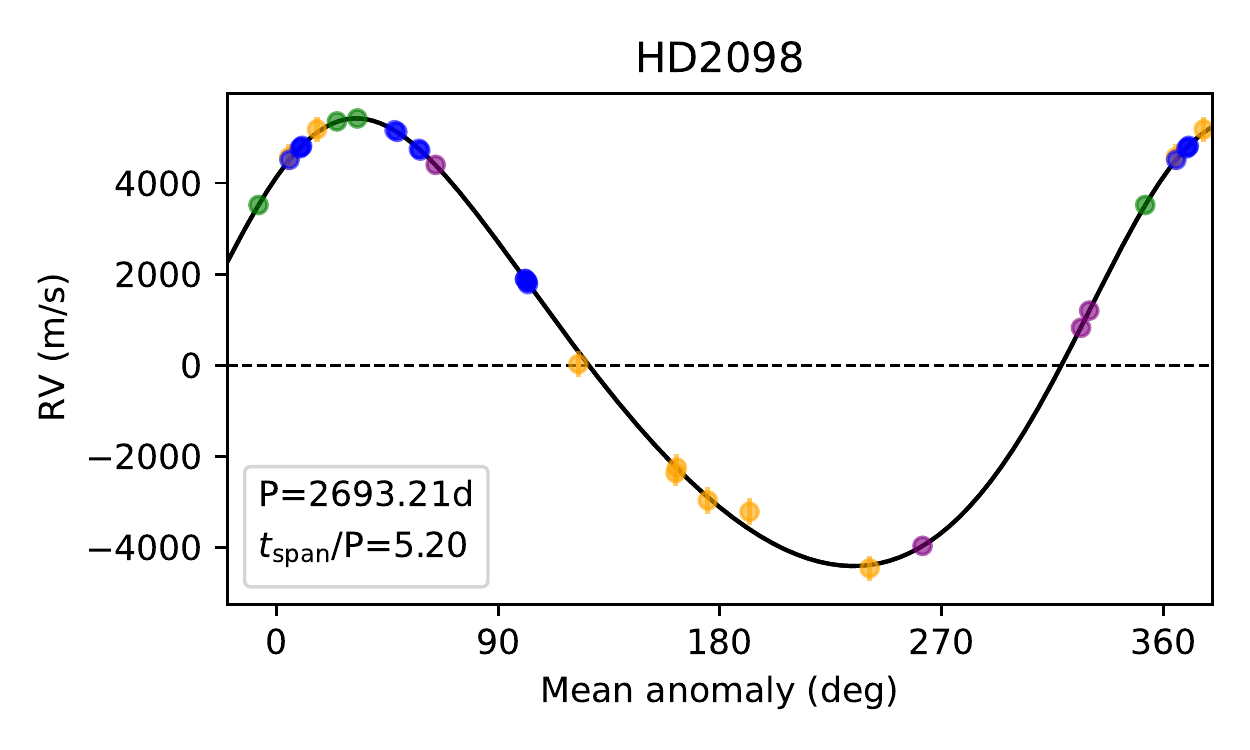}&
		\includegraphics[width=0.22\linewidth]{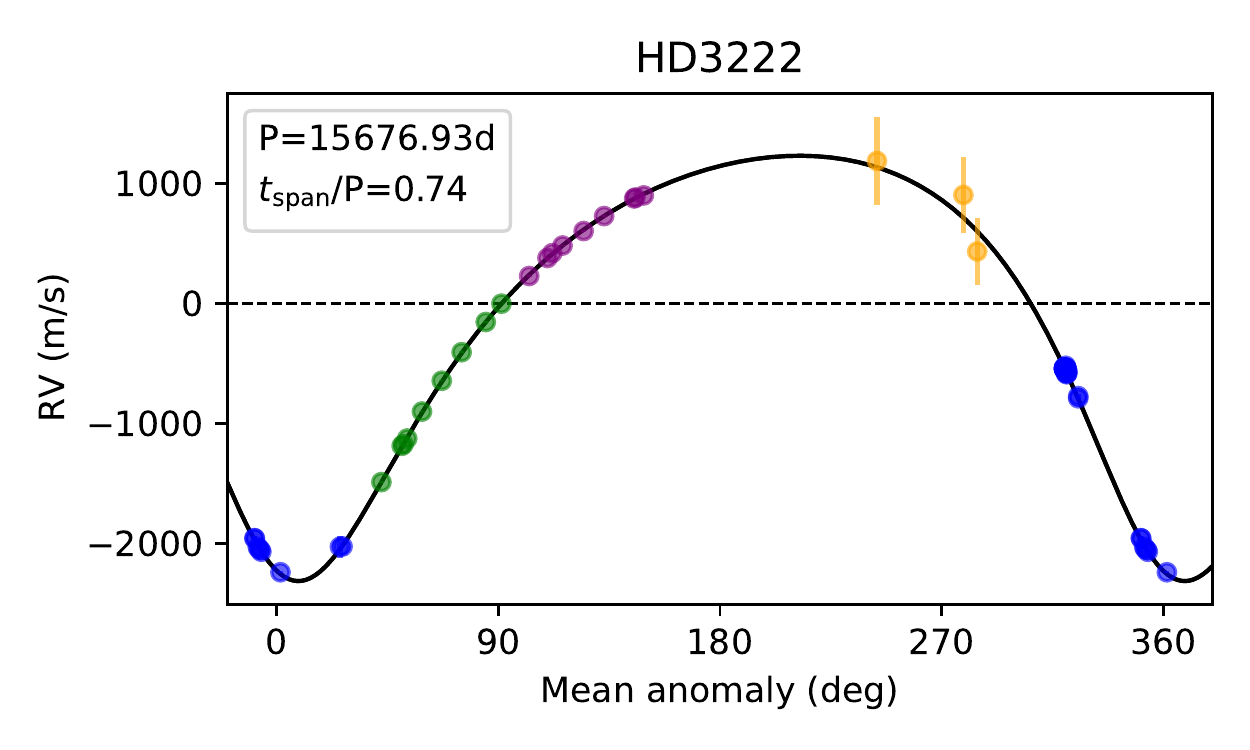}&
		\includegraphics[width=0.22\linewidth]{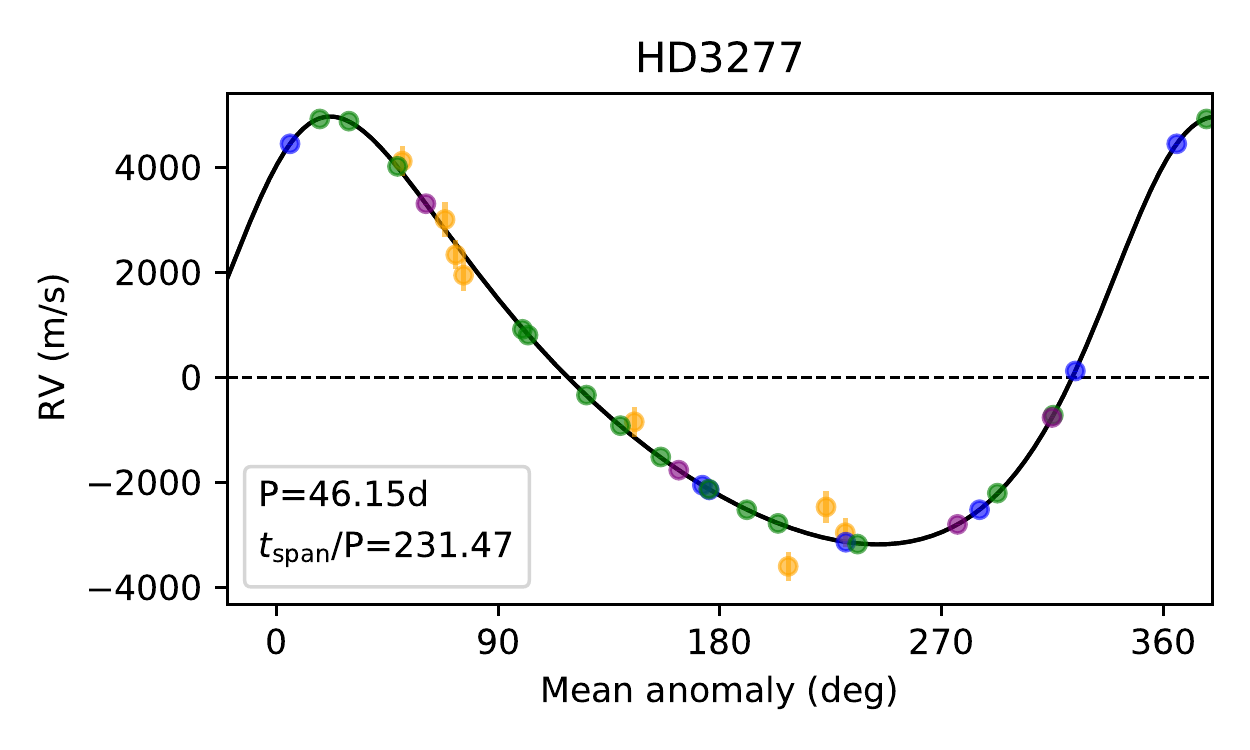}&
		\includegraphics[width=0.22\linewidth]{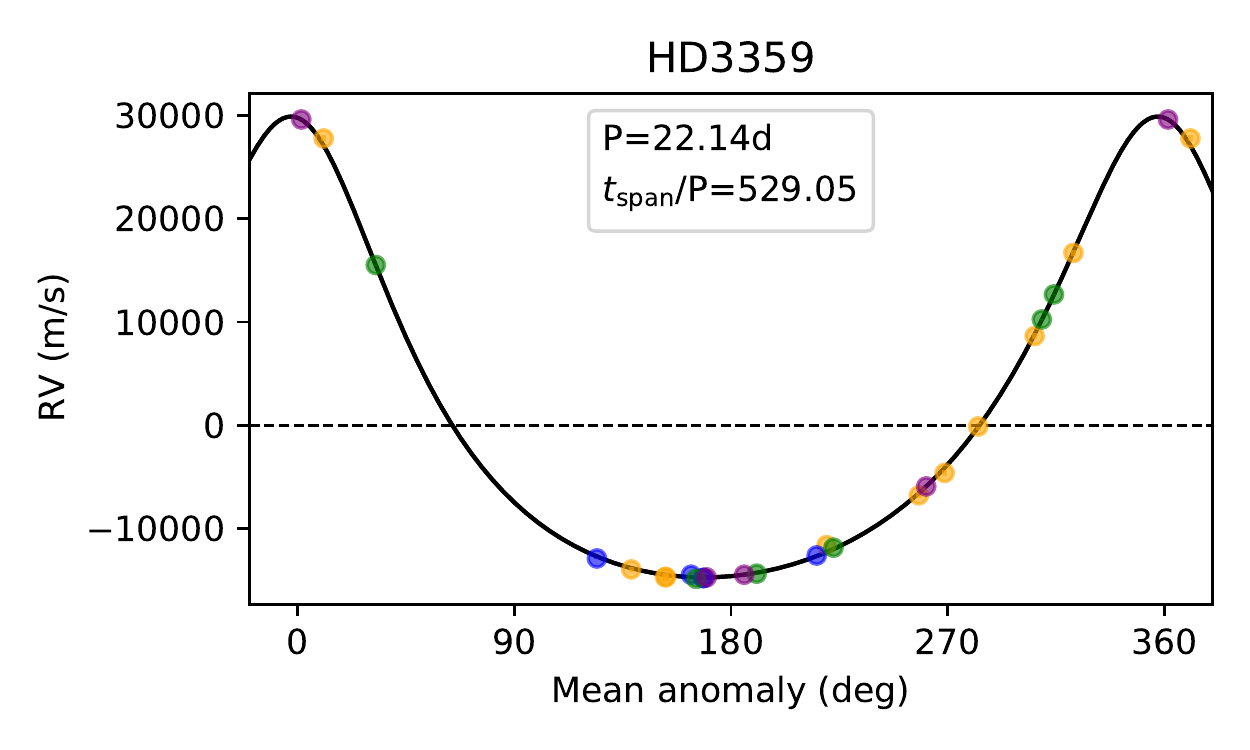}\\

		\includegraphics[width=0.22\linewidth]{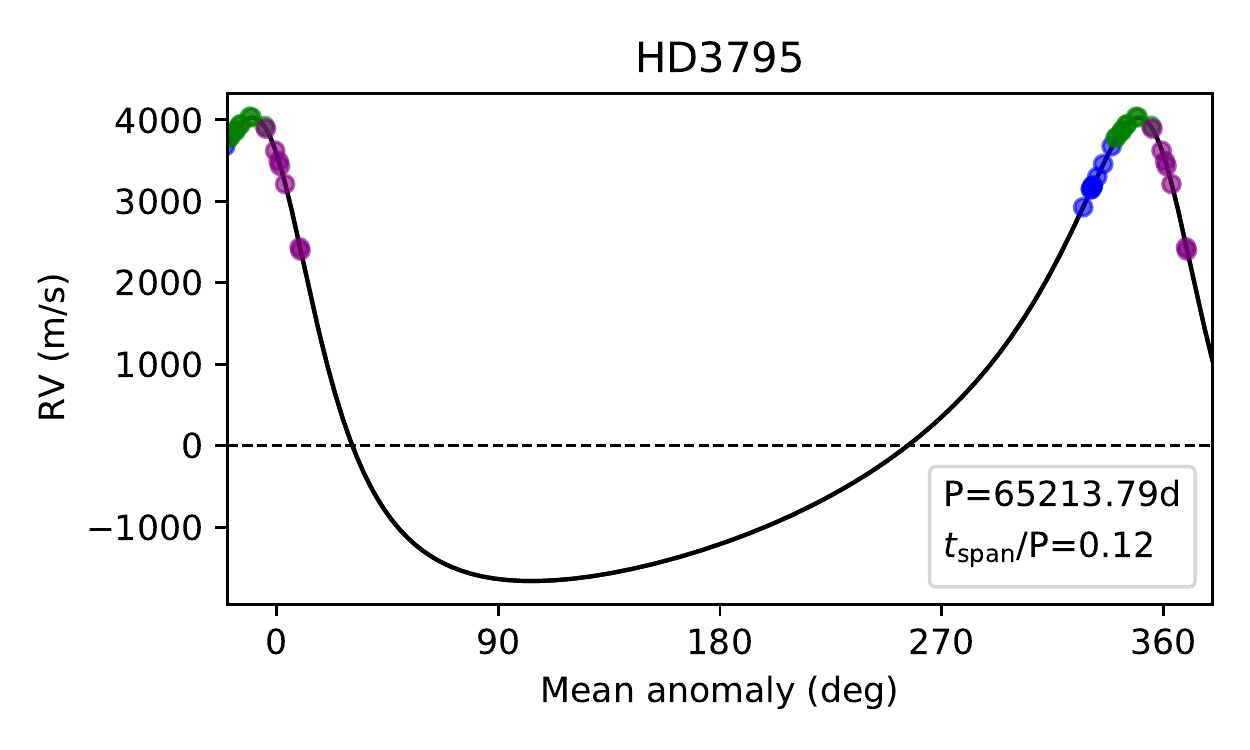}&
		\includegraphics[width=0.22\linewidth]{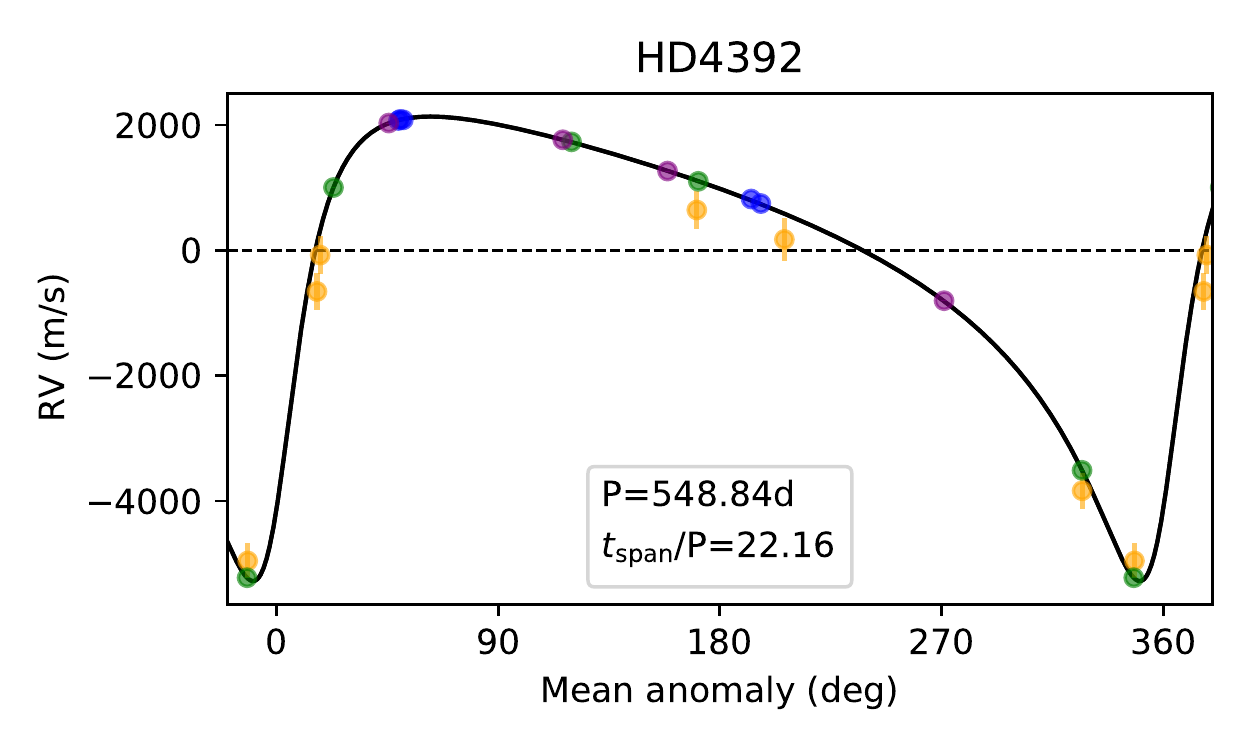}&
		\includegraphics[width=0.22\linewidth]{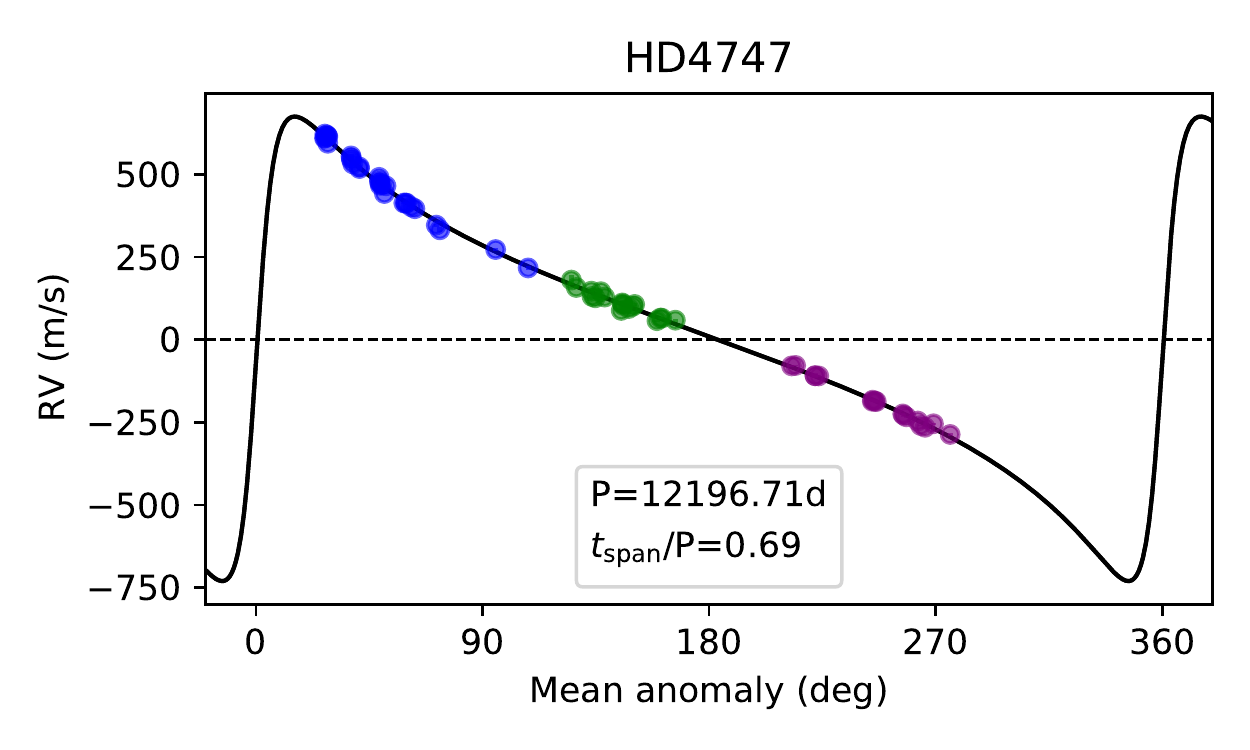}&
		\includegraphics[width=0.22\linewidth]{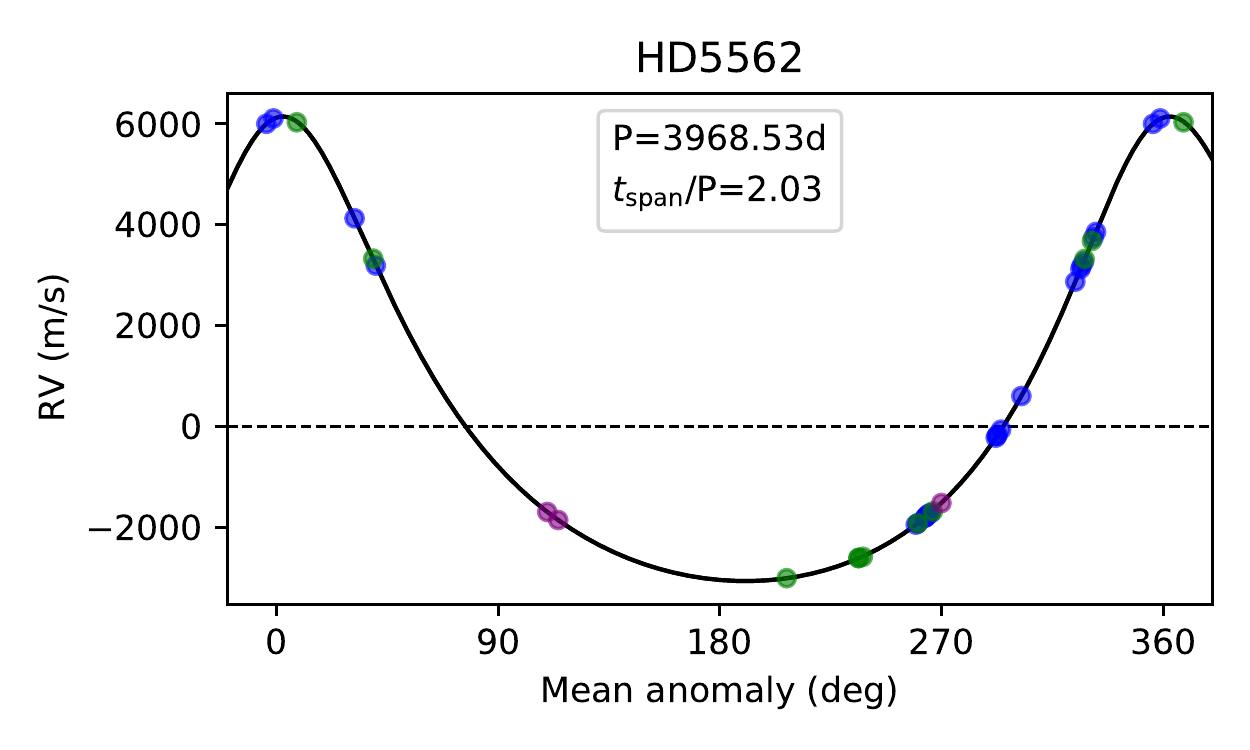}\\

		\includegraphics[width=0.22\linewidth]{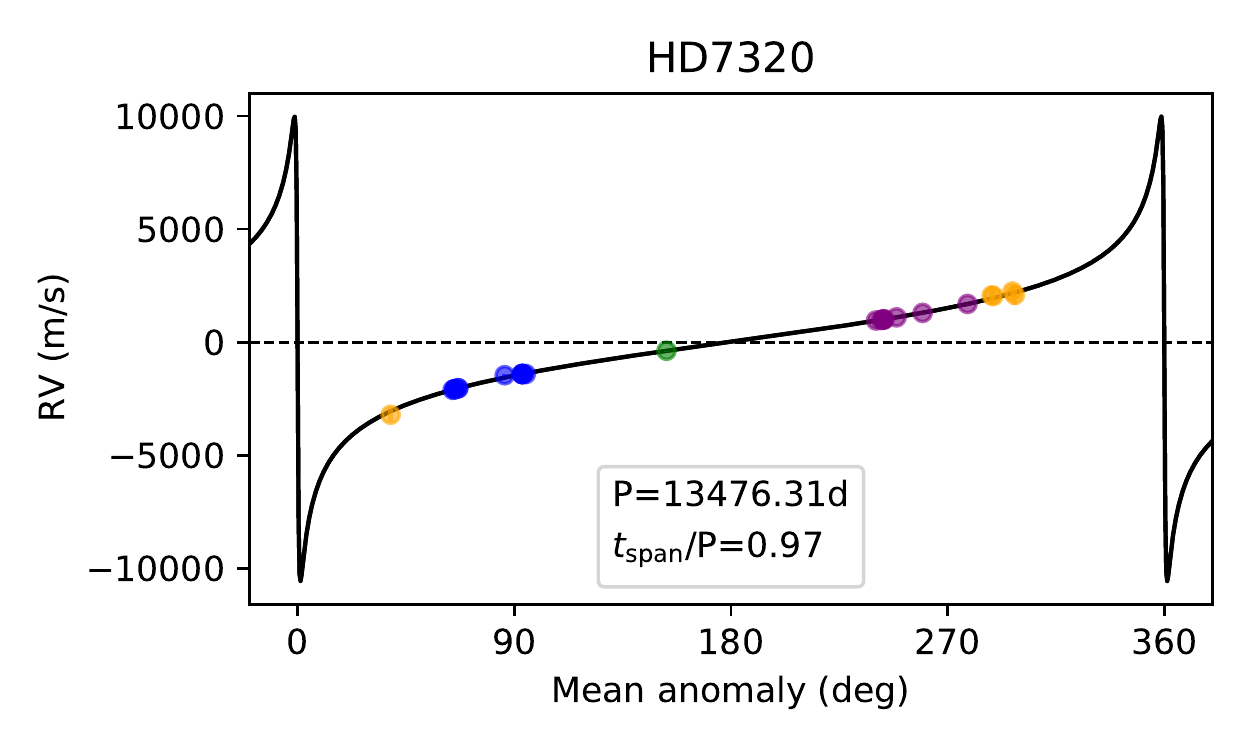}&
		\includegraphics[width=0.22\linewidth]{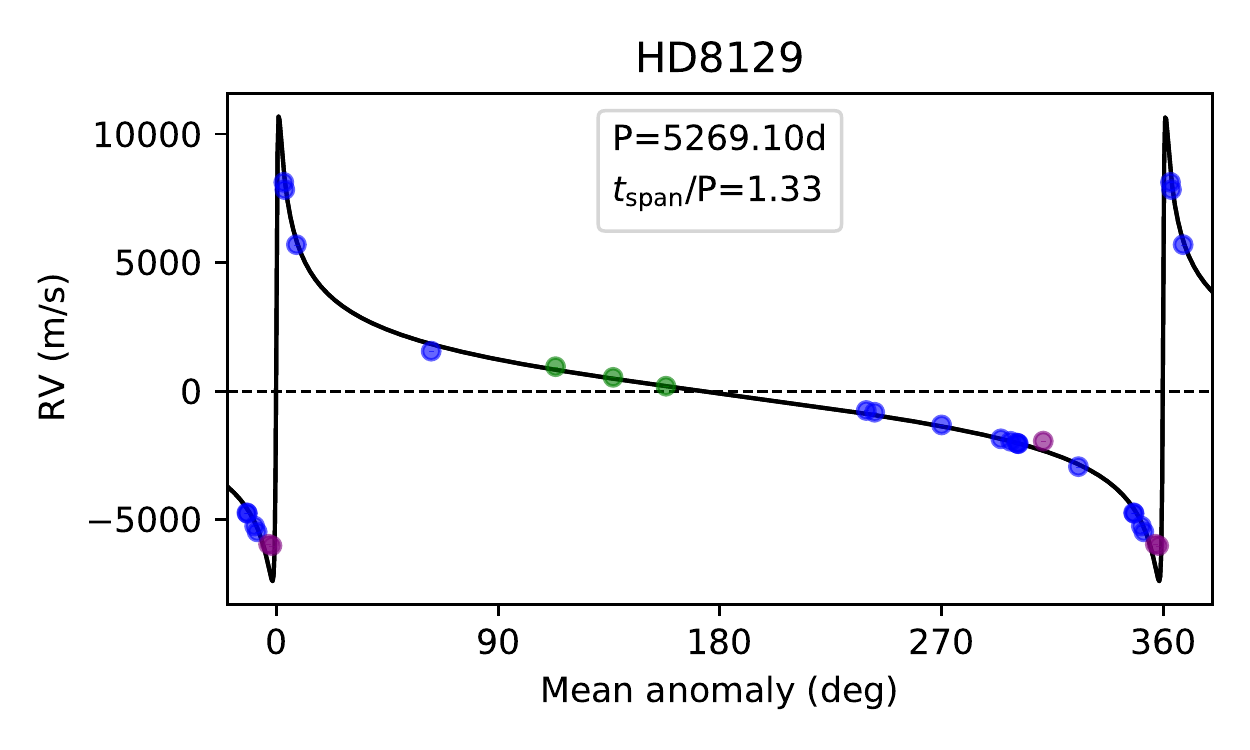}&
		\includegraphics[width=0.22\linewidth]{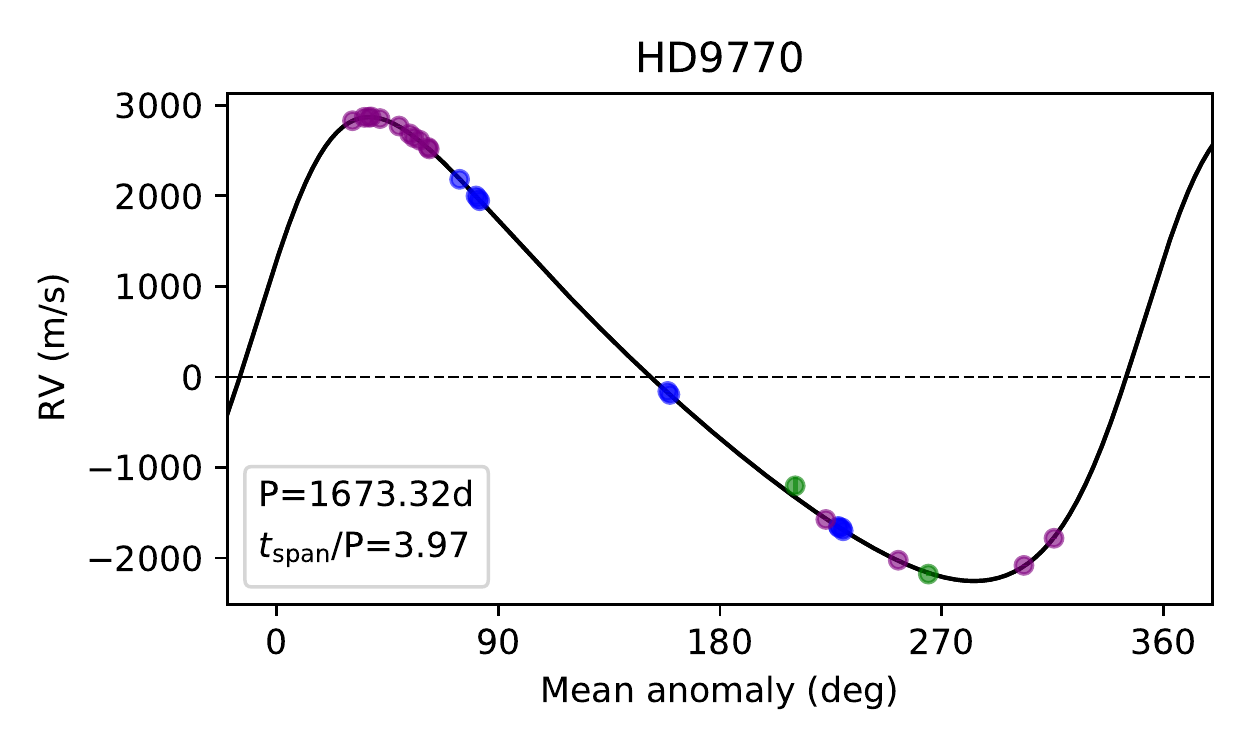}&
		\includegraphics[width=0.22\linewidth]{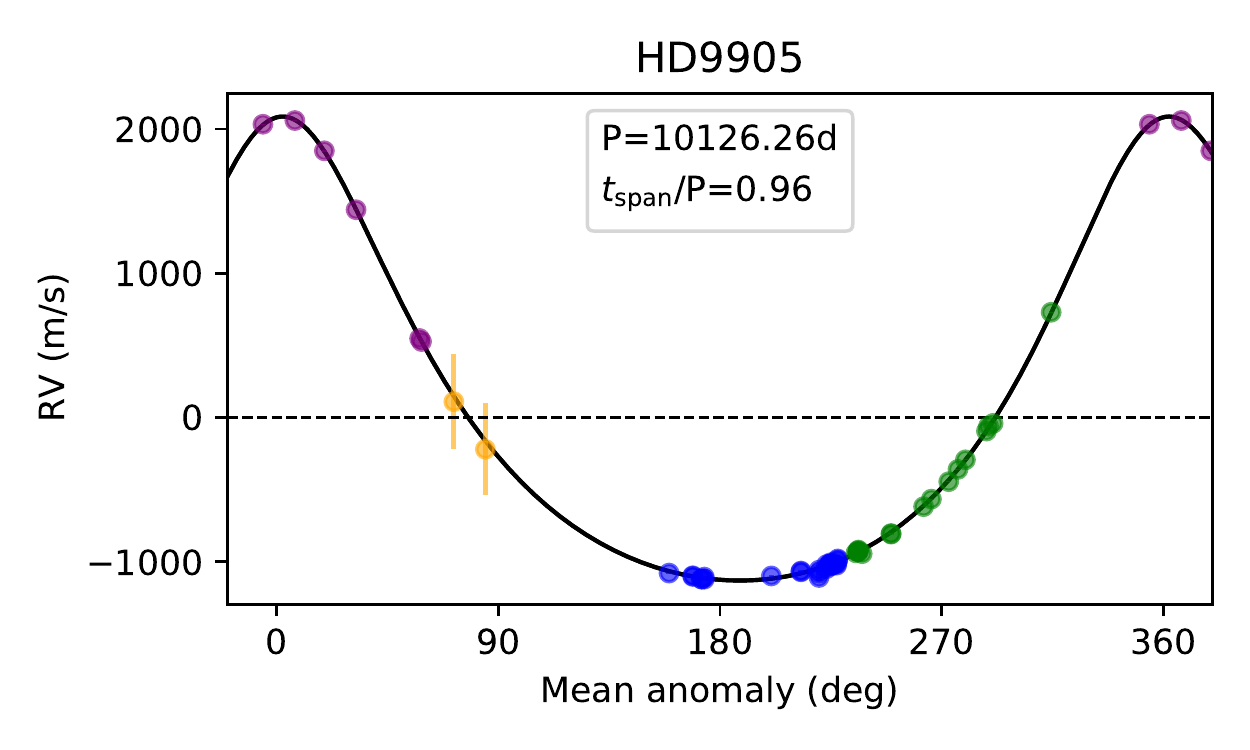}\\

		\includegraphics[width=0.22\linewidth]{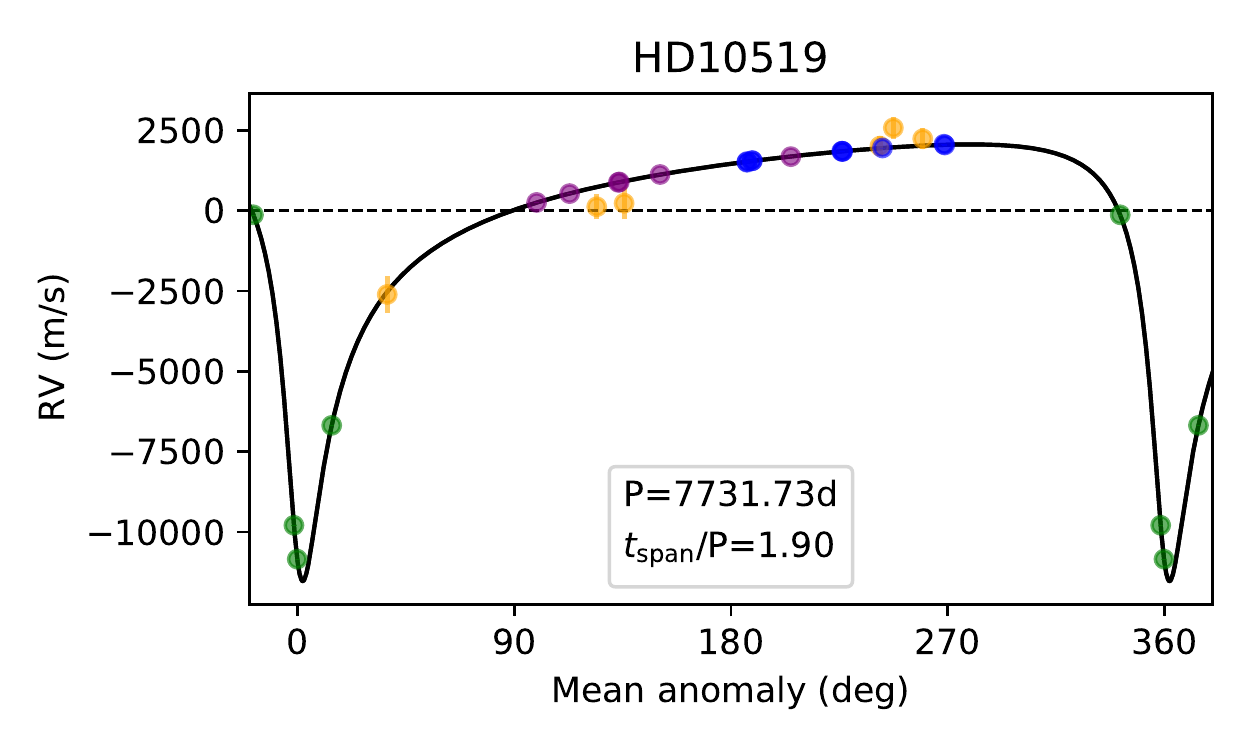}&
		\includegraphics[width=0.22\linewidth]{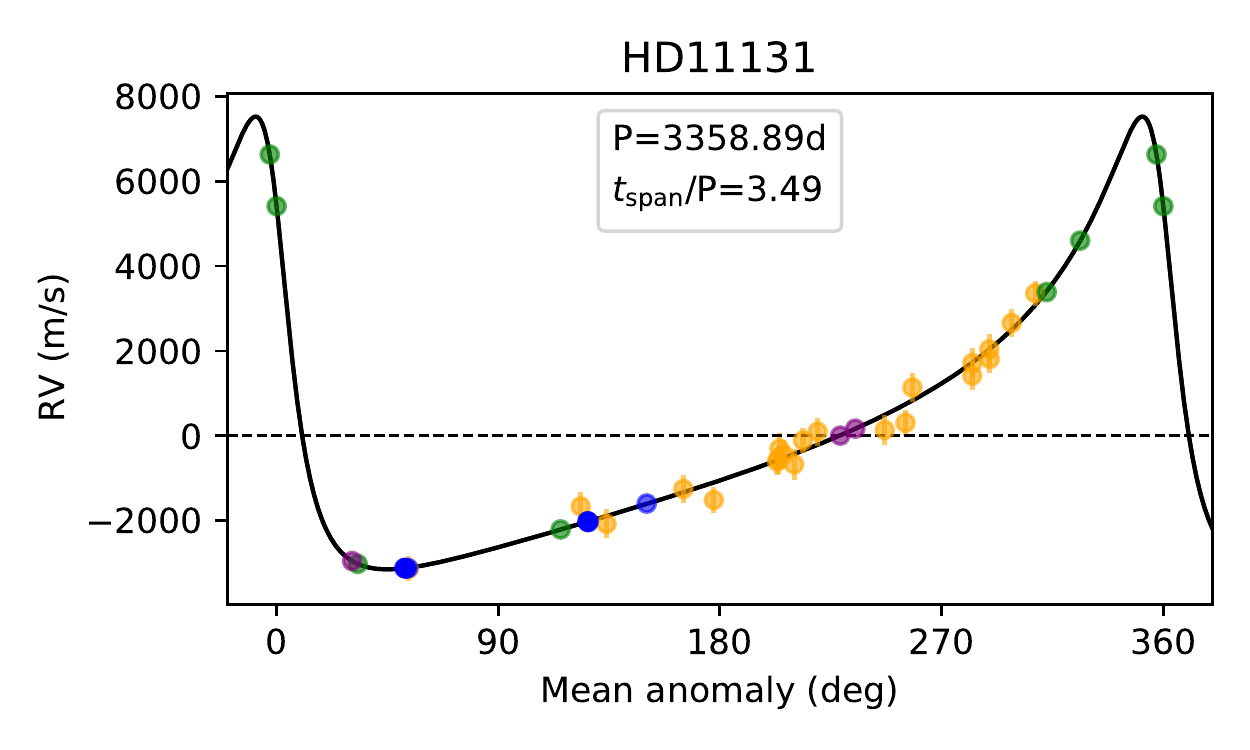}&
		\includegraphics[width=0.22\linewidth]{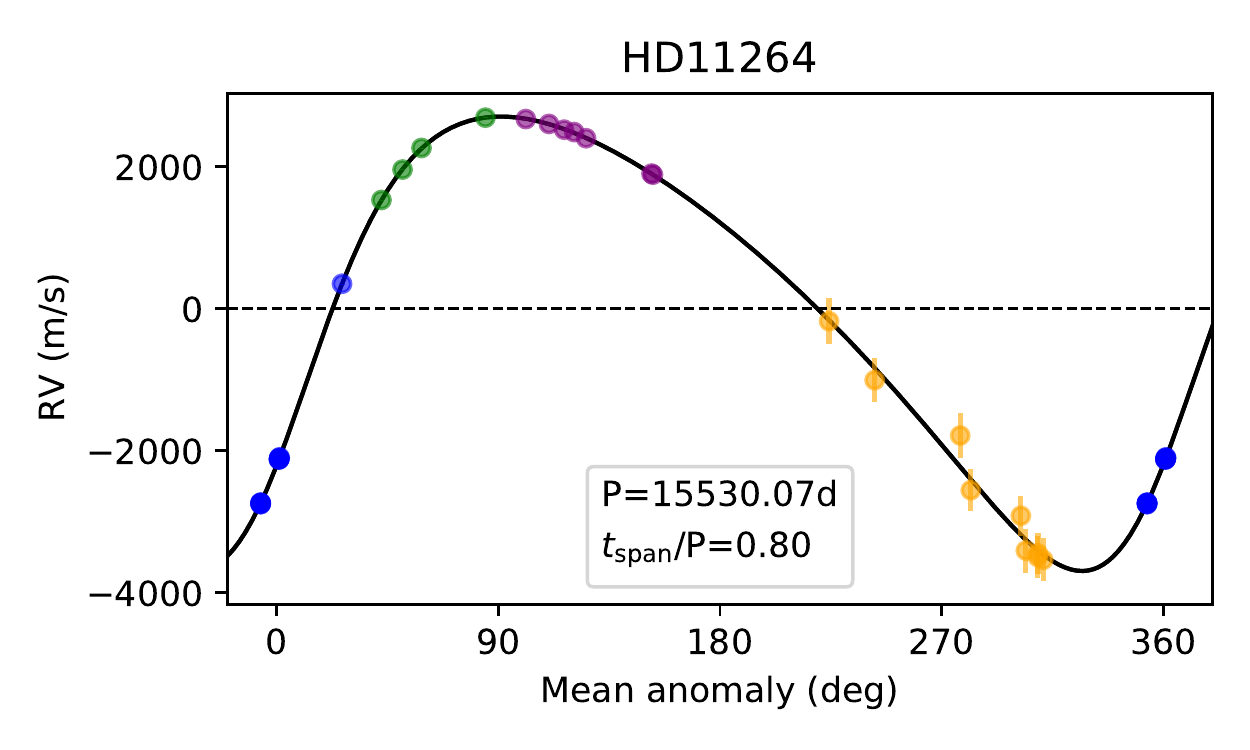}&
		\includegraphics[width=0.22\linewidth]{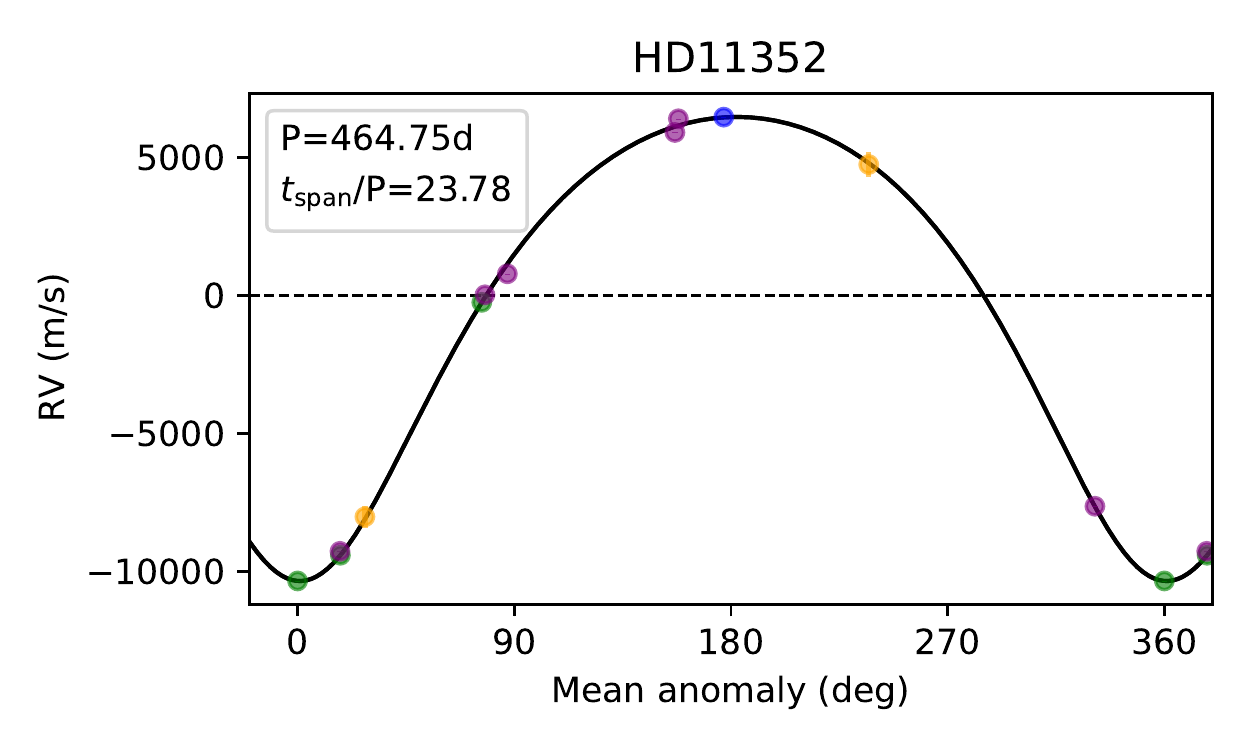}\\

		\includegraphics[width=0.22\linewidth]{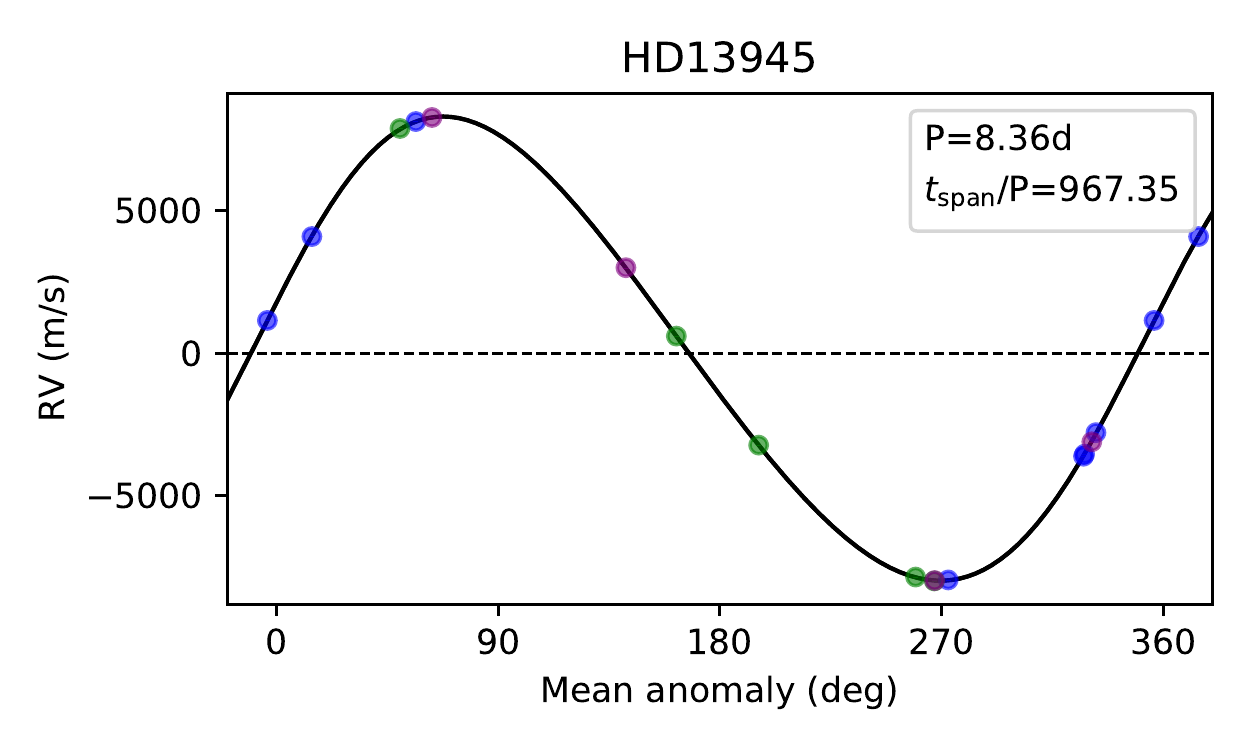}&
		\includegraphics[width=0.22\linewidth]{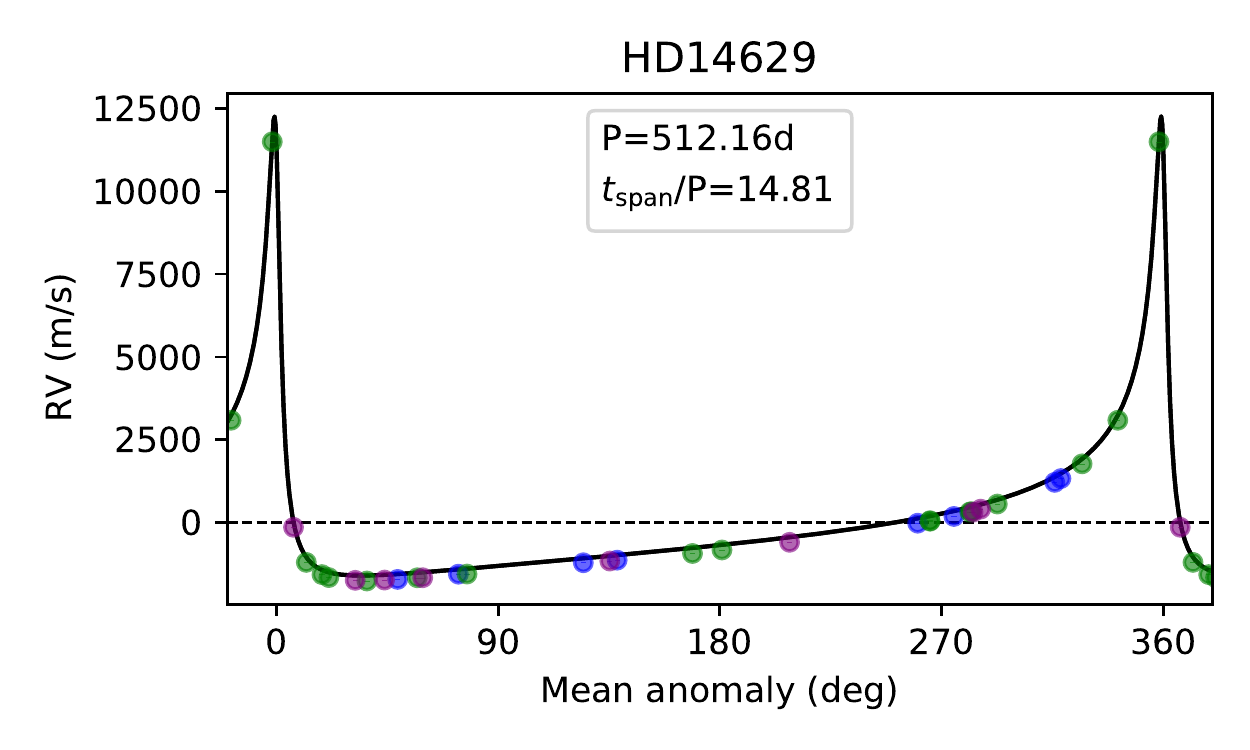}&
		\includegraphics[width=0.22\linewidth]{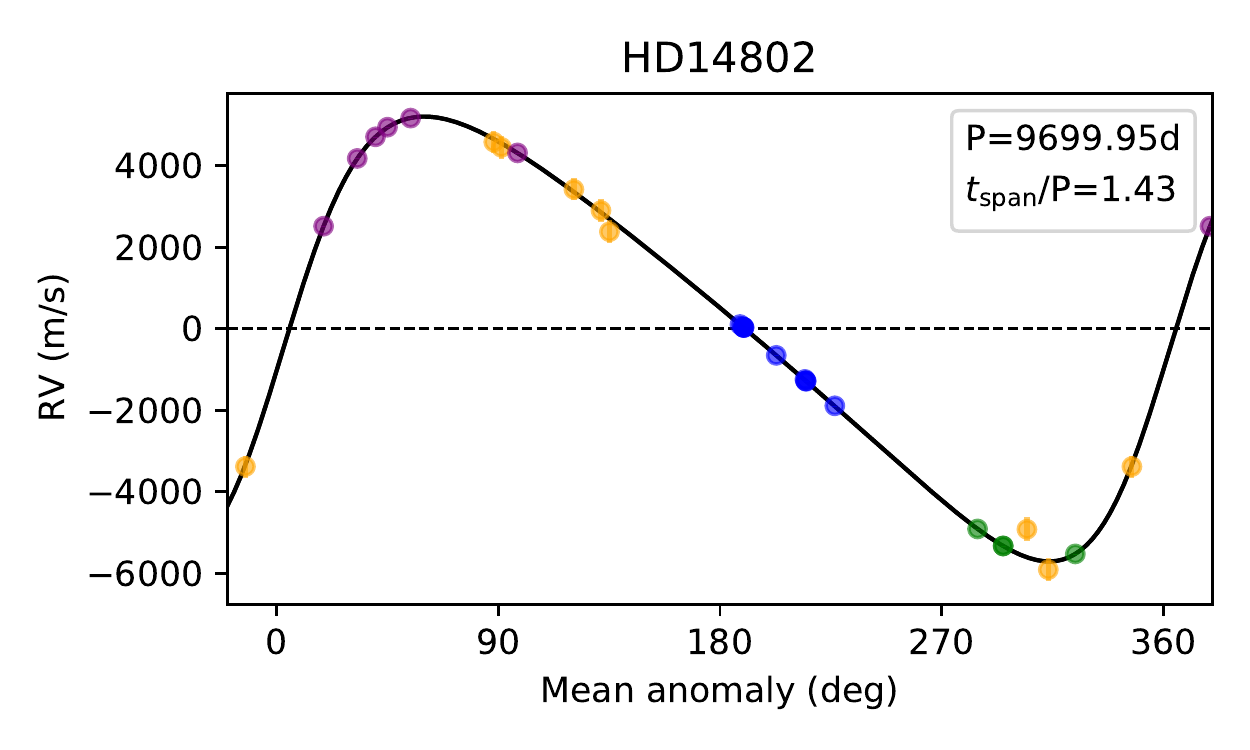}&
		\includegraphics[width=0.22\linewidth]{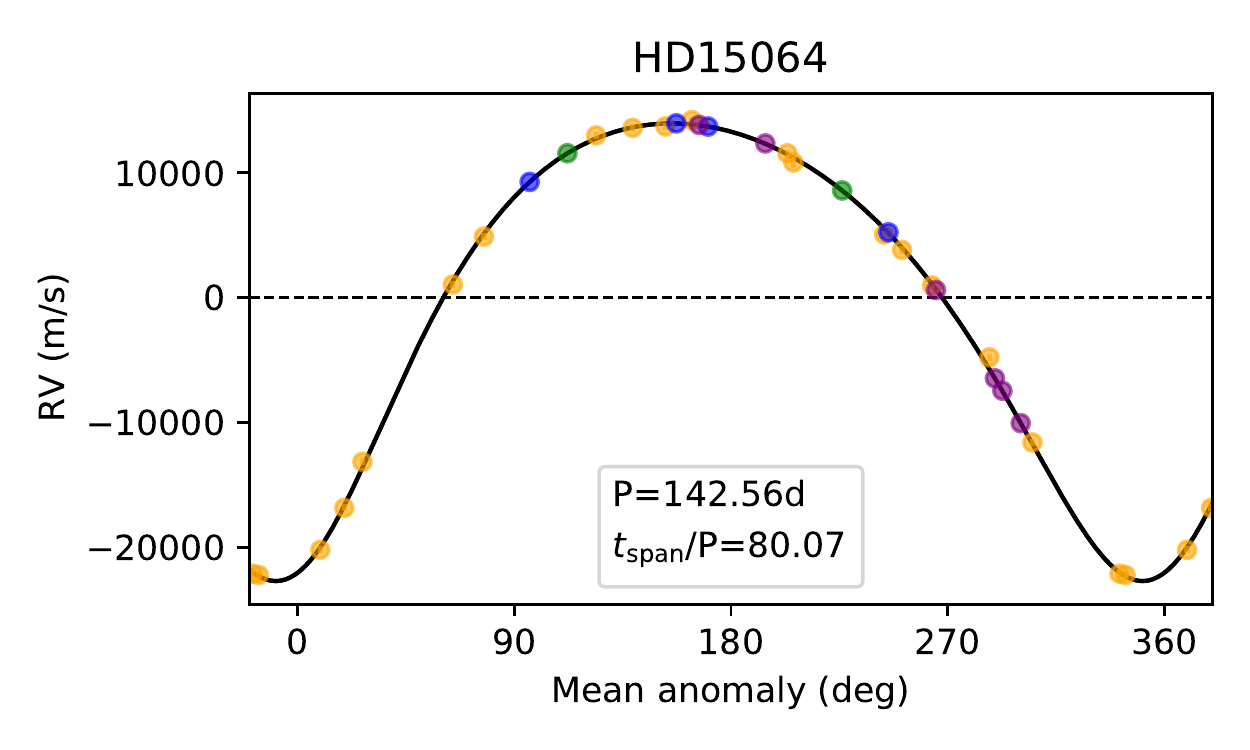}\\

		\includegraphics[width=0.22\linewidth]{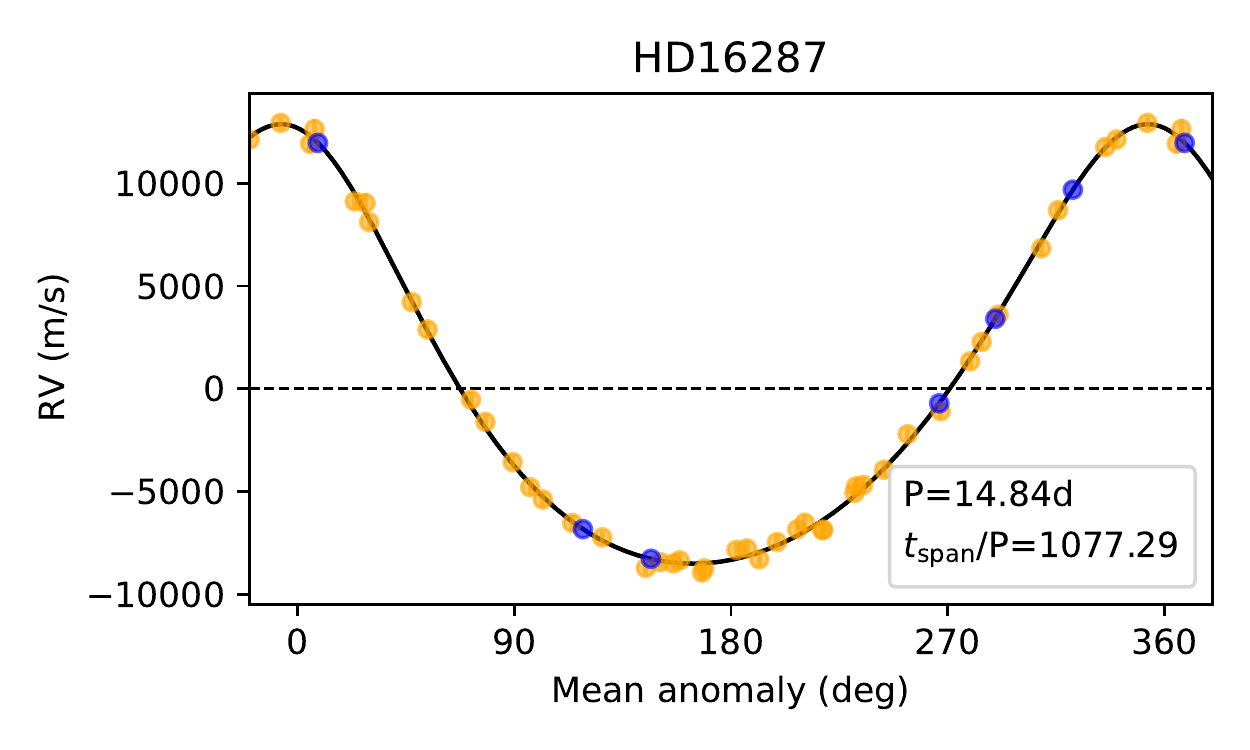}&
		\includegraphics[width=0.22\linewidth]{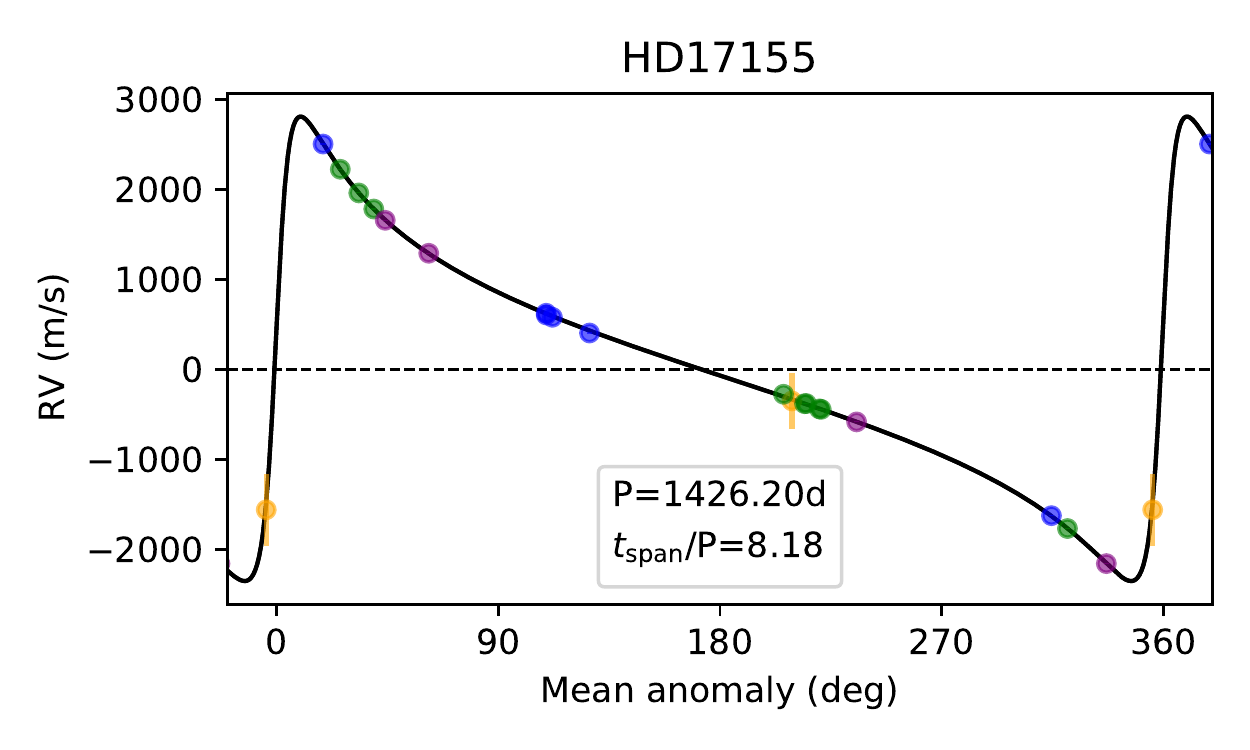}&
		\includegraphics[width=0.22\linewidth]{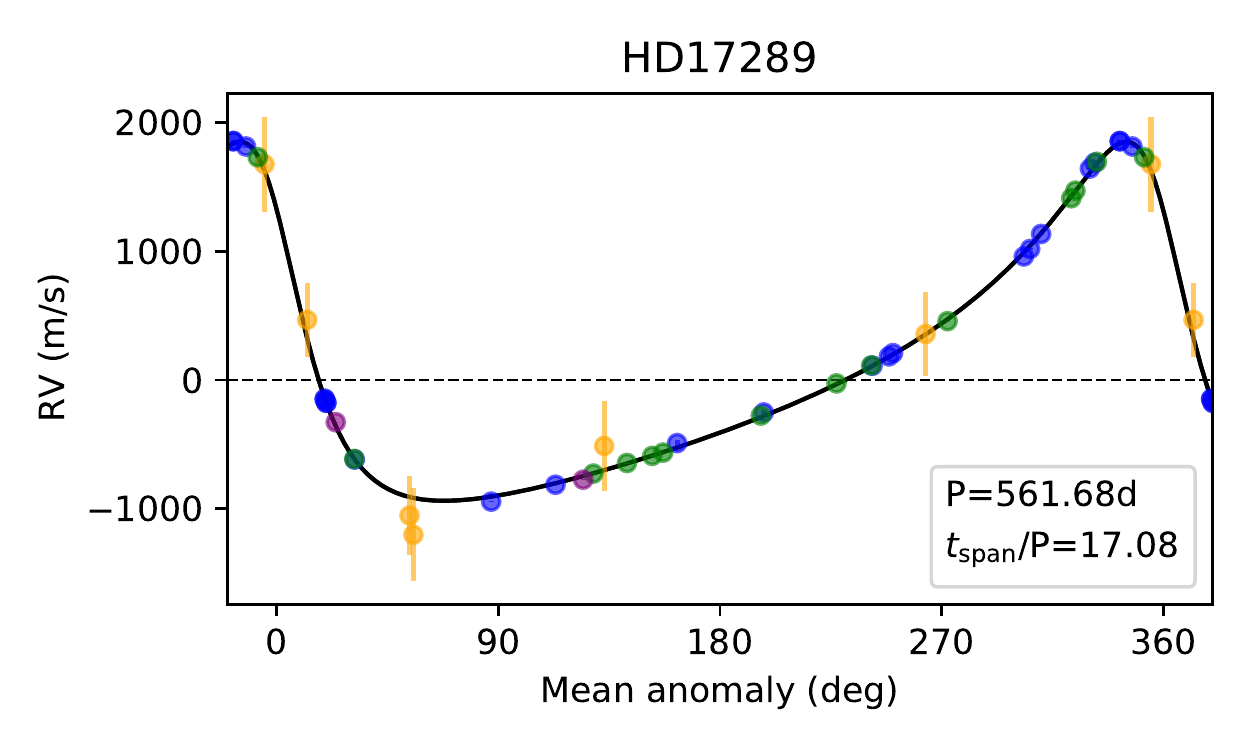}&
		\includegraphics[width=0.22\linewidth]{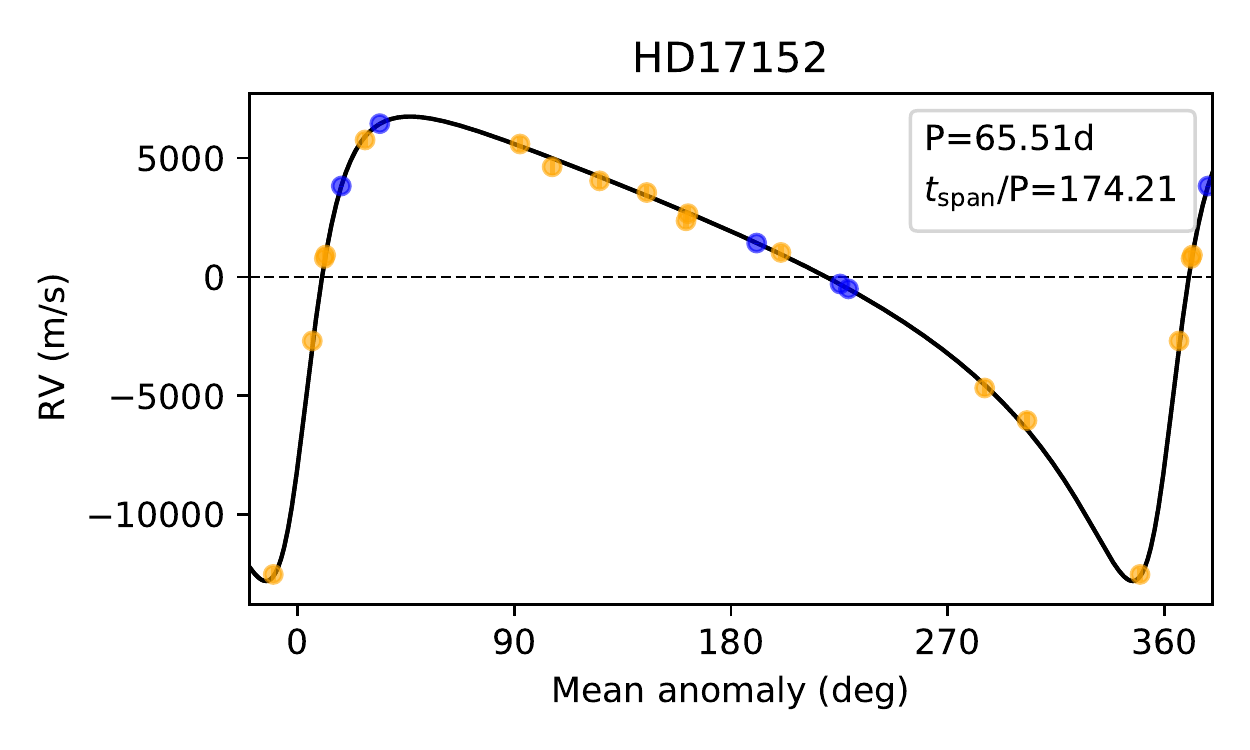}\\

		\includegraphics[width=0.22\linewidth]{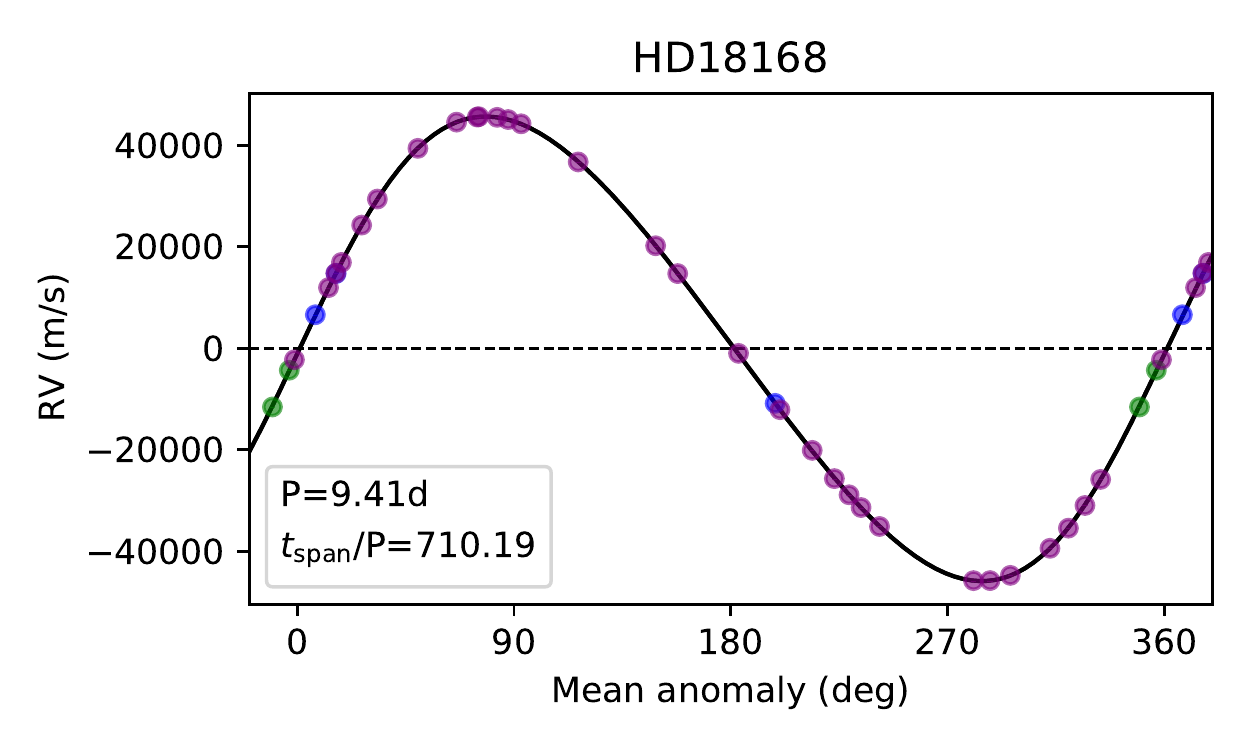}&
		\includegraphics[width=0.22\linewidth]{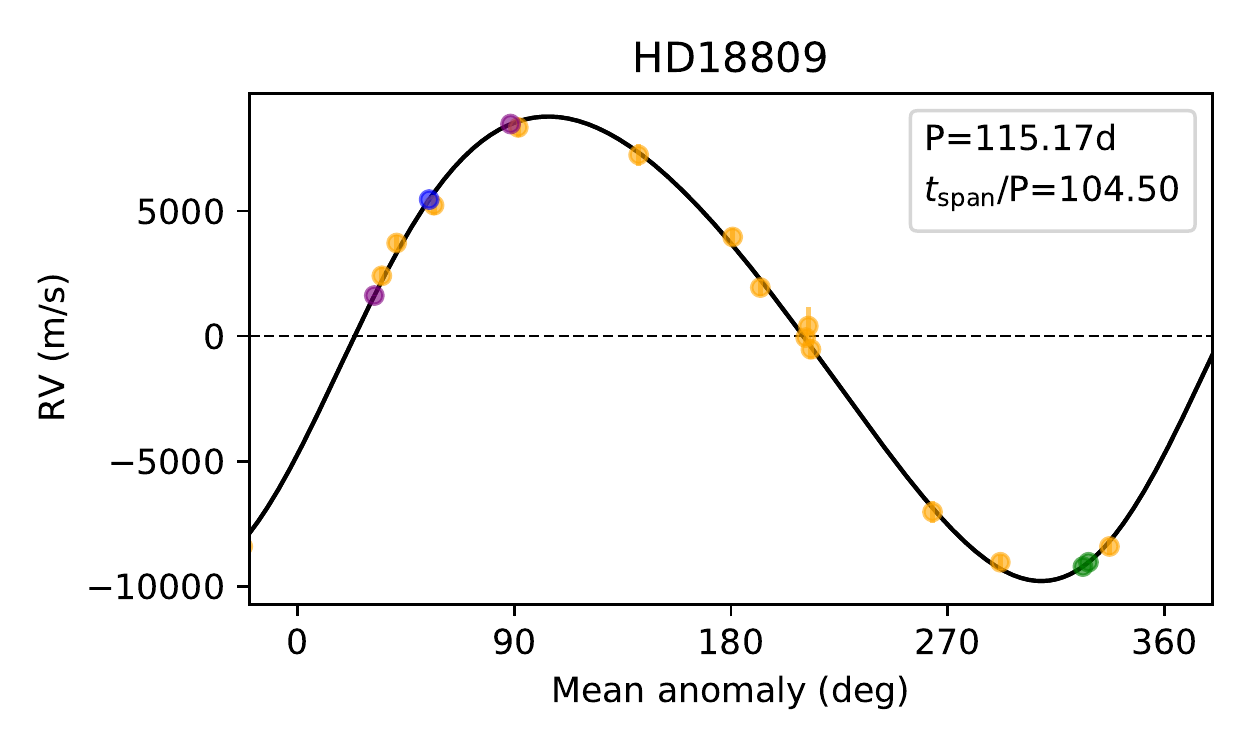}&
		\includegraphics[width=0.22\linewidth]{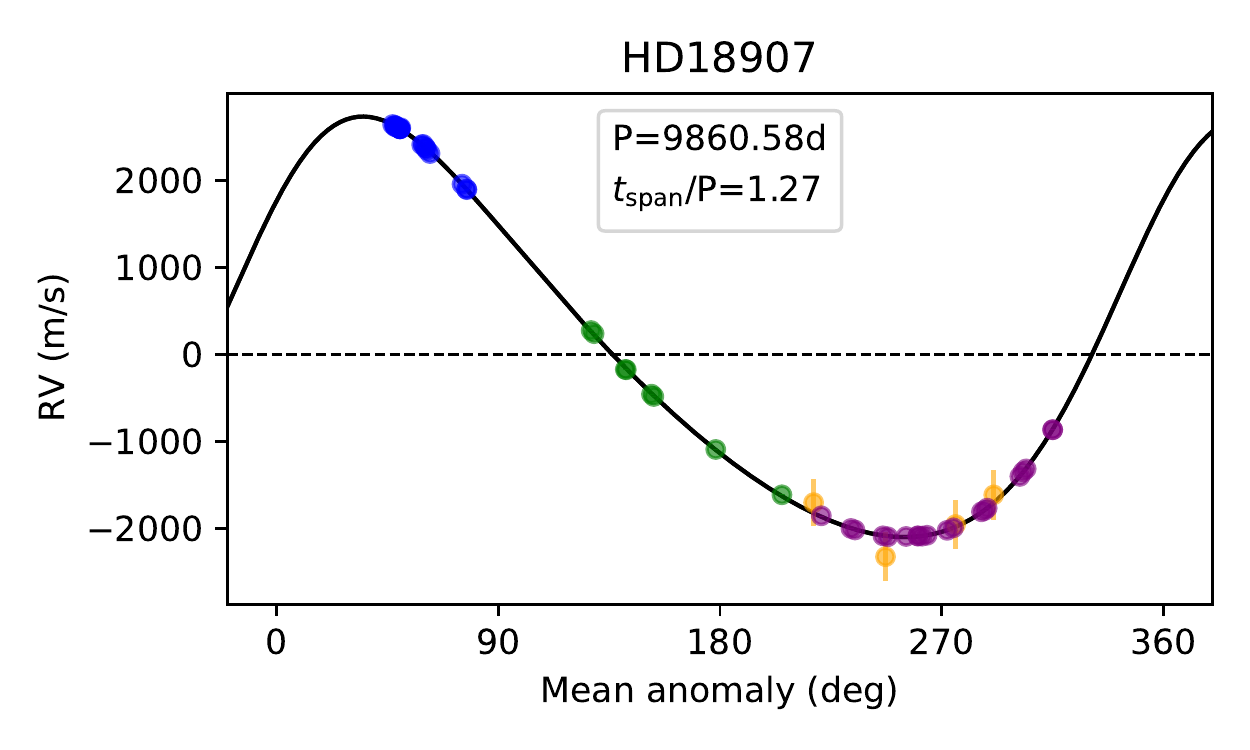}&
		\includegraphics[width=0.22\linewidth]{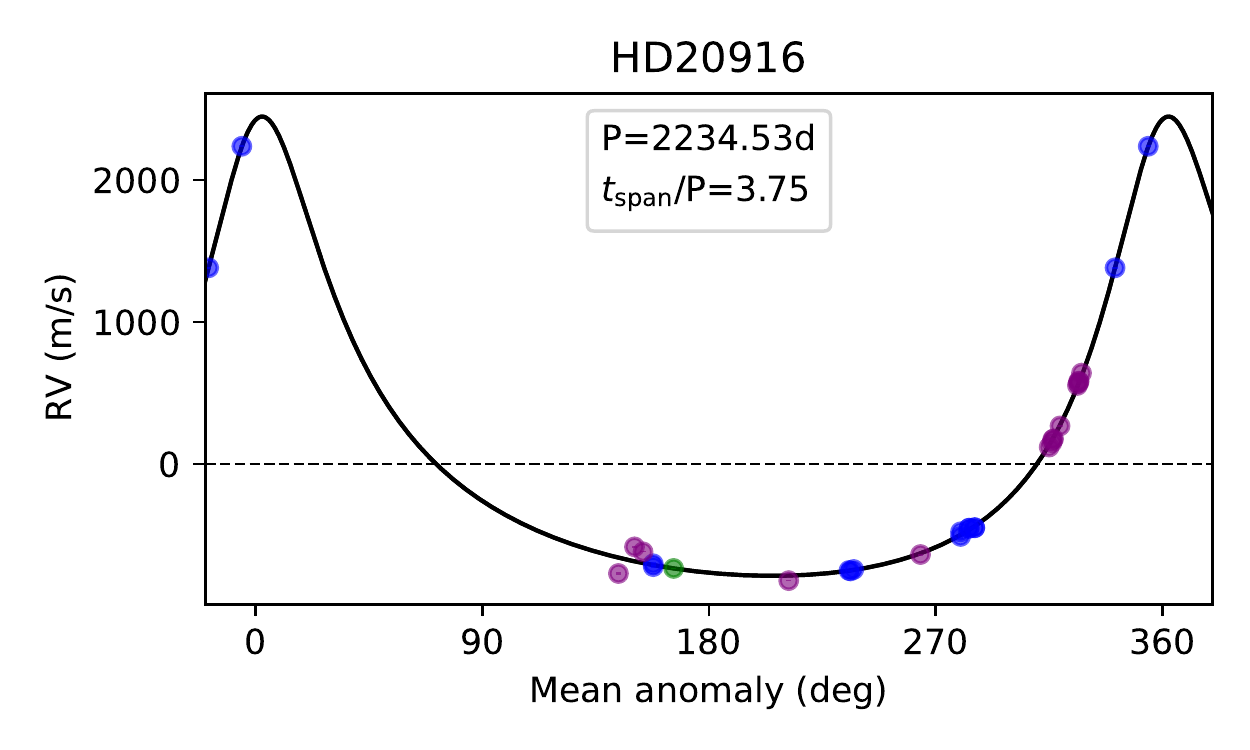}\\

		\includegraphics[width=0.22\linewidth]{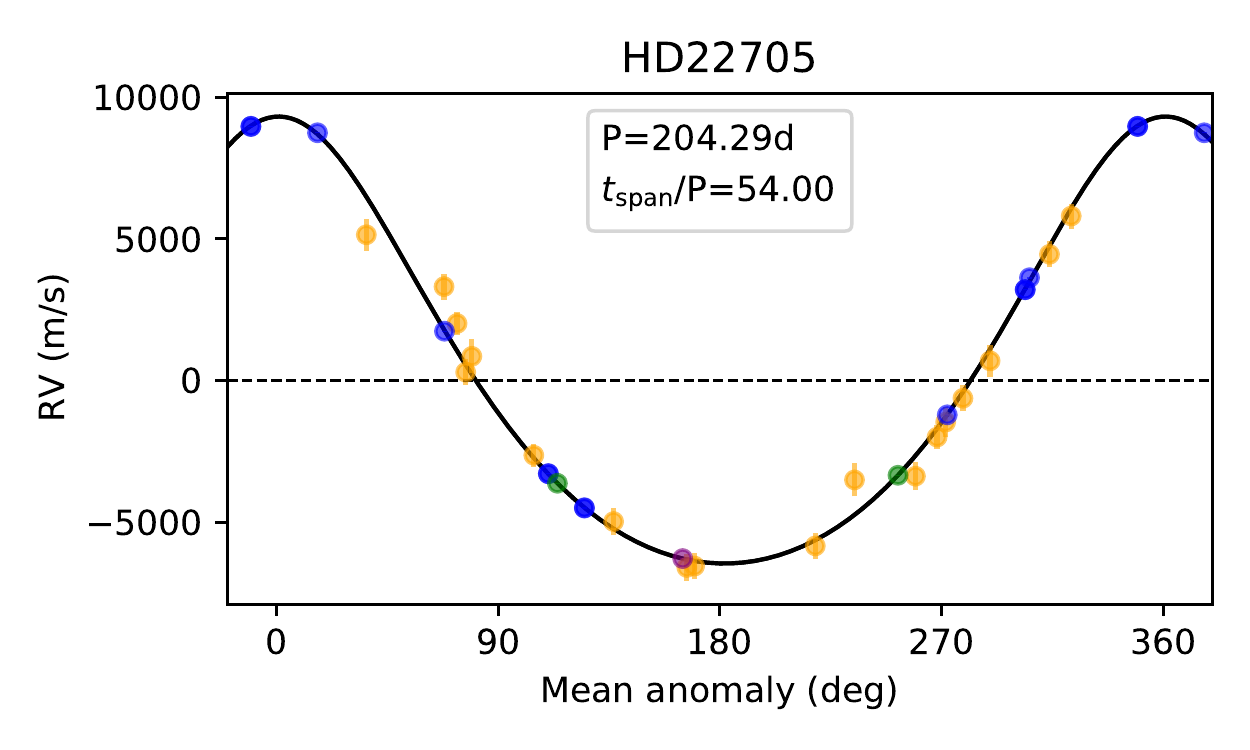}&
		\includegraphics[width=0.22\linewidth]{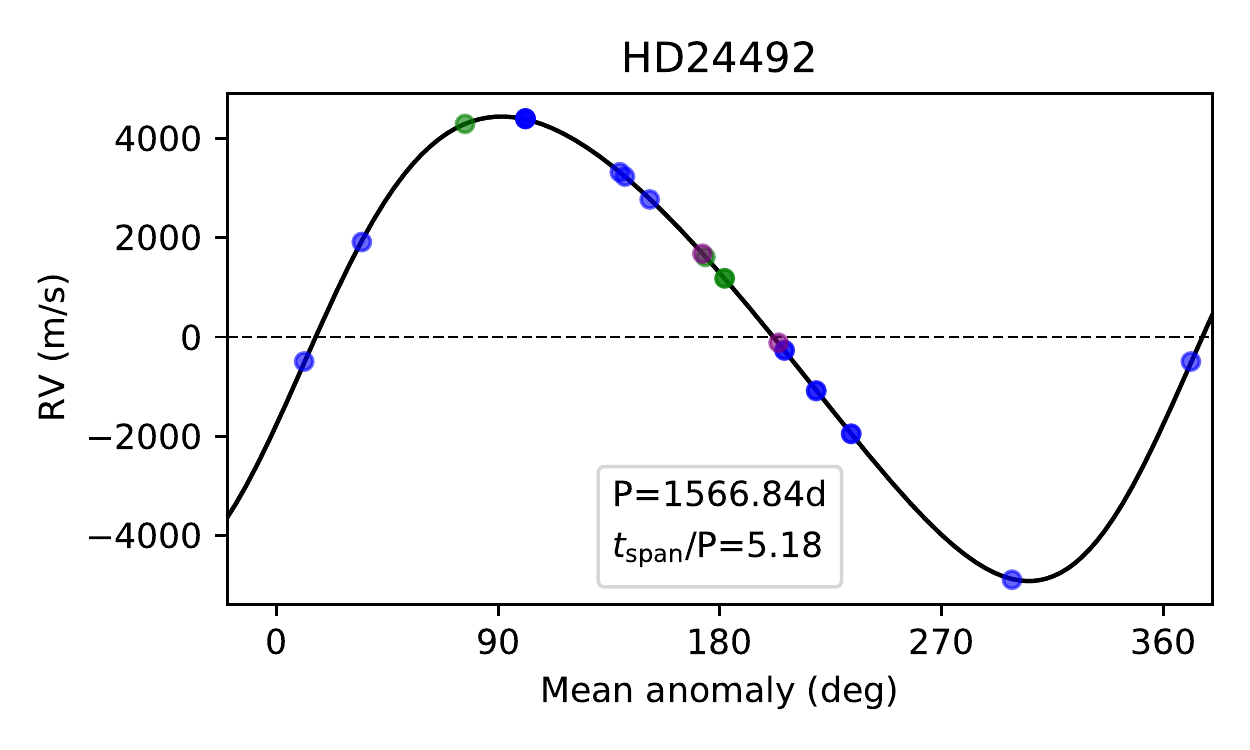}&
		\includegraphics[width=0.22\linewidth]{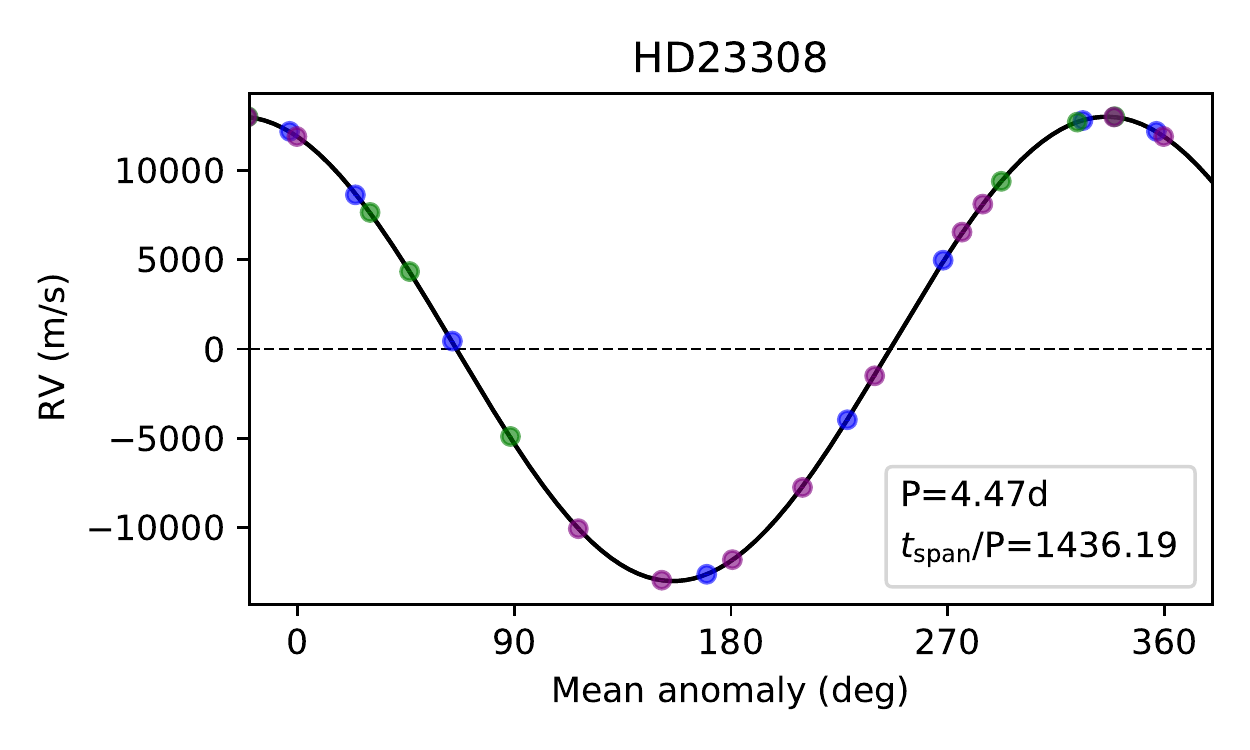}&
		\includegraphics[width=0.22\linewidth]{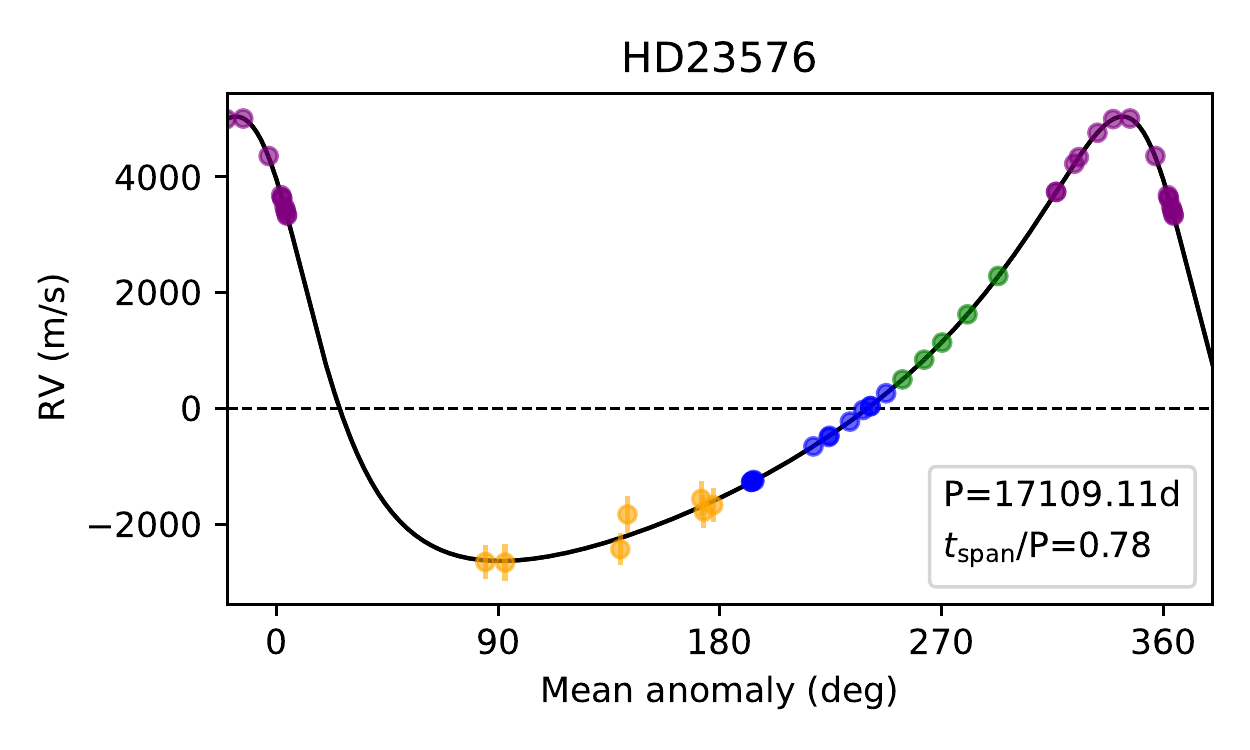}\\

		\includegraphics[width=0.22\linewidth]{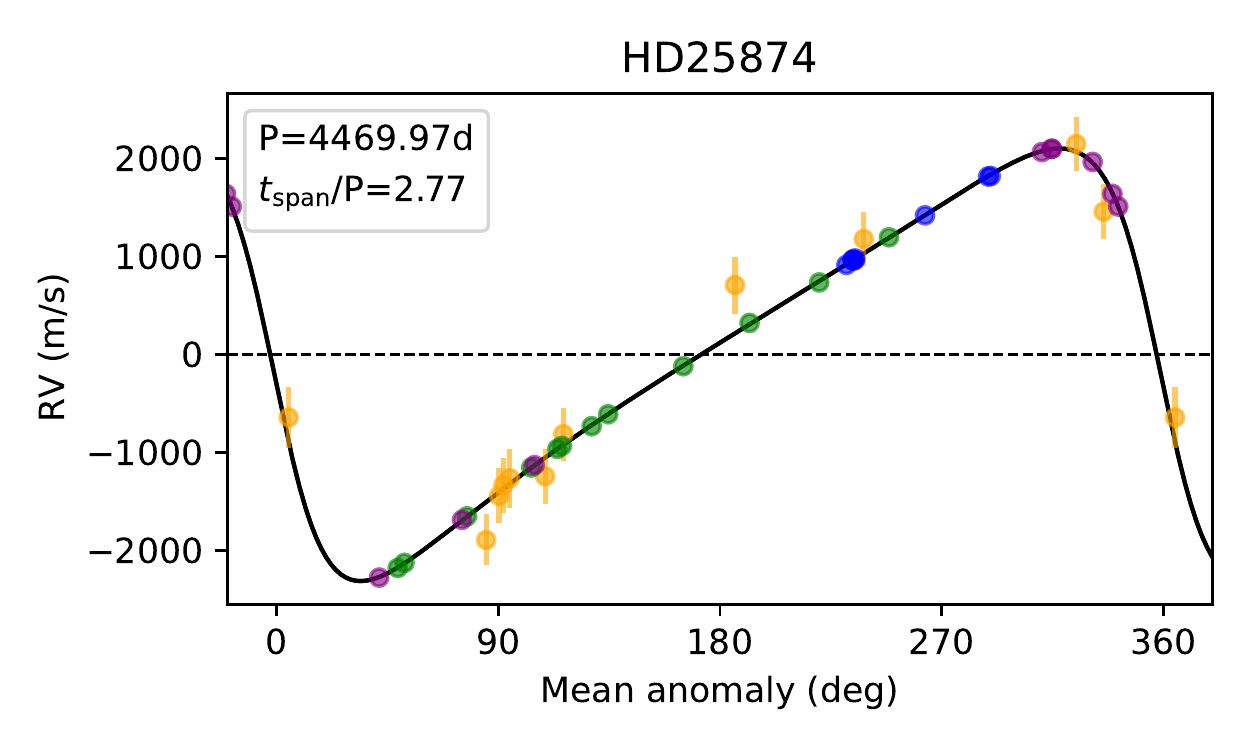}&
		\includegraphics[width=0.22\linewidth]{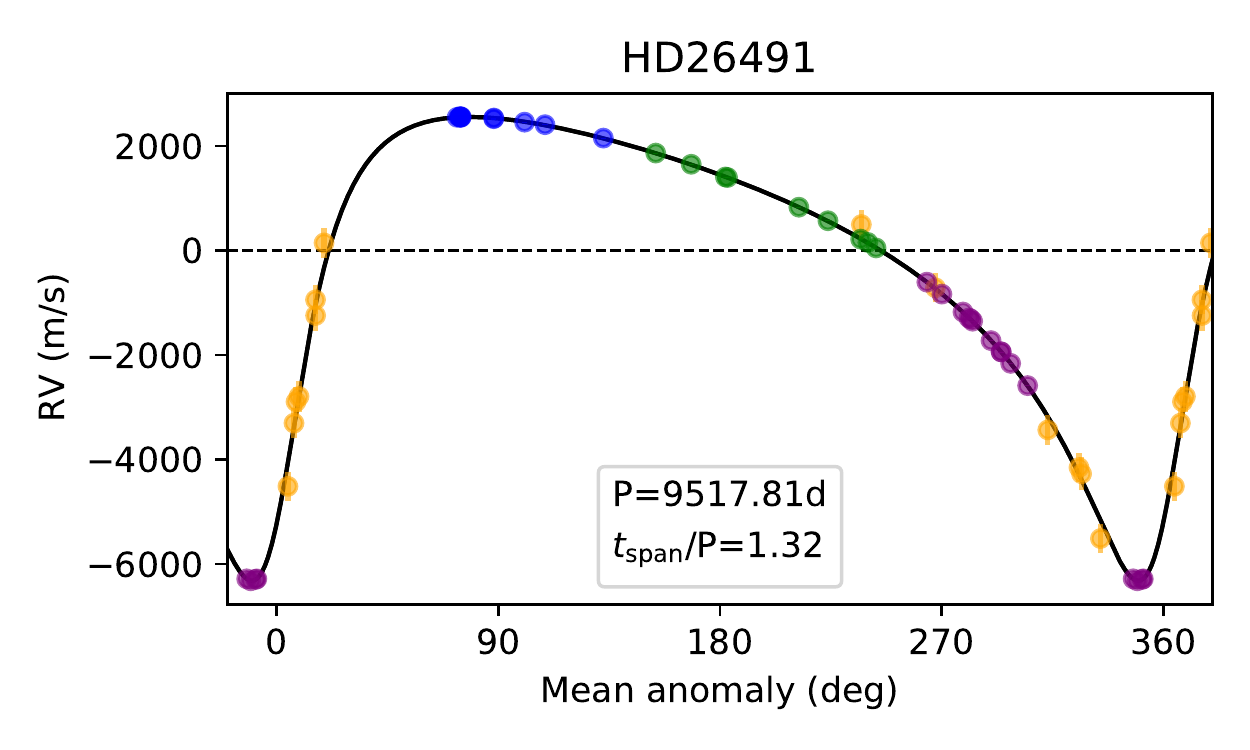}&
		\includegraphics[width=0.22\linewidth]{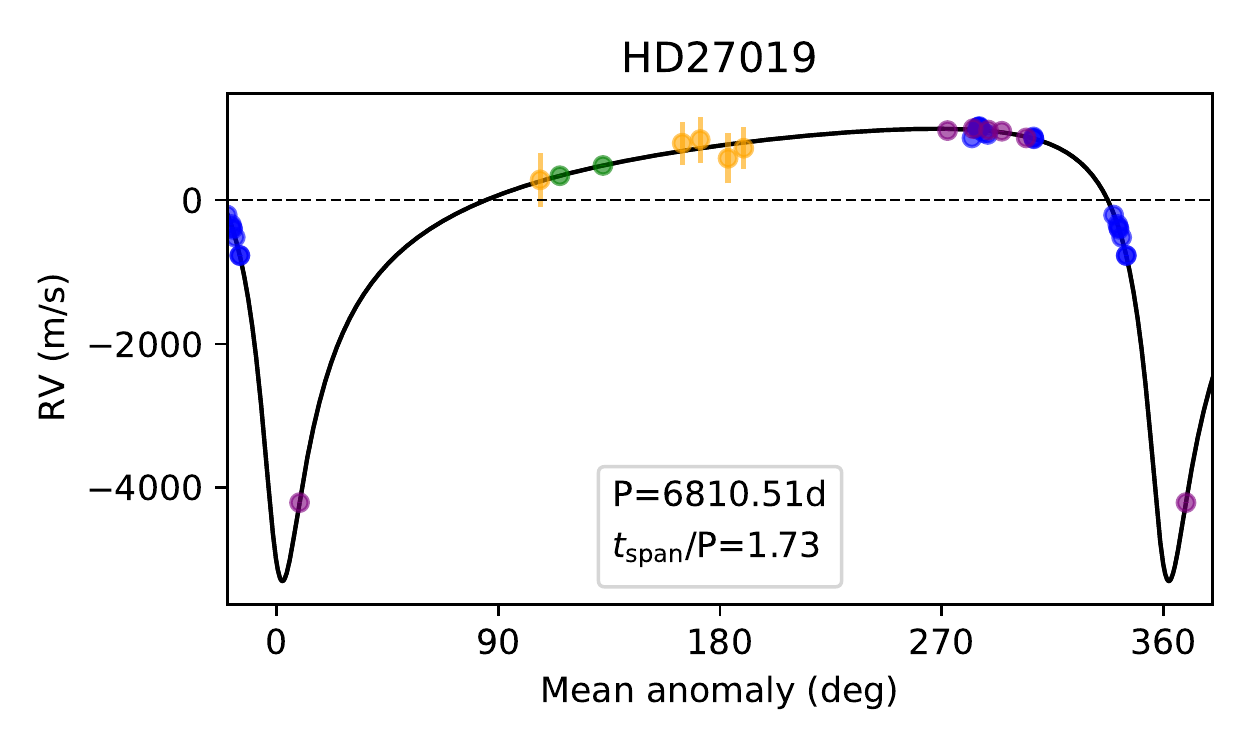}&
		\includegraphics[width=0.22\linewidth]{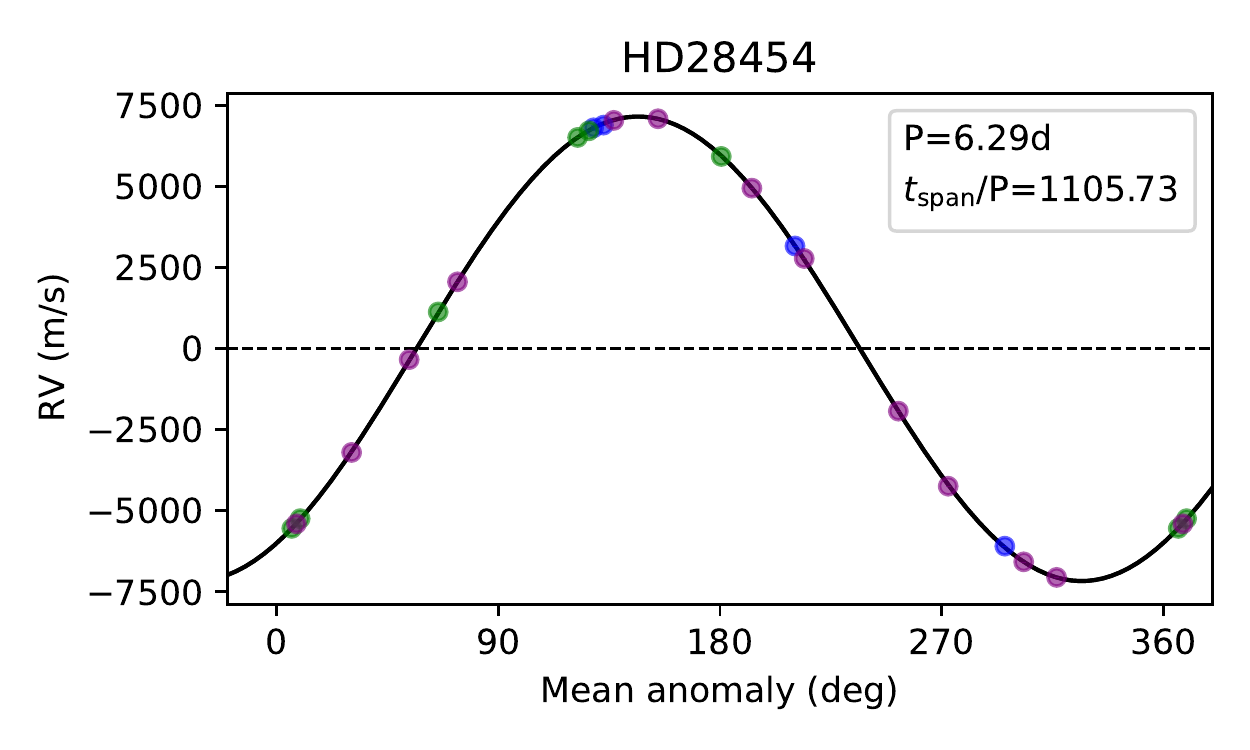}\\

		\includegraphics[width=0.22\linewidth]{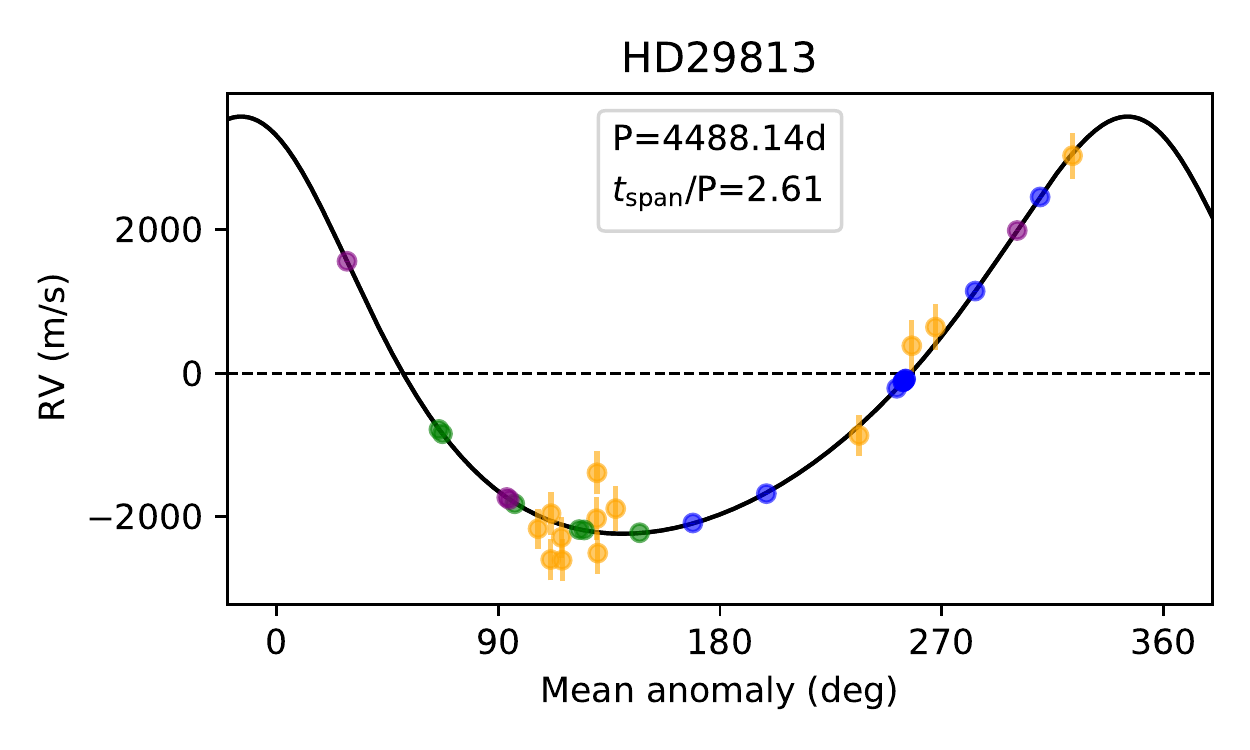}&
		\includegraphics[width=0.22\linewidth]{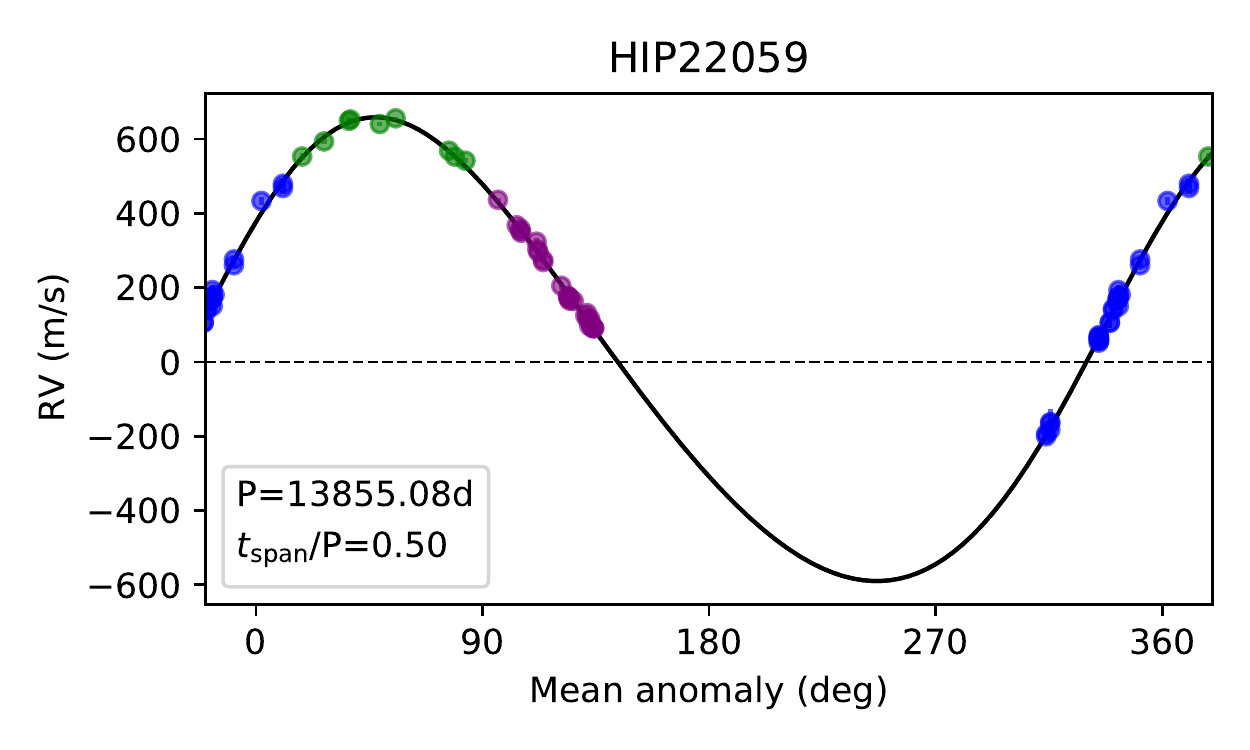}&
		\includegraphics[width=0.22\linewidth]{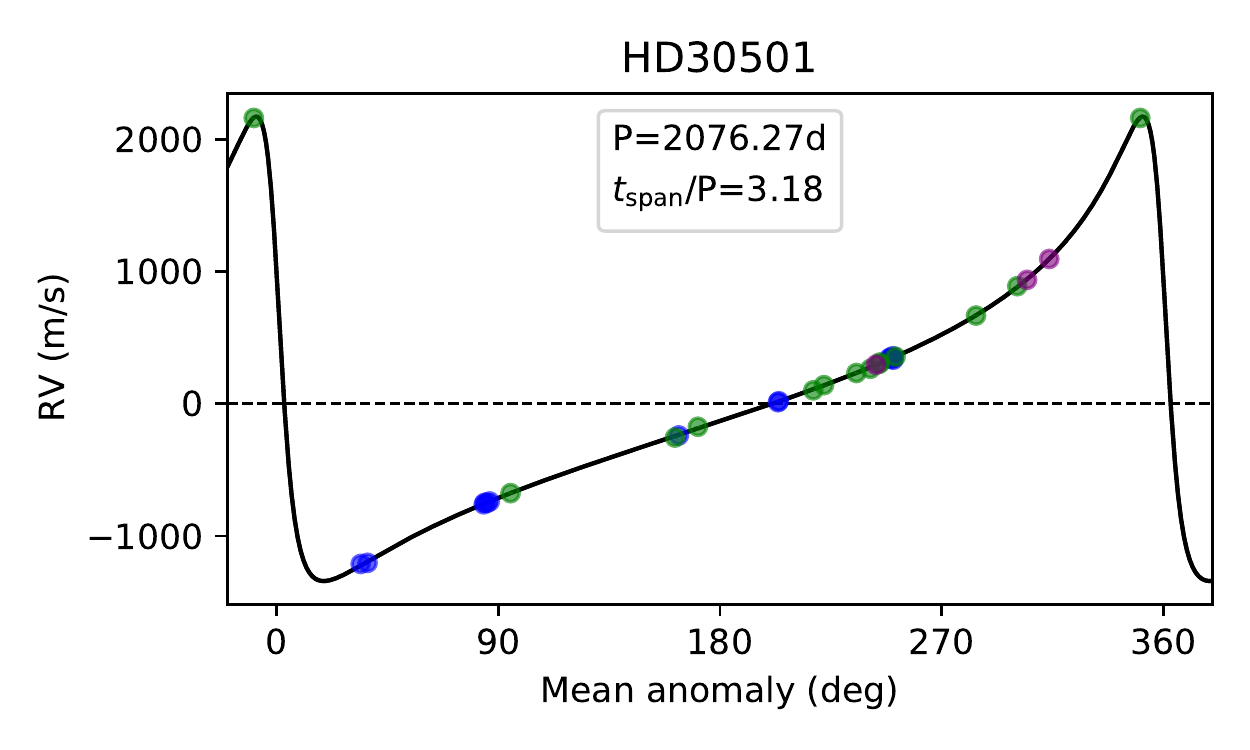}&
		\includegraphics[width=0.22\linewidth]{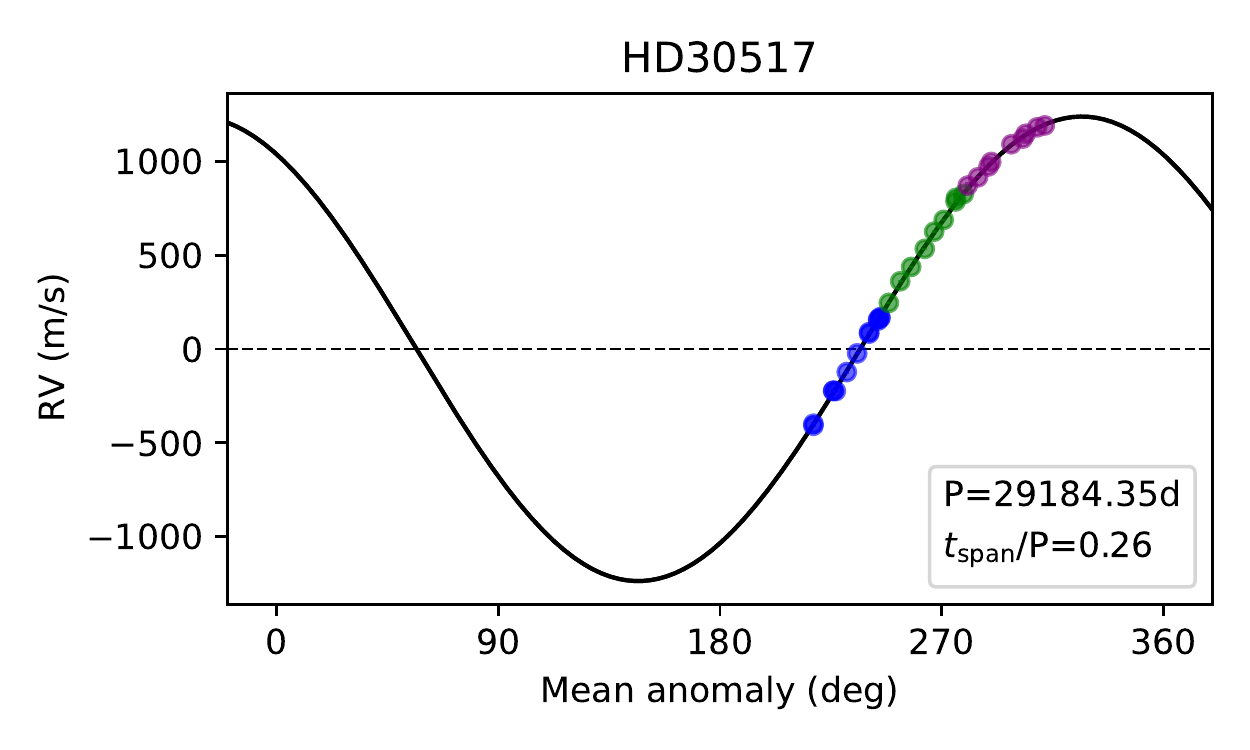}\\

		\includegraphics[width=0.22\linewidth]{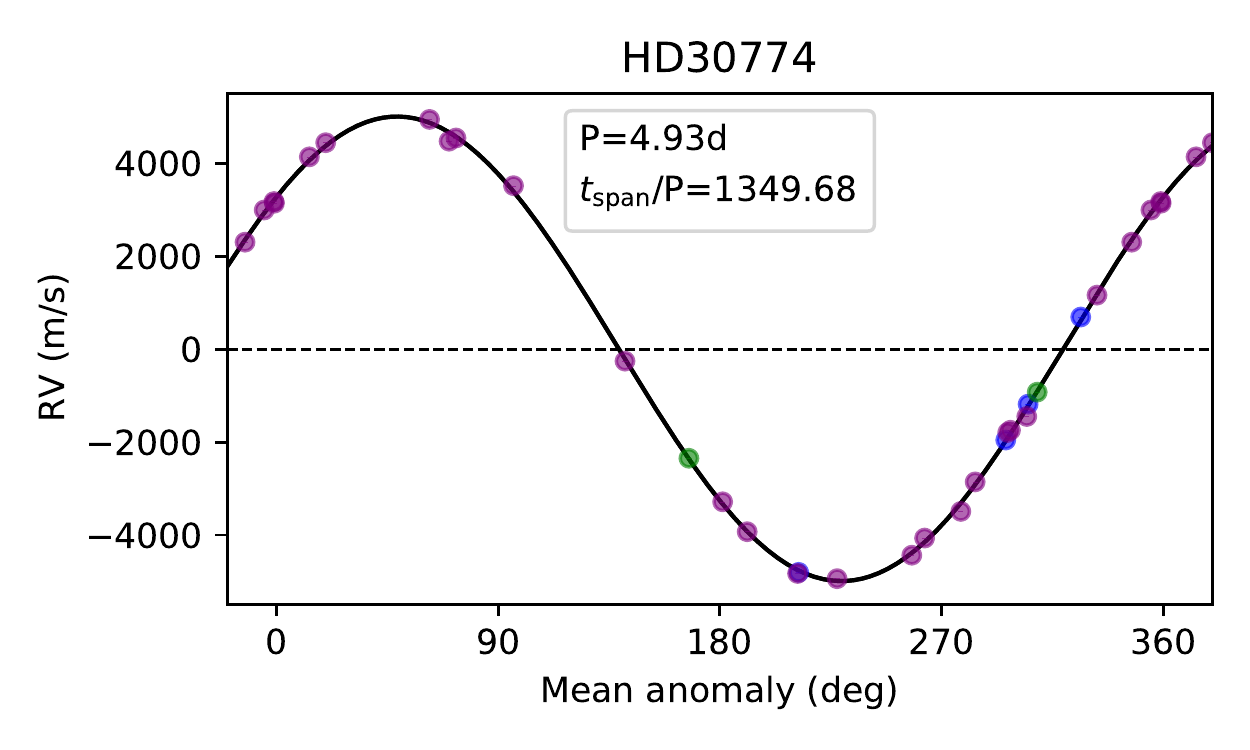}&
		\includegraphics[width=0.22\linewidth]{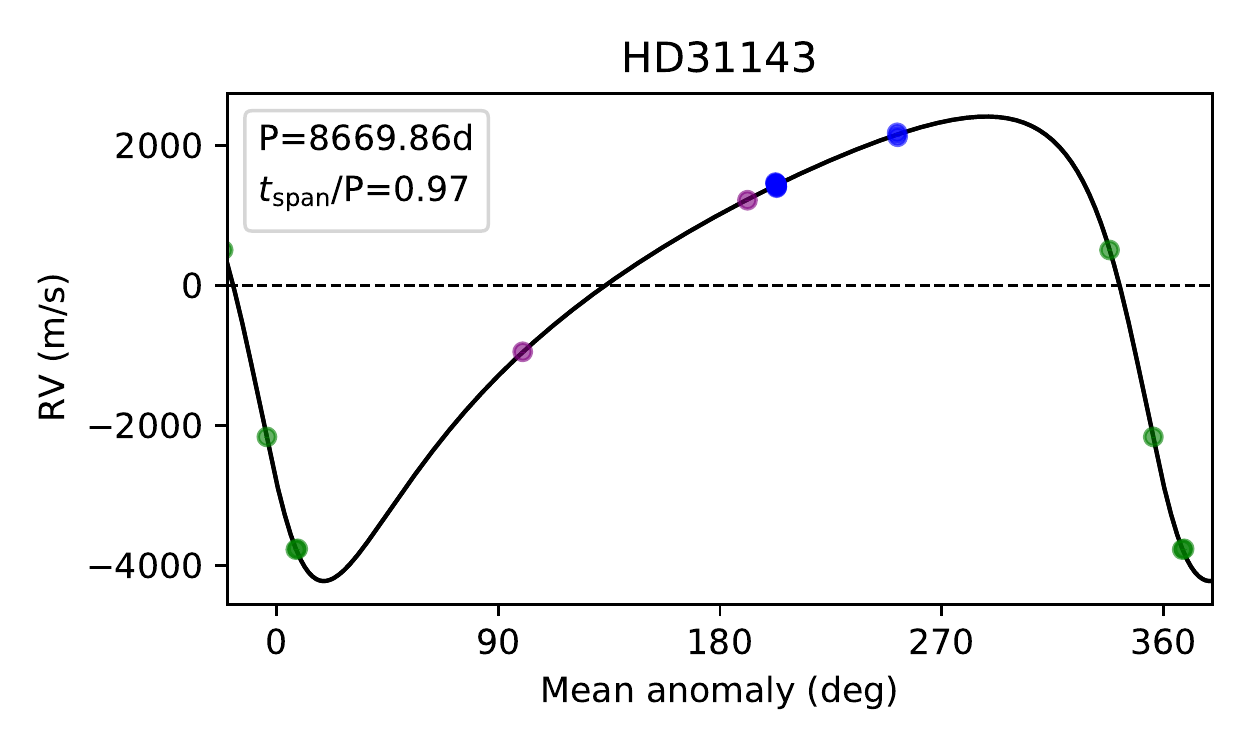}&
		\includegraphics[width=0.22\linewidth]{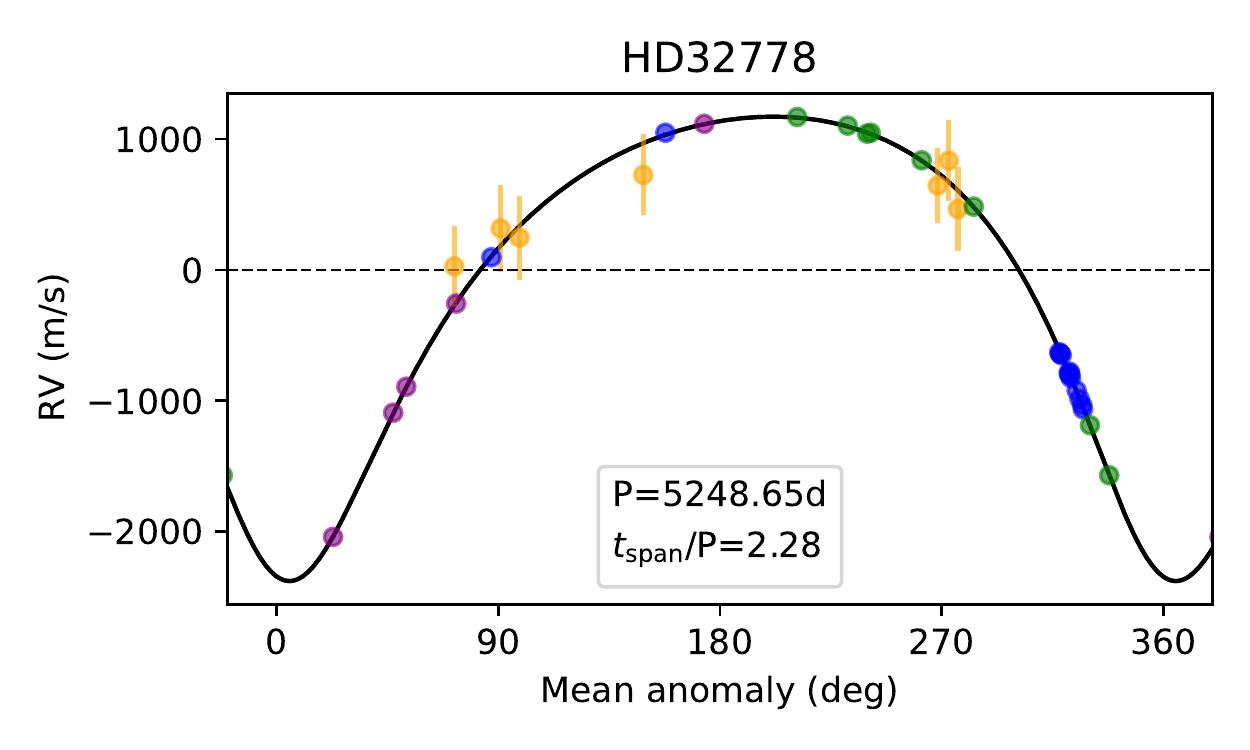}&
		\includegraphics[width=0.22\linewidth]{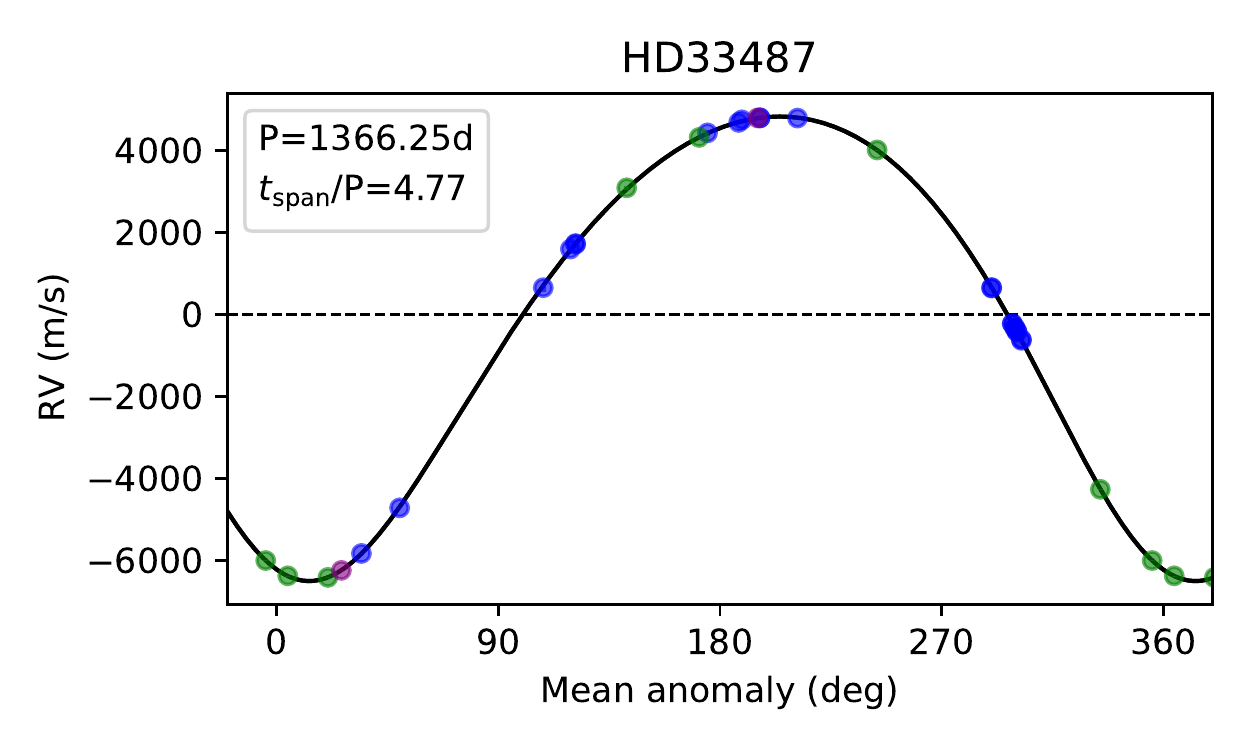}\\

		\includegraphics[width=0.22\linewidth]{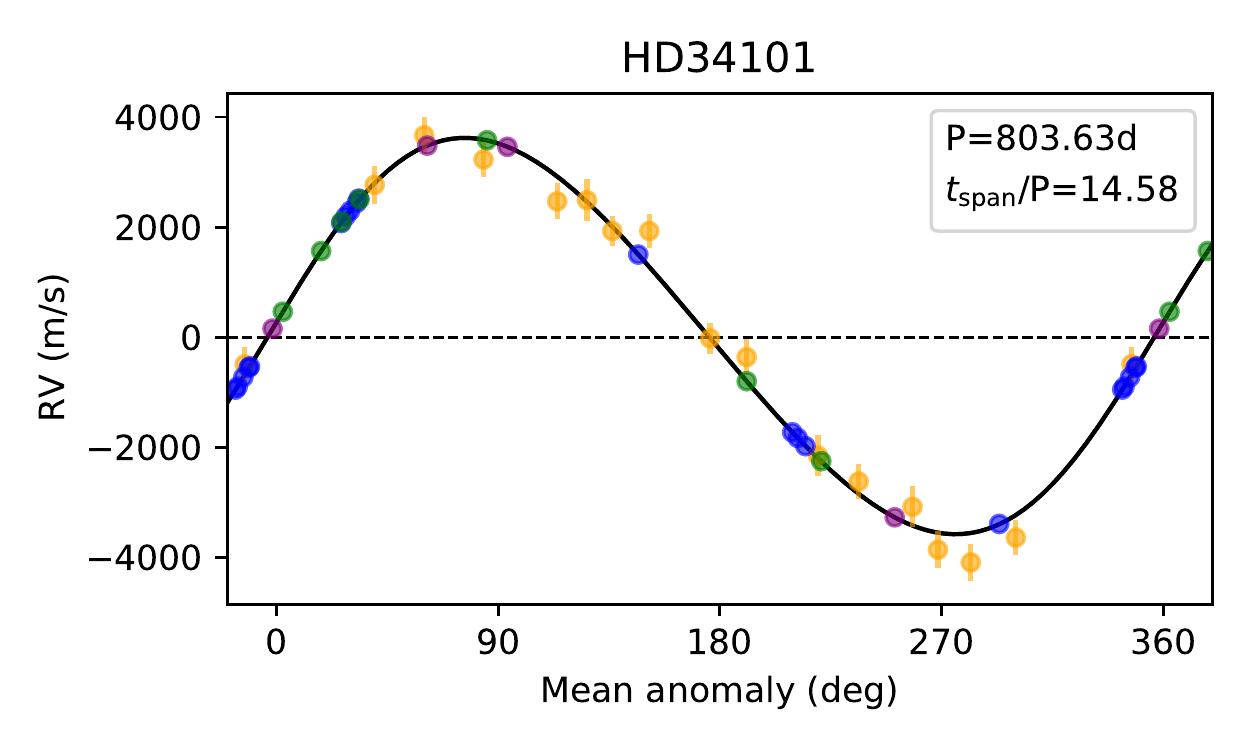}&
		\includegraphics[width=0.22\linewidth]{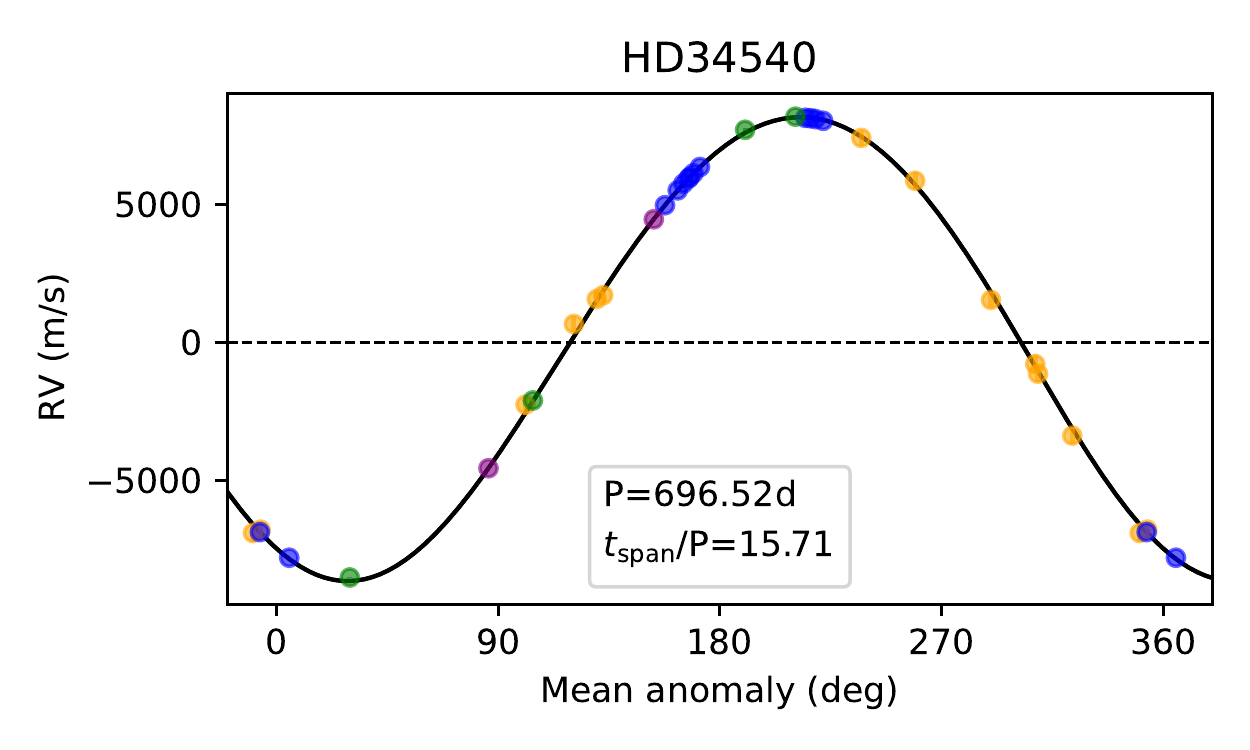}&
		\includegraphics[width=0.22\linewidth]{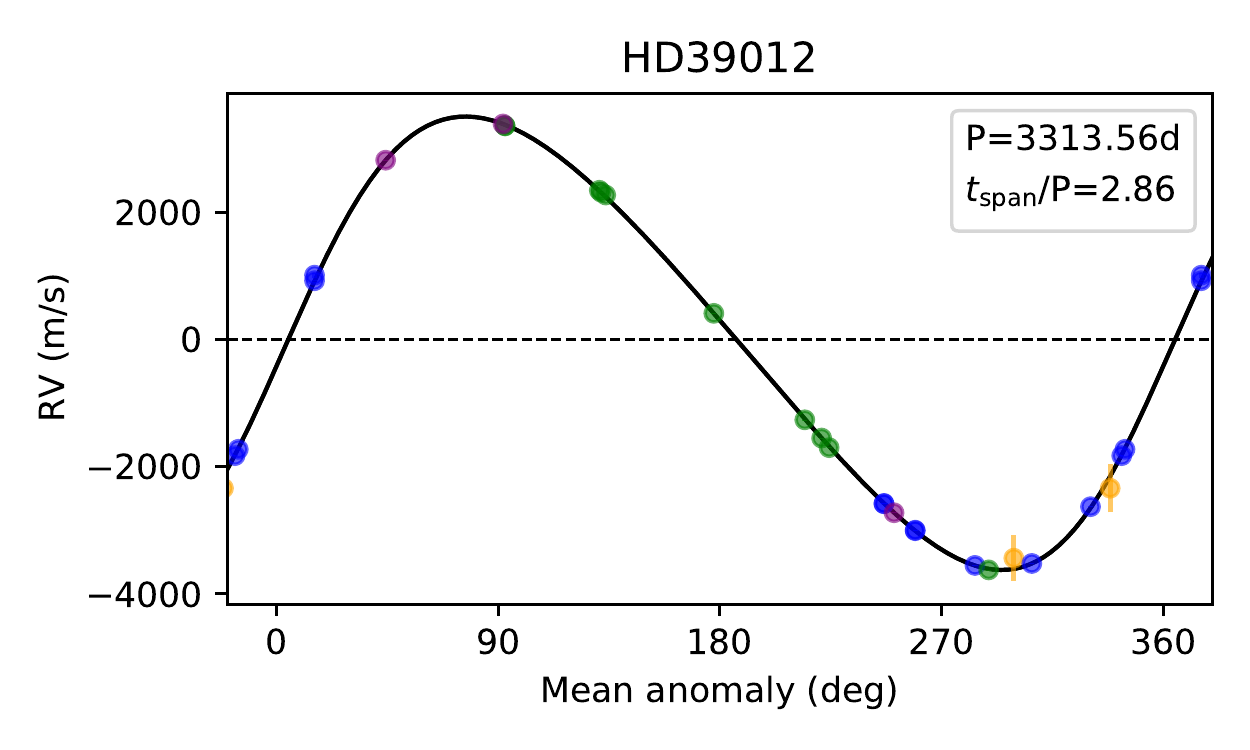}&
		\includegraphics[width=0.22\linewidth]{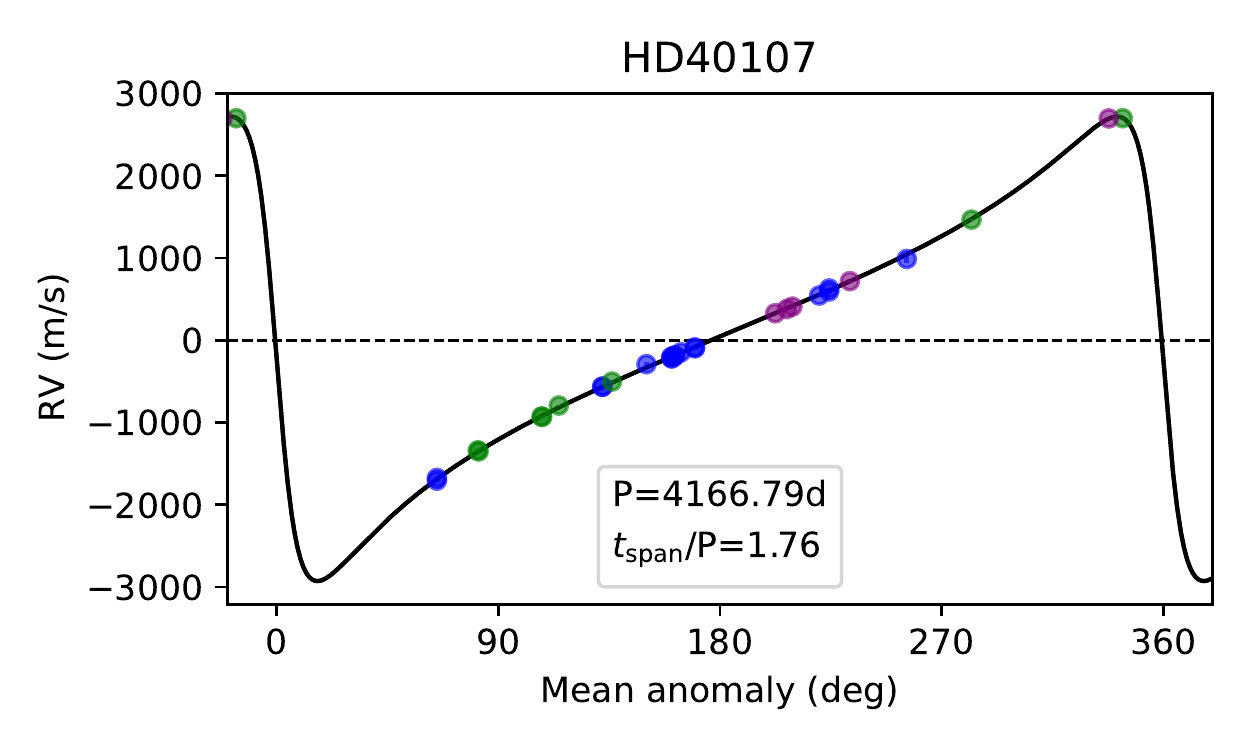}\\

		\includegraphics[width=0.22\linewidth]{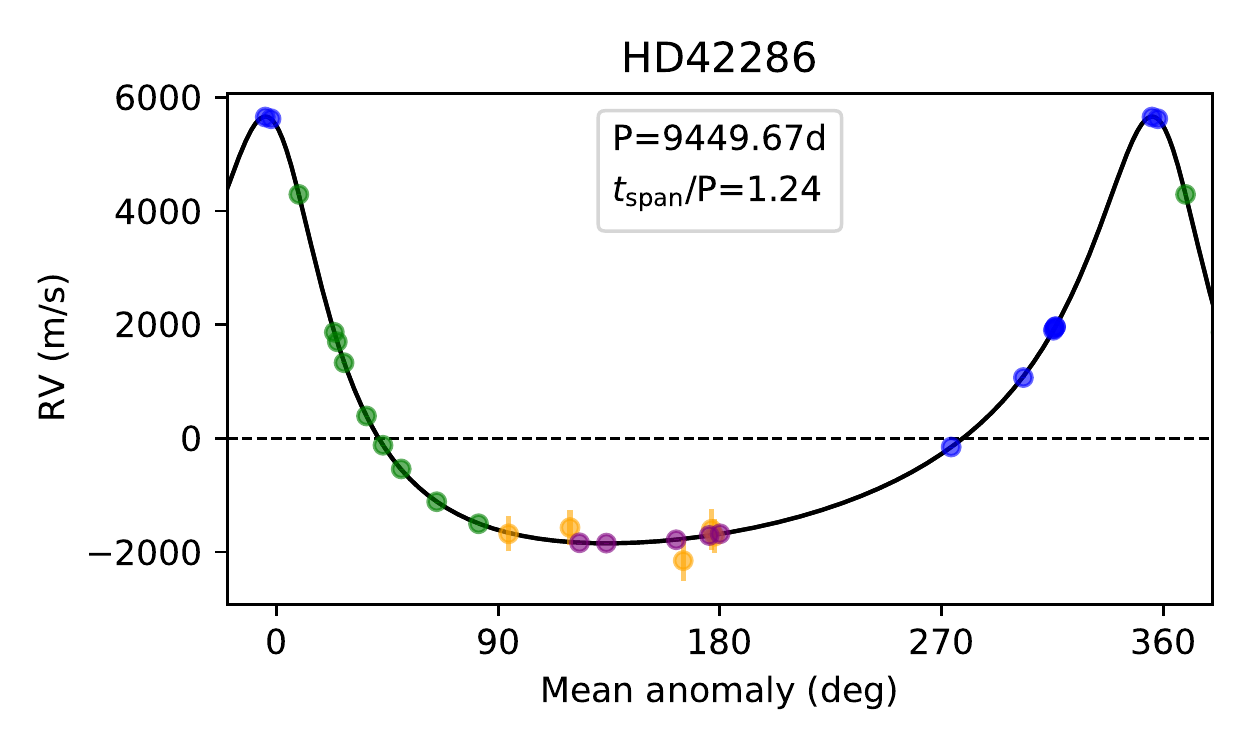}&
		\includegraphics[width=0.22\linewidth]{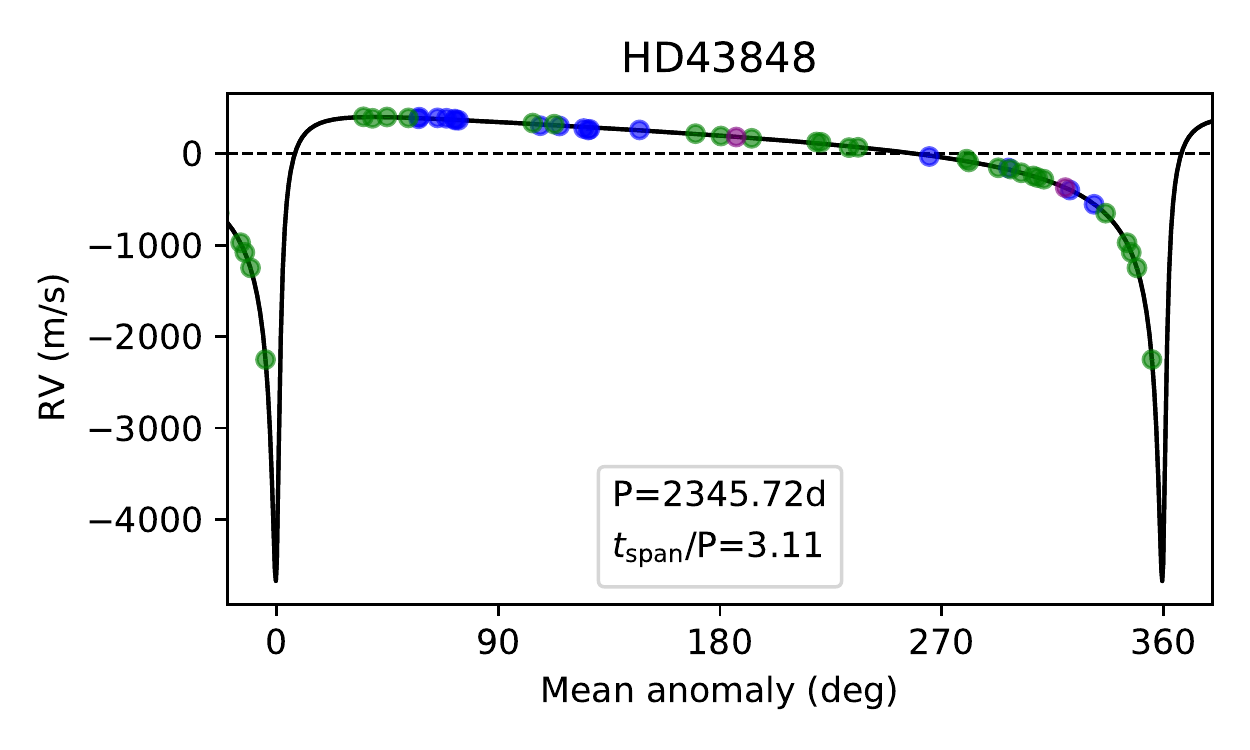}&
		\includegraphics[width=0.22\linewidth]{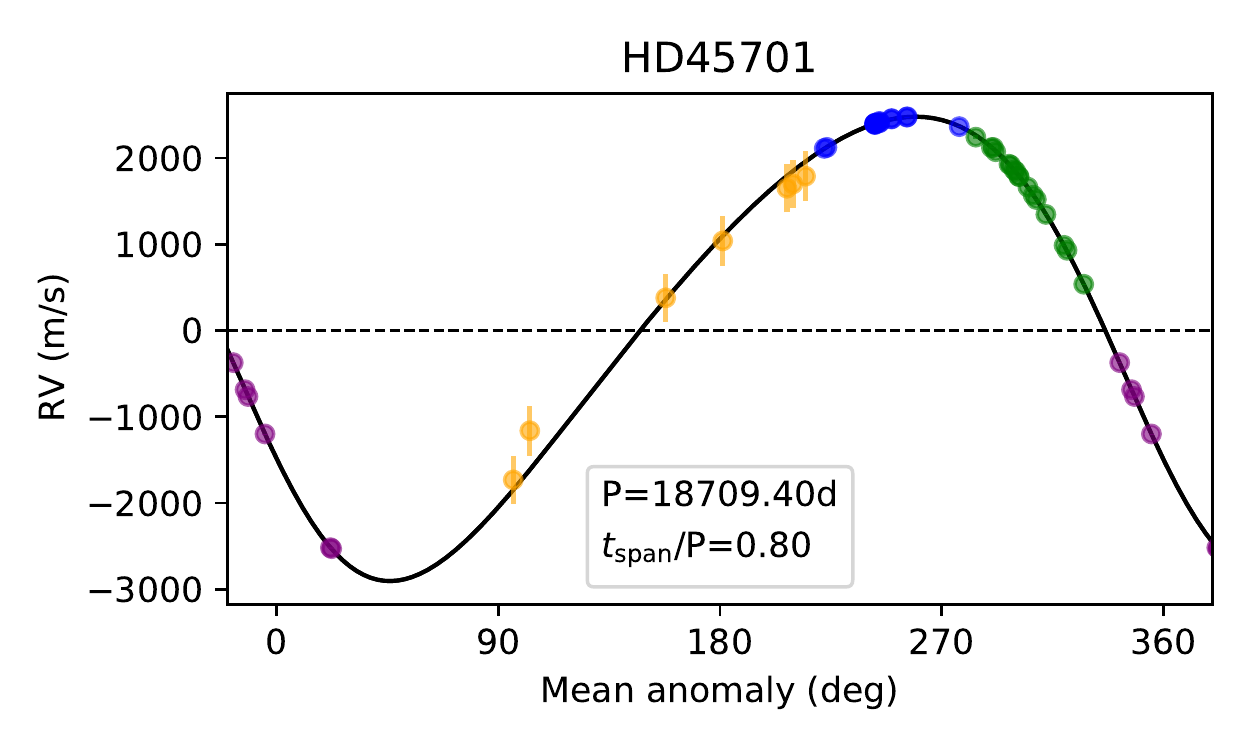}&
		\includegraphics[width=0.22\linewidth]{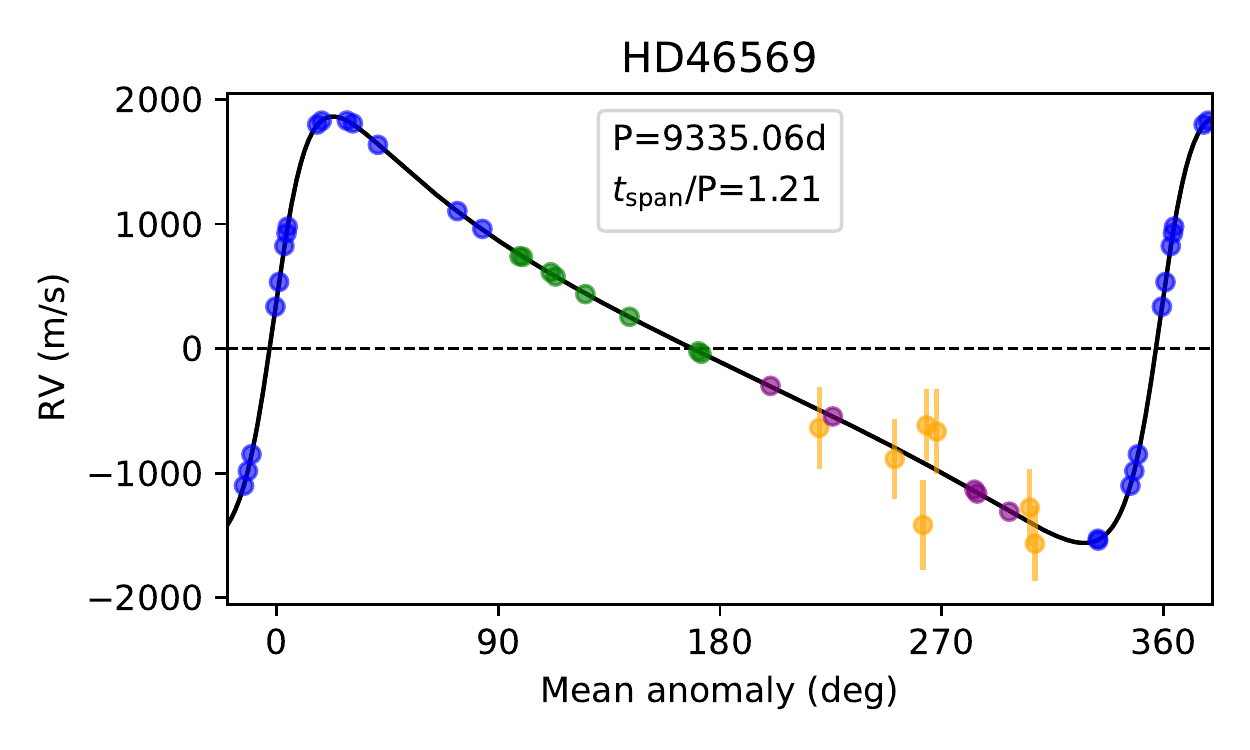}\\

		\includegraphics[width=0.22\linewidth]{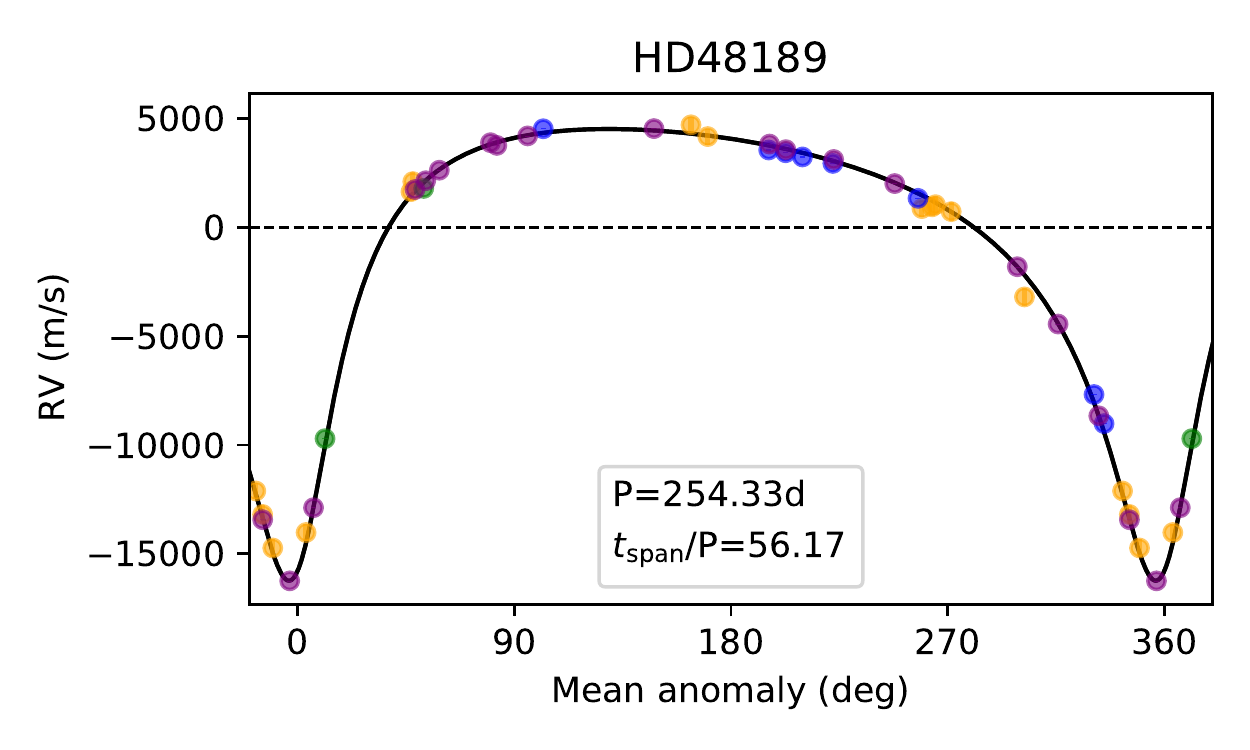}&
		\includegraphics[width=0.22\linewidth]{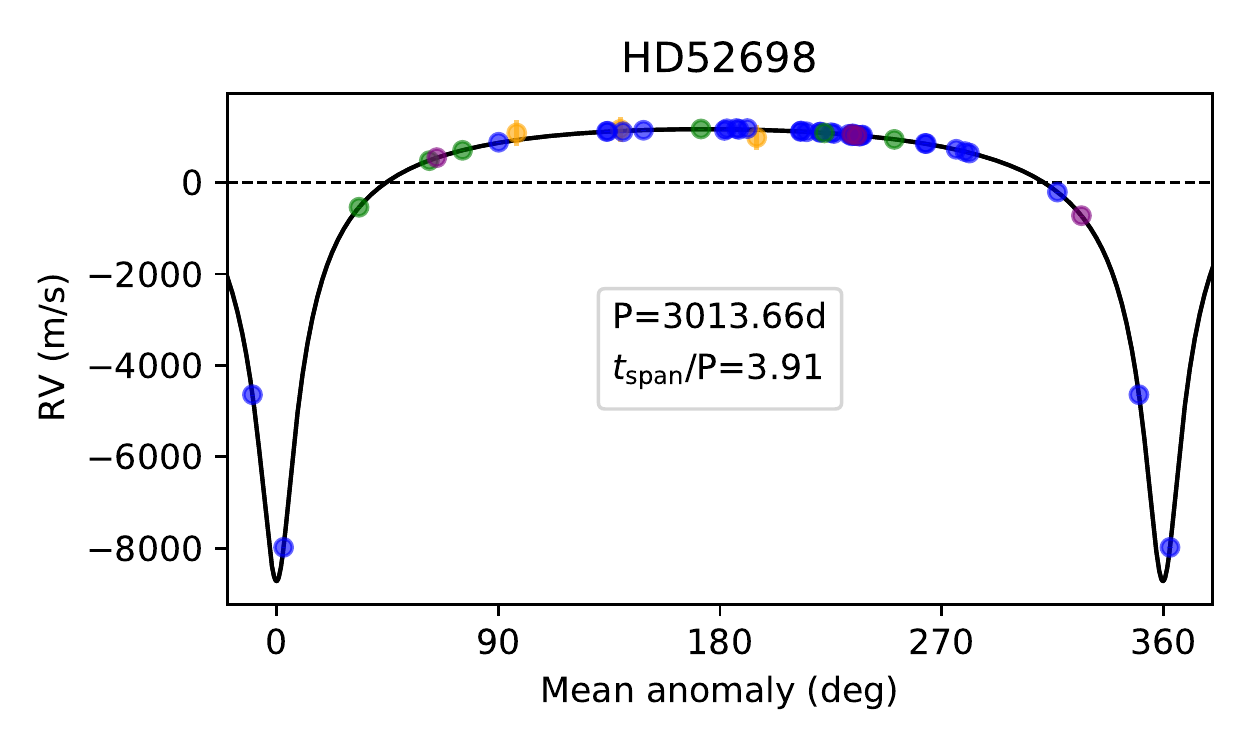}&
		\includegraphics[width=0.22\linewidth]{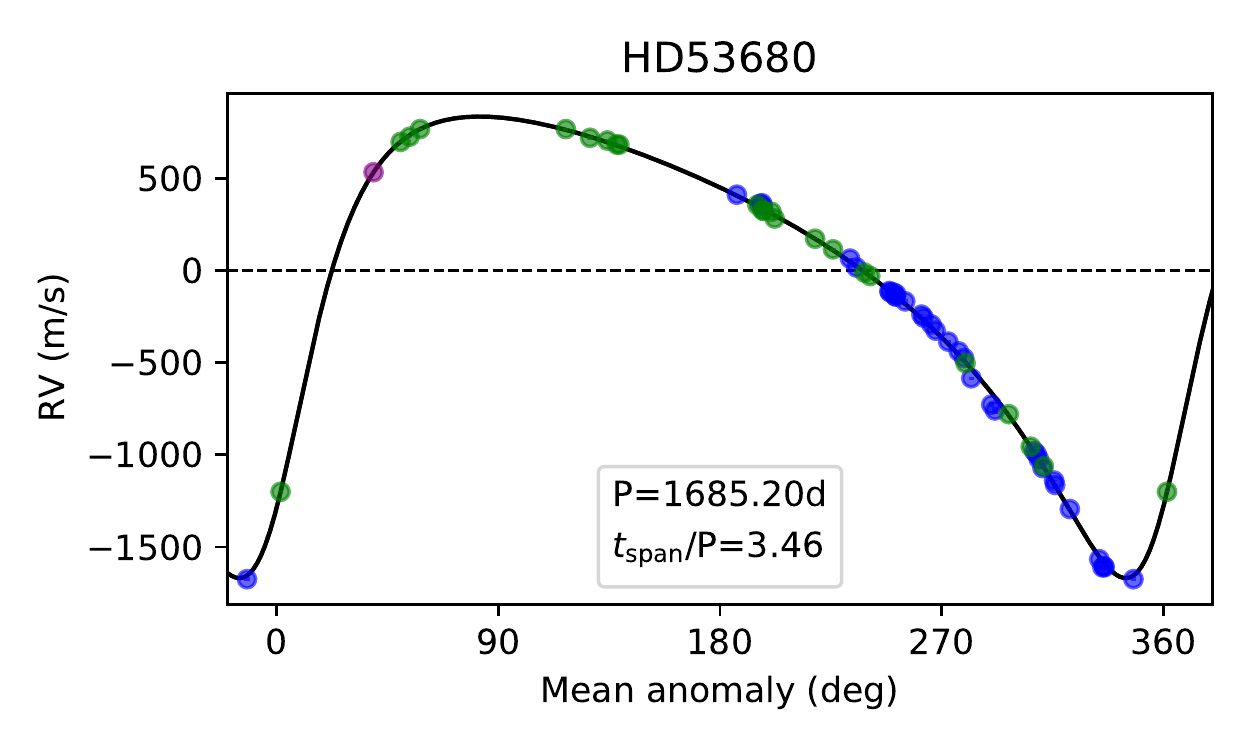}&
		\includegraphics[width=0.22\linewidth]{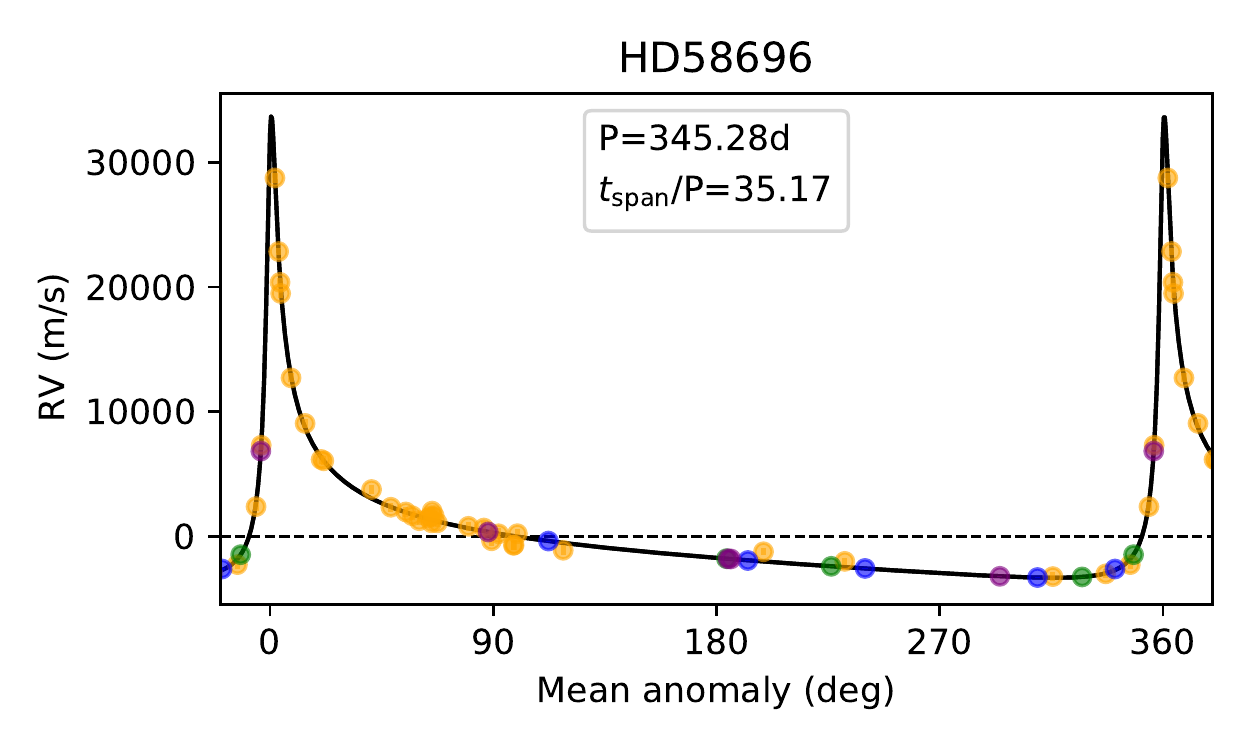}\\

		\includegraphics[width=0.22\linewidth]{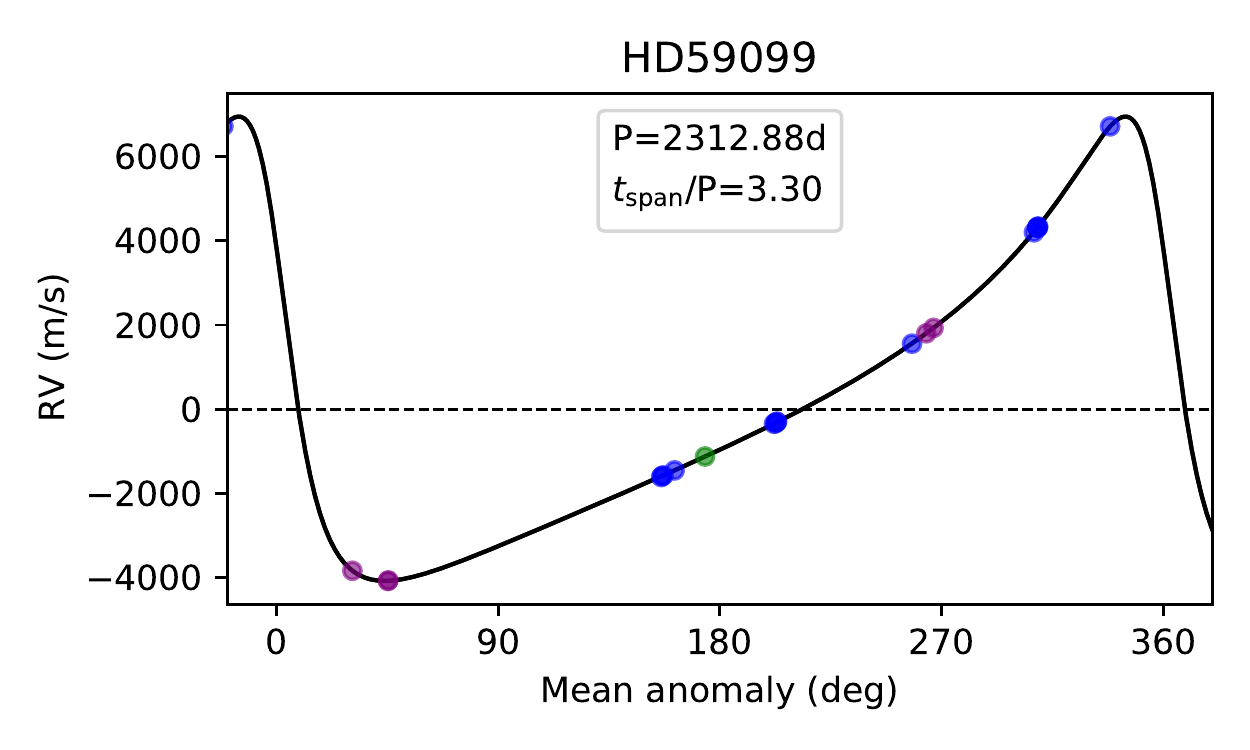}&
		\includegraphics[width=0.22\linewidth]{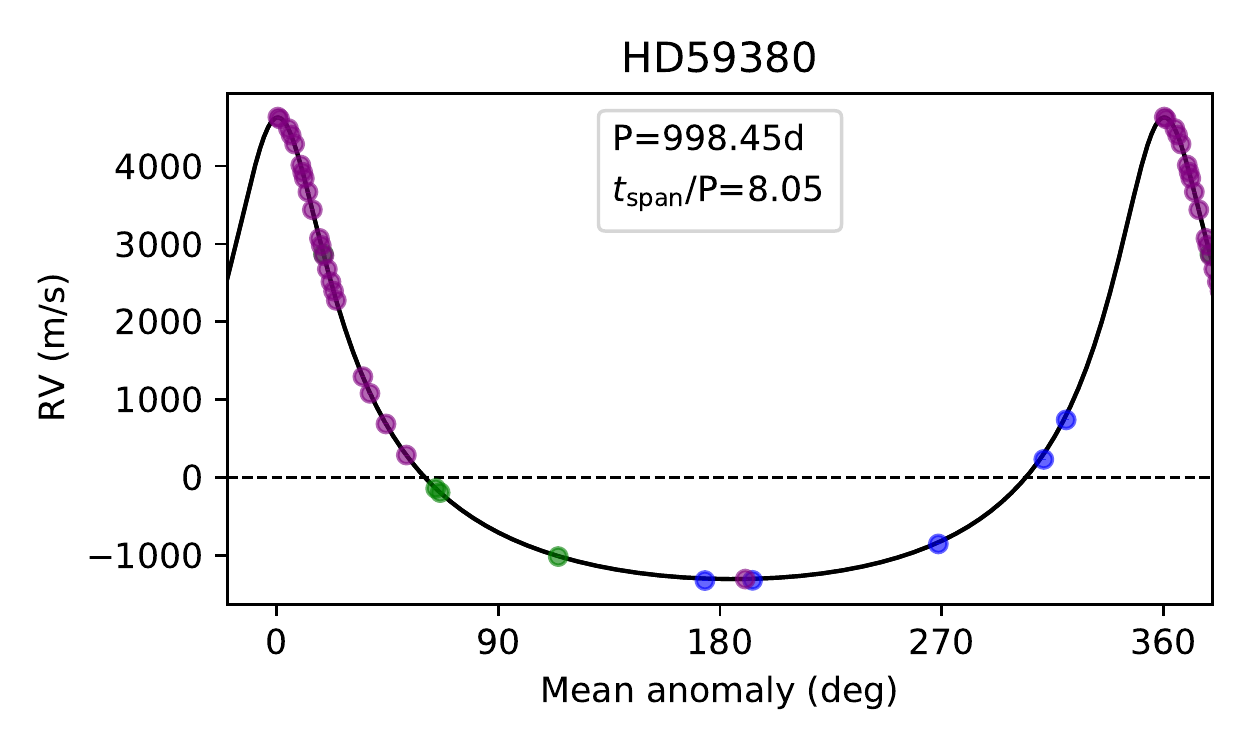}&
		\includegraphics[width=0.22\linewidth]{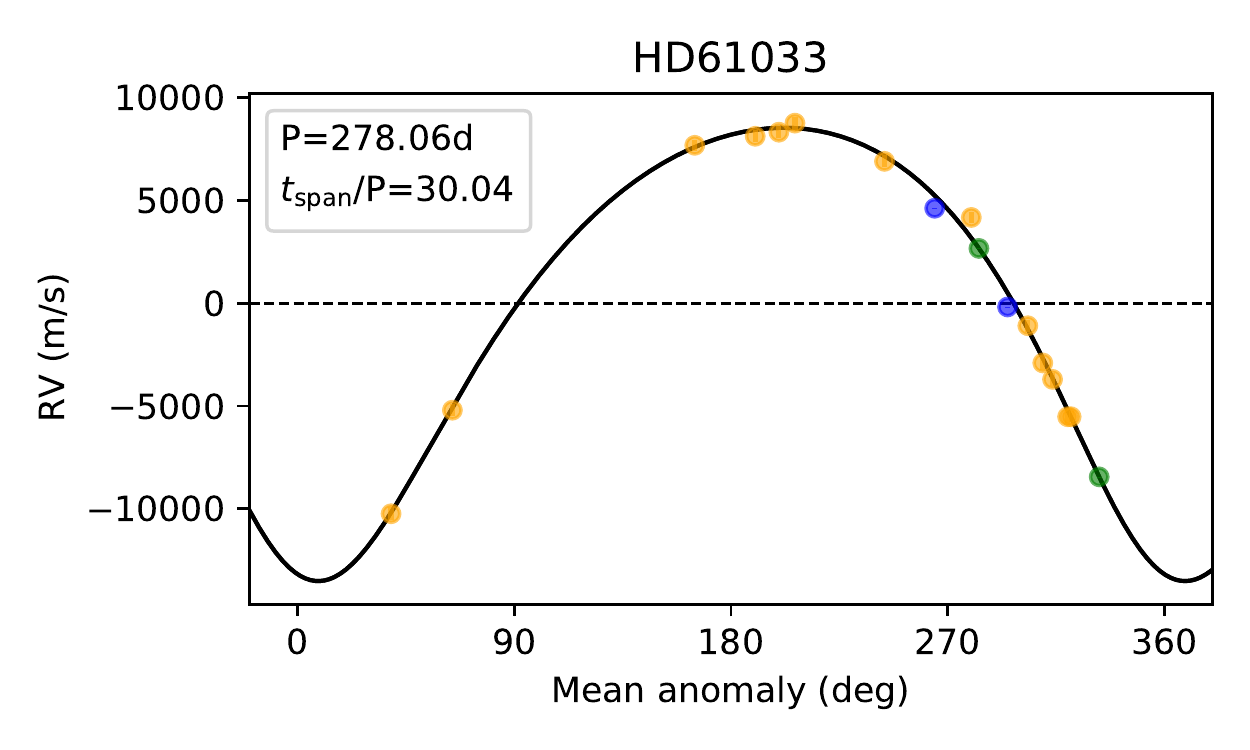}&
		\includegraphics[width=0.22\linewidth]{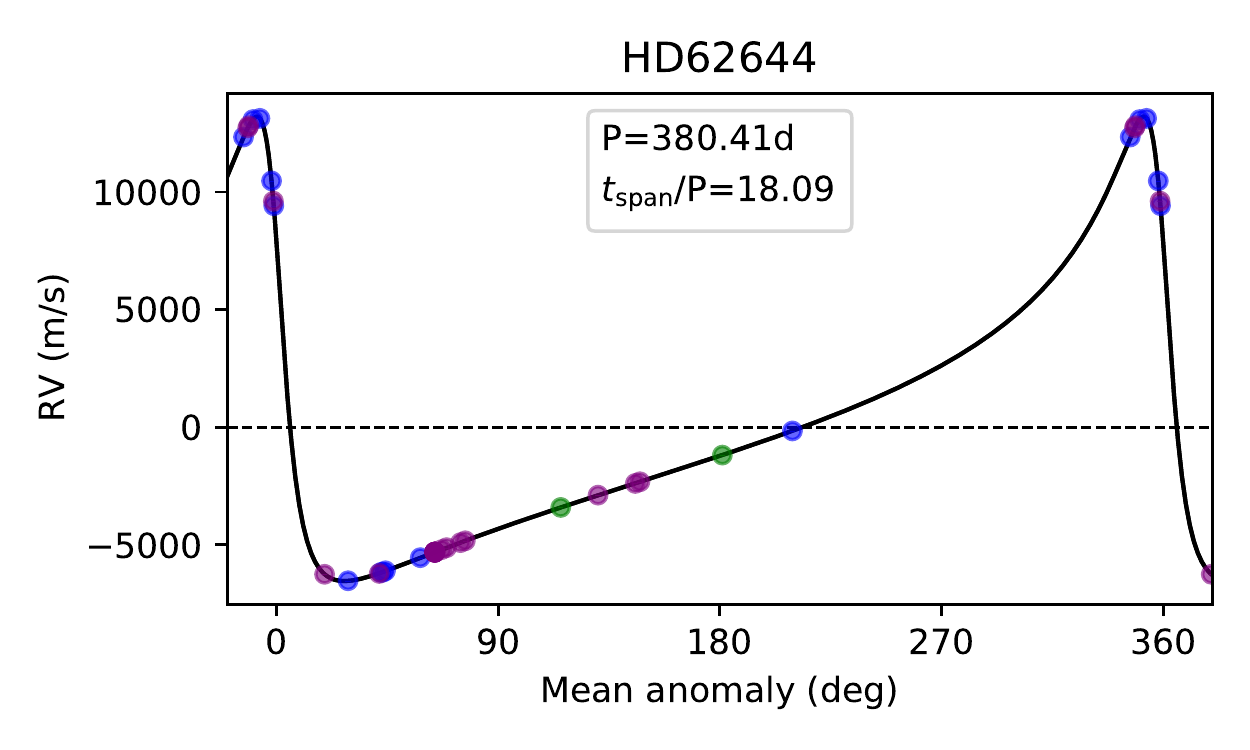}\\

		\includegraphics[width=0.22\linewidth]{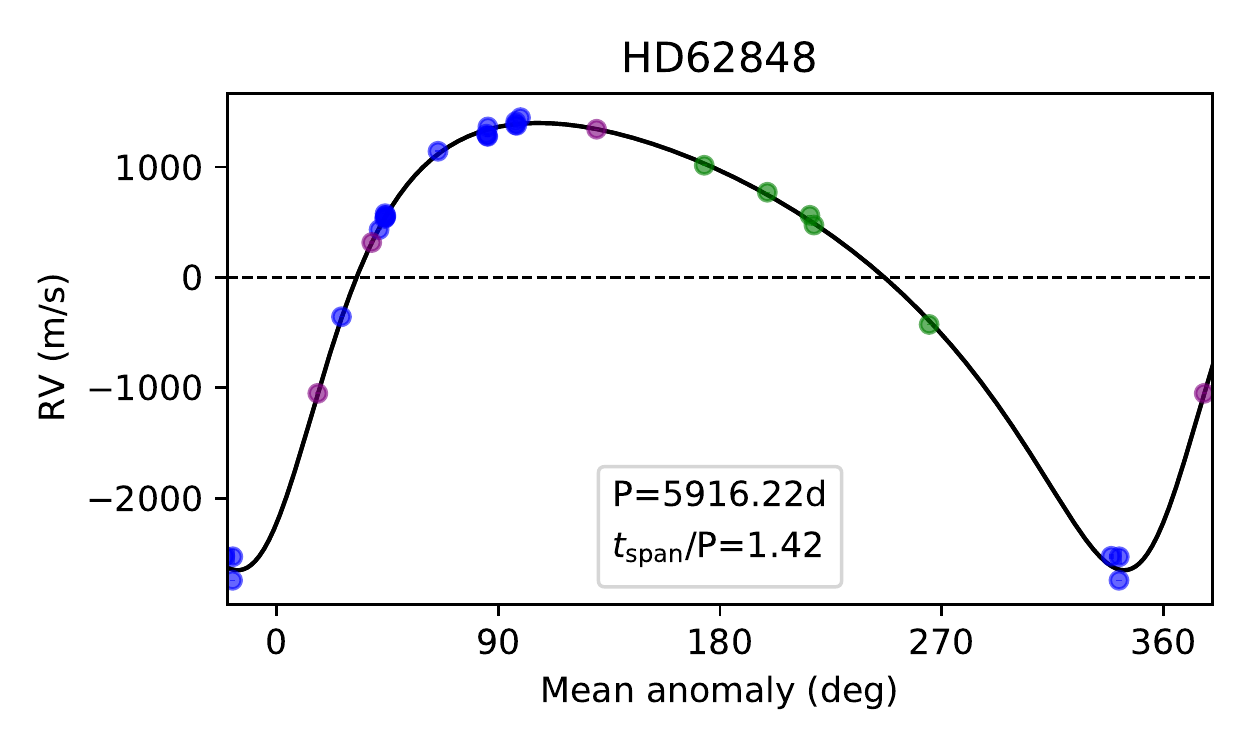}&
		\includegraphics[width=0.22\linewidth]{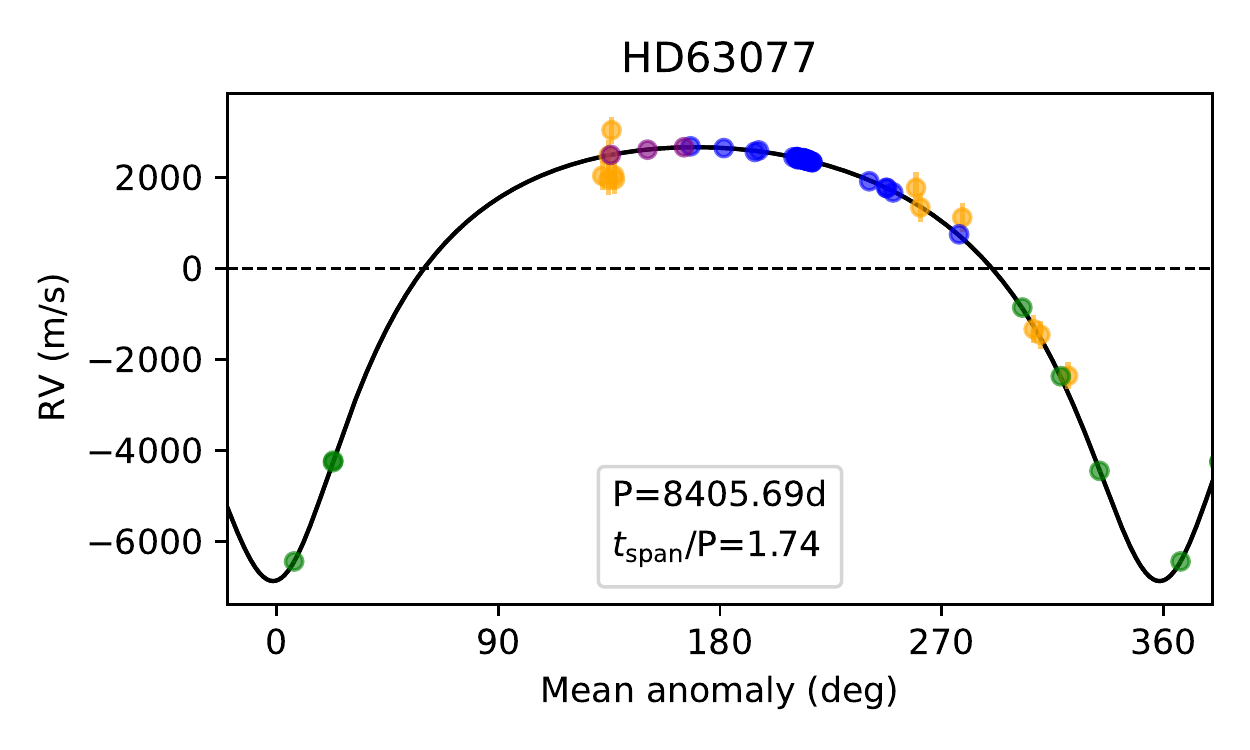}&
		\includegraphics[width=0.22\linewidth]{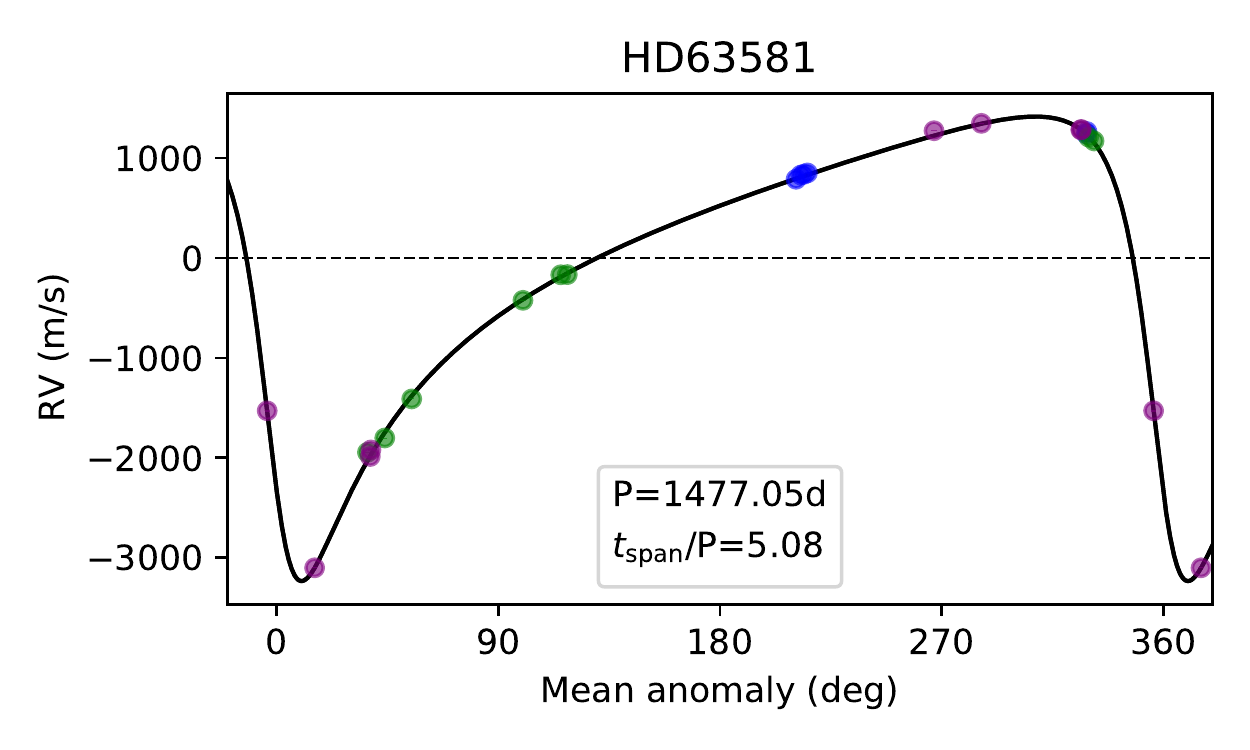}&
		\includegraphics[width=0.22\linewidth]{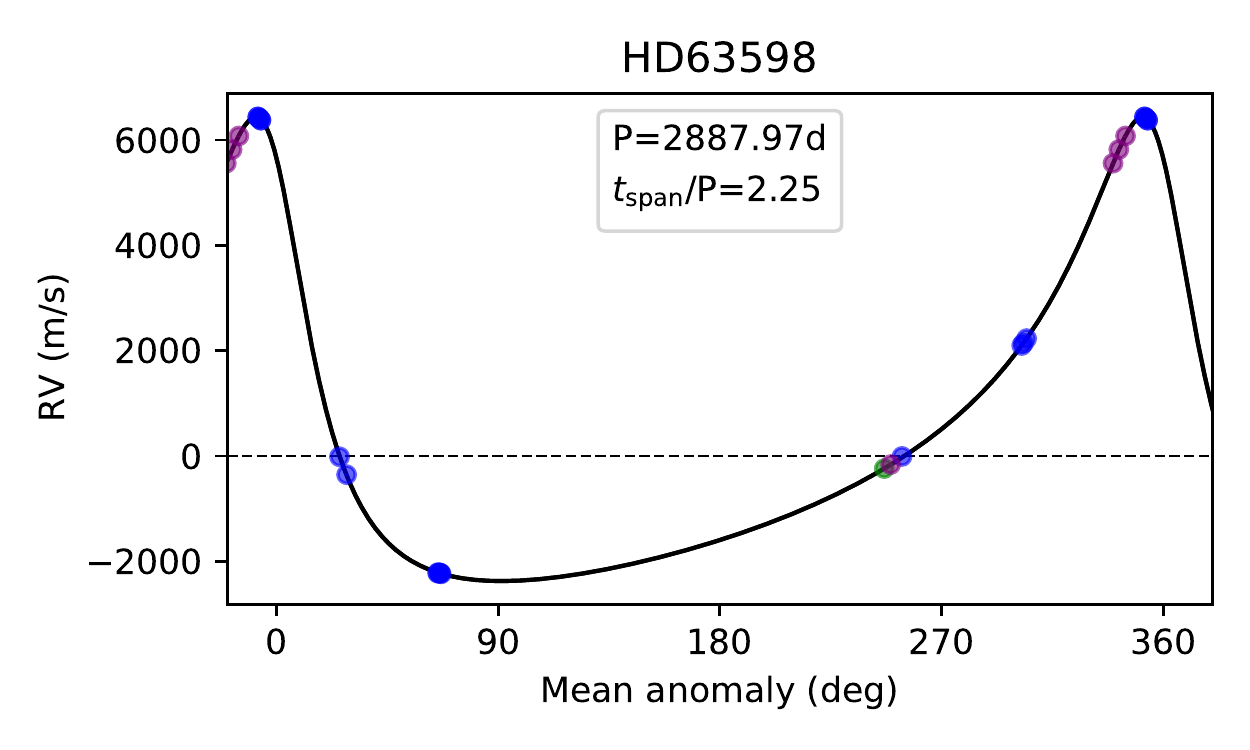}\\

		\includegraphics[width=0.22\linewidth]{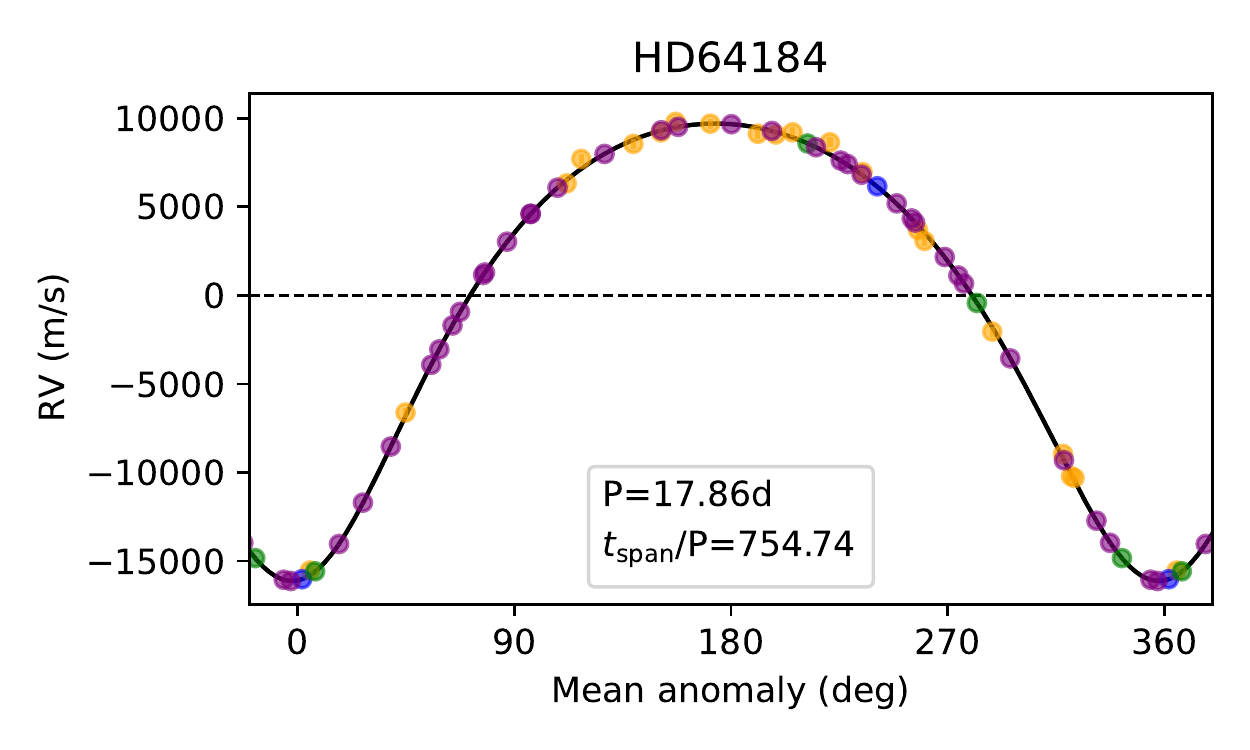}&
		\includegraphics[width=0.22\linewidth]{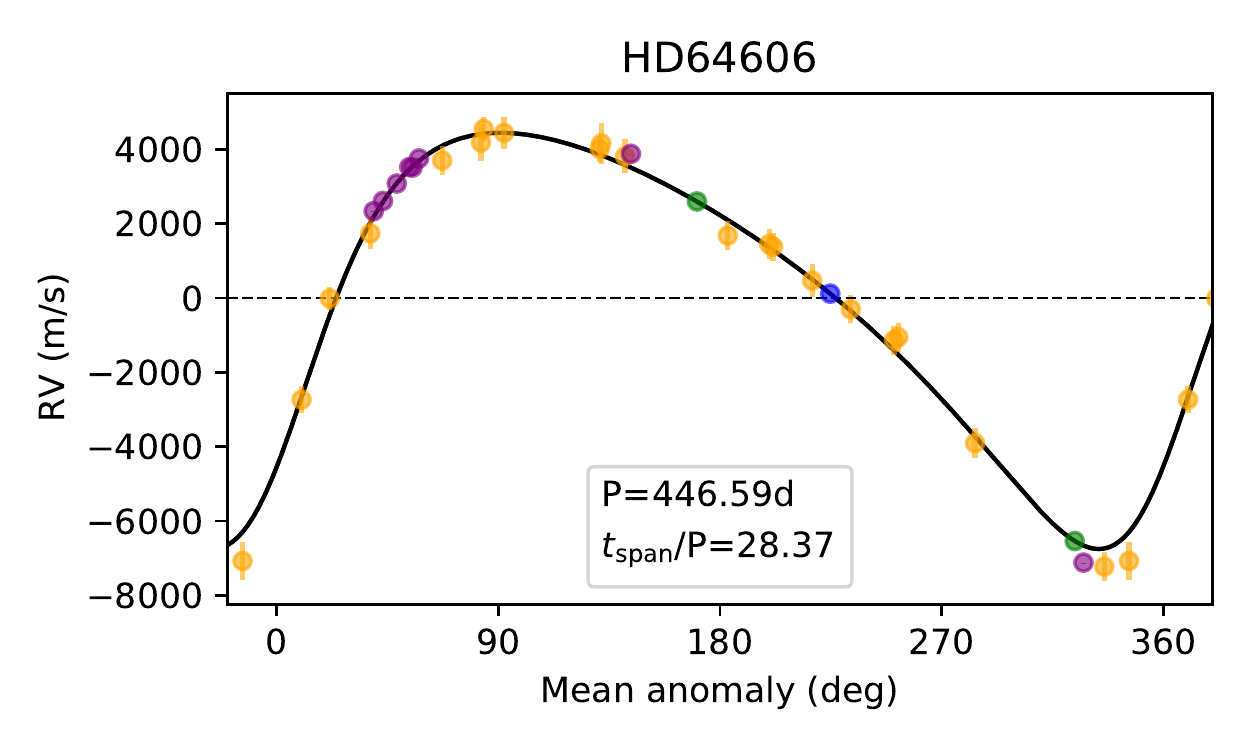}&
		\includegraphics[width=0.22\linewidth]{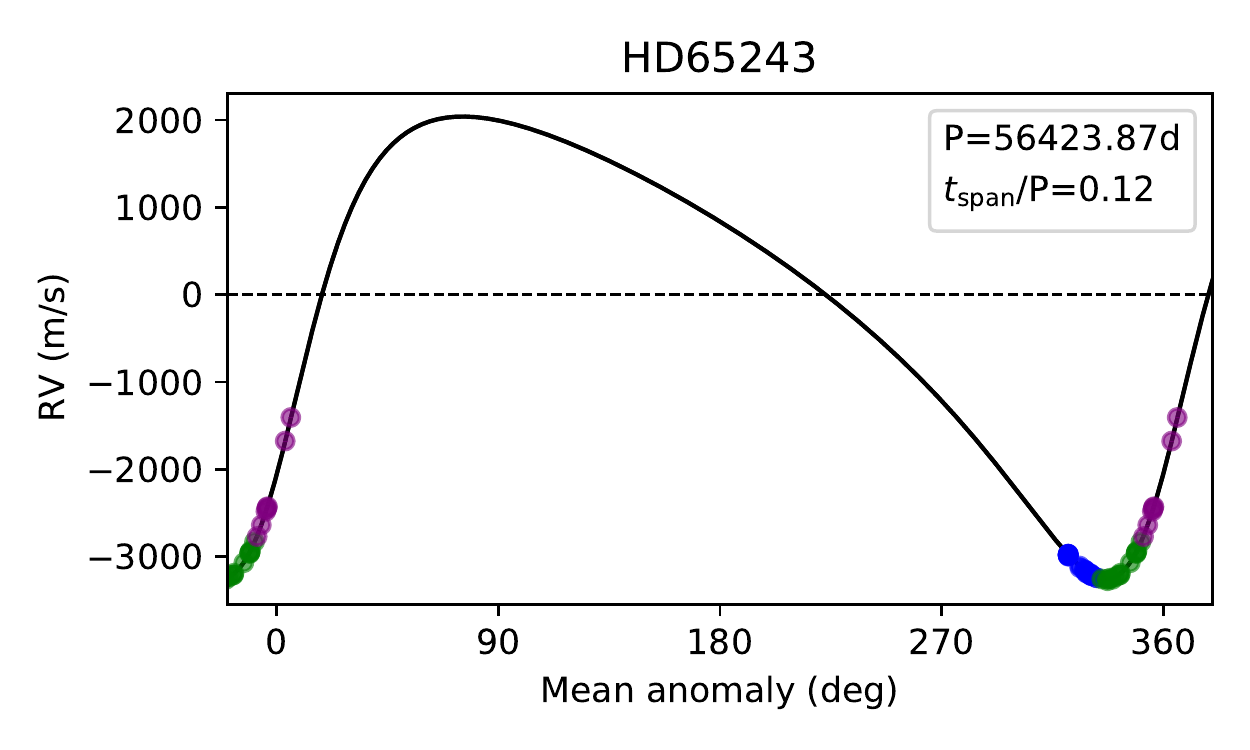}&
		\includegraphics[width=0.22\linewidth]{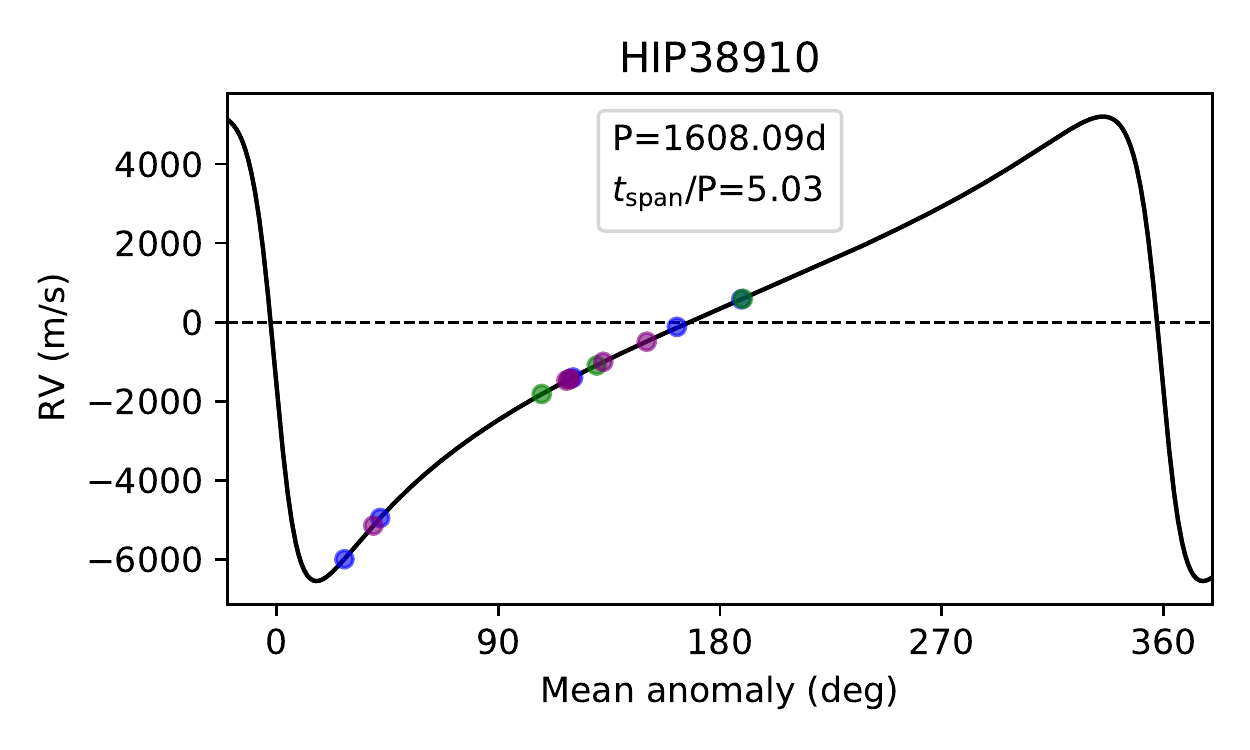}\\

		\includegraphics[width=0.22\linewidth]{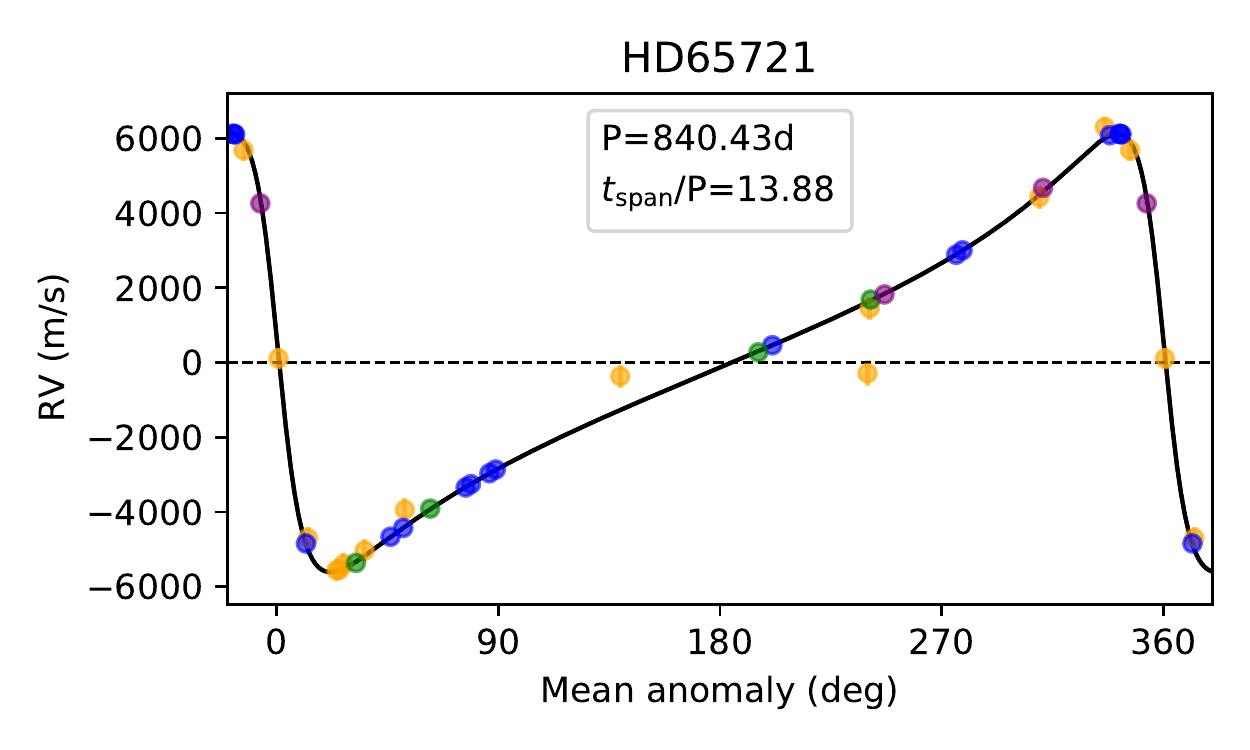}&
		\includegraphics[width=0.22\linewidth]{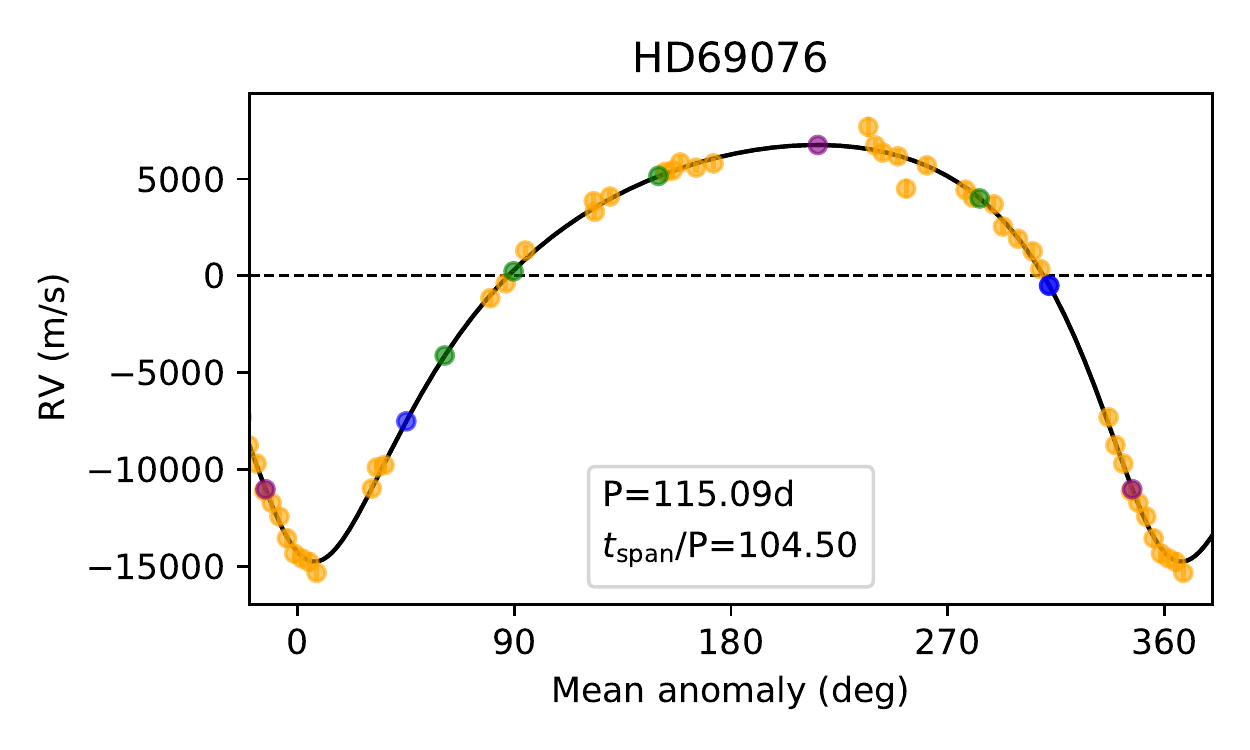}&
		\includegraphics[width=0.22\linewidth]{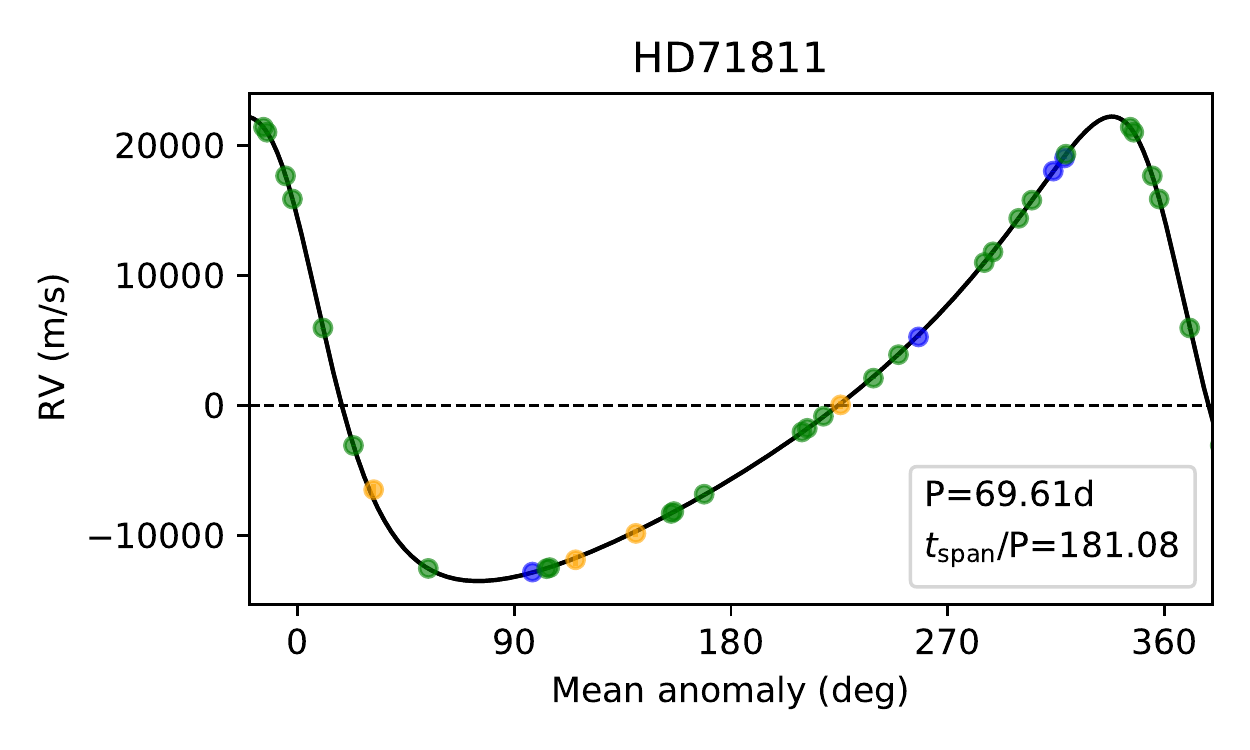}&
		\includegraphics[width=0.22\linewidth]{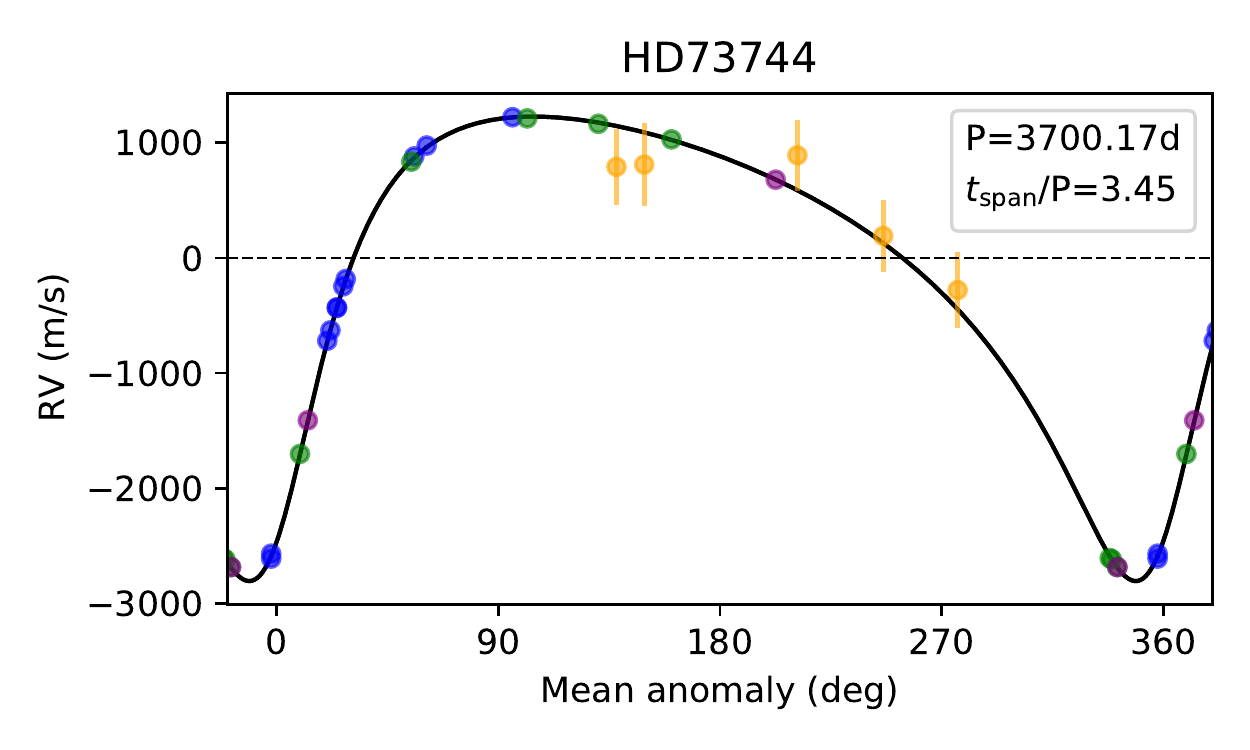}\\

		\includegraphics[width=0.22\linewidth]{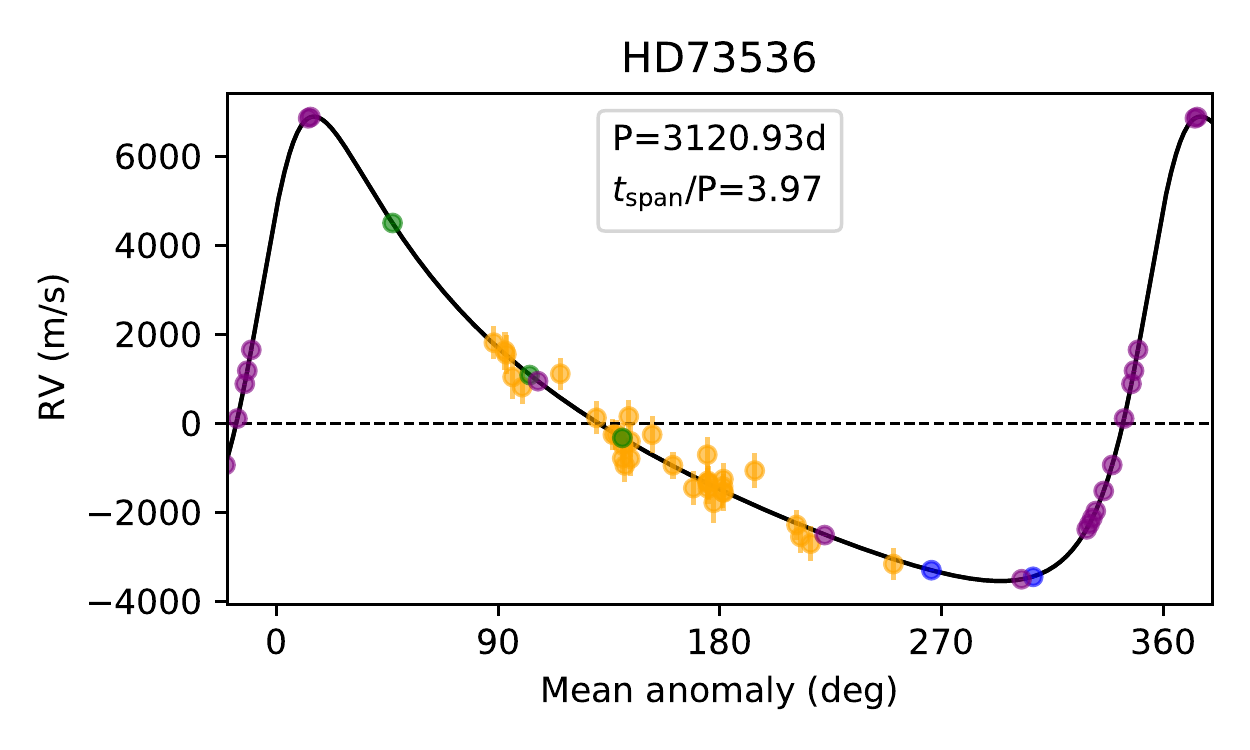}&
		\includegraphics[width=0.22\linewidth]{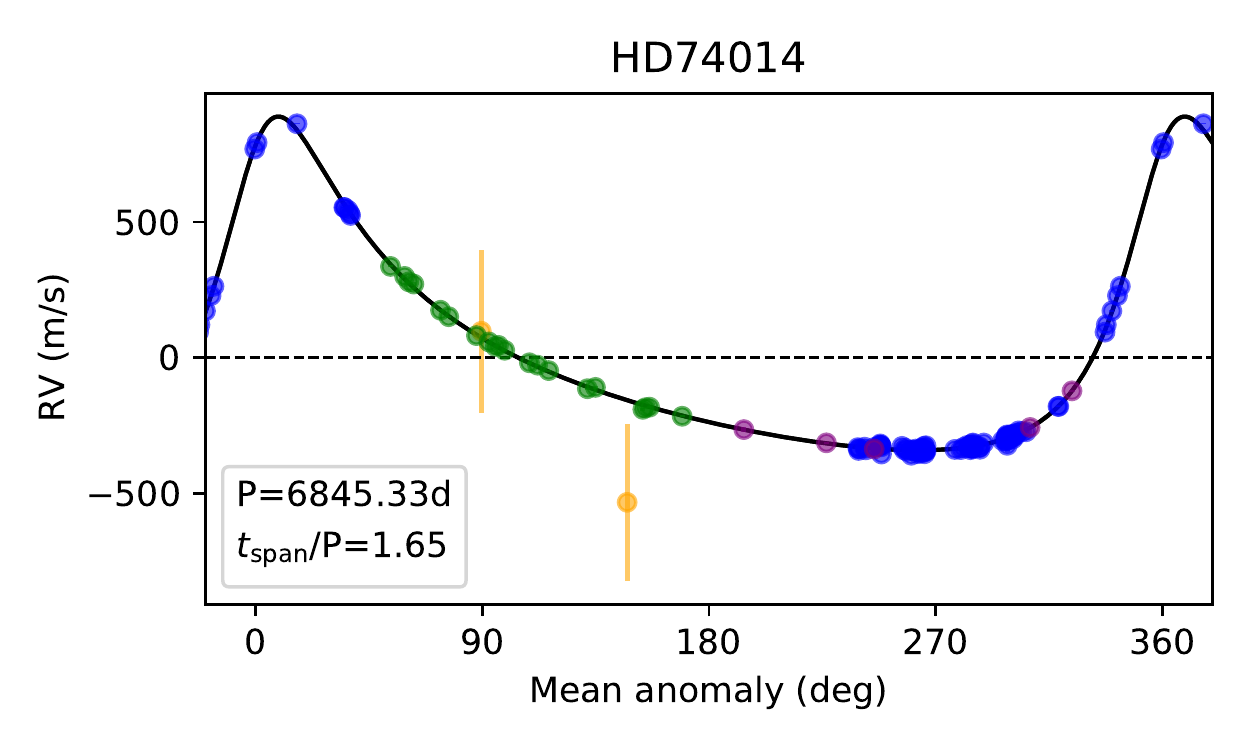}&
		\includegraphics[width=0.22\linewidth]{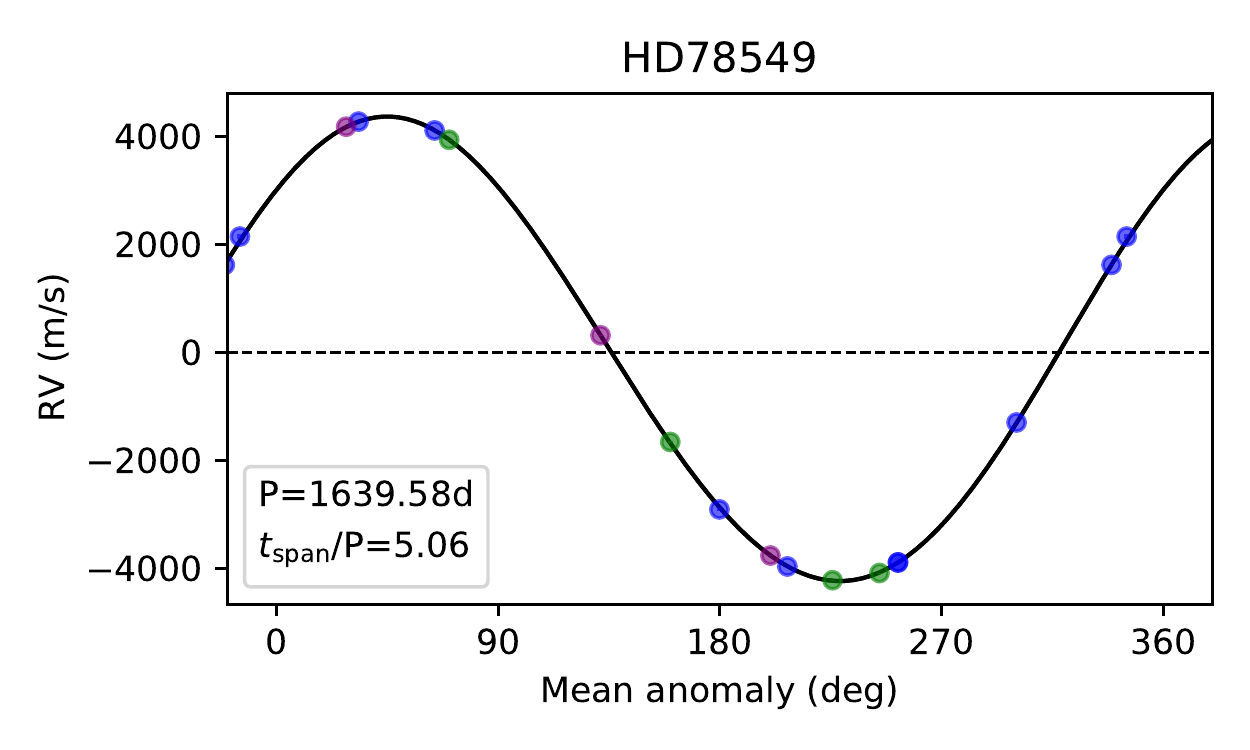}&
		\includegraphics[width=0.22\linewidth]{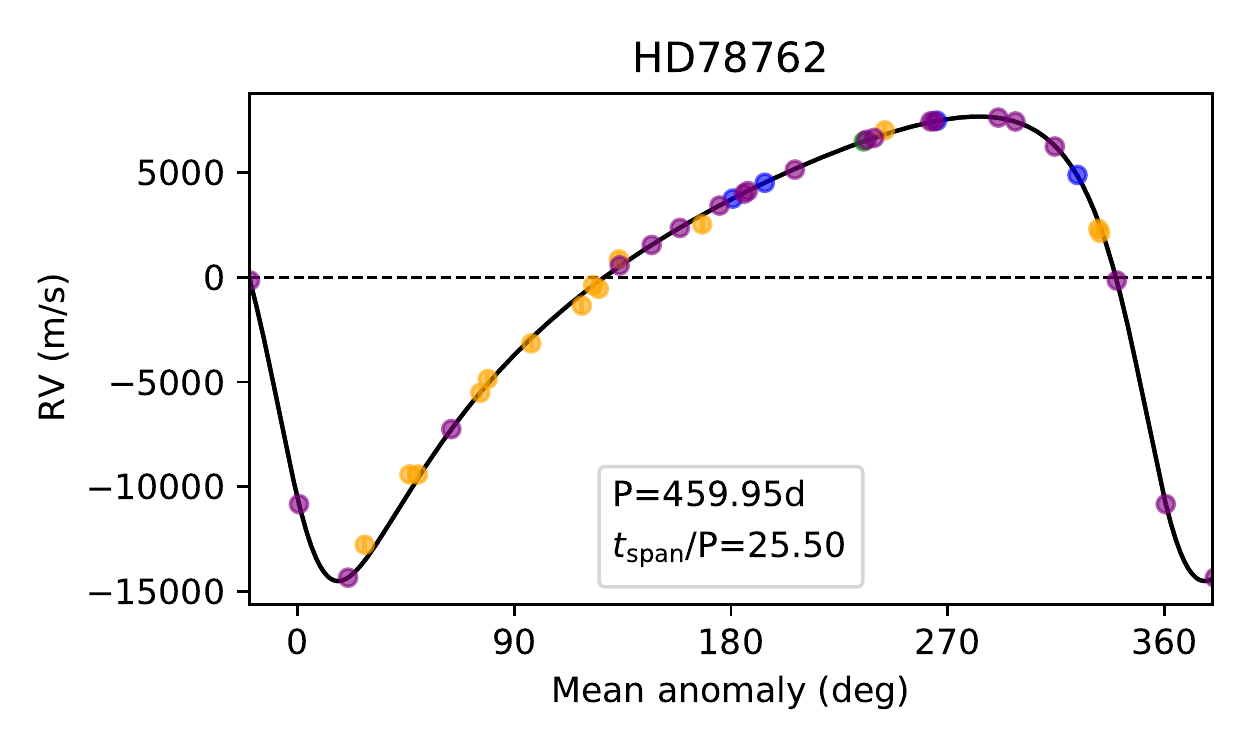}\\

		\includegraphics[width=0.22\linewidth]{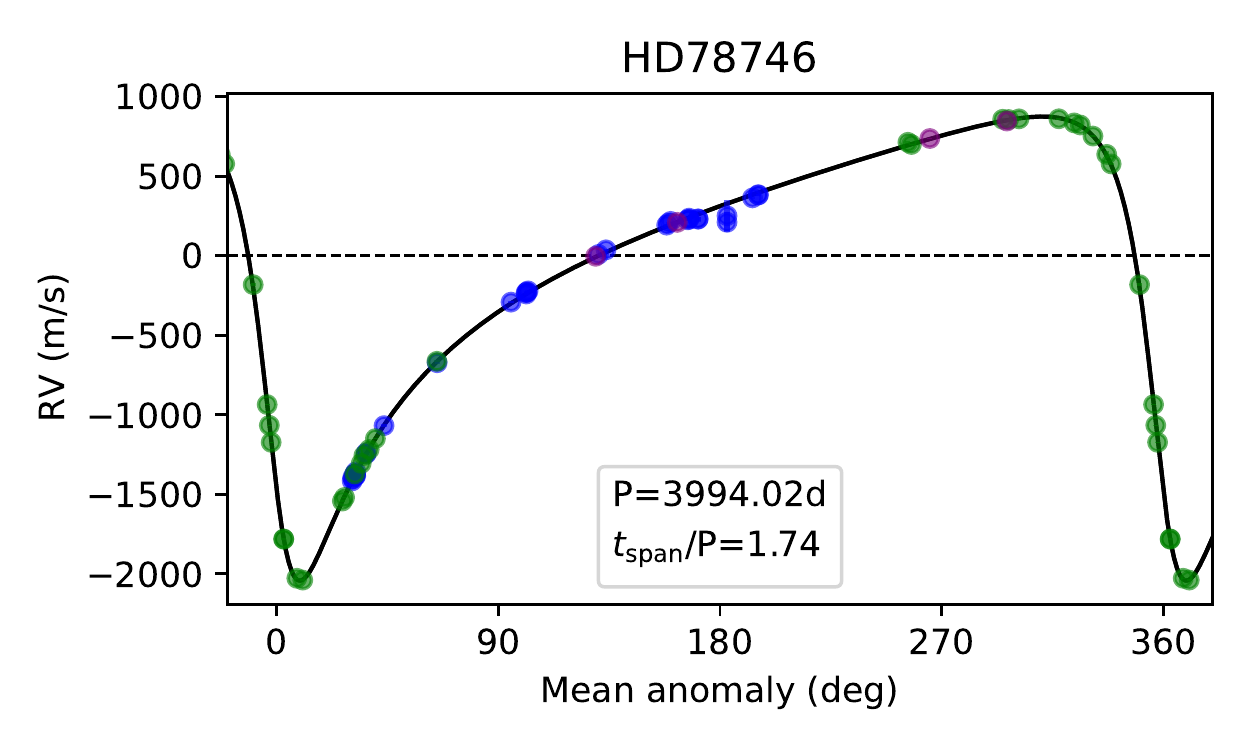}&
		\includegraphics[width=0.22\linewidth]{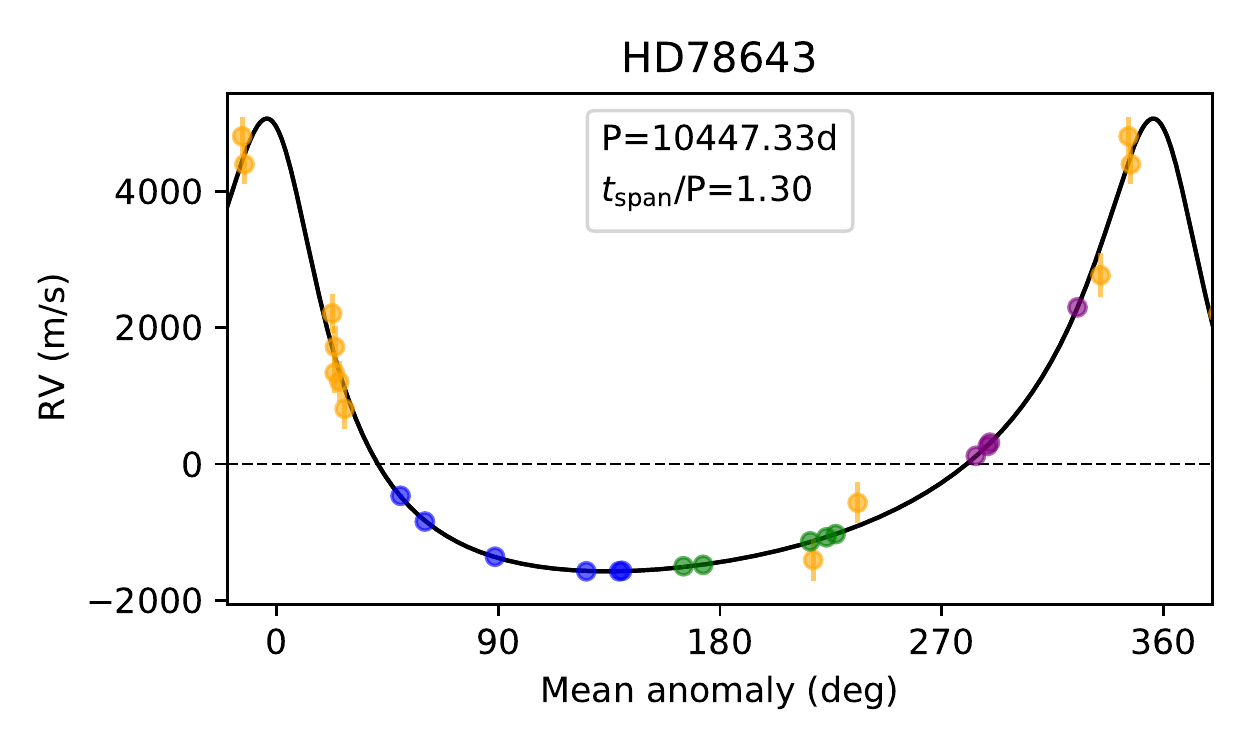}&
		\includegraphics[width=0.22\linewidth]{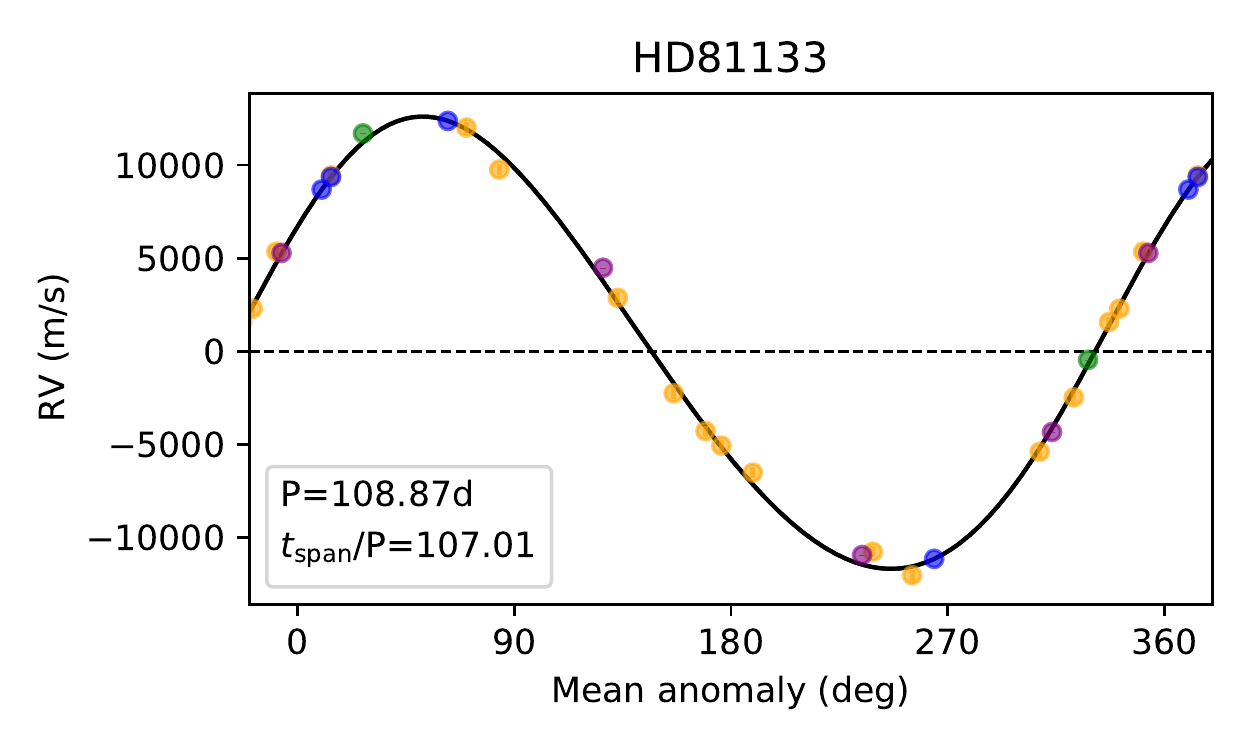}&
		\includegraphics[width=0.22\linewidth]{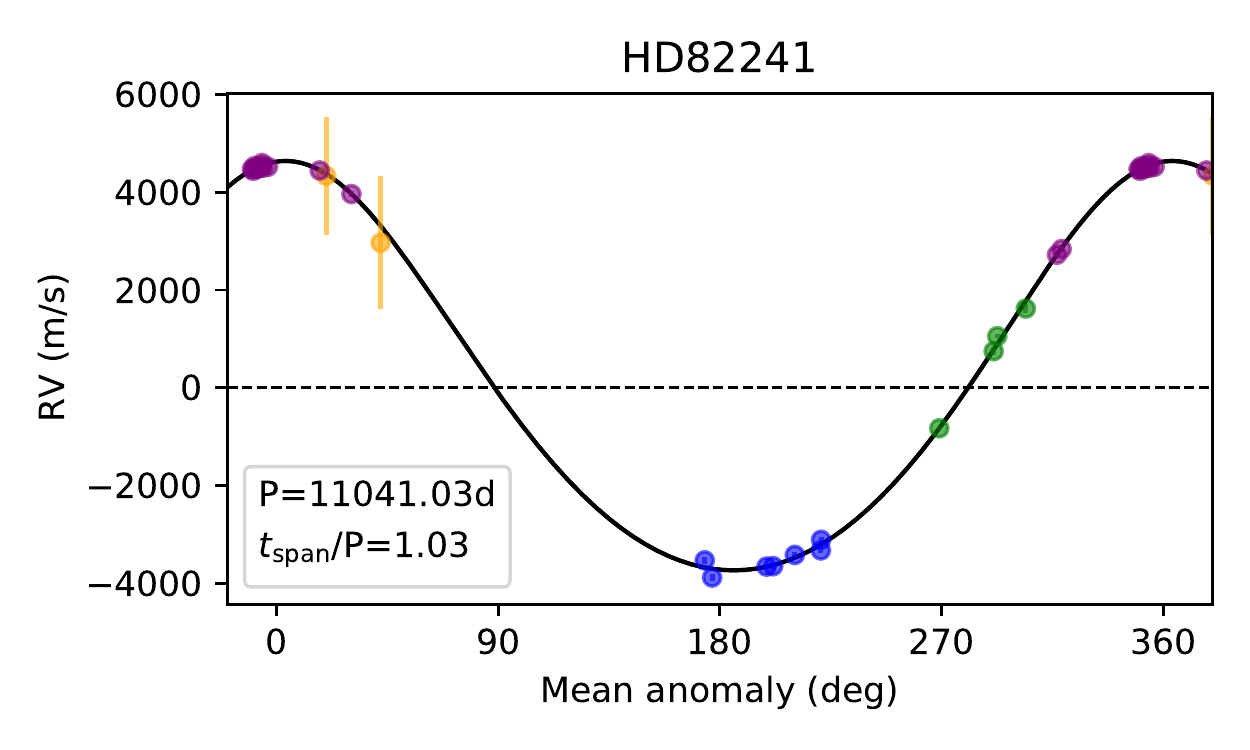}\\

		\includegraphics[width=0.22\linewidth]{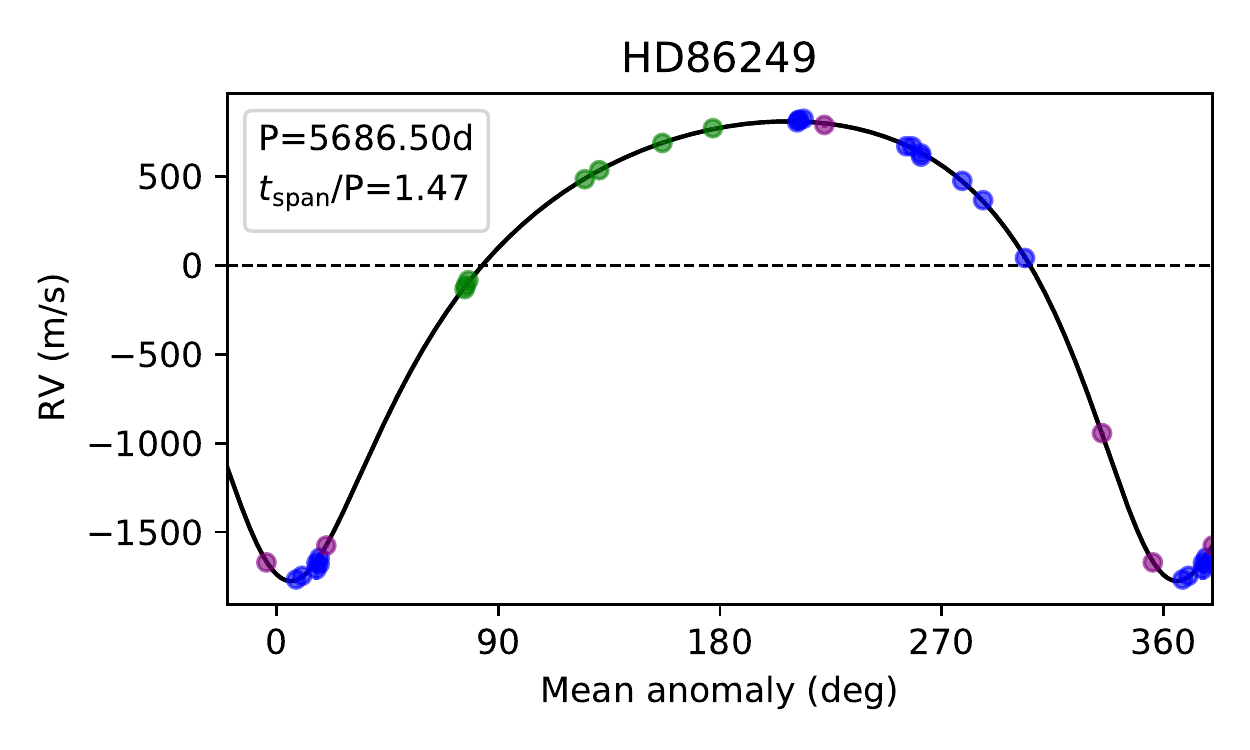}&
		\includegraphics[width=0.22\linewidth]{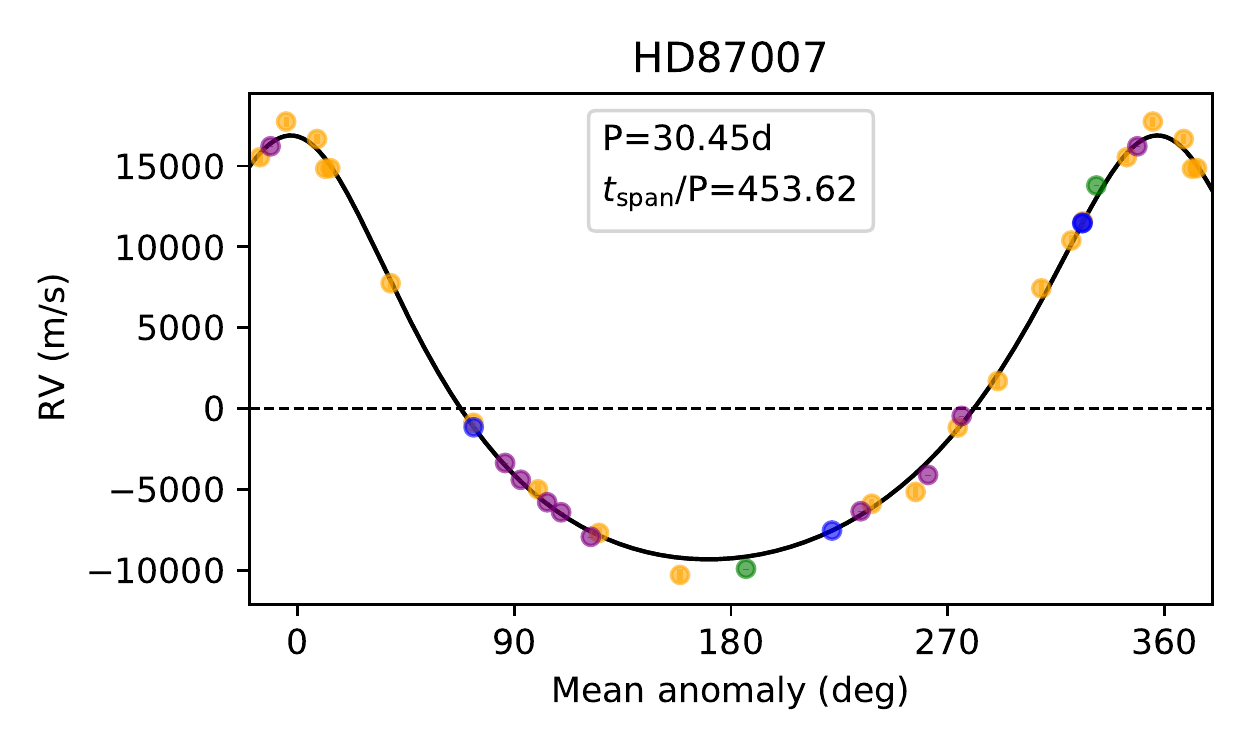}&
		\includegraphics[width=0.22\linewidth]{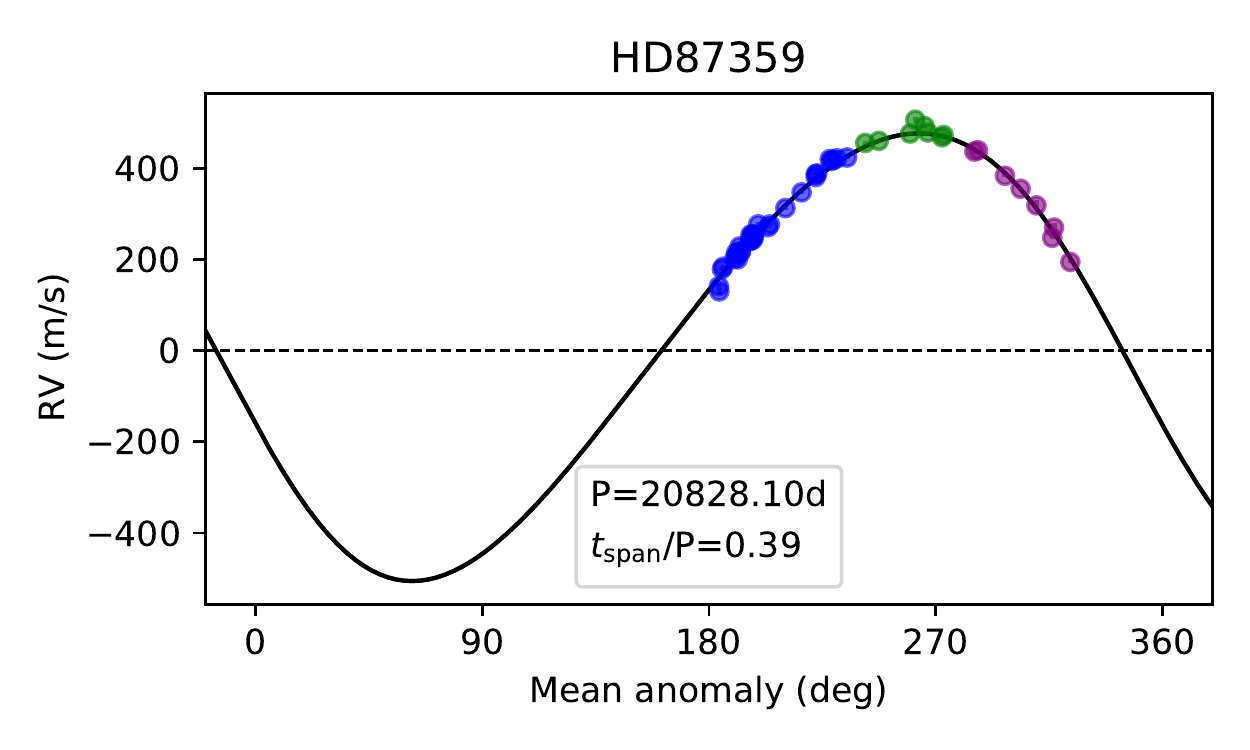}&
		\includegraphics[width=0.22\linewidth]{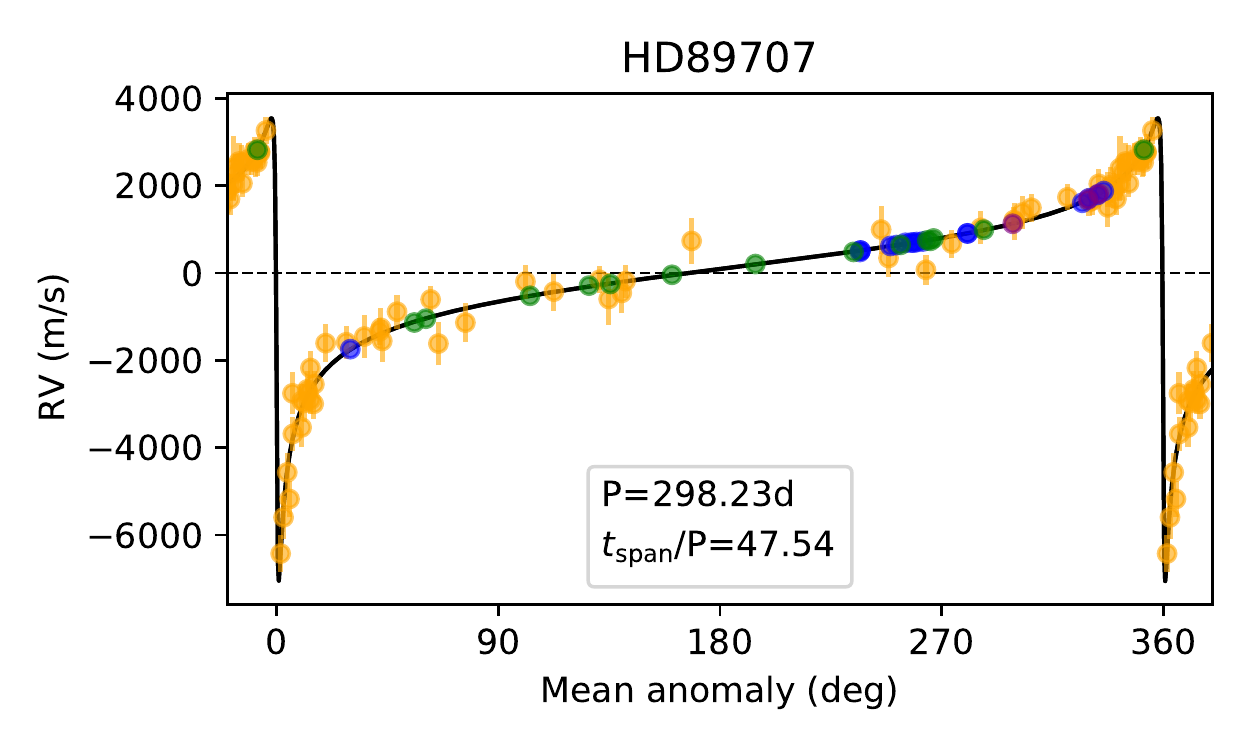}\\

		\includegraphics[width=0.22\linewidth]{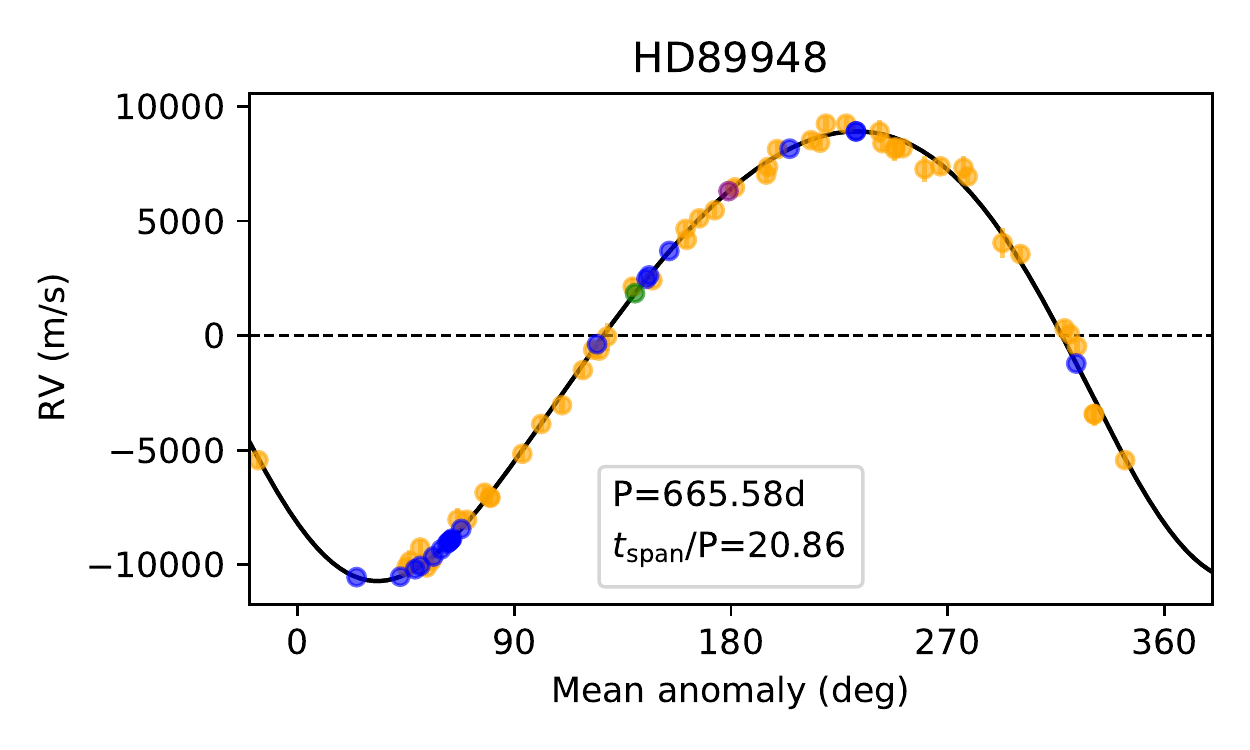}&
		\includegraphics[width=0.22\linewidth]{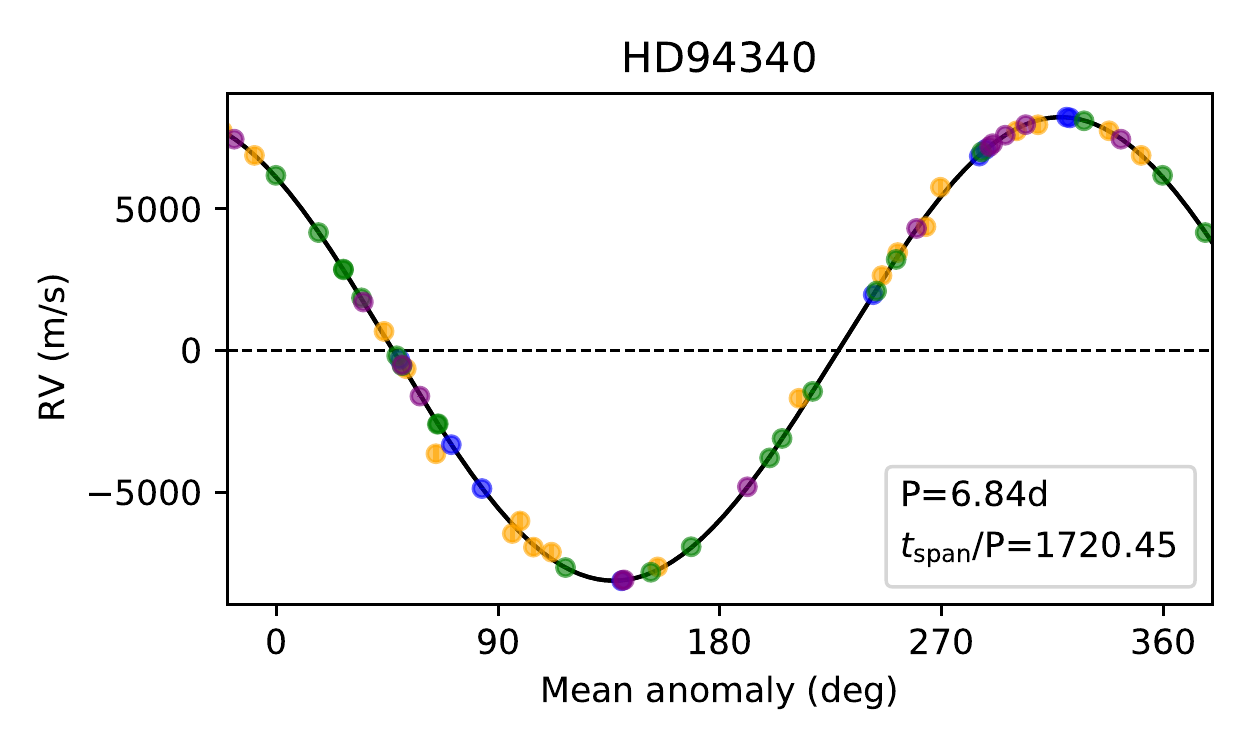}&
		\includegraphics[width=0.22\linewidth]{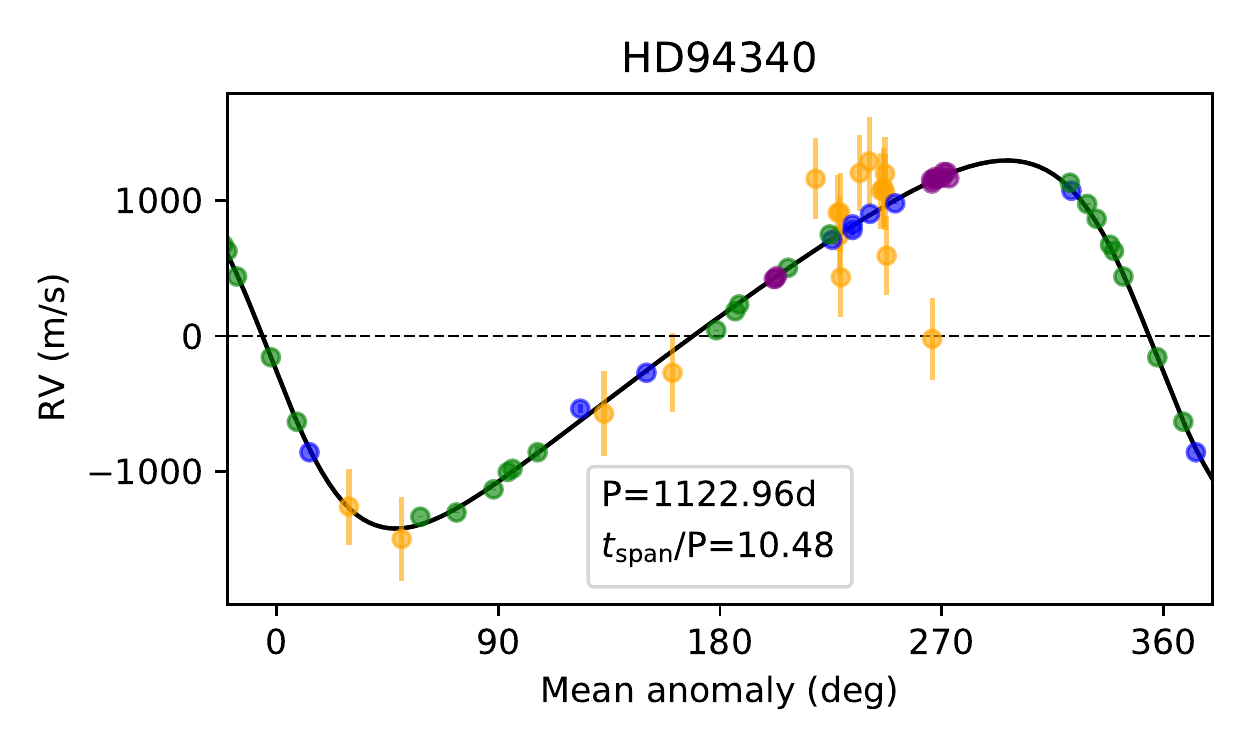}&
		\includegraphics[width=0.22\linewidth]{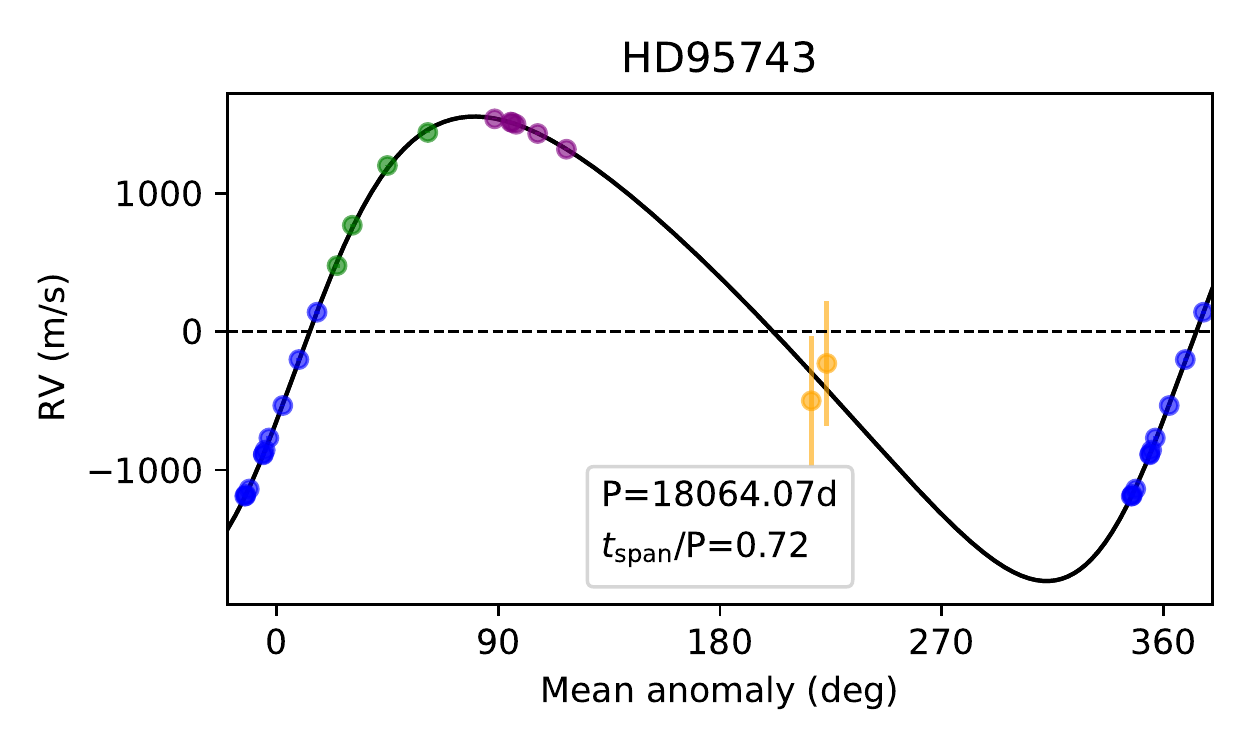}\\

		\includegraphics[width=0.22\linewidth]{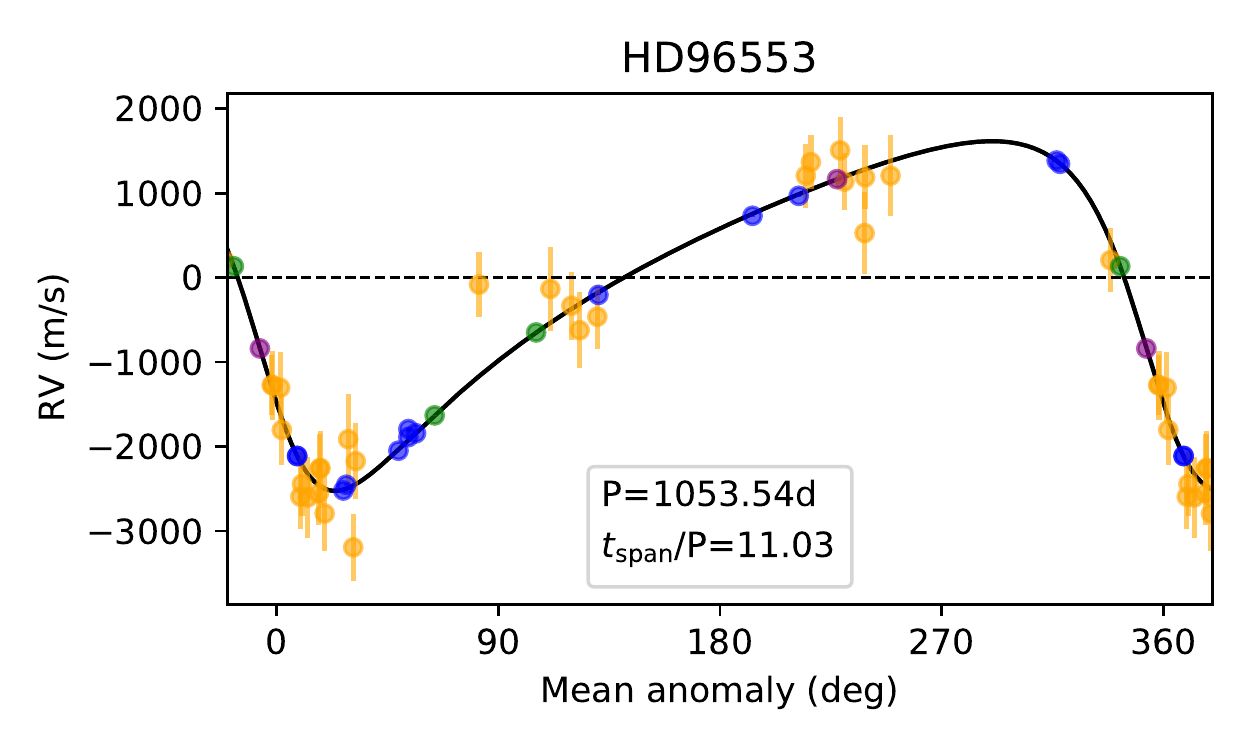}&
		\includegraphics[width=0.22\linewidth]{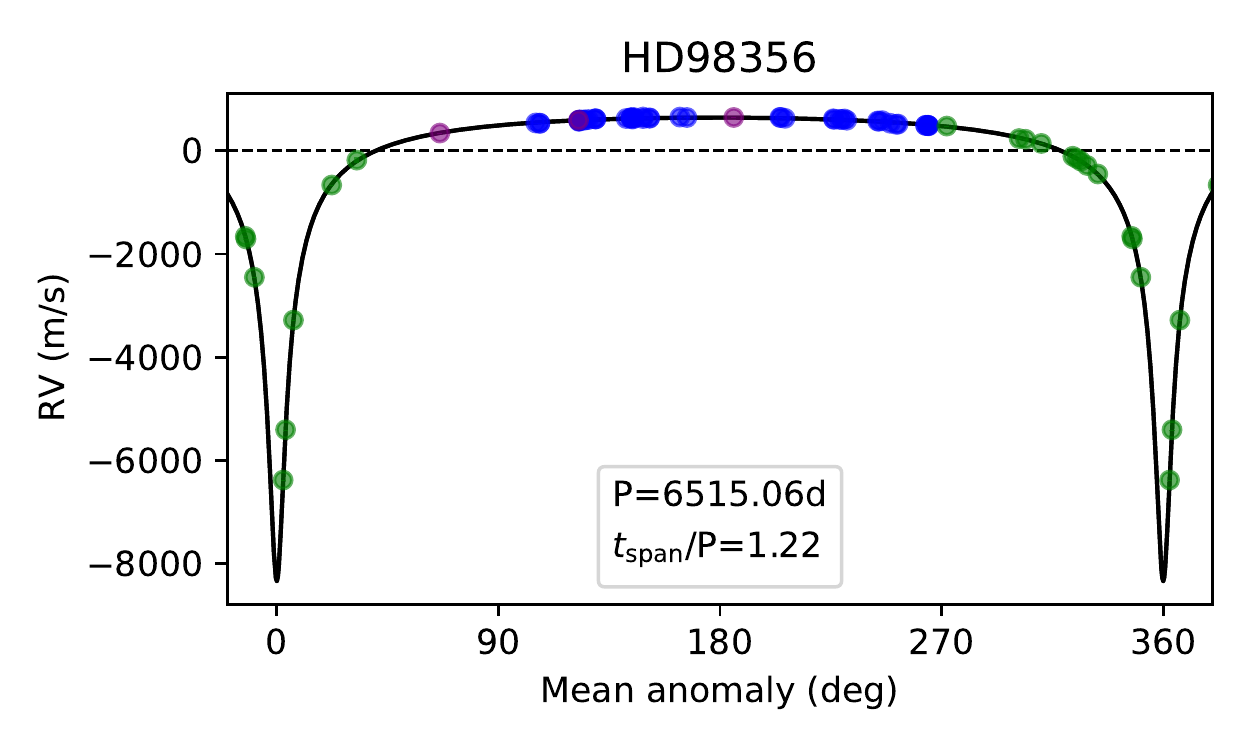}&
		\includegraphics[width=0.22\linewidth]{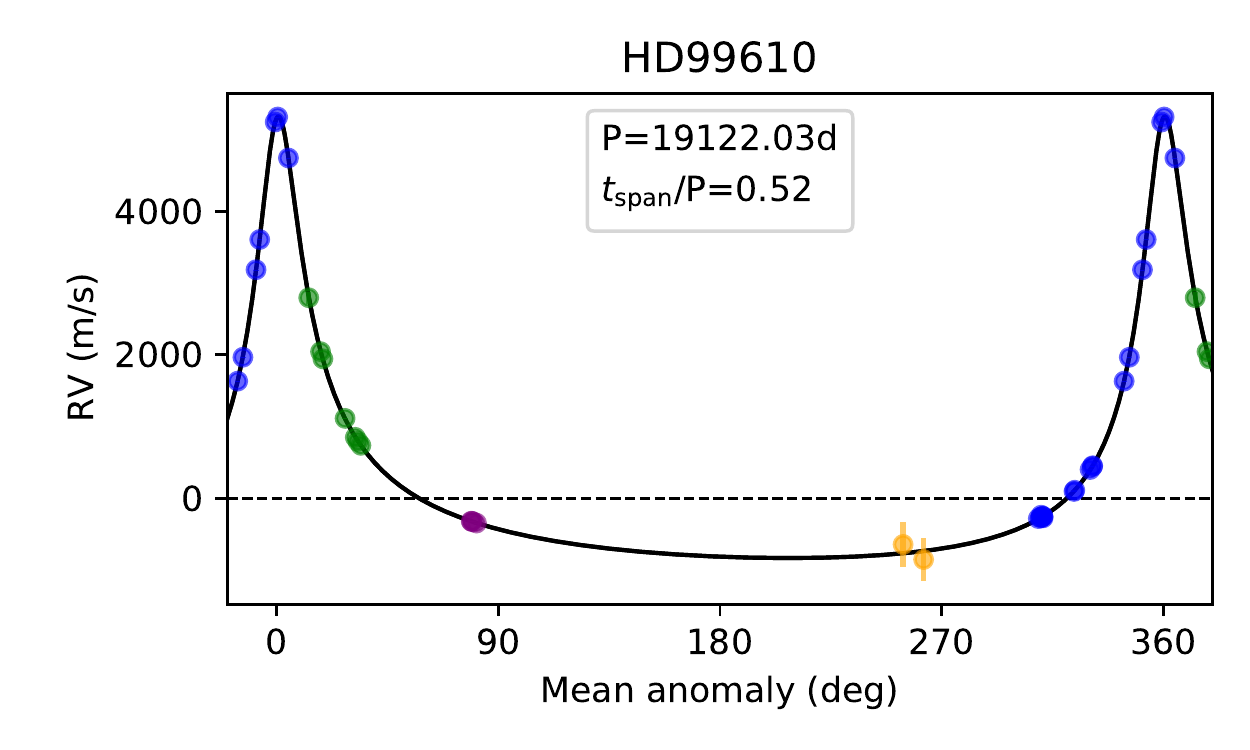}&
		\includegraphics[width=0.22\linewidth]{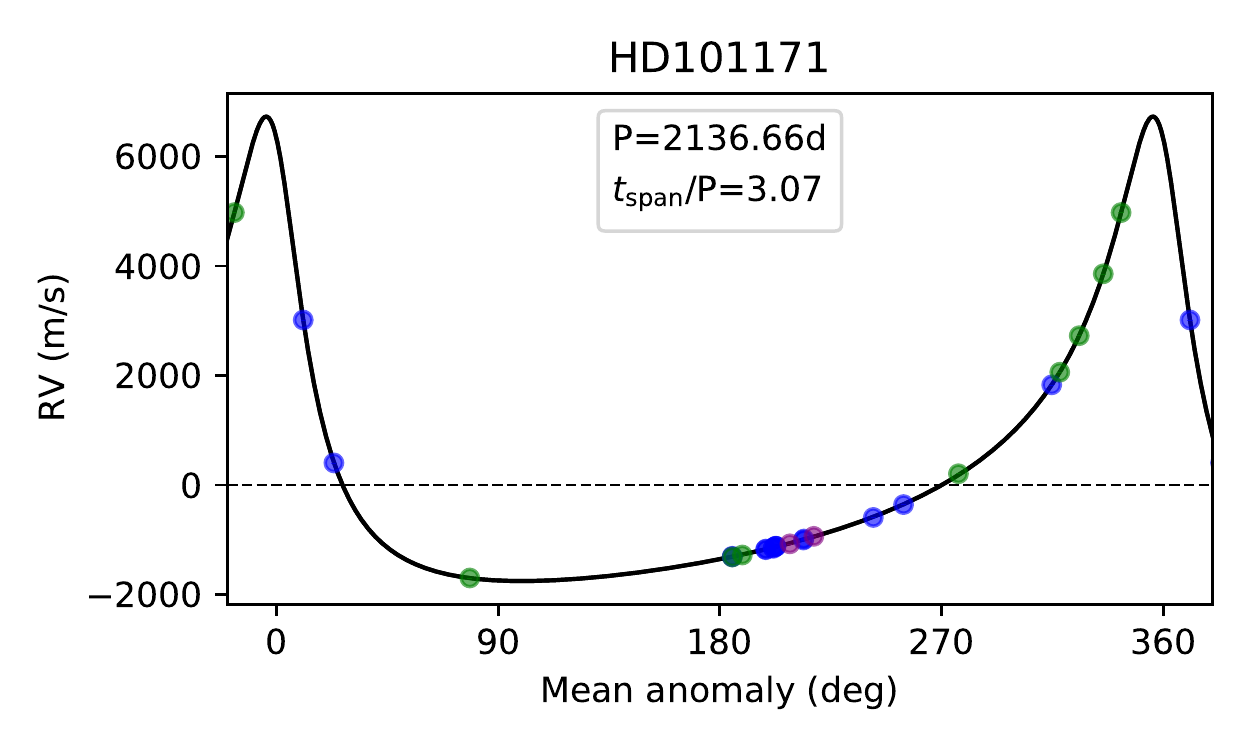}\\

		\includegraphics[width=0.22\linewidth]{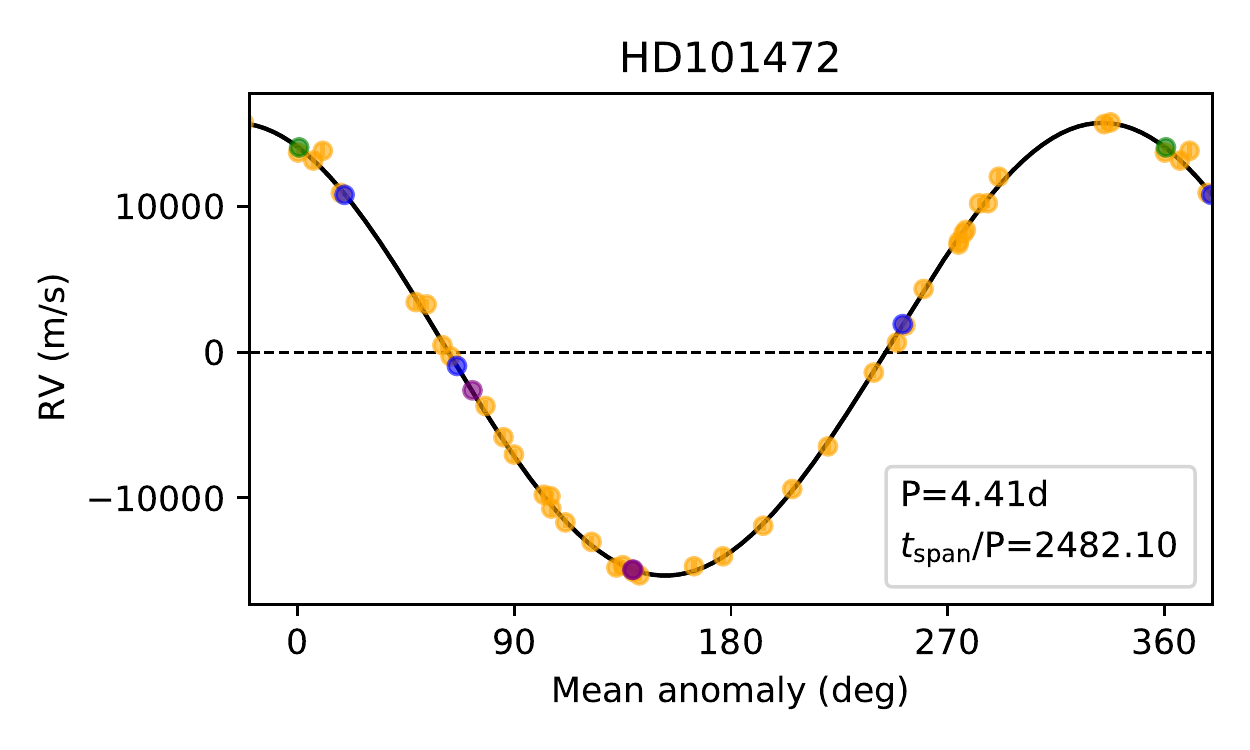}&
		\includegraphics[width=0.22\linewidth]{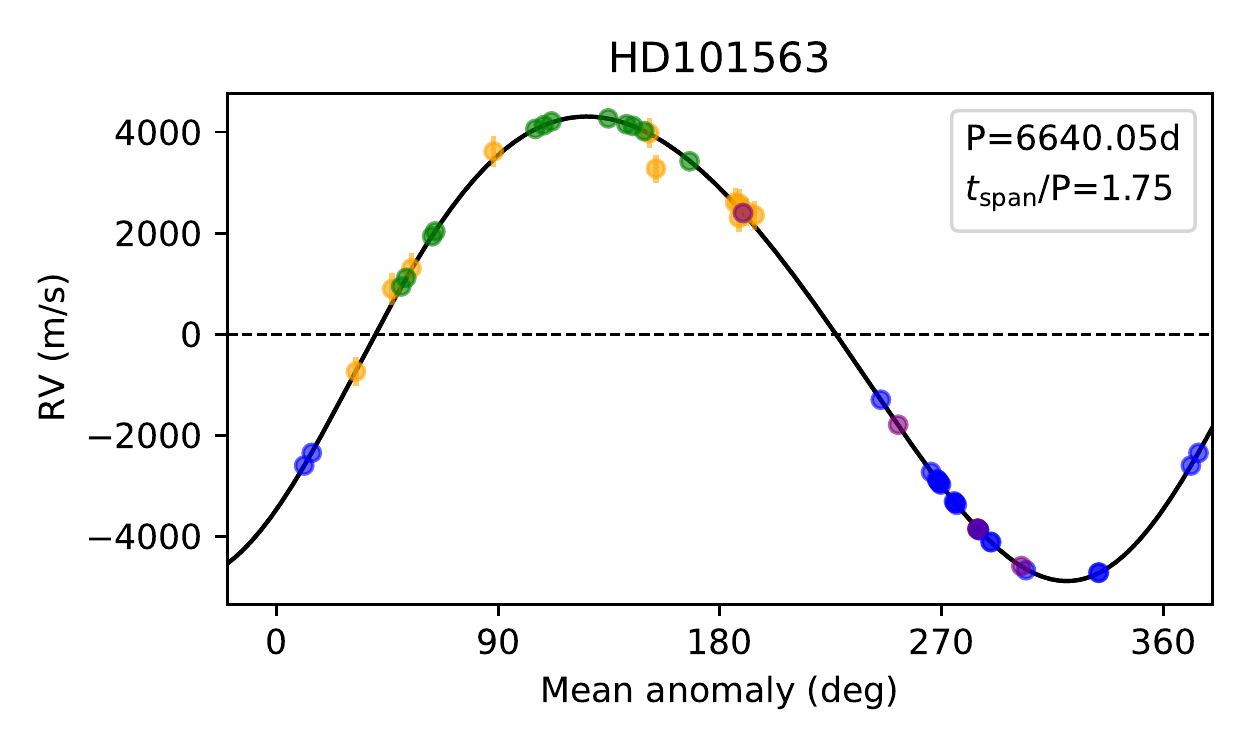}&
		\includegraphics[width=0.22\linewidth]{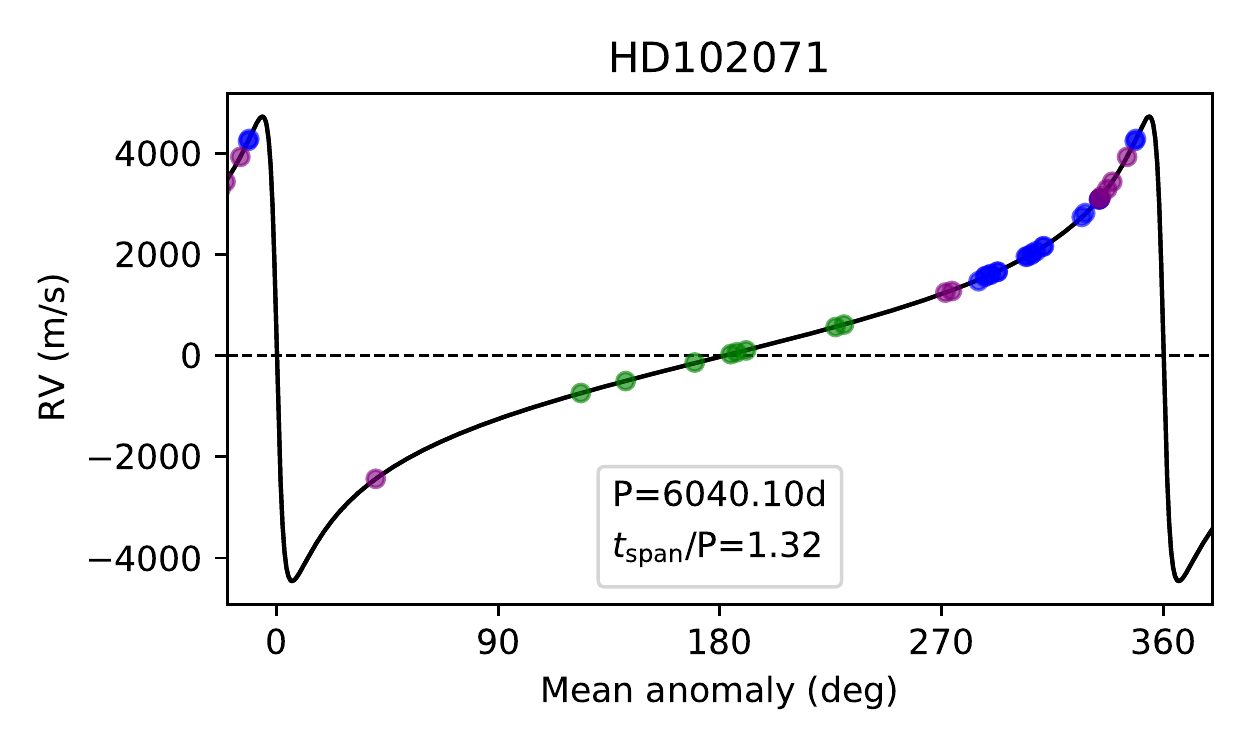}&
		\includegraphics[width=0.22\linewidth]{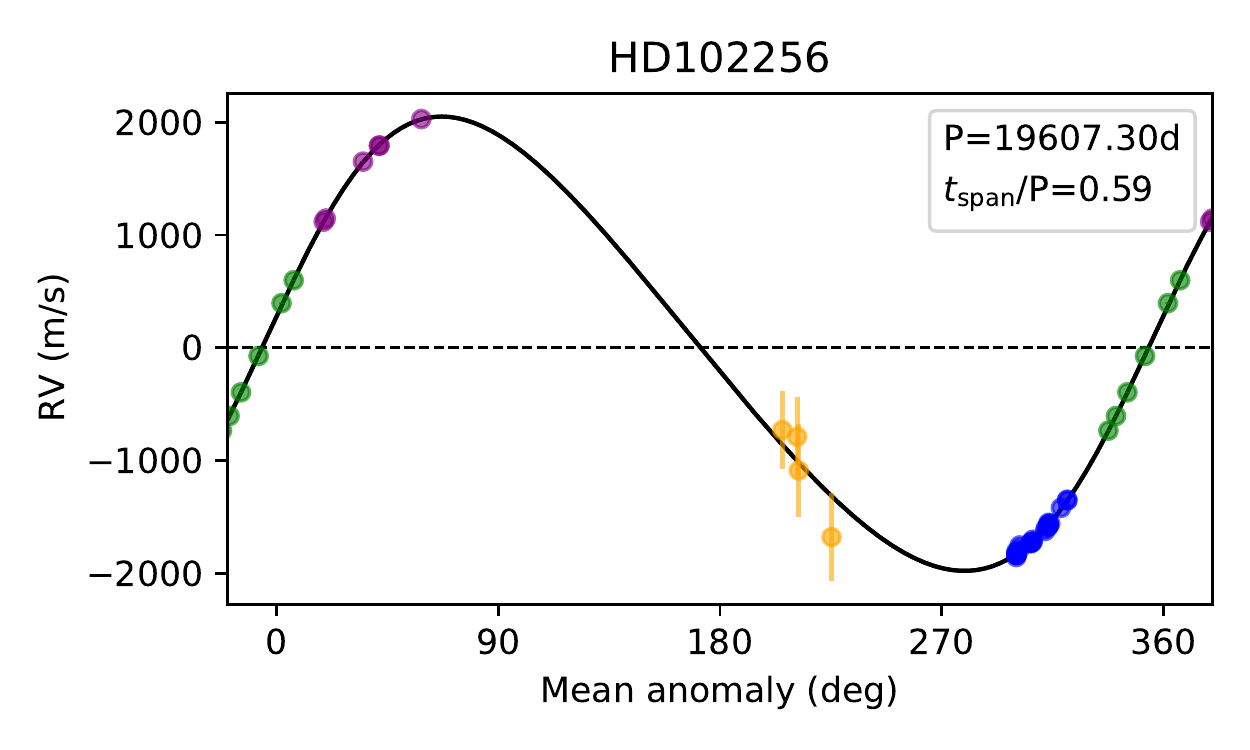}\\

		\includegraphics[width=0.22\linewidth]{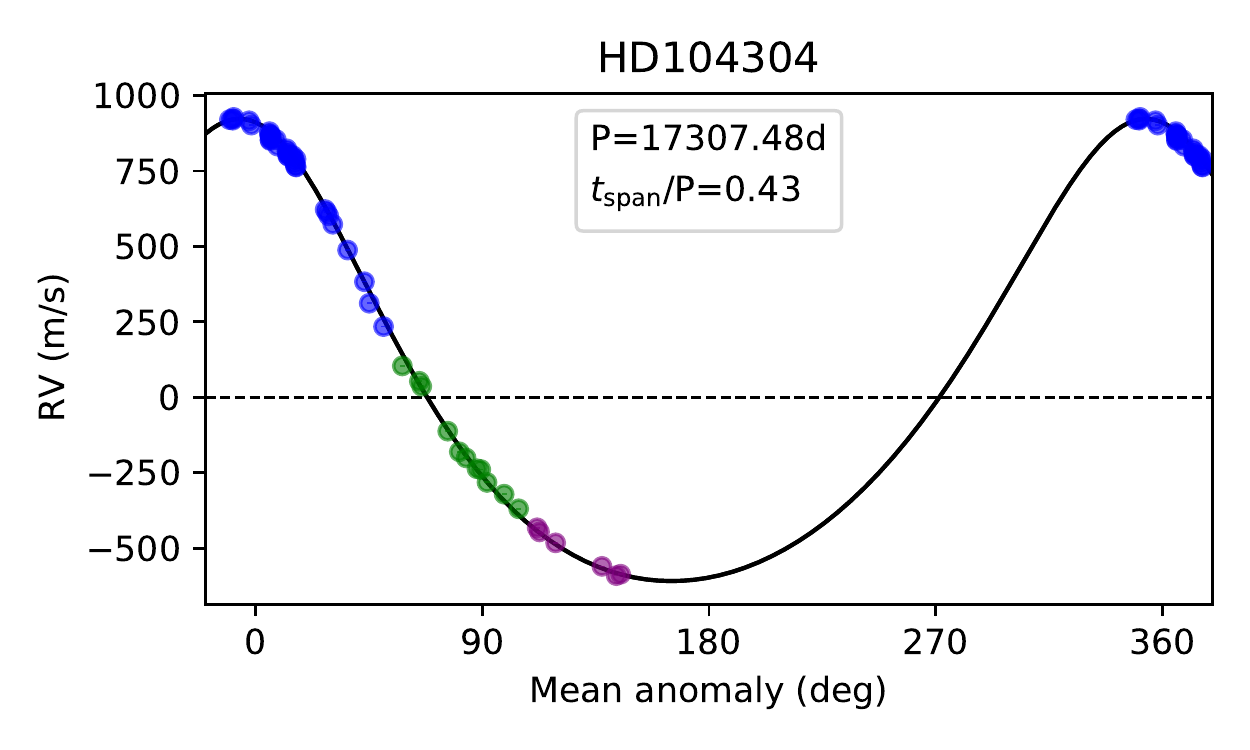}&
		\includegraphics[width=0.22\linewidth]{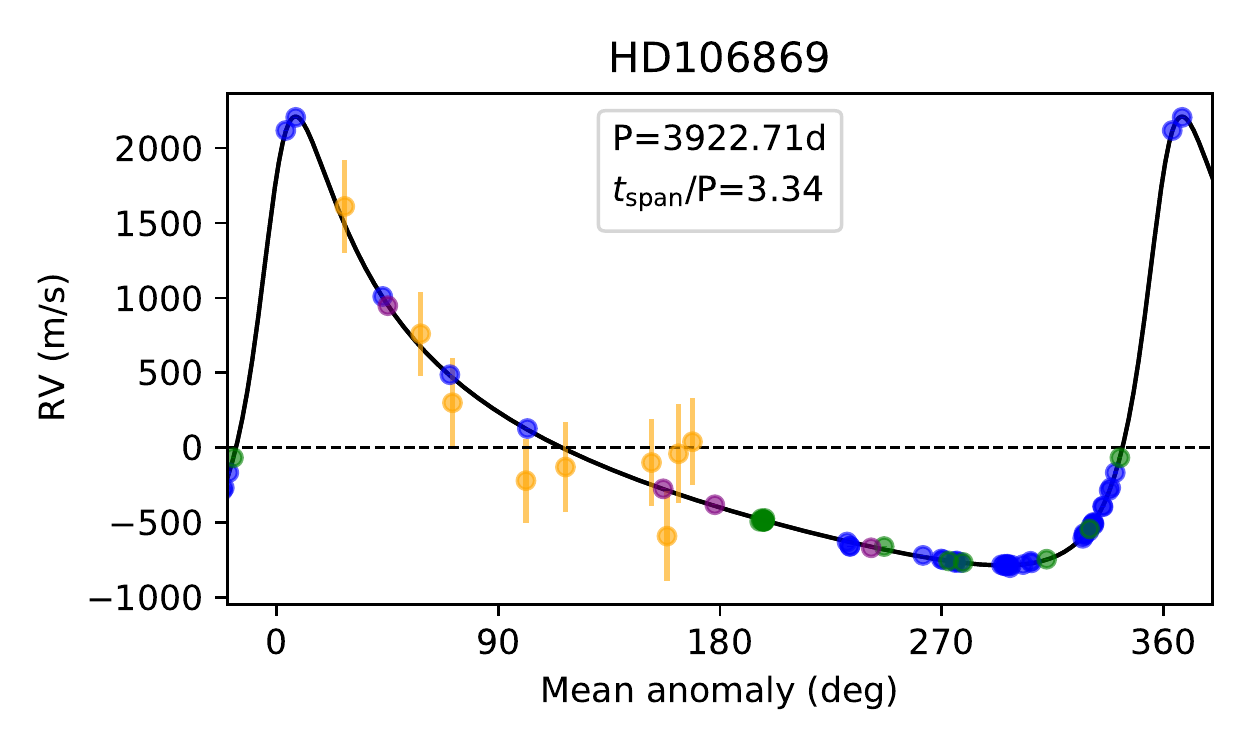}&
		\includegraphics[width=0.22\linewidth]{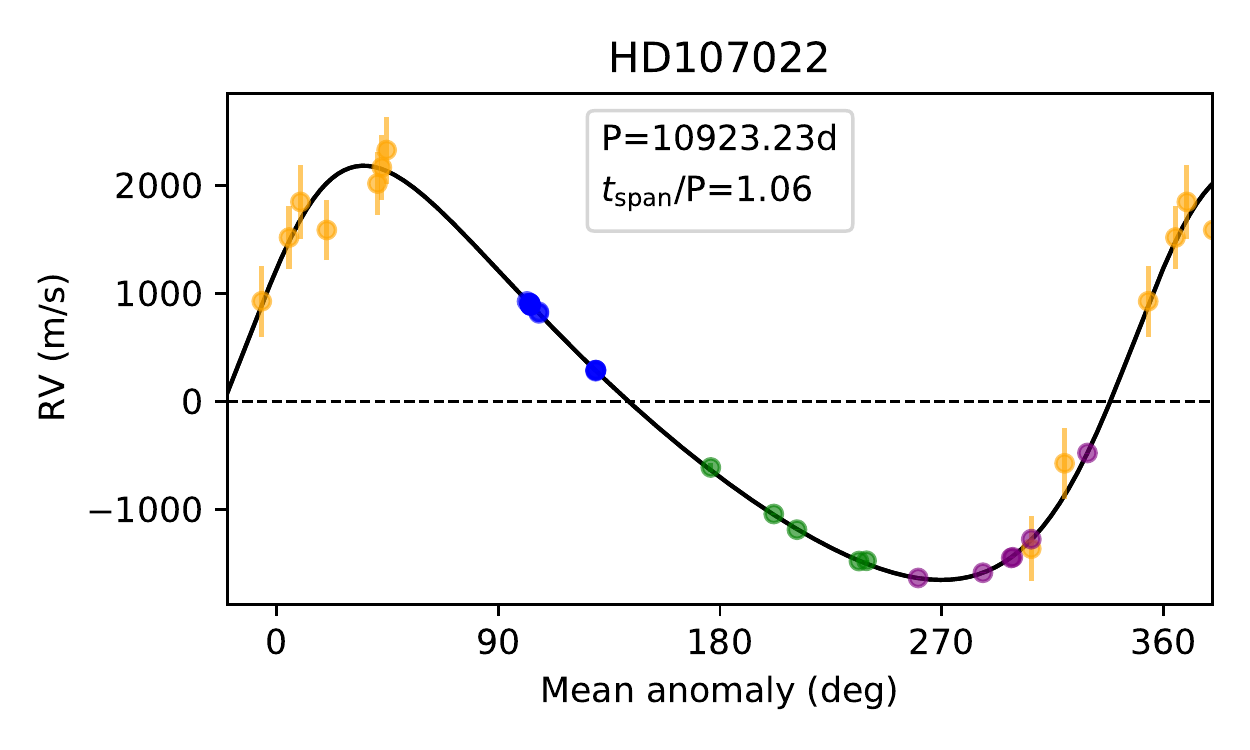}&
		\includegraphics[width=0.22\linewidth]{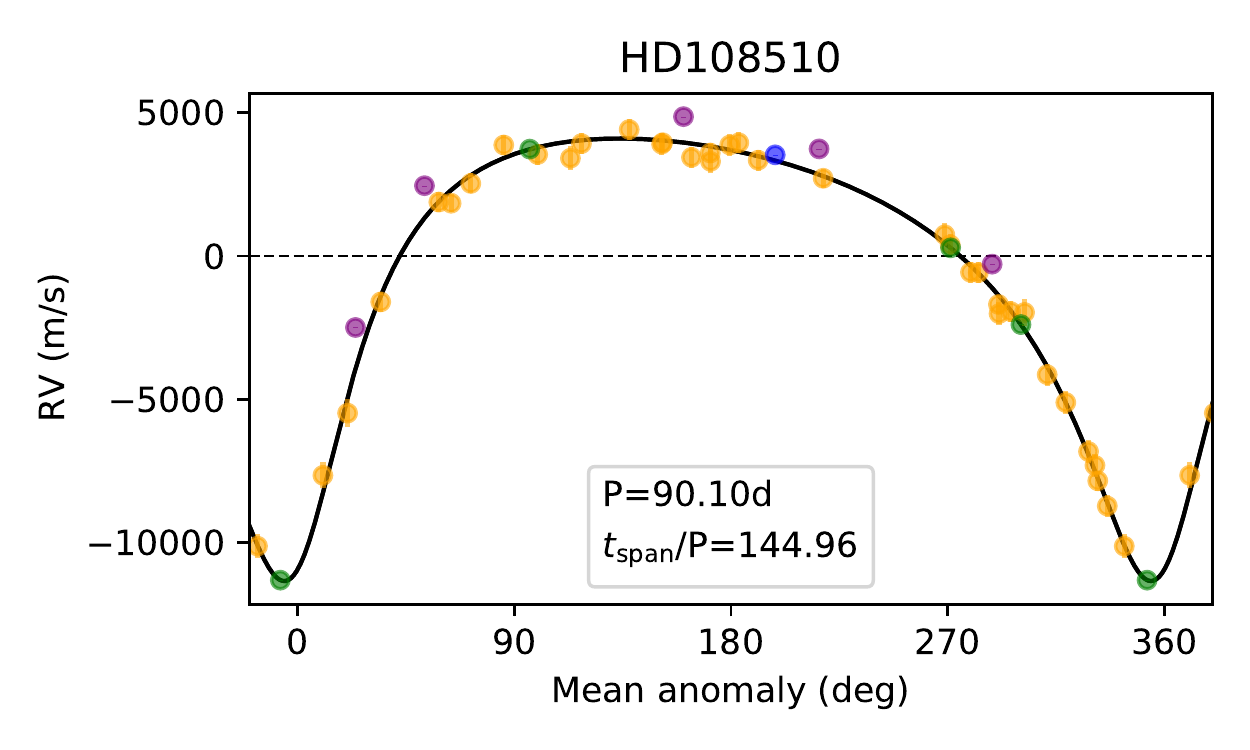}\\

		\includegraphics[width=0.22\linewidth]{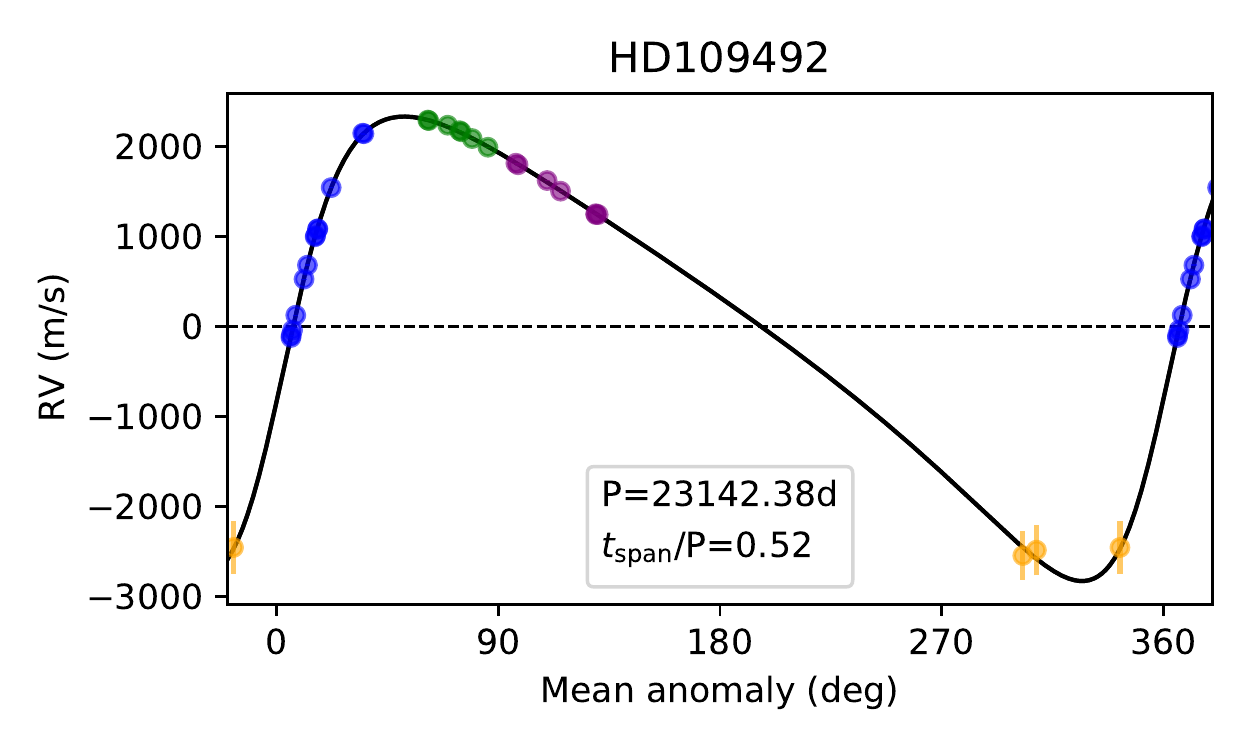}&
		\includegraphics[width=0.22\linewidth]{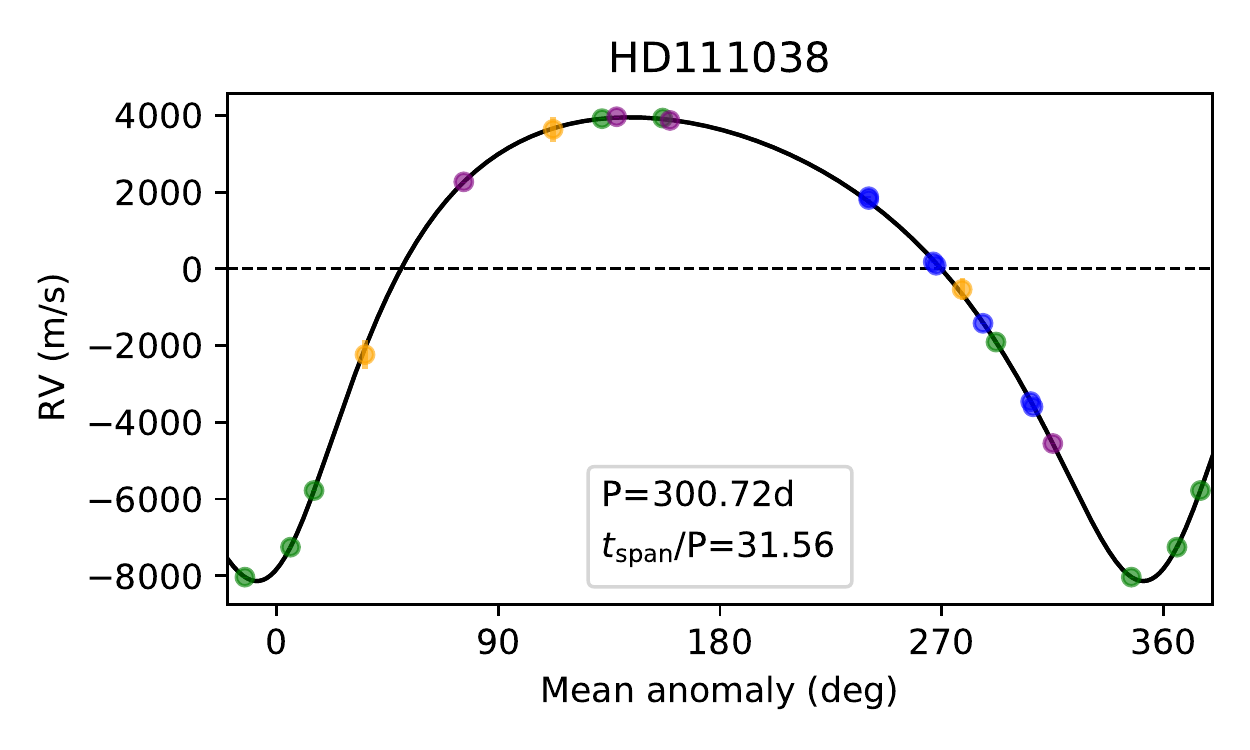}&
		\includegraphics[width=0.22\linewidth]{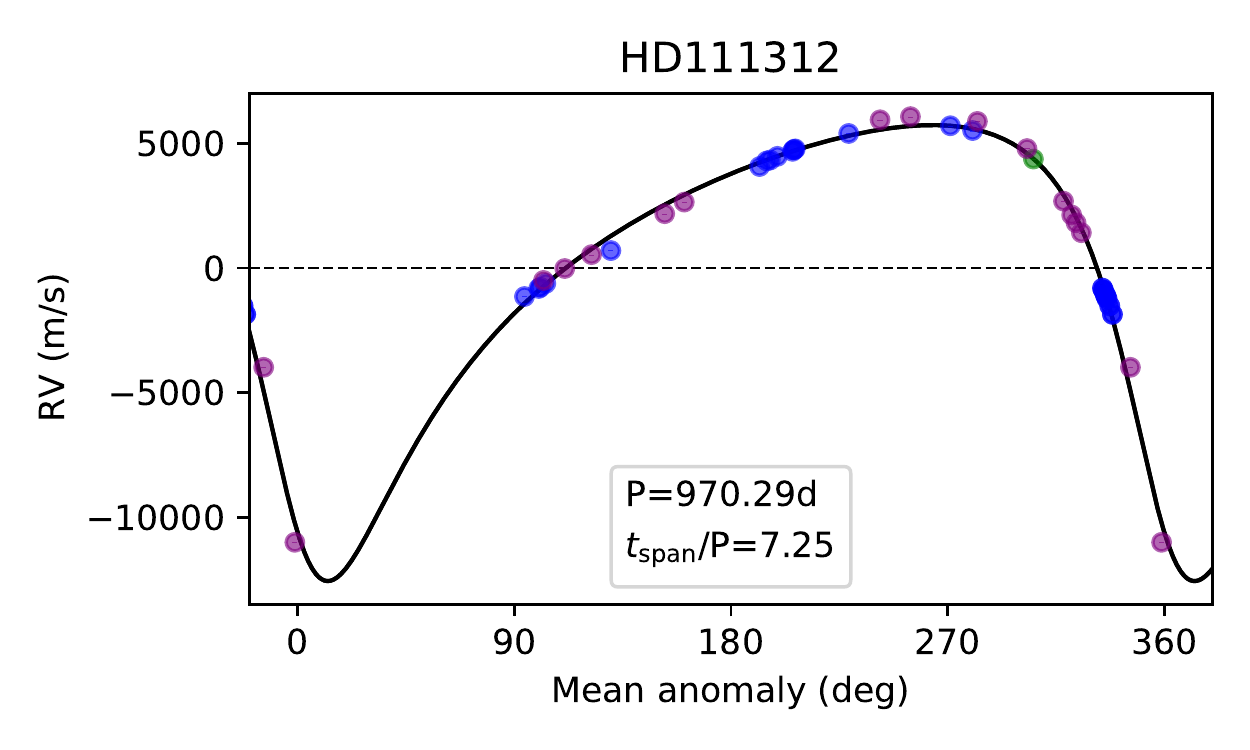}&
		\includegraphics[width=0.22\linewidth]{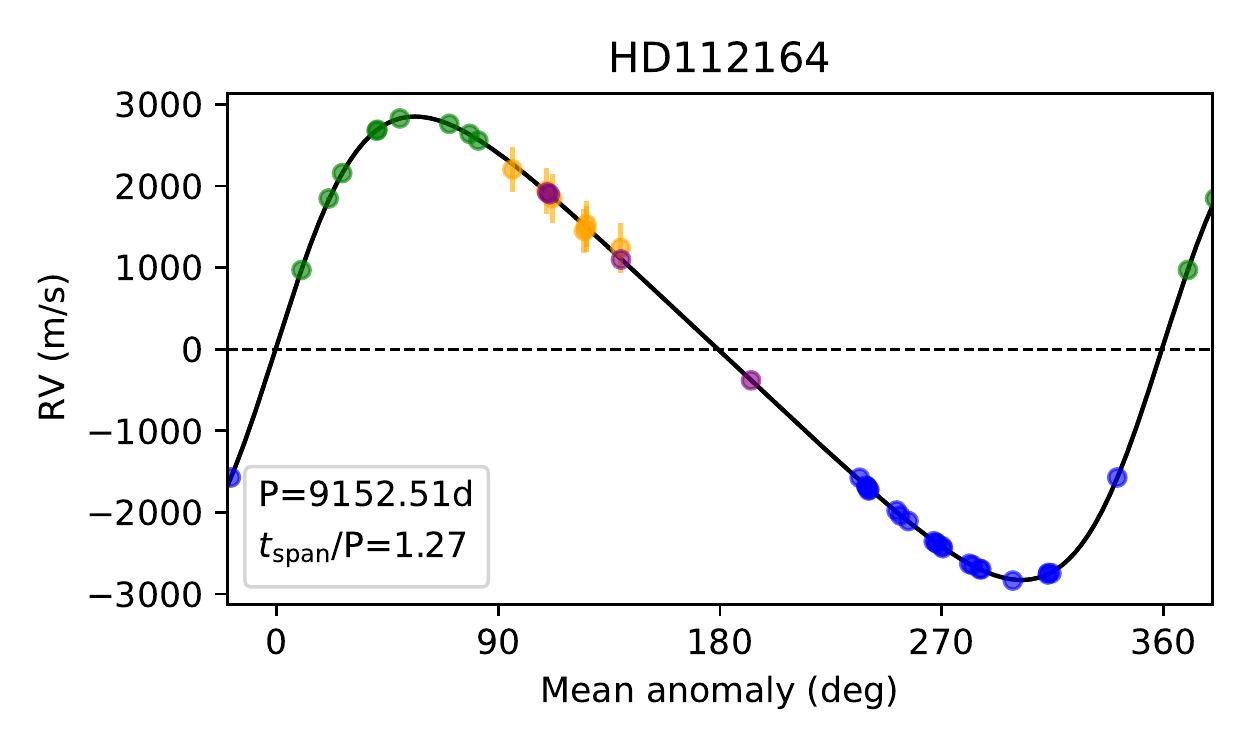}\\

		\includegraphics[width=0.22\linewidth]{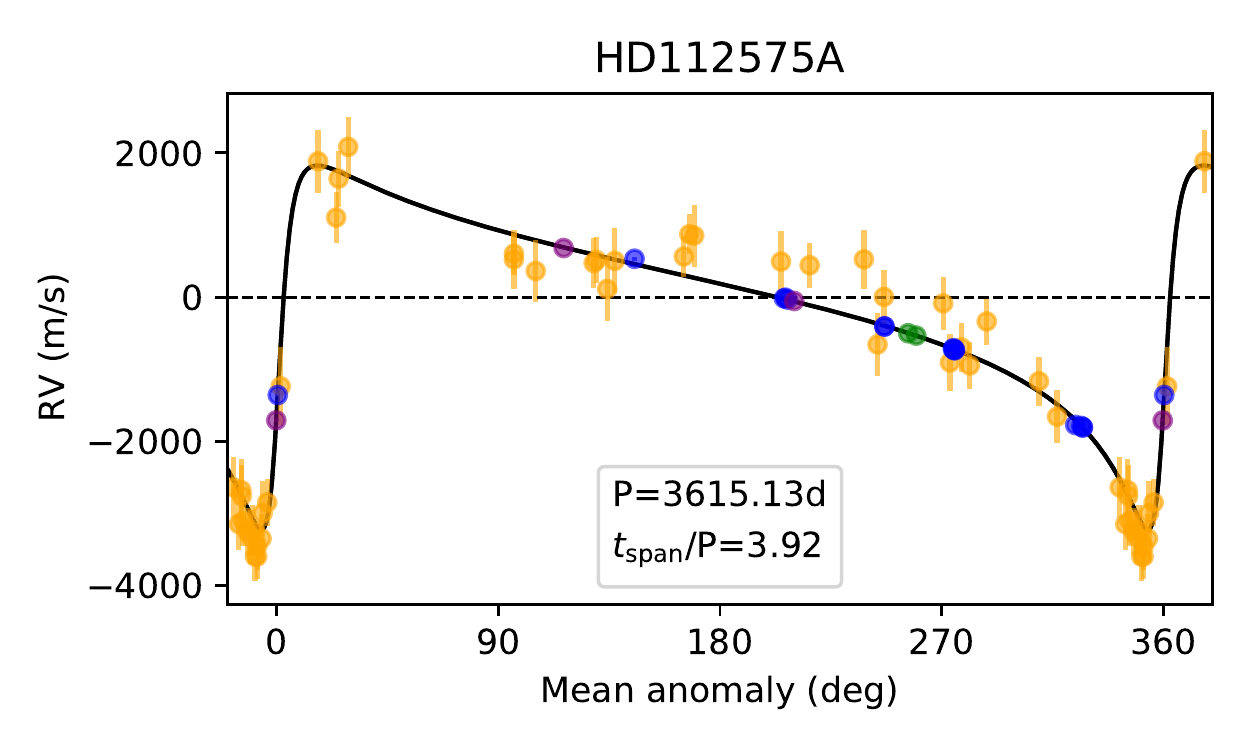}&
		\includegraphics[width=0.22\linewidth]{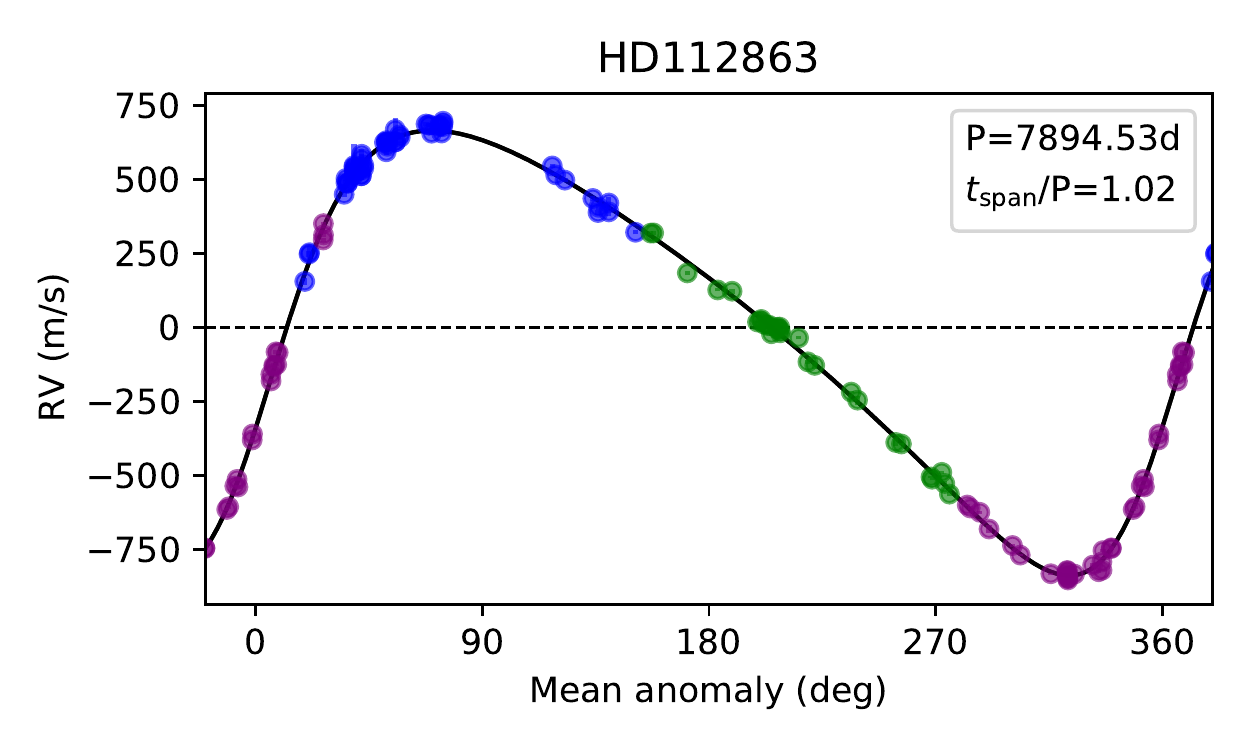}&
		\includegraphics[width=0.22\linewidth]{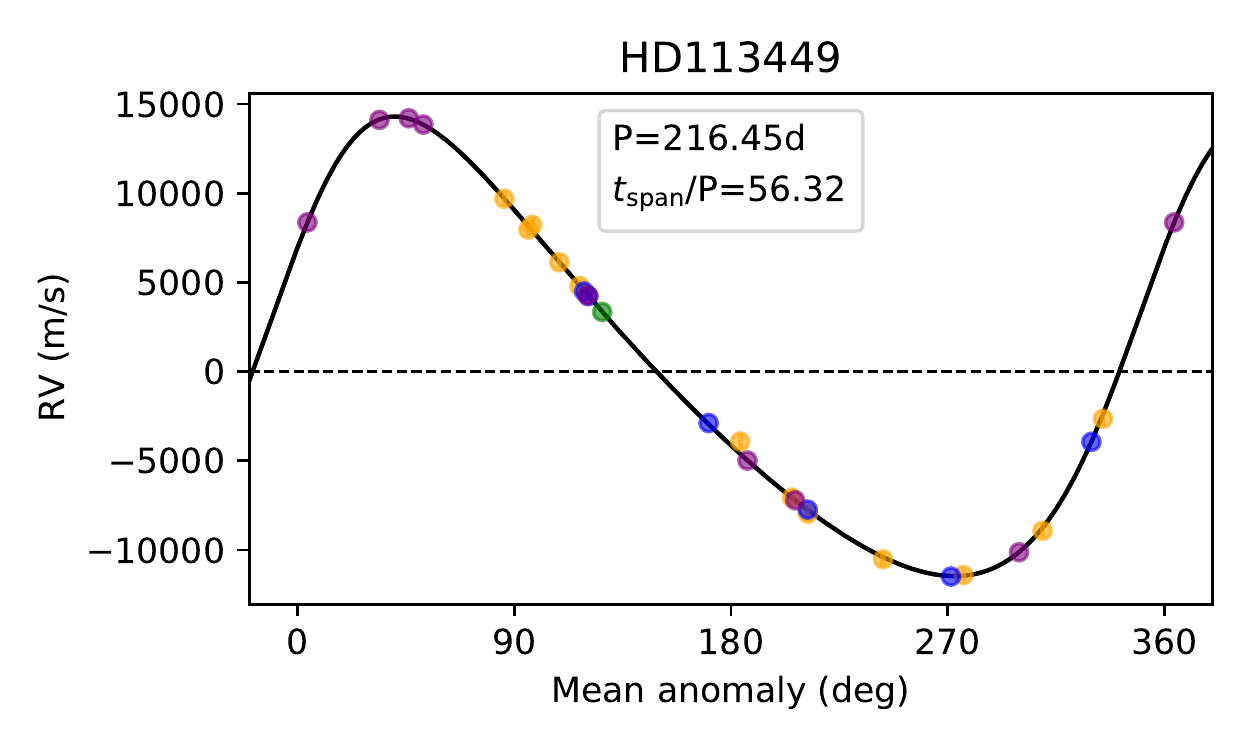}&
		\includegraphics[width=0.22\linewidth]{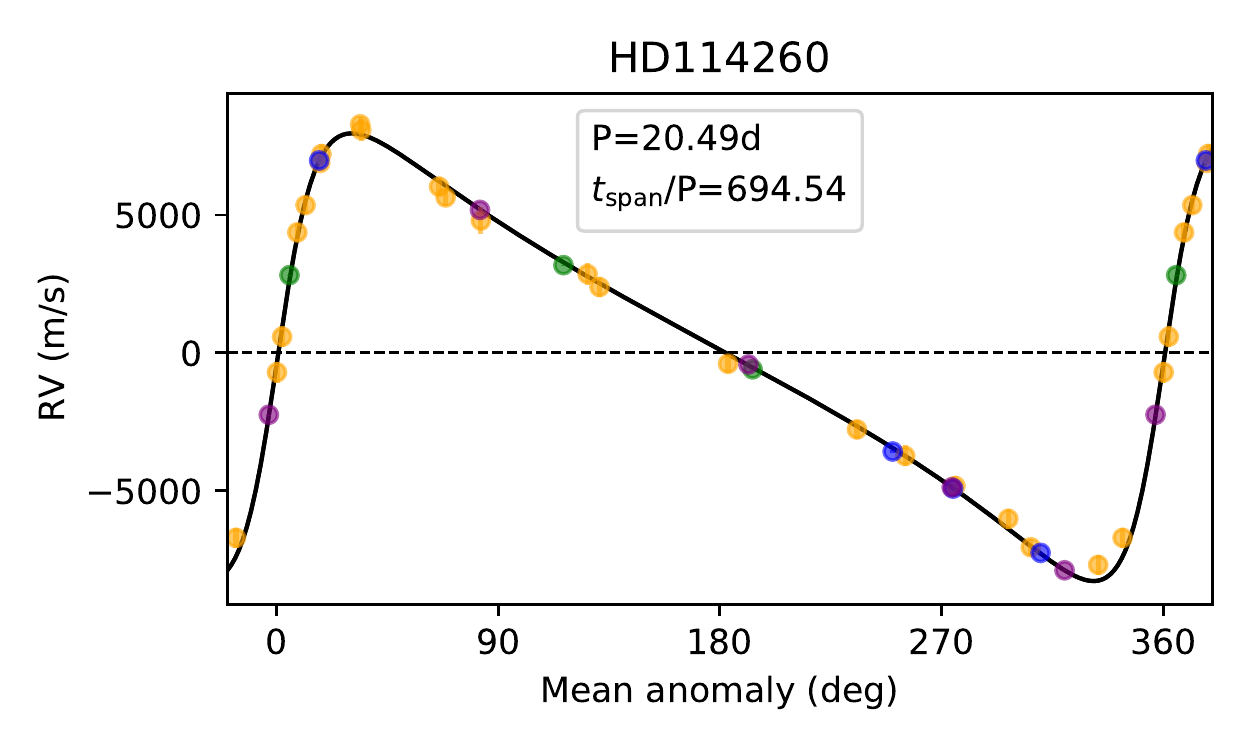}\\

		\includegraphics[width=0.22\linewidth]{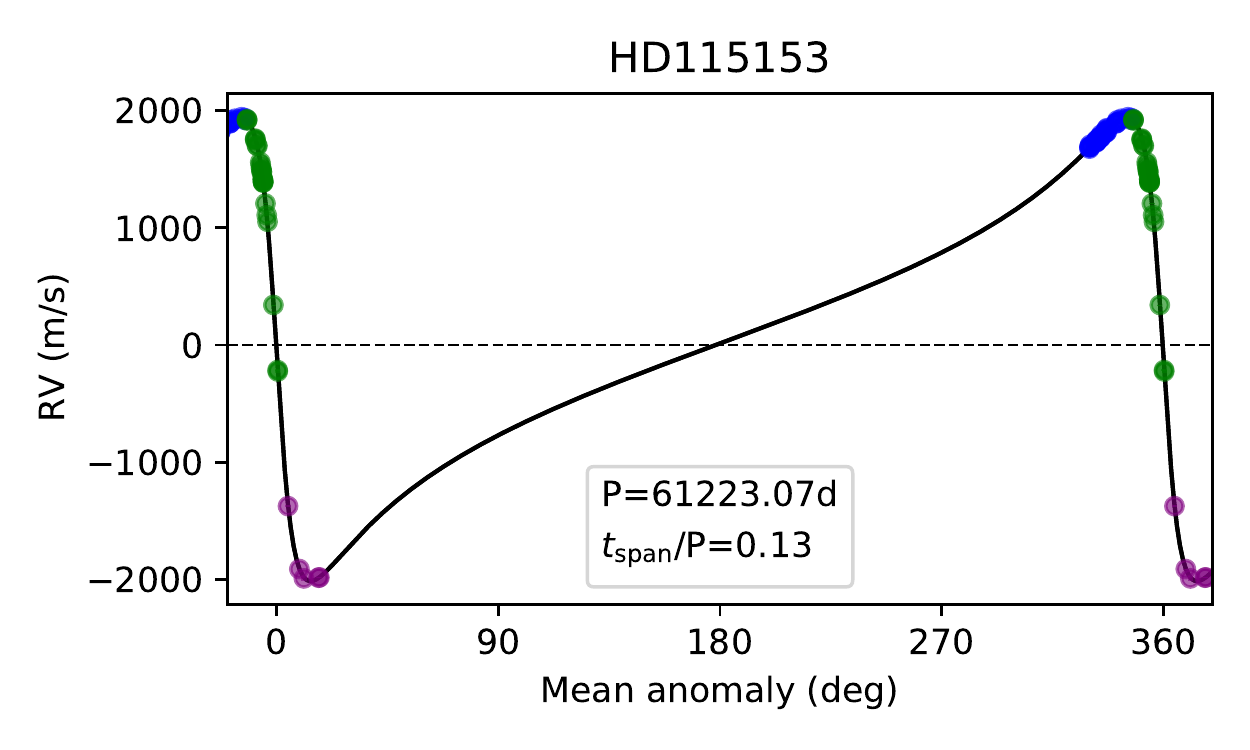}&
		\includegraphics[width=0.22\linewidth]{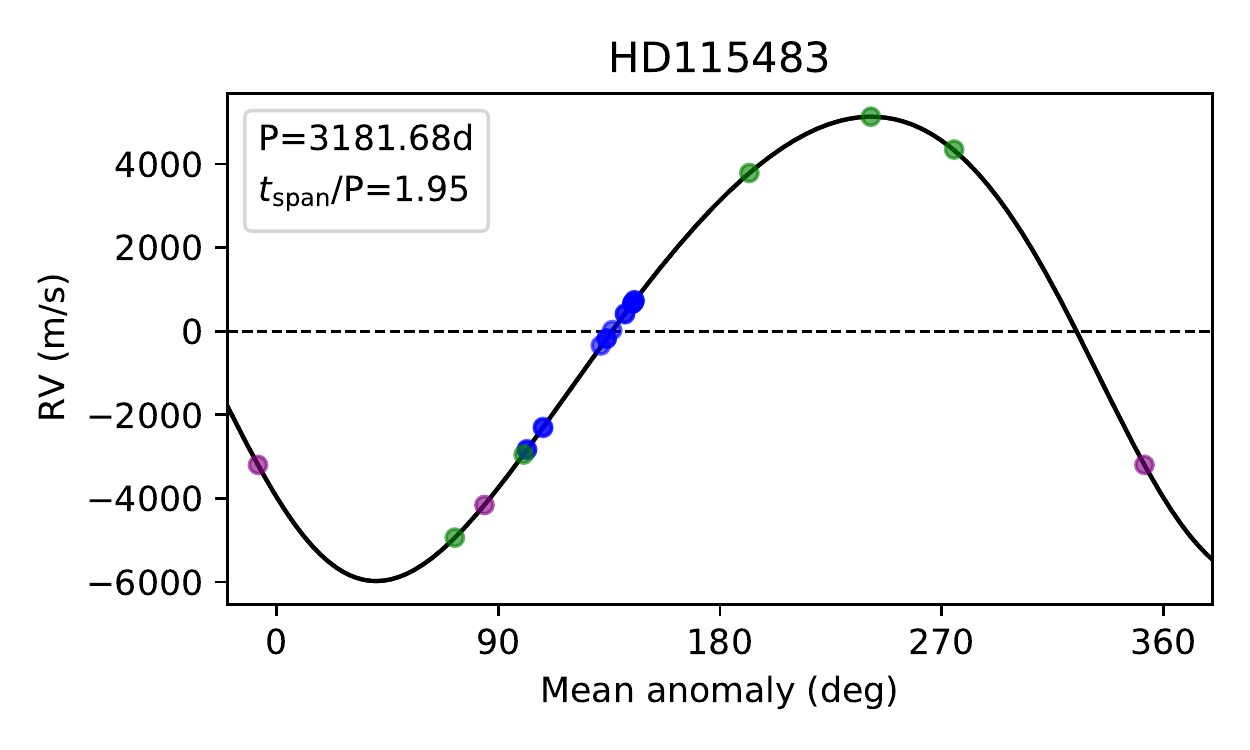}&
		\includegraphics[width=0.22\linewidth]{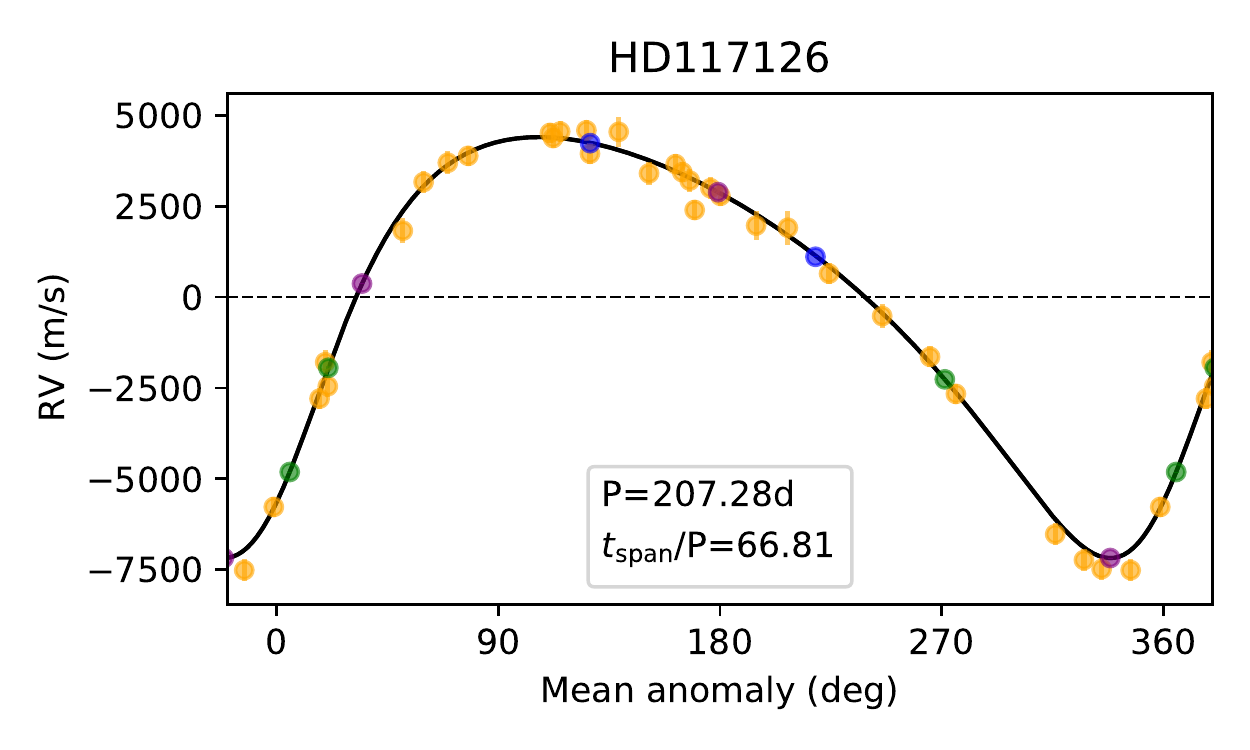}&
		\includegraphics[width=0.22\linewidth]{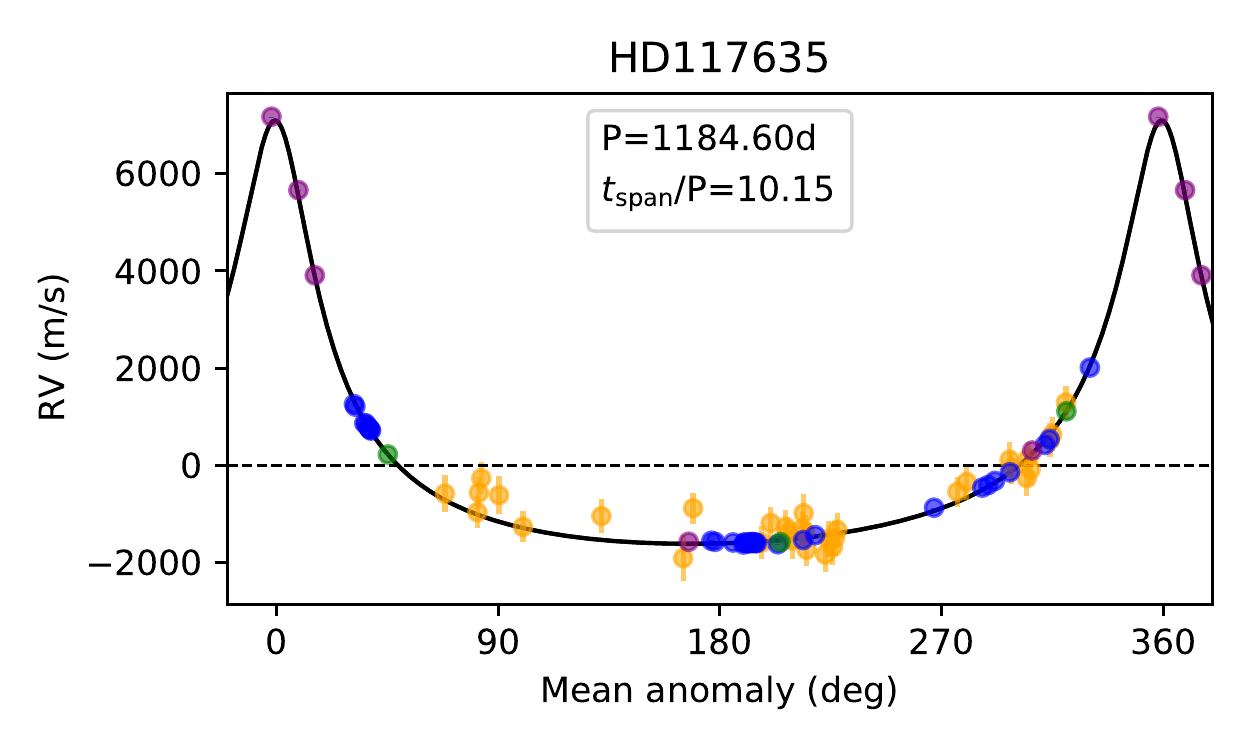}\\

		\includegraphics[width=0.22\linewidth]{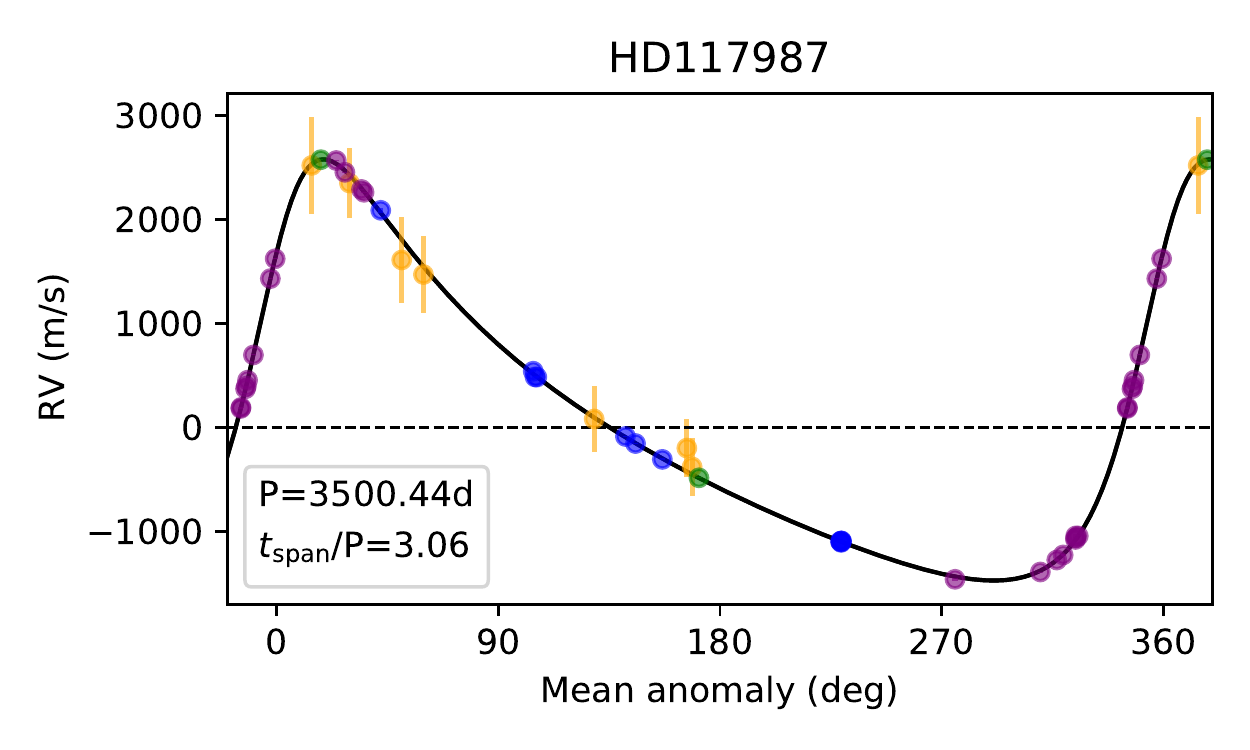}&
		\includegraphics[width=0.22\linewidth]{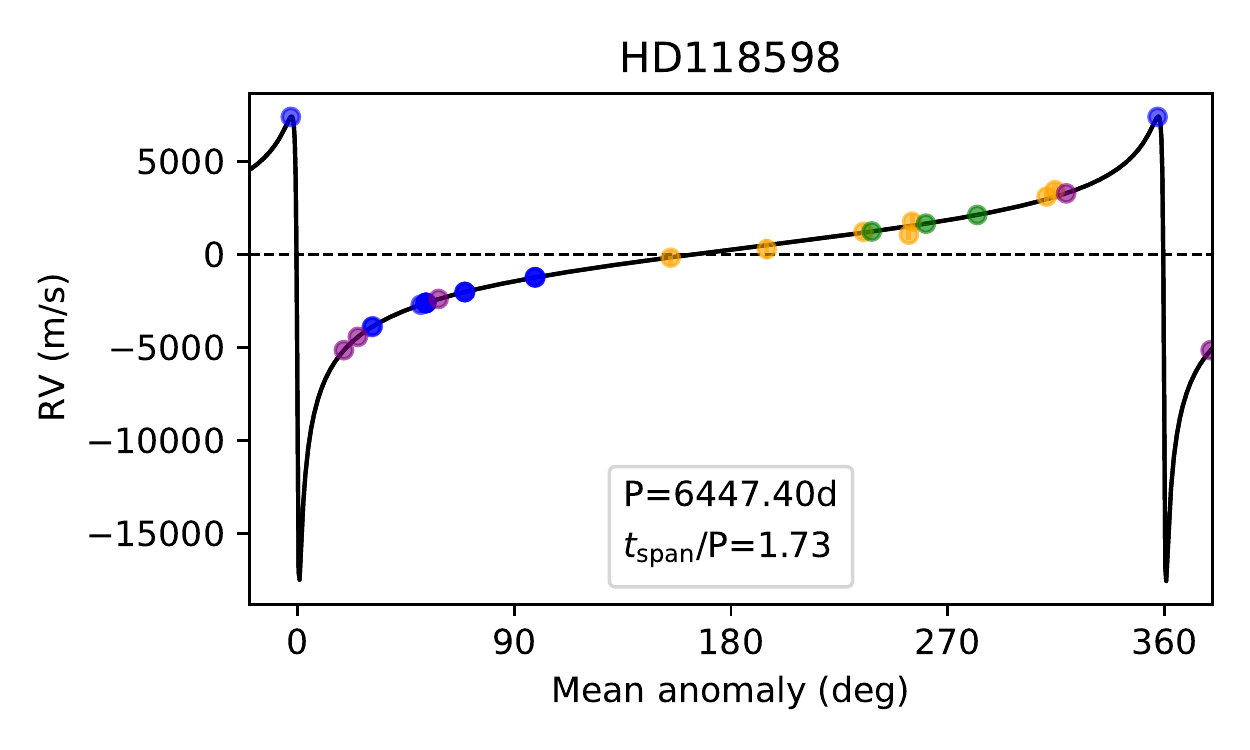}&
		\includegraphics[width=0.22\linewidth]{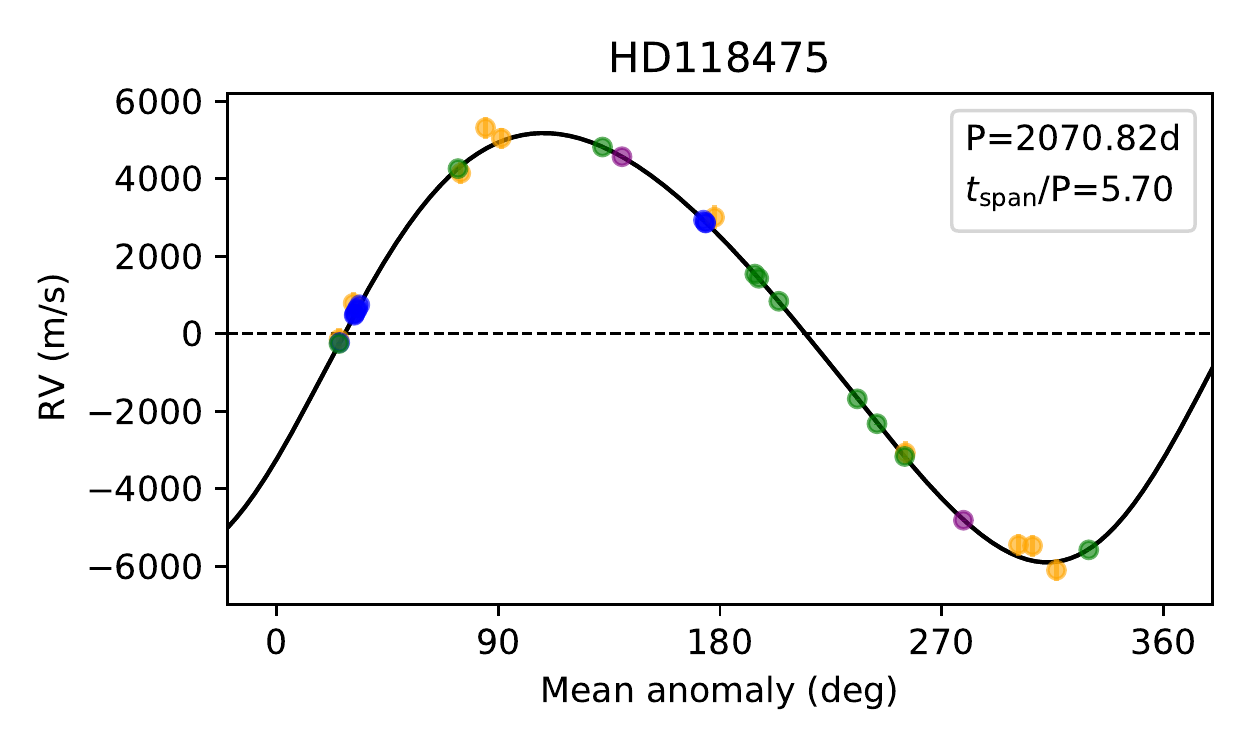}&
		\includegraphics[width=0.22\linewidth]{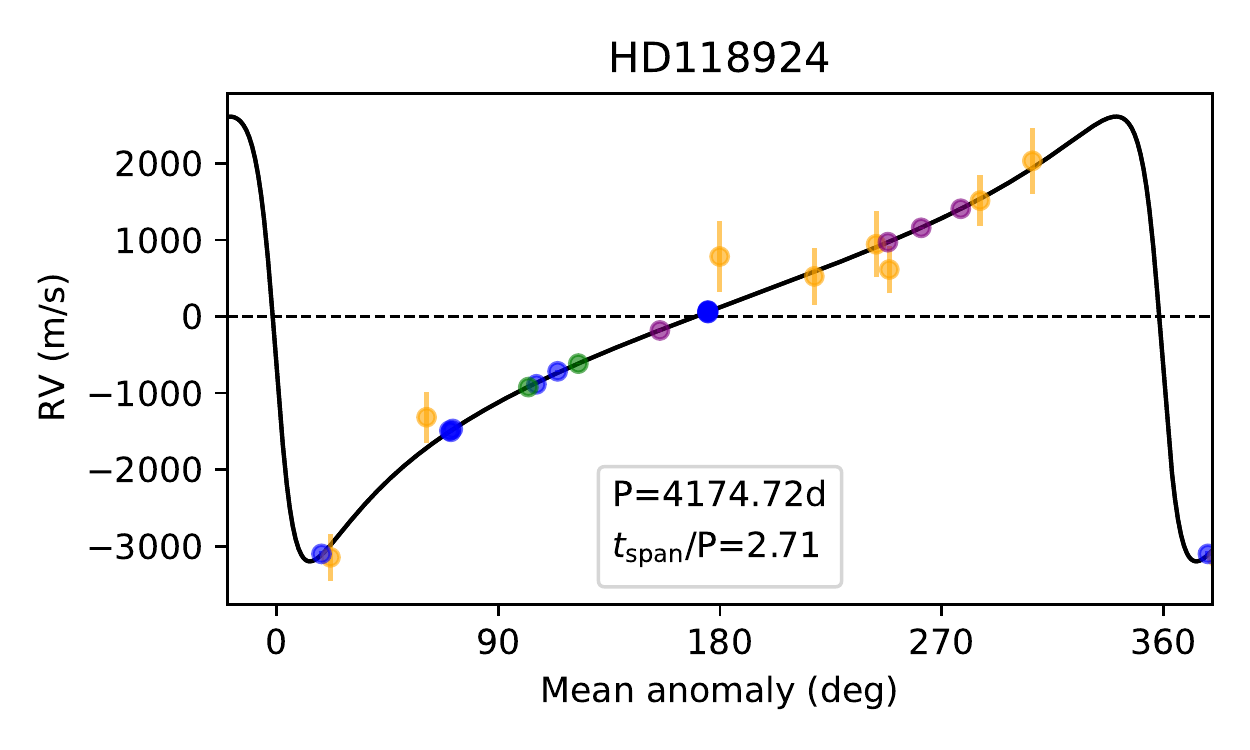}\\

		\includegraphics[width=0.22\linewidth]{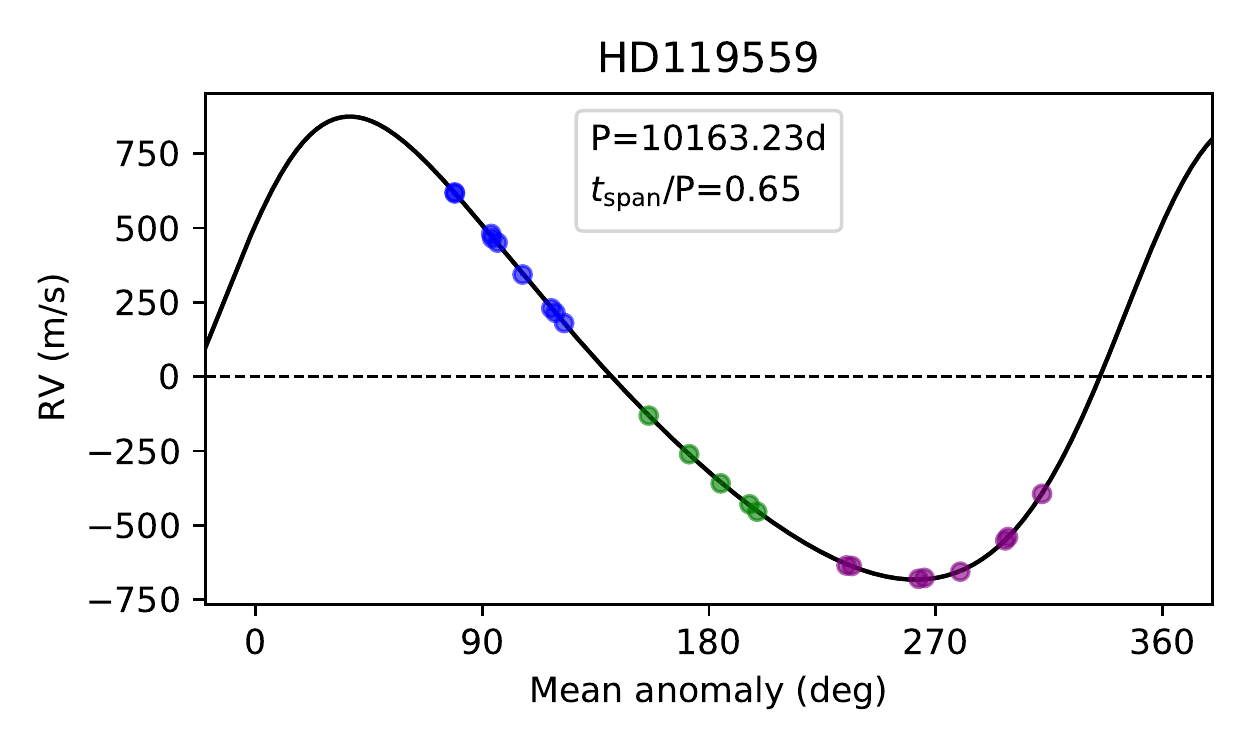}&
		\includegraphics[width=0.22\linewidth]{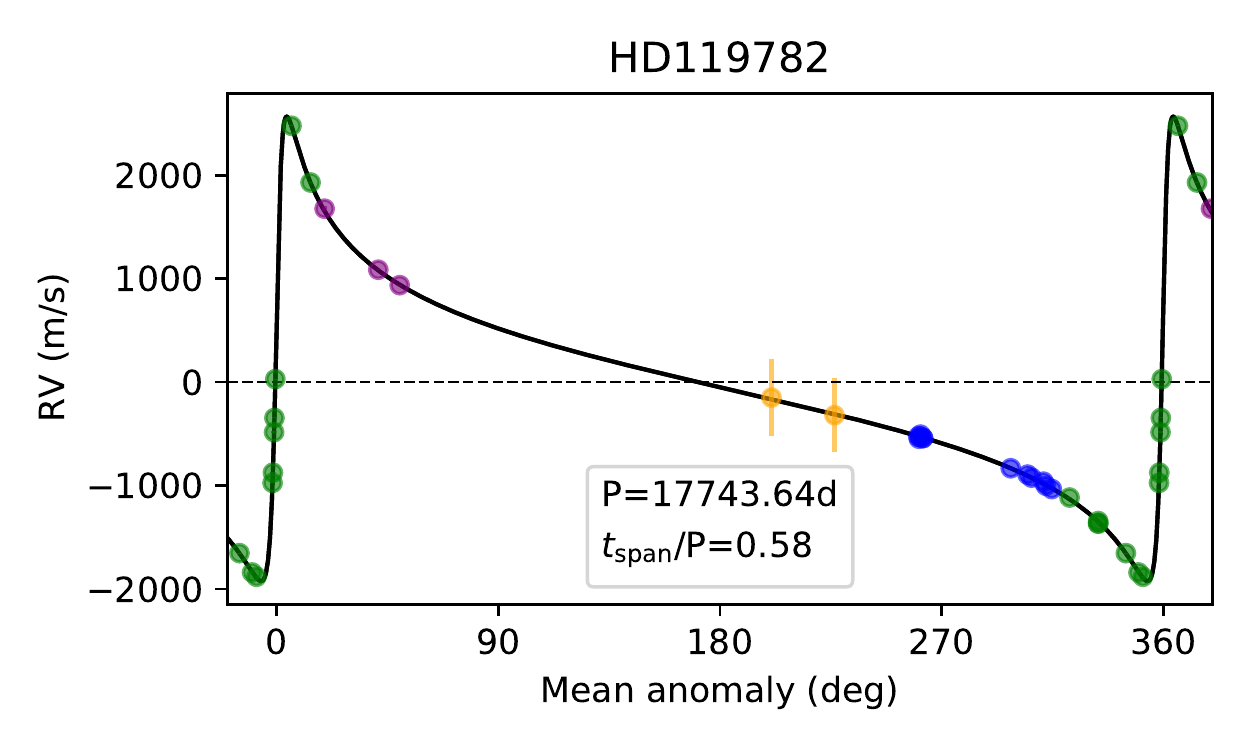}&
		\includegraphics[width=0.22\linewidth]{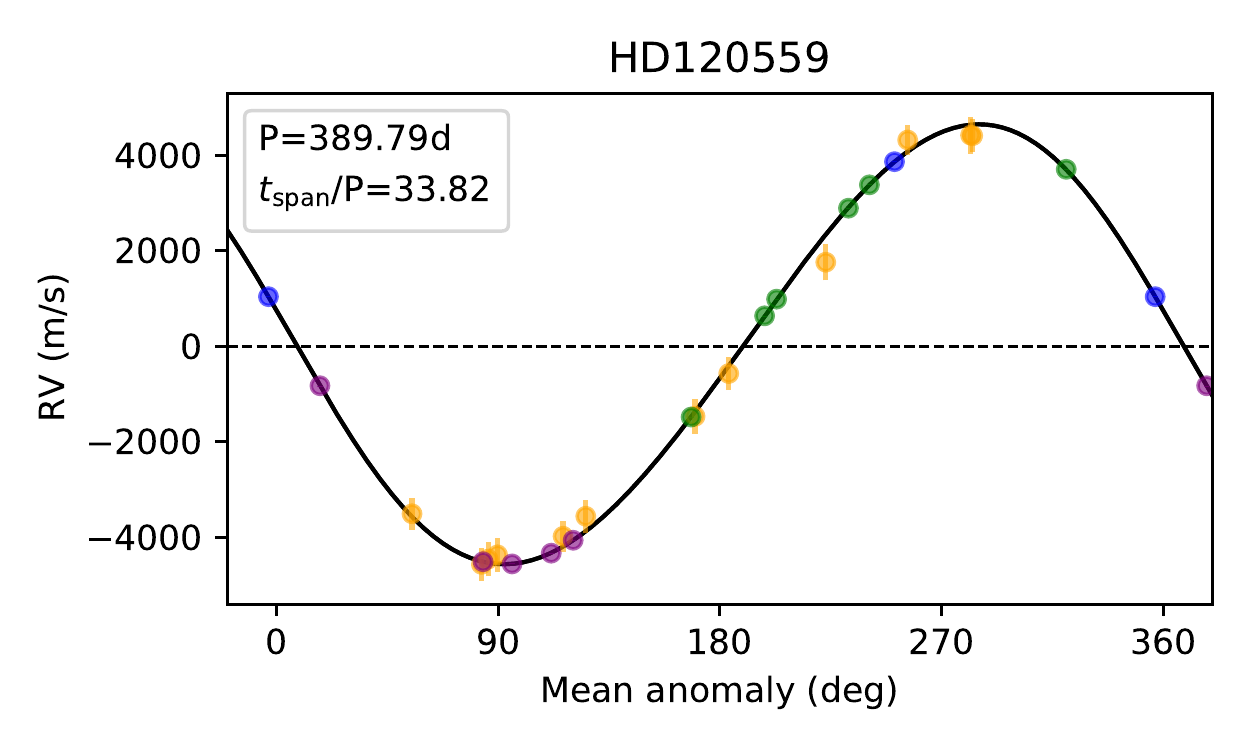}&
		\includegraphics[width=0.22\linewidth]{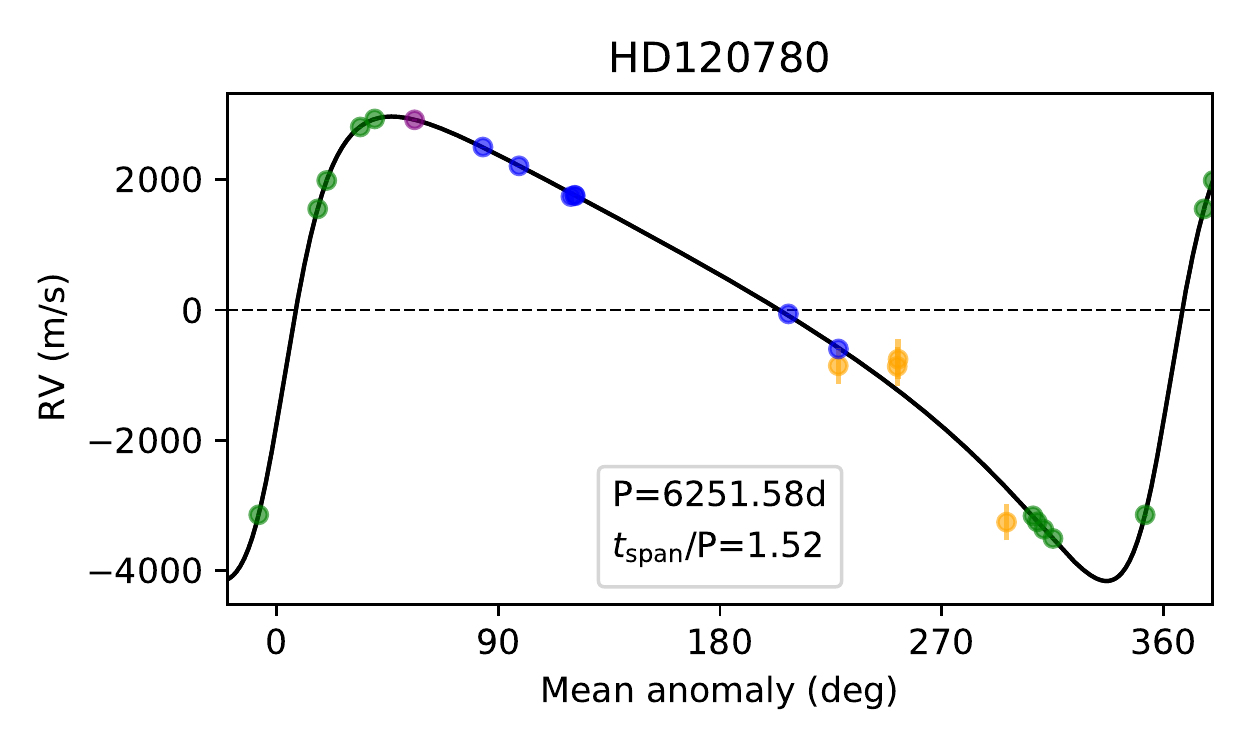}\\

		\includegraphics[width=0.22\linewidth]{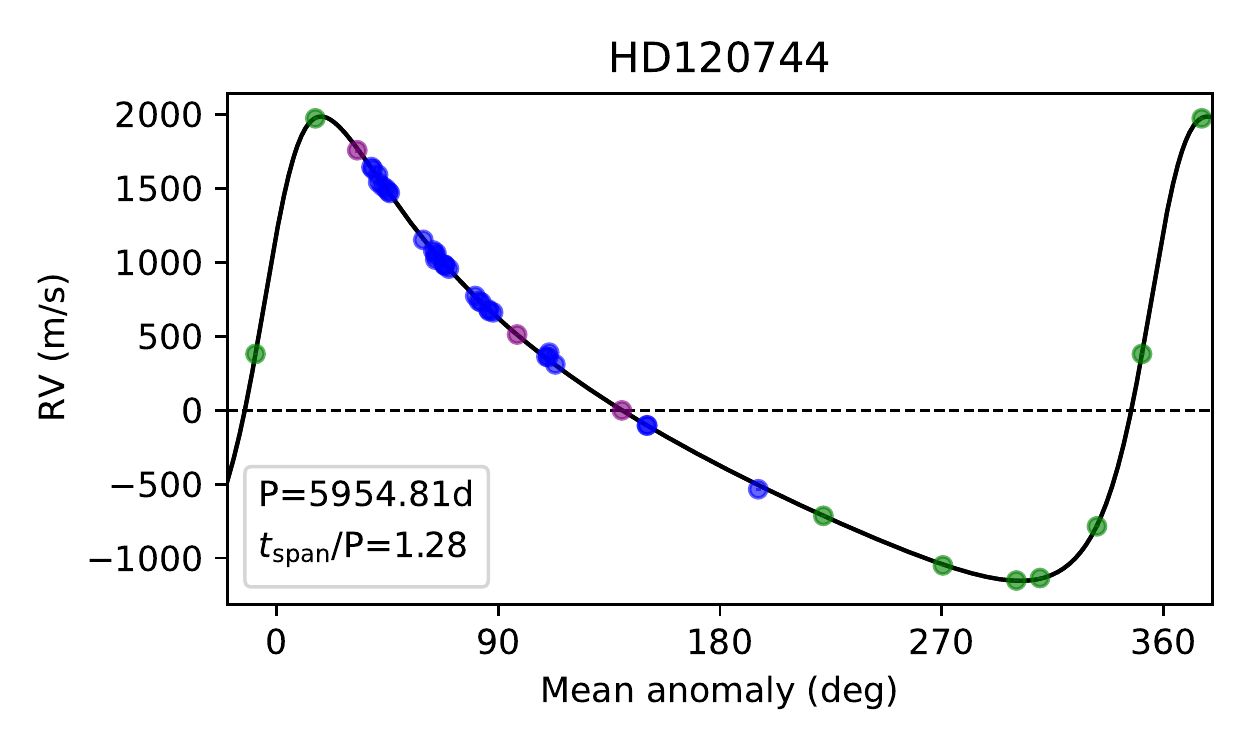}&
		\includegraphics[width=0.22\linewidth]{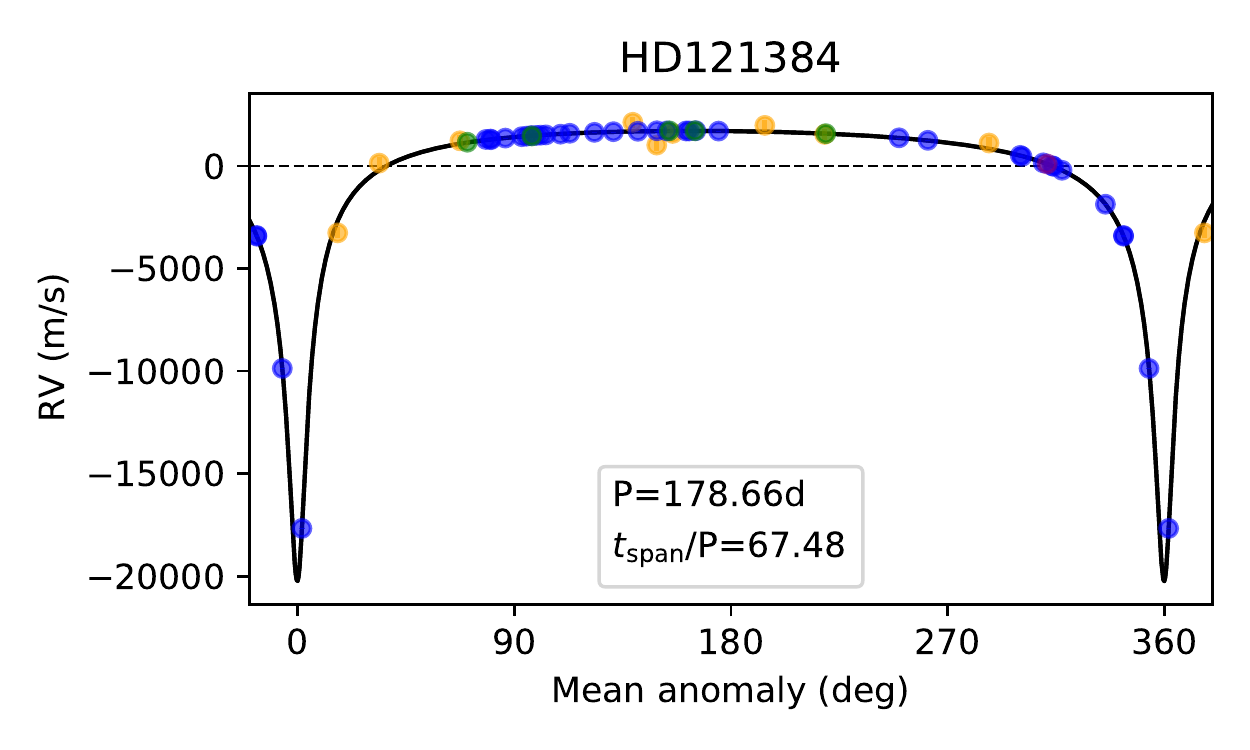}&
		\includegraphics[width=0.22\linewidth]{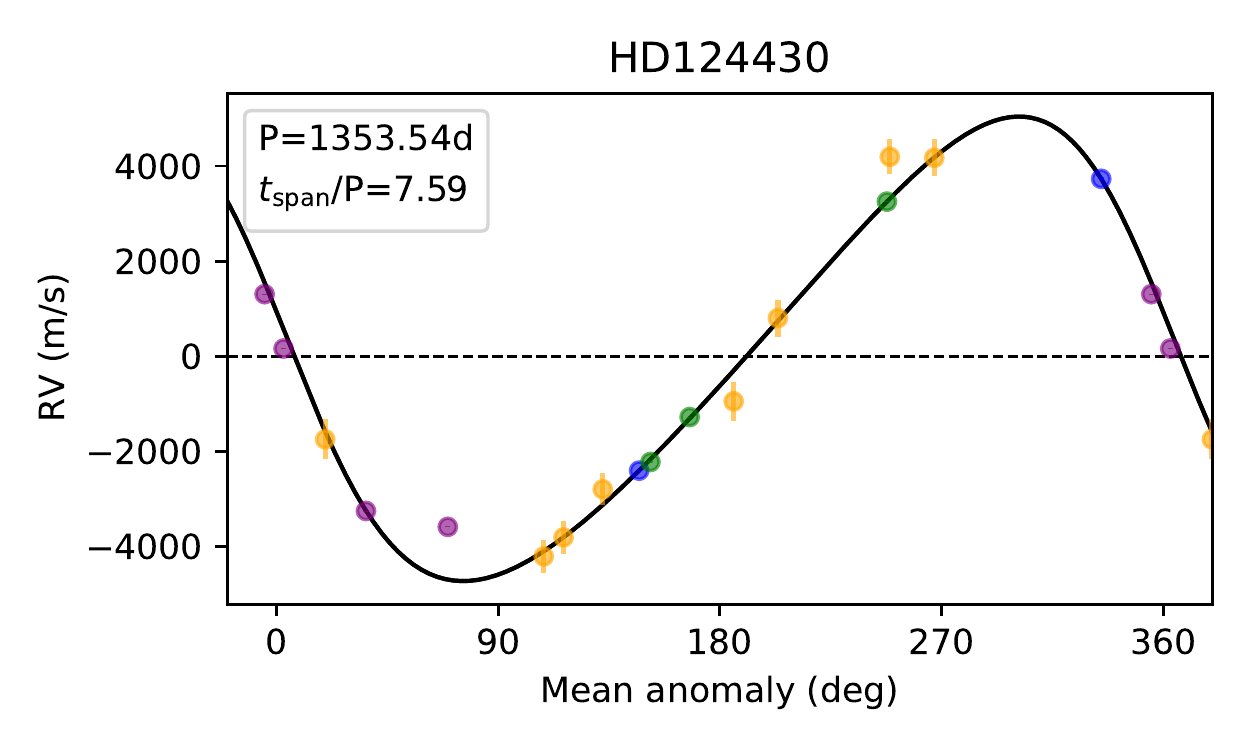}&
		\includegraphics[width=0.22\linewidth]{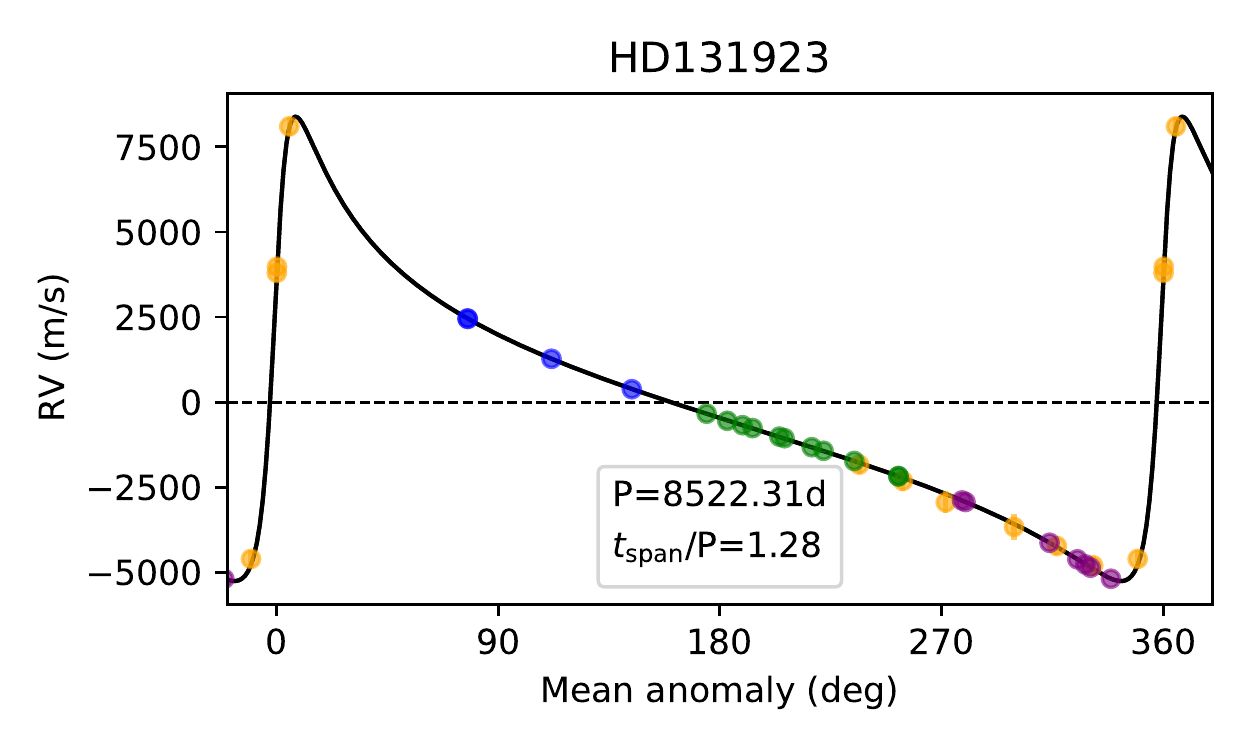}\\

		\includegraphics[width=0.22\linewidth]{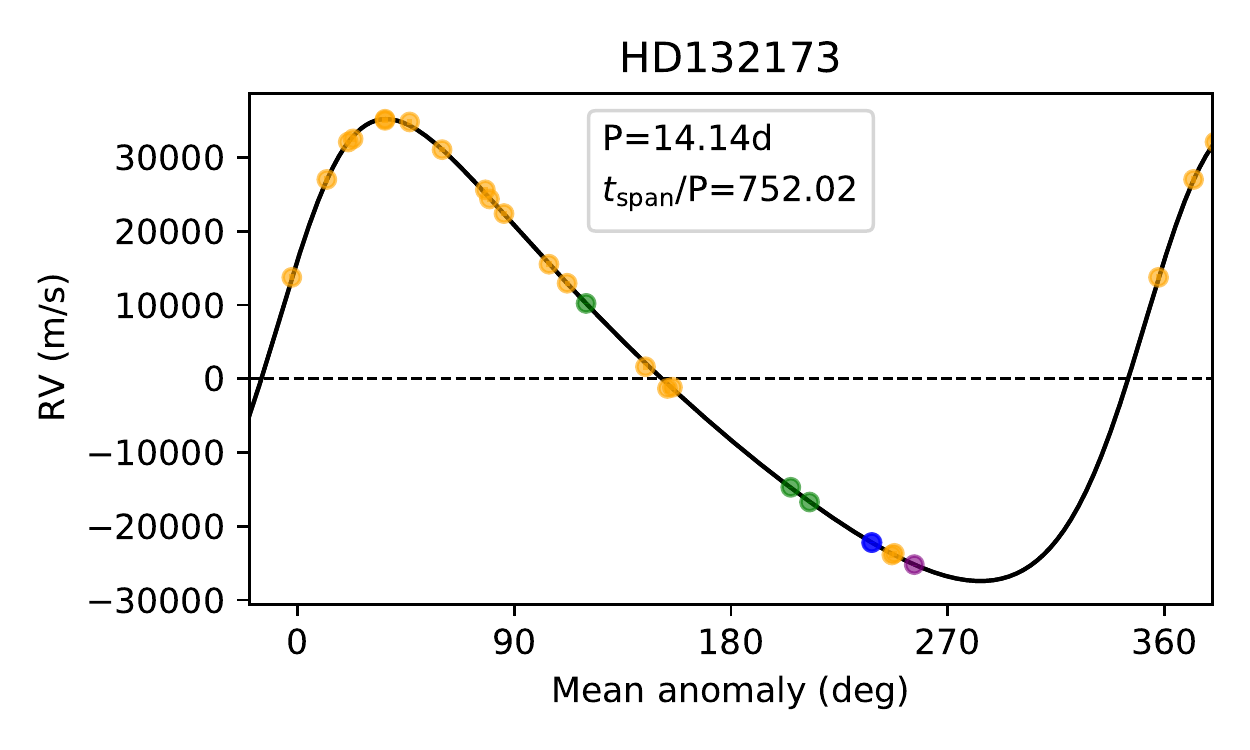}&
		\includegraphics[width=0.22\linewidth]{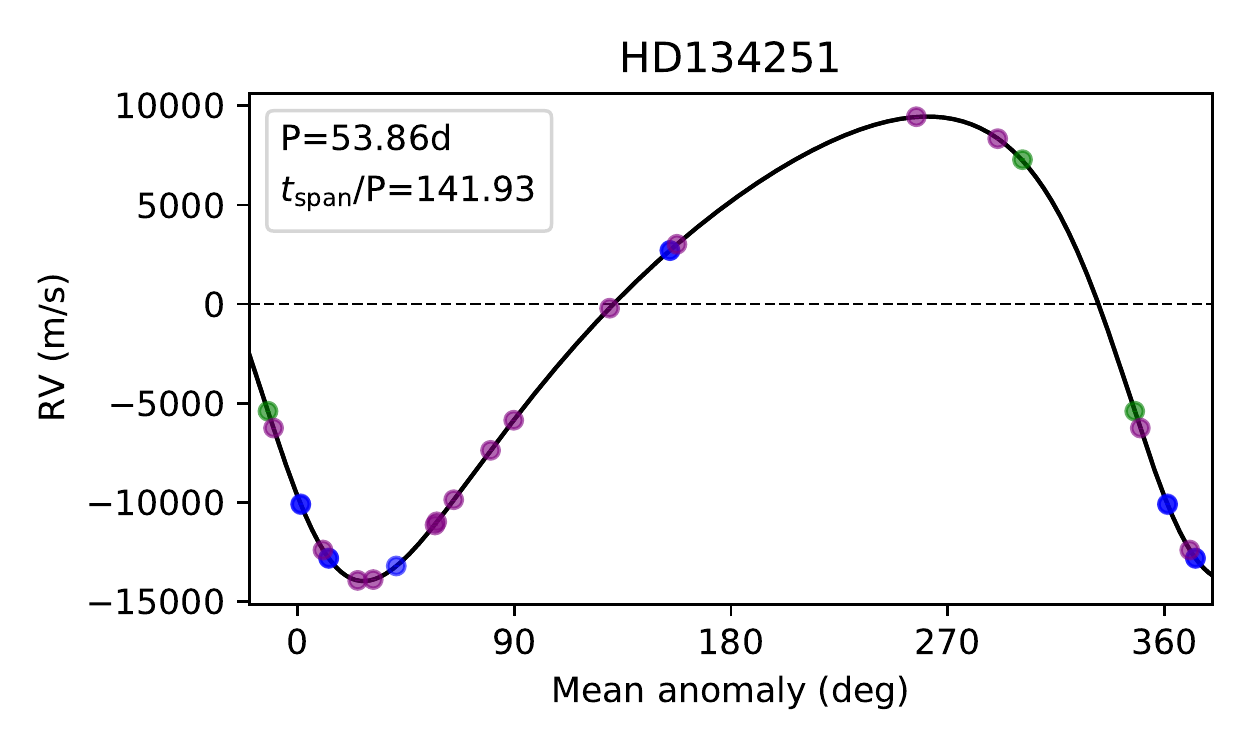}&
		\includegraphics[width=0.22\linewidth]{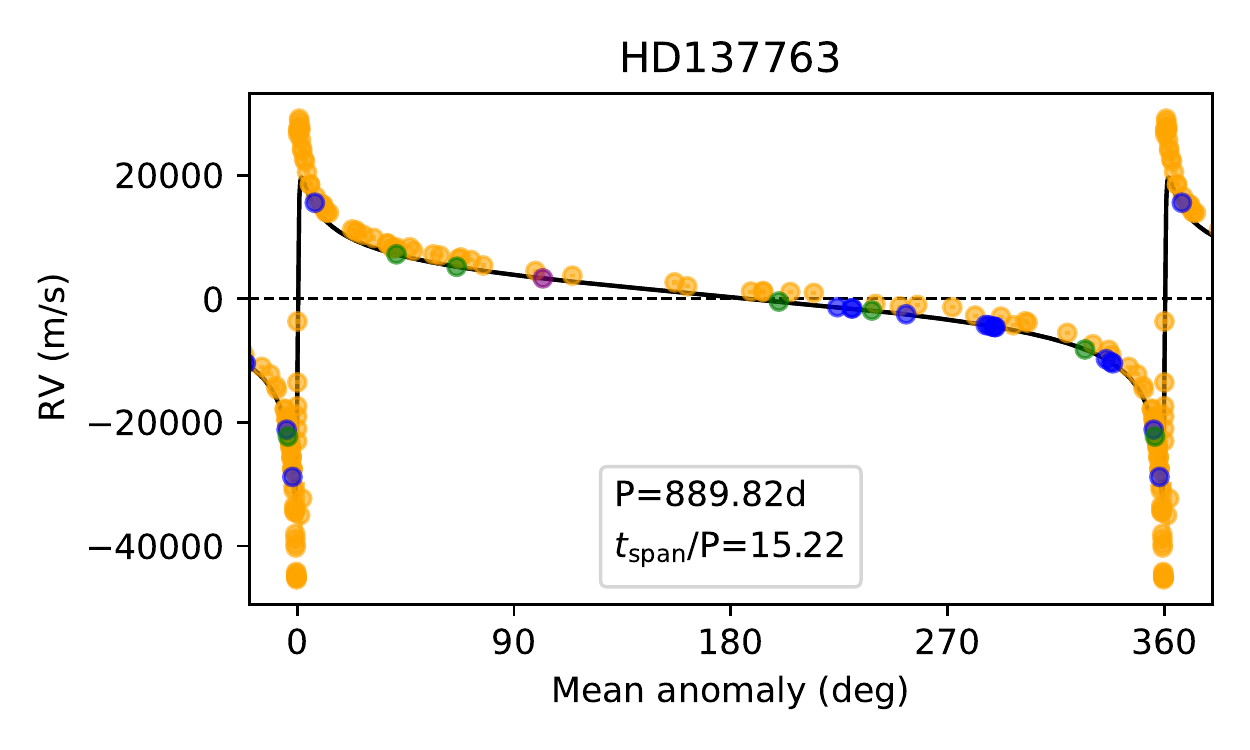}&
		\includegraphics[width=0.22\linewidth]{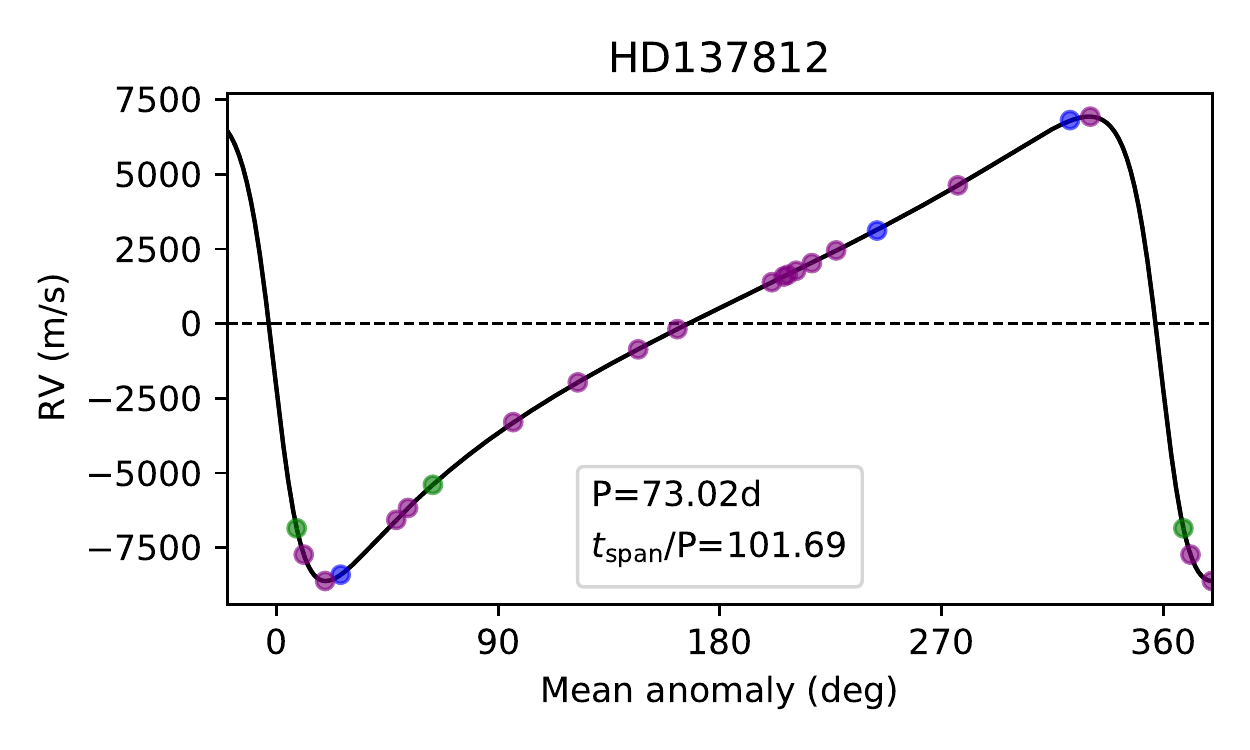}\\

		\includegraphics[width=0.22\linewidth]{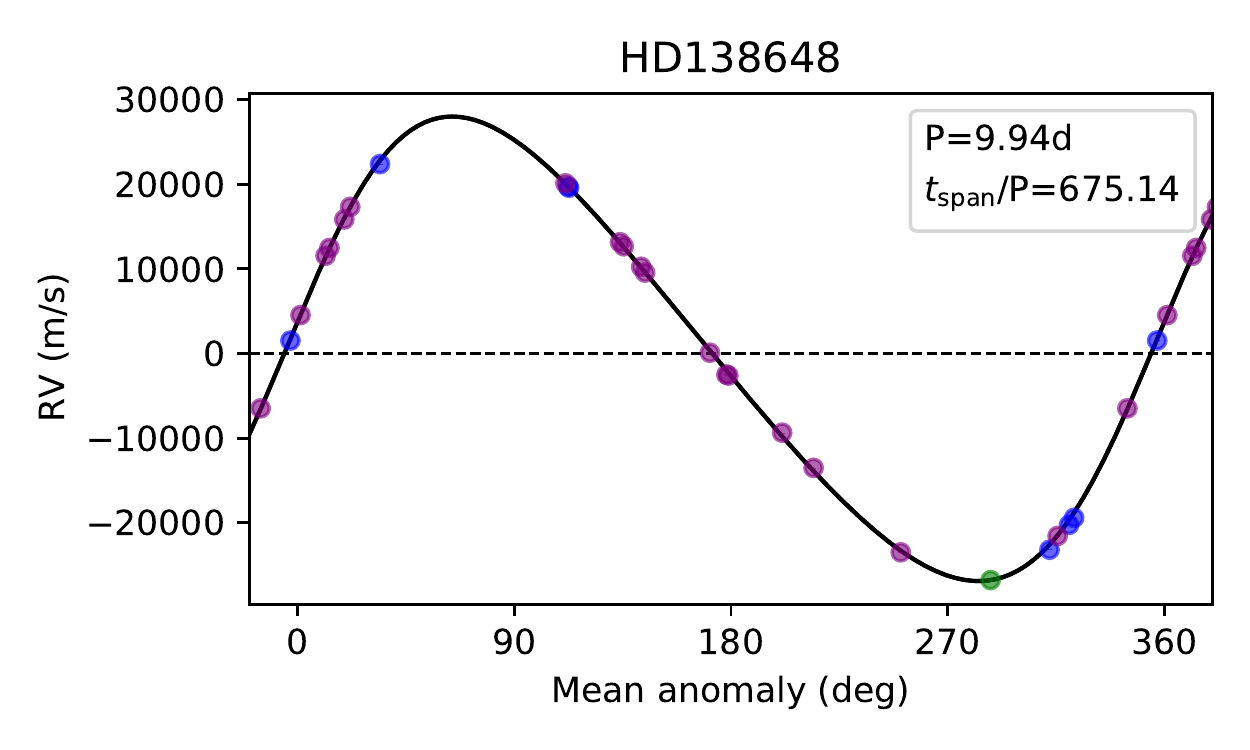}&
		\includegraphics[width=0.22\linewidth]{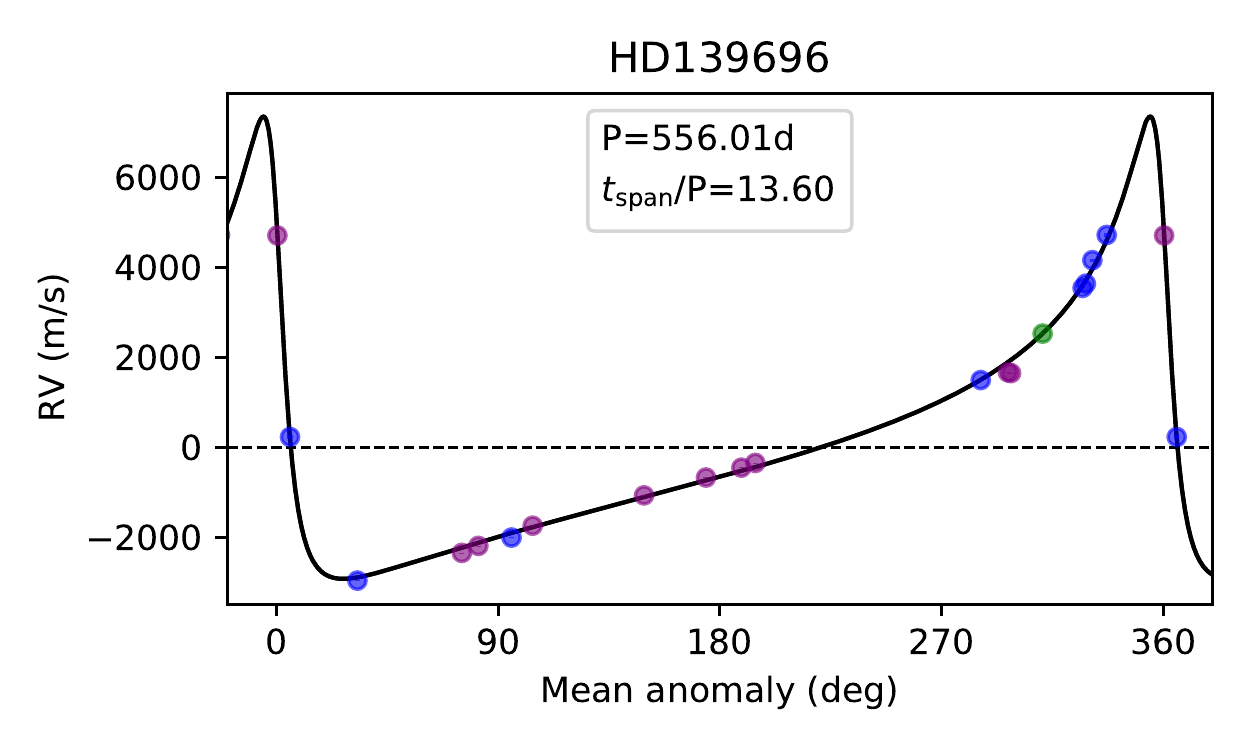}&
		\includegraphics[width=0.22\linewidth]{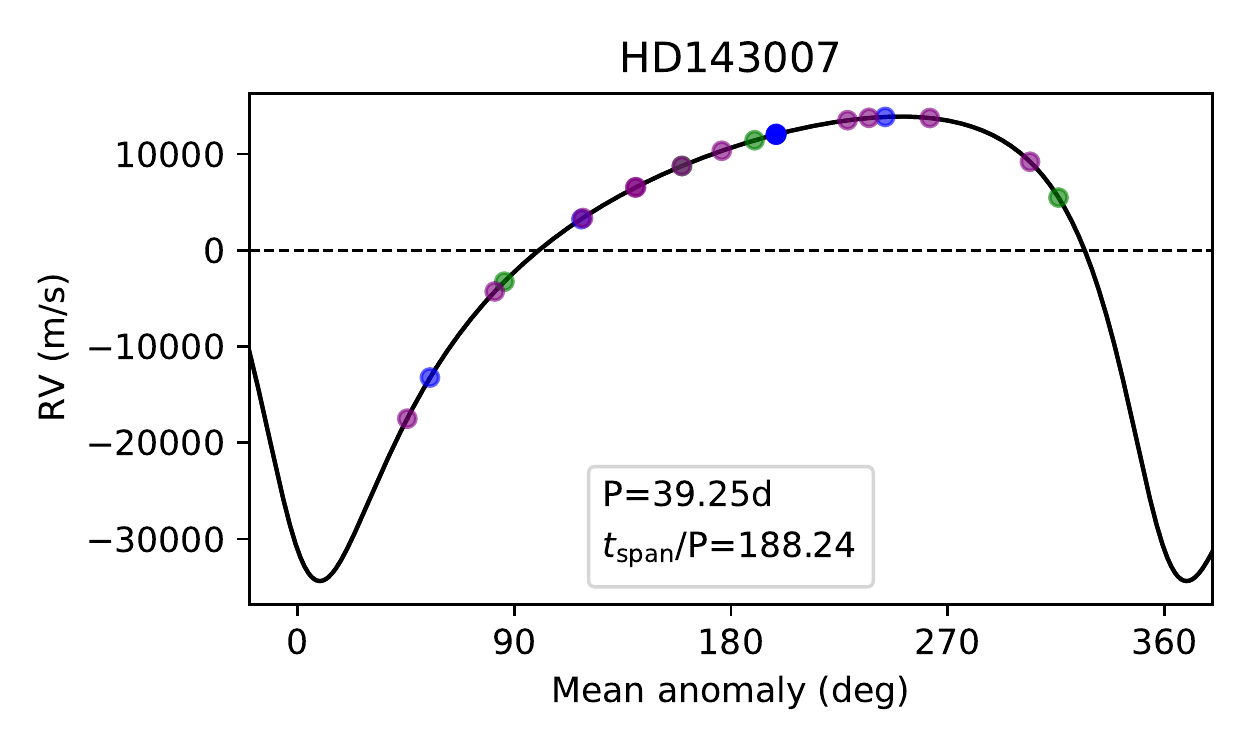}&
		\includegraphics[width=0.22\linewidth]{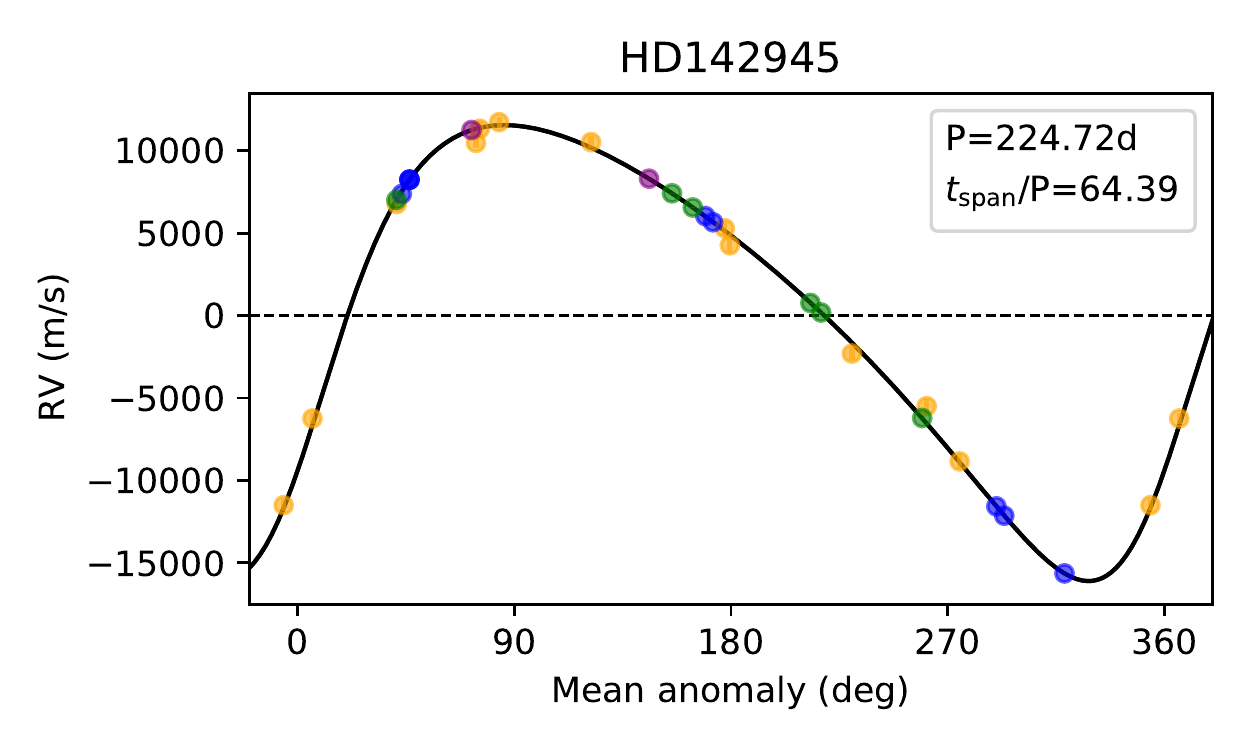}\\

		\includegraphics[width=0.22\linewidth]{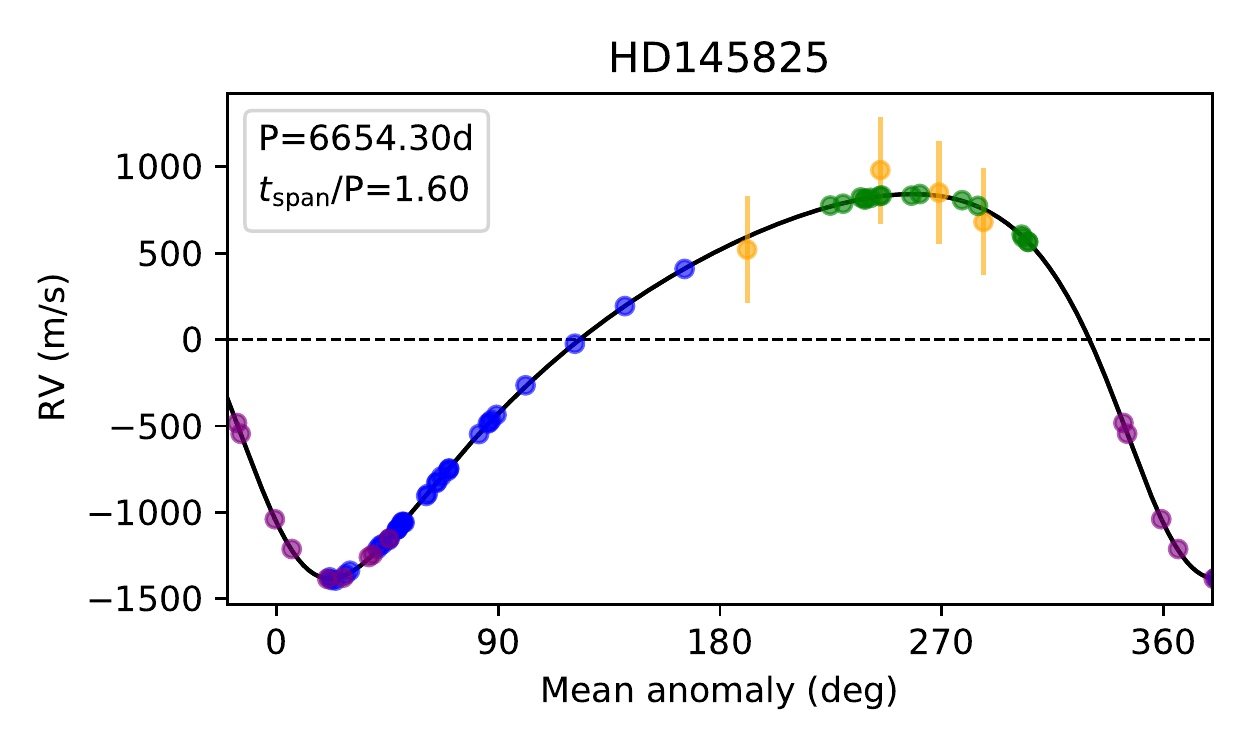}&
		\includegraphics[width=0.22\linewidth]{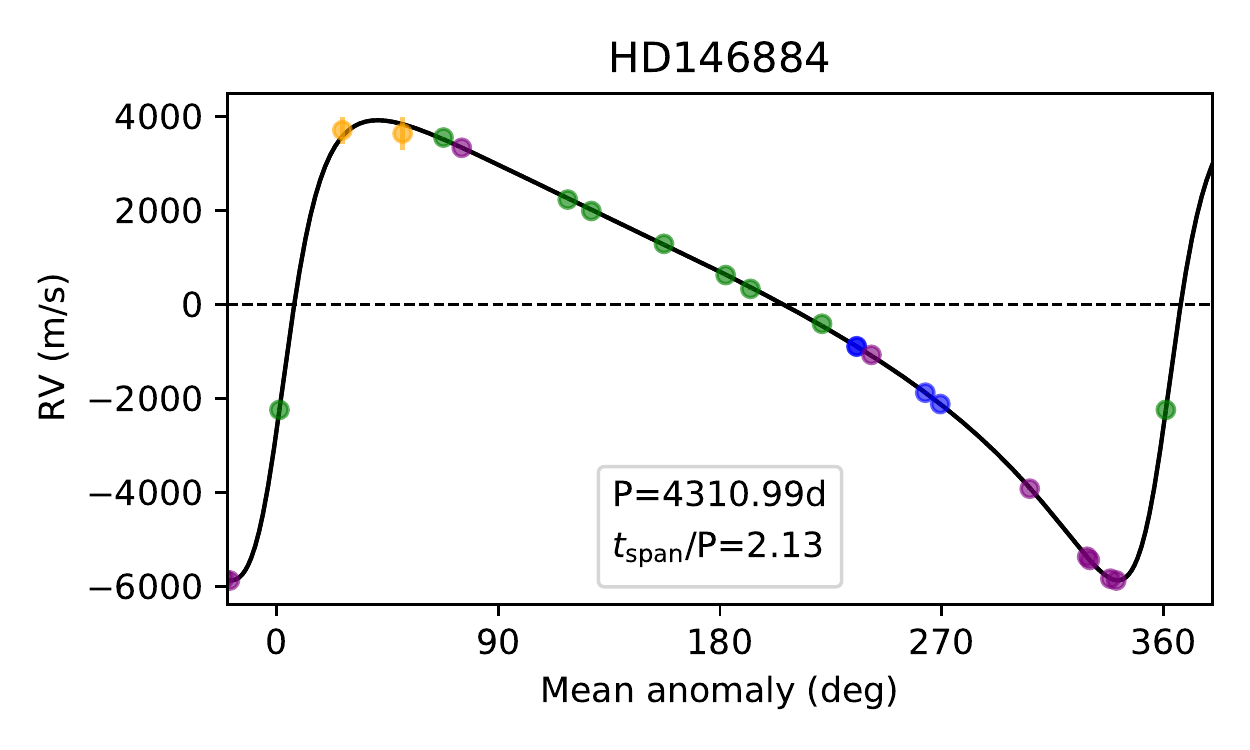}&
		\includegraphics[width=0.22\linewidth]{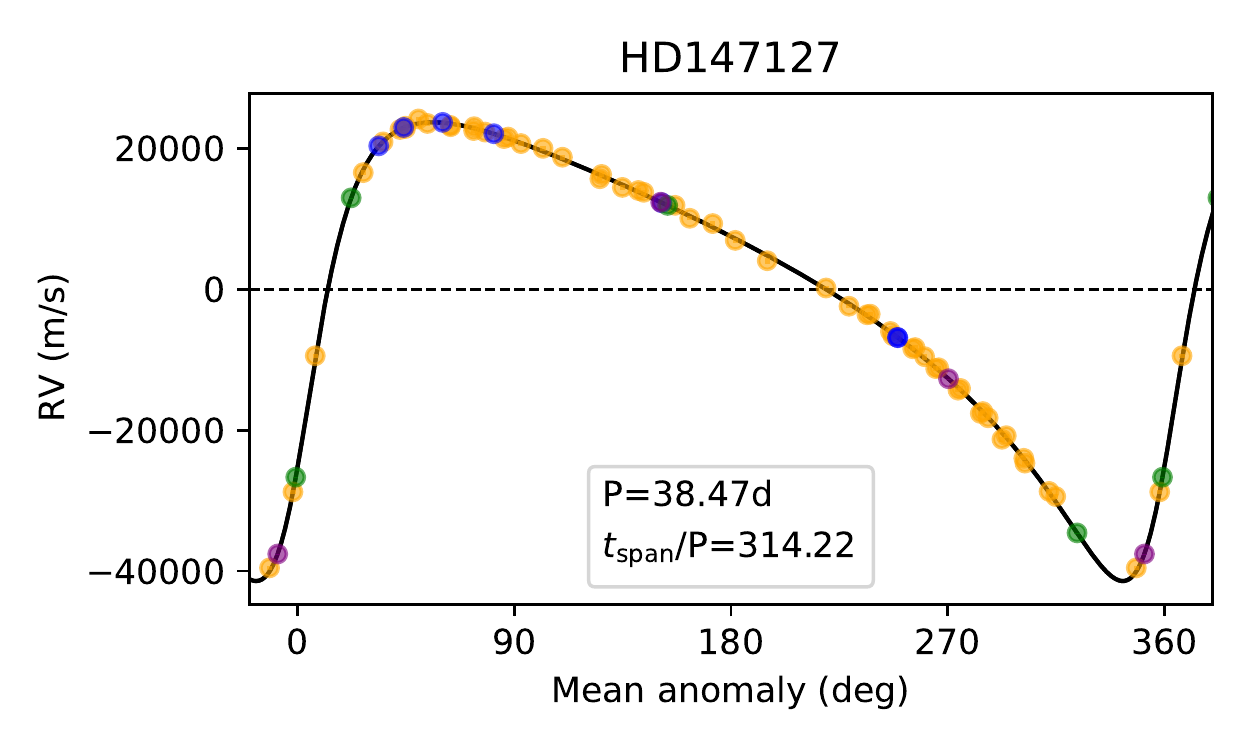}&
		\includegraphics[width=0.22\linewidth]{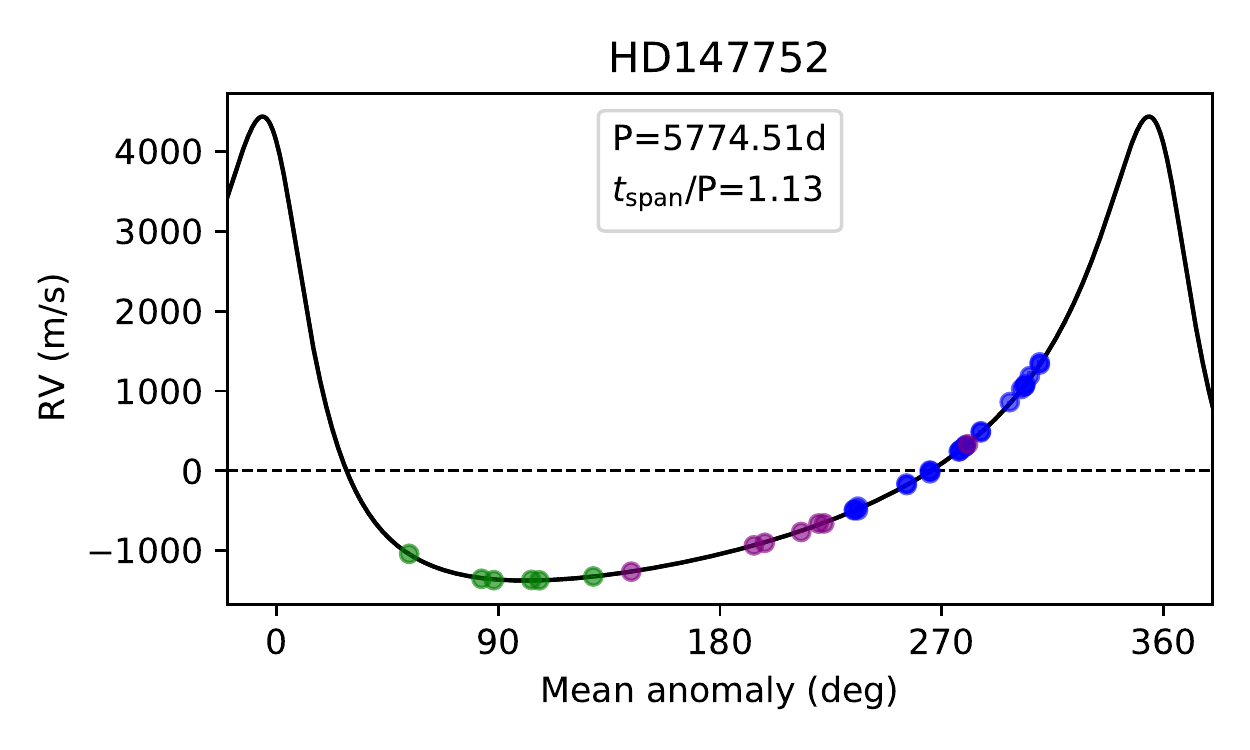}\\

		\includegraphics[width=0.22\linewidth]{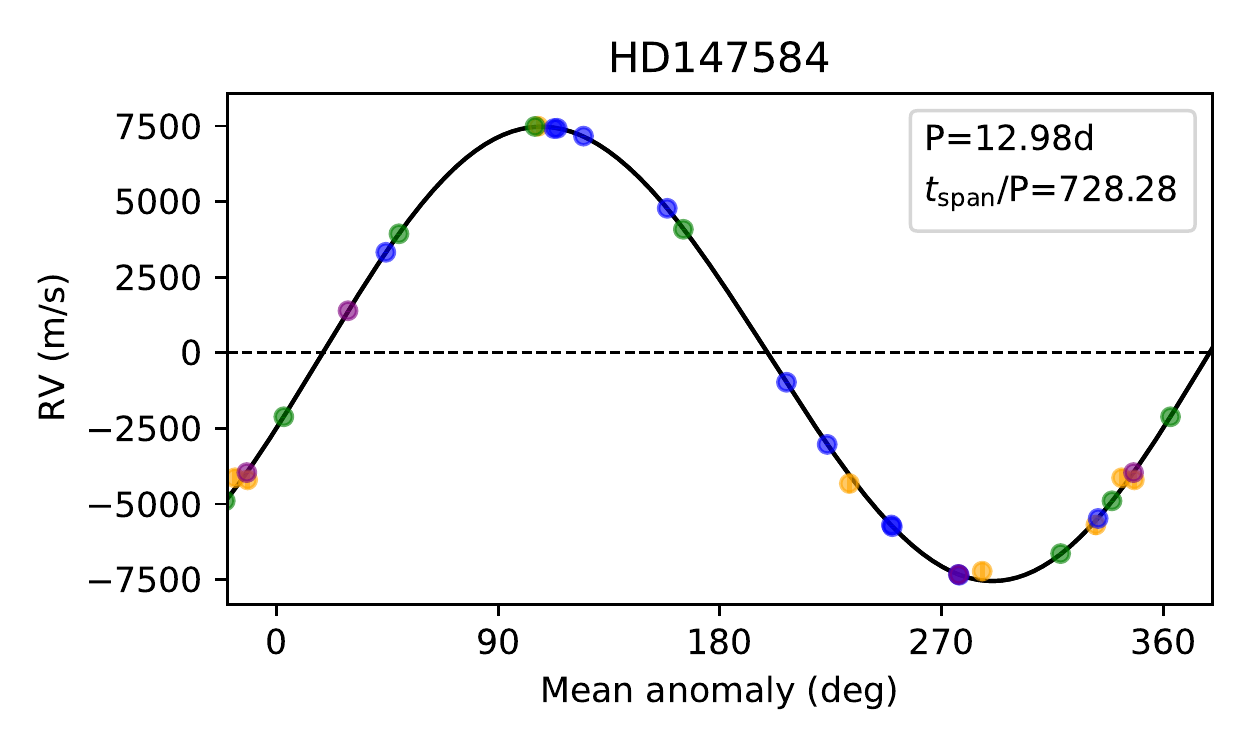}&
		\includegraphics[width=0.22\linewidth]{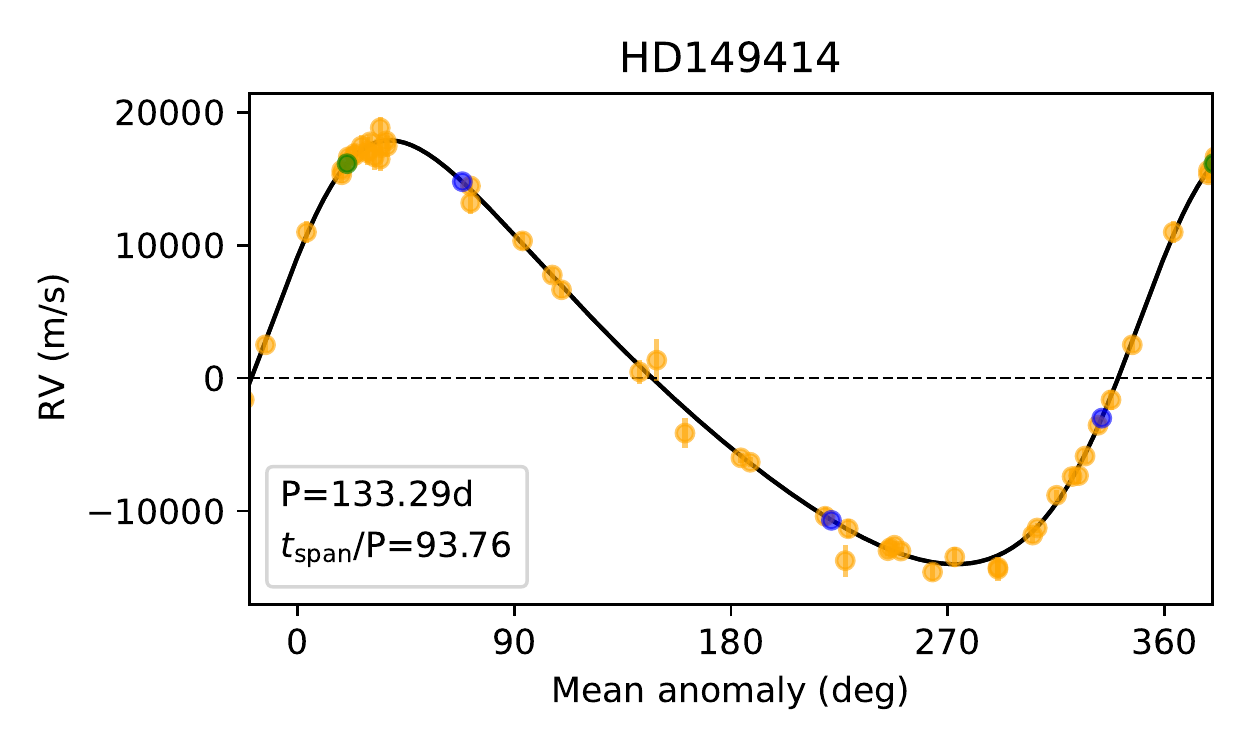}&
		\includegraphics[width=0.22\linewidth]{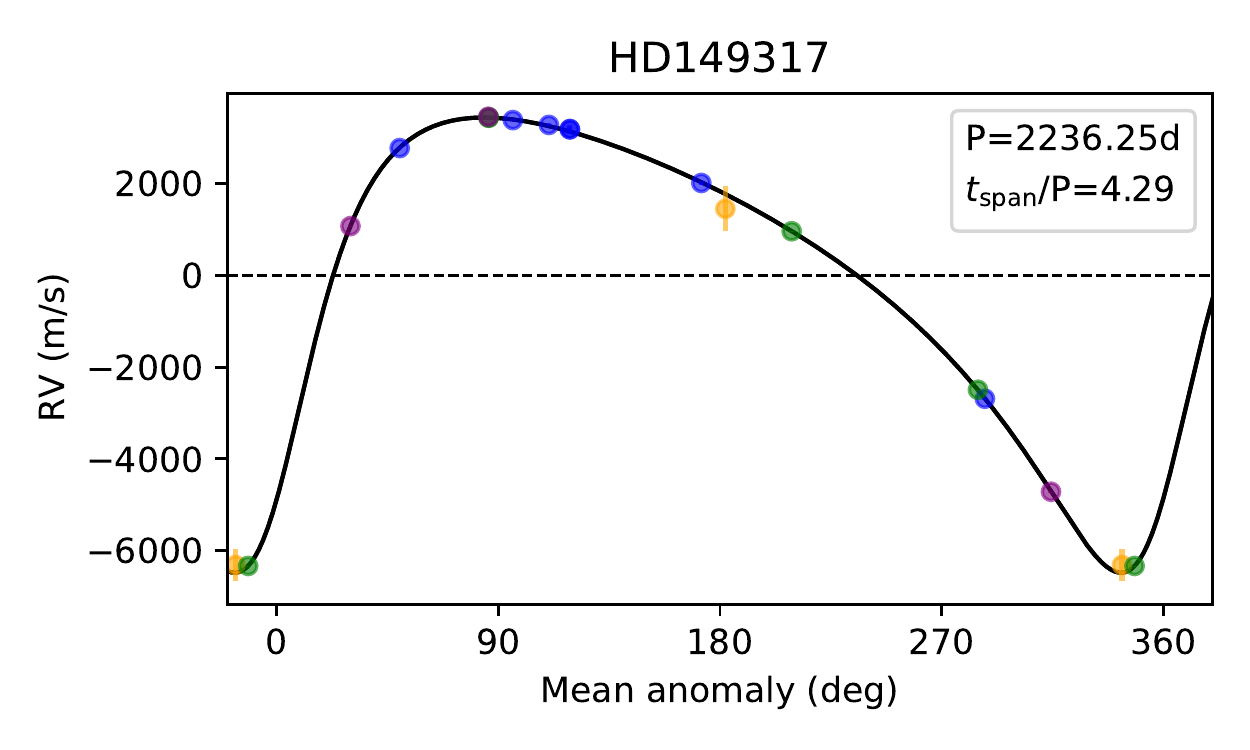}&
		\includegraphics[width=0.22\linewidth]{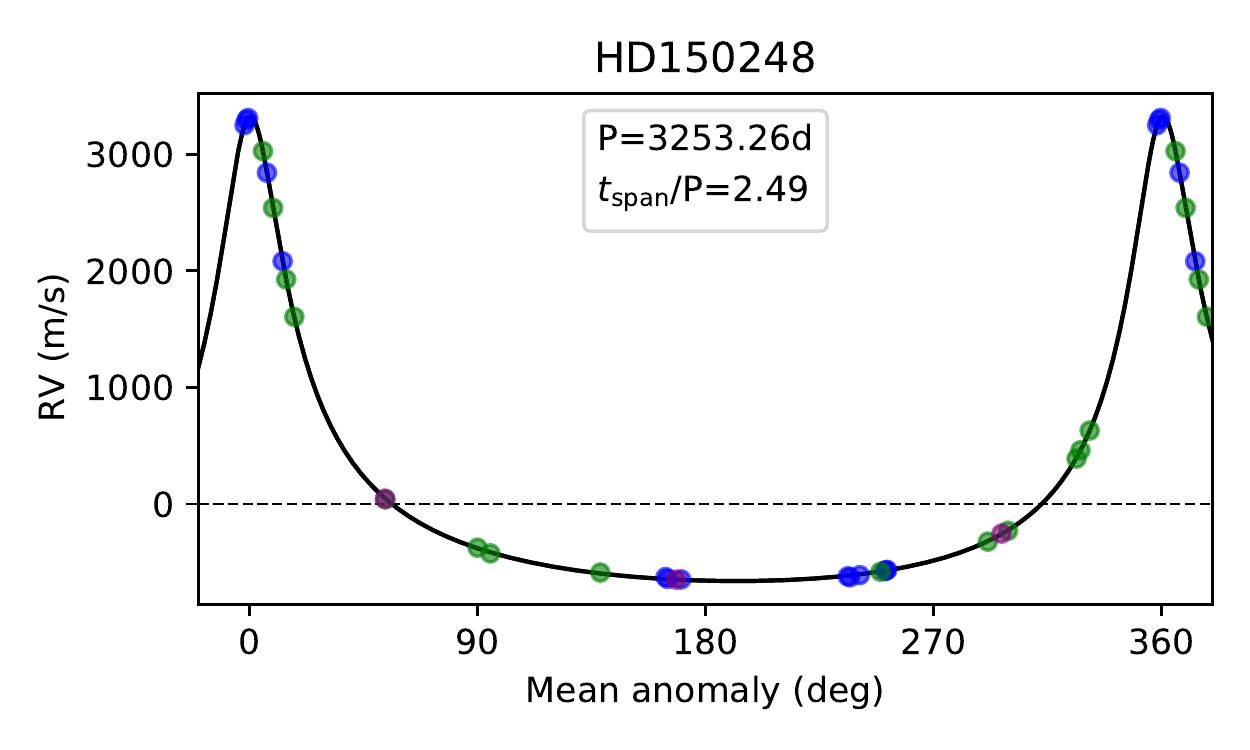}\\

		\includegraphics[width=0.22\linewidth]{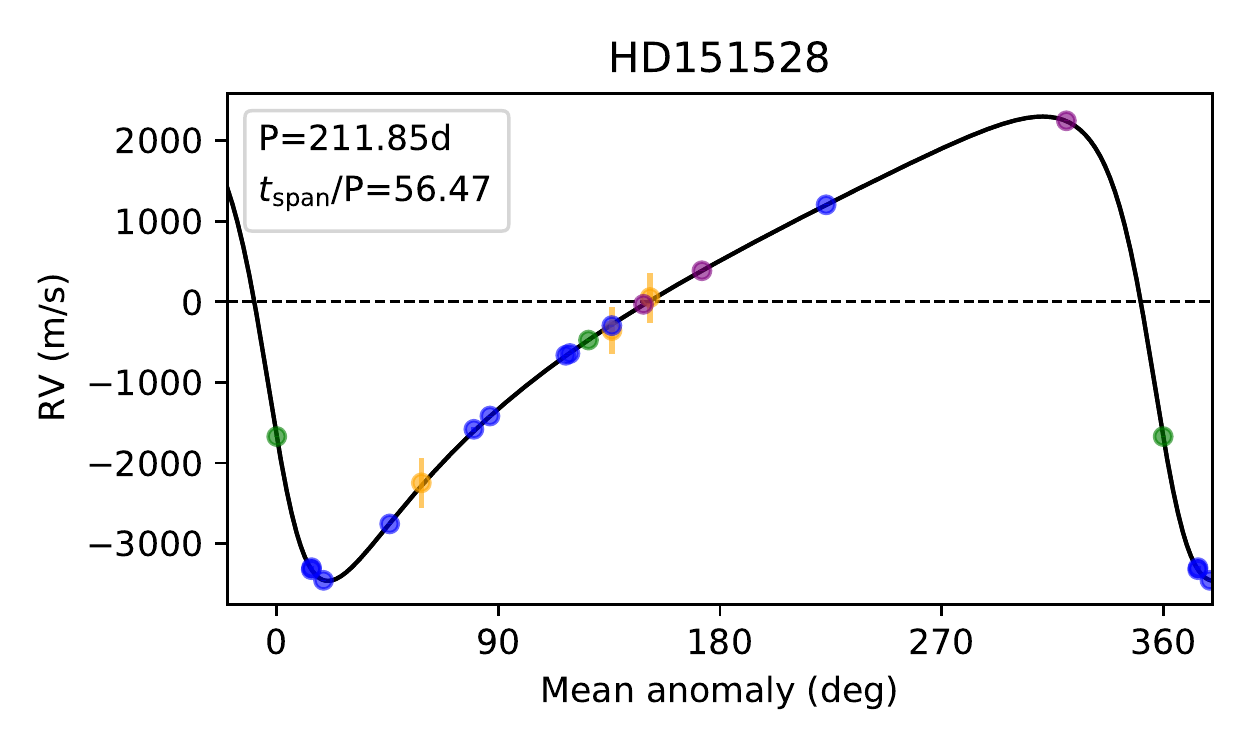}&
		\includegraphics[width=0.22\linewidth]{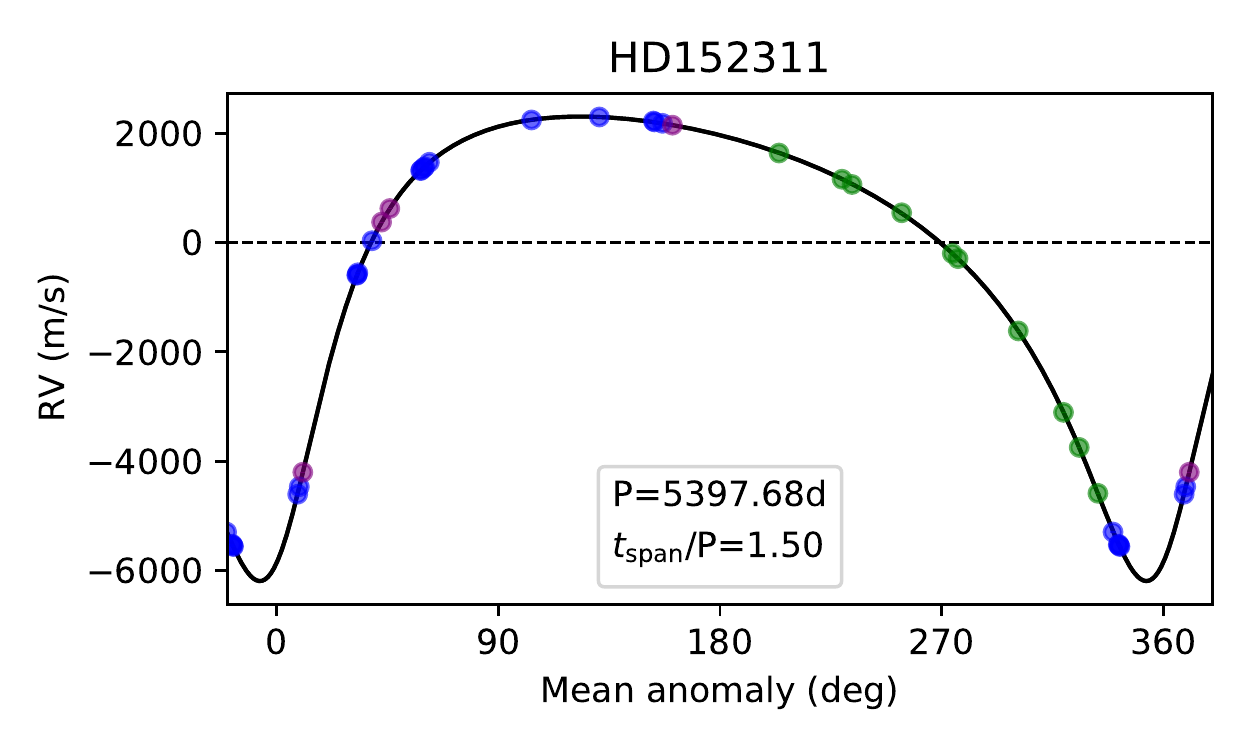}&
		\includegraphics[width=0.22\linewidth]{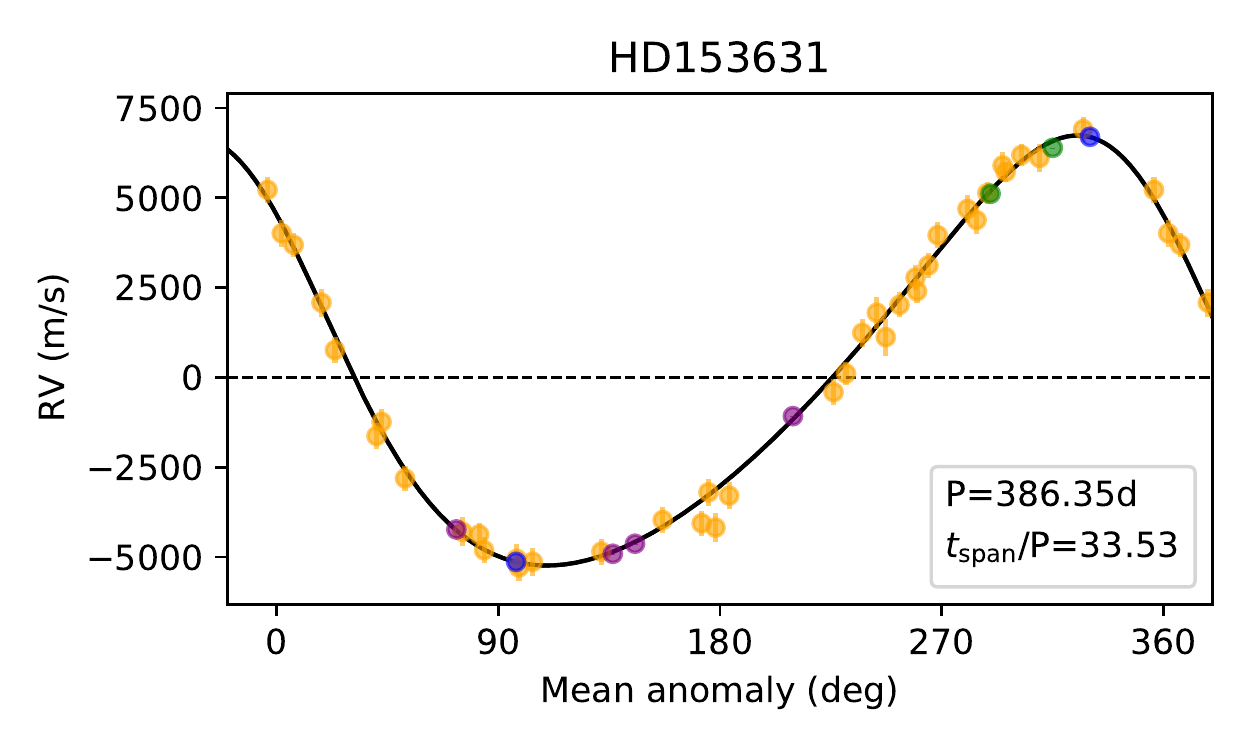}&
		\includegraphics[width=0.22\linewidth]{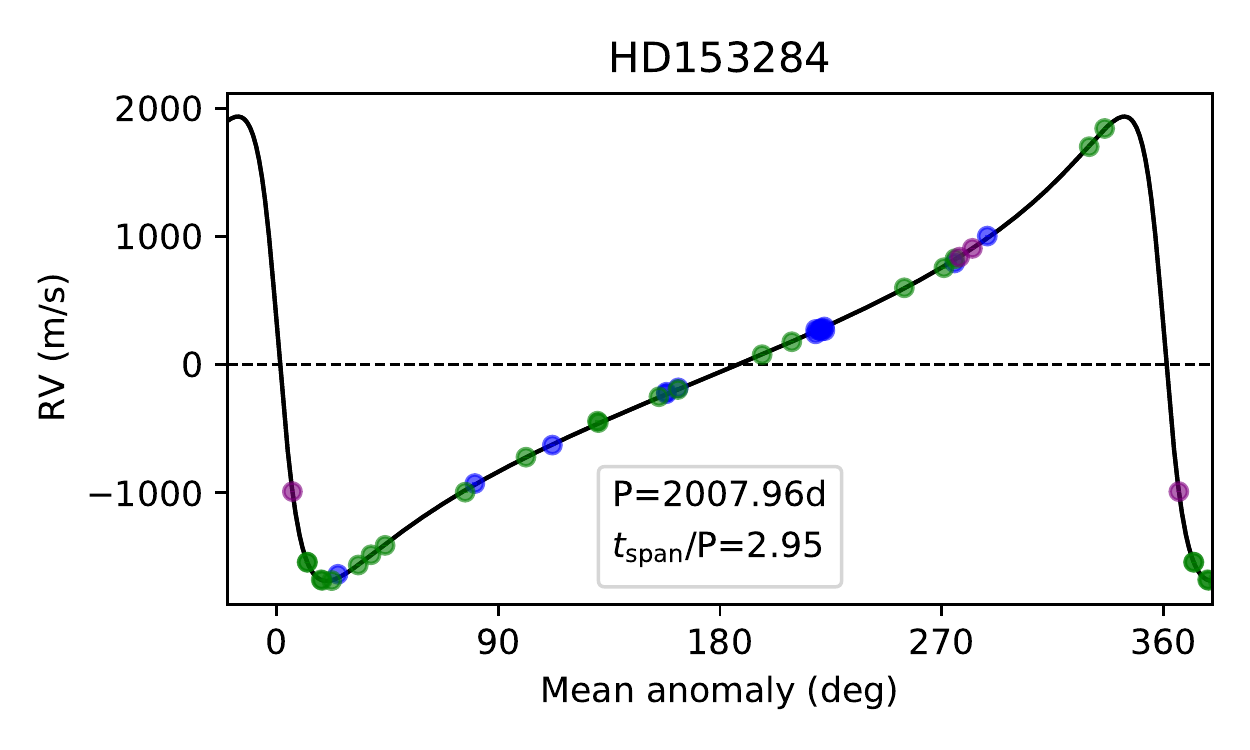}\\

		\includegraphics[width=0.22\linewidth]{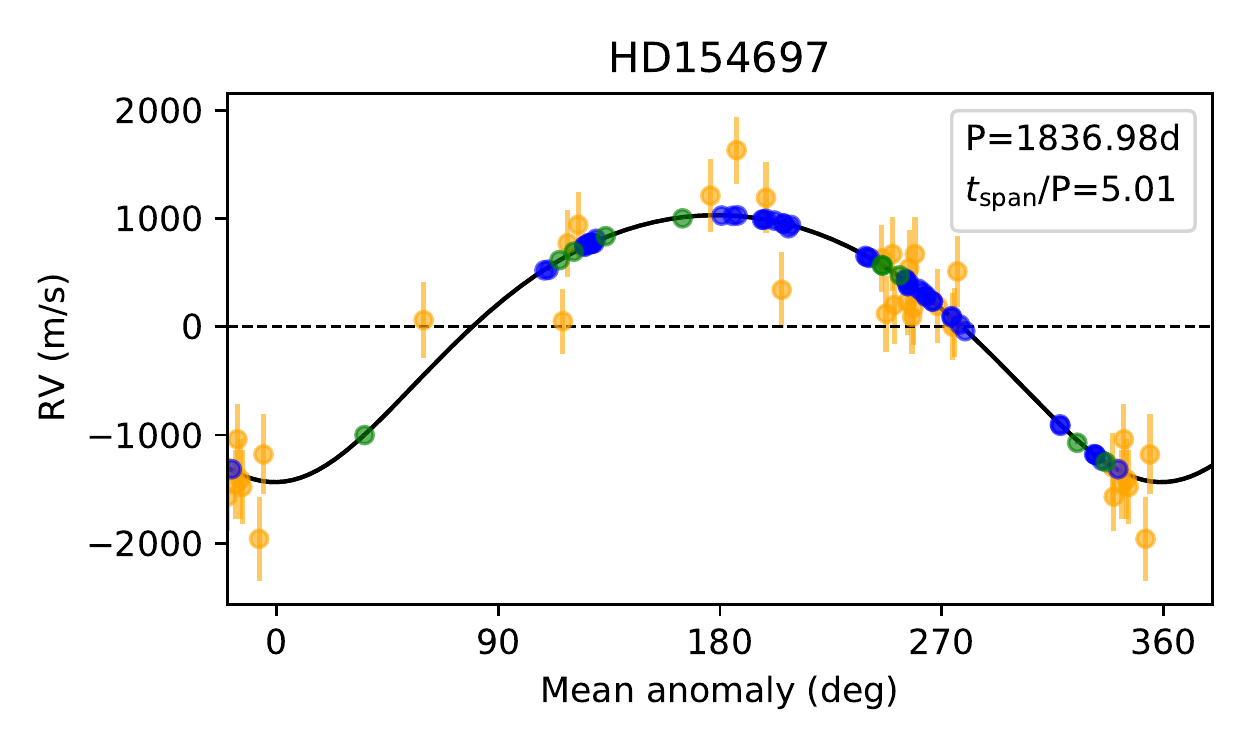}&
		\includegraphics[width=0.22\linewidth]{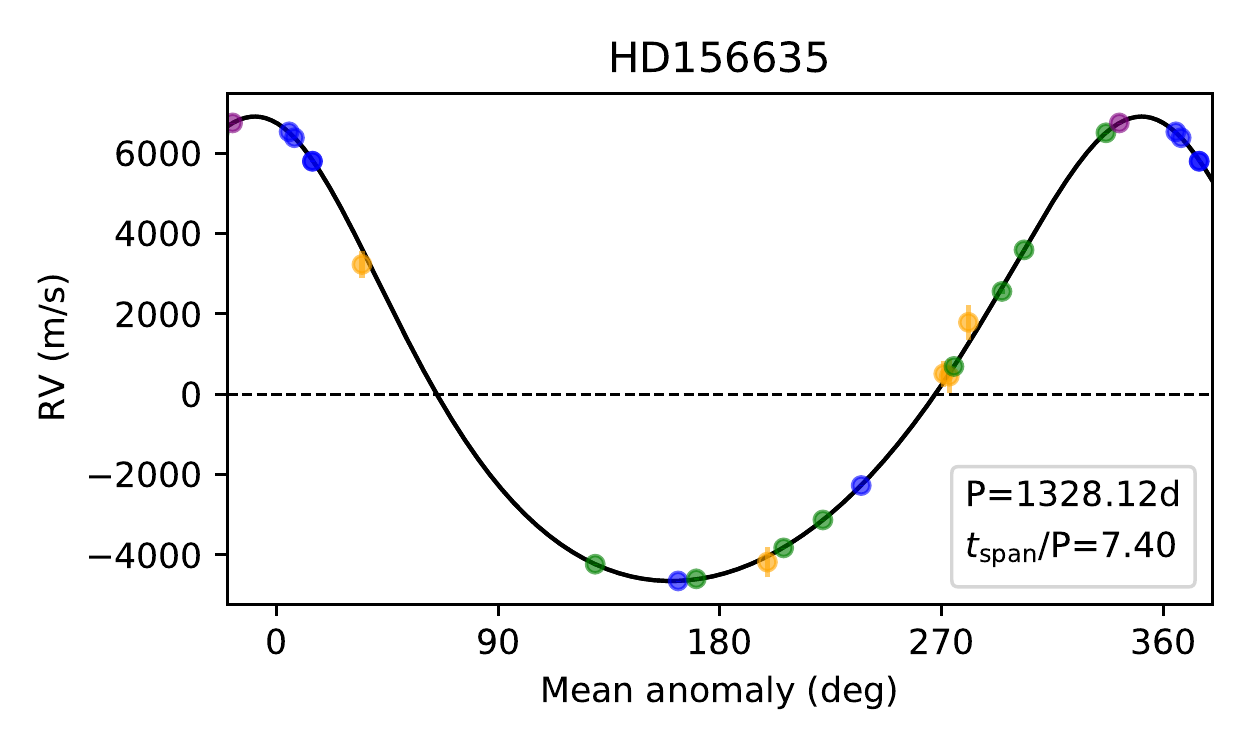}&
		\includegraphics[width=0.22\linewidth]{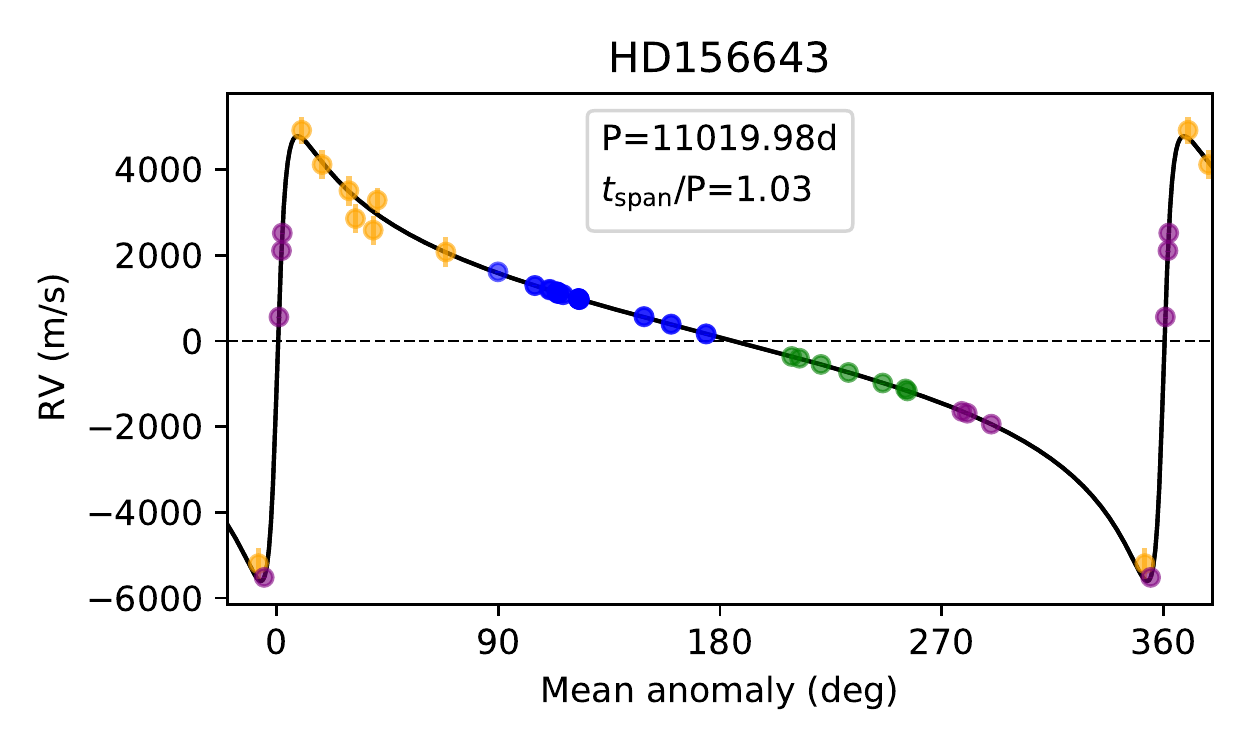}&
		\includegraphics[width=0.22\linewidth]{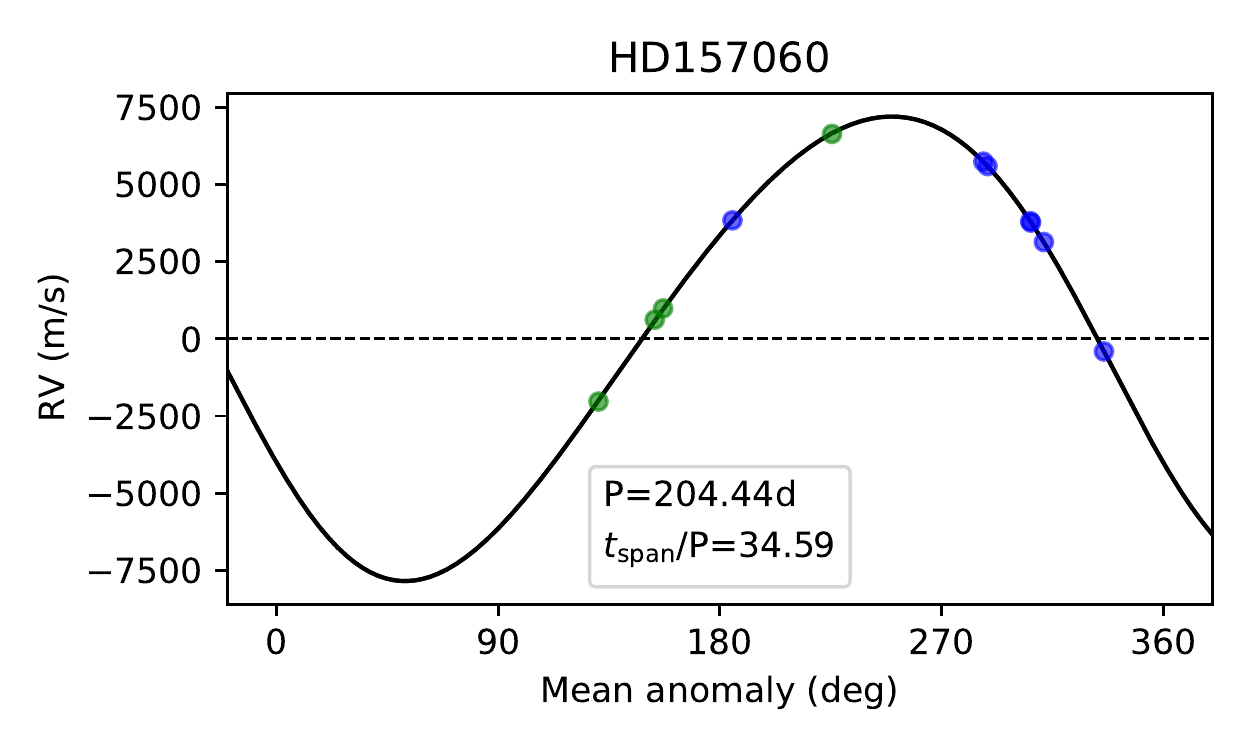}\\

		\includegraphics[width=0.22\linewidth]{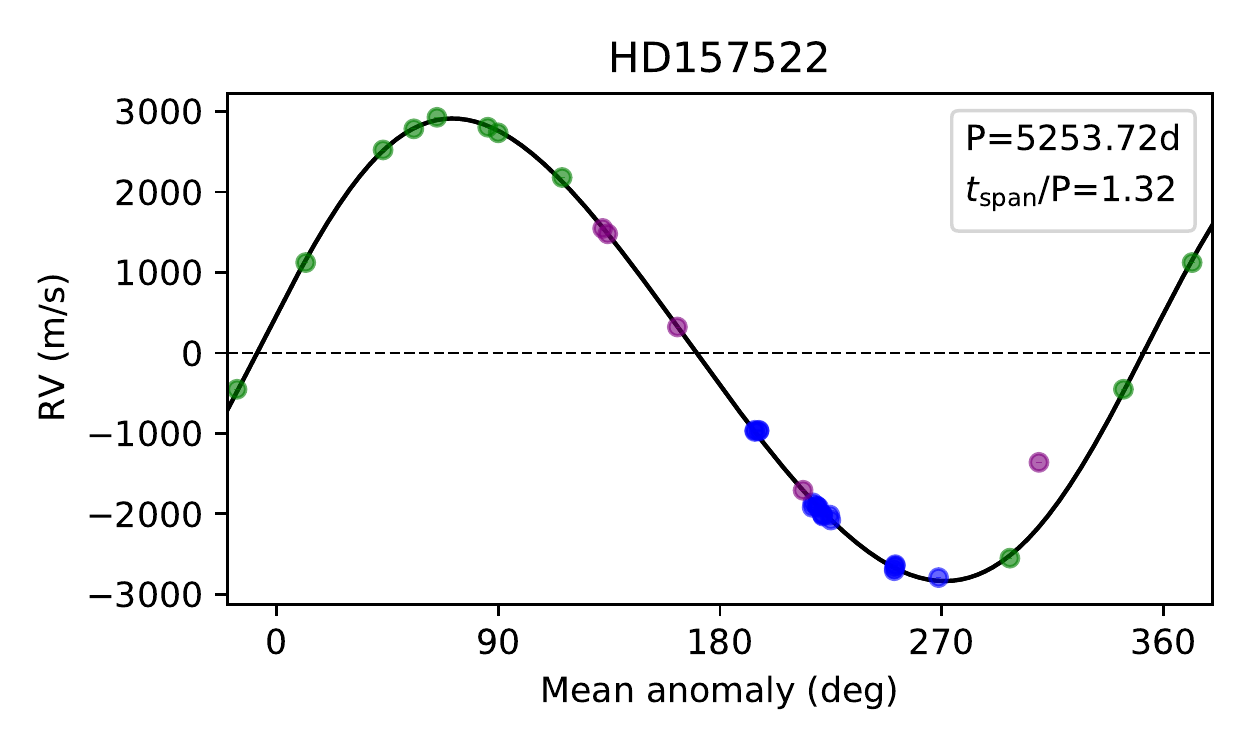}&
		\includegraphics[width=0.22\linewidth]{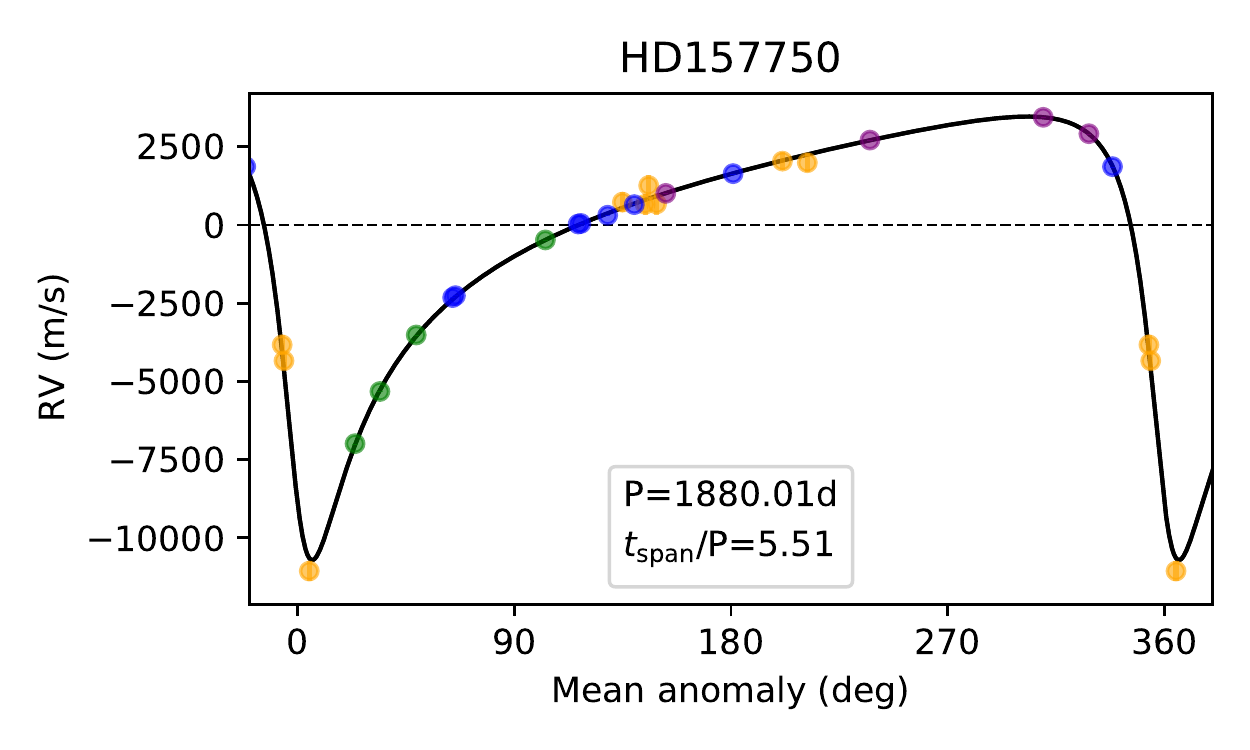}&
		\includegraphics[width=0.22\linewidth]{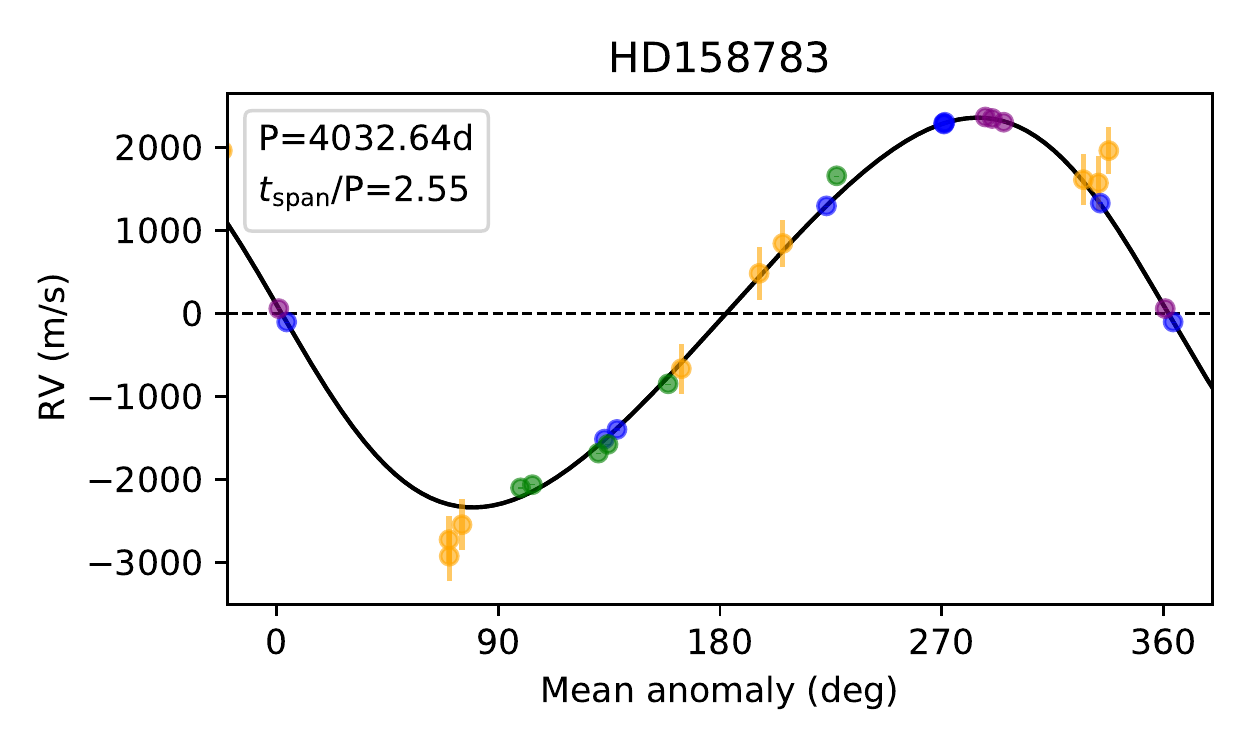}&
		\includegraphics[width=0.22\linewidth]{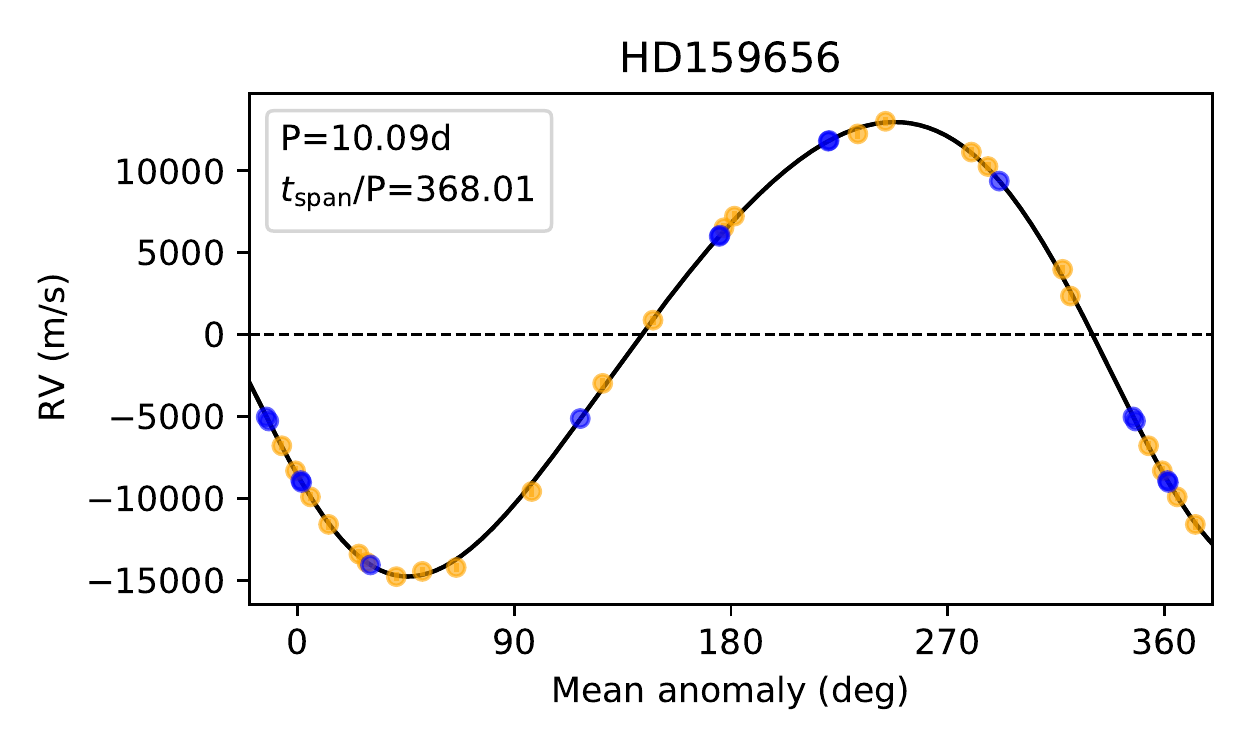}\\

		\includegraphics[width=0.22\linewidth]{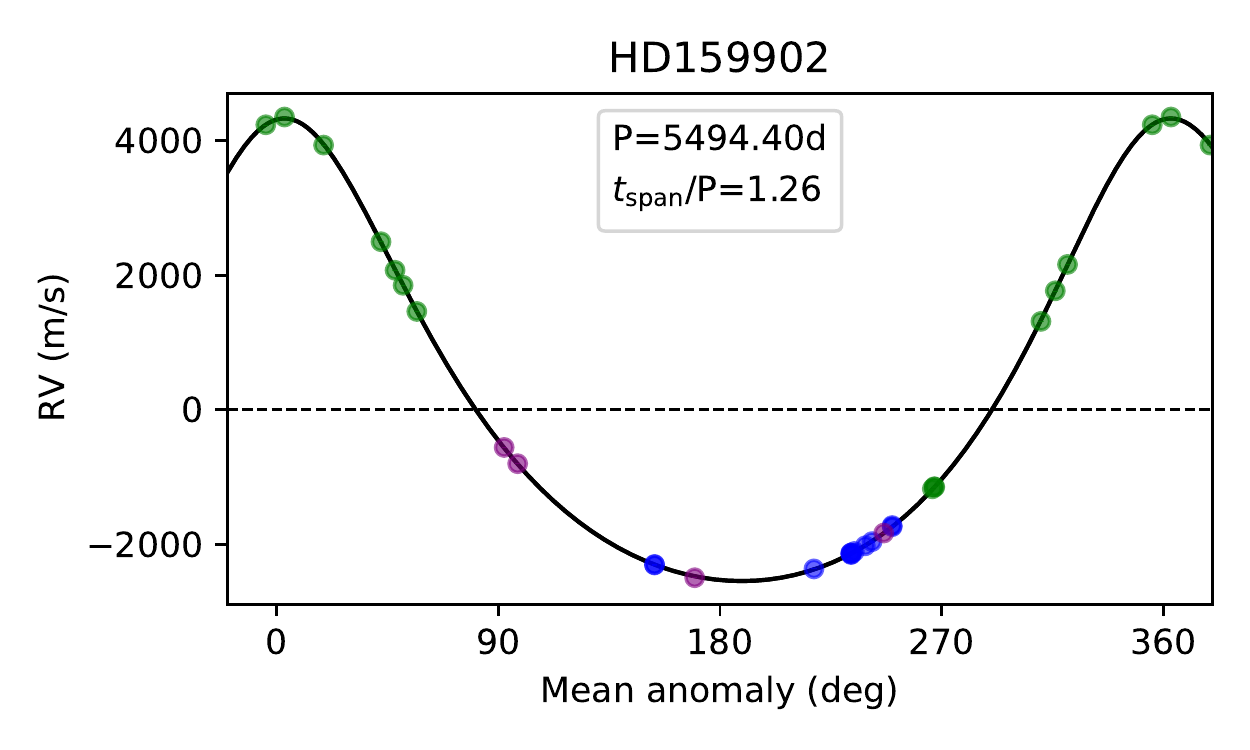}&
		\includegraphics[width=0.22\linewidth]{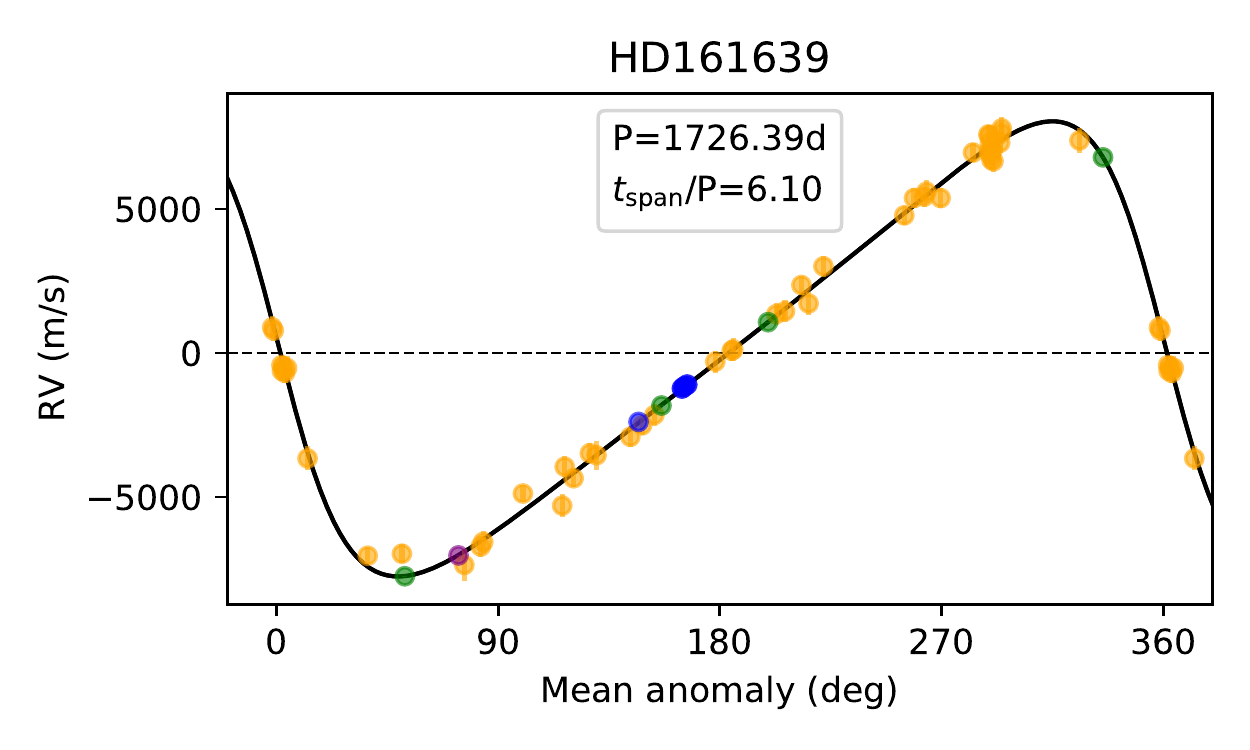}&
		\includegraphics[width=0.22\linewidth]{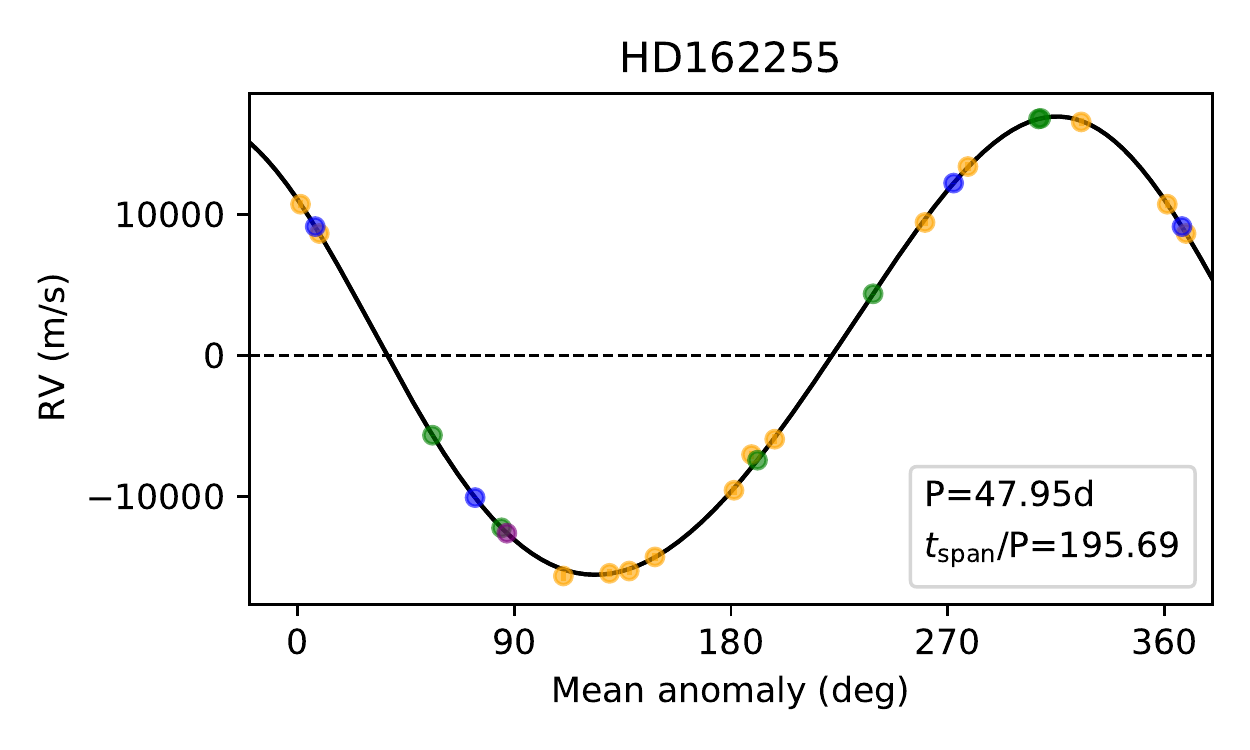}&
		\includegraphics[width=0.22\linewidth]{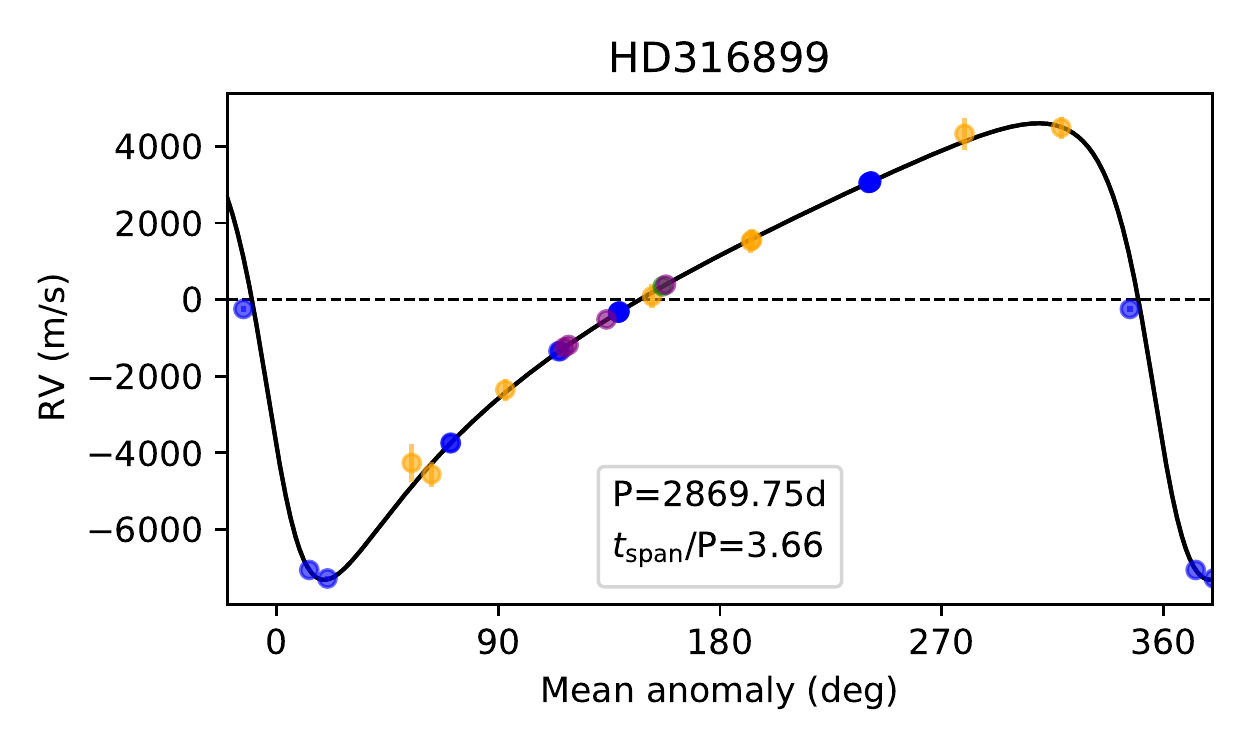}\\

		\includegraphics[width=0.22\linewidth]{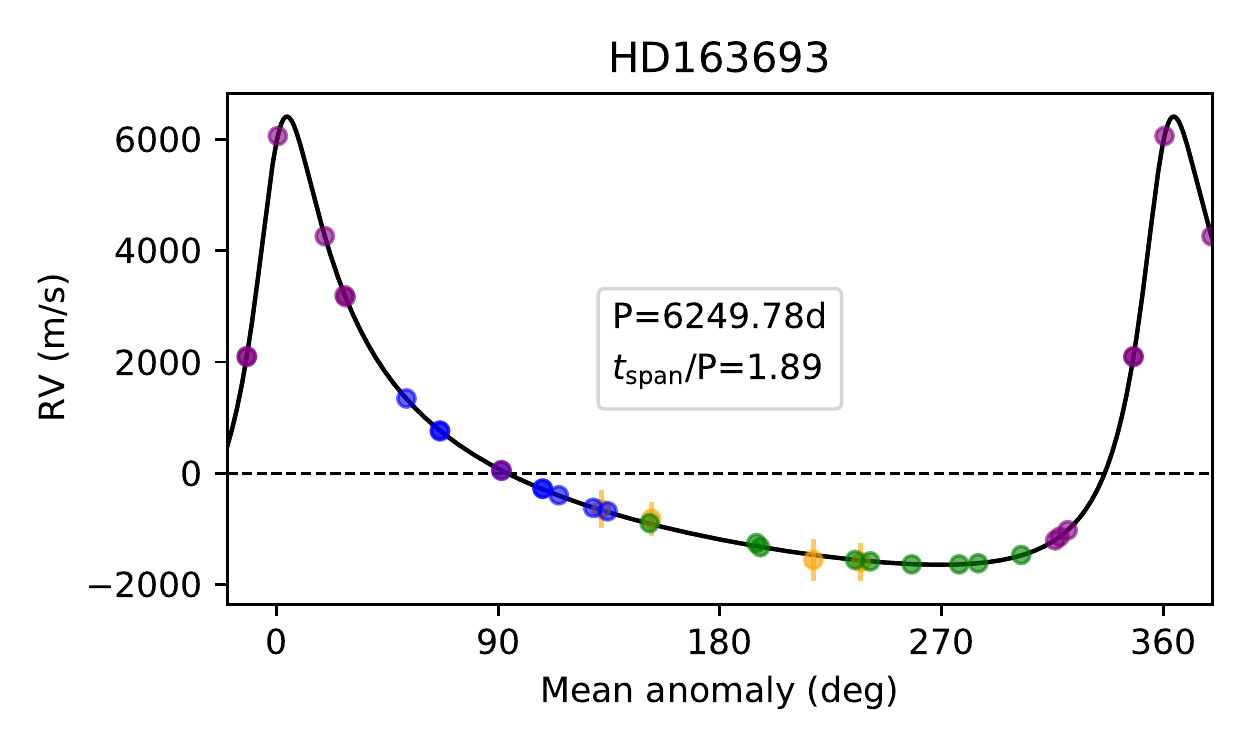}&
		\includegraphics[width=0.22\linewidth]{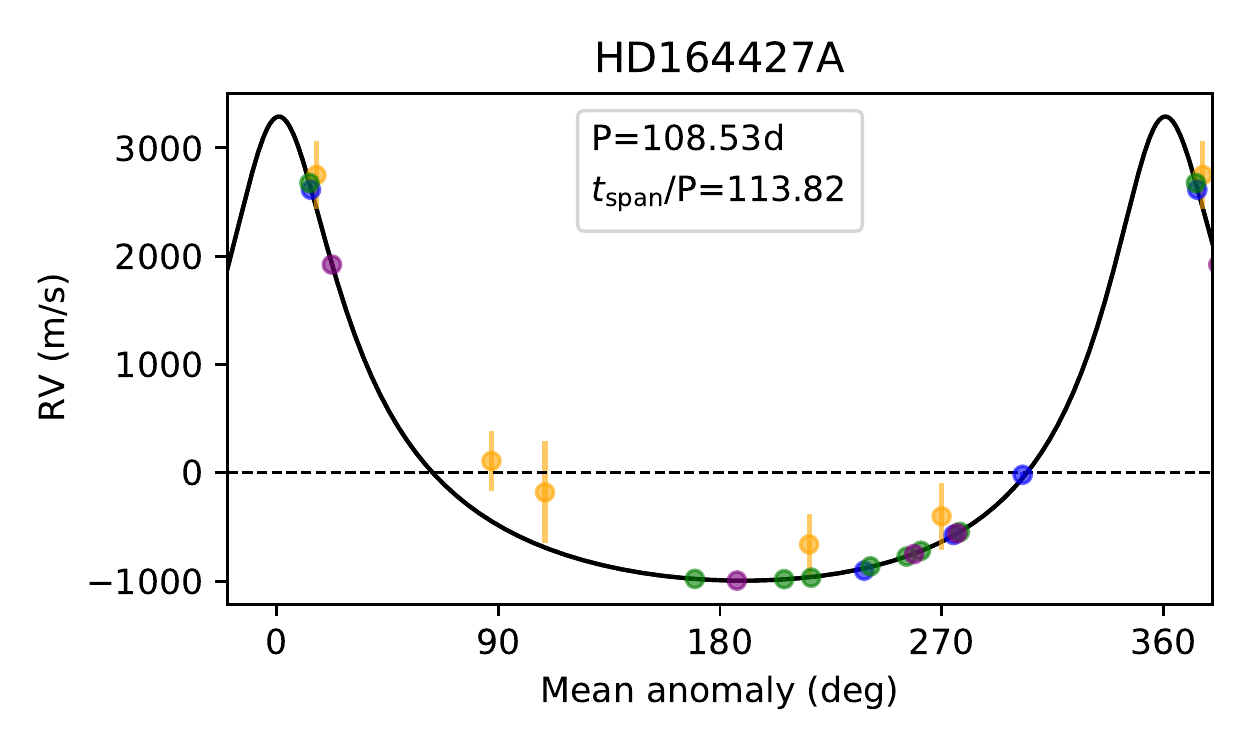}&
		\includegraphics[width=0.22\linewidth]{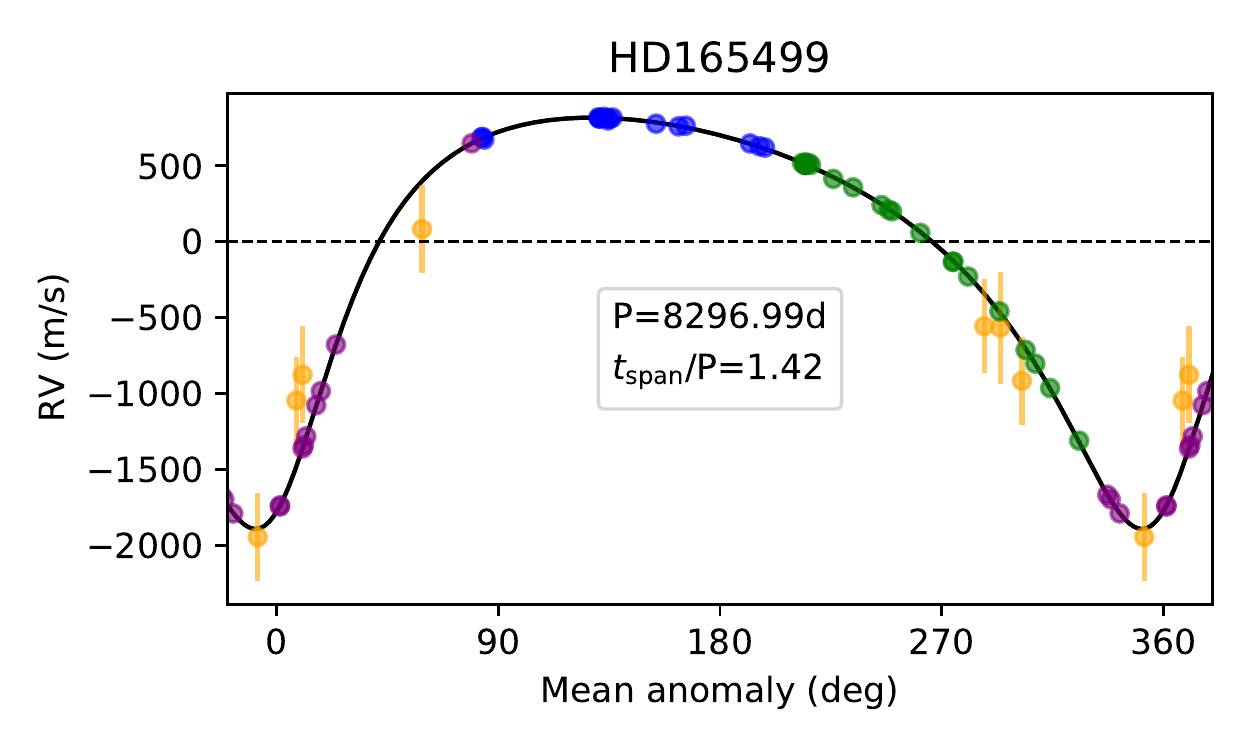}&
		\includegraphics[width=0.22\linewidth]{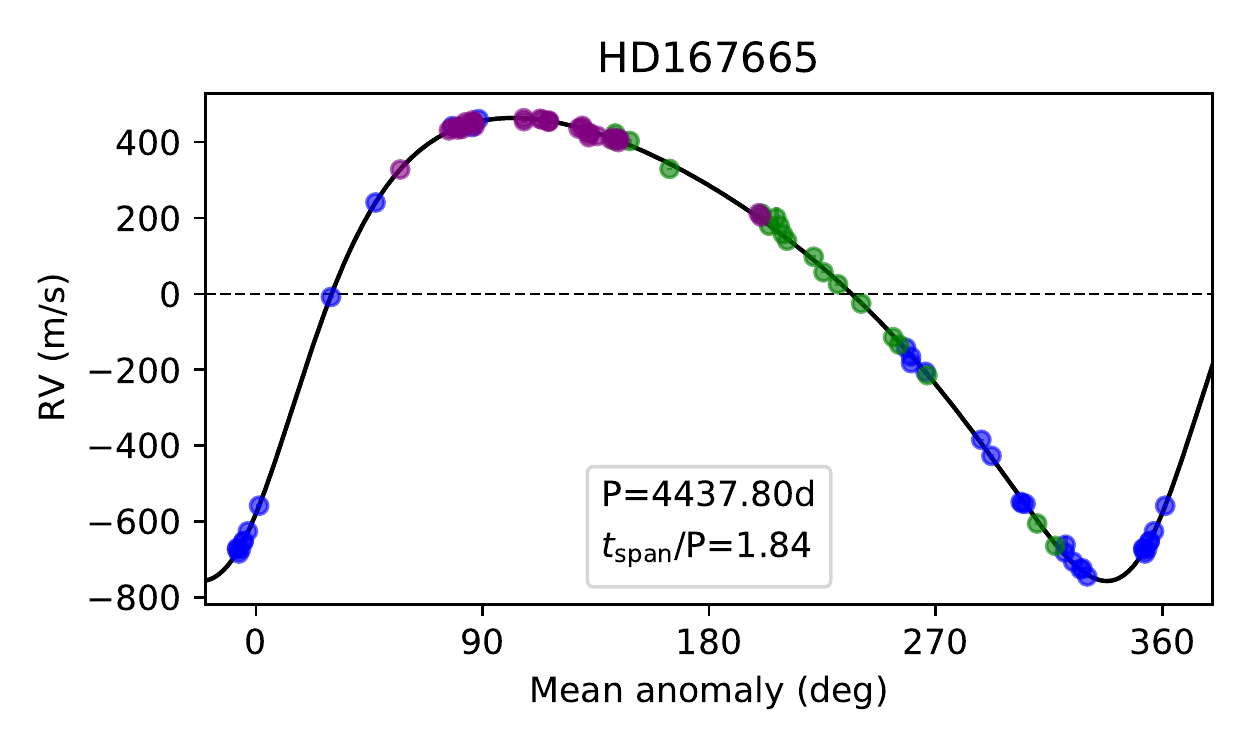}\\

		\includegraphics[width=0.22\linewidth]{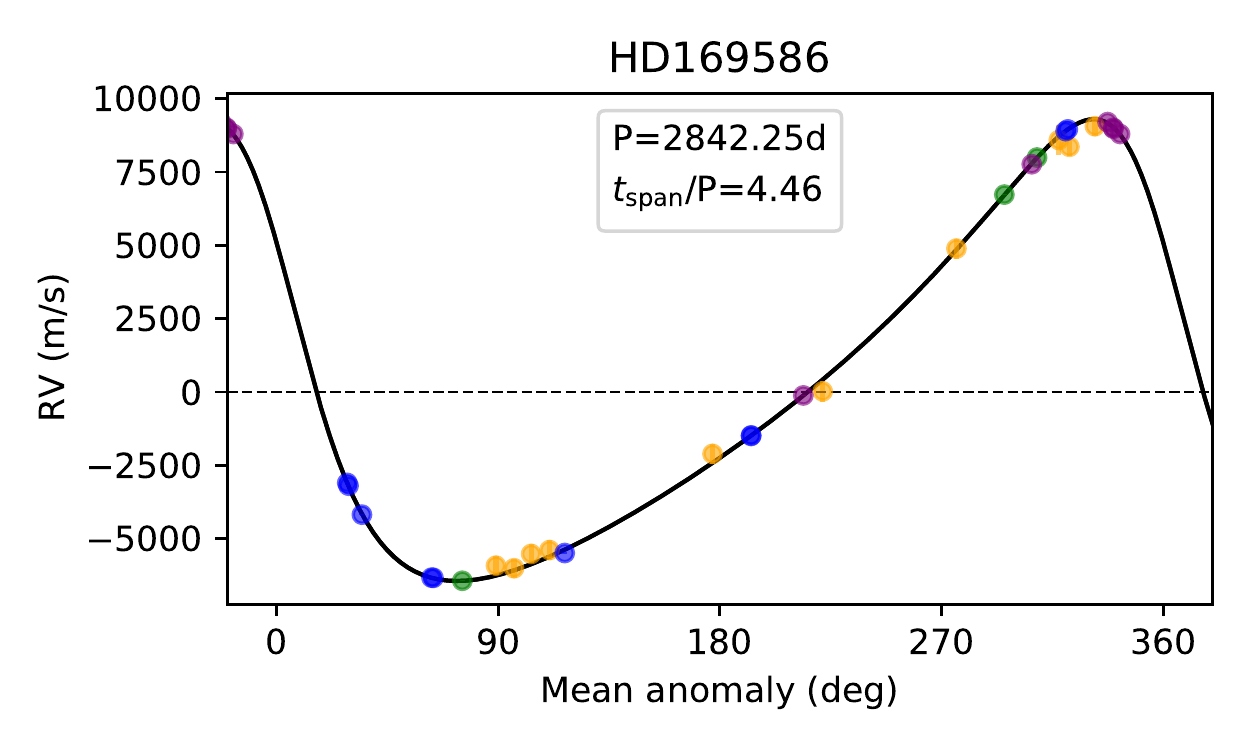}&
		\includegraphics[width=0.22\linewidth]{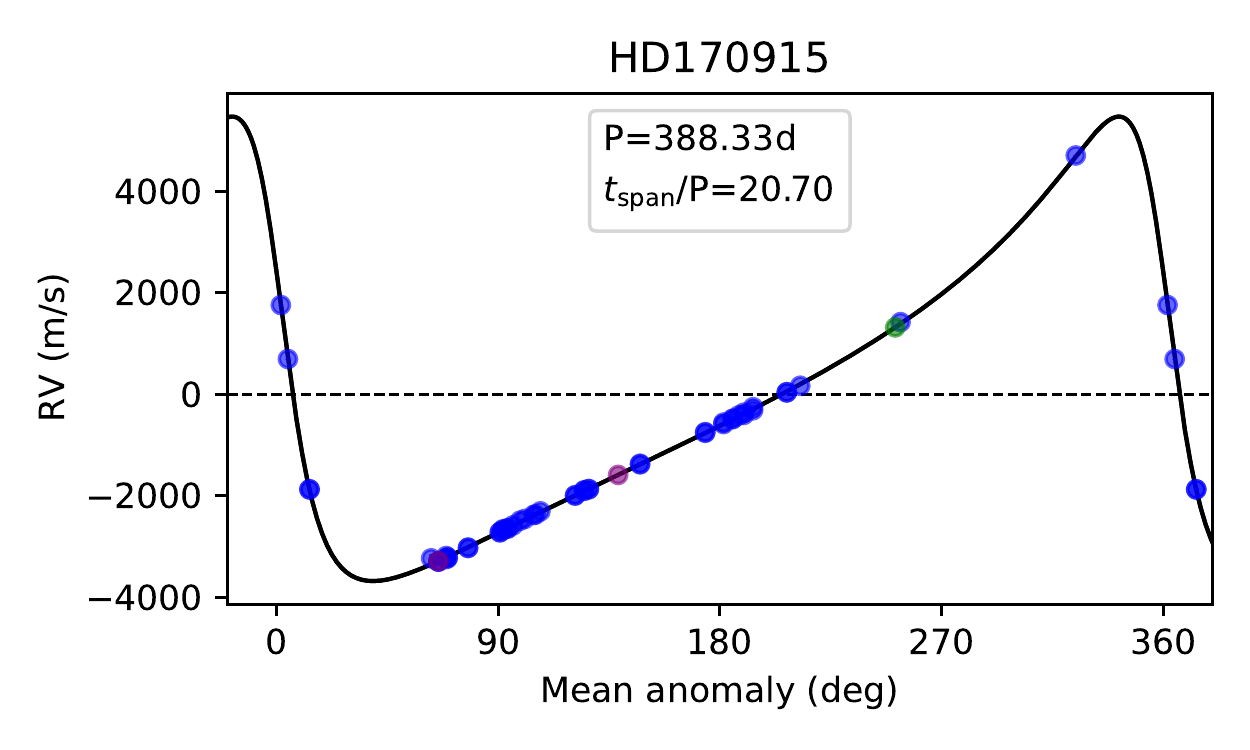}&
		\includegraphics[width=0.22\linewidth]{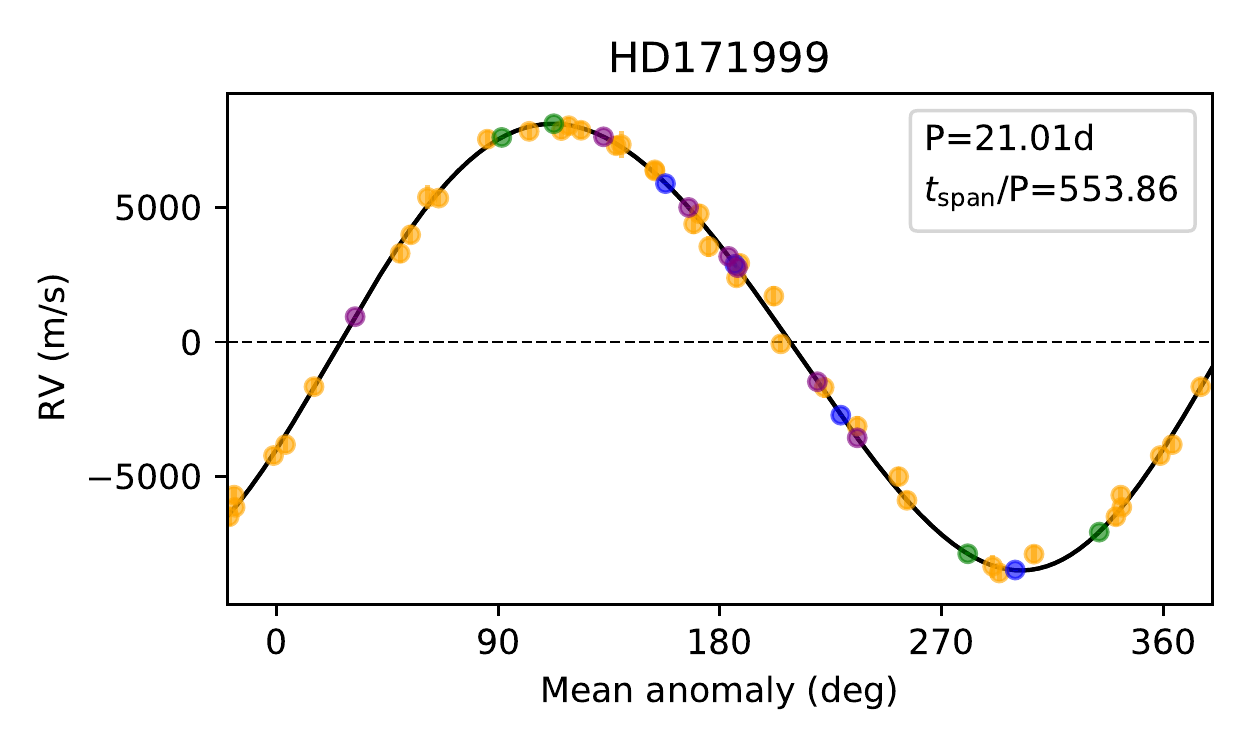}&
		\includegraphics[width=0.22\linewidth]{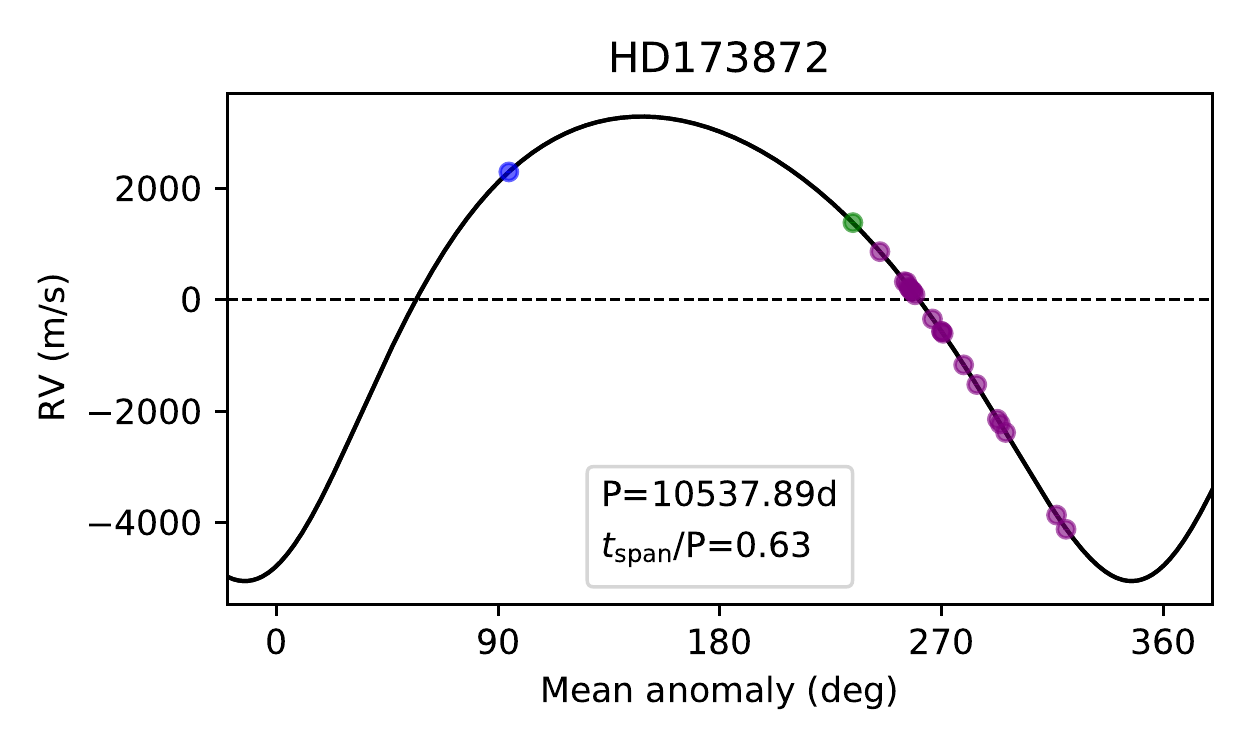}\\

		\includegraphics[width=0.22\linewidth]{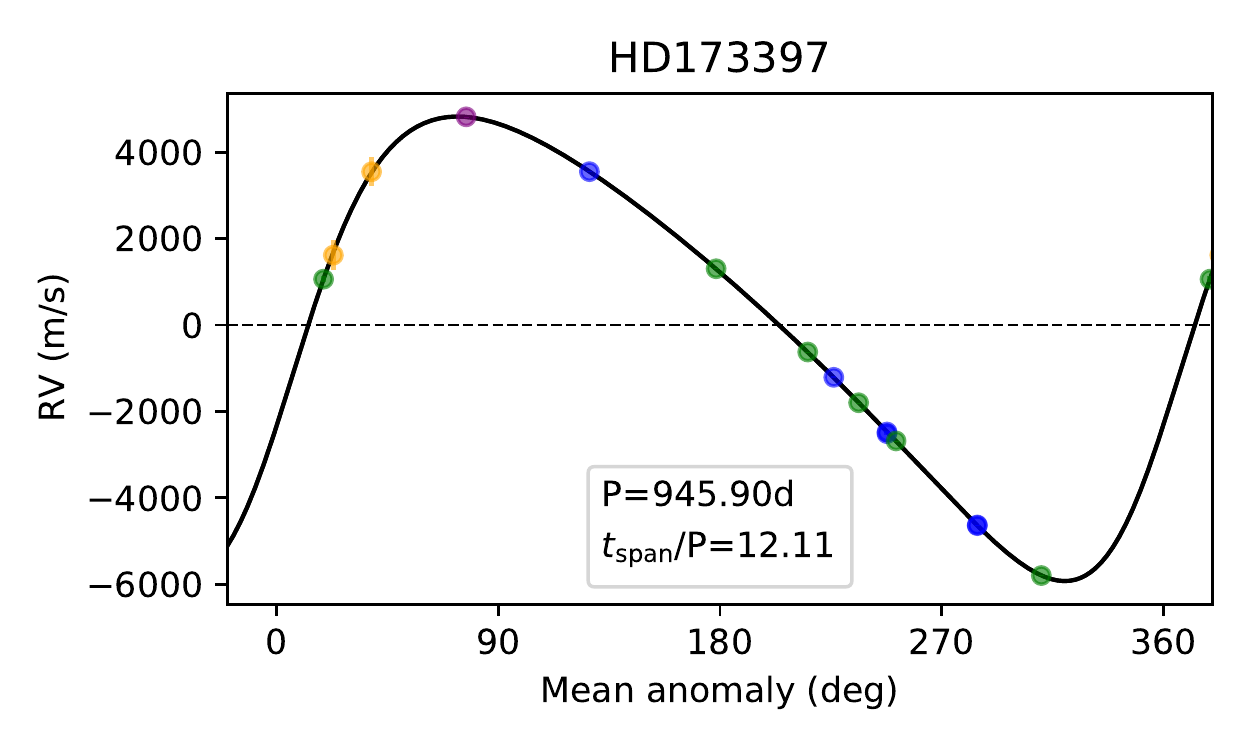}&
		\includegraphics[width=0.22\linewidth]{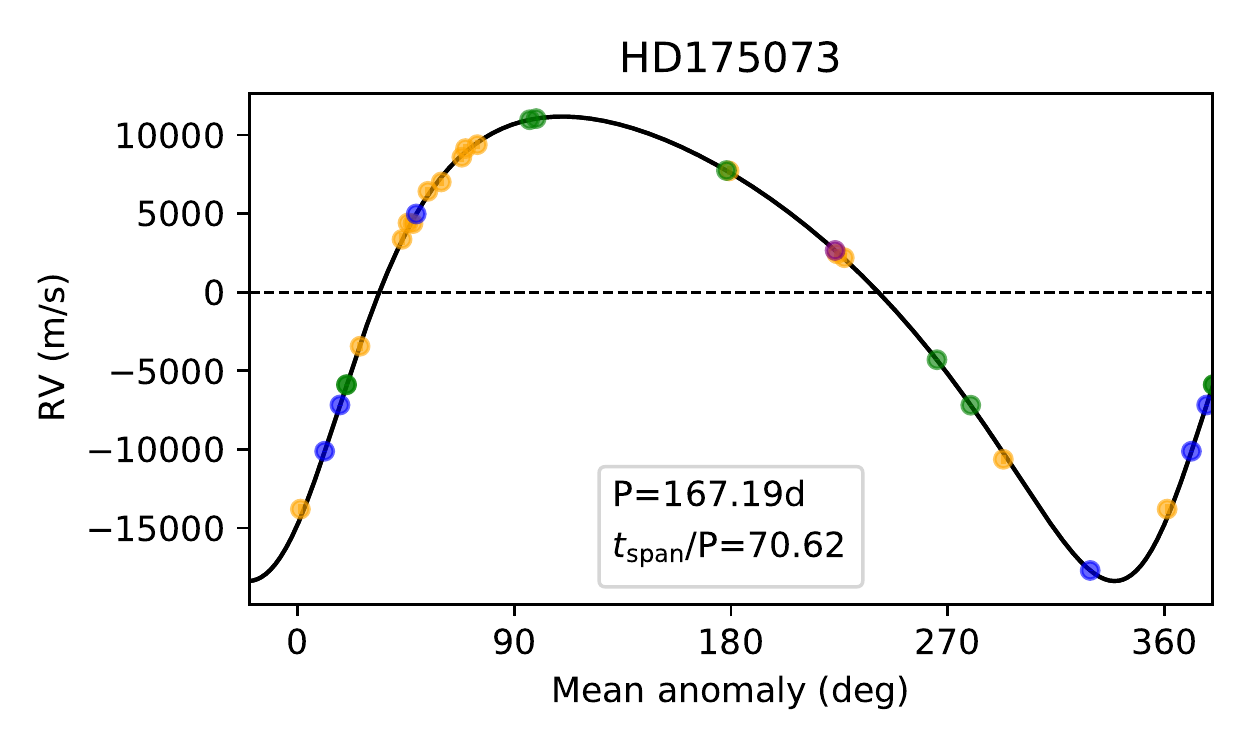}&
		\includegraphics[width=0.22\linewidth]{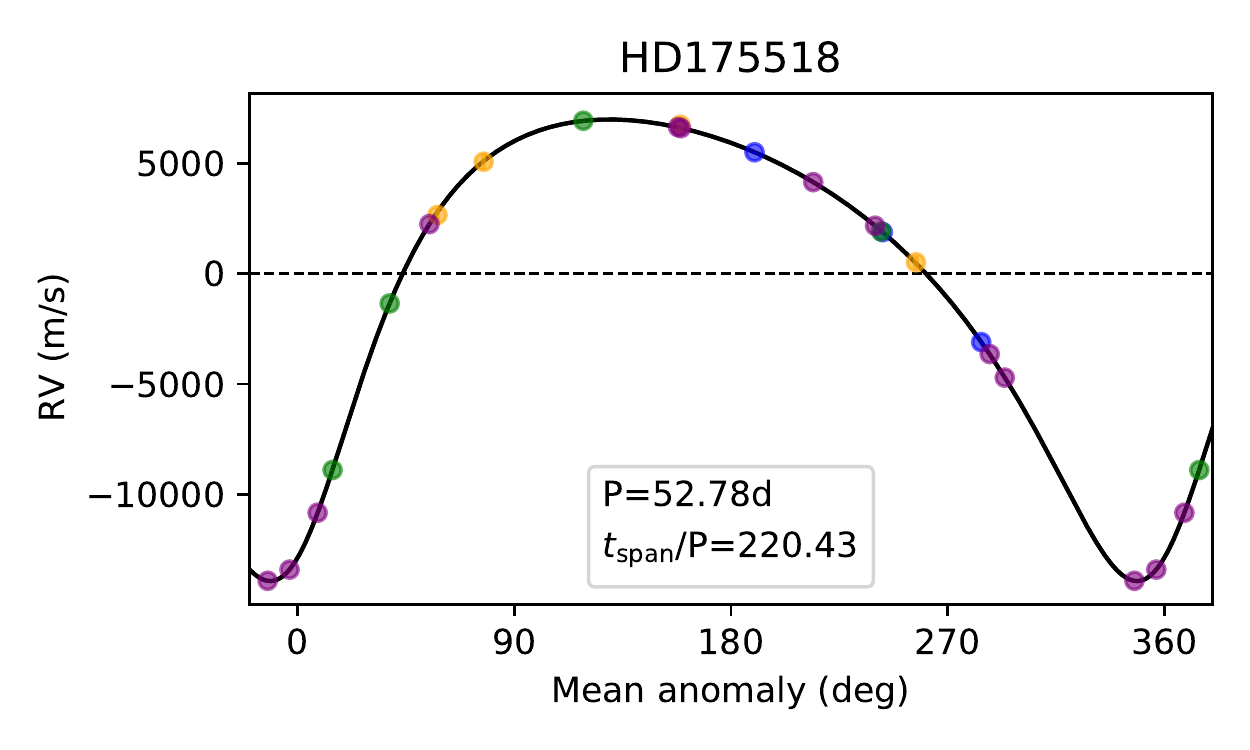}&
		\includegraphics[width=0.22\linewidth]{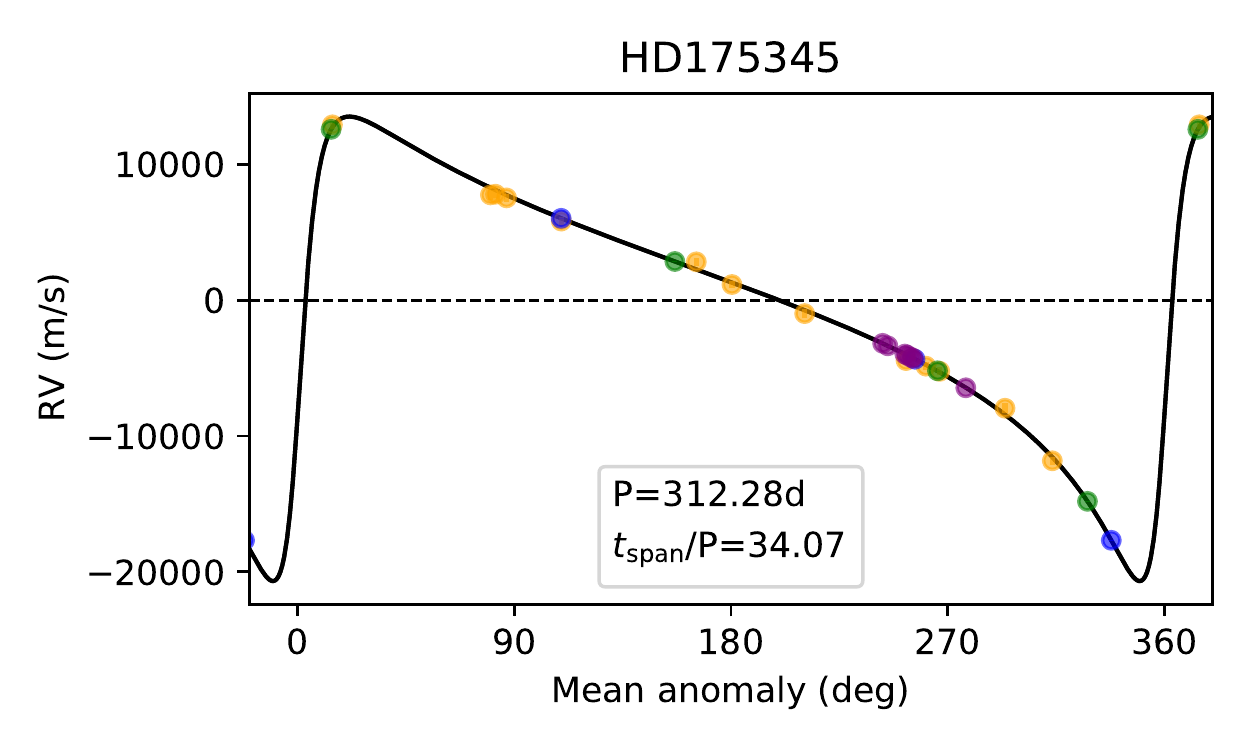}\\

		\includegraphics[width=0.22\linewidth]{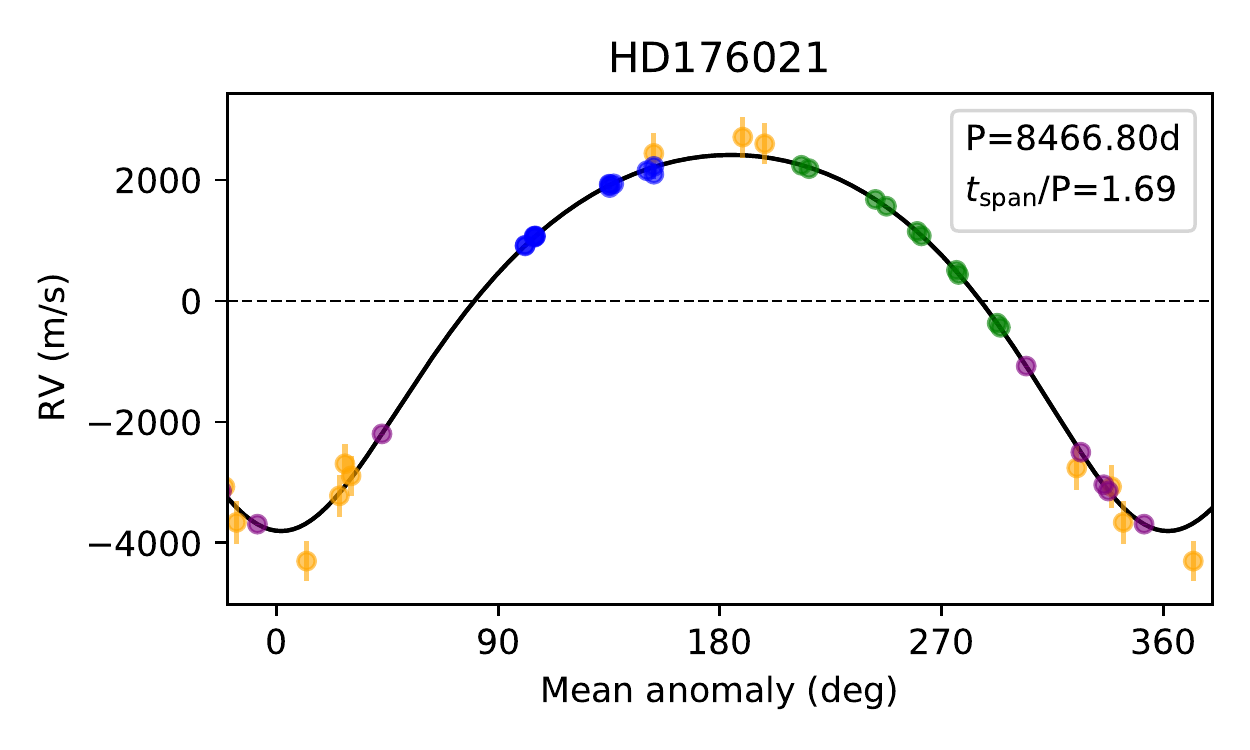}&
		\includegraphics[width=0.22\linewidth]{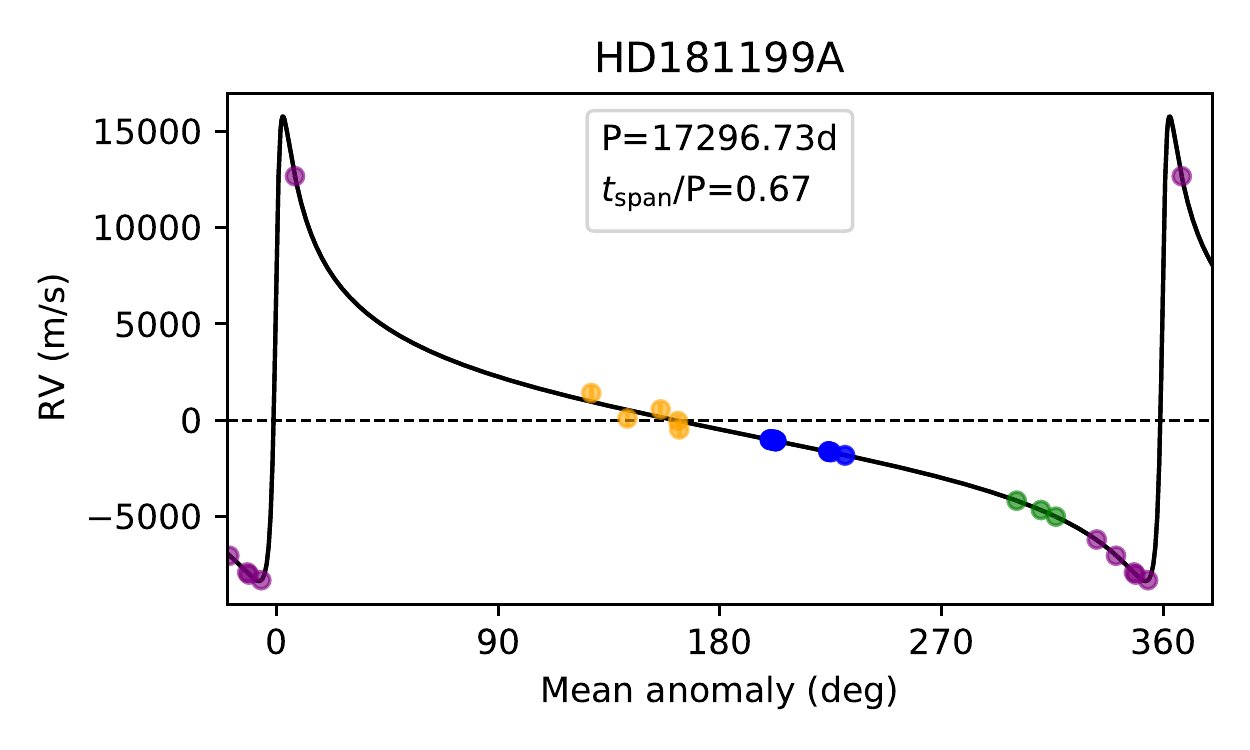}&
		\includegraphics[width=0.22\linewidth]{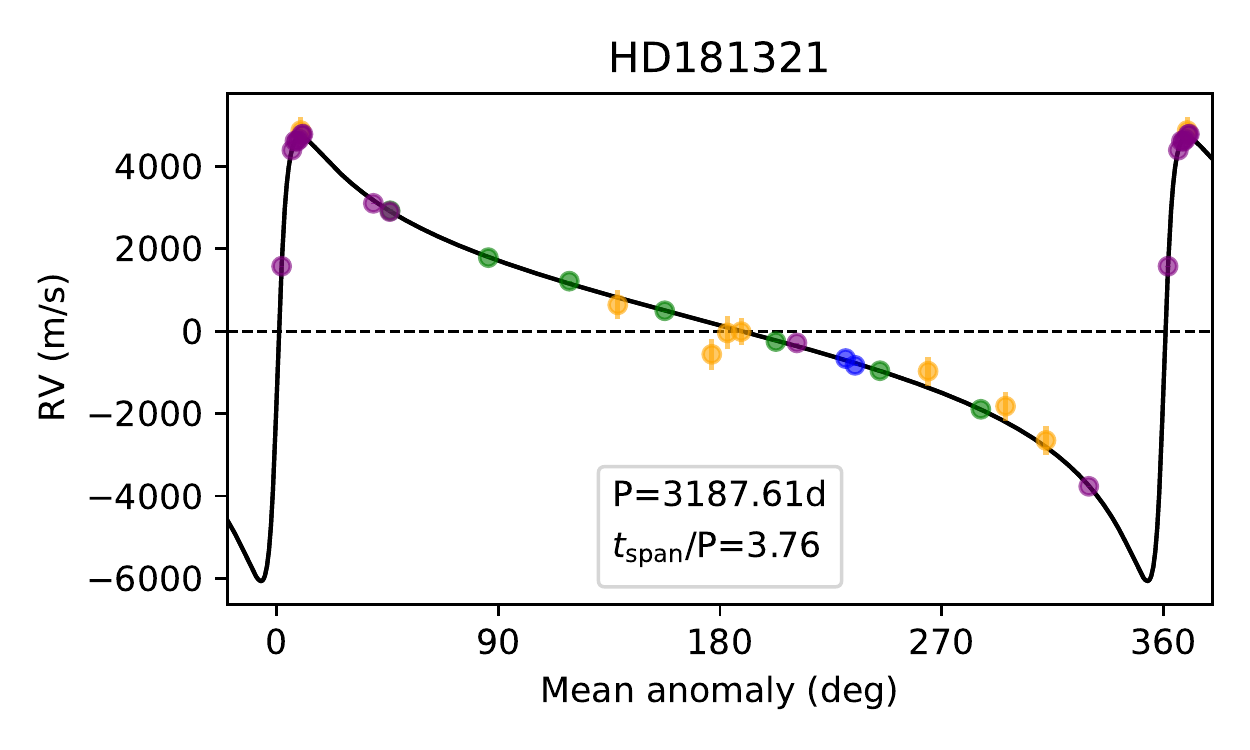}&
		\includegraphics[width=0.22\linewidth]{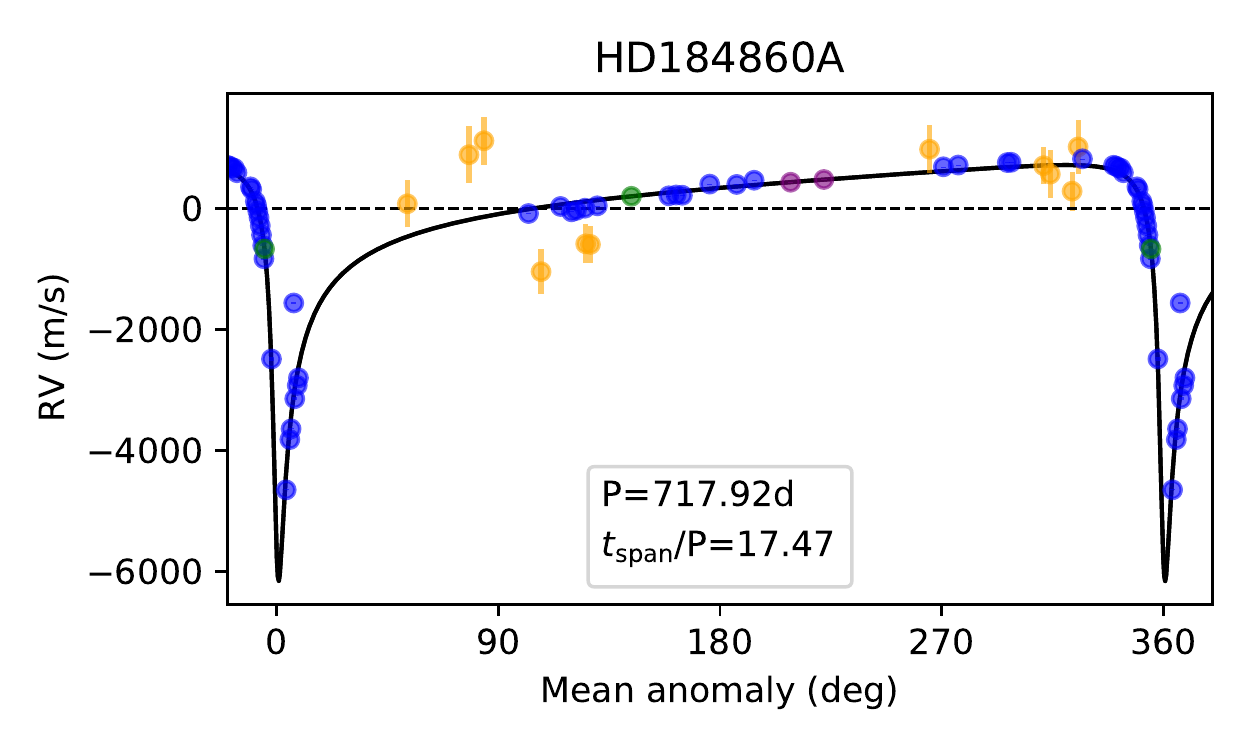}\\

		\includegraphics[width=0.22\linewidth]{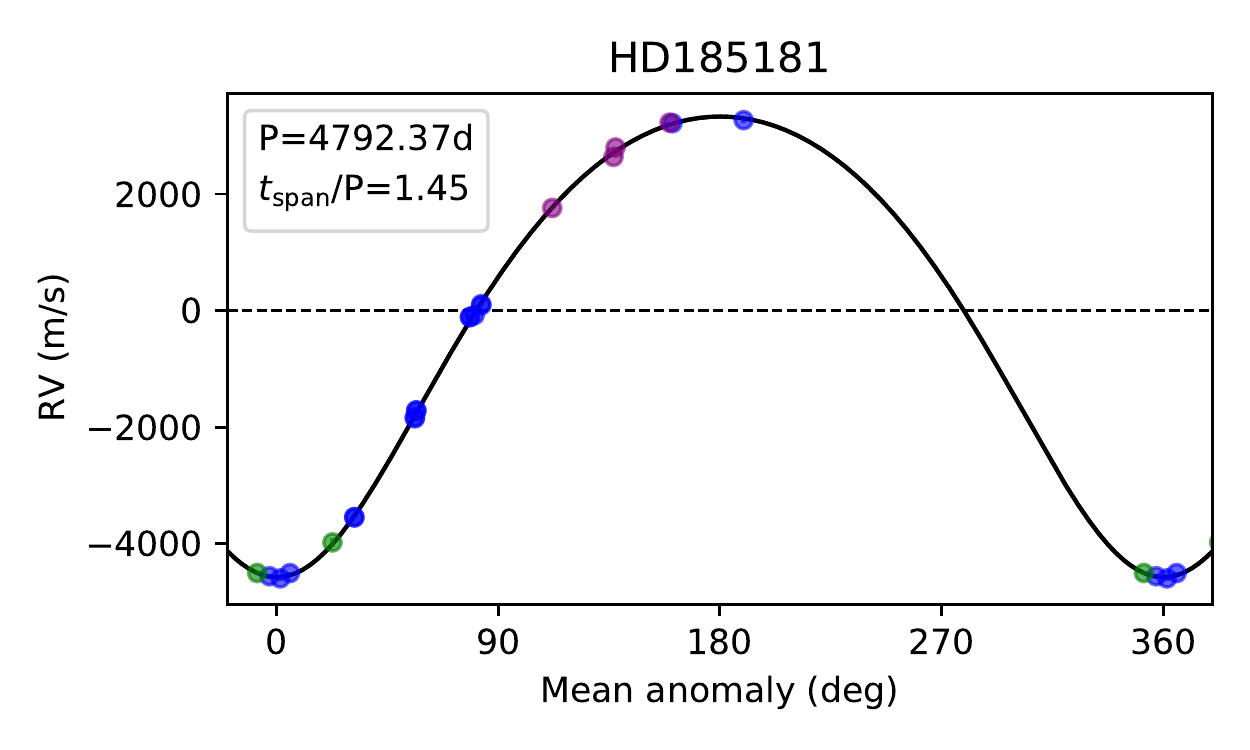}&
		\includegraphics[width=0.22\linewidth]{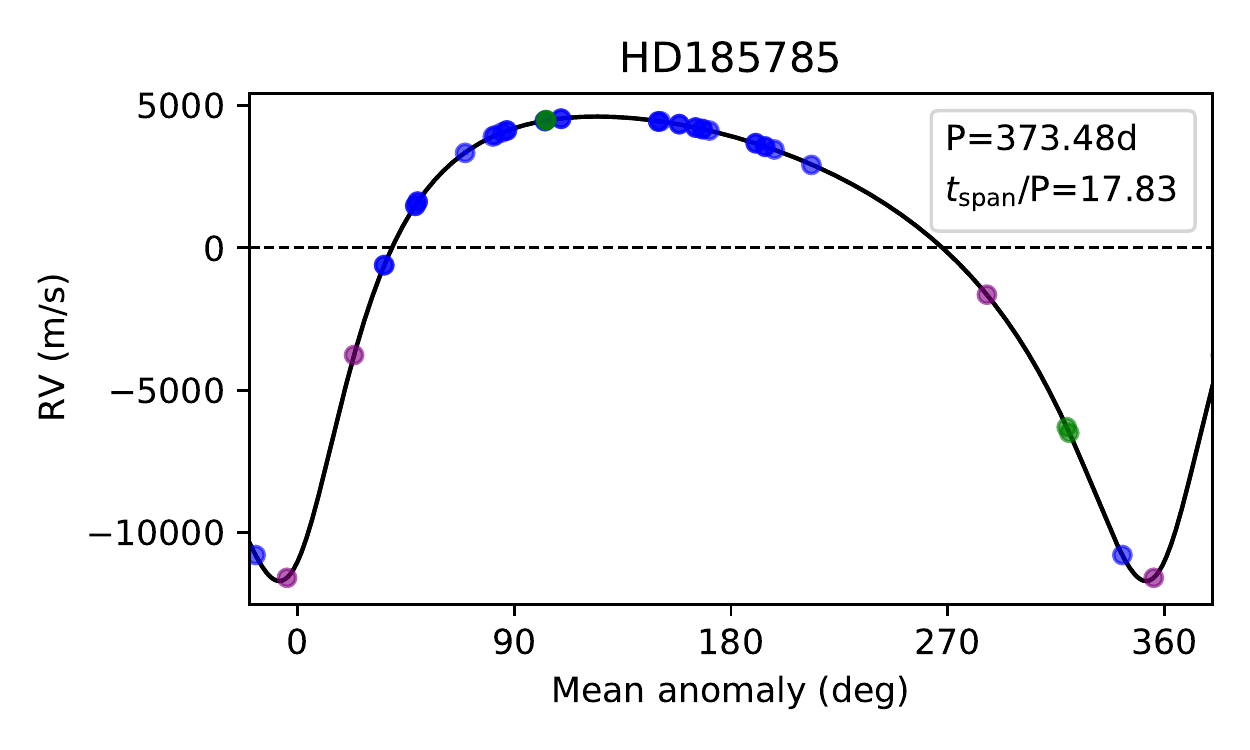}&
		\includegraphics[width=0.22\linewidth]{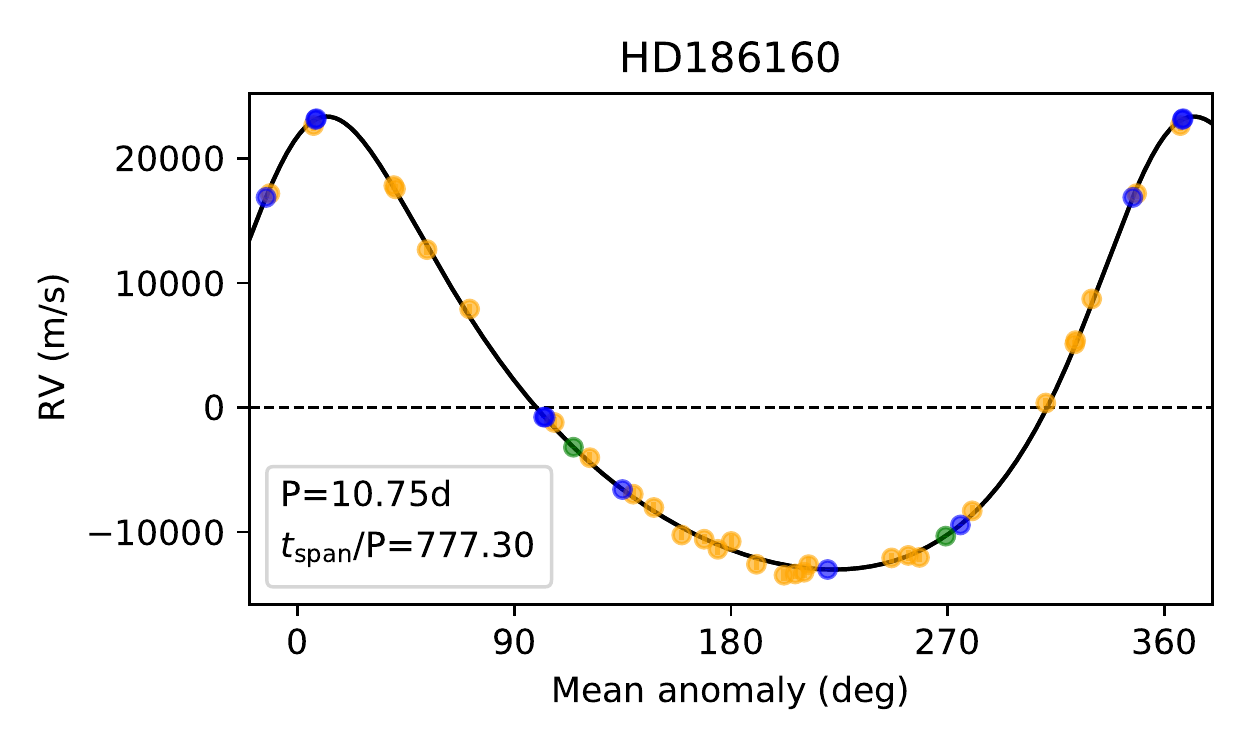}&
		\includegraphics[width=0.22\linewidth]{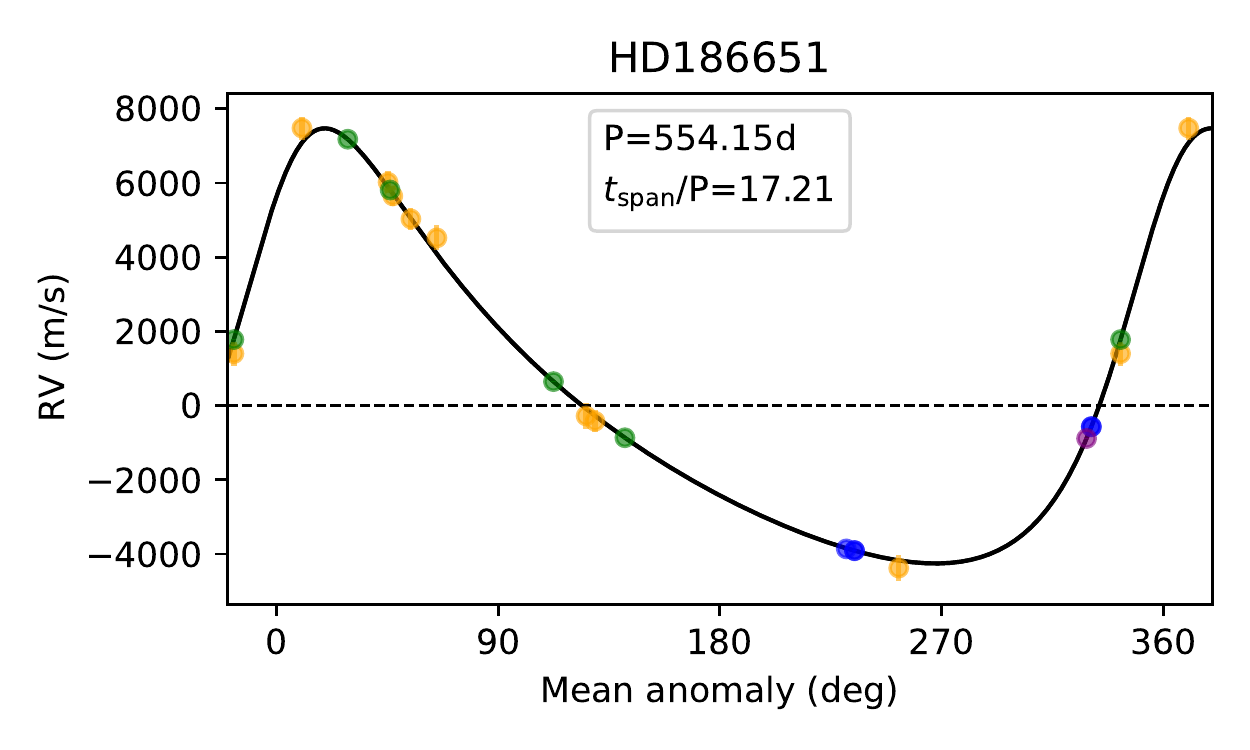}\\

		\includegraphics[width=0.22\linewidth]{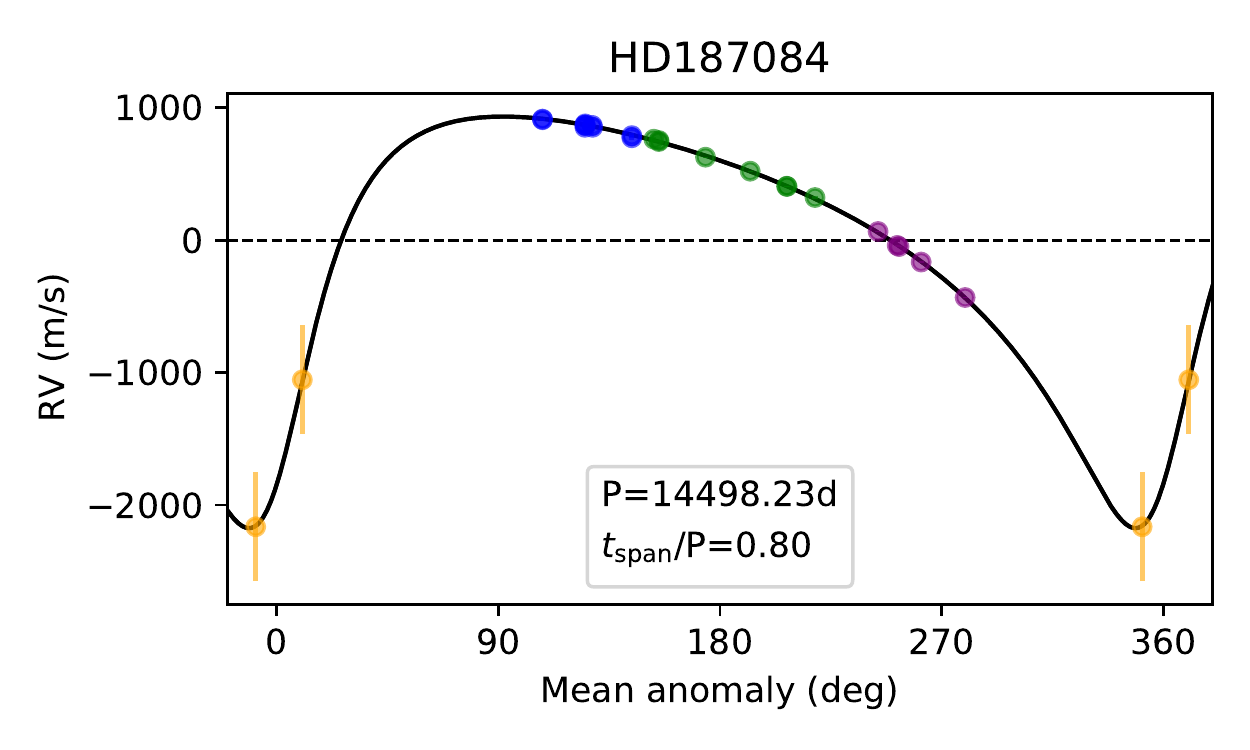}&
		\includegraphics[width=0.22\linewidth]{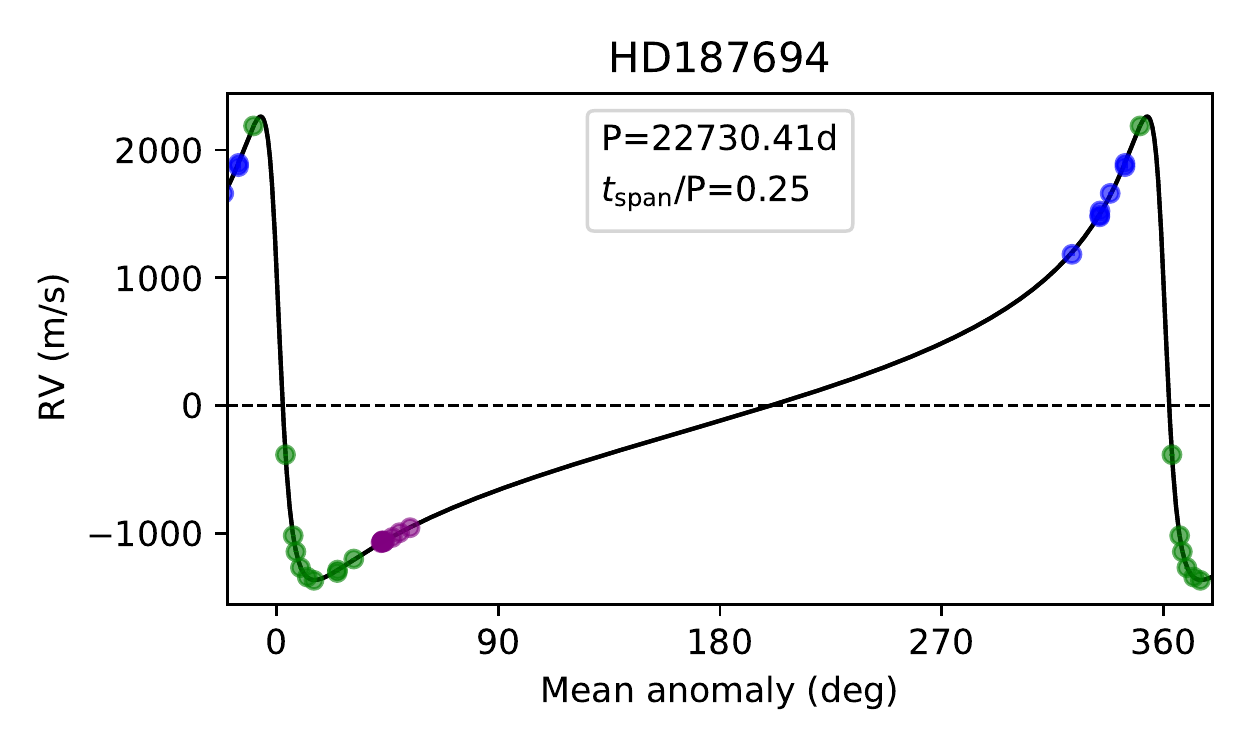}&
		\includegraphics[width=0.22\linewidth]{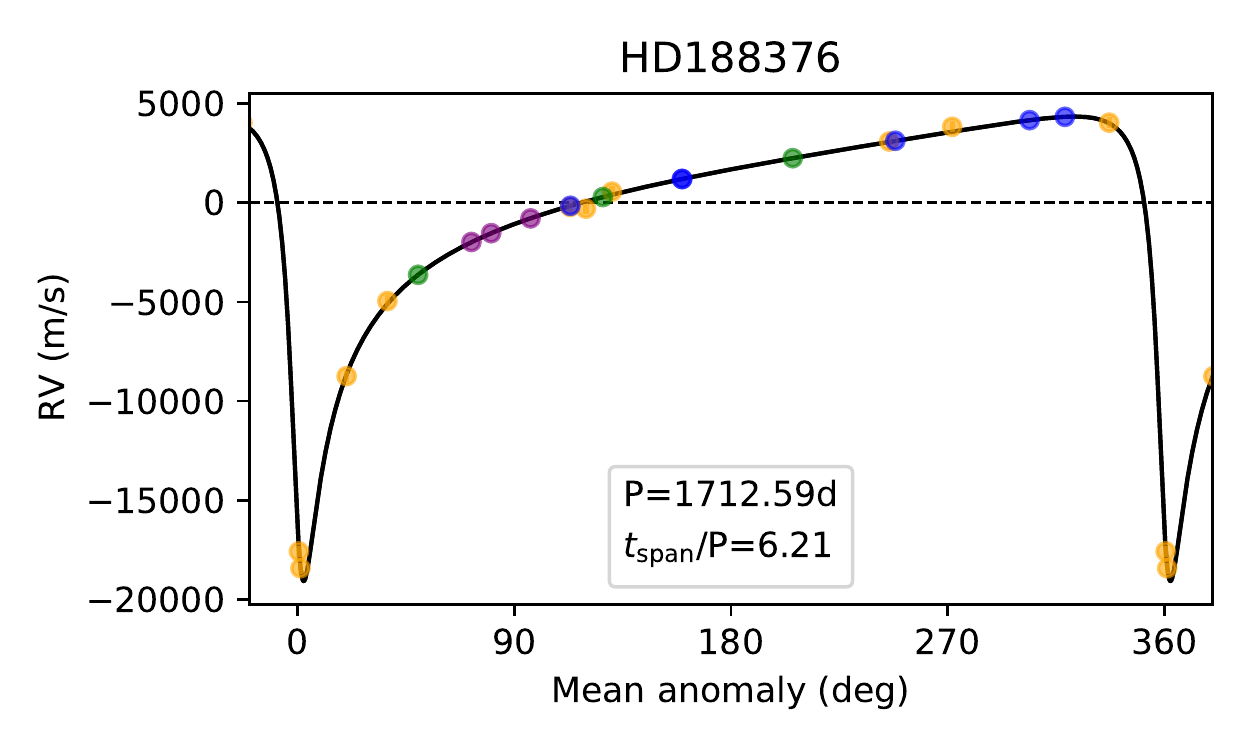}&
		\includegraphics[width=0.22\linewidth]{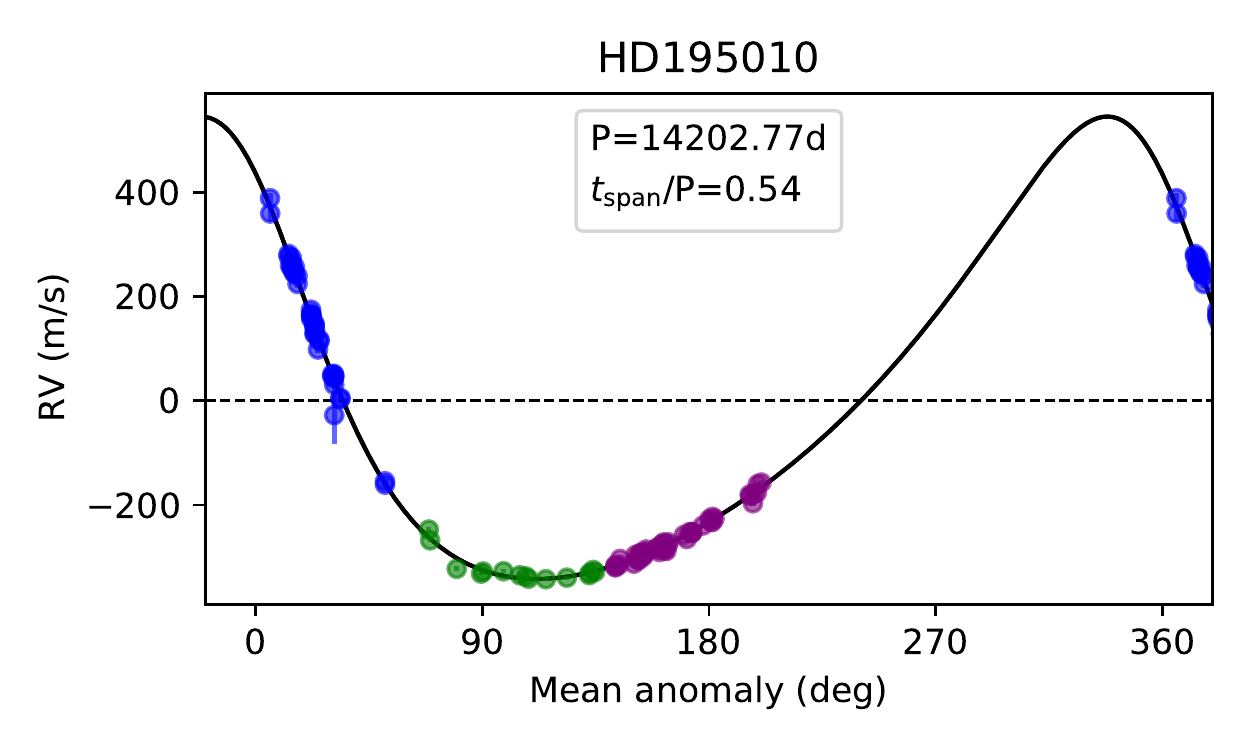}\\

		\includegraphics[width=0.22\linewidth]{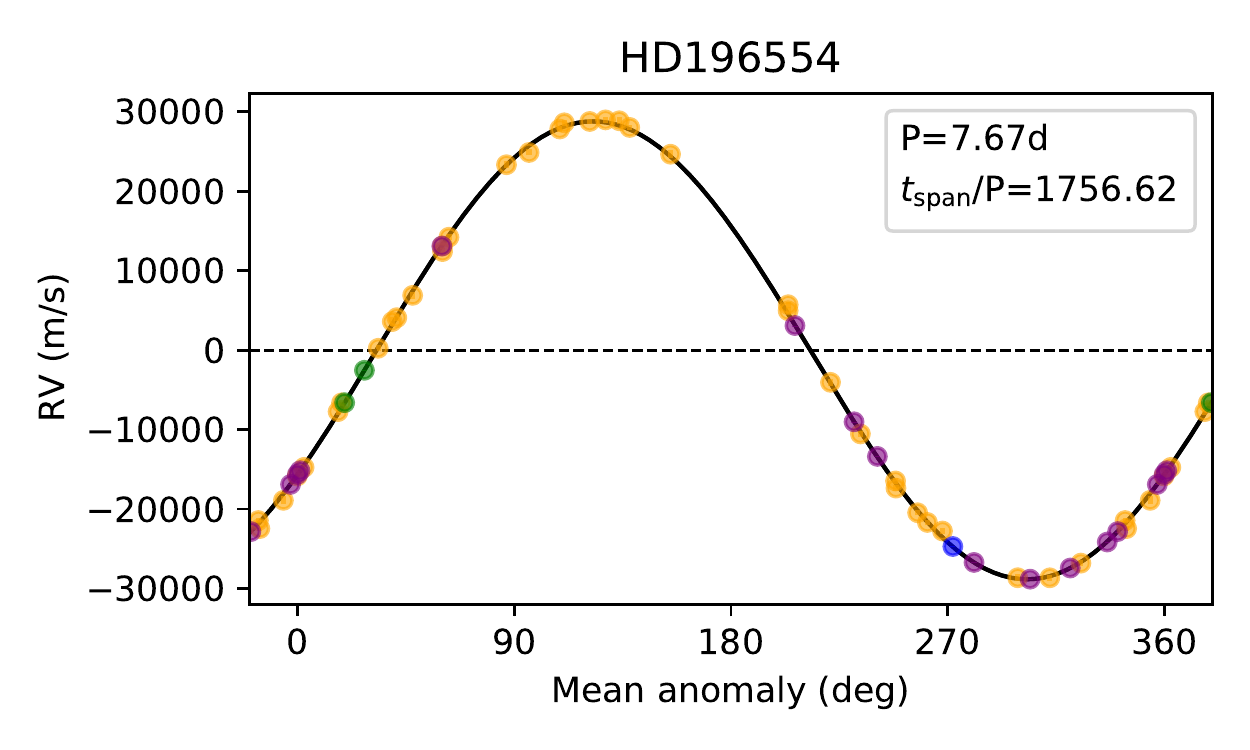}&
		\includegraphics[width=0.22\linewidth]{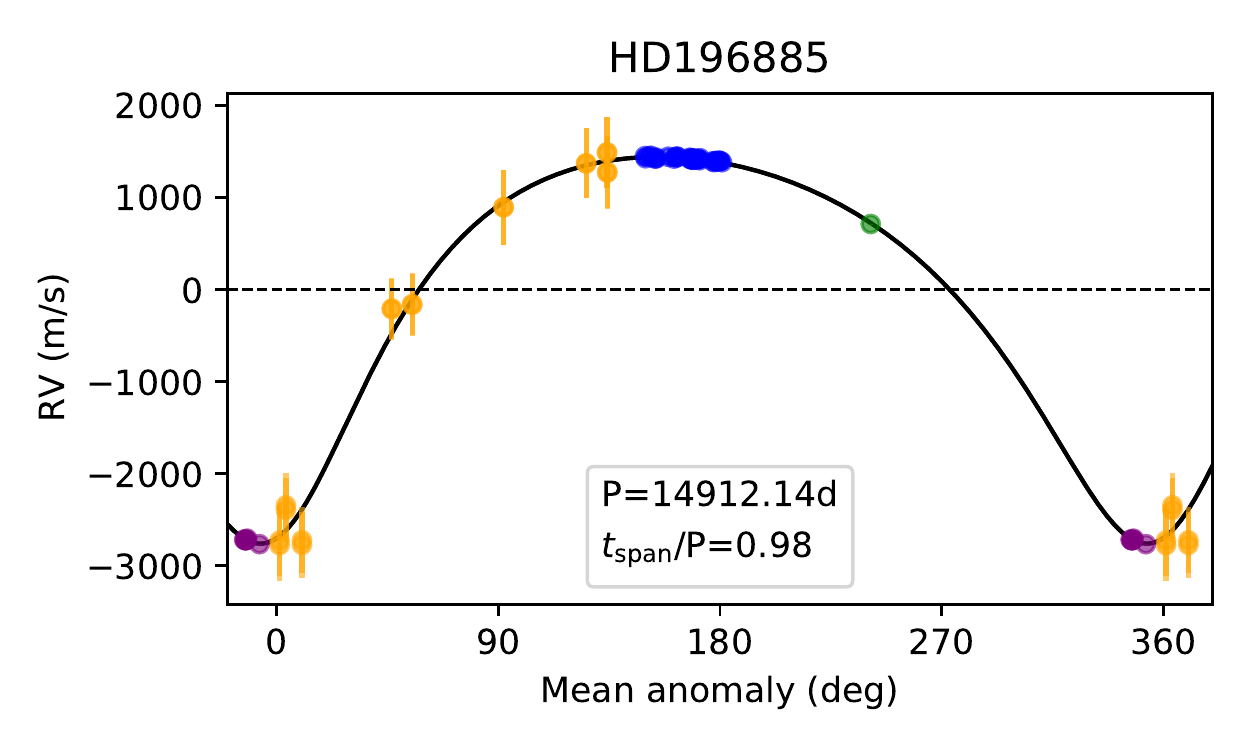}&
		\includegraphics[width=0.22\linewidth]{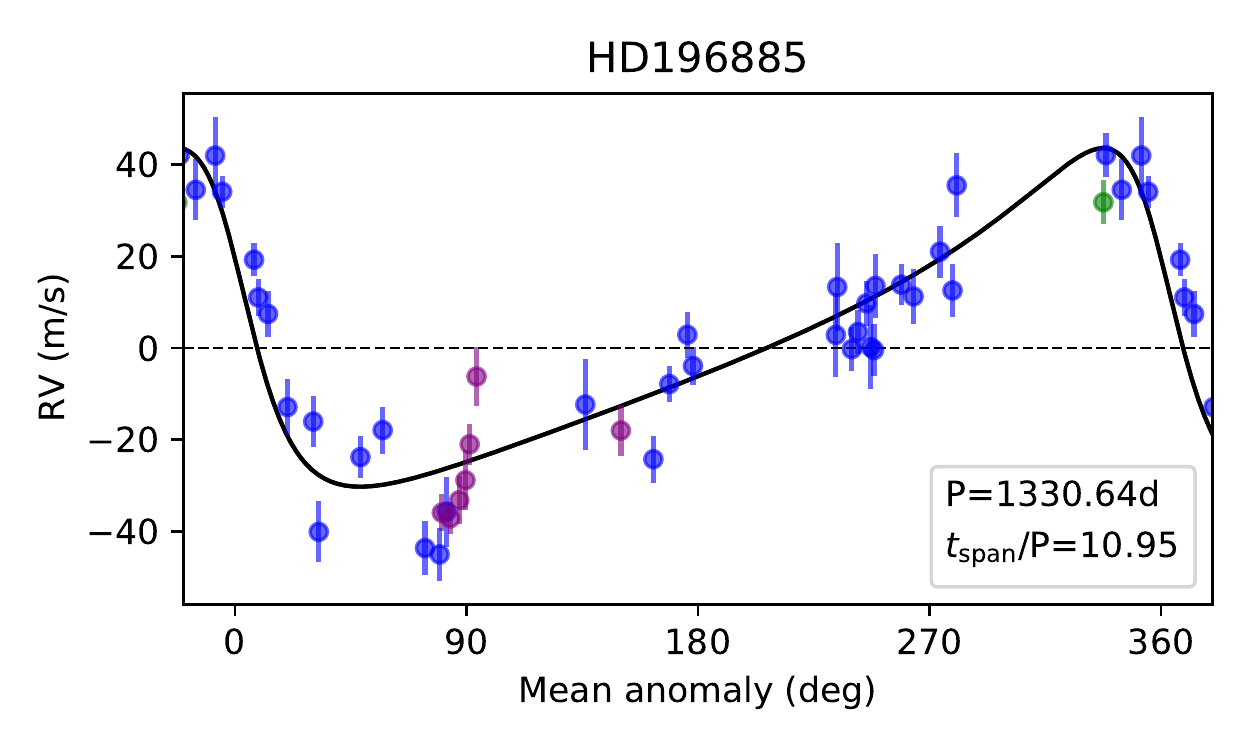}&
		\includegraphics[width=0.22\linewidth]{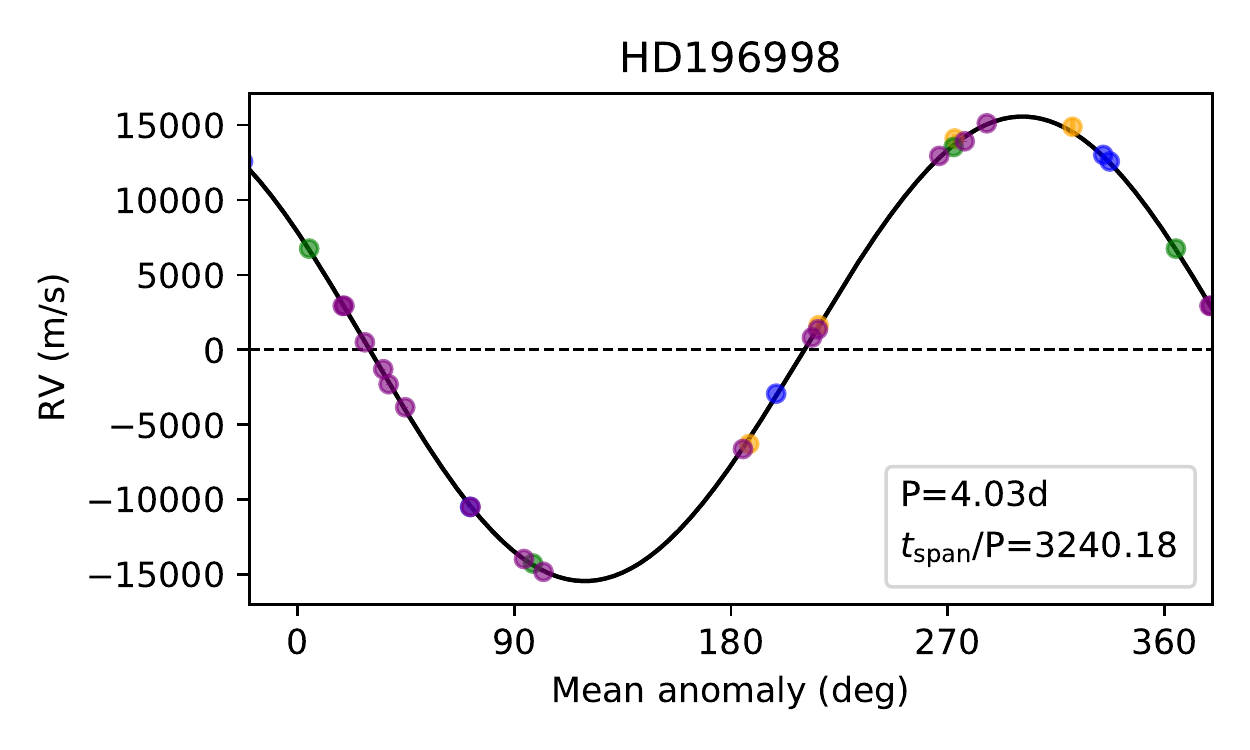}\\

		\includegraphics[width=0.22\linewidth]{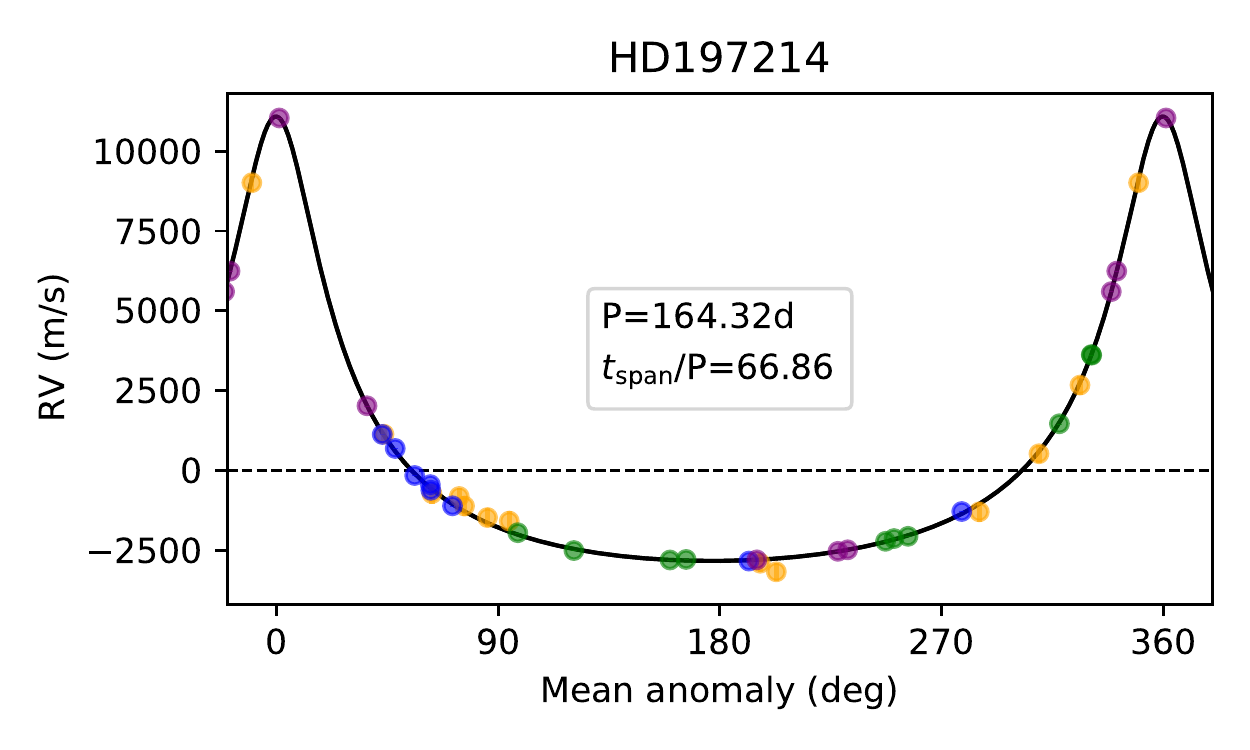}&
		\includegraphics[width=0.22\linewidth]{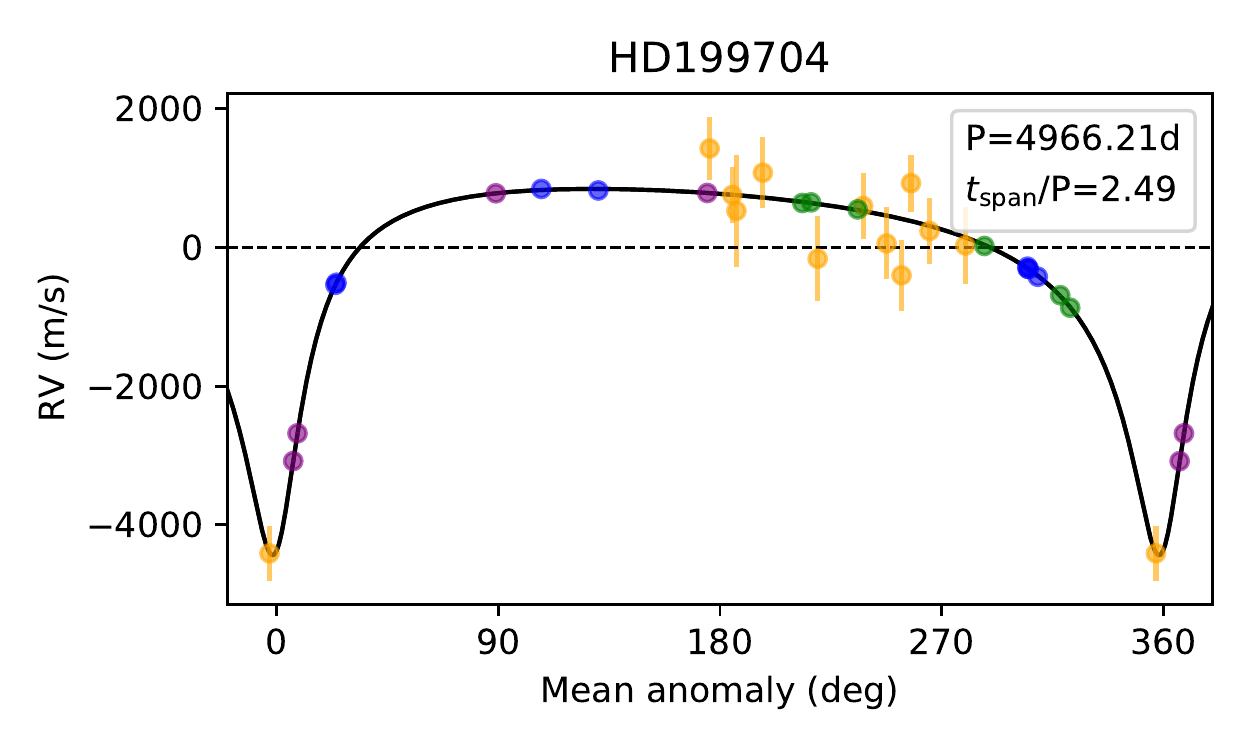}&
		\includegraphics[width=0.22\linewidth]{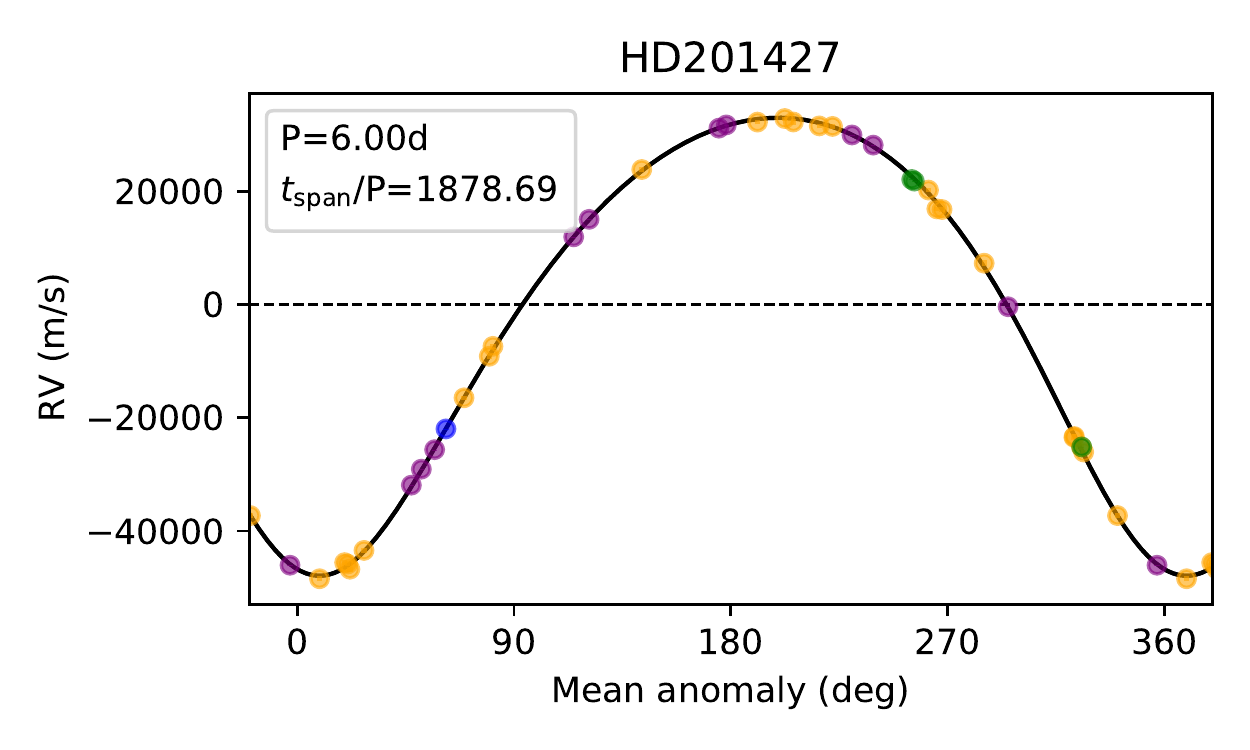}&
		\includegraphics[width=0.22\linewidth]{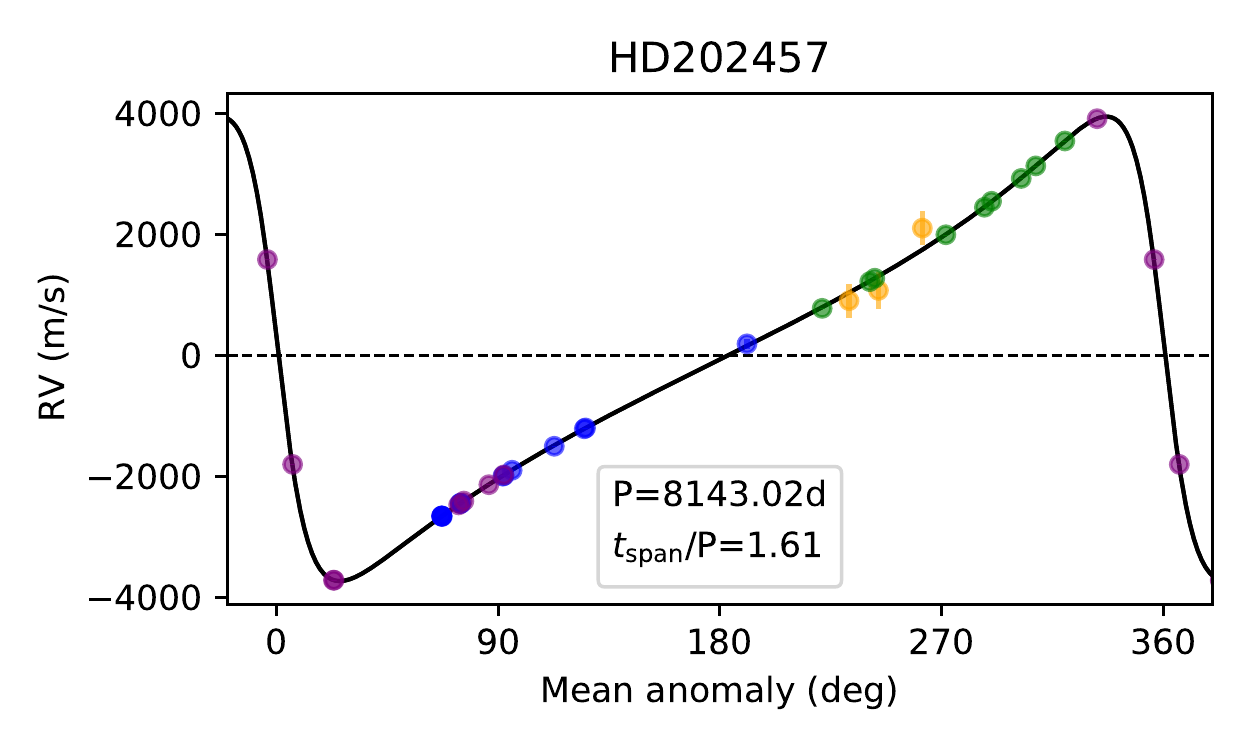}\\

		\includegraphics[width=0.22\linewidth]{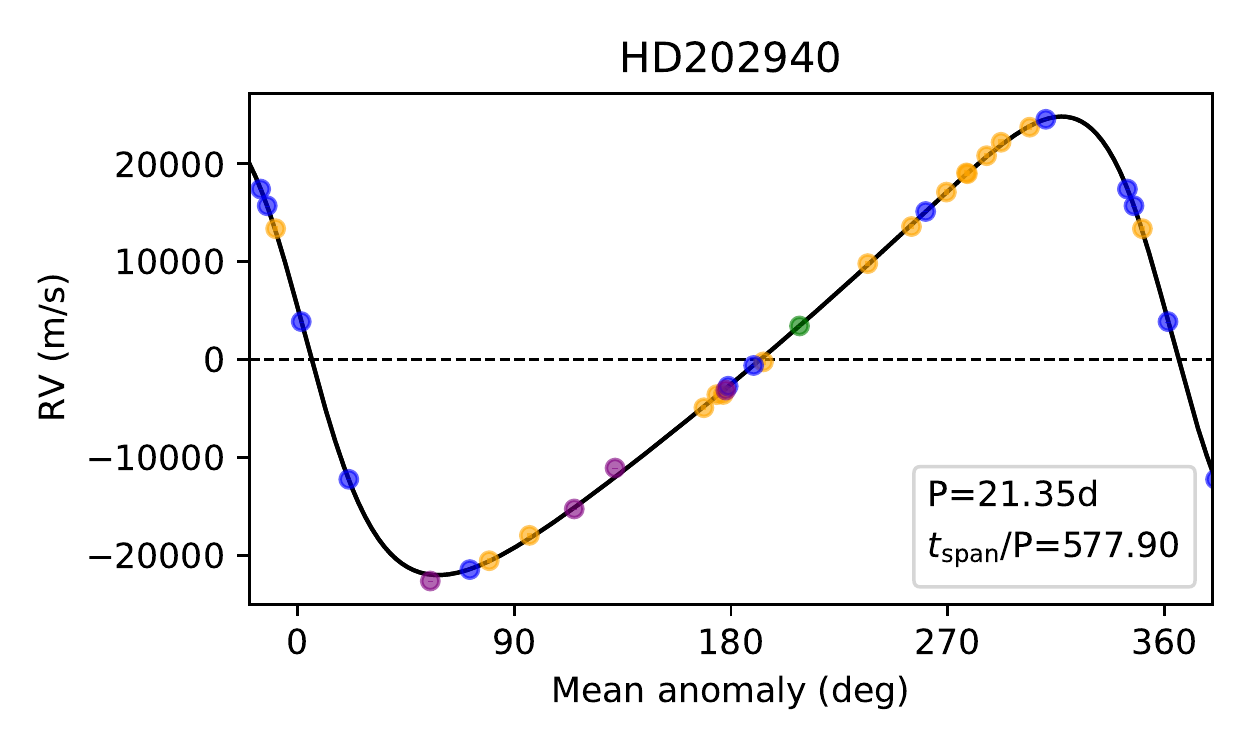}&
		\includegraphics[width=0.22\linewidth]{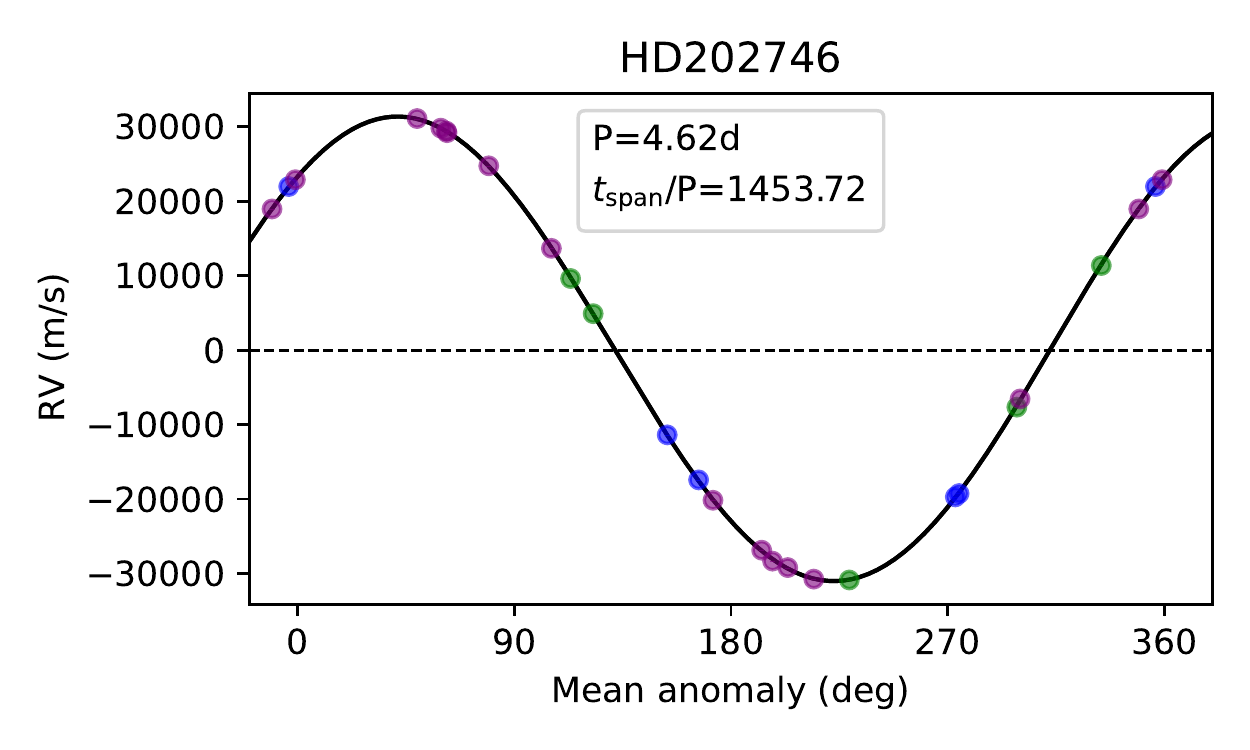}&
		\includegraphics[width=0.22\linewidth]{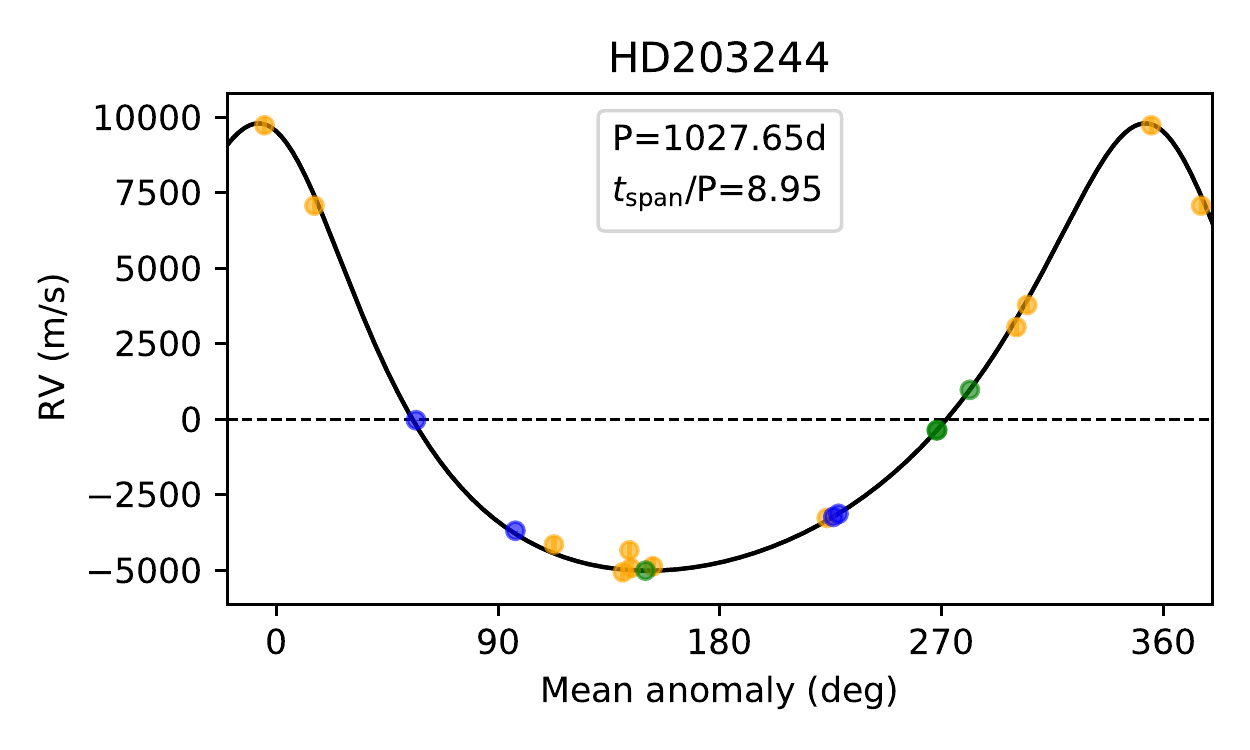}&
		\includegraphics[width=0.22\linewidth]{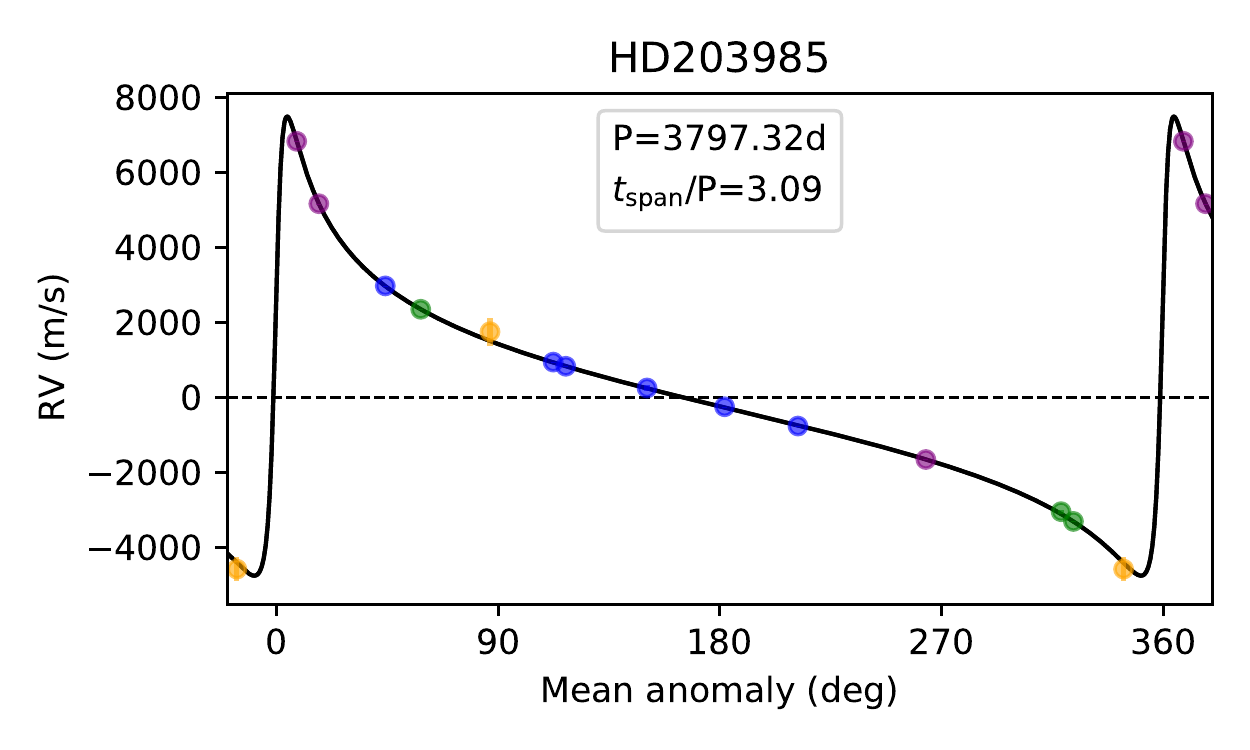}\\

		\includegraphics[width=0.22\linewidth]{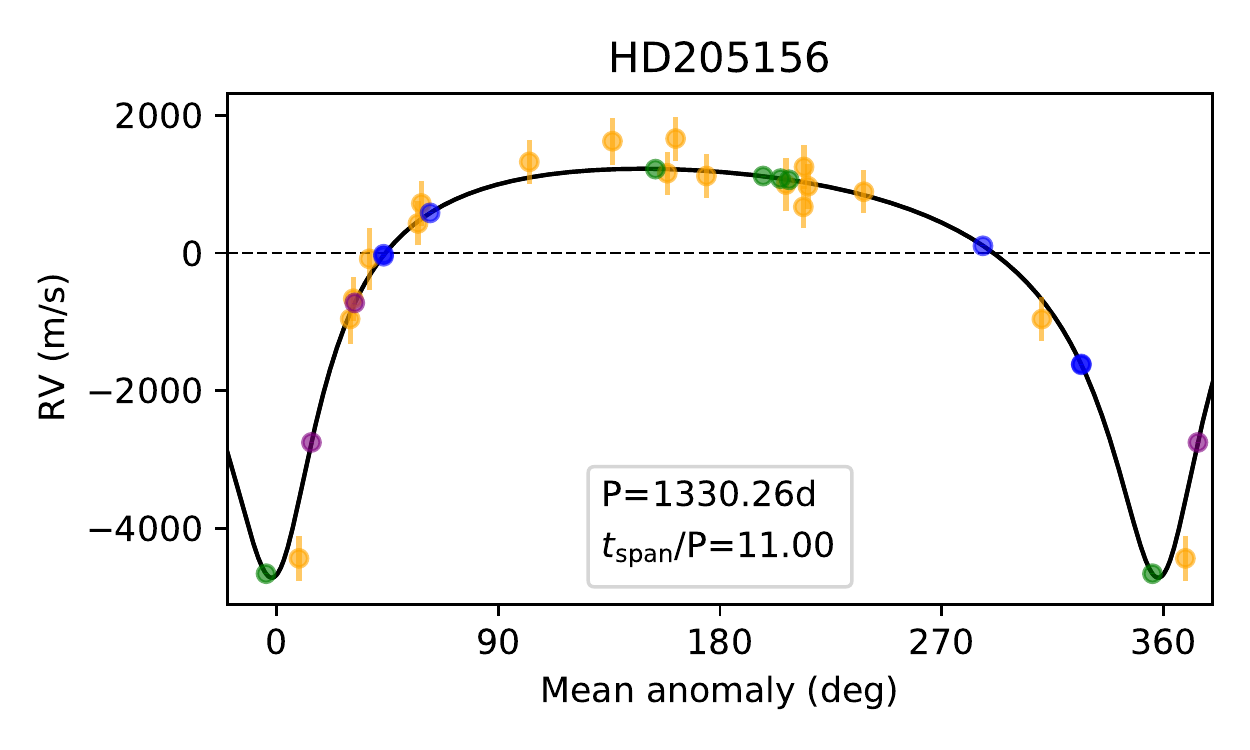}&
		\includegraphics[width=0.22\linewidth]{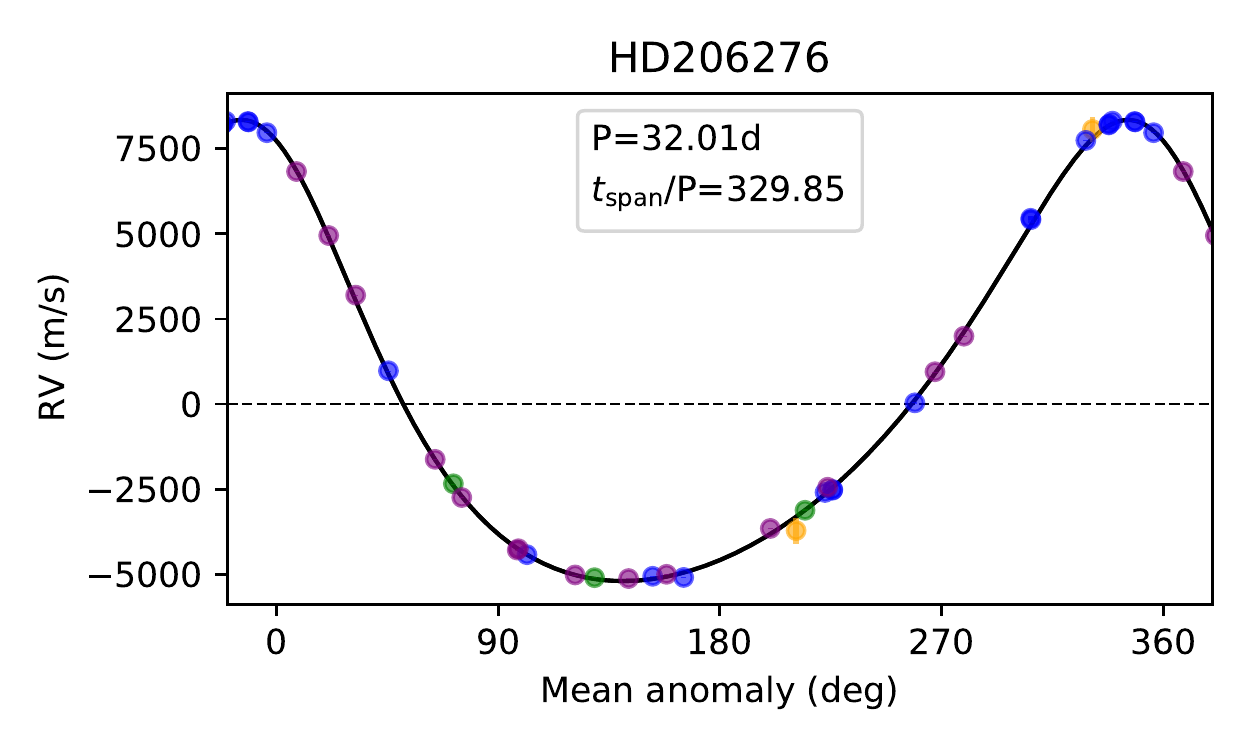}&
		\includegraphics[width=0.22\linewidth]{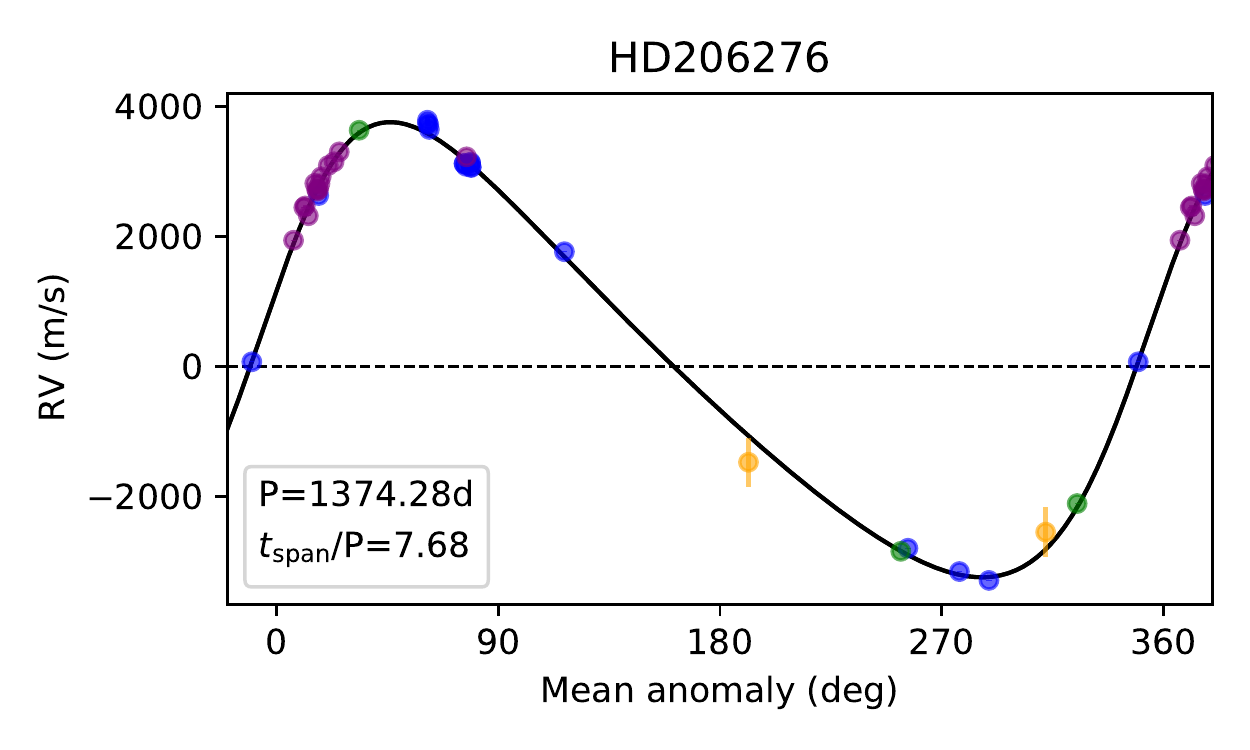}&
		\includegraphics[width=0.22\linewidth]{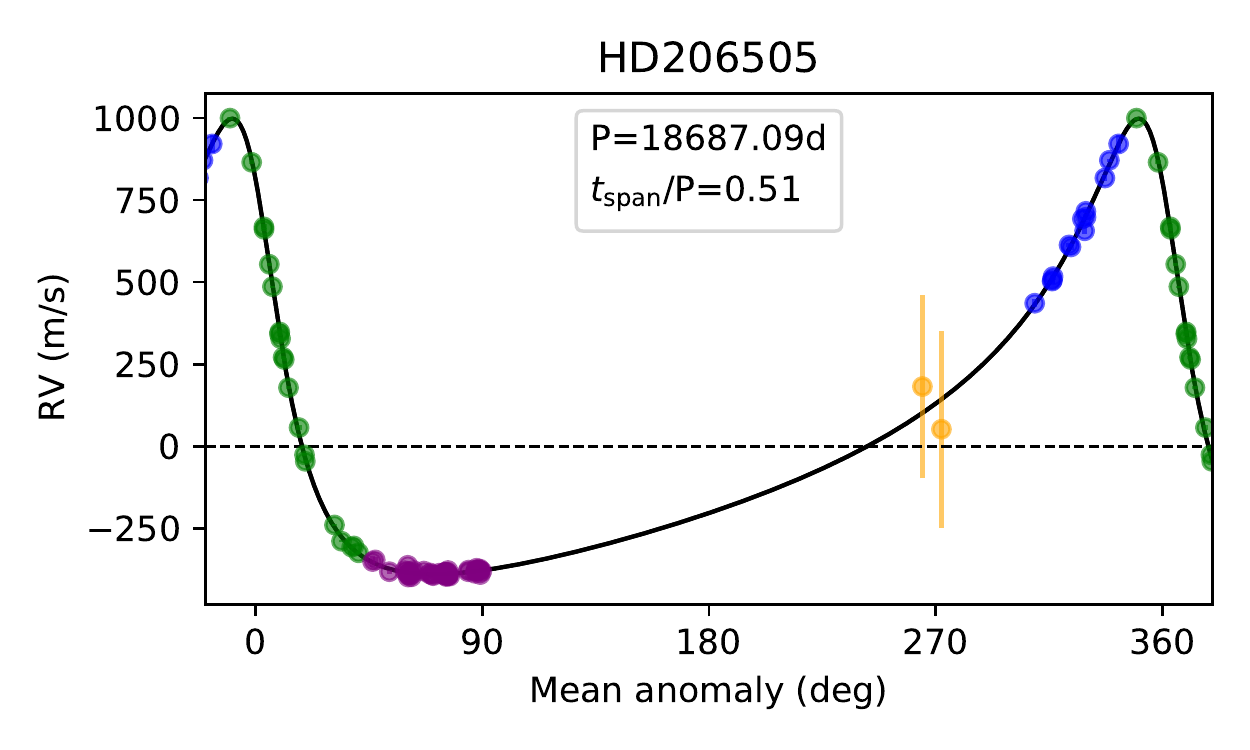}\\

		\includegraphics[width=0.22\linewidth]{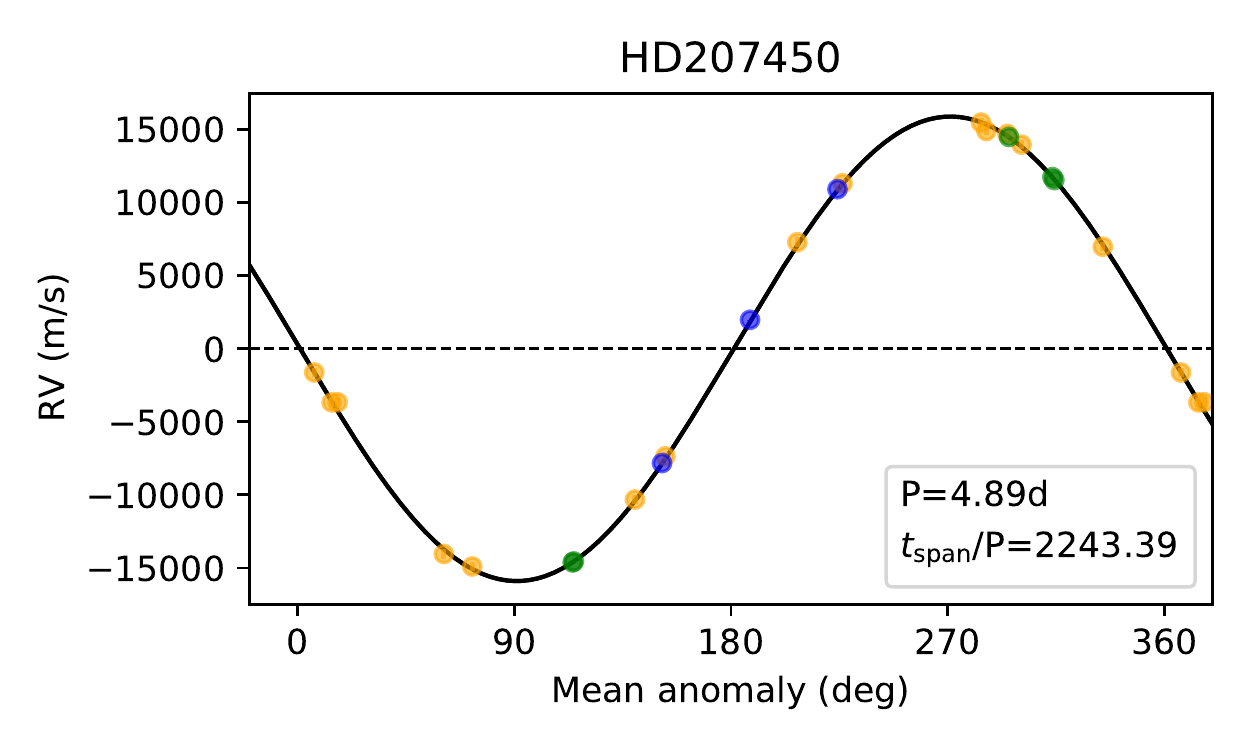}&
		\includegraphics[width=0.22\linewidth]{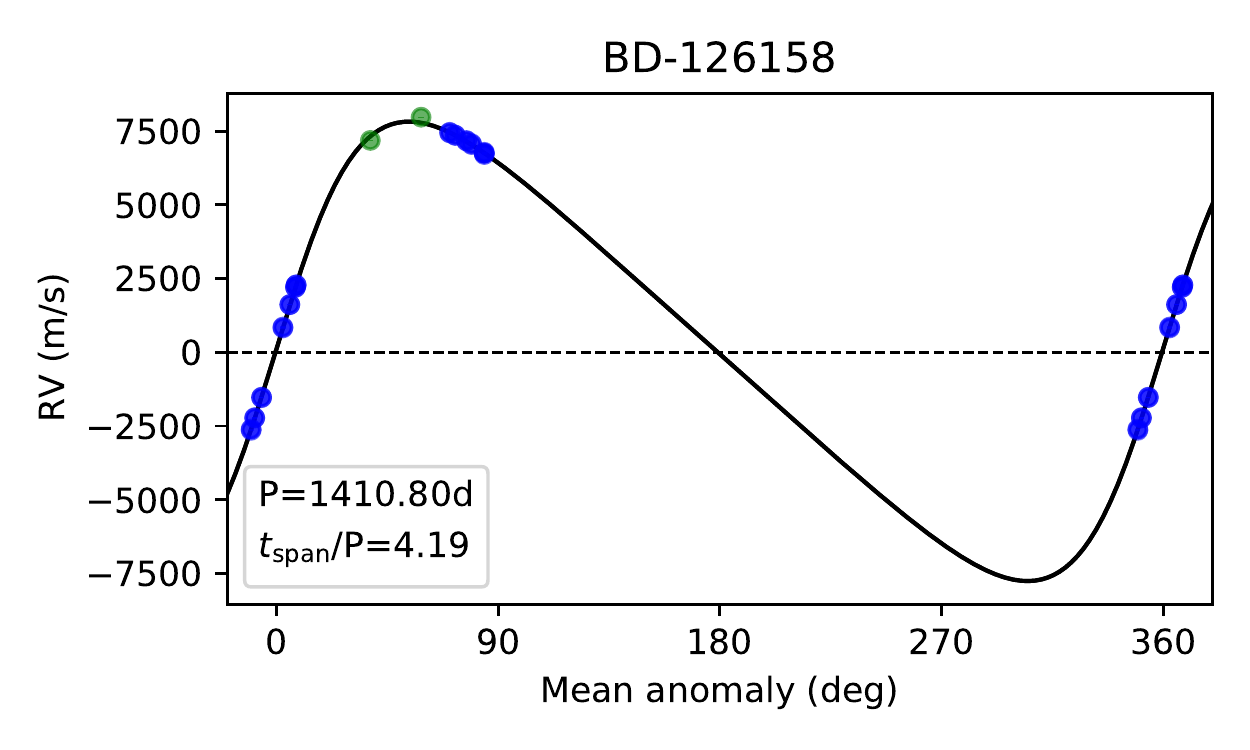}&
		\includegraphics[width=0.22\linewidth]{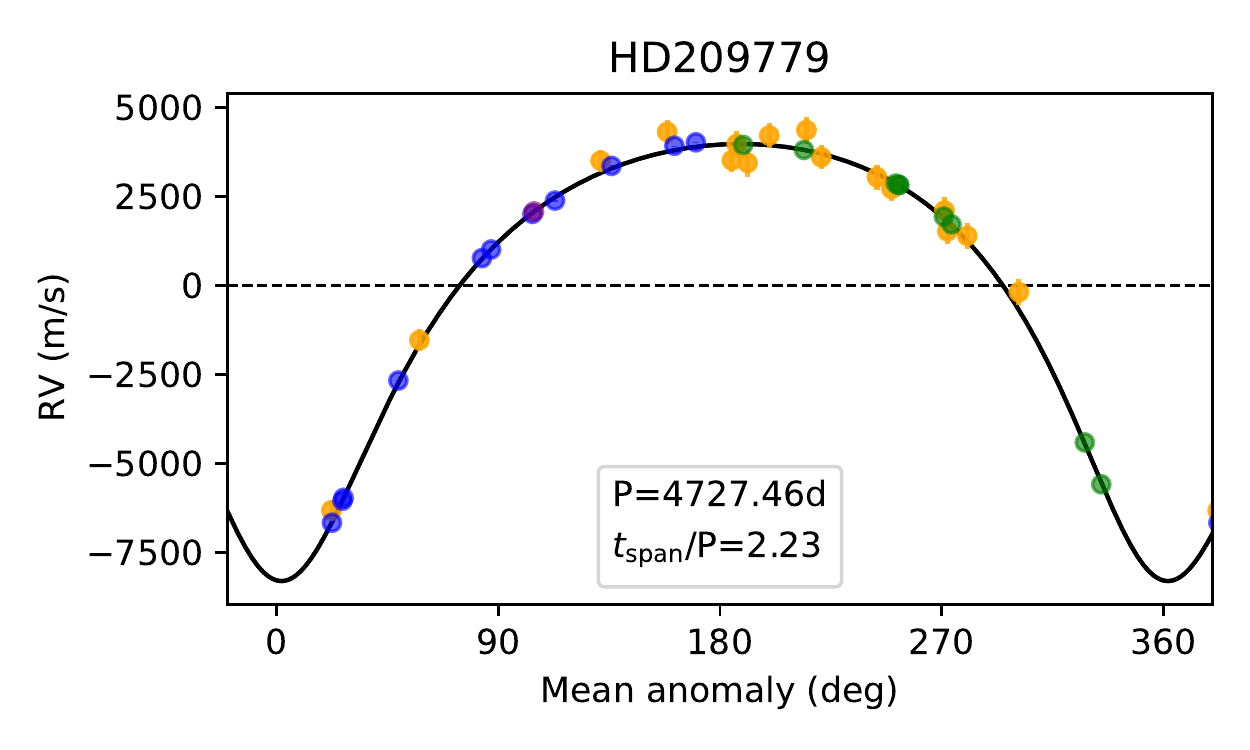}&
		\includegraphics[width=0.22\linewidth]{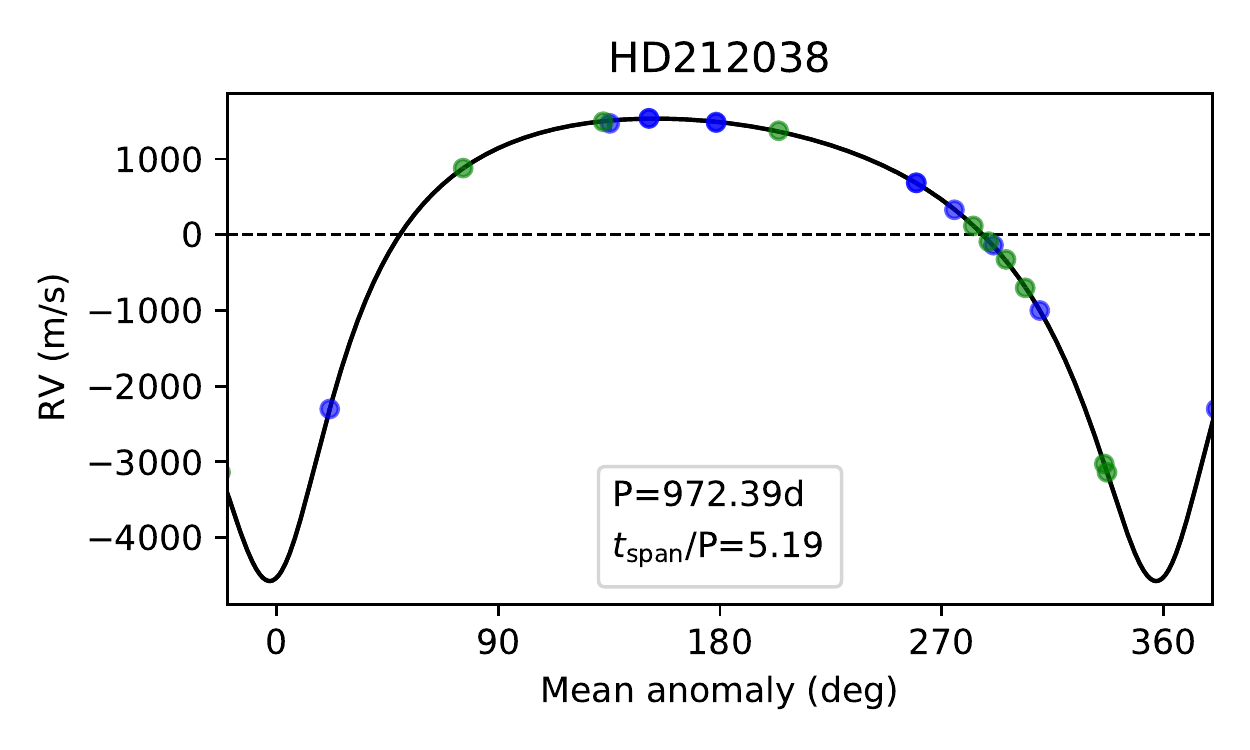}\\

		\includegraphics[width=0.22\linewidth]{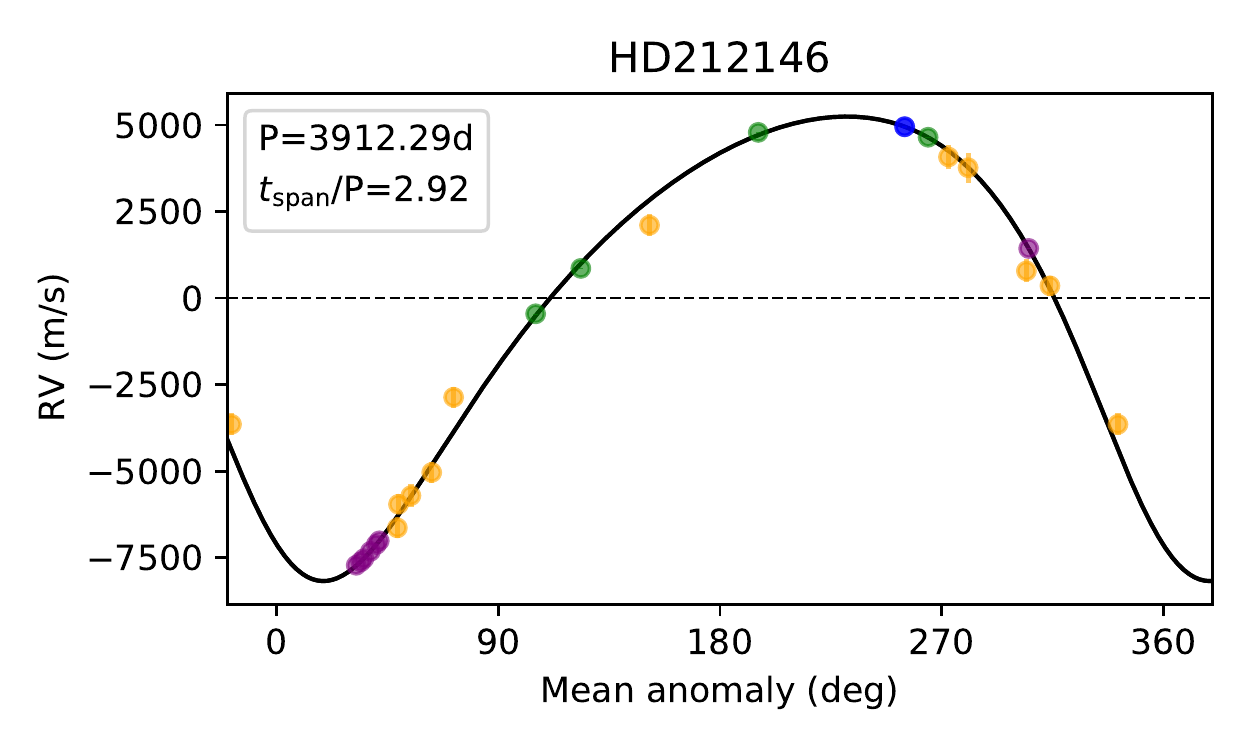}&
		\includegraphics[width=0.22\linewidth]{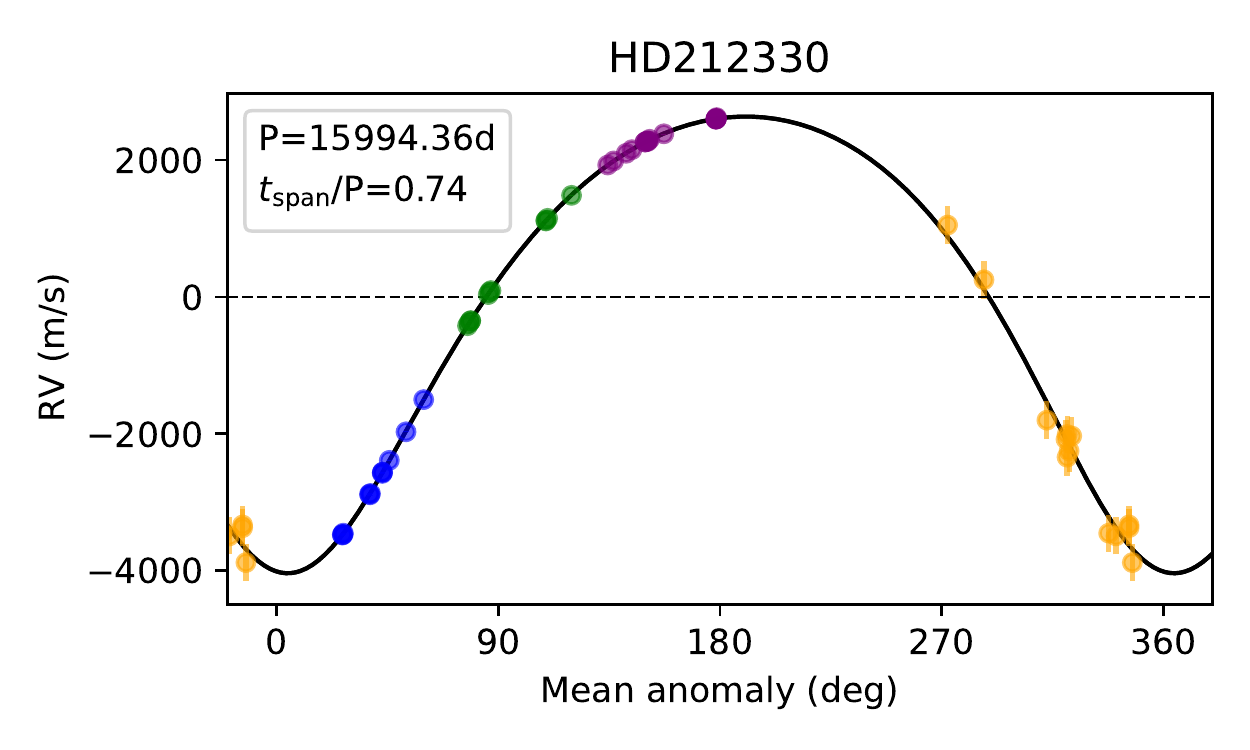}&
		\includegraphics[width=0.22\linewidth]{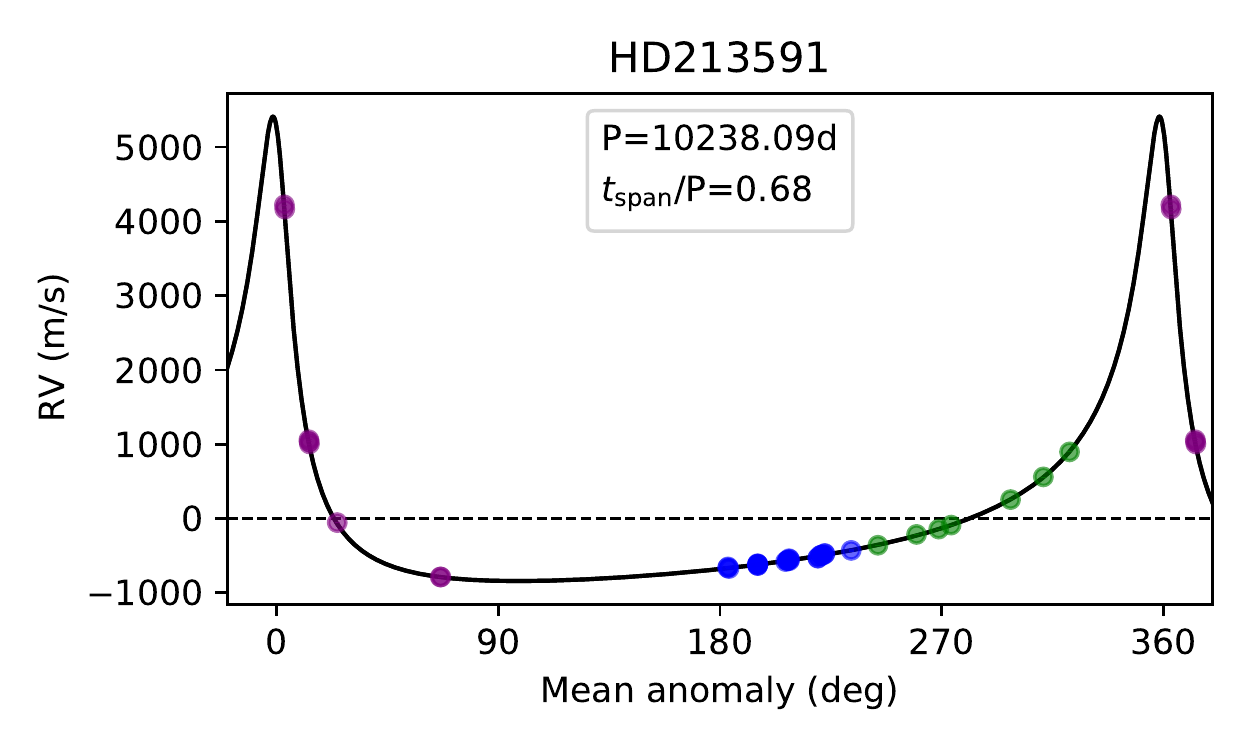}&
		\includegraphics[width=0.22\linewidth]{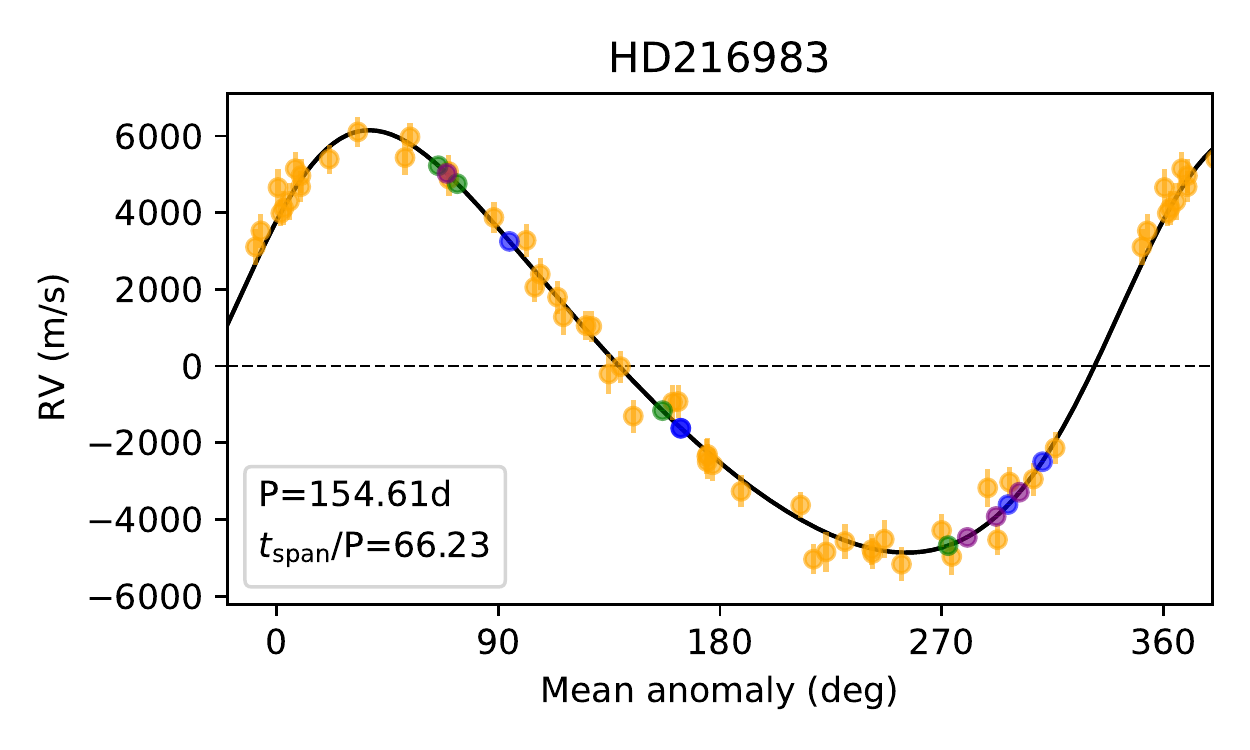}\\

		\includegraphics[width=0.22\linewidth]{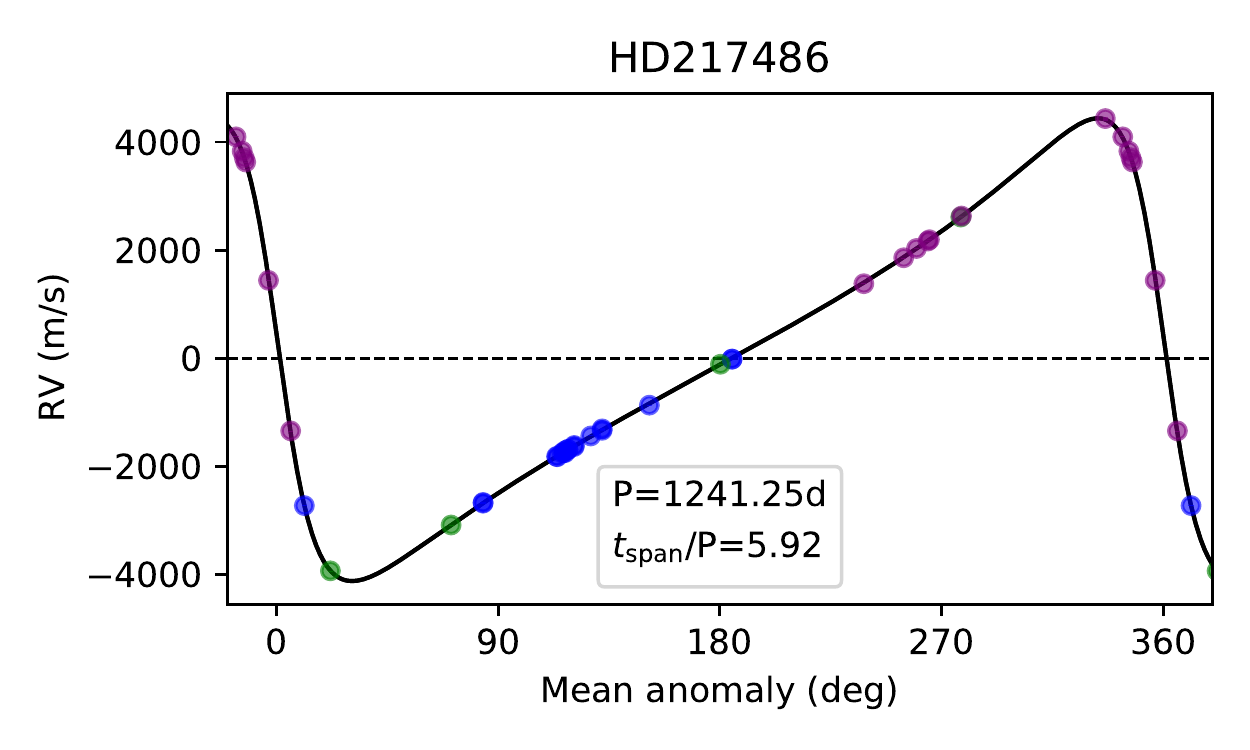}&
		\includegraphics[width=0.22\linewidth]{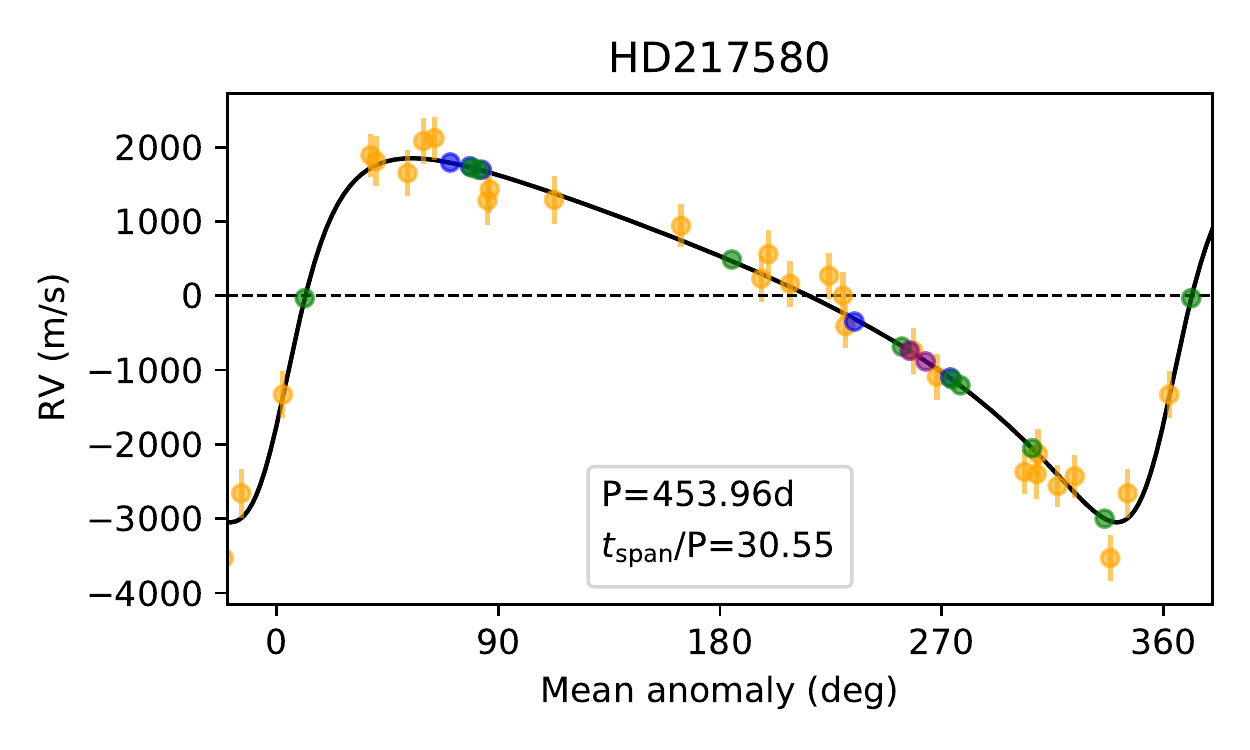}&
		\includegraphics[width=0.22\linewidth]{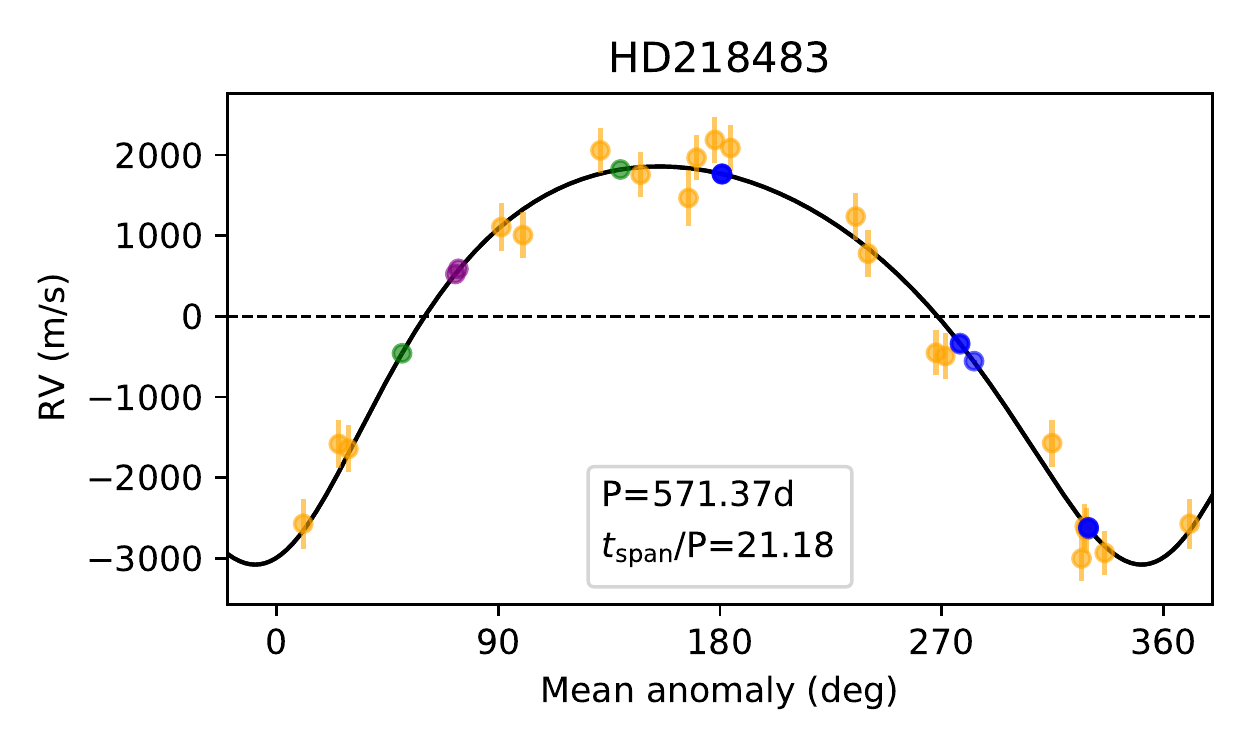}&
		\includegraphics[width=0.22\linewidth]{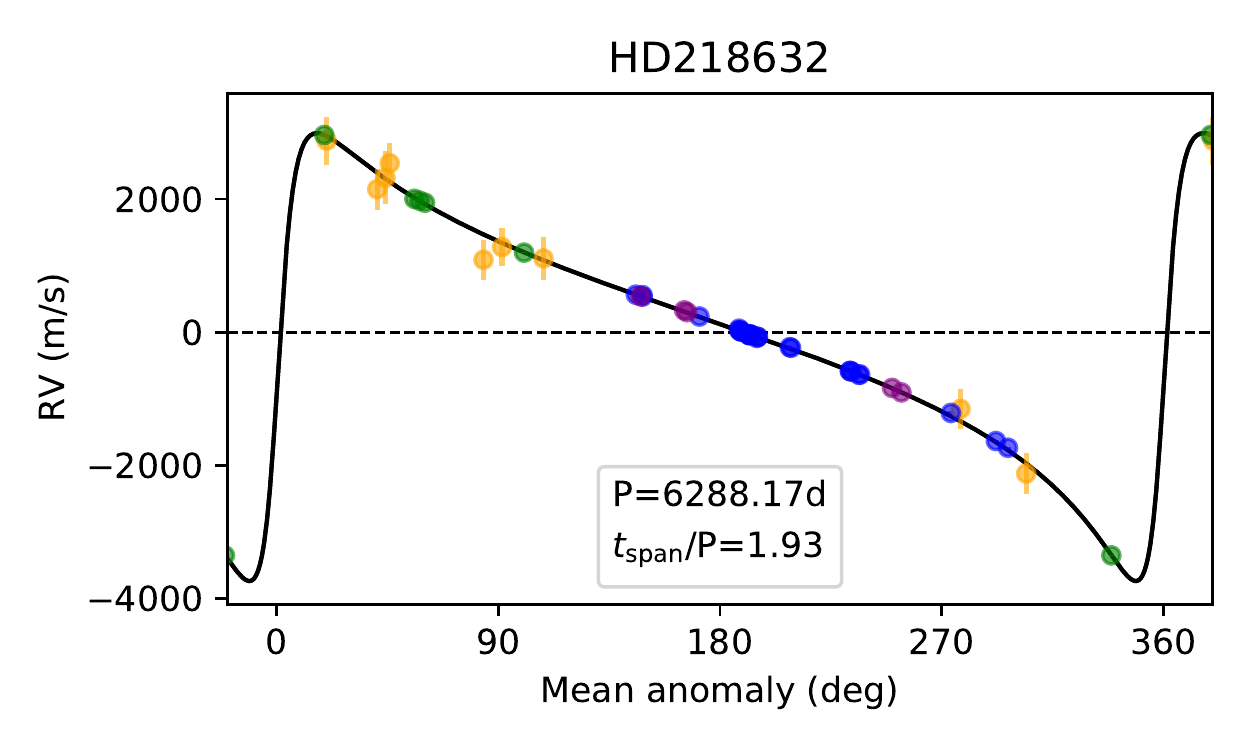}\\

		\includegraphics[width=0.22\linewidth]{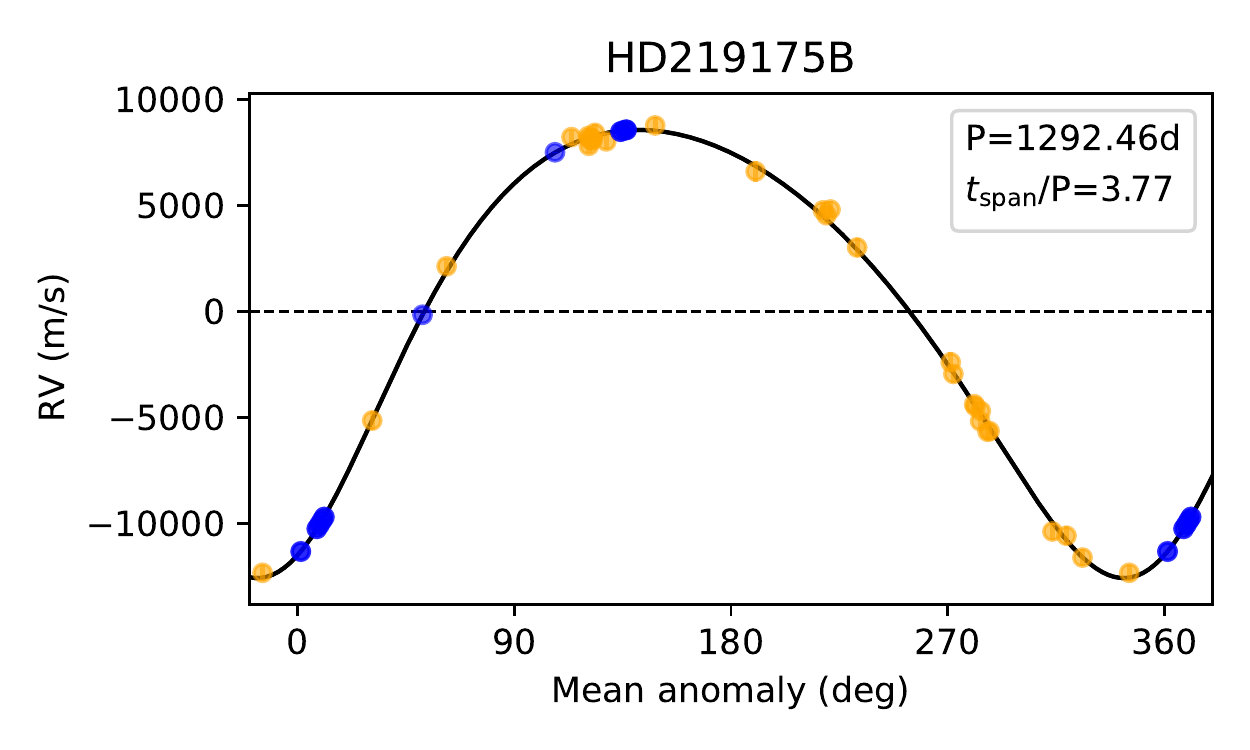}&
		\includegraphics[width=0.22\linewidth]{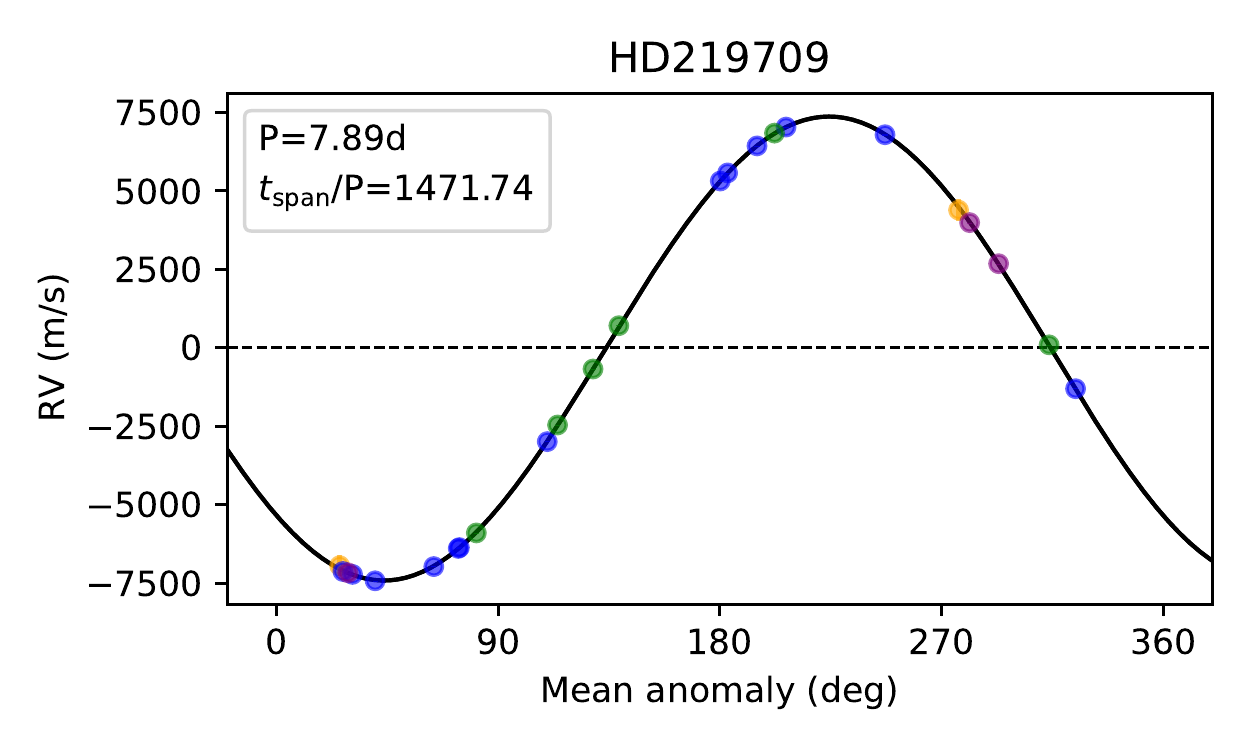}&
		\includegraphics[width=0.22\linewidth]{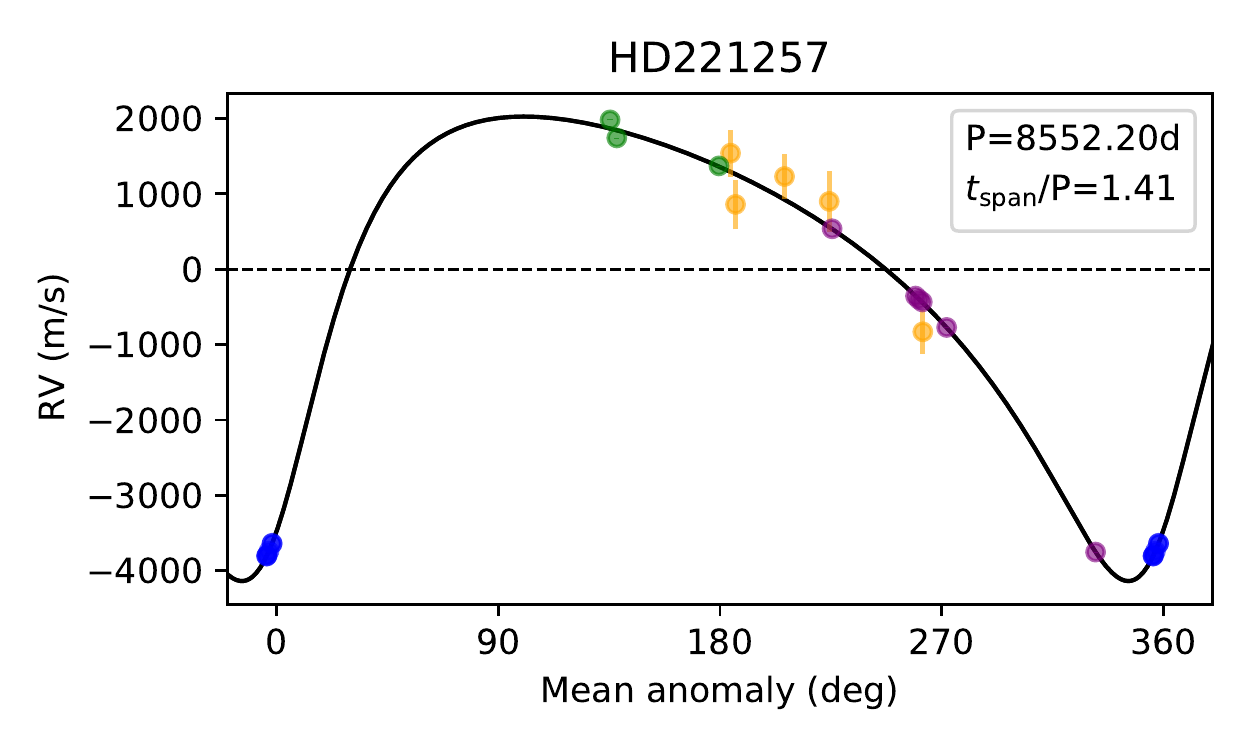}&
		\includegraphics[width=0.22\linewidth]{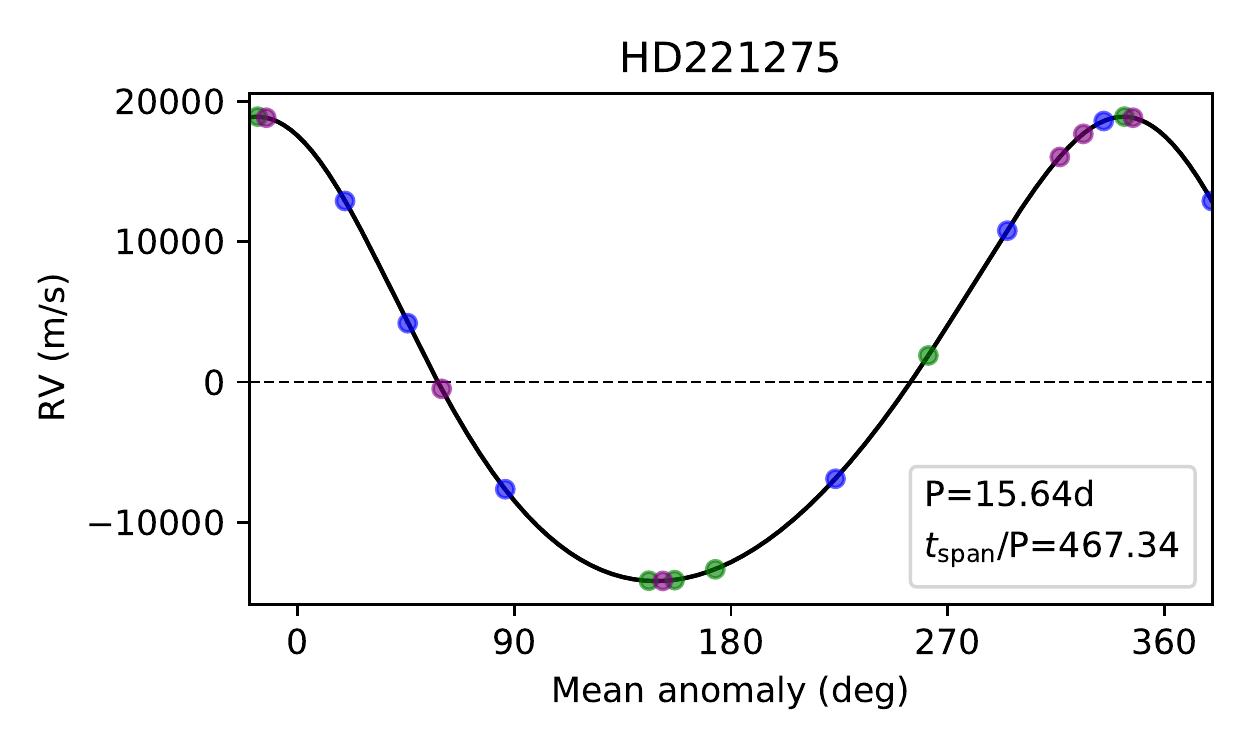}\\

		\includegraphics[width=0.22\linewidth]{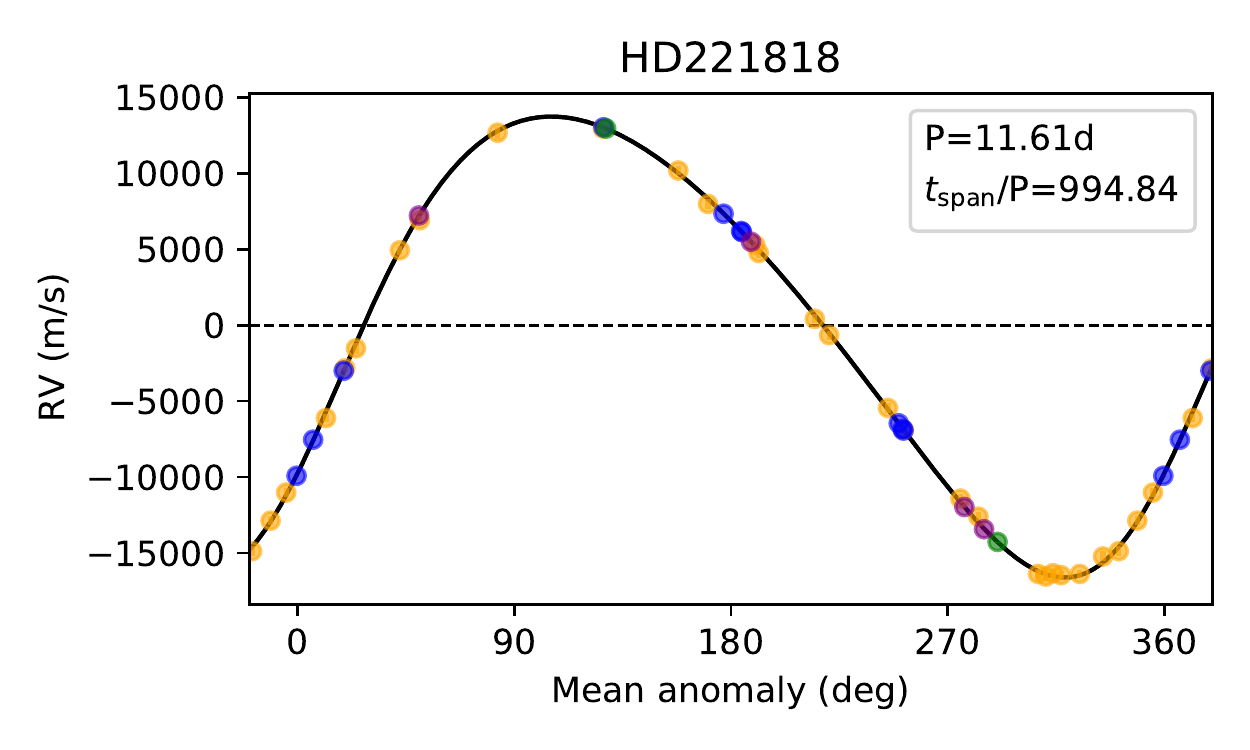}&
		\includegraphics[width=0.22\linewidth]{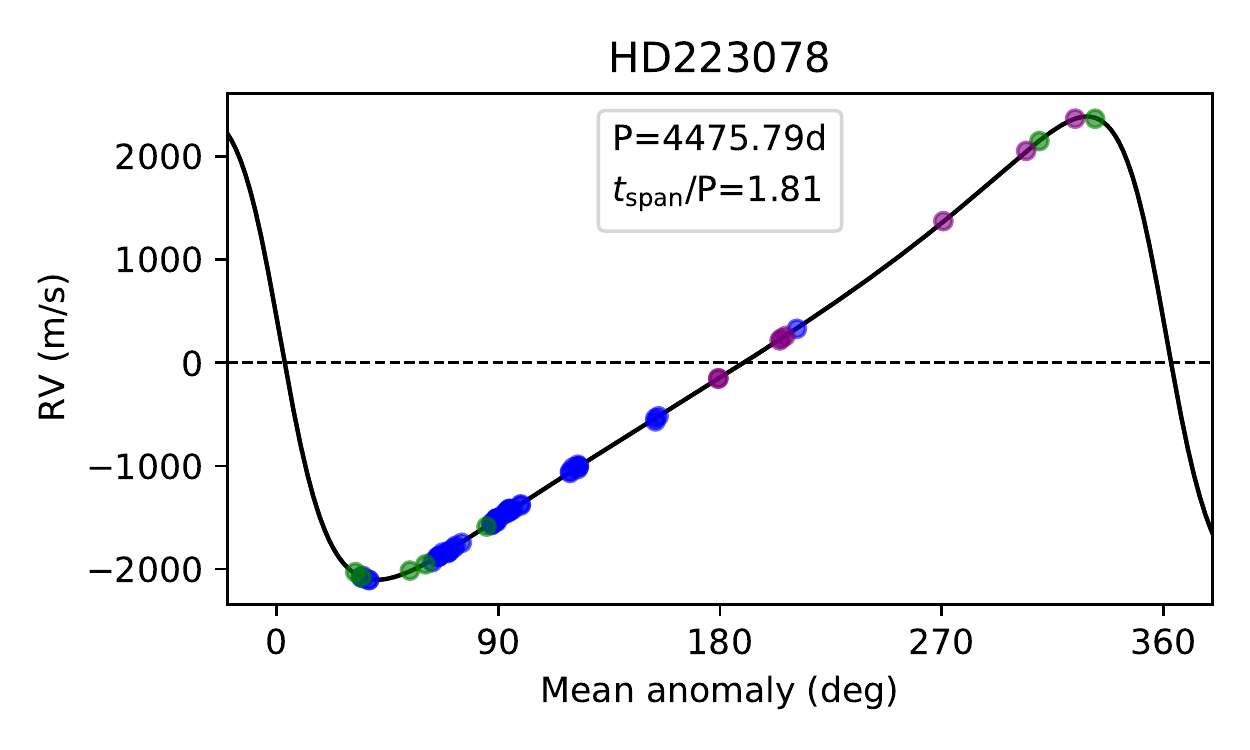}&
		\includegraphics[width=0.22\linewidth]{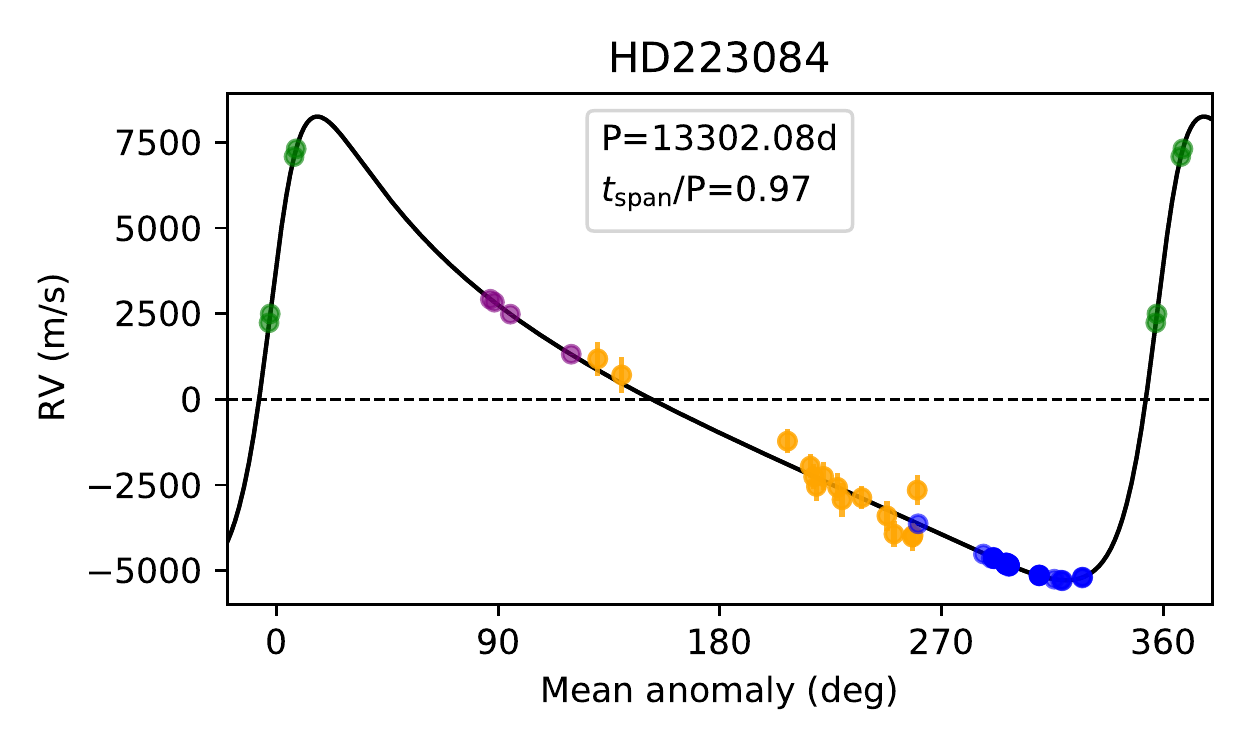}&
		\includegraphics[width=0.22\linewidth]{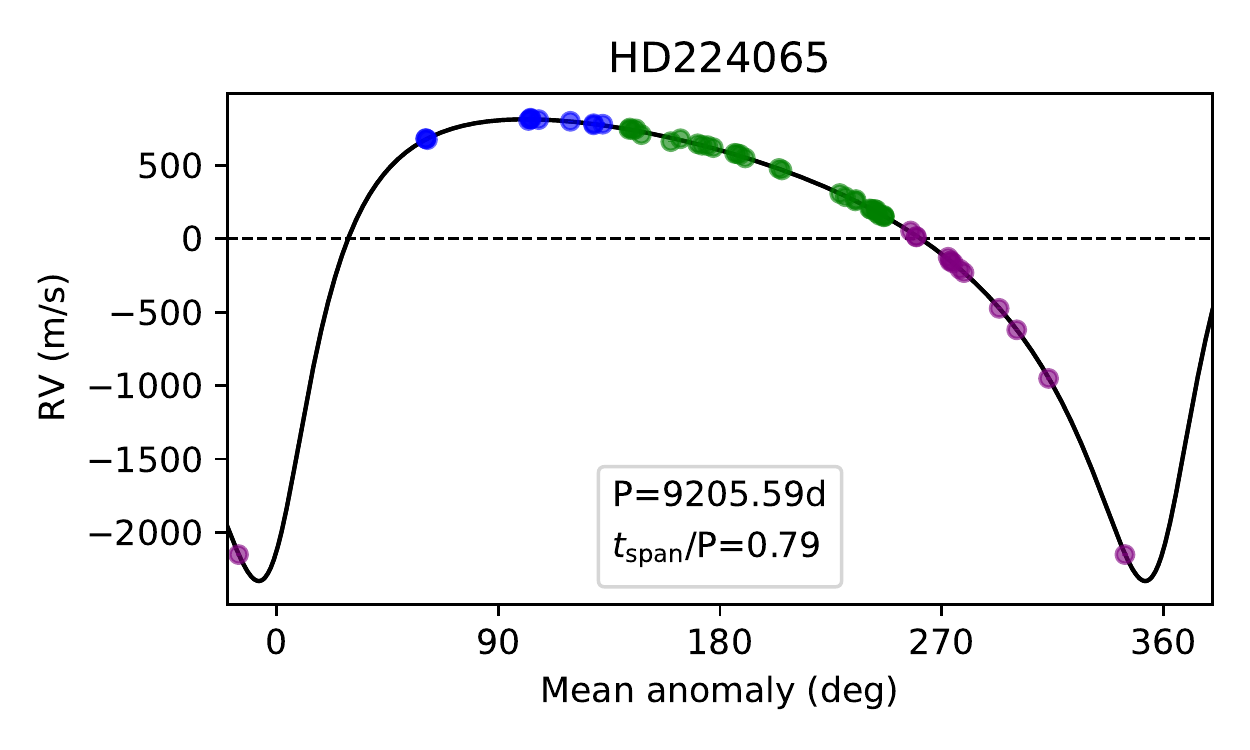}\\

		\includegraphics[width=0.22\linewidth]{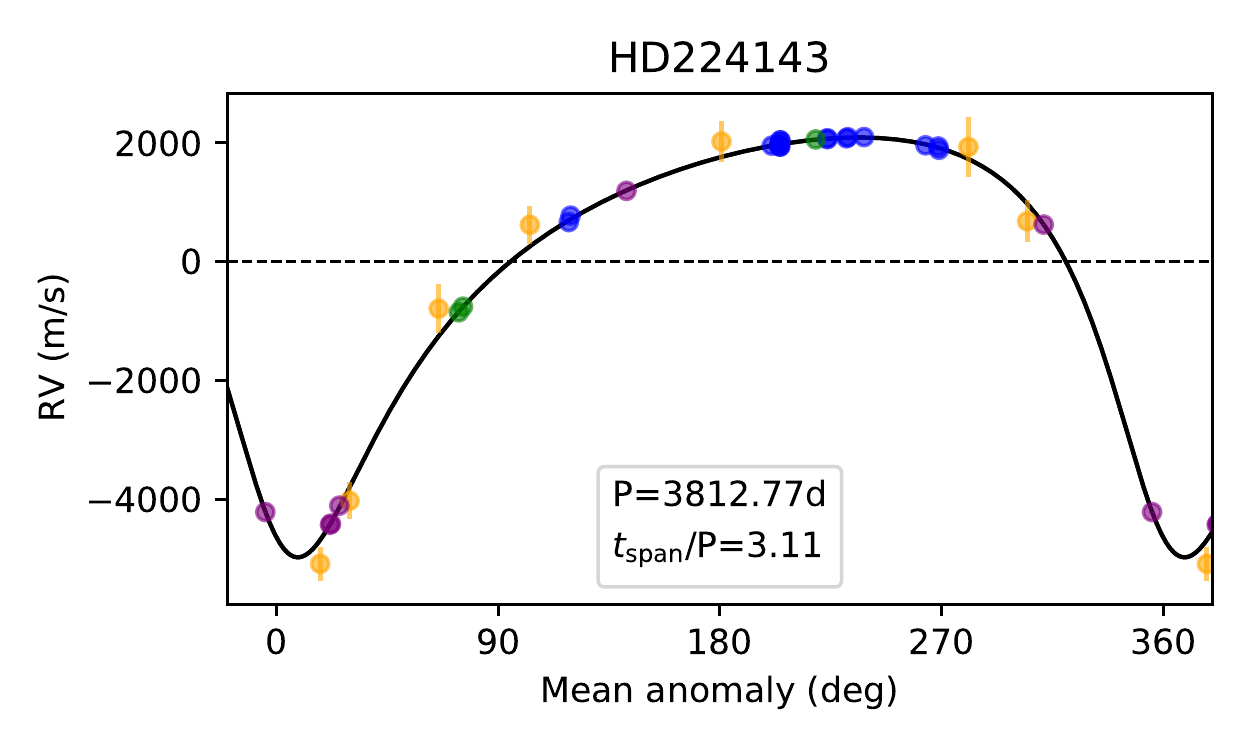}&
    	\end{longtable}
    	\captionof{figure}{Phase folded radial velocity curves for the companions detected around each star in the binary sample as described in Sect.\ref{sec:rv}. In each plot, the CORAVEL, CORALIE98, CORALIE07 and CORALIE14 measurements are shown in orange, blue, green and purple respectively over the phase folded model radial velocity curve. Companion orbital period and orbit completeness, indicated as the ratio between observational span and period, are noted in each plot's box.}}\label{fig:app-phasefolded}
    \clearpage
    \twocolumn
    
    \clearpage
    \onecolumn
    \section{Detection limits maps} \label{app:detection-limits}
    	{
    	\centering
    	\begin{longtable}{c c c c}
    		\includegraphics[width=0.22\linewidth]{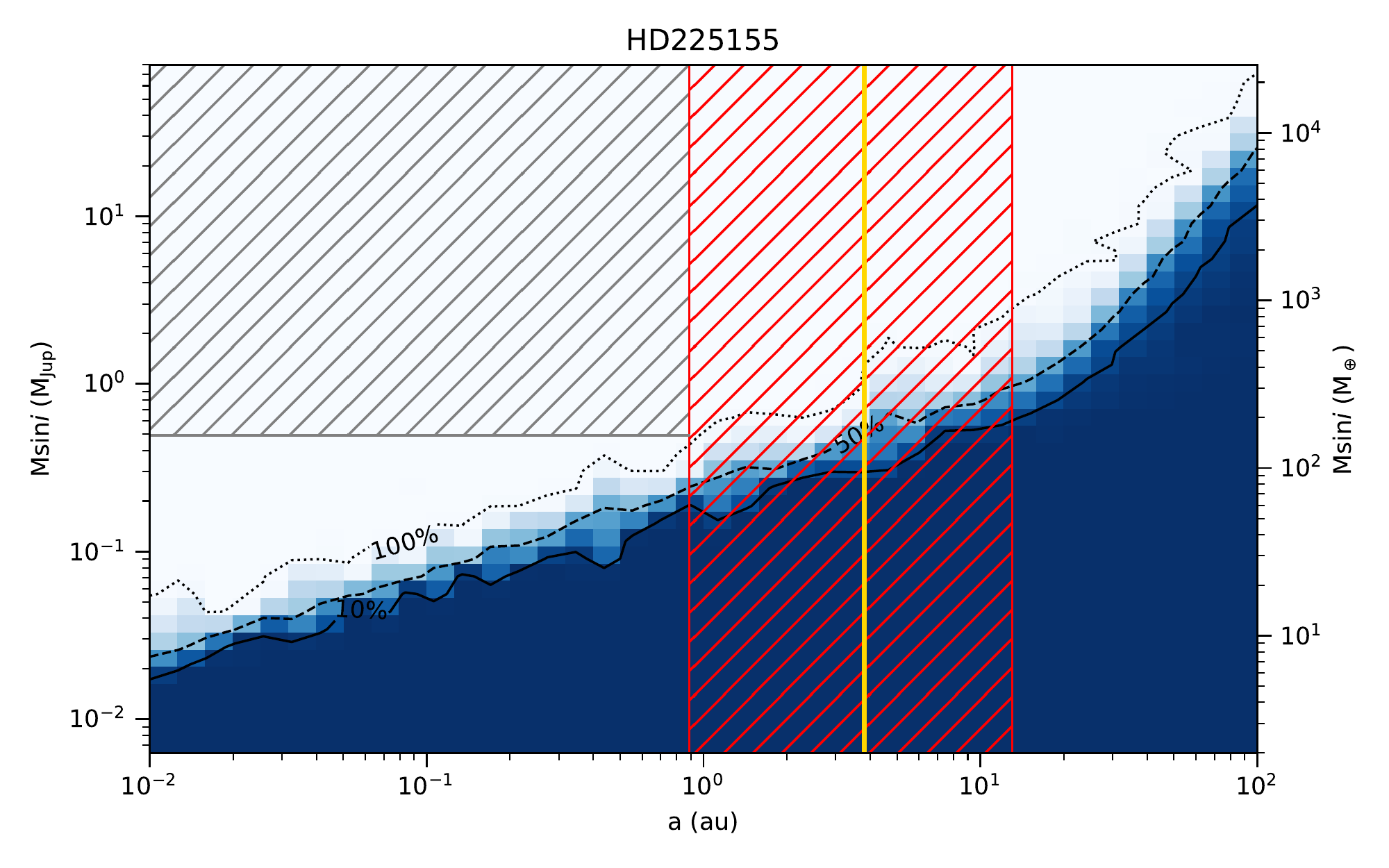}&
    		\includegraphics[width=0.22\linewidth]{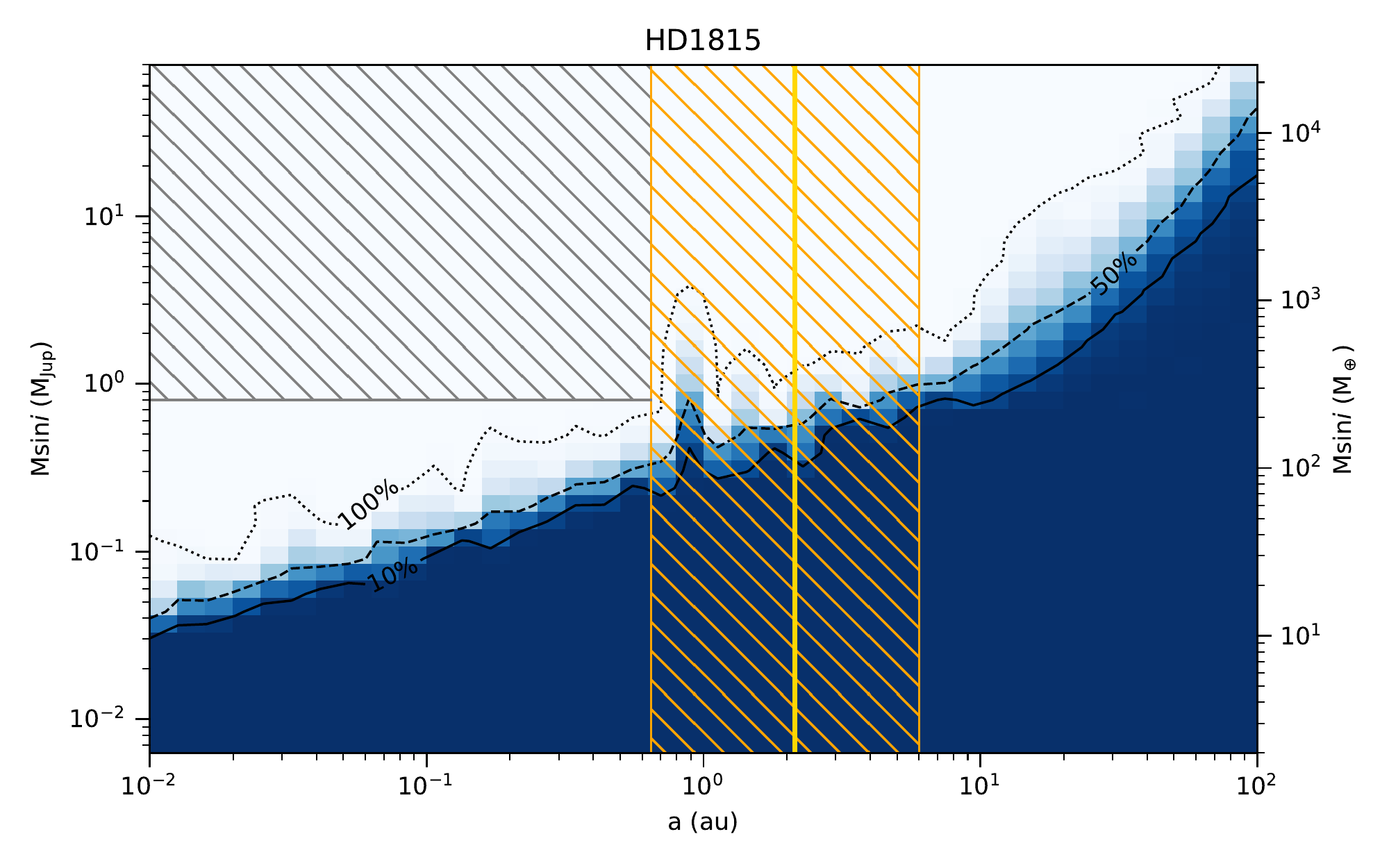}&
    		\includegraphics[width=0.22\linewidth]{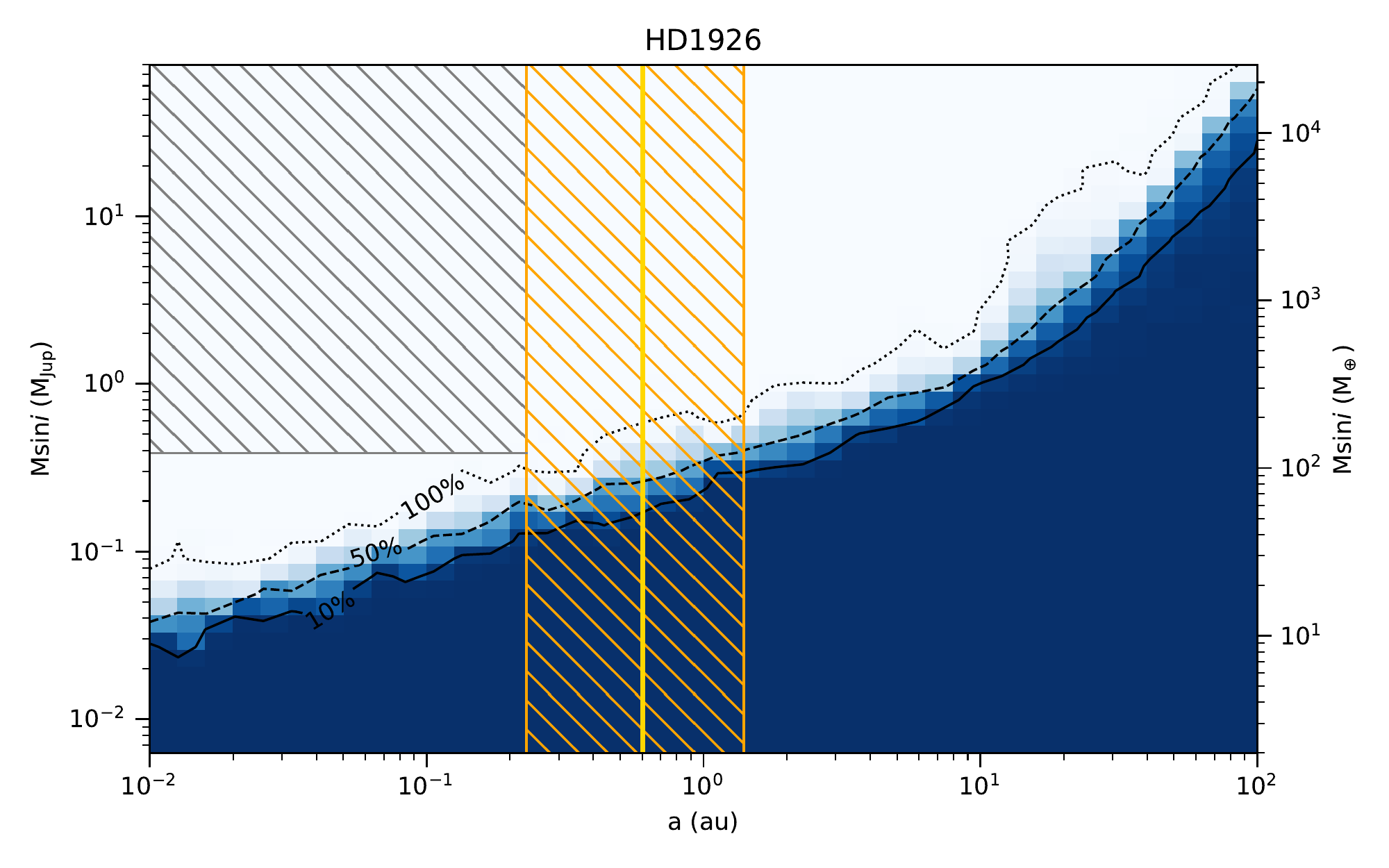}&
    		\includegraphics[width=0.22\linewidth]{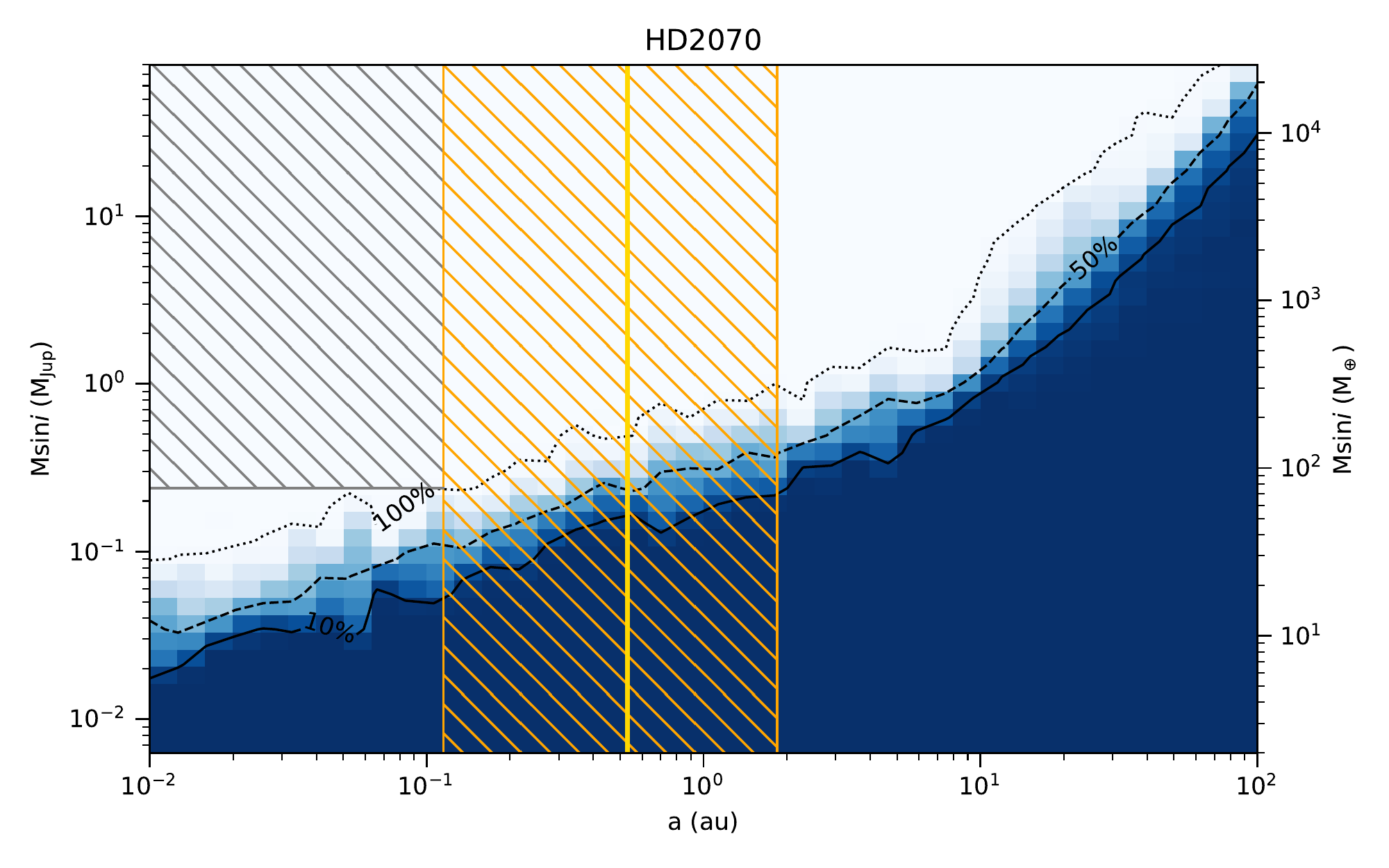}\\
    
    		\includegraphics[width=0.22\linewidth]{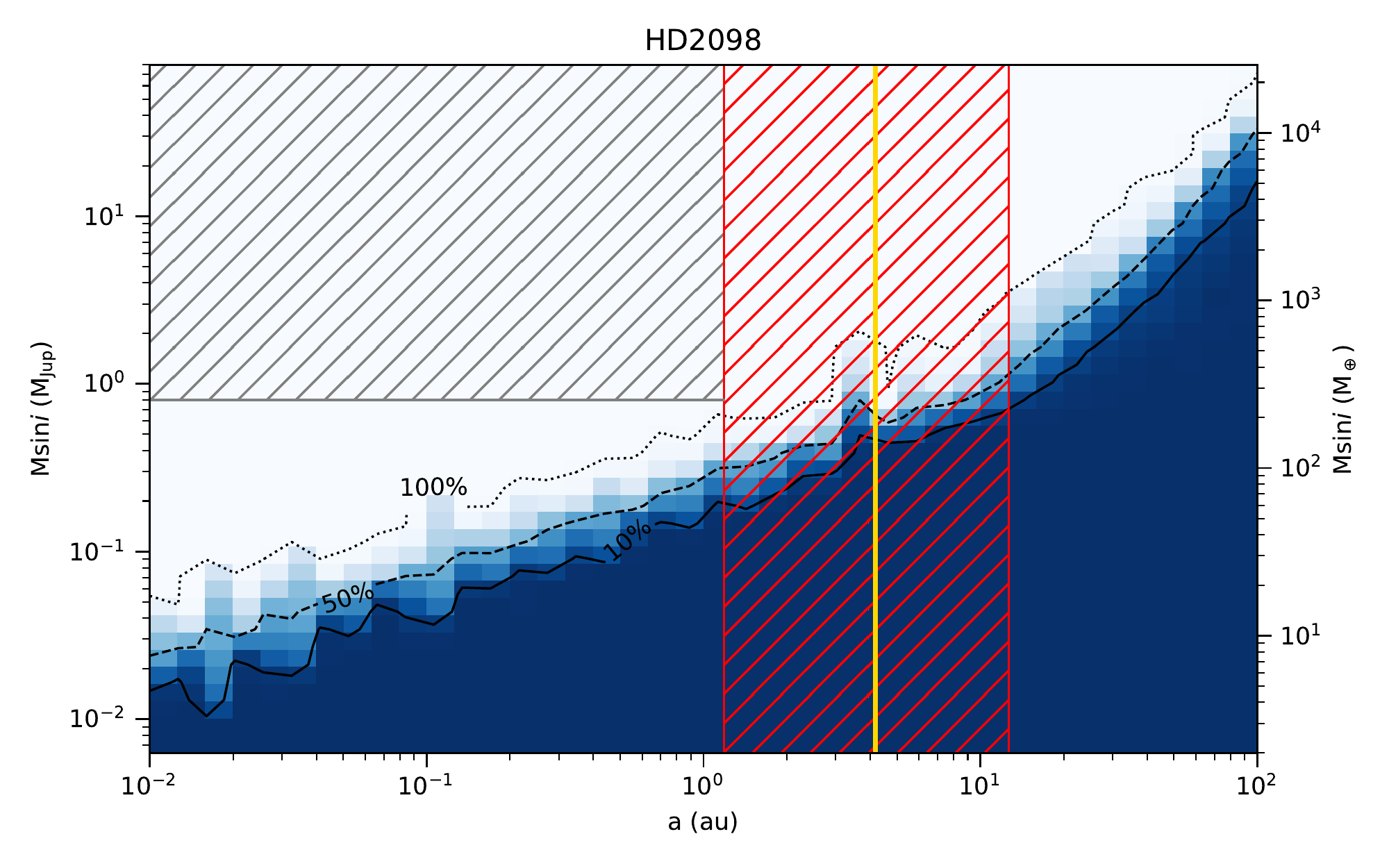}&
    		\includegraphics[width=0.22\linewidth]{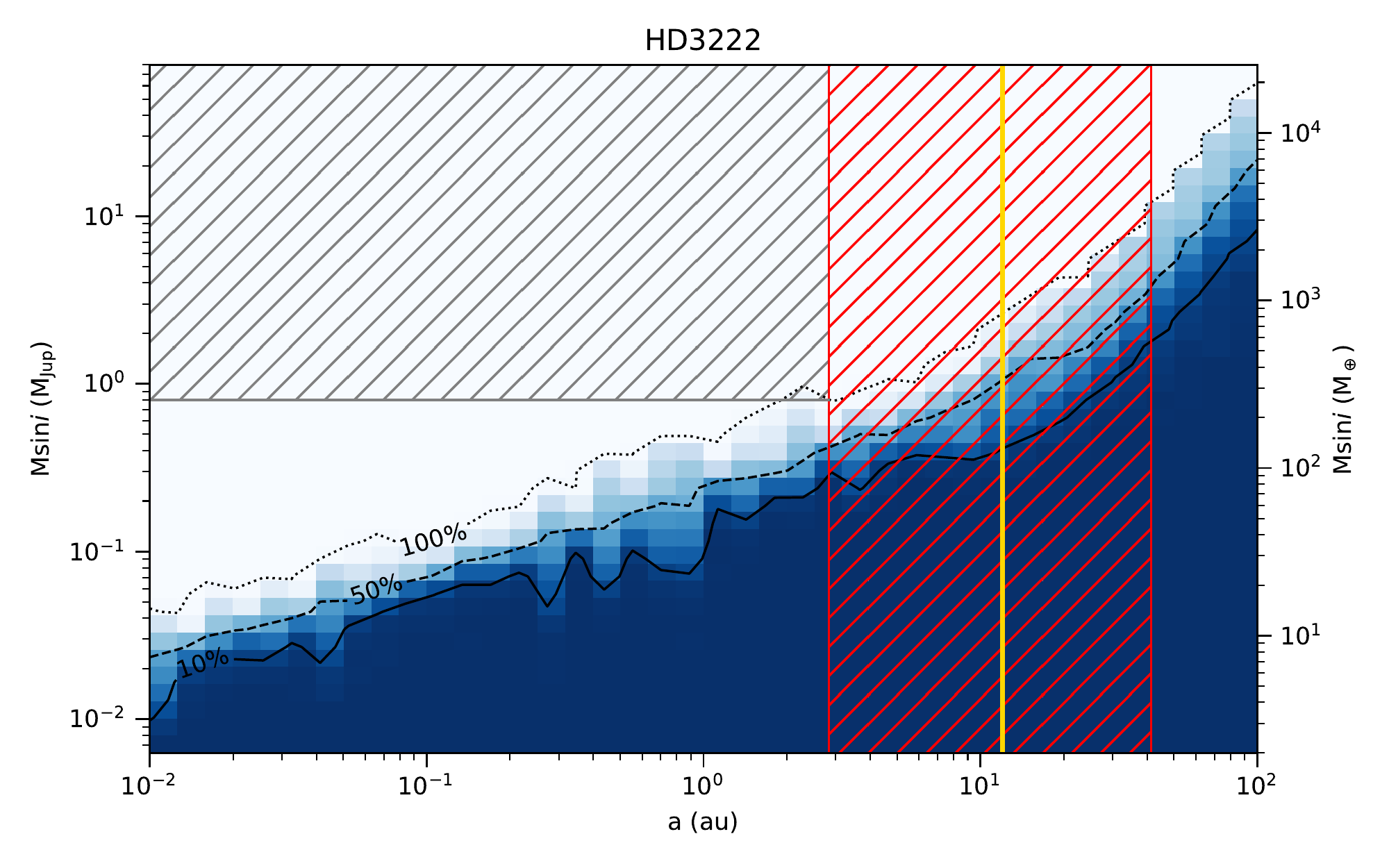}&
    		\includegraphics[width=0.22\linewidth]{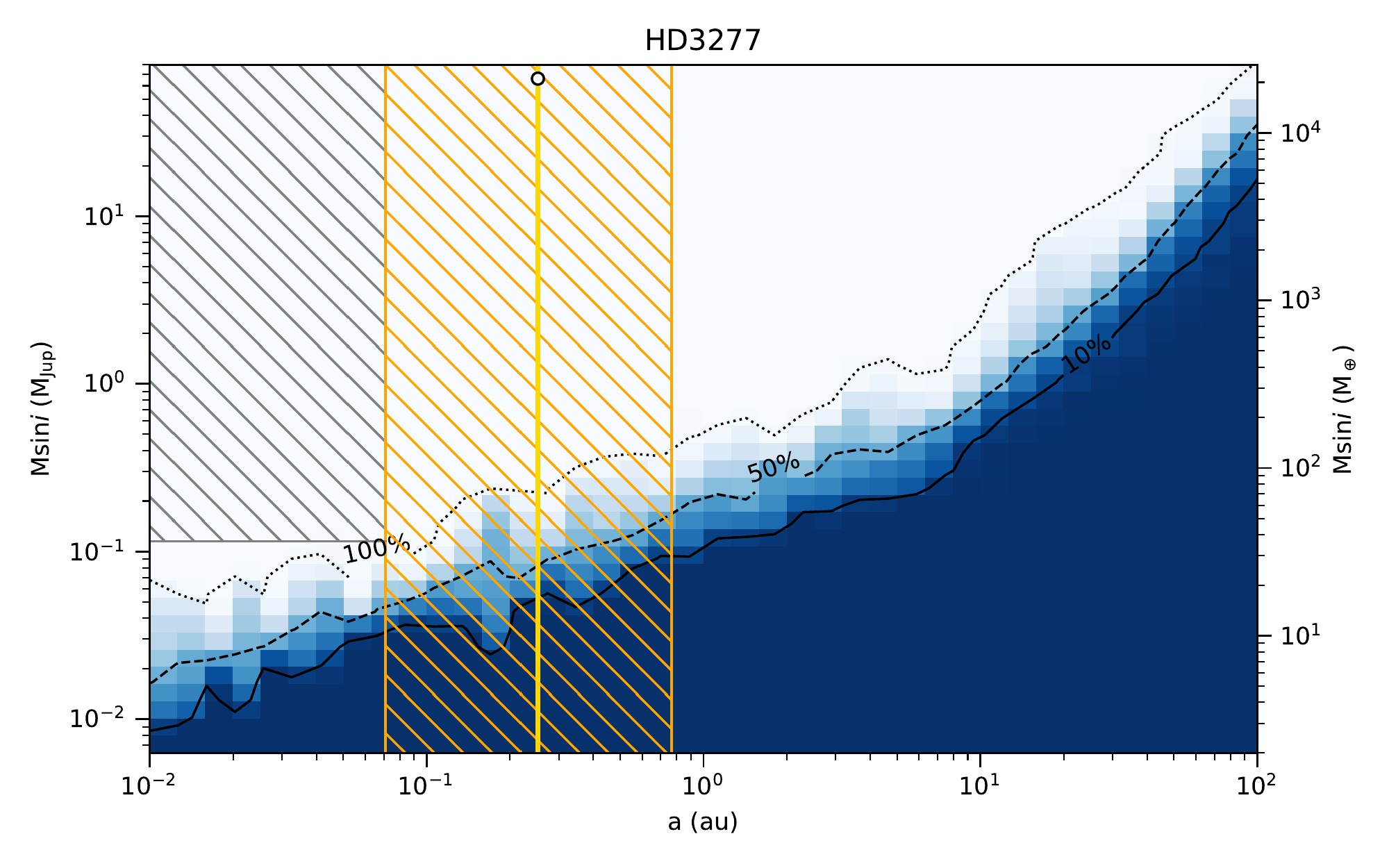}&
    		\includegraphics[width=0.22\linewidth]{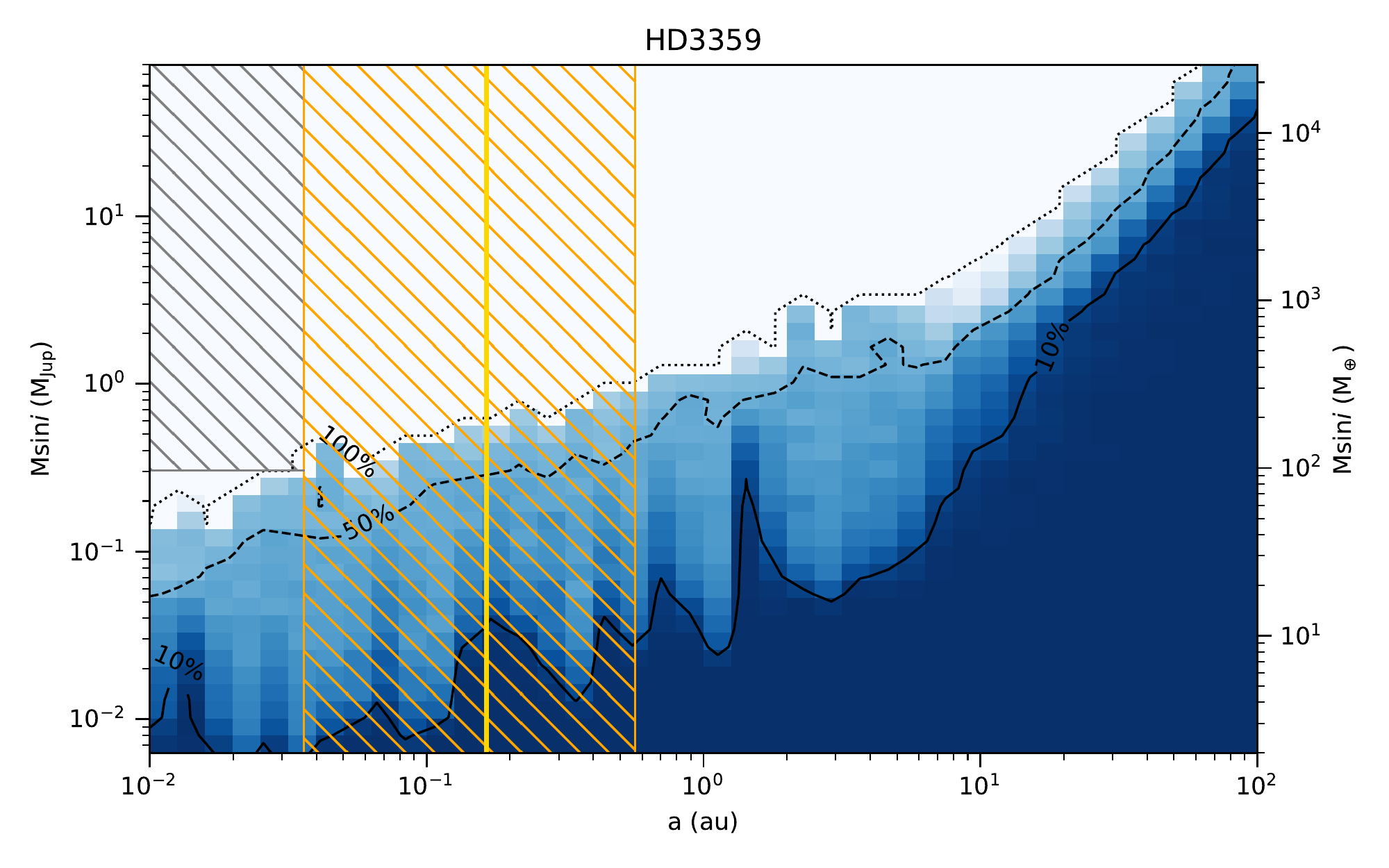}\\
    
    		\includegraphics[width=0.22\linewidth]{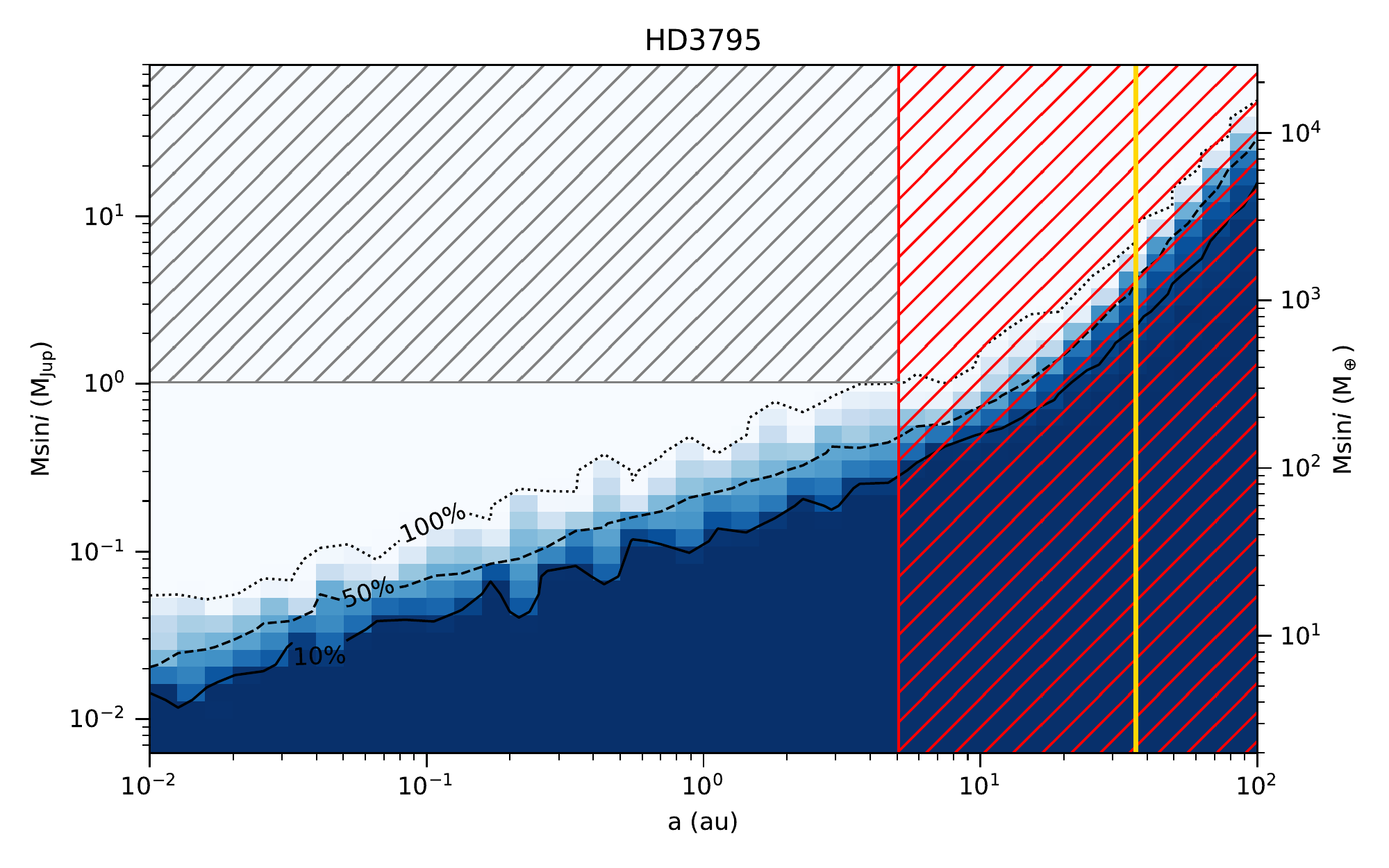}&
    		\includegraphics[width=0.22\linewidth]{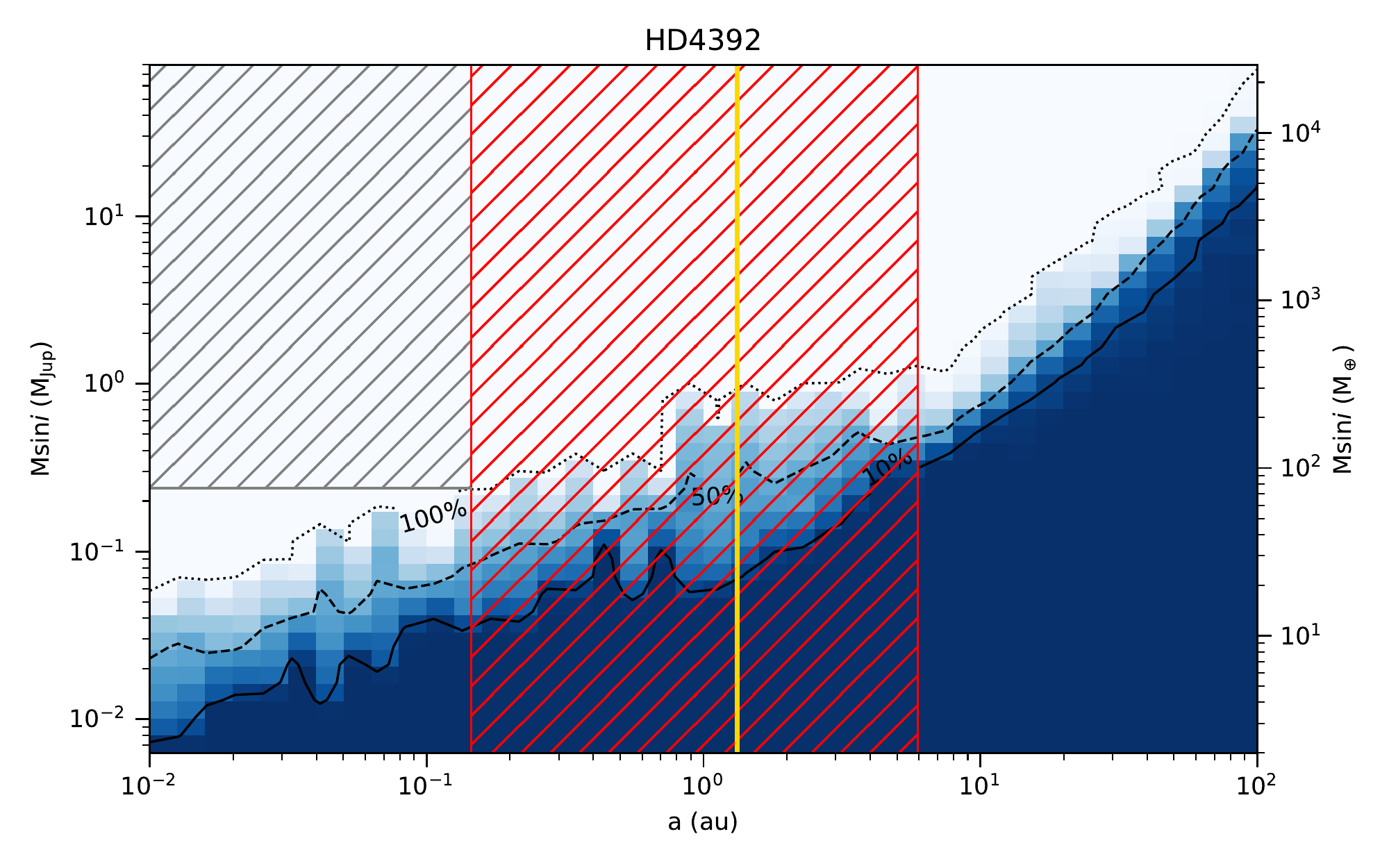}&
    		\includegraphics[width=0.22\linewidth]{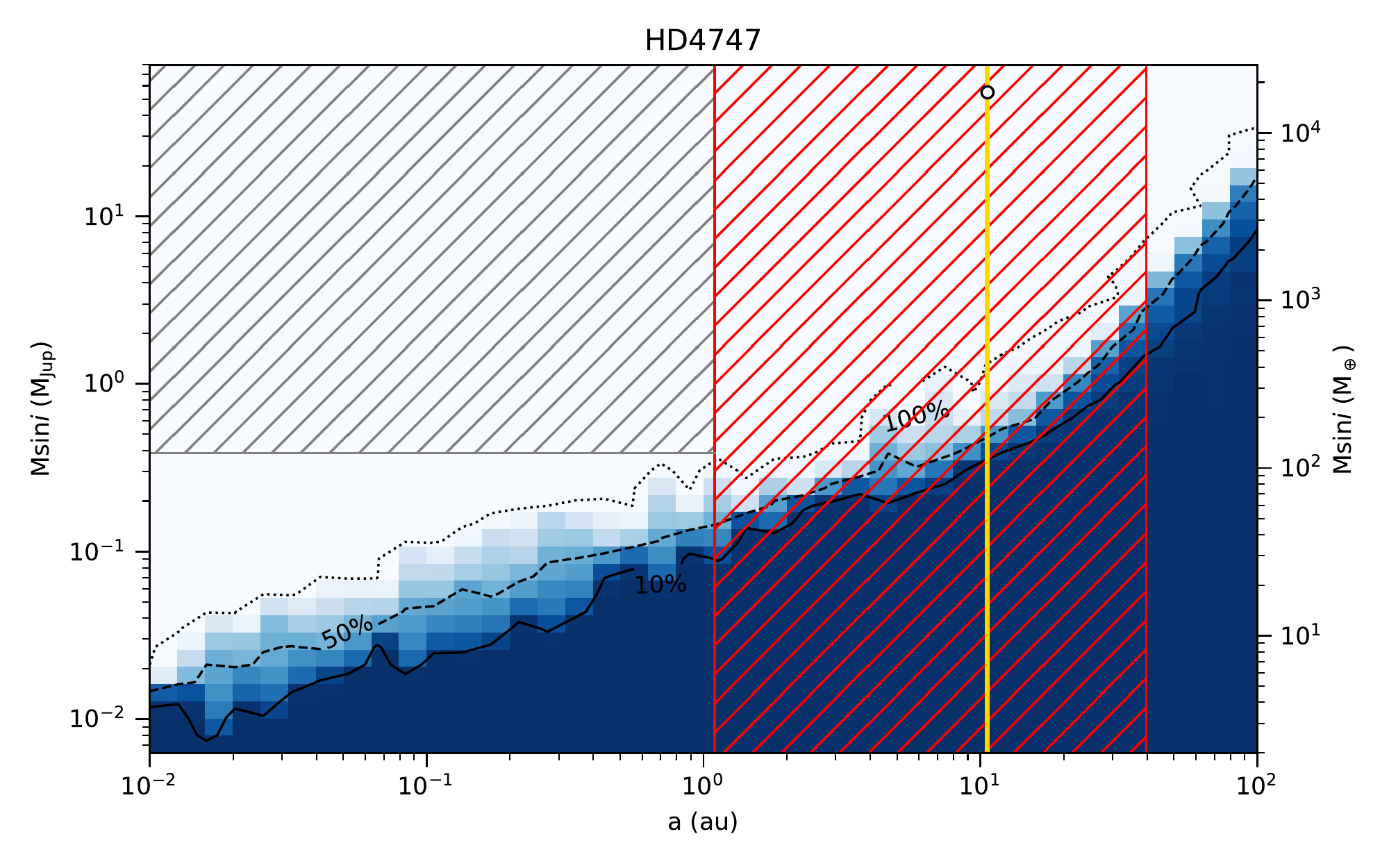}&
    		\includegraphics[width=0.22\linewidth]{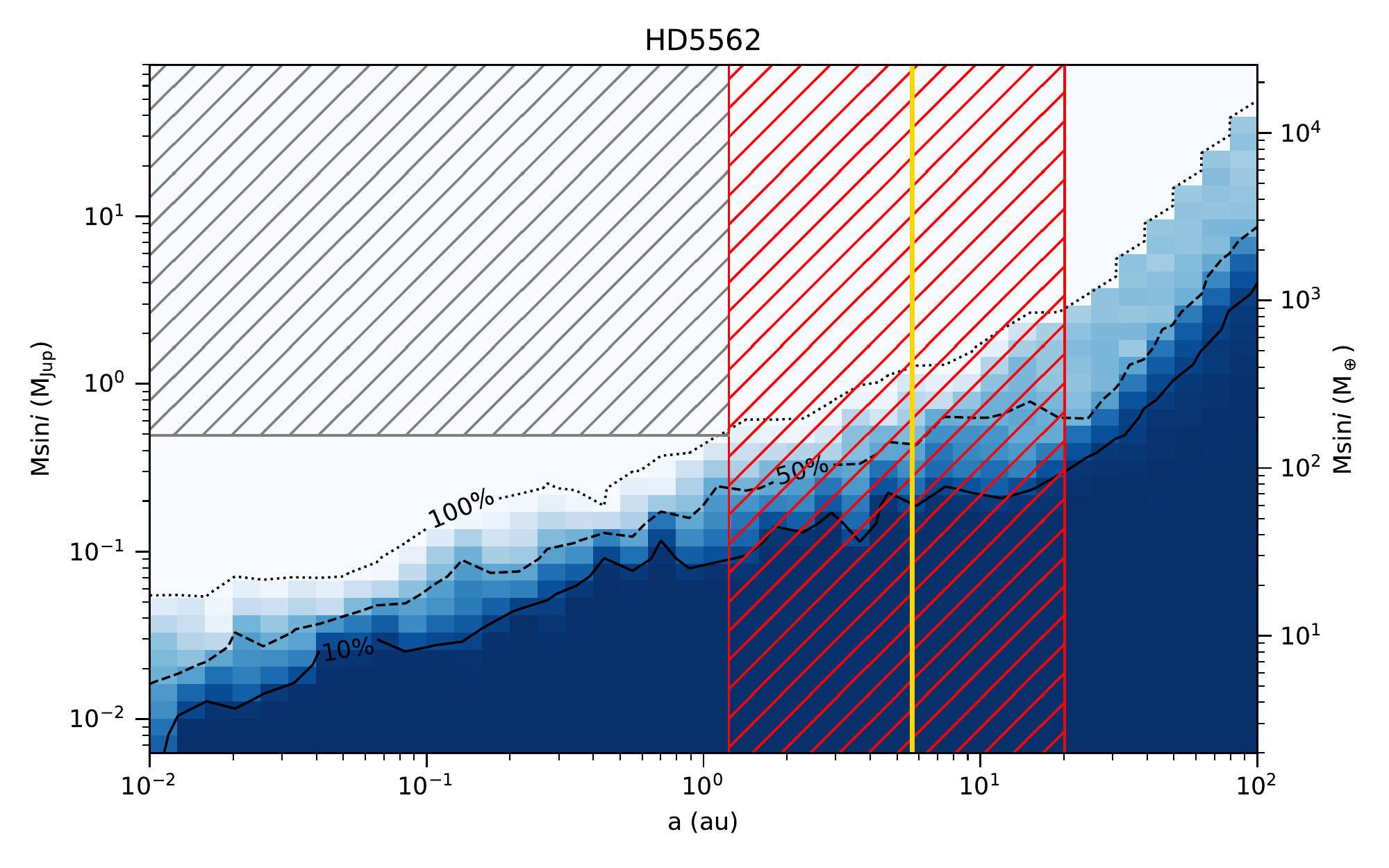}\\
    
    		\includegraphics[width=0.22\linewidth]{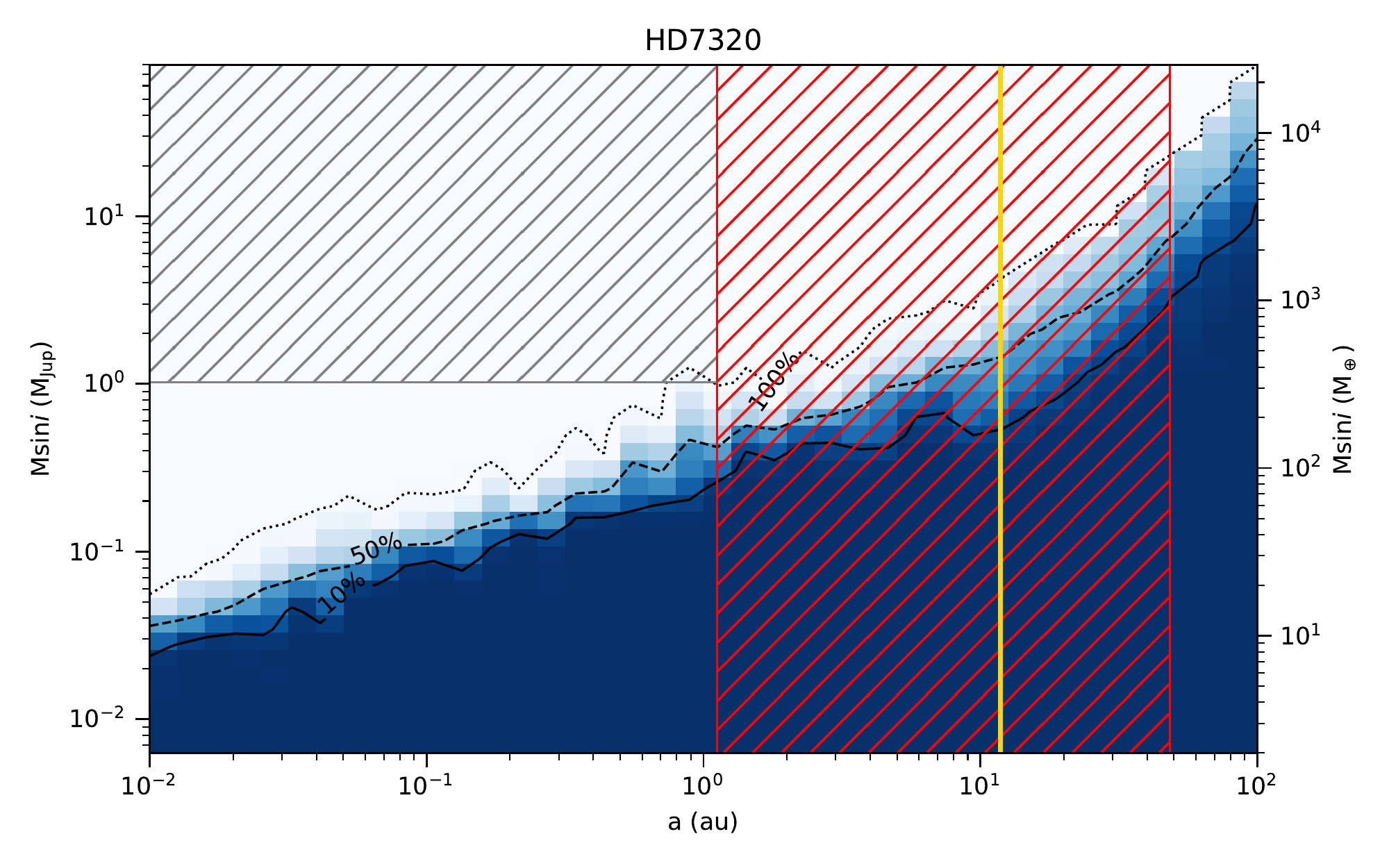}&
    		\includegraphics[width=0.22\linewidth]{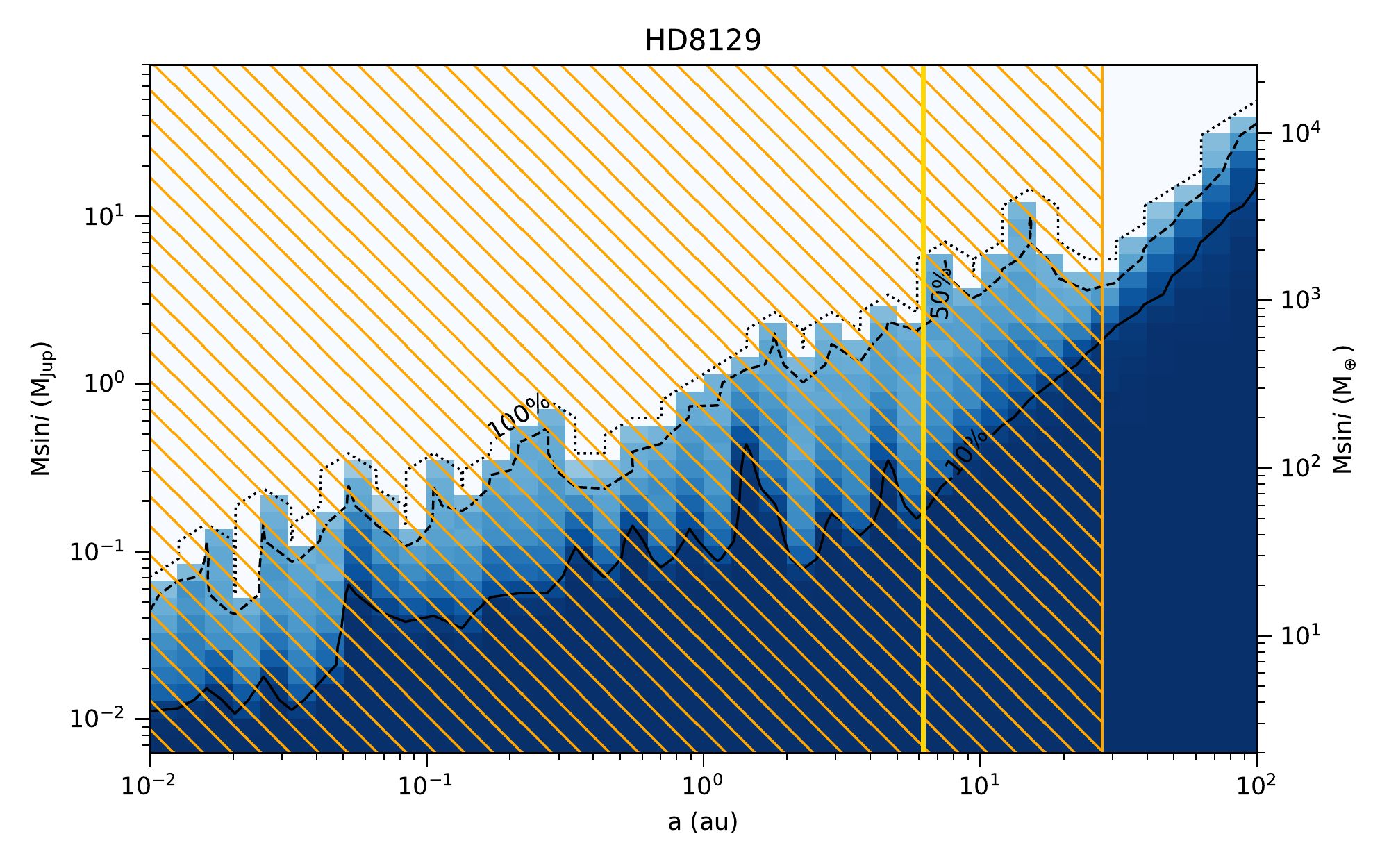}&
    		\includegraphics[width=0.22\linewidth]{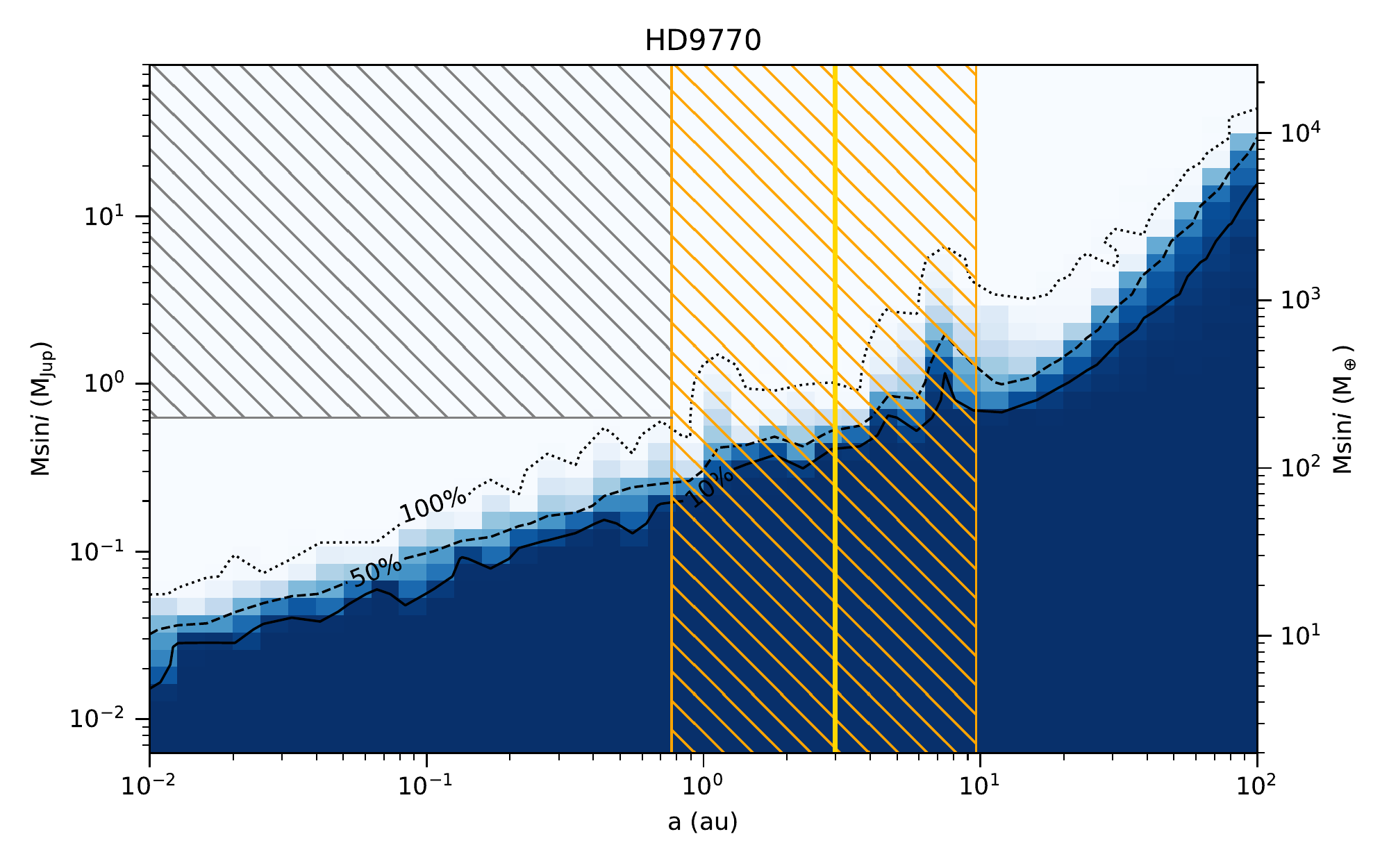}&
    		\includegraphics[width=0.22\linewidth]{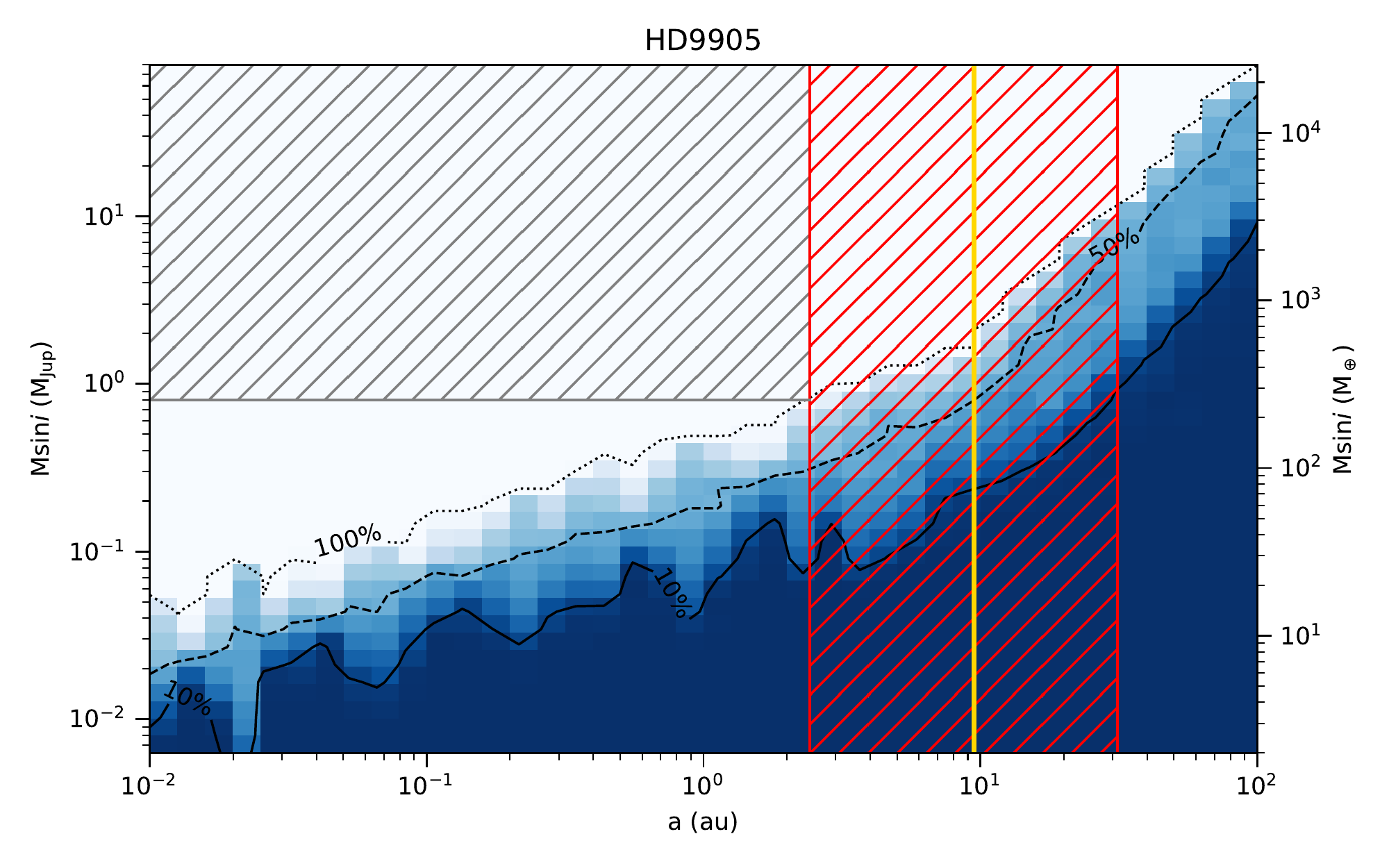}\\
    
    		\includegraphics[width=0.22\linewidth]{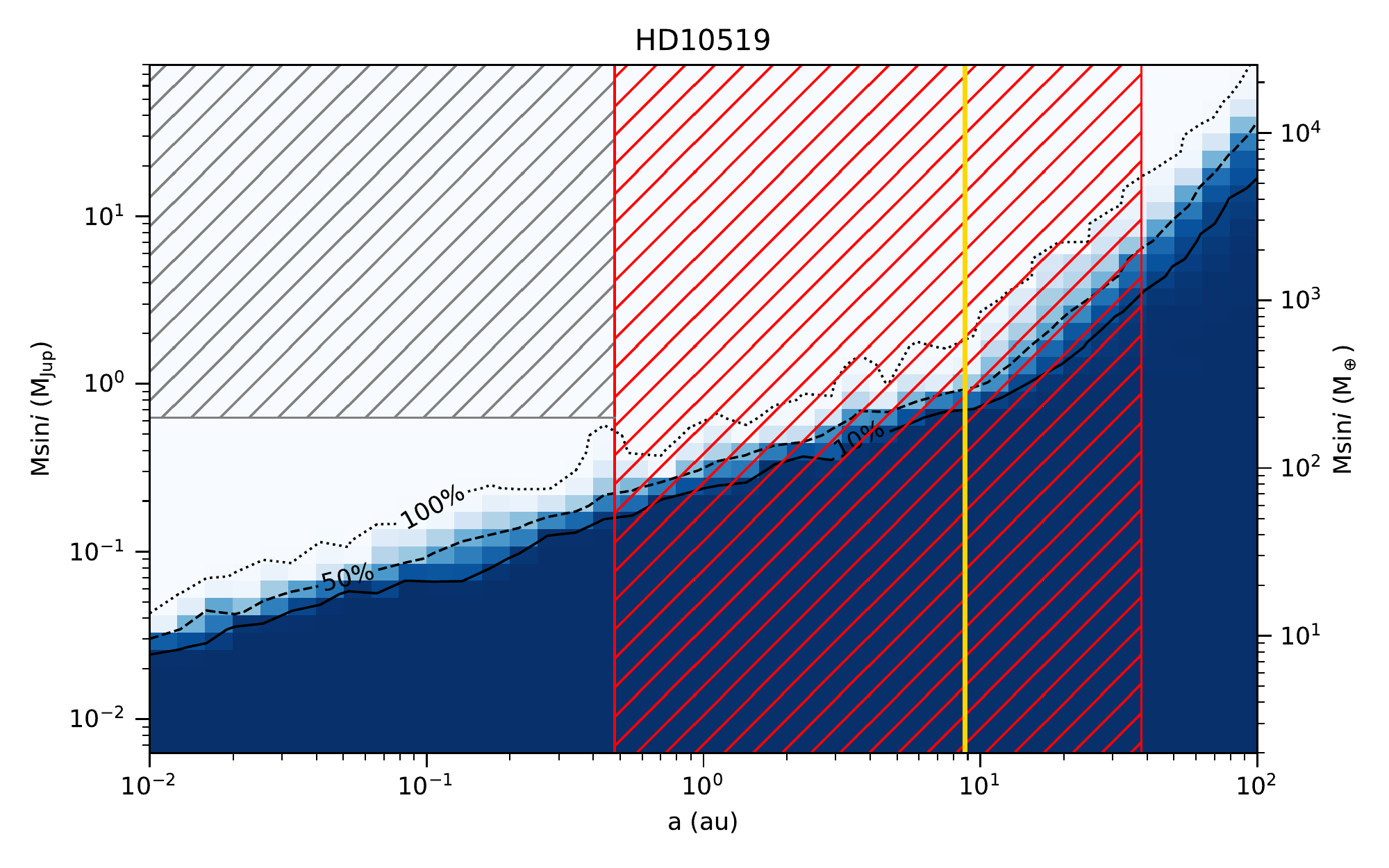}&
    		\includegraphics[width=0.22\linewidth]{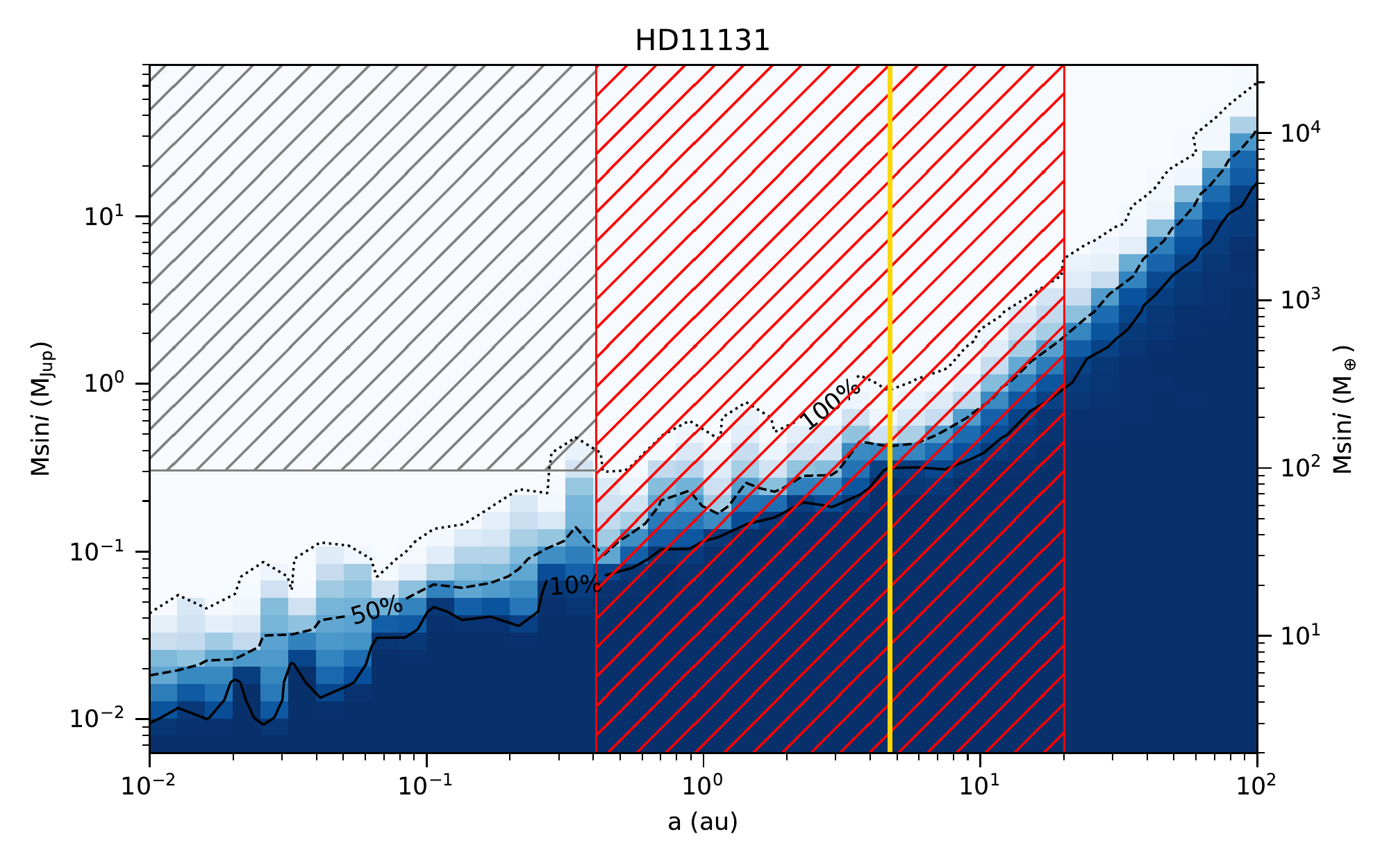}&
    		\includegraphics[width=0.22\linewidth]{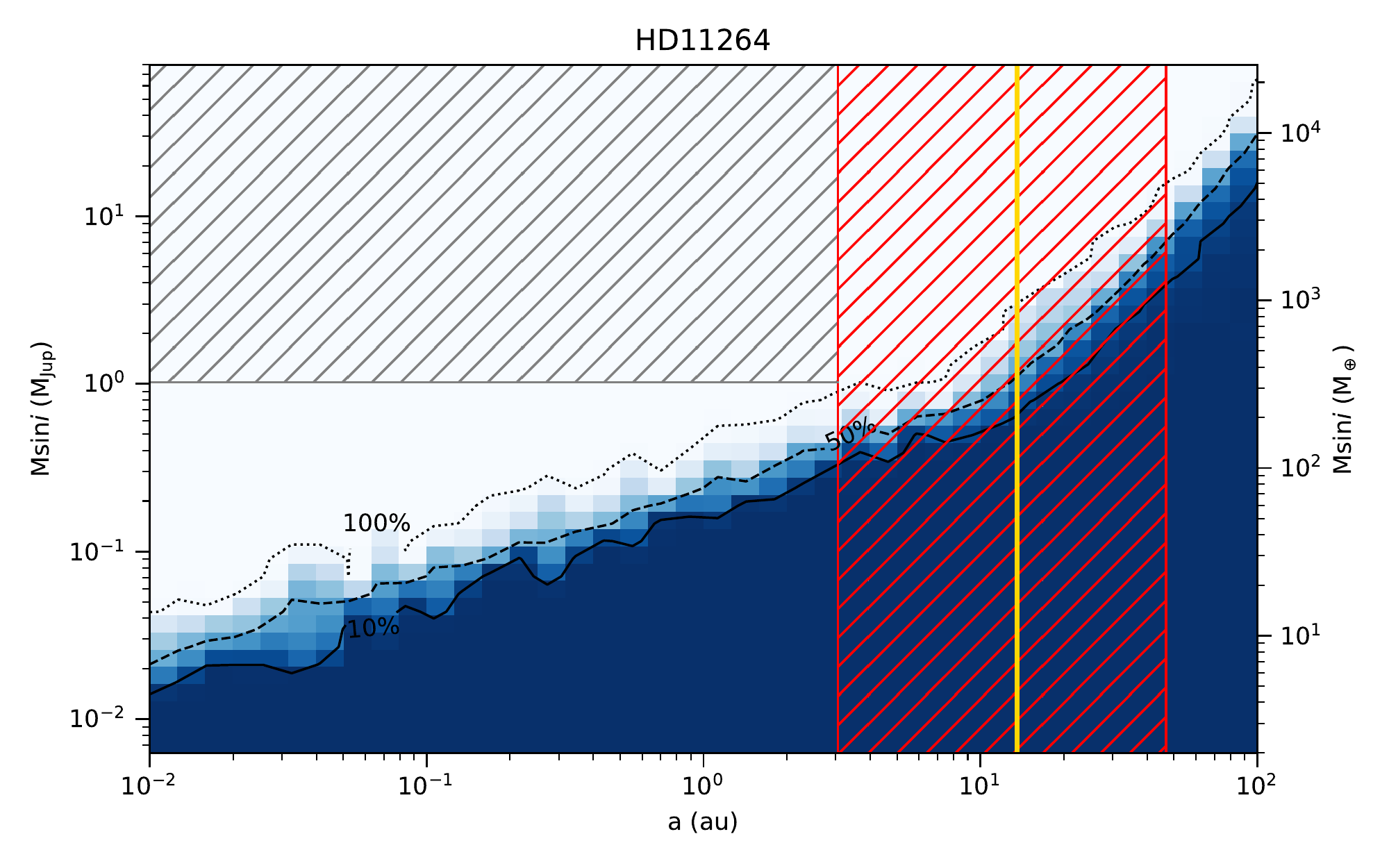}&
    		\includegraphics[width=0.22\linewidth]{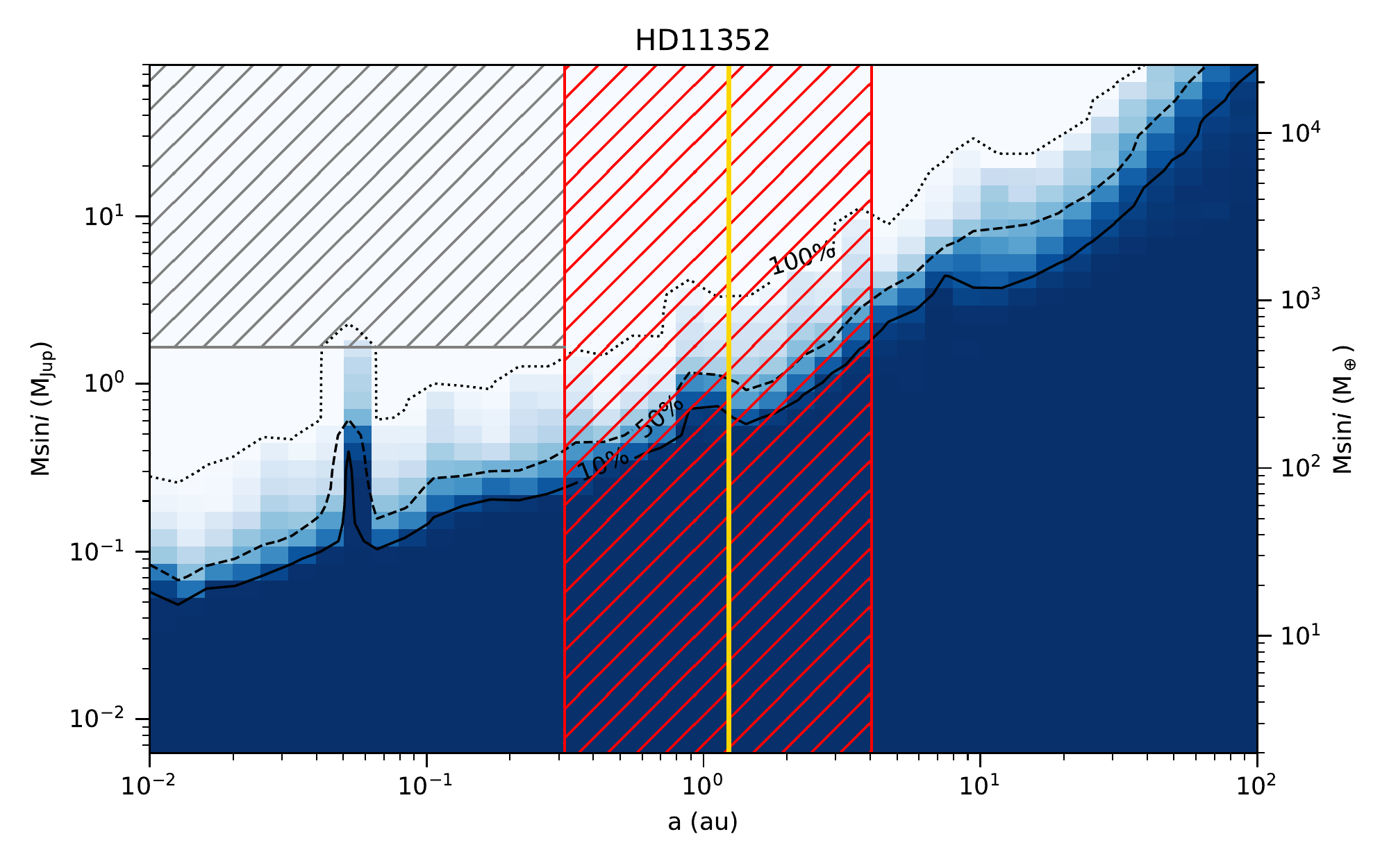}\\
    
    		\includegraphics[width=0.22\linewidth]{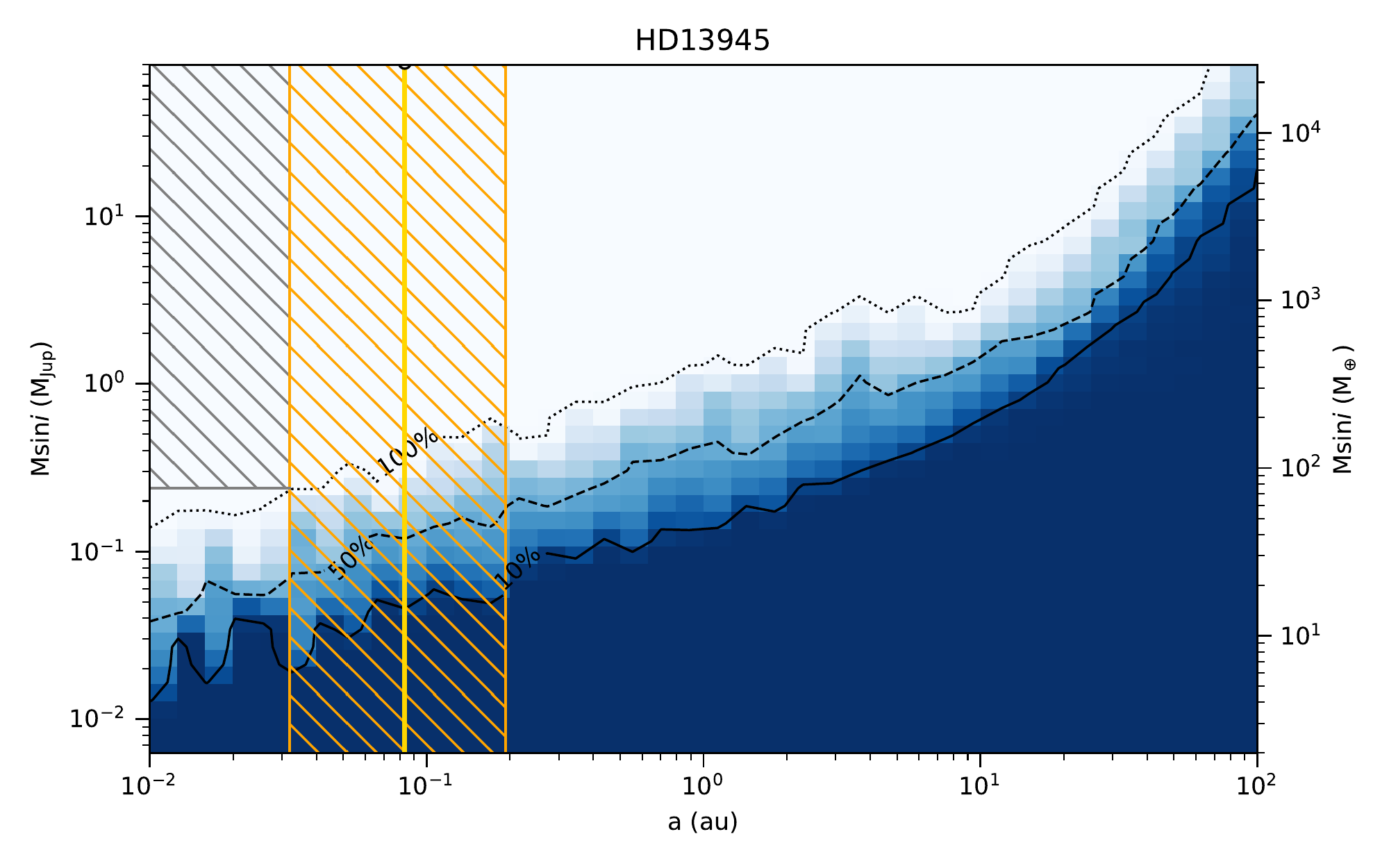}&
    		\includegraphics[width=0.22\linewidth]{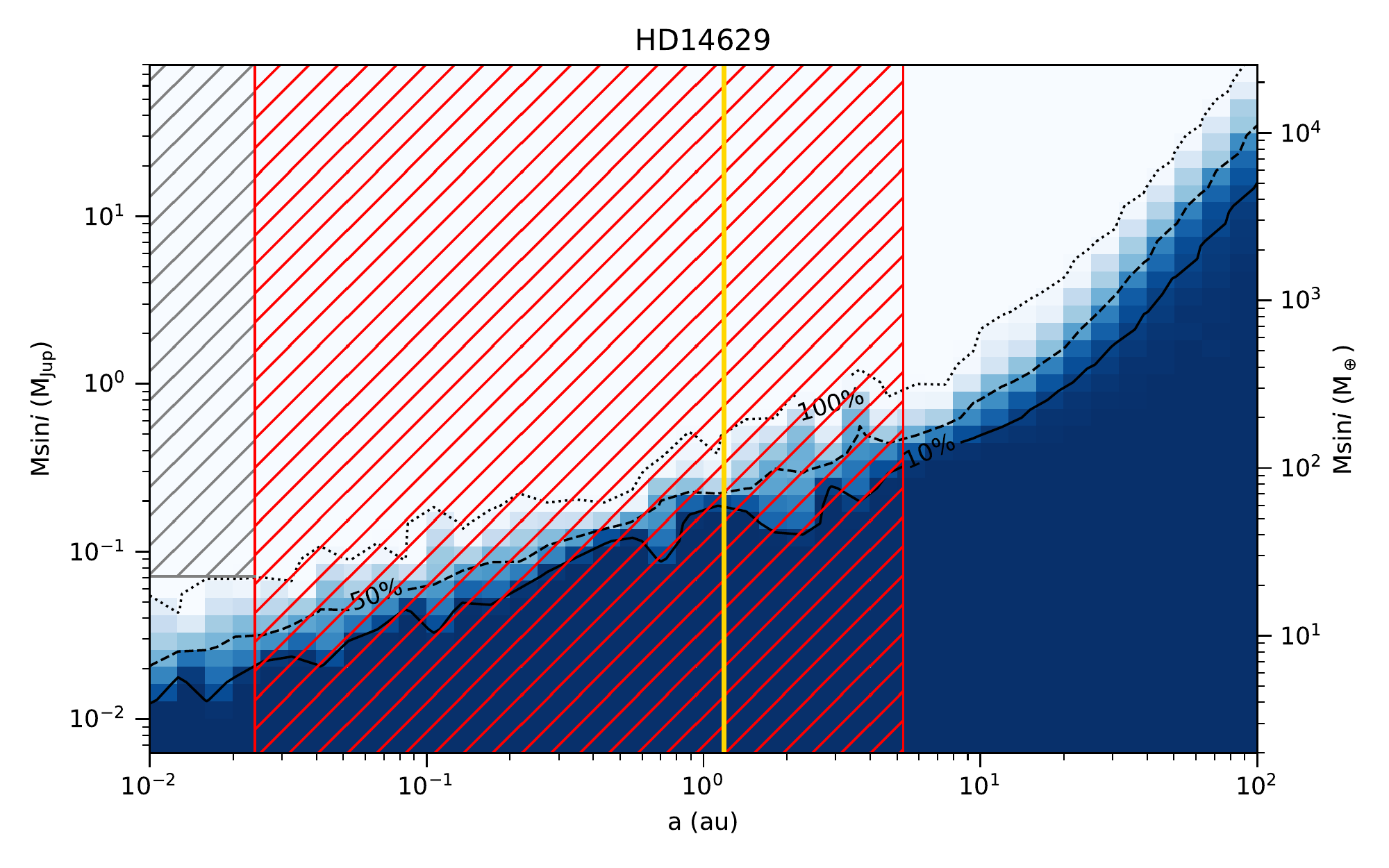}&
    		\includegraphics[width=0.22\linewidth]{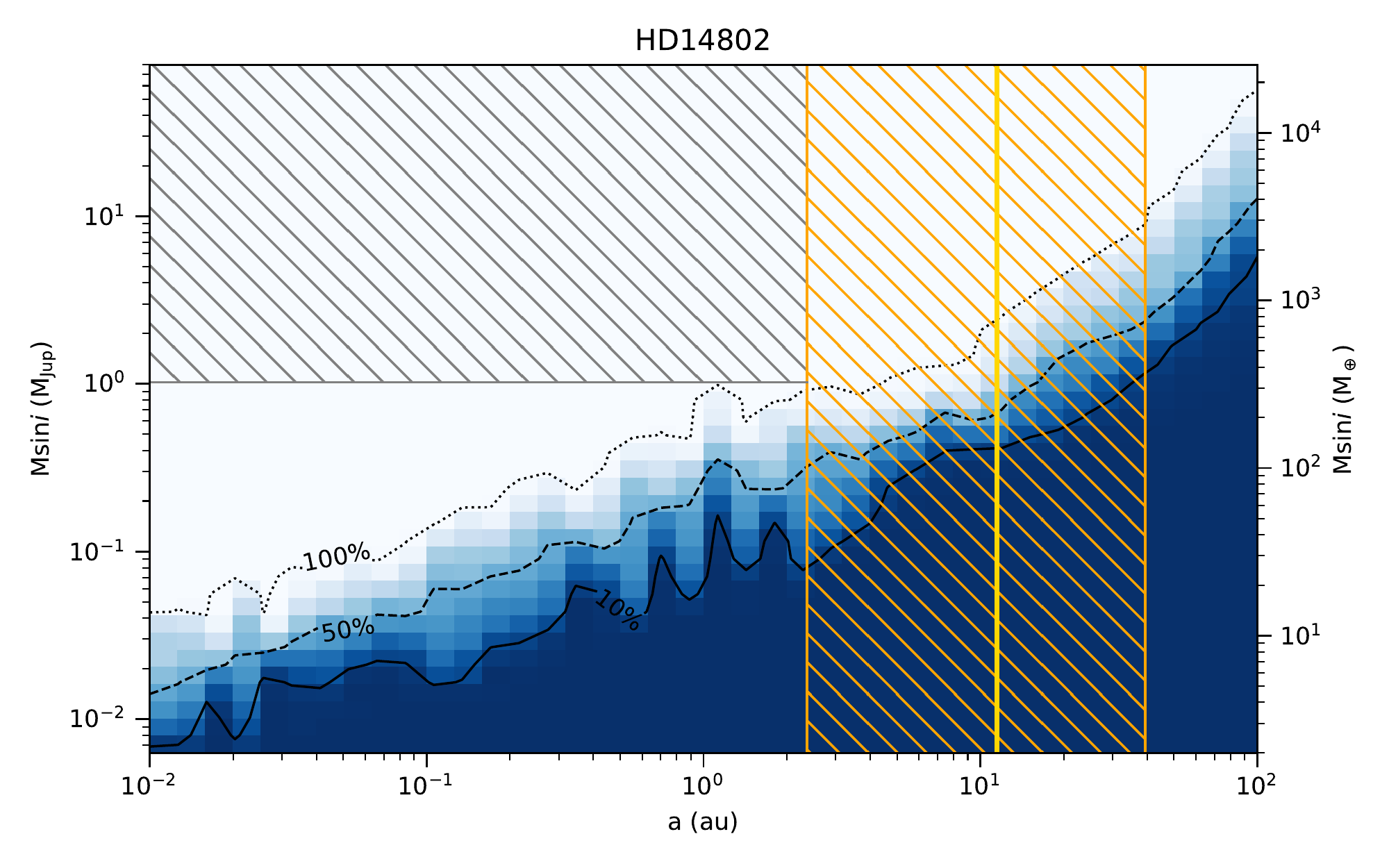}&
    		\includegraphics[width=0.22\linewidth]{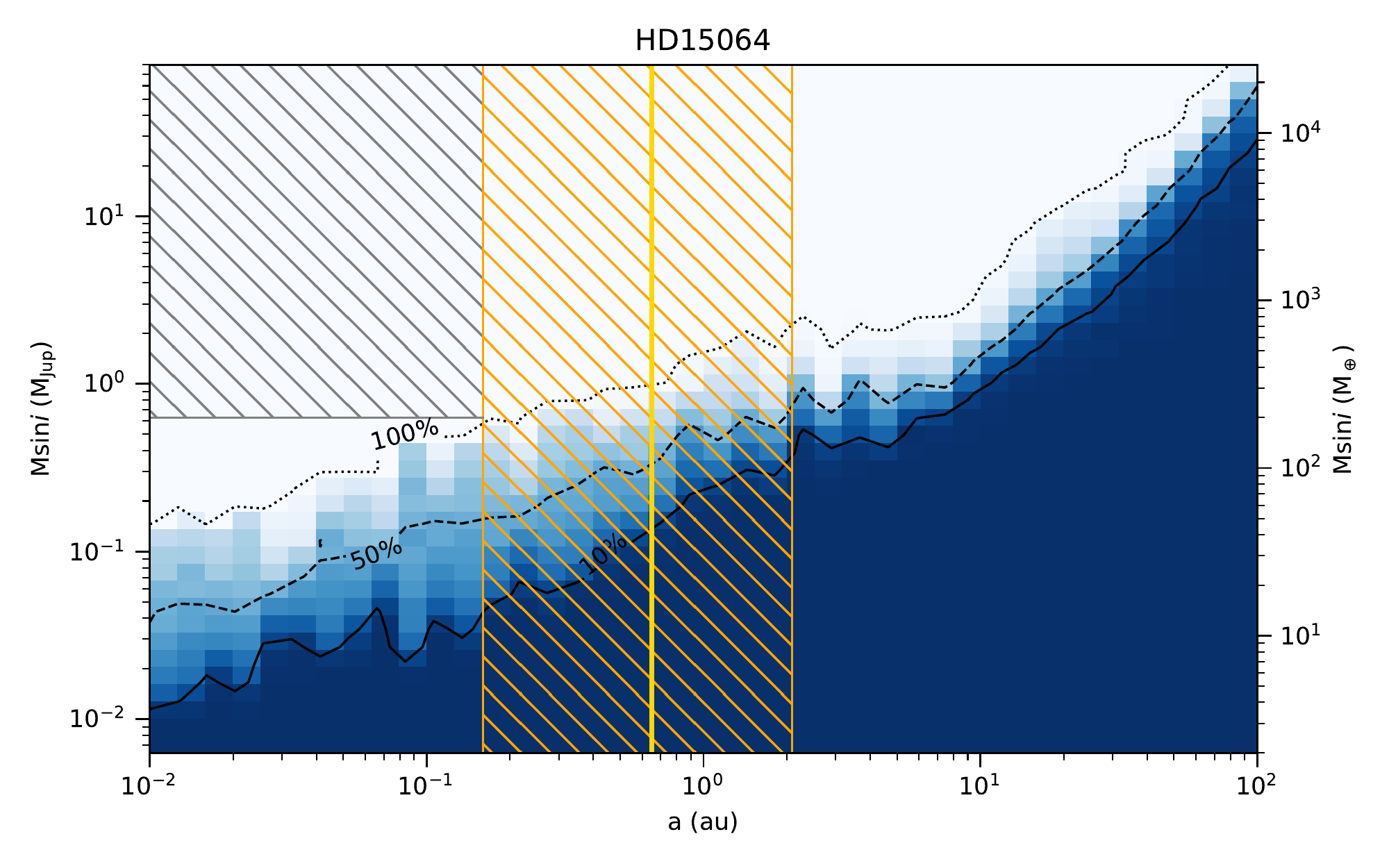}\\
    
    		\includegraphics[width=0.22\linewidth]{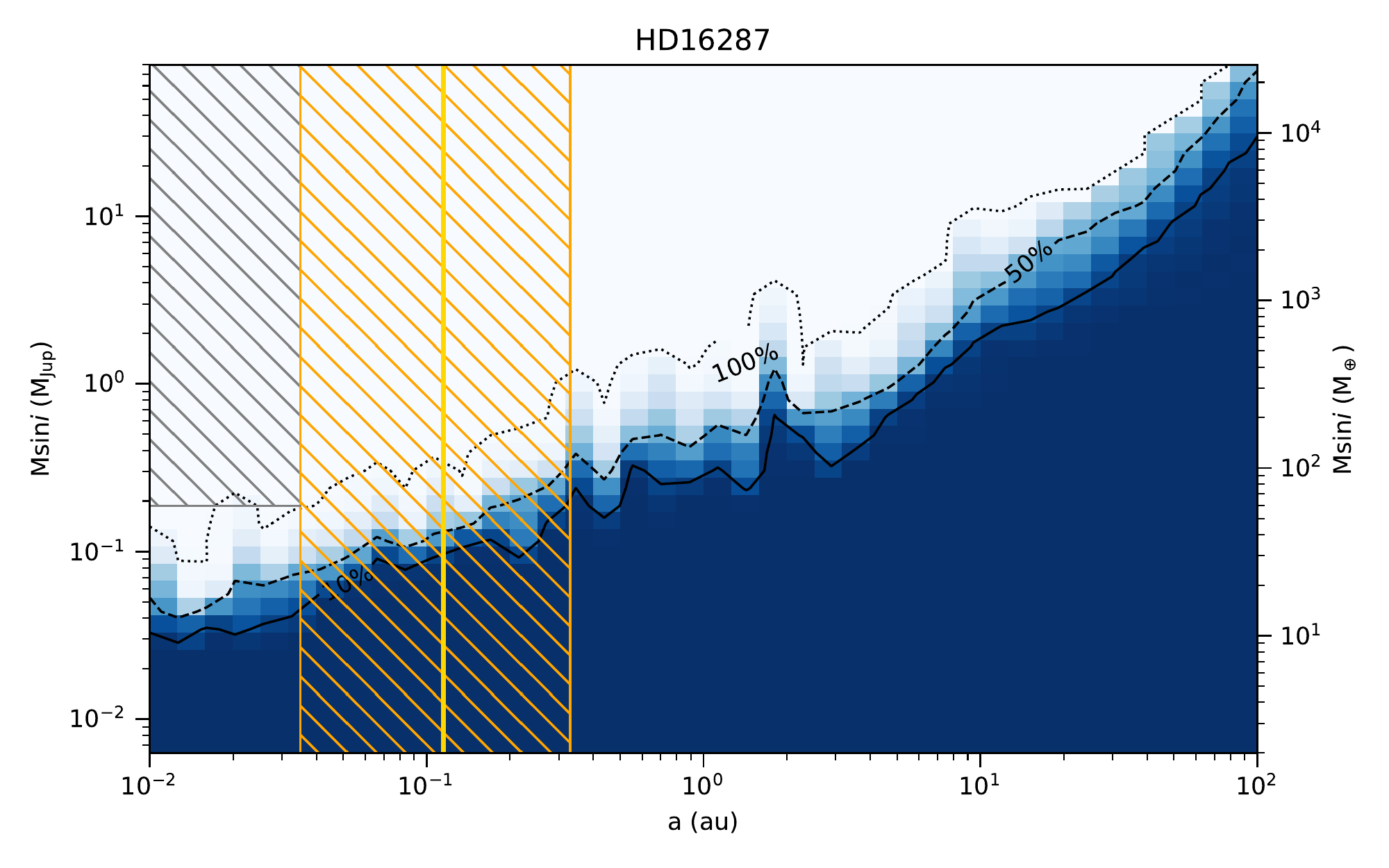}&
    		\includegraphics[width=0.22\linewidth]{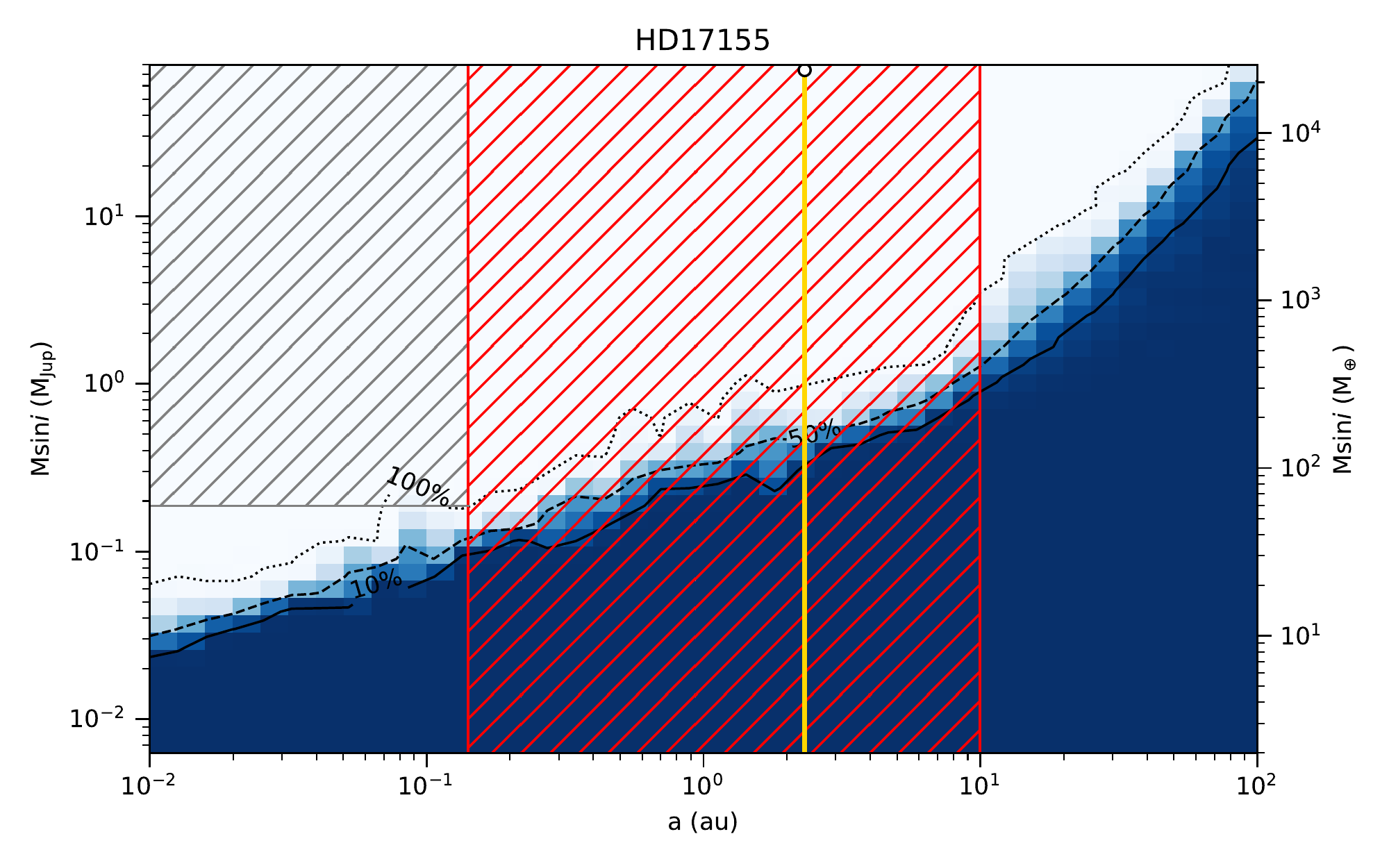}&
    		\includegraphics[width=0.22\linewidth]{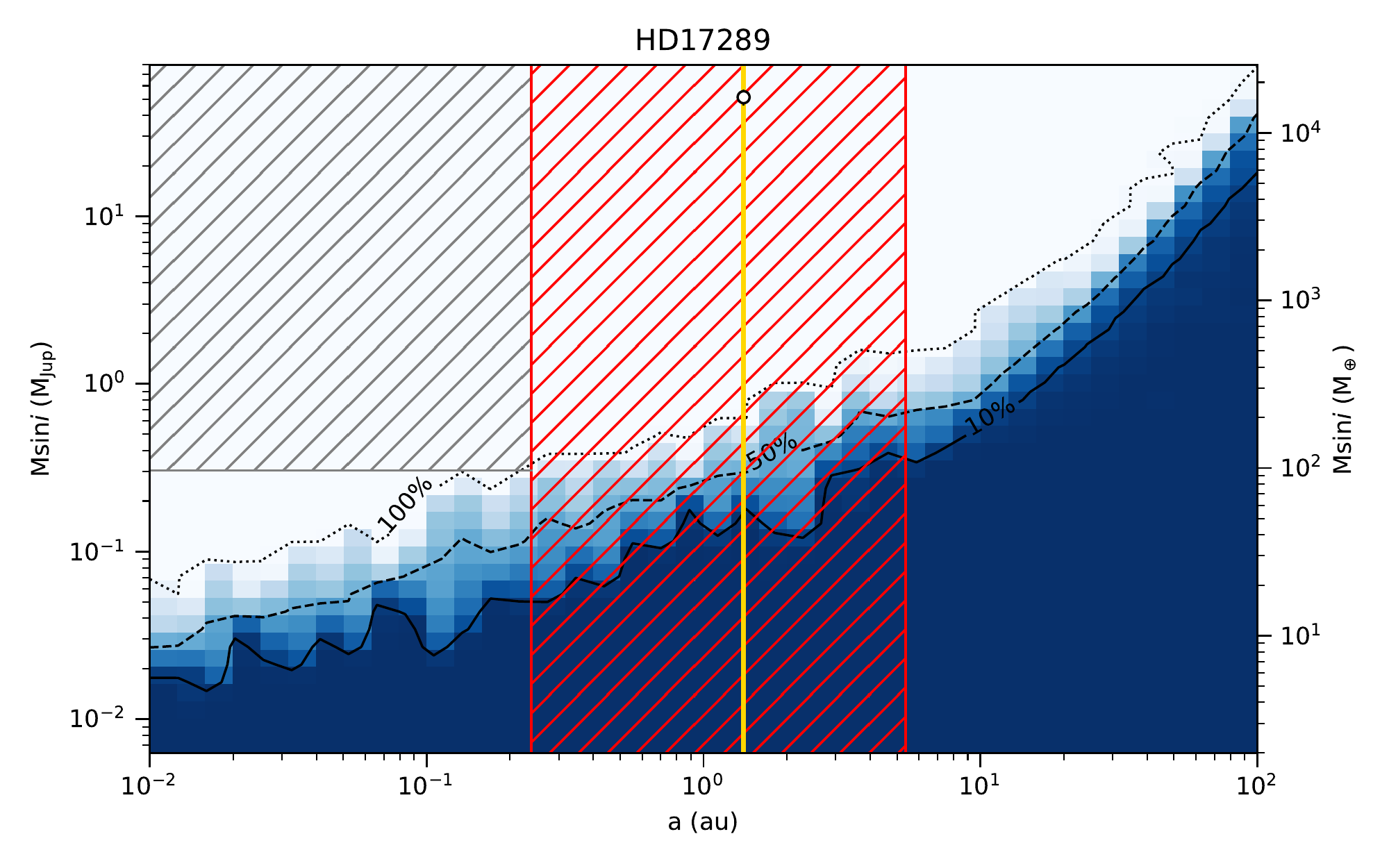}&
    		\includegraphics[width=0.22\linewidth]{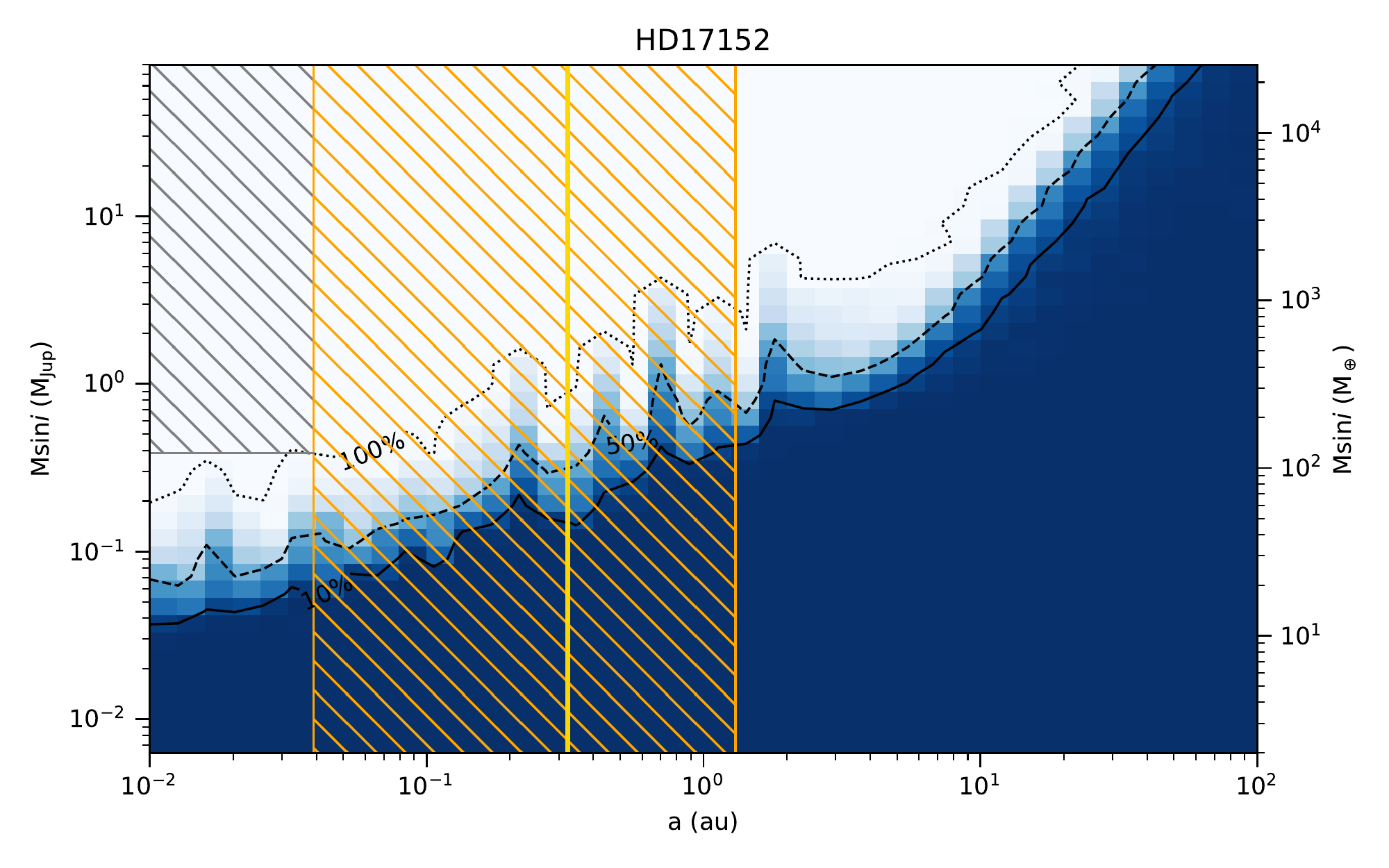}\\
    
    		\includegraphics[width=0.22\linewidth]{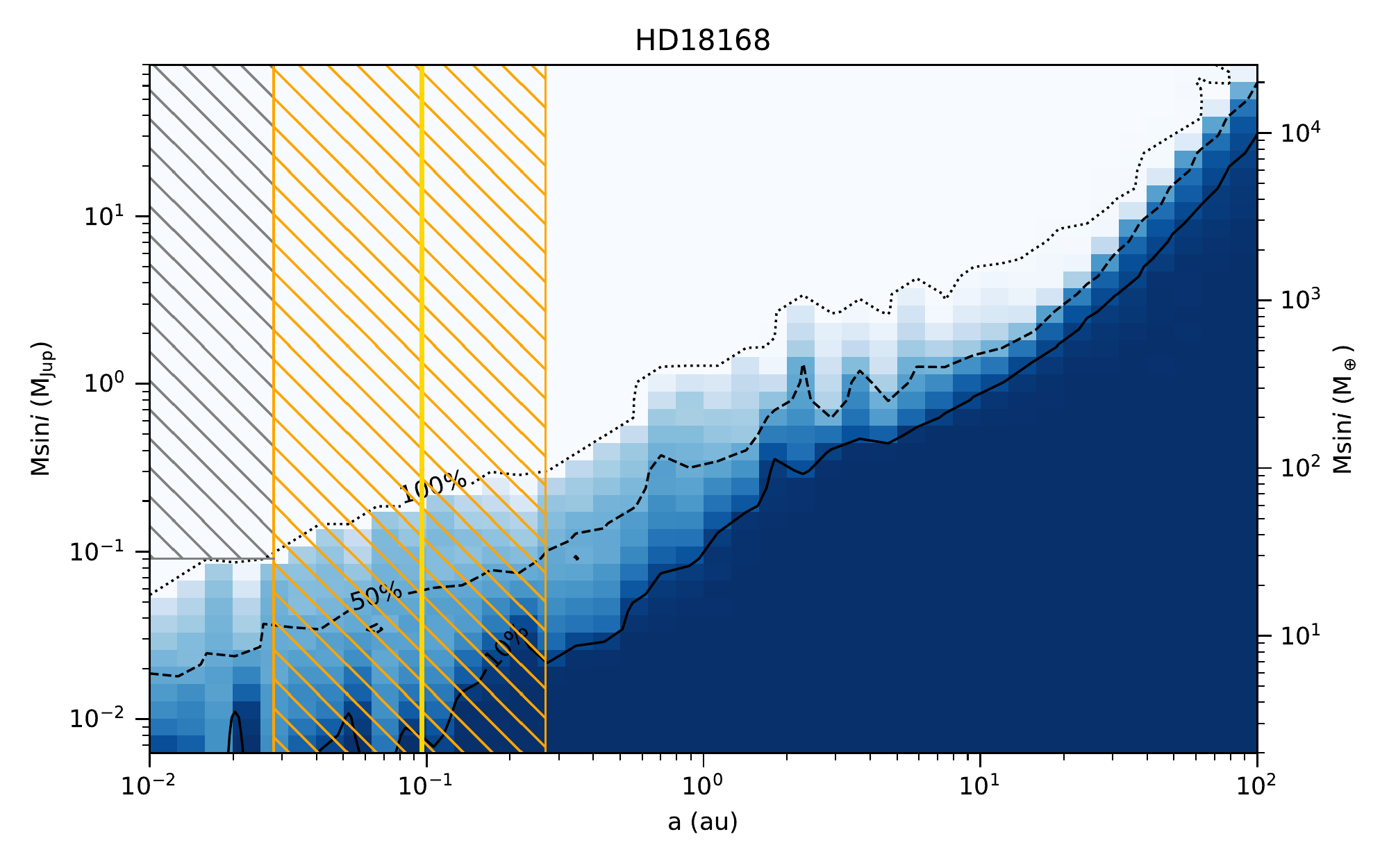}&
    		\includegraphics[width=0.22\linewidth]{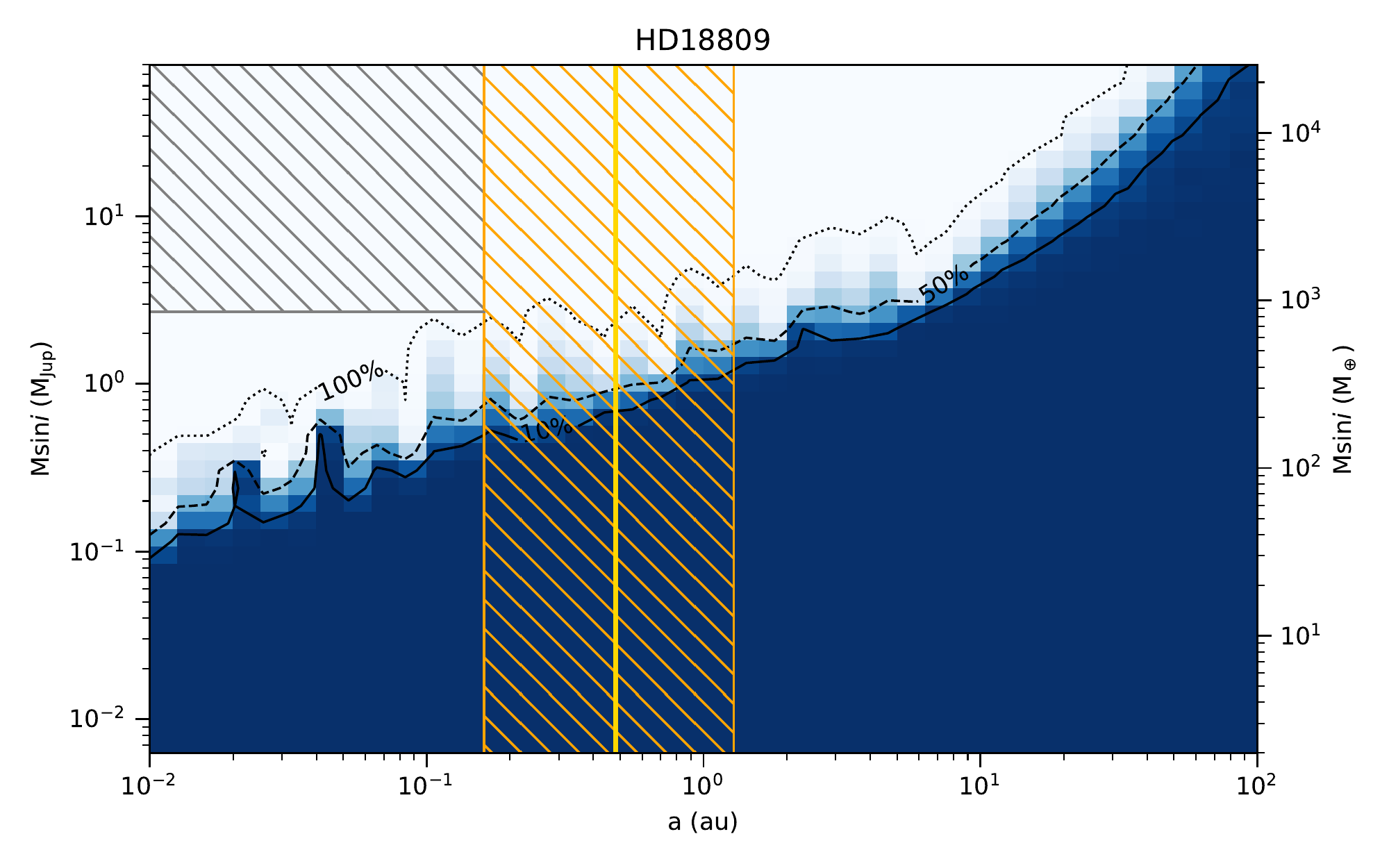}&
    		\includegraphics[width=0.22\linewidth]{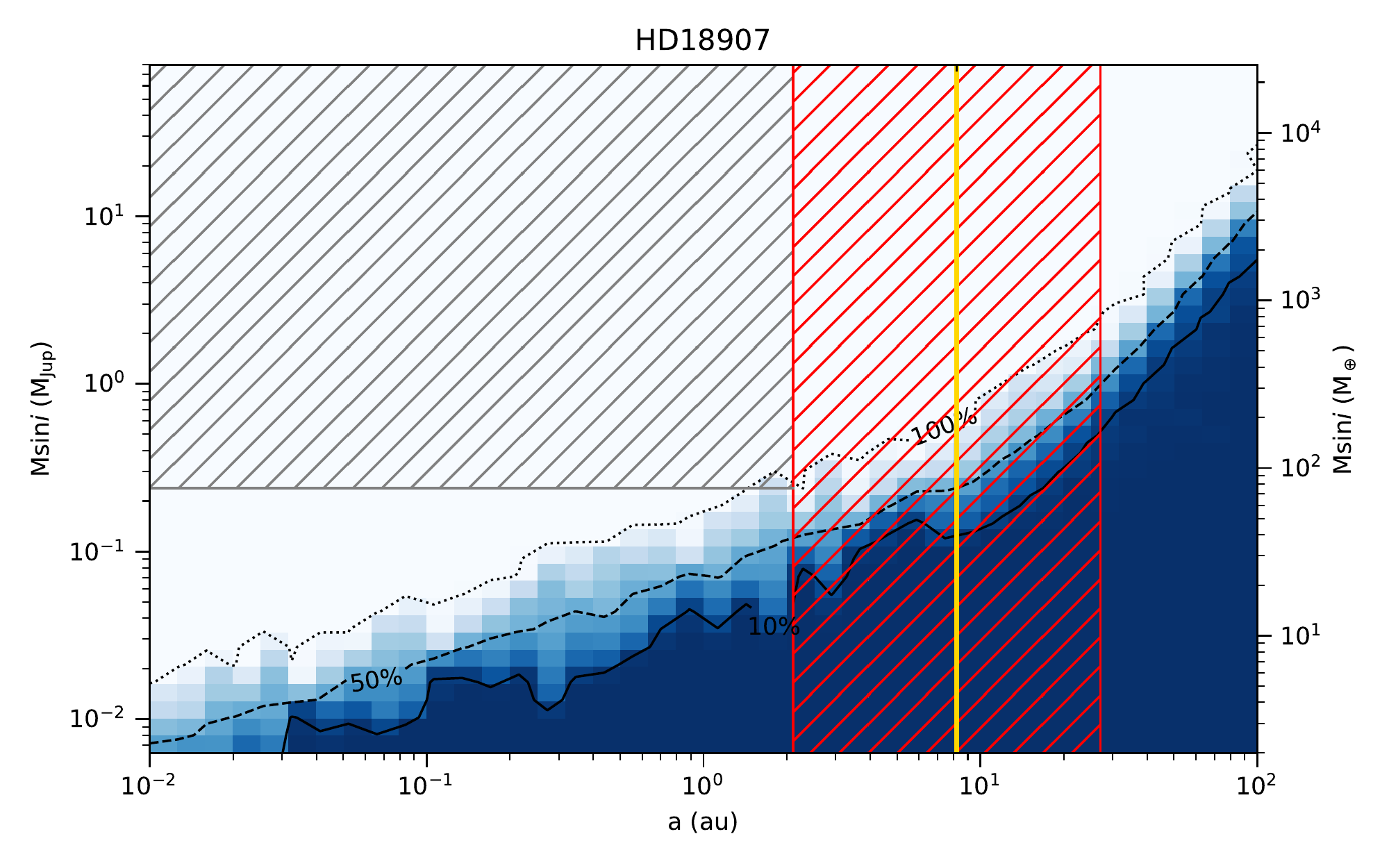}&
    		\includegraphics[width=0.22\linewidth]{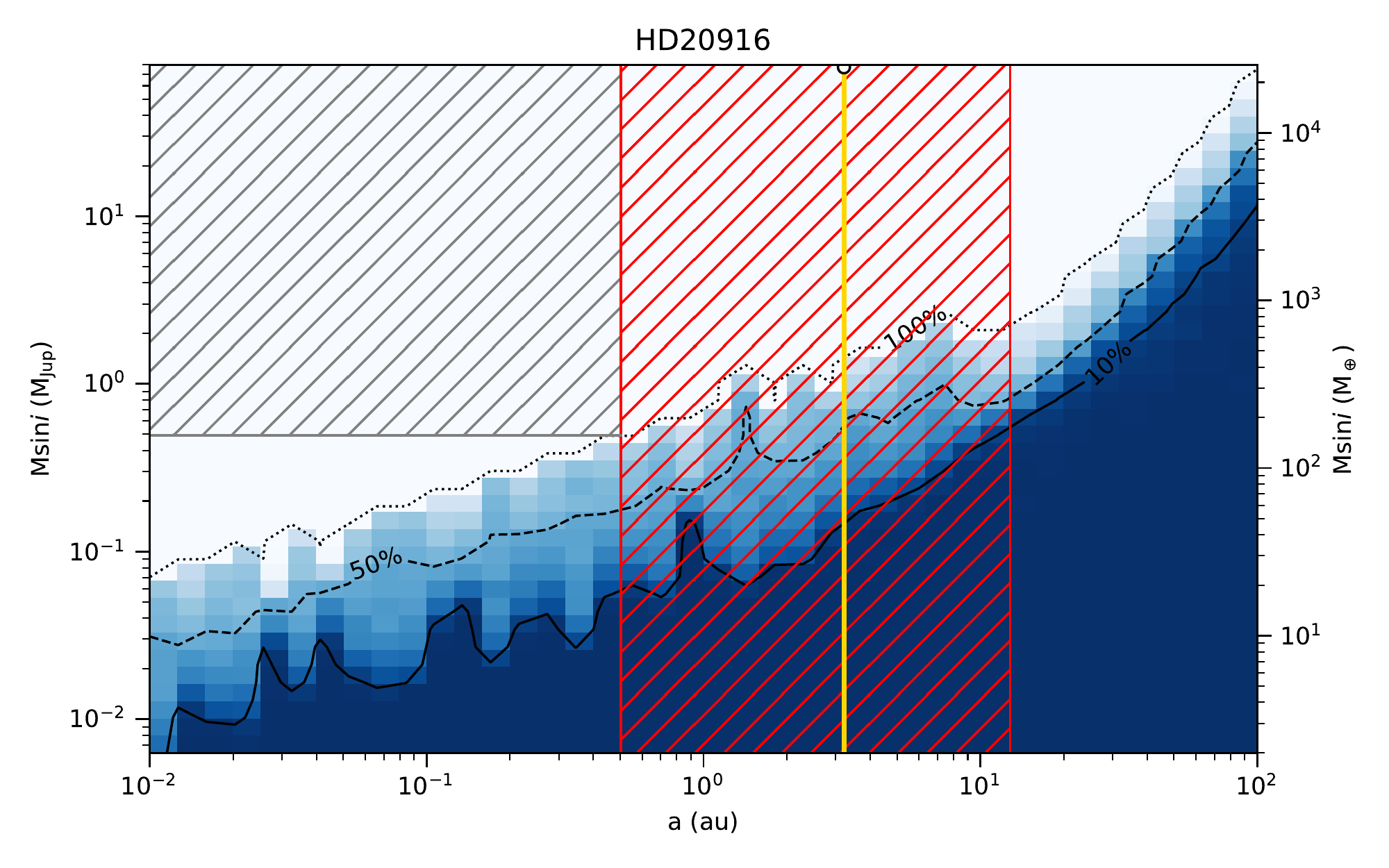}\\
    
    		\includegraphics[width=0.22\linewidth]{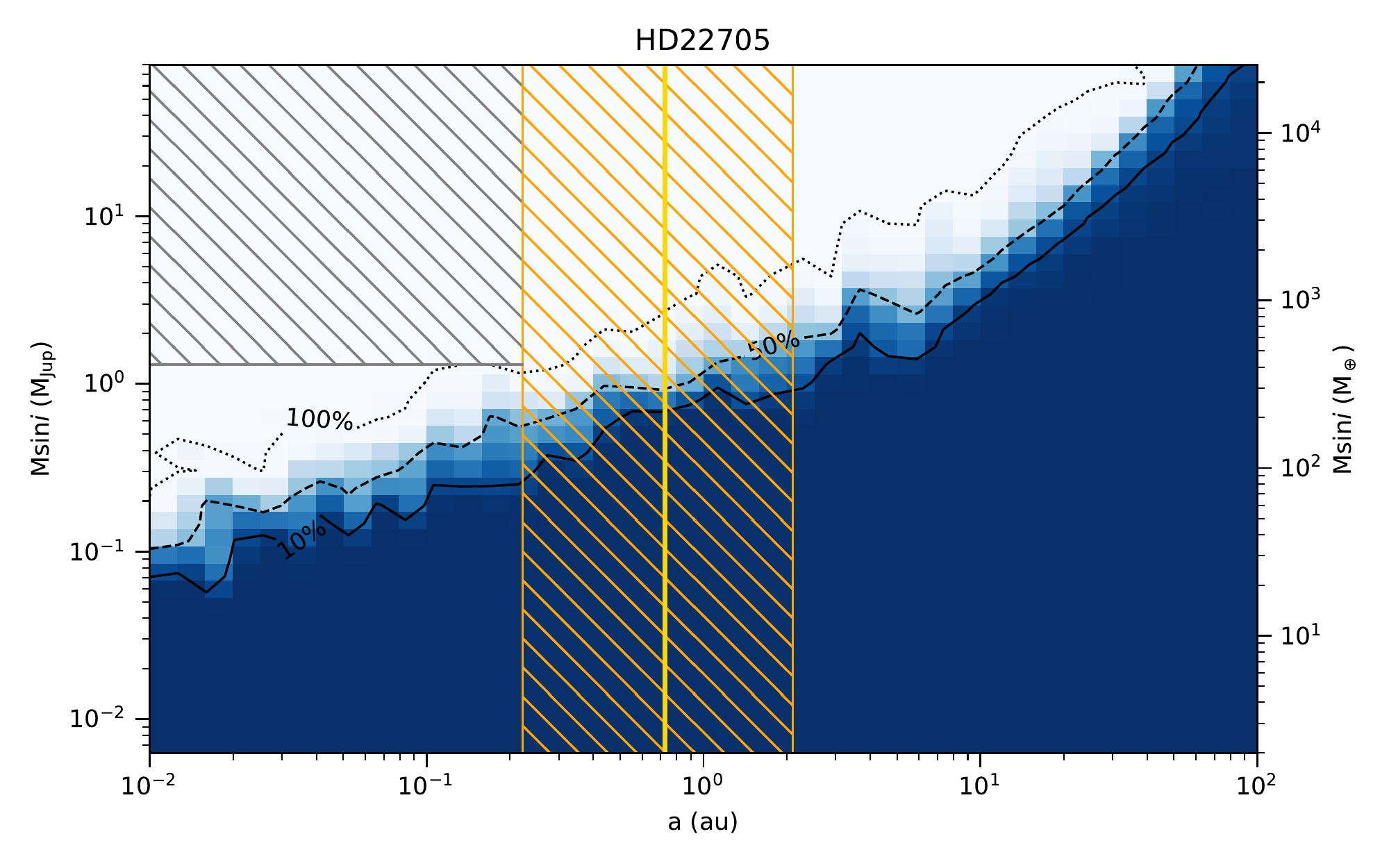}&
    		\includegraphics[width=0.22\linewidth]{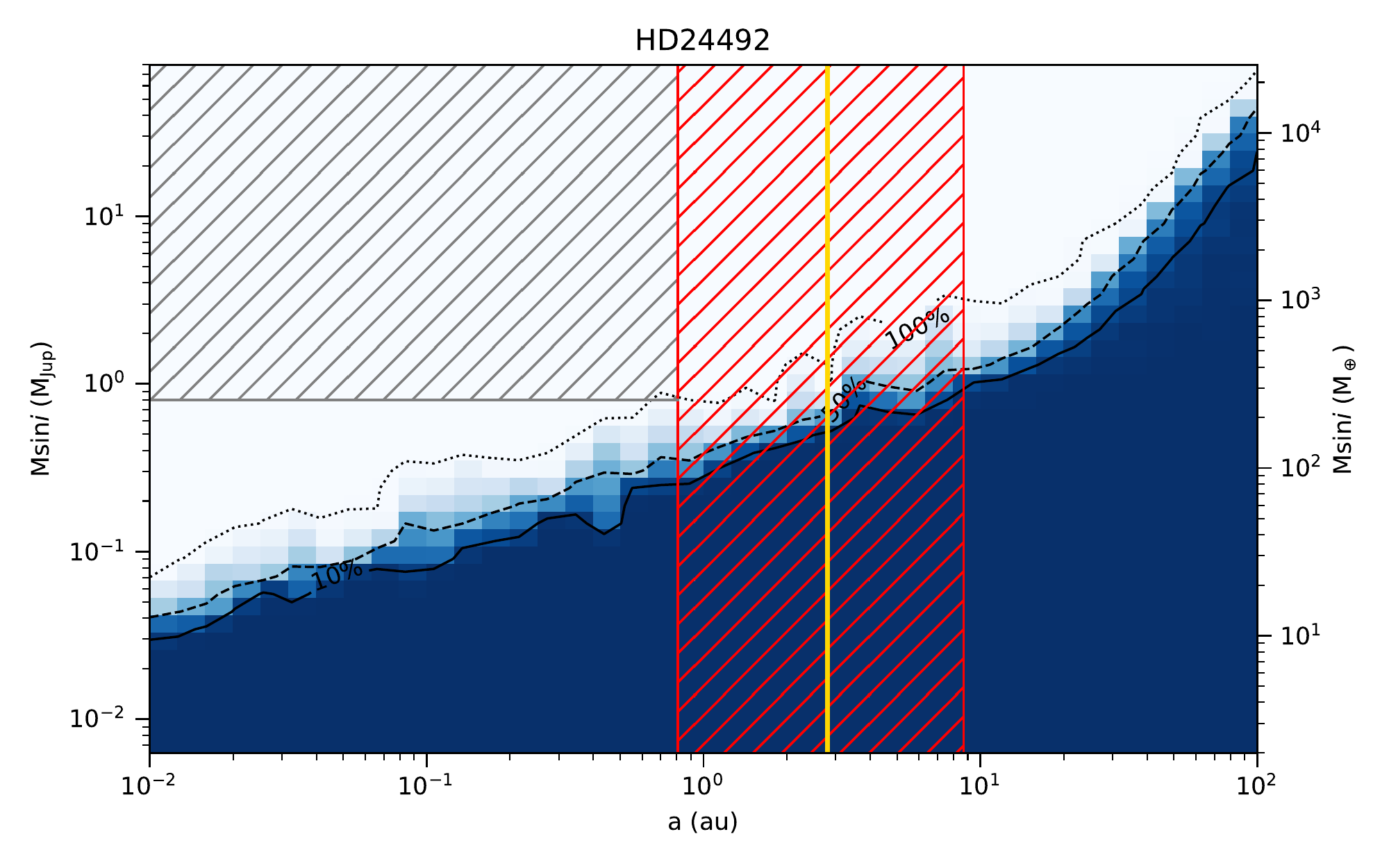}&
    		\includegraphics[width=0.22\linewidth]{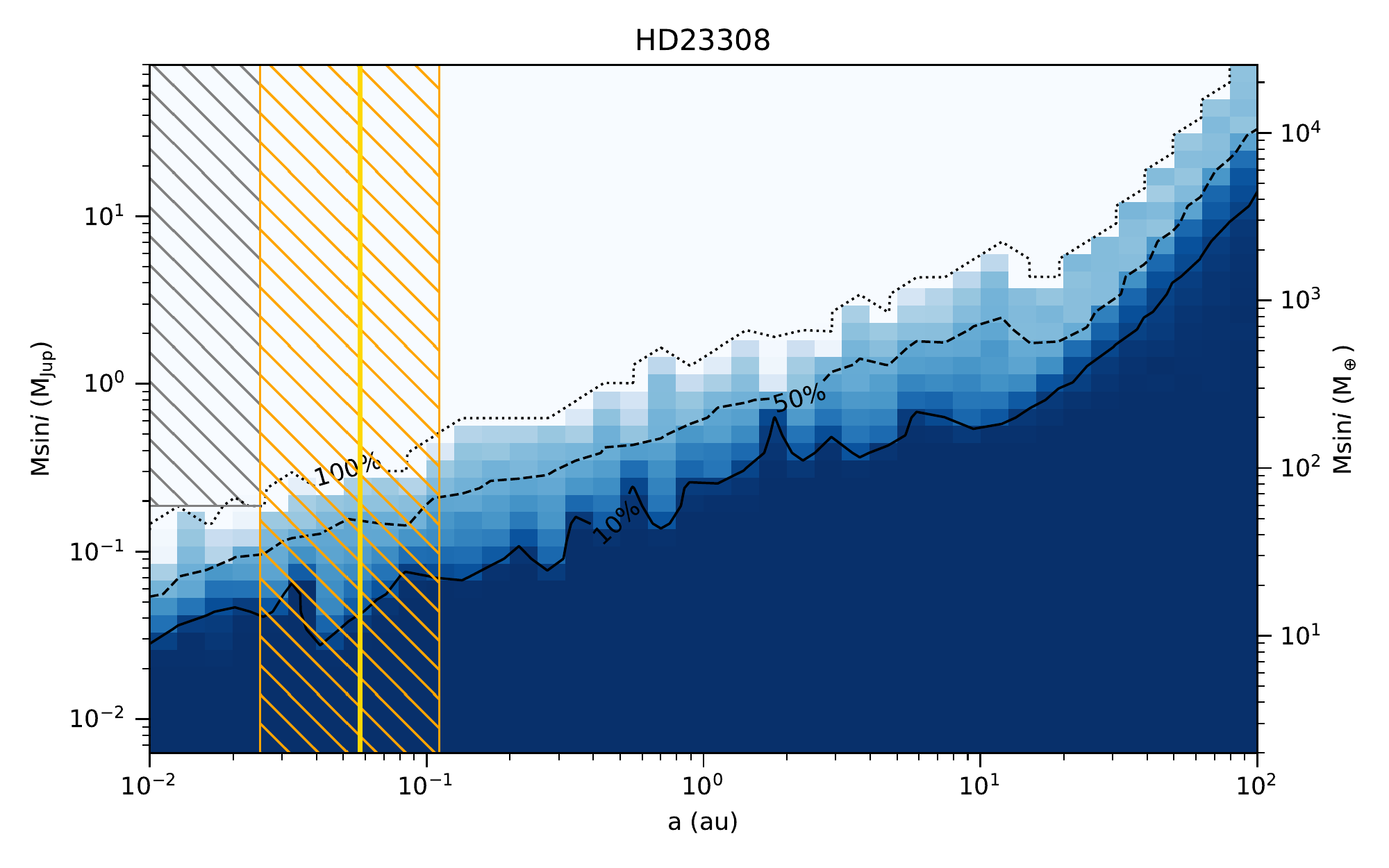}&
    		\includegraphics[width=0.22\linewidth]{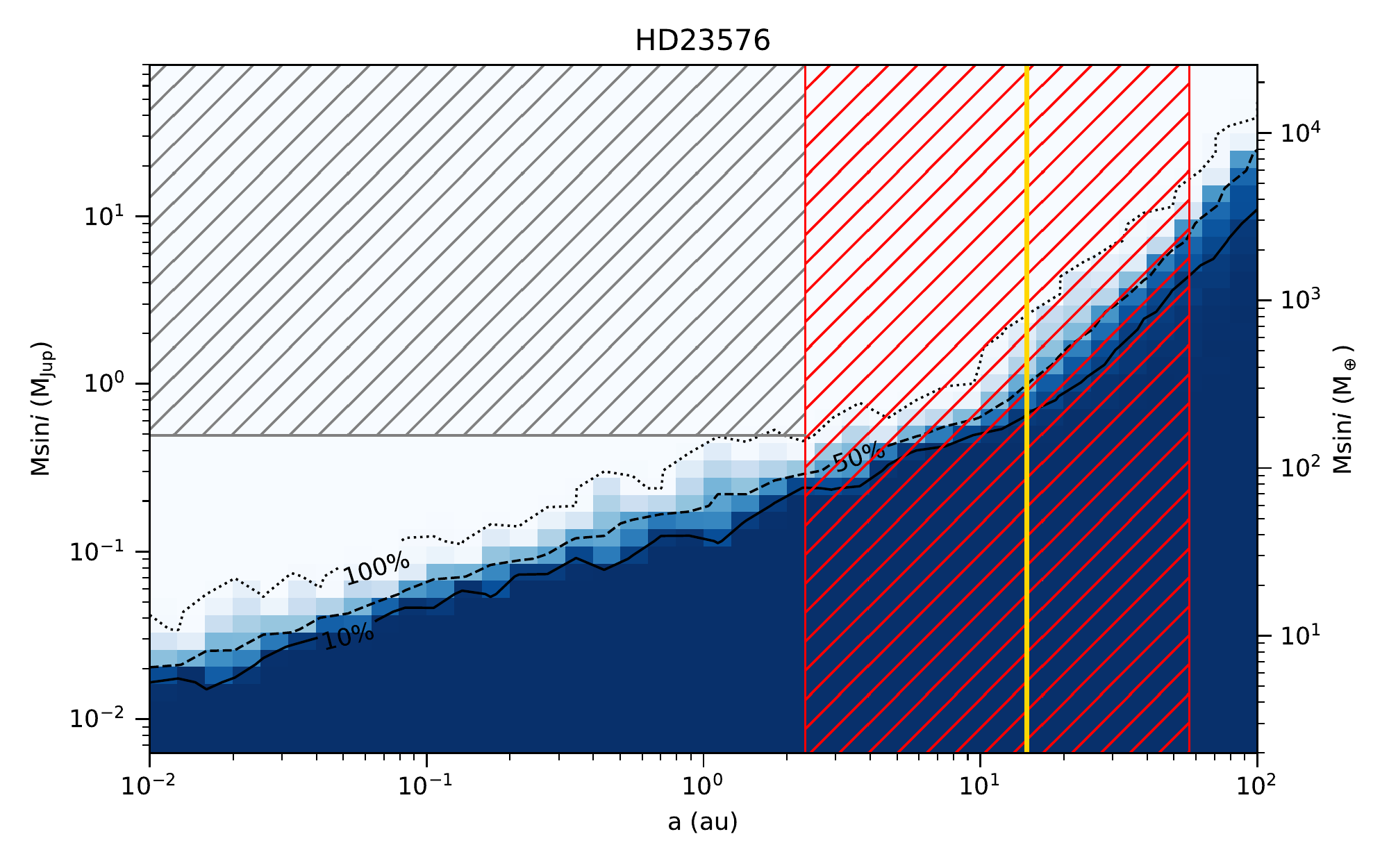}\\
    
    		\includegraphics[width=0.22\linewidth]{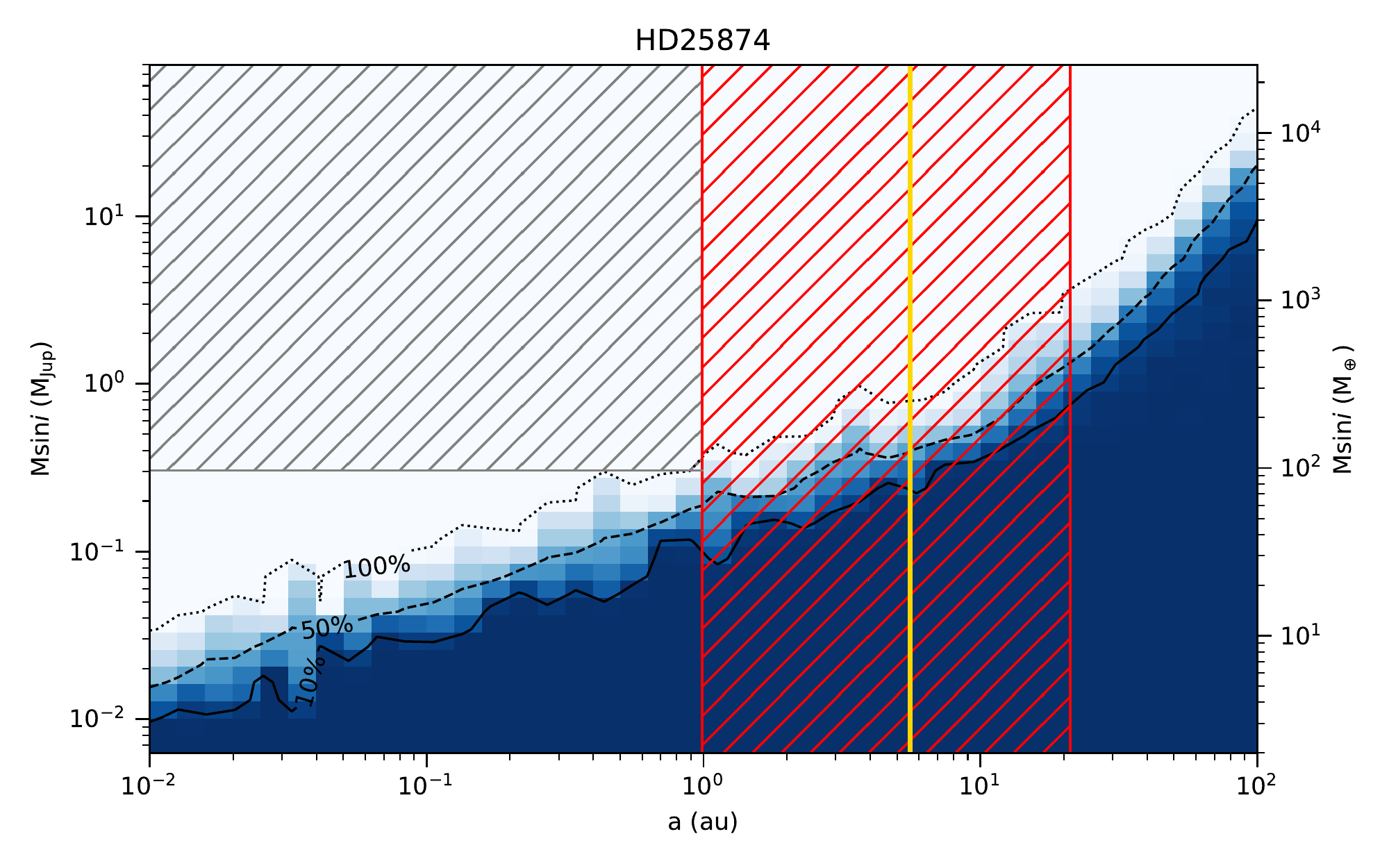}&
    		\includegraphics[width=0.22\linewidth]{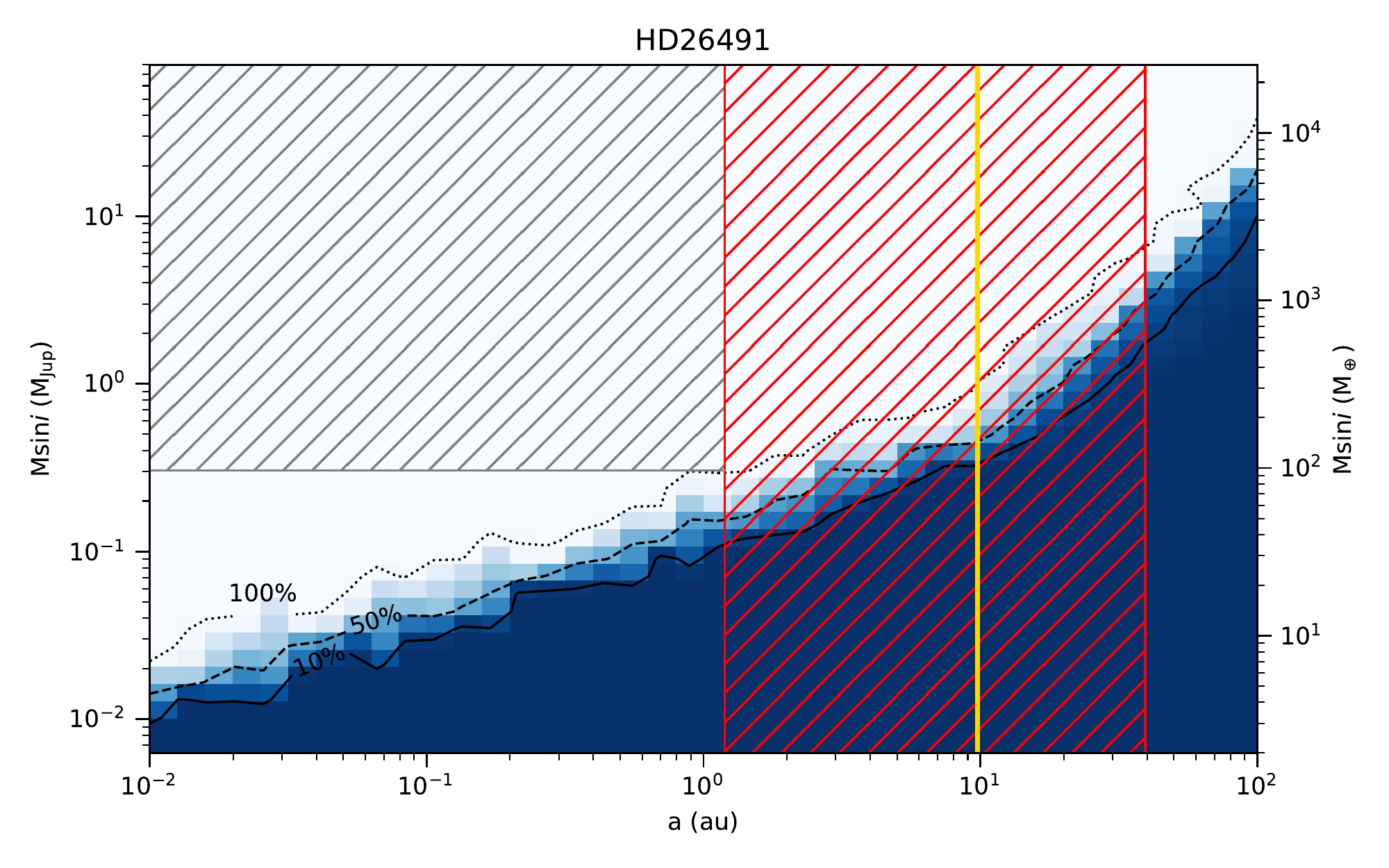}&
    		\includegraphics[width=0.22\linewidth]{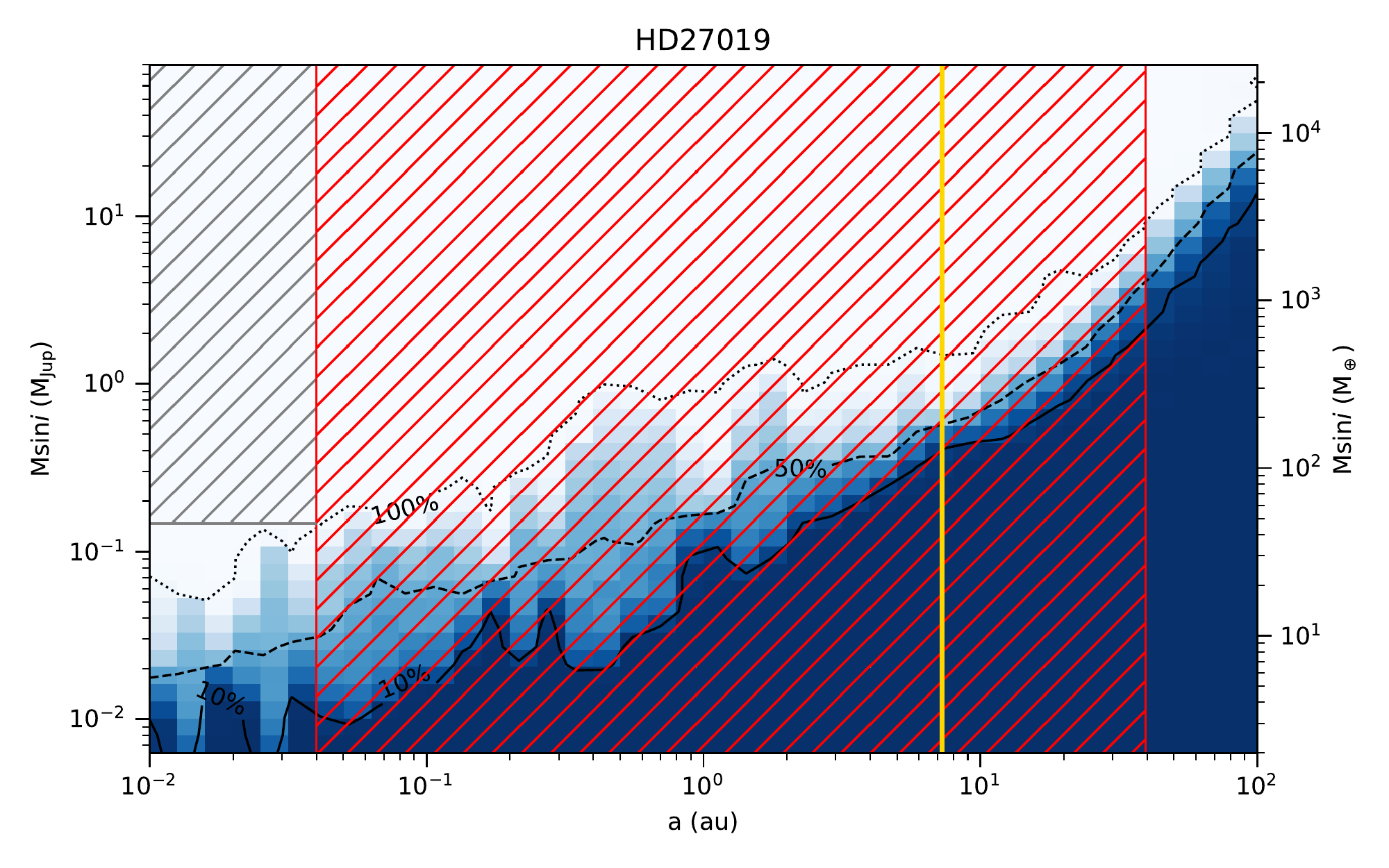}&
    		\includegraphics[width=0.22\linewidth]{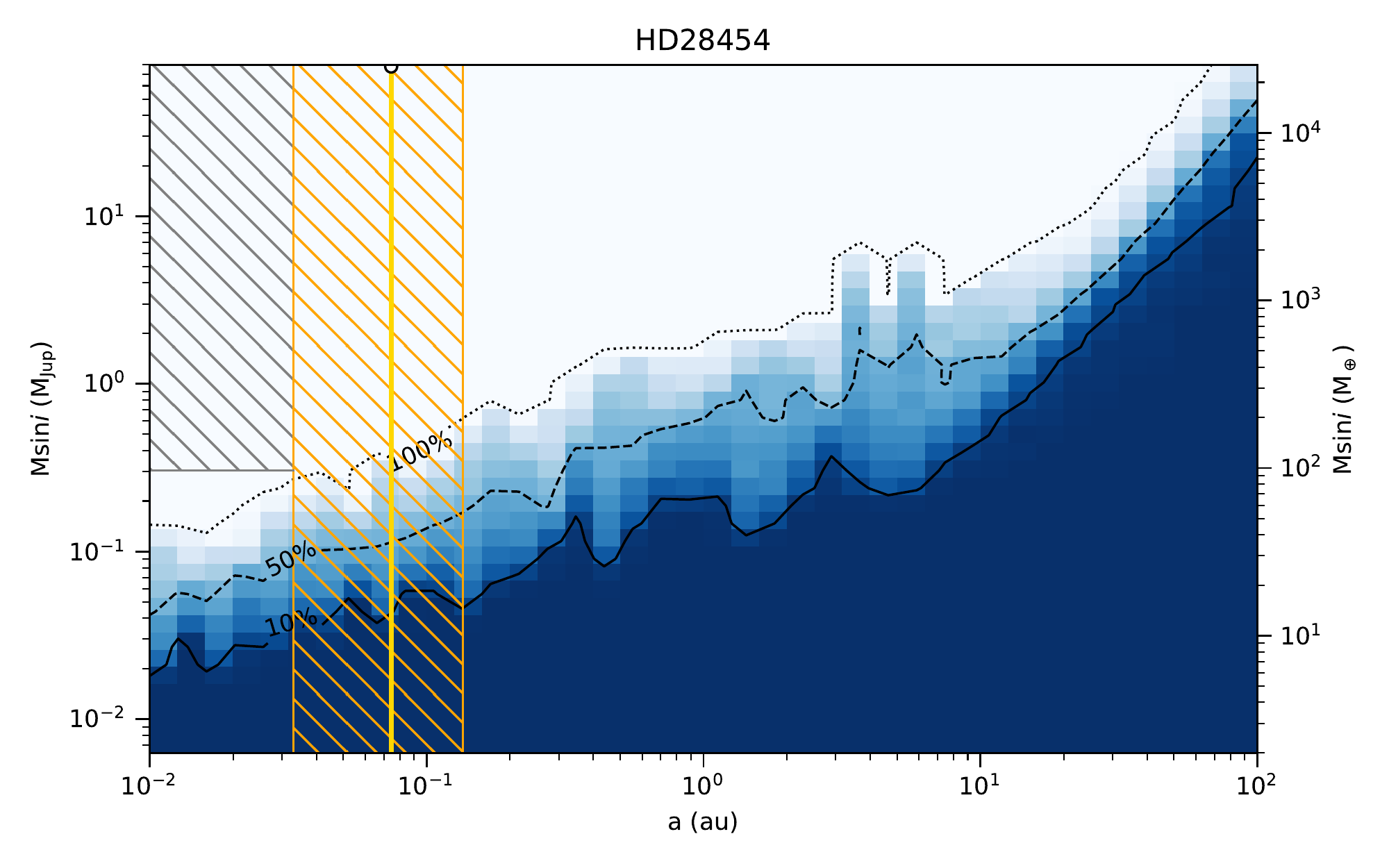}\\
    
    		\includegraphics[width=0.22\linewidth]{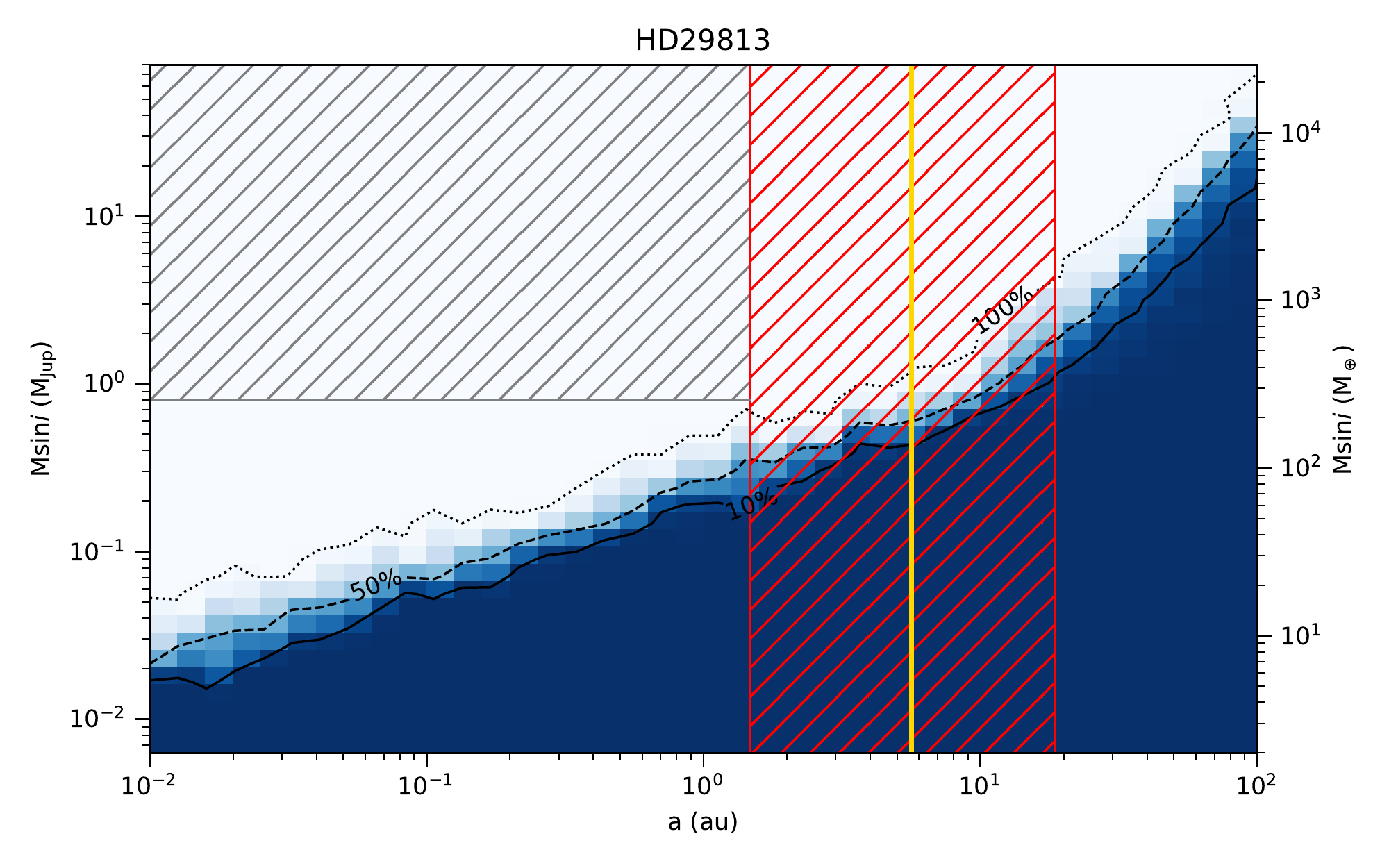}&
    		\includegraphics[width=0.22\linewidth]{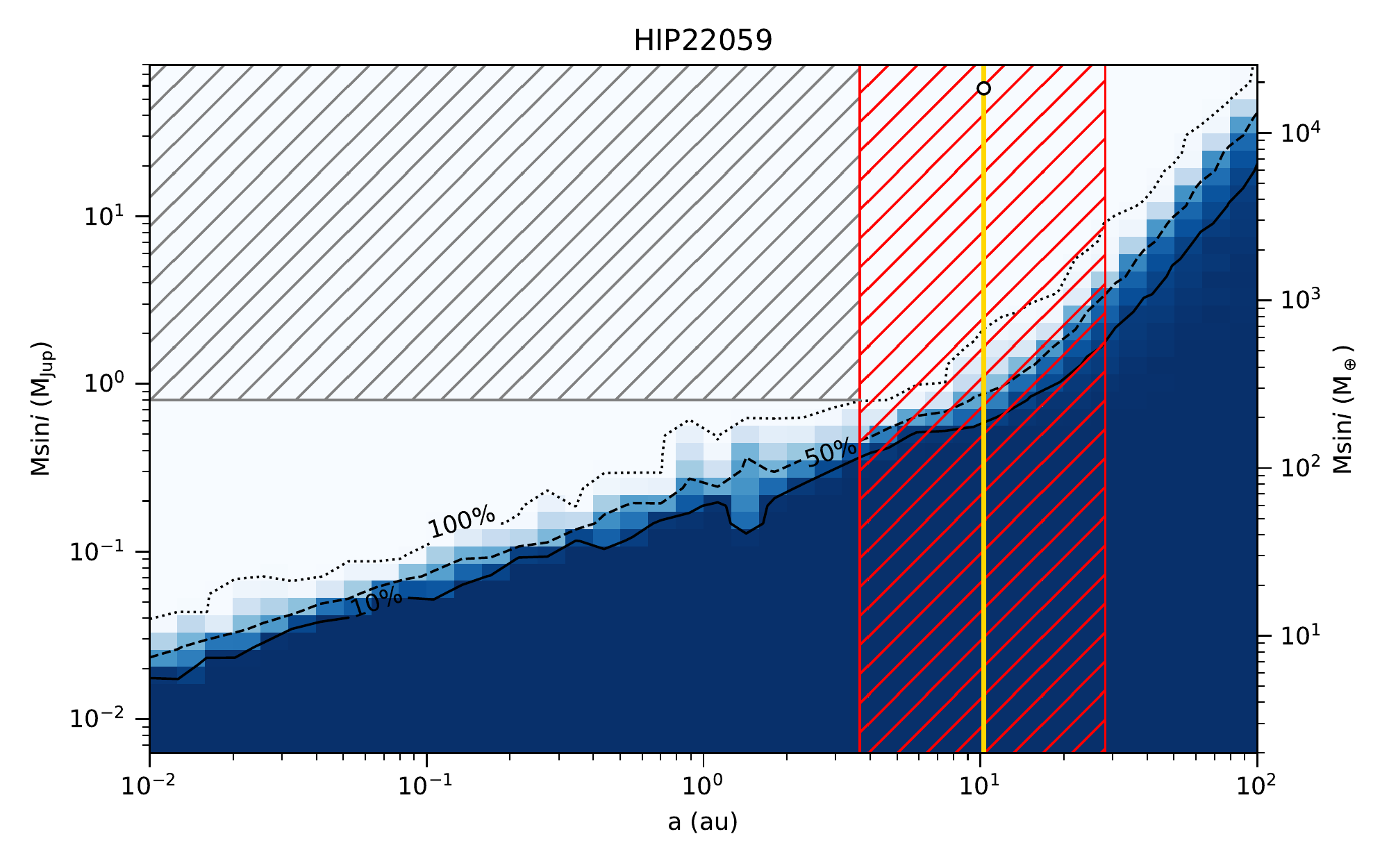}&
    		\includegraphics[width=0.22\linewidth]{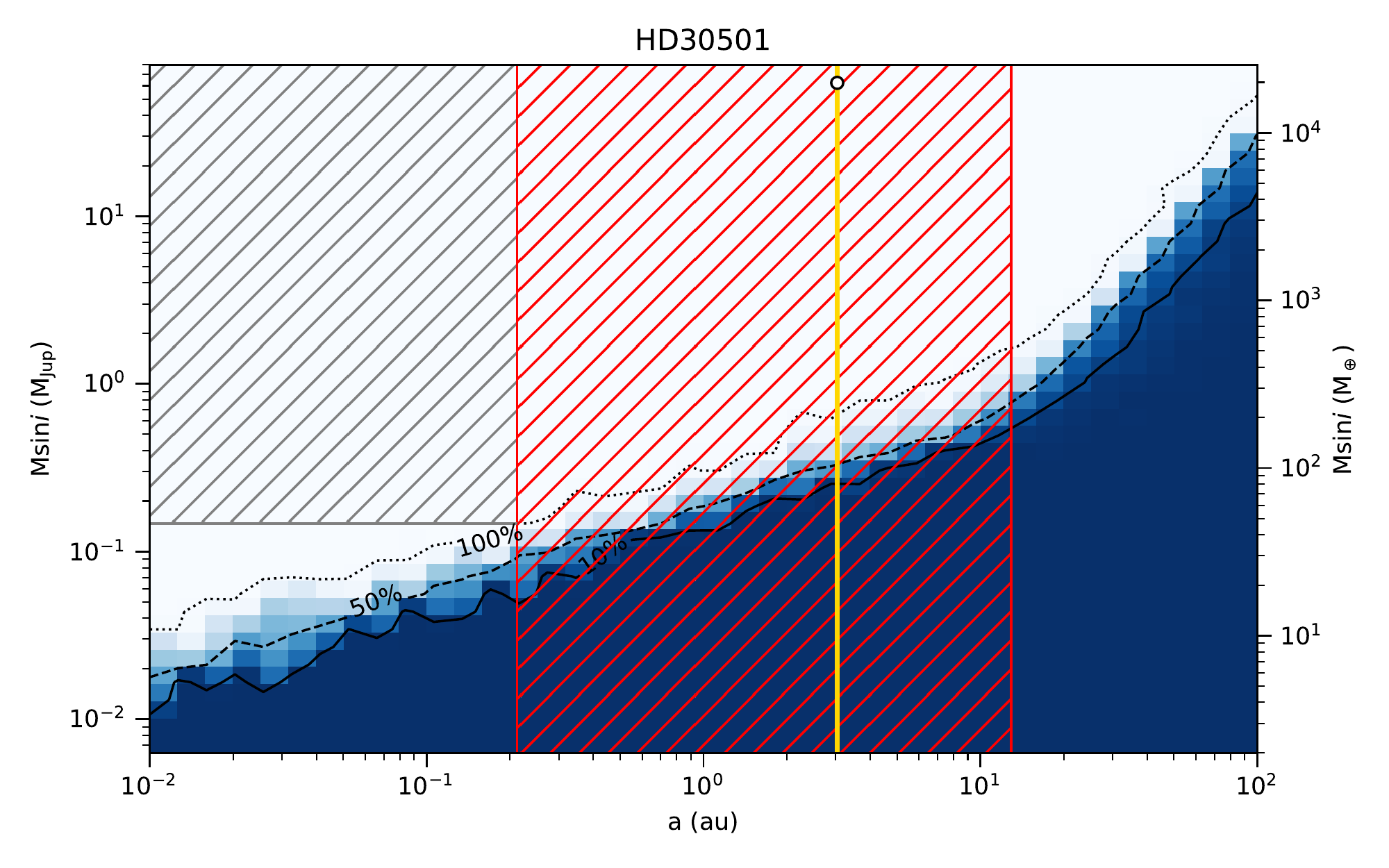}&
    		\includegraphics[width=0.22\linewidth]{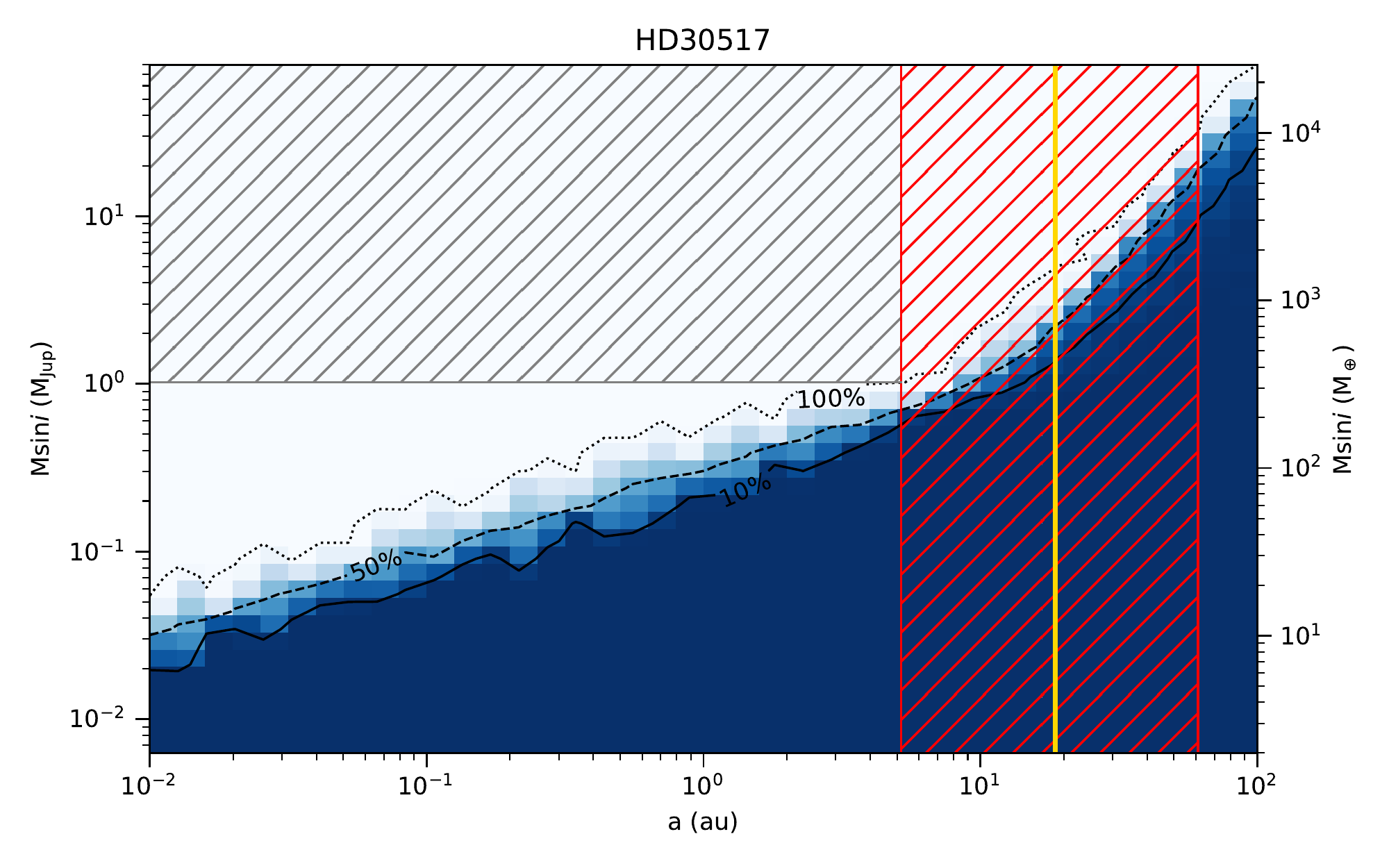}\\
    
    		\includegraphics[width=0.22\linewidth]{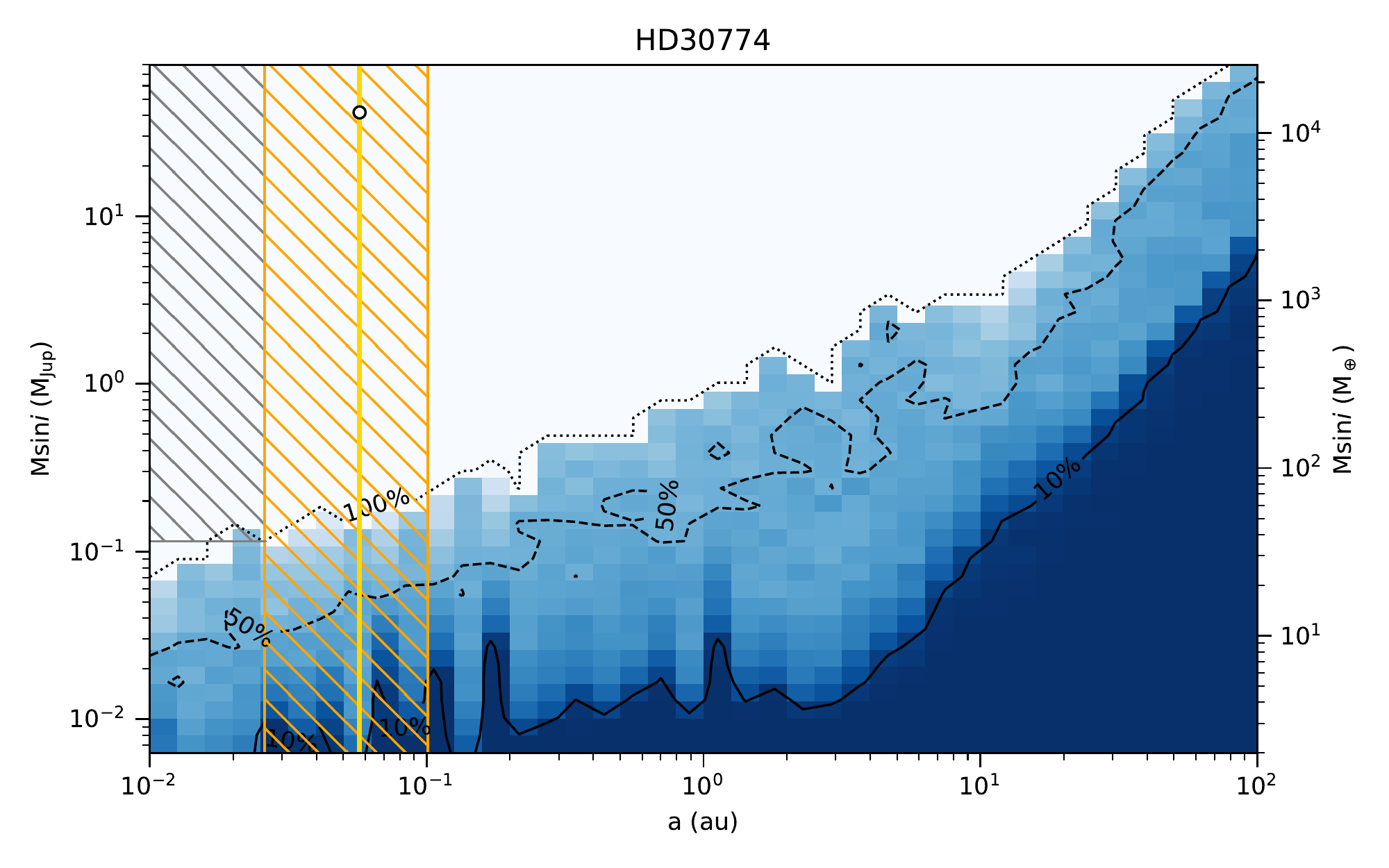}&
    		\includegraphics[width=0.22\linewidth]{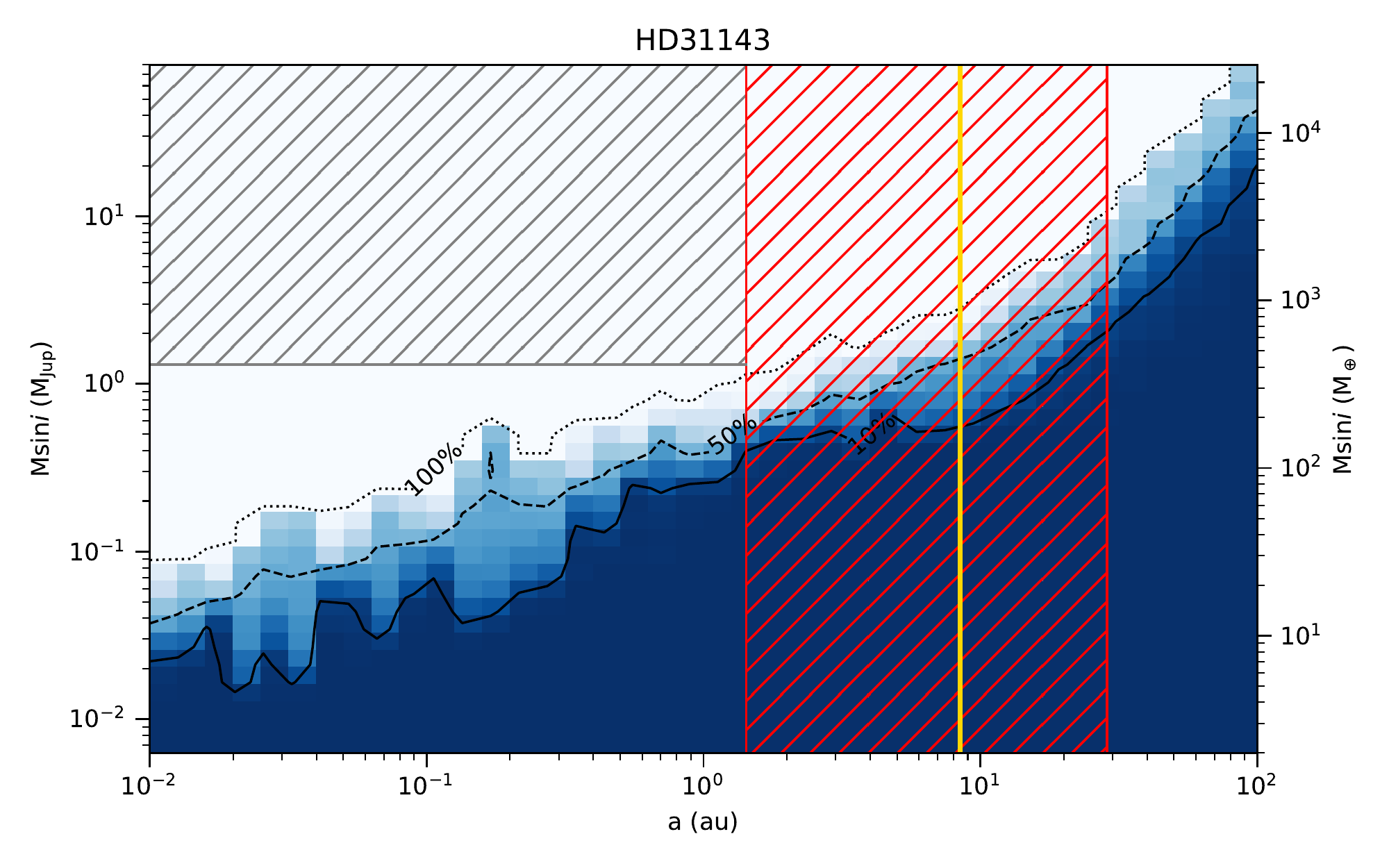}&
    		\includegraphics[width=0.22\linewidth]{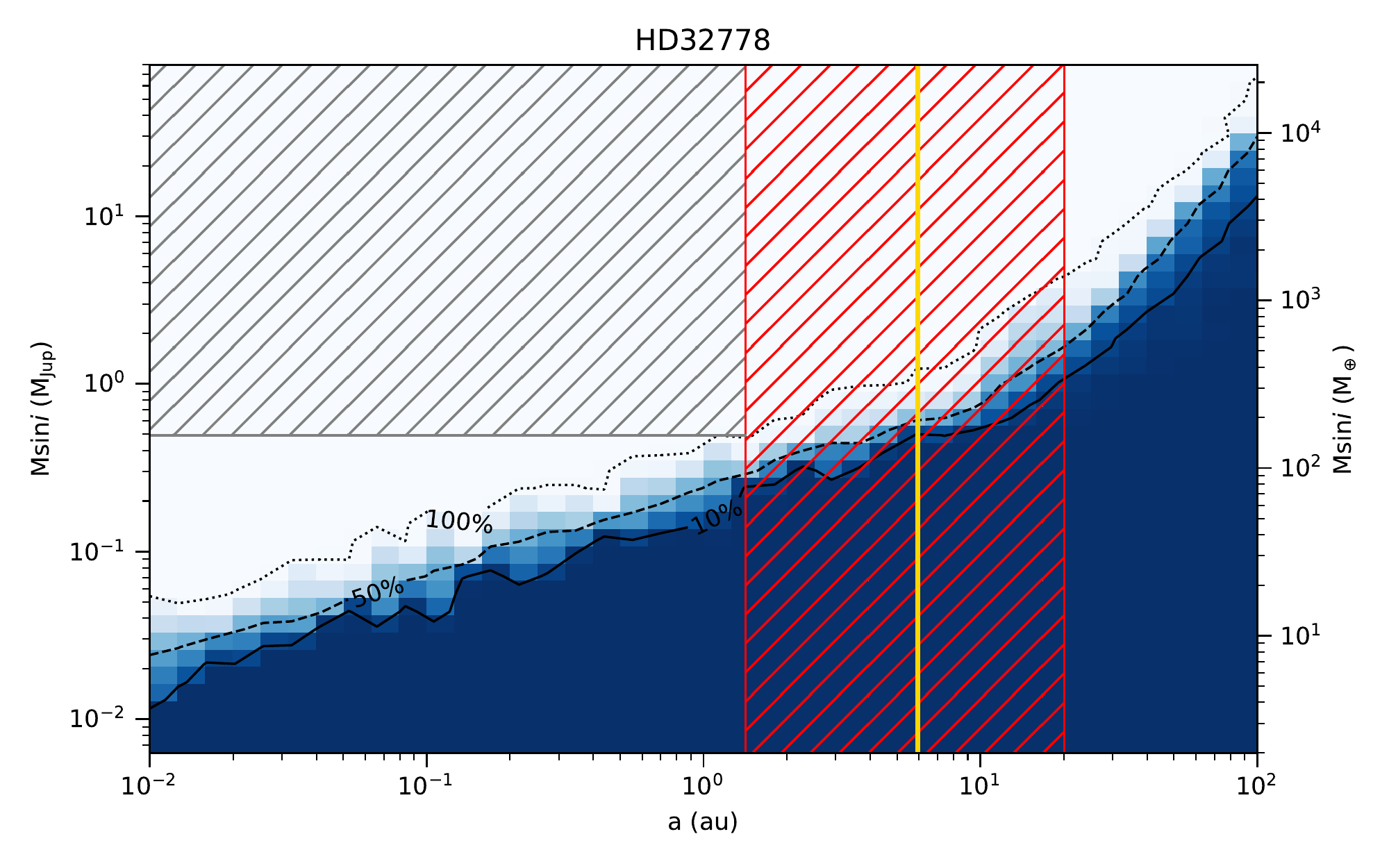}&
    		\includegraphics[width=0.22\linewidth]{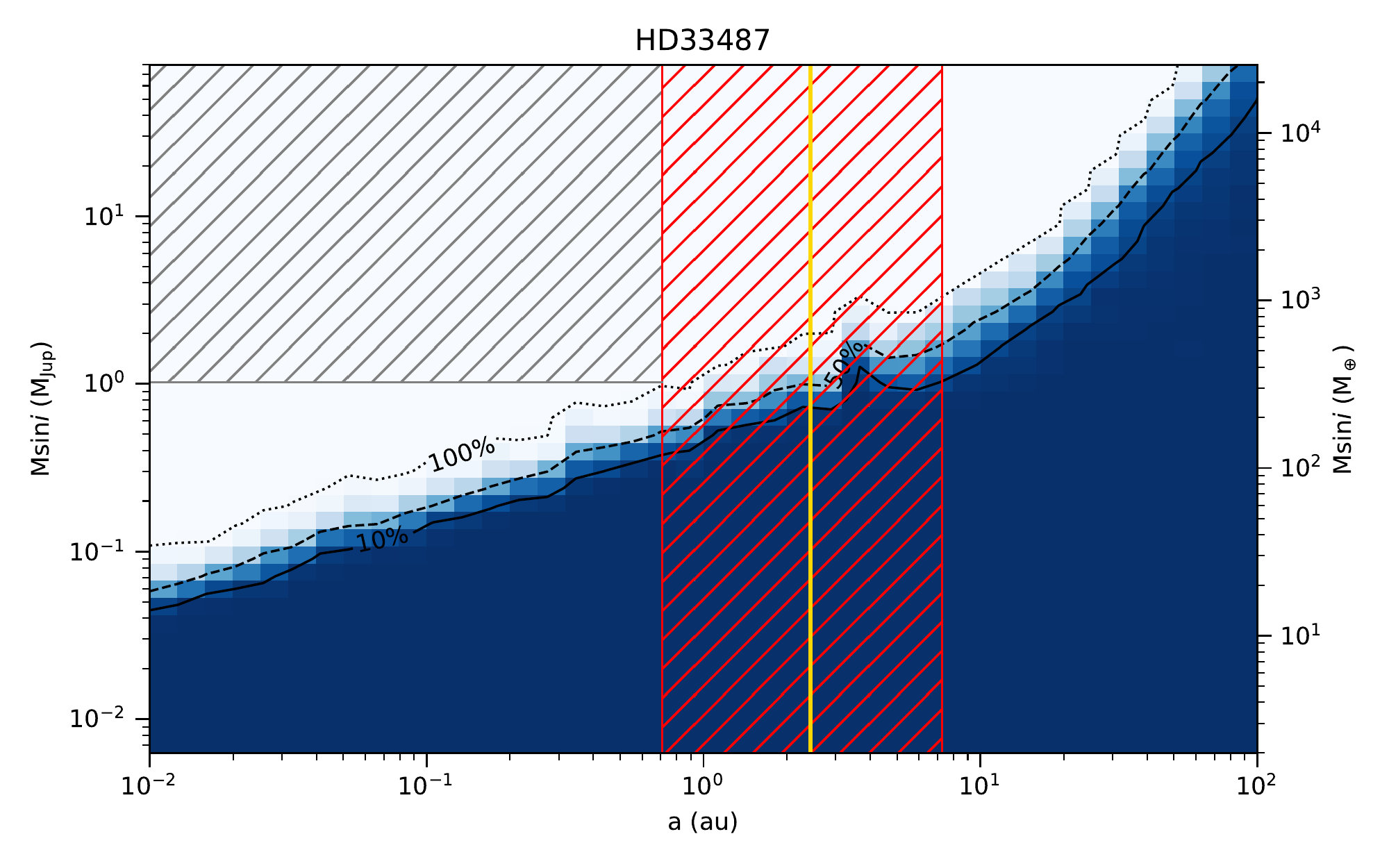}\\
    
    		\includegraphics[width=0.22\linewidth]{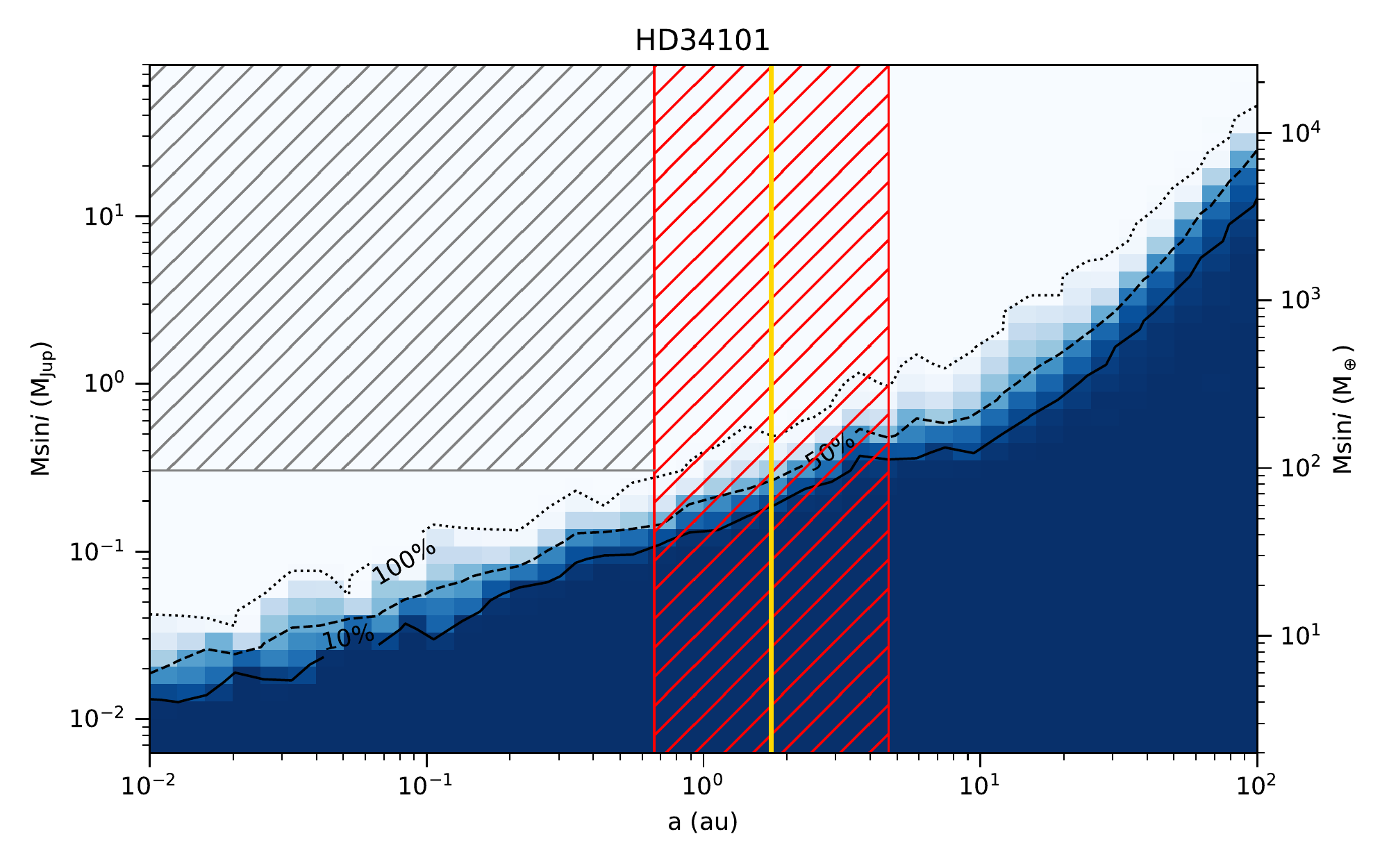}&
    		\includegraphics[width=0.22\linewidth]{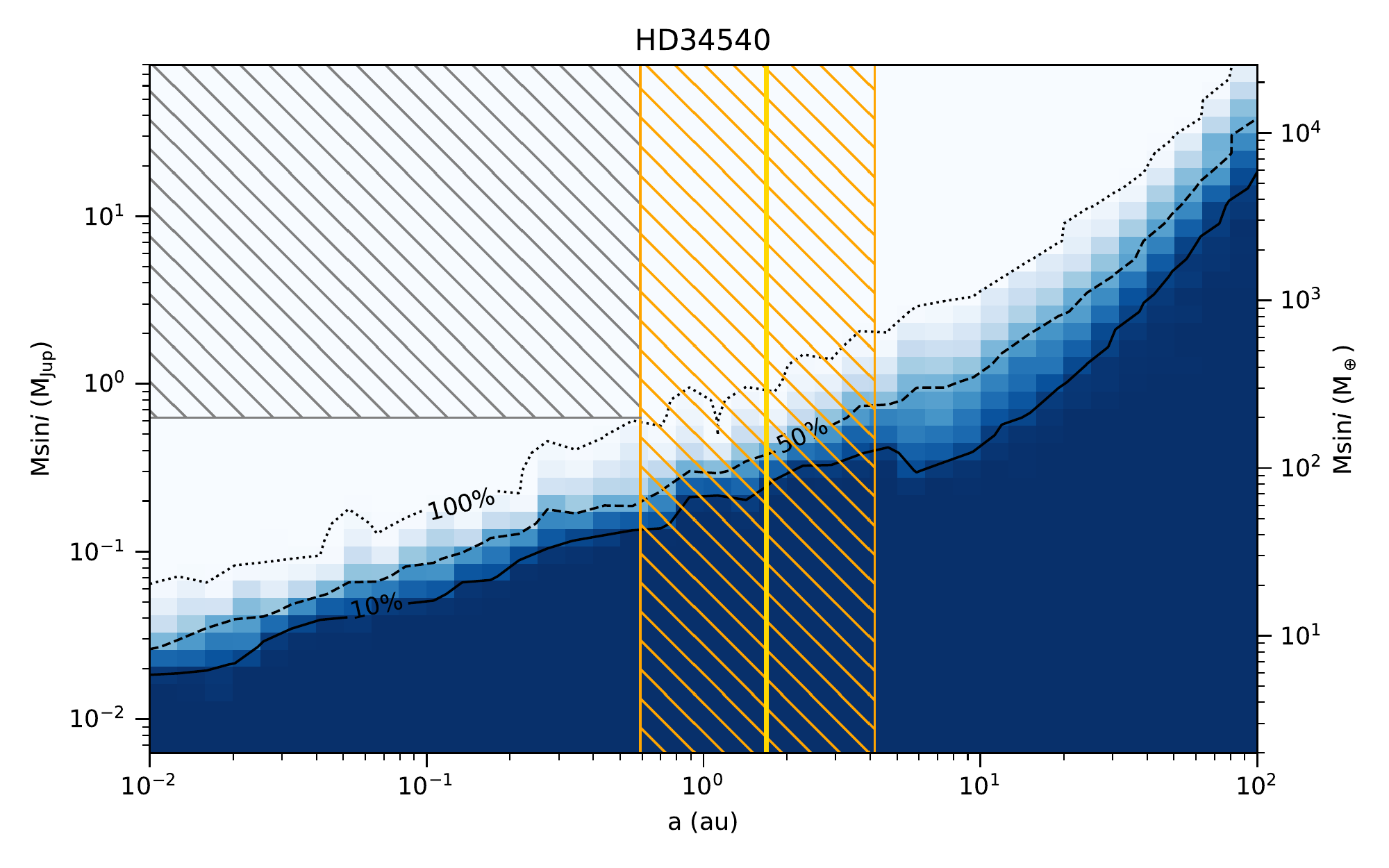}&
    		\includegraphics[width=0.22\linewidth]{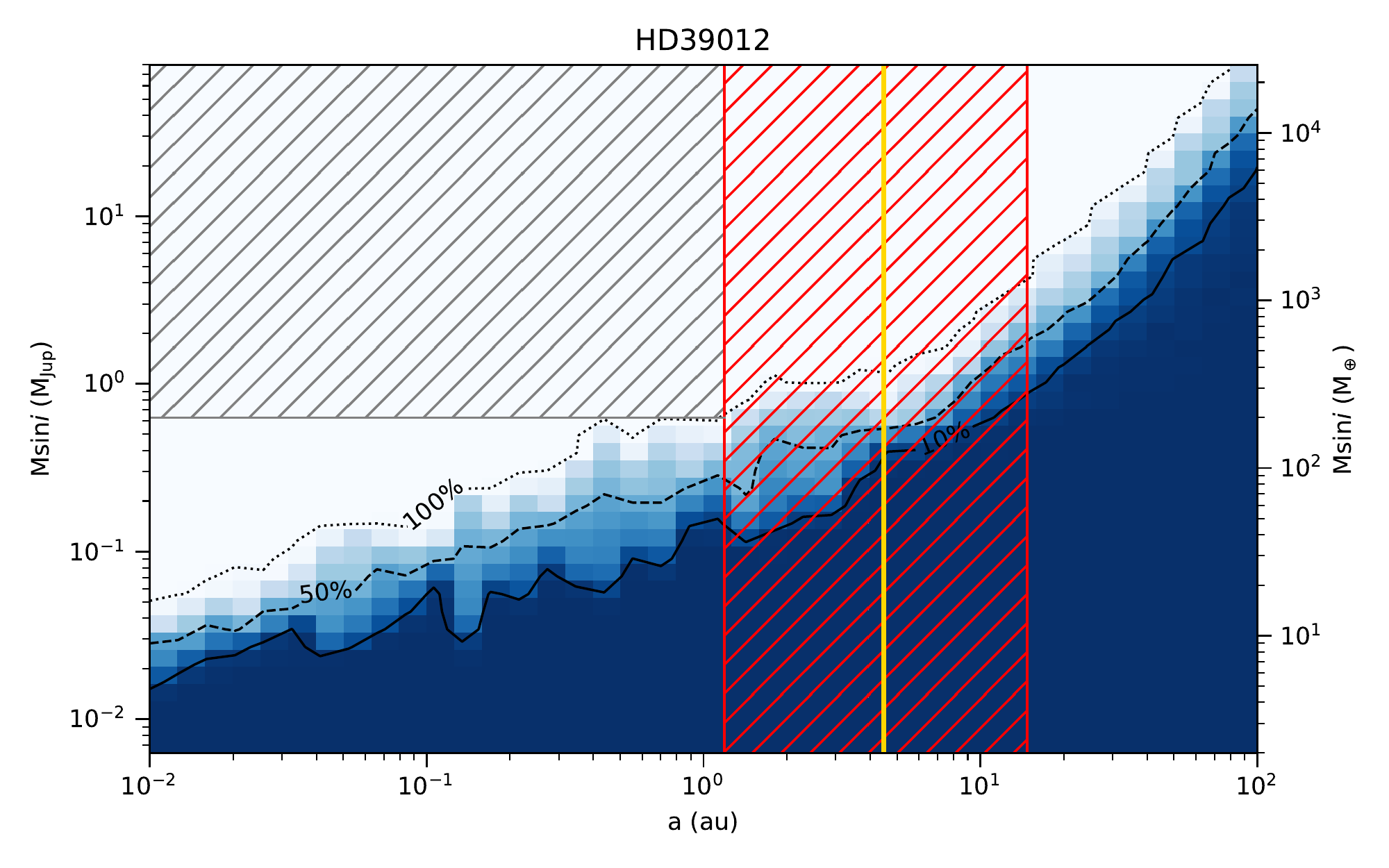}&
    		\includegraphics[width=0.22\linewidth]{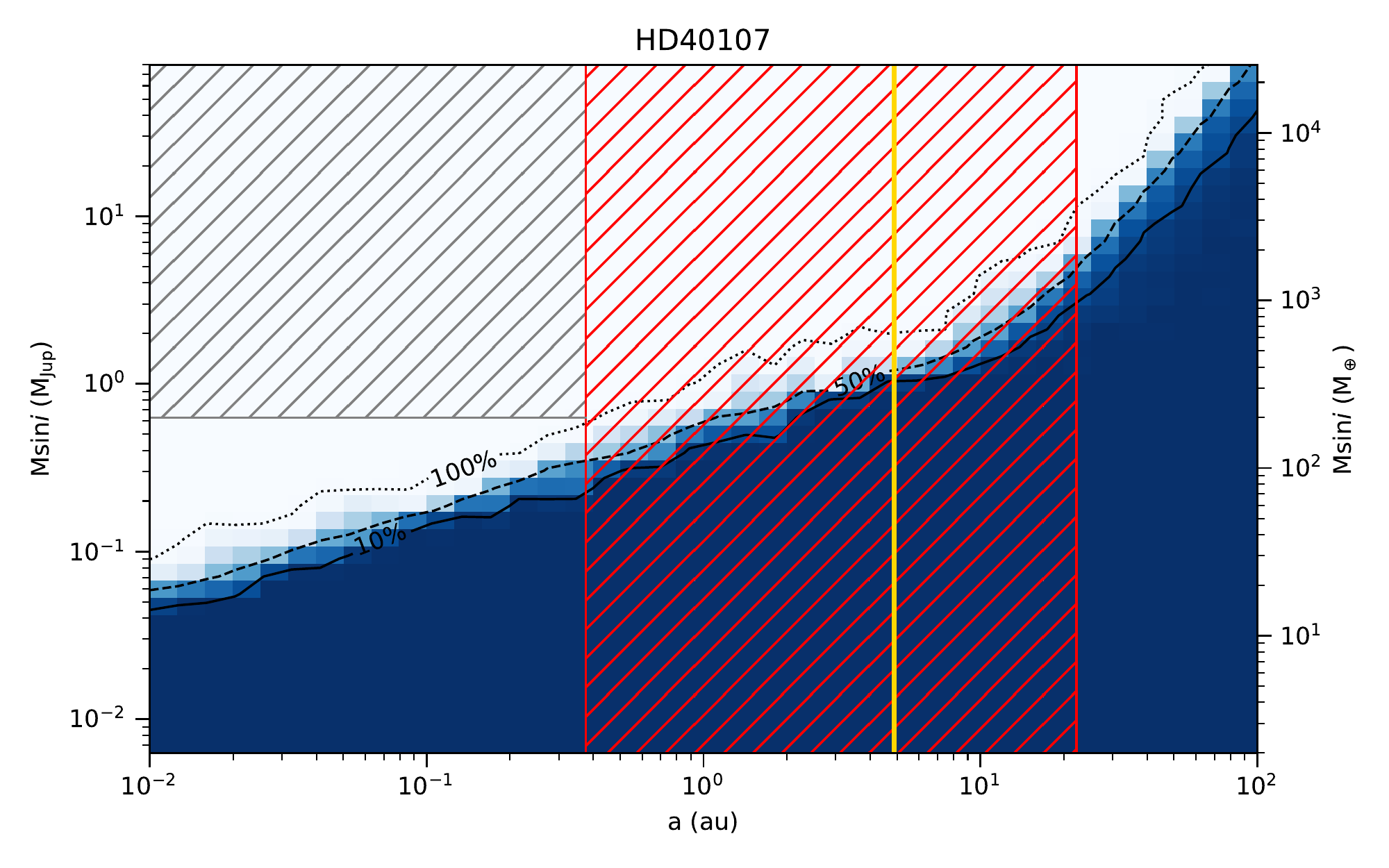}\\
    
    		\includegraphics[width=0.22\linewidth]{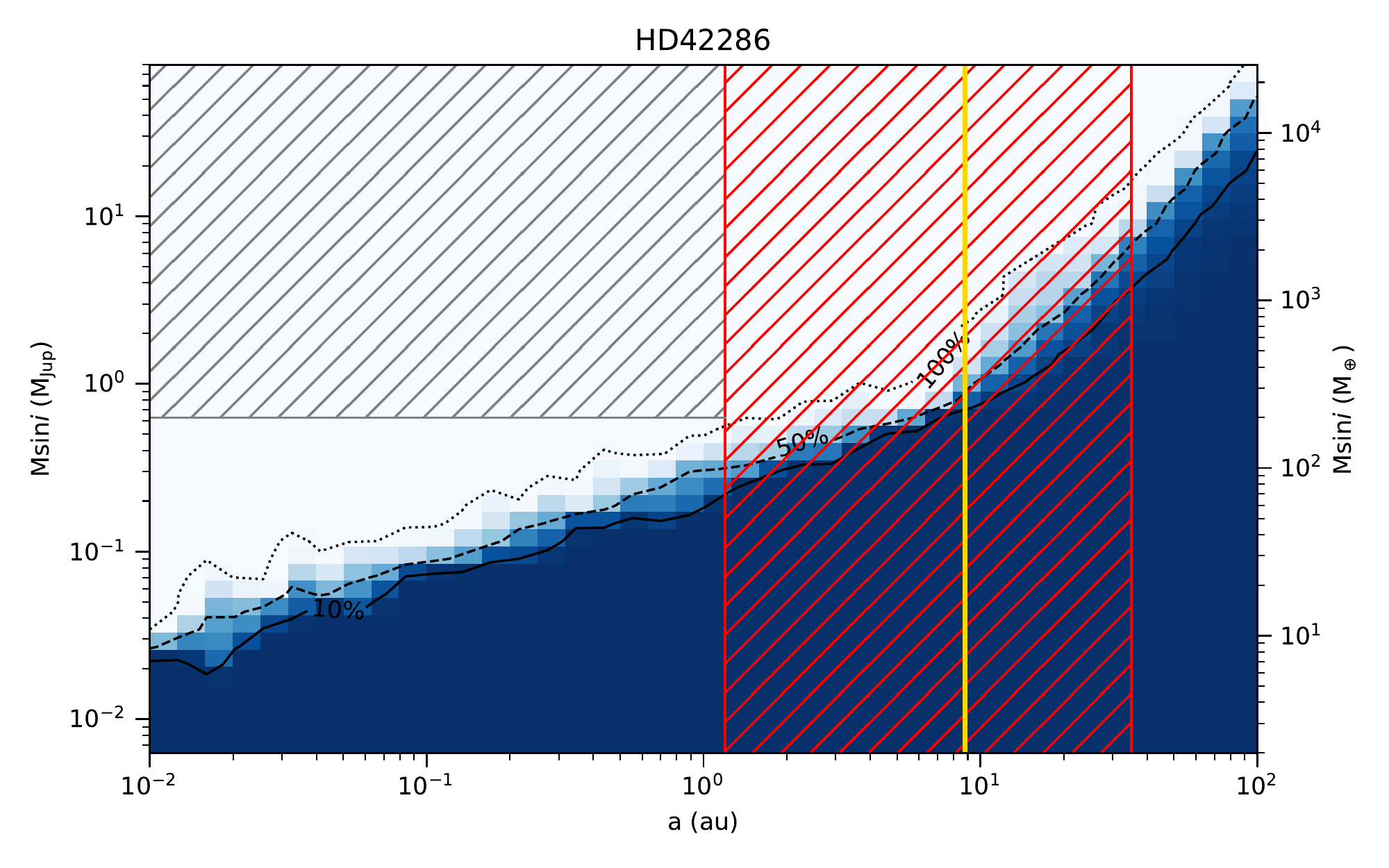}&
    		\includegraphics[width=0.22\linewidth]{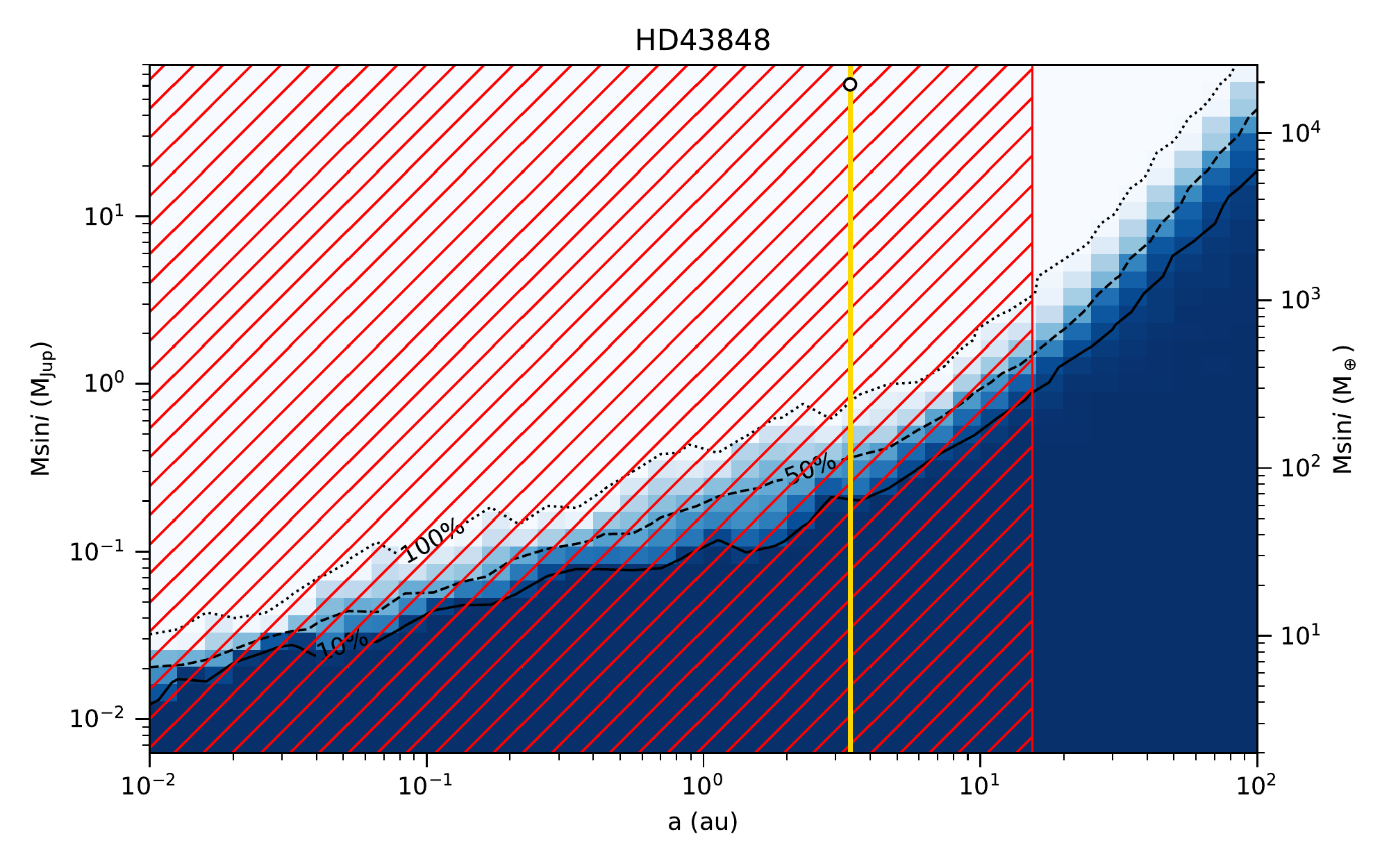}&
    		\includegraphics[width=0.22\linewidth]{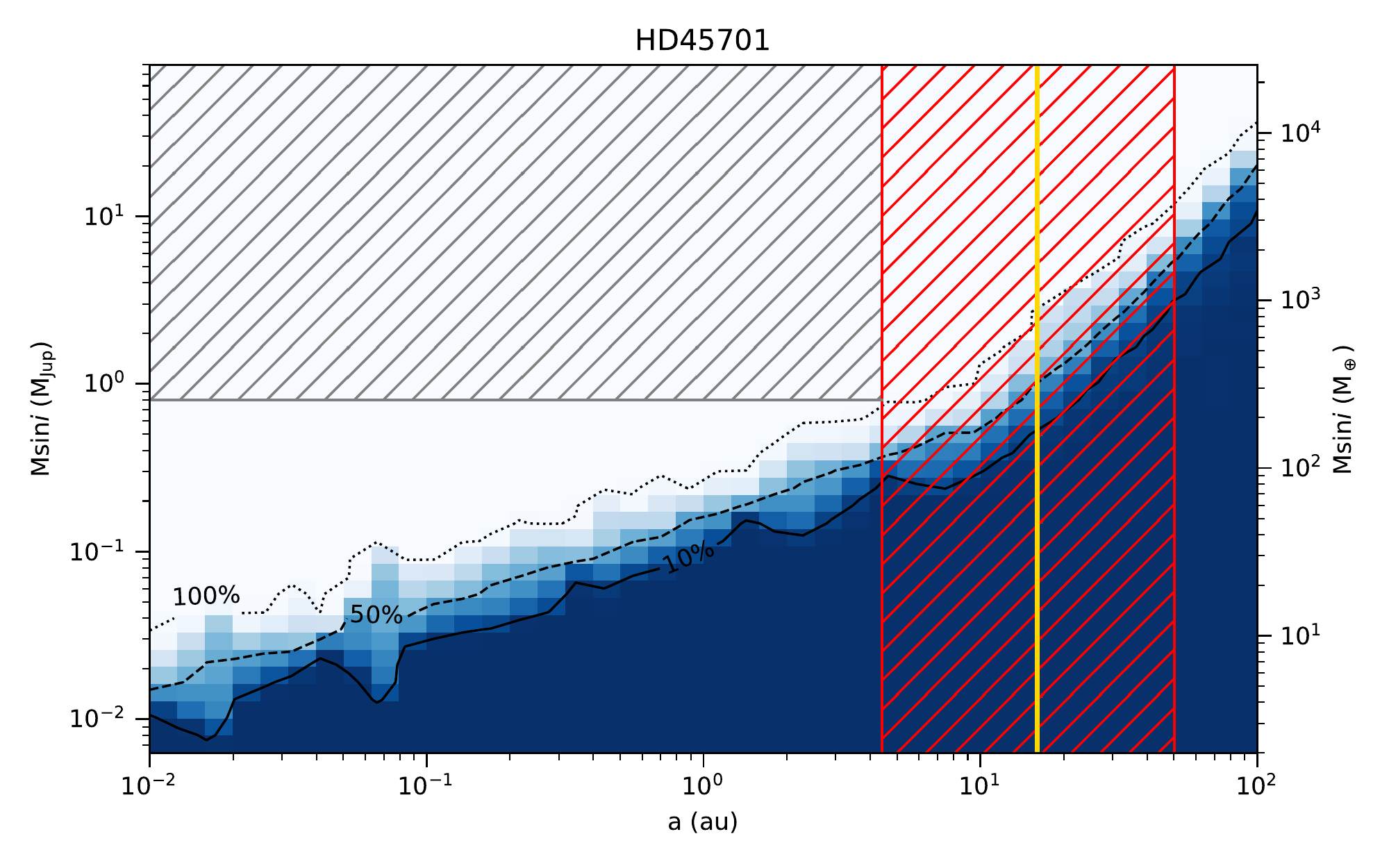}&
    		\includegraphics[width=0.22\linewidth]{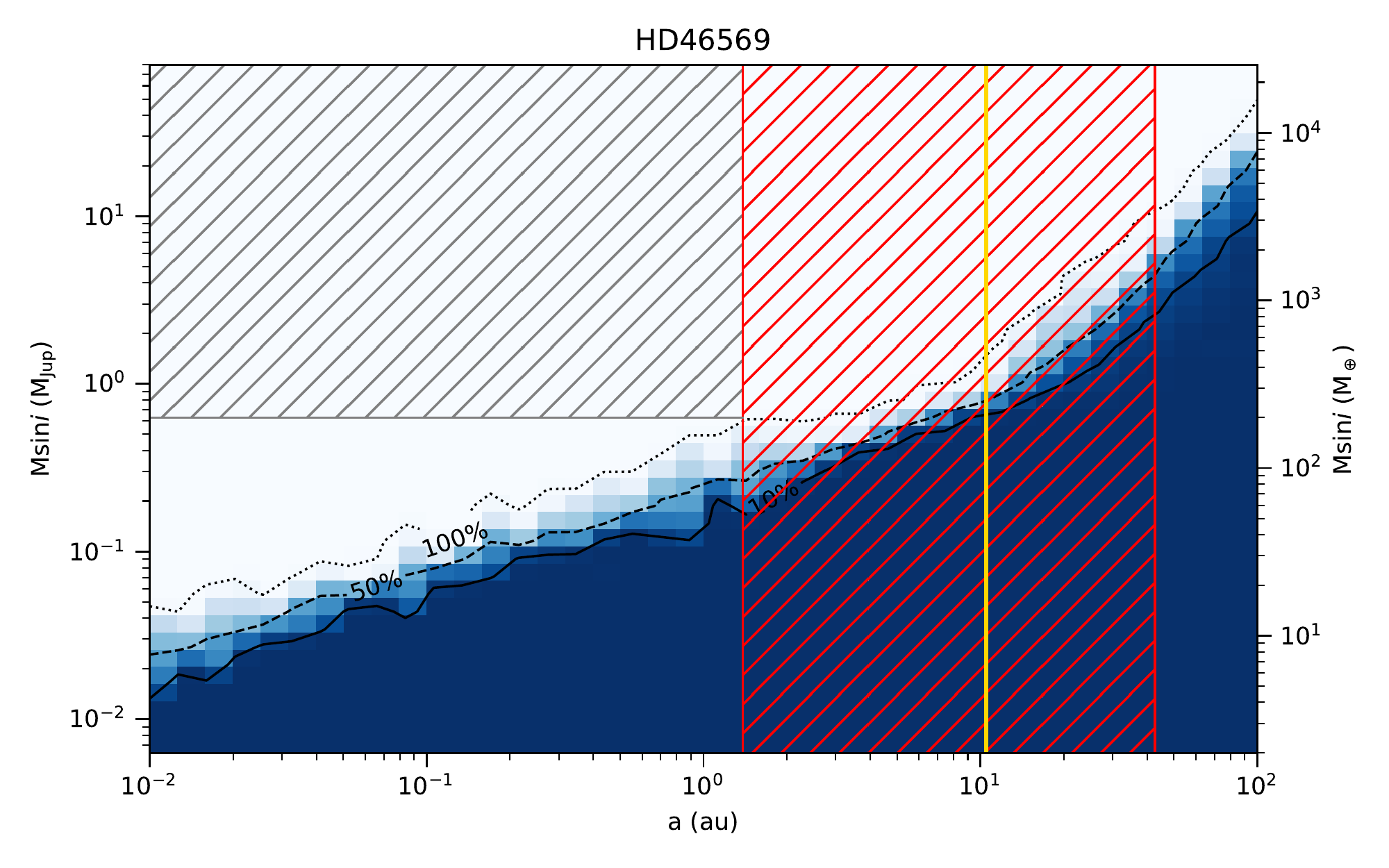}\\
    
    		\includegraphics[width=0.22\linewidth]{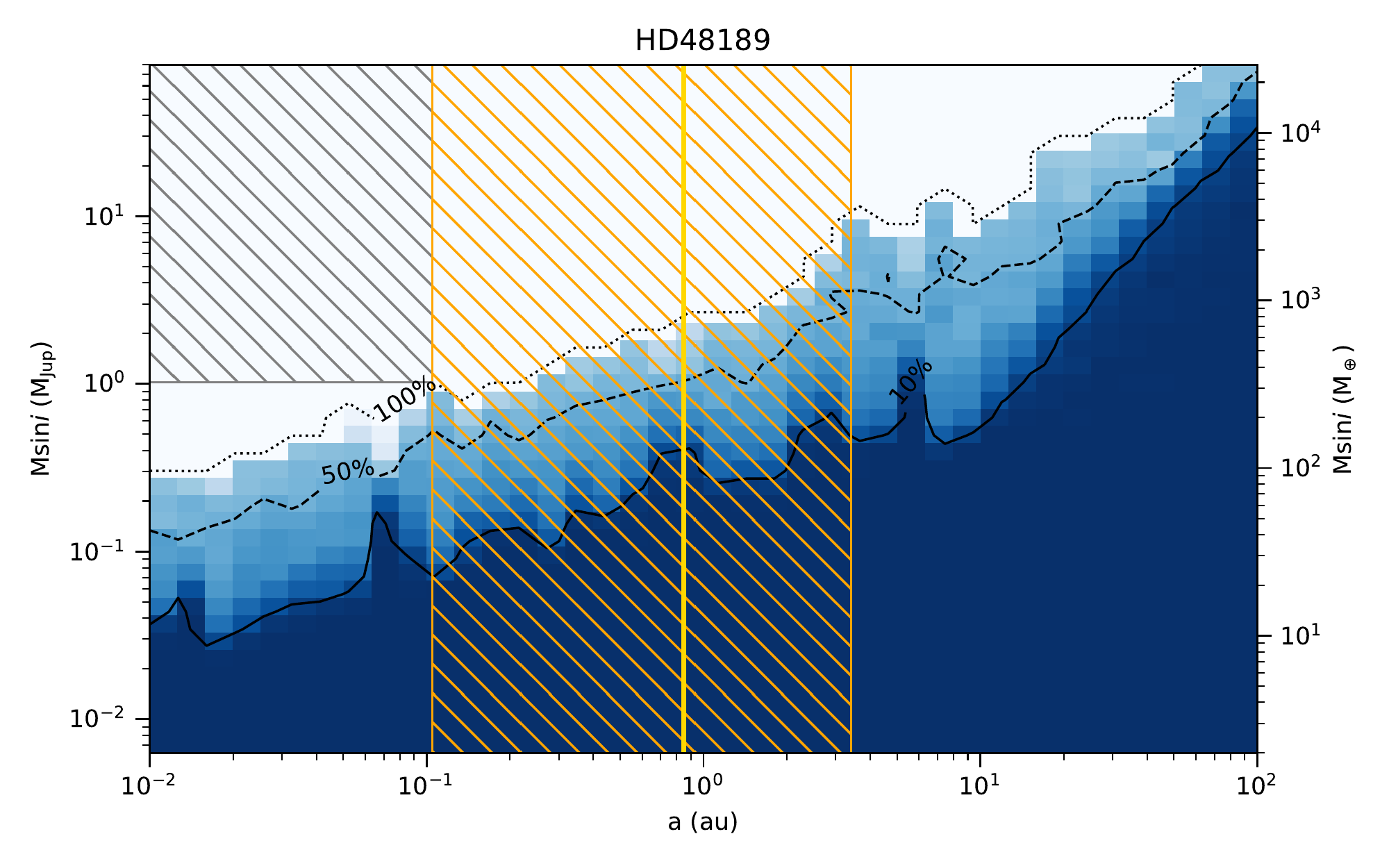}&
    		\includegraphics[width=0.22\linewidth]{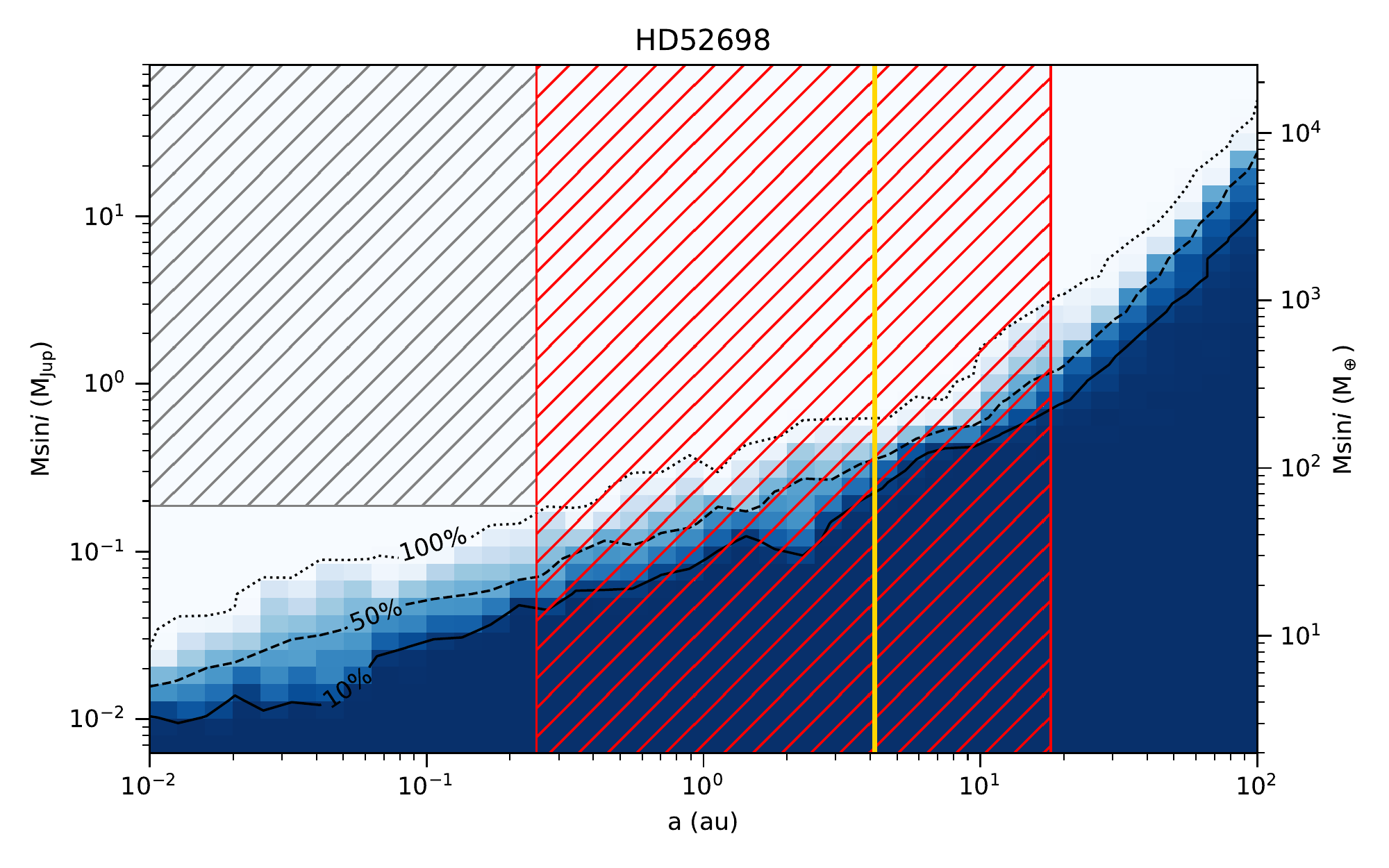}&
    		\includegraphics[width=0.22\linewidth]{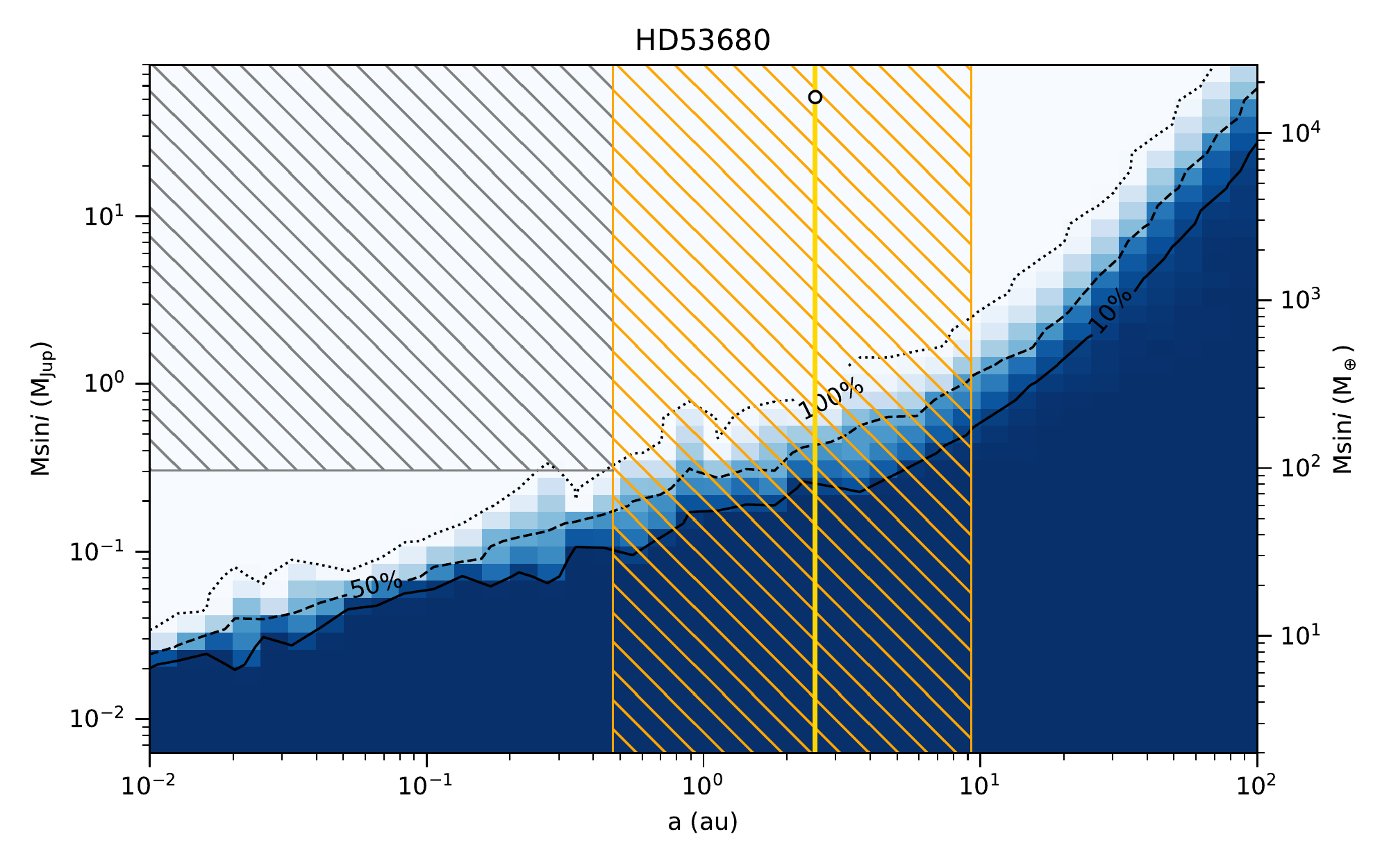}&
    		\includegraphics[width=0.22\linewidth]{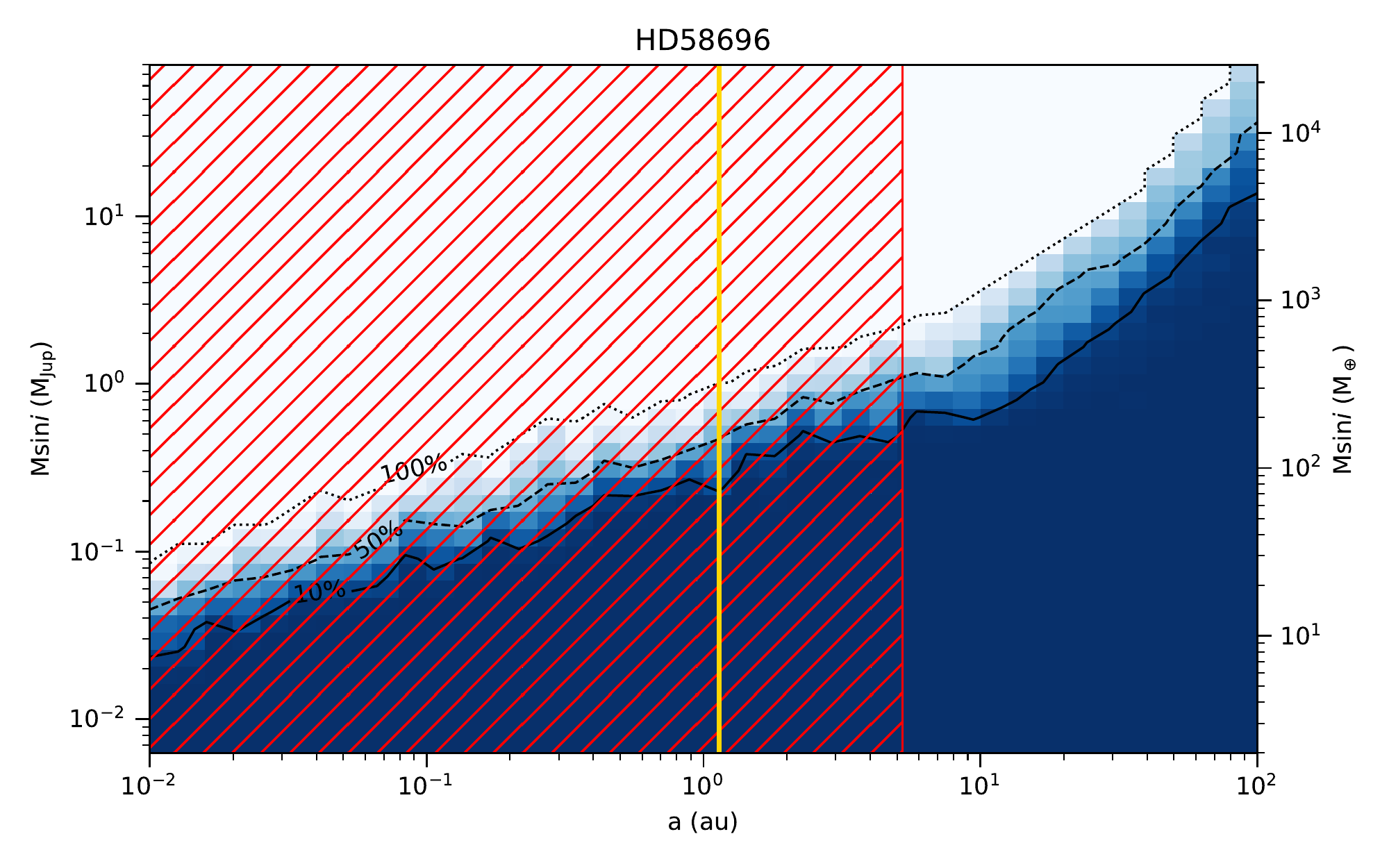}\\
    
    		\includegraphics[width=0.22\linewidth]{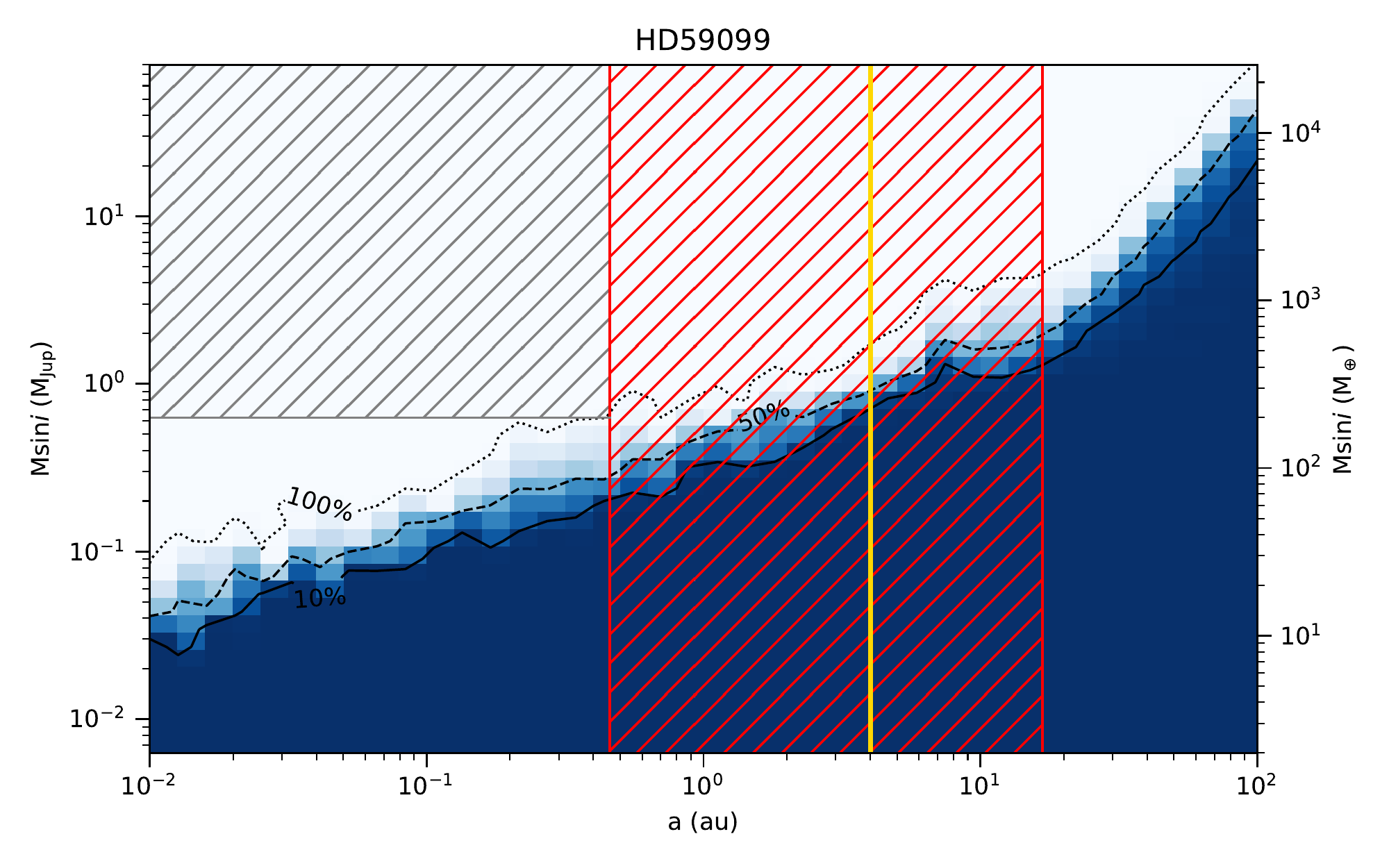}&
    		\includegraphics[width=0.22\linewidth]{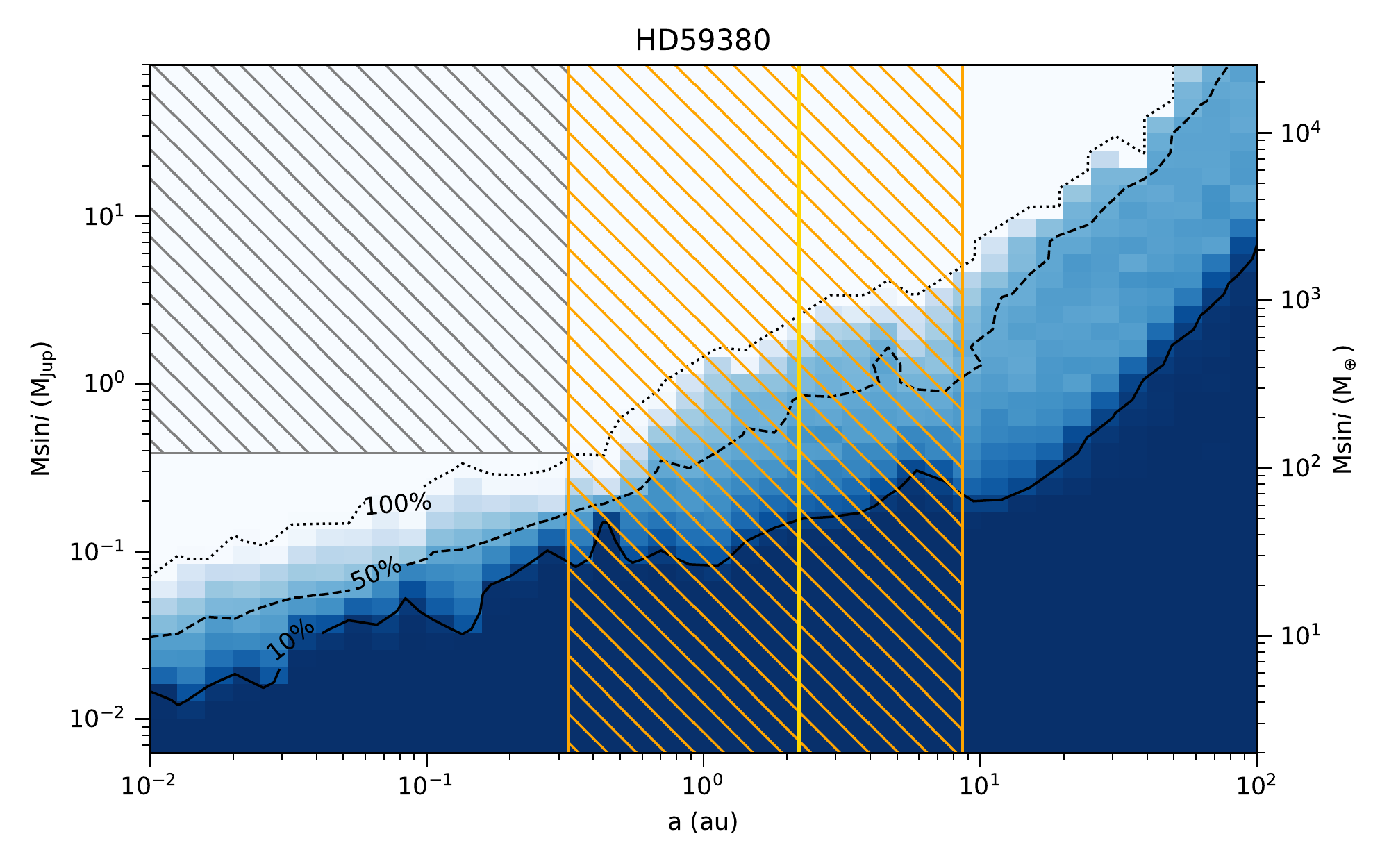}&
    		\includegraphics[width=0.22\linewidth]{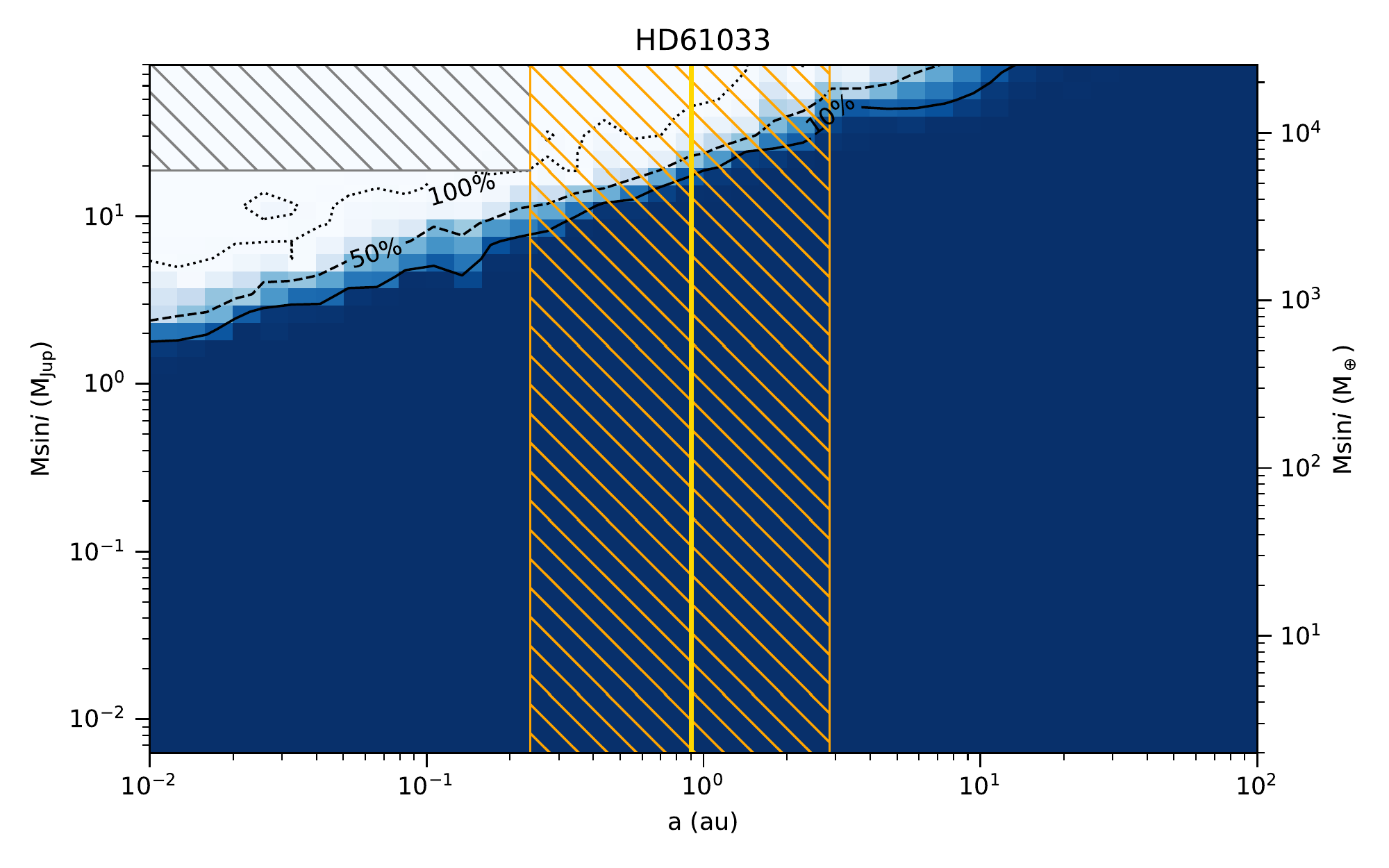}&
    		\includegraphics[width=0.22\linewidth]{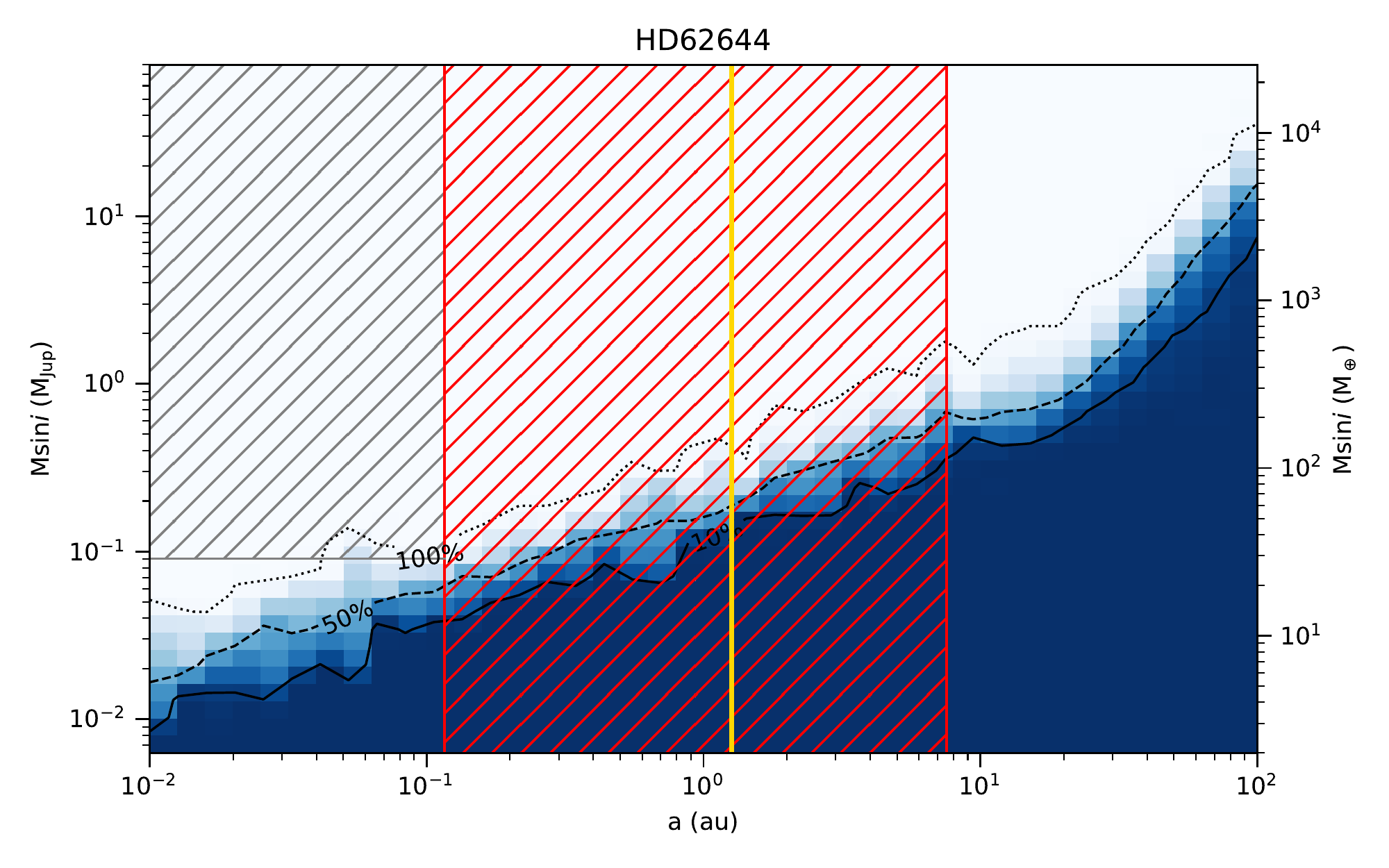}\\
    
    		\includegraphics[width=0.22\linewidth]{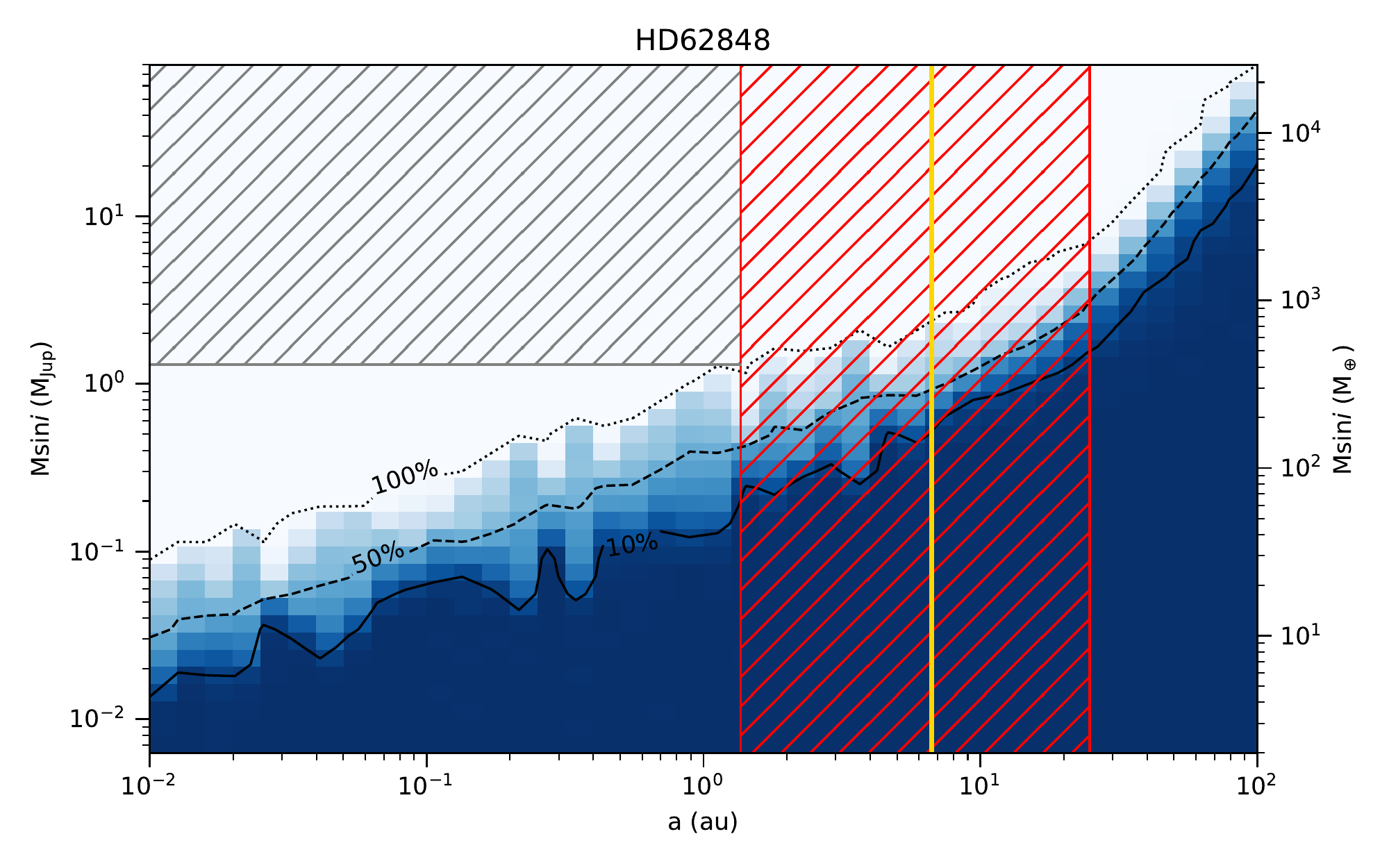}&
    		\includegraphics[width=0.22\linewidth]{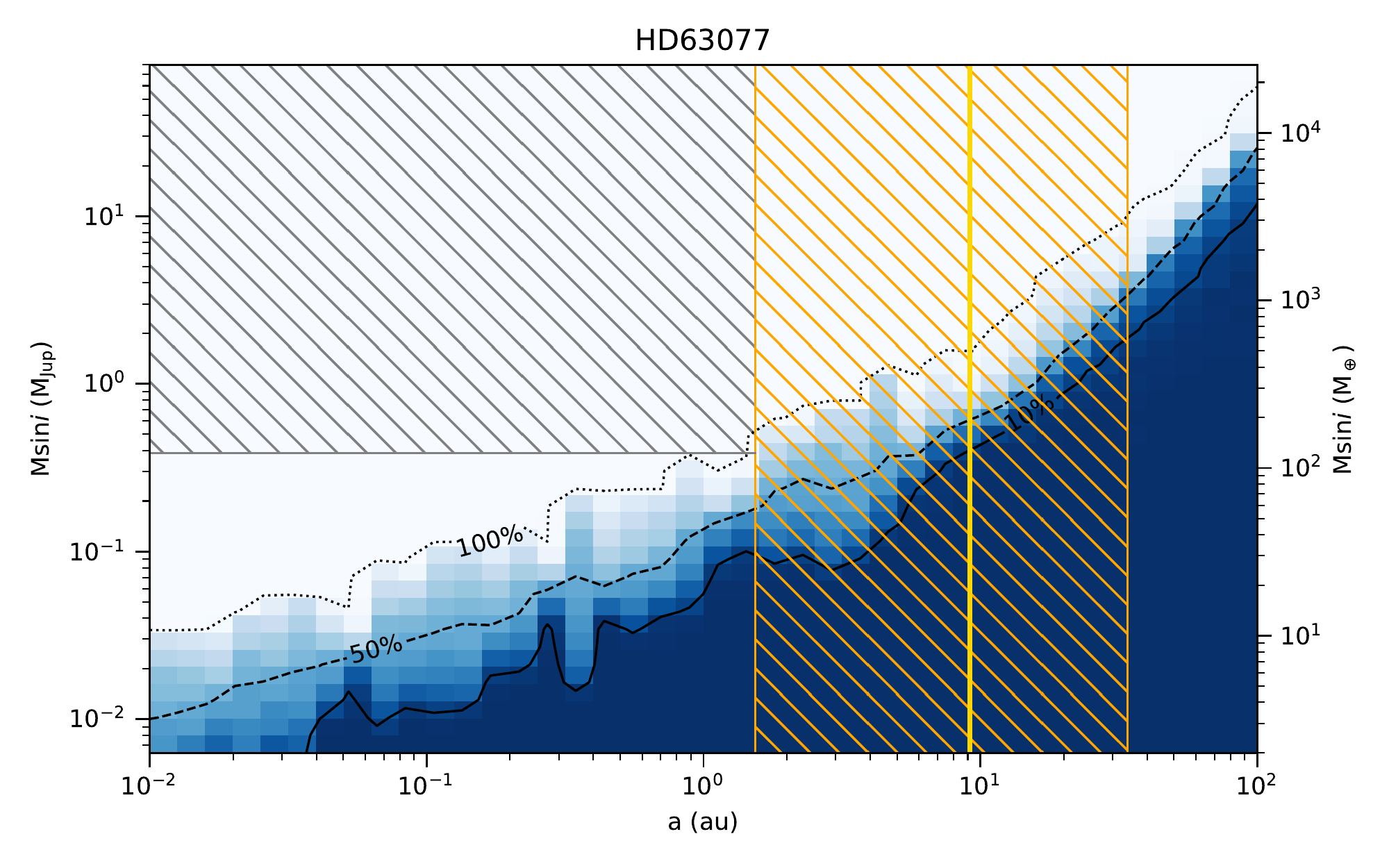}&
    		\includegraphics[width=0.22\linewidth]{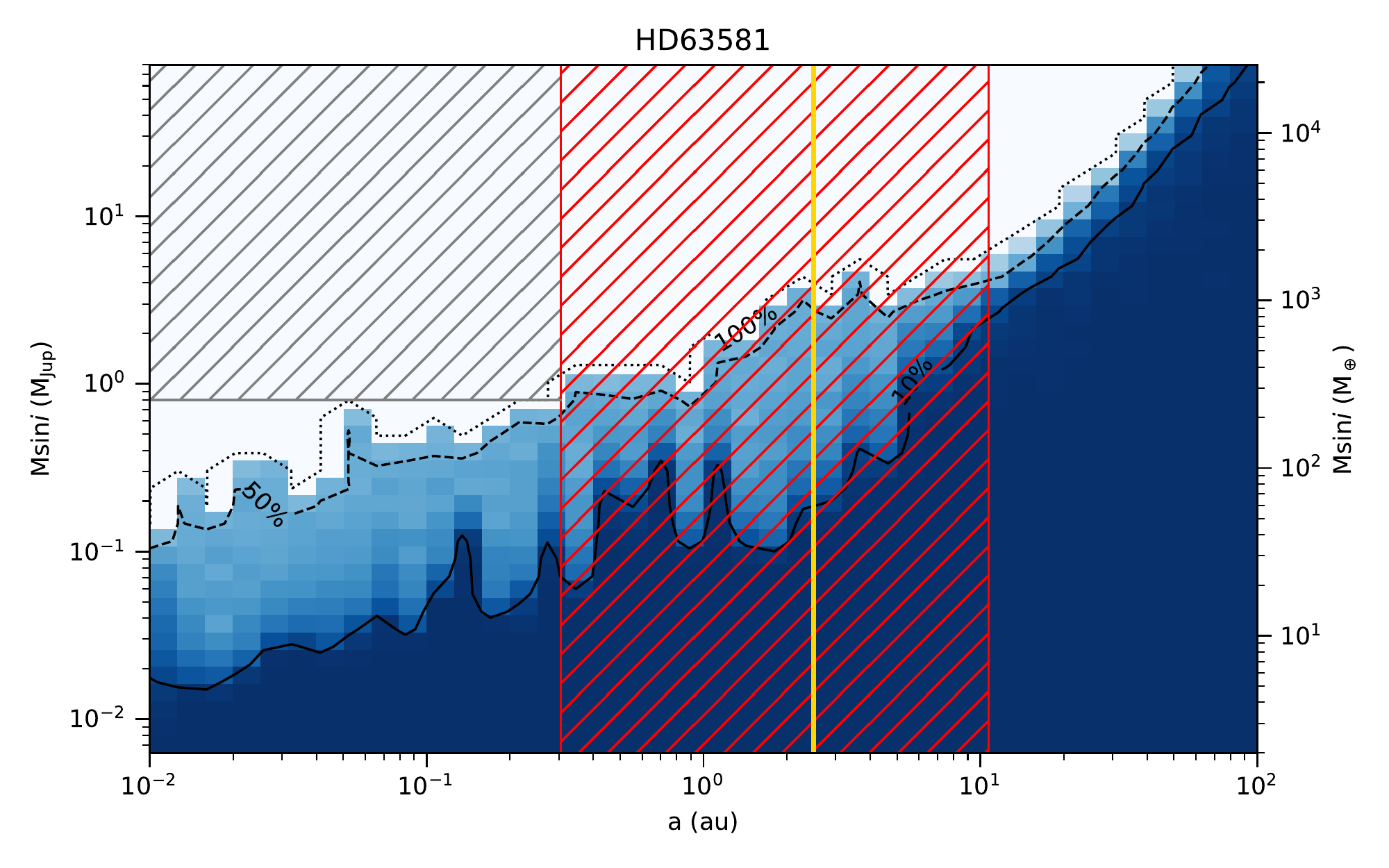}&
    		\includegraphics[width=0.22\linewidth]{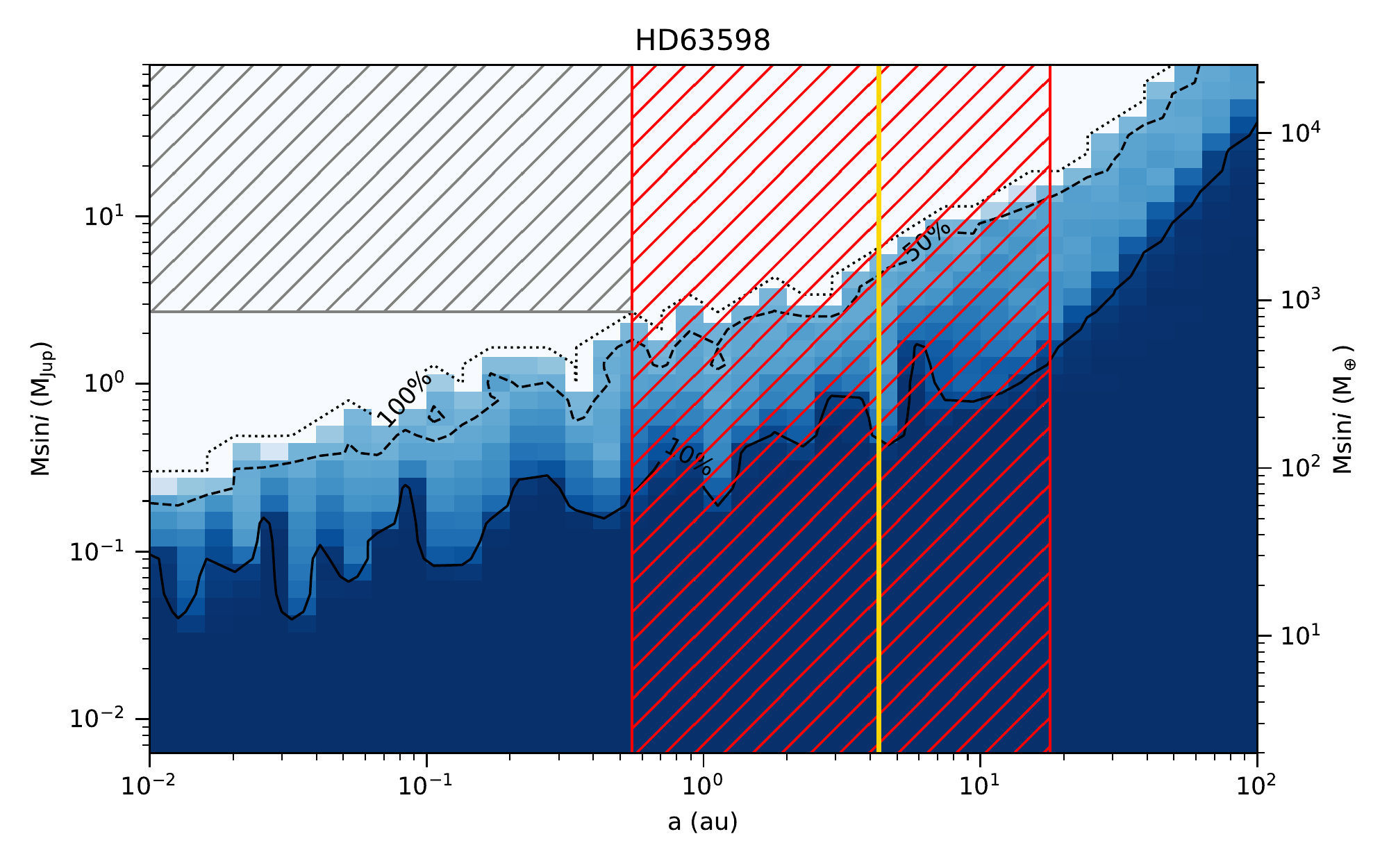}\\
    
    		\includegraphics[width=0.22\linewidth]{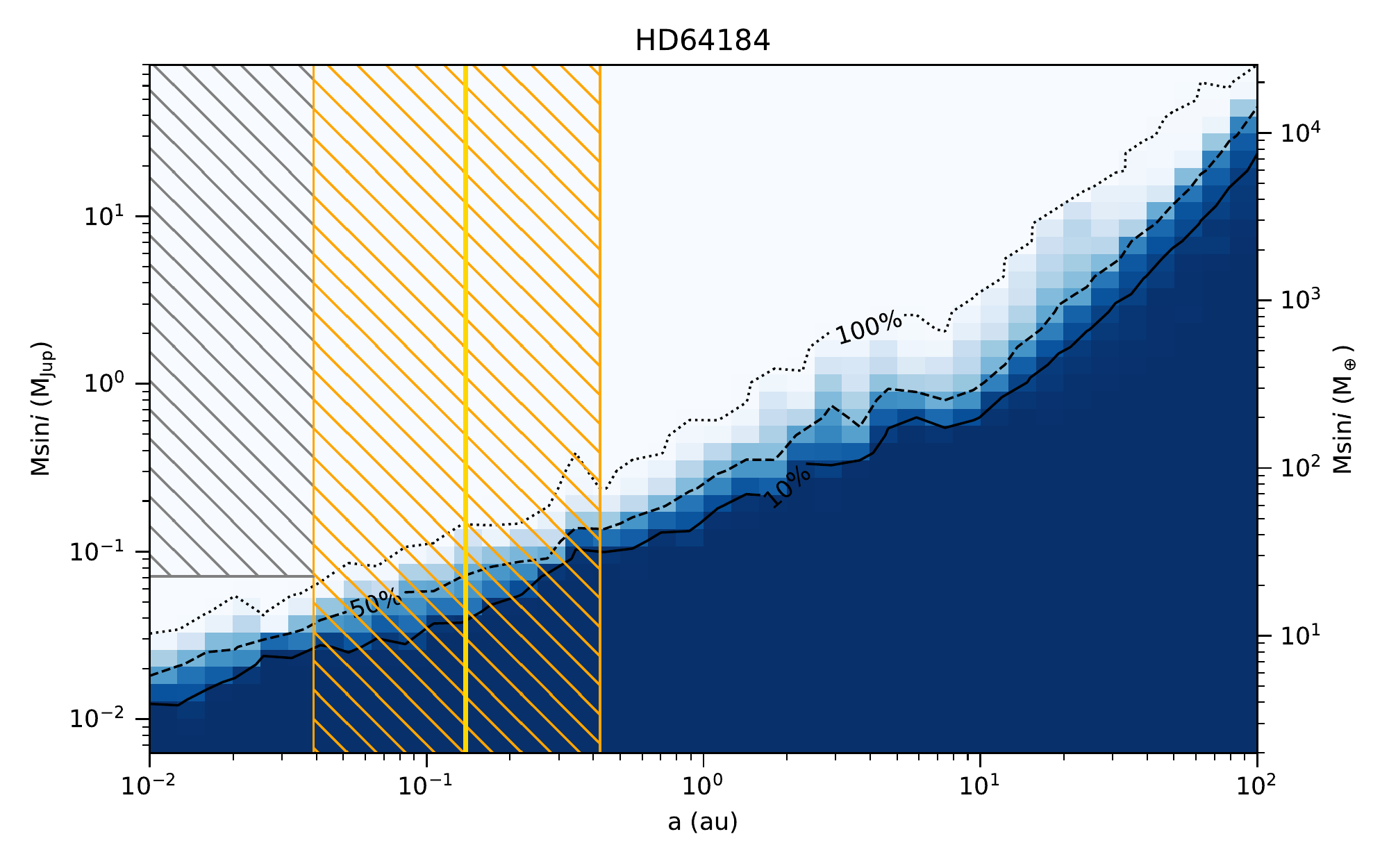}&
    		\includegraphics[width=0.22\linewidth]{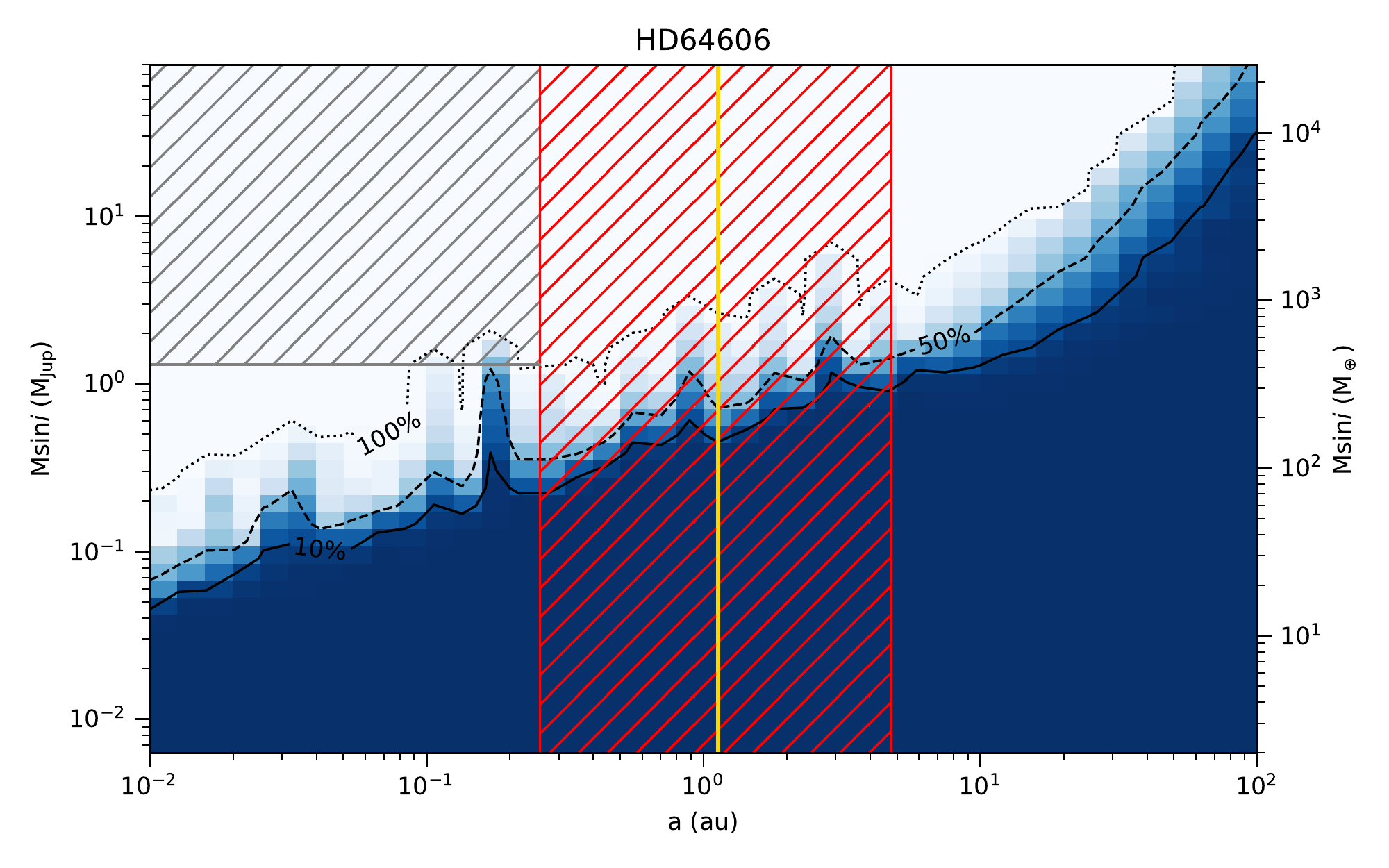}&
    		\includegraphics[width=0.22\linewidth]{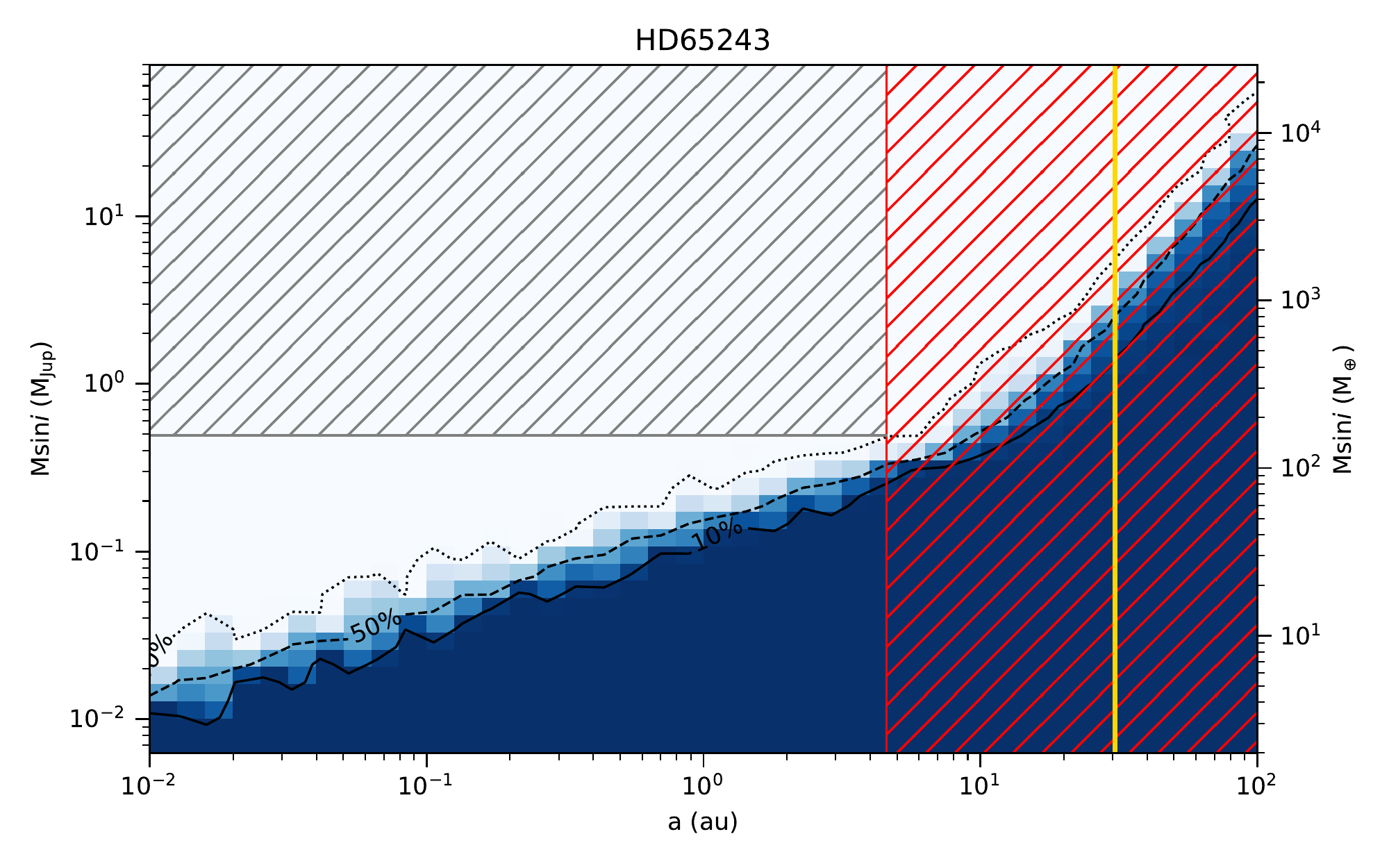}&
    		\includegraphics[width=0.22\linewidth]{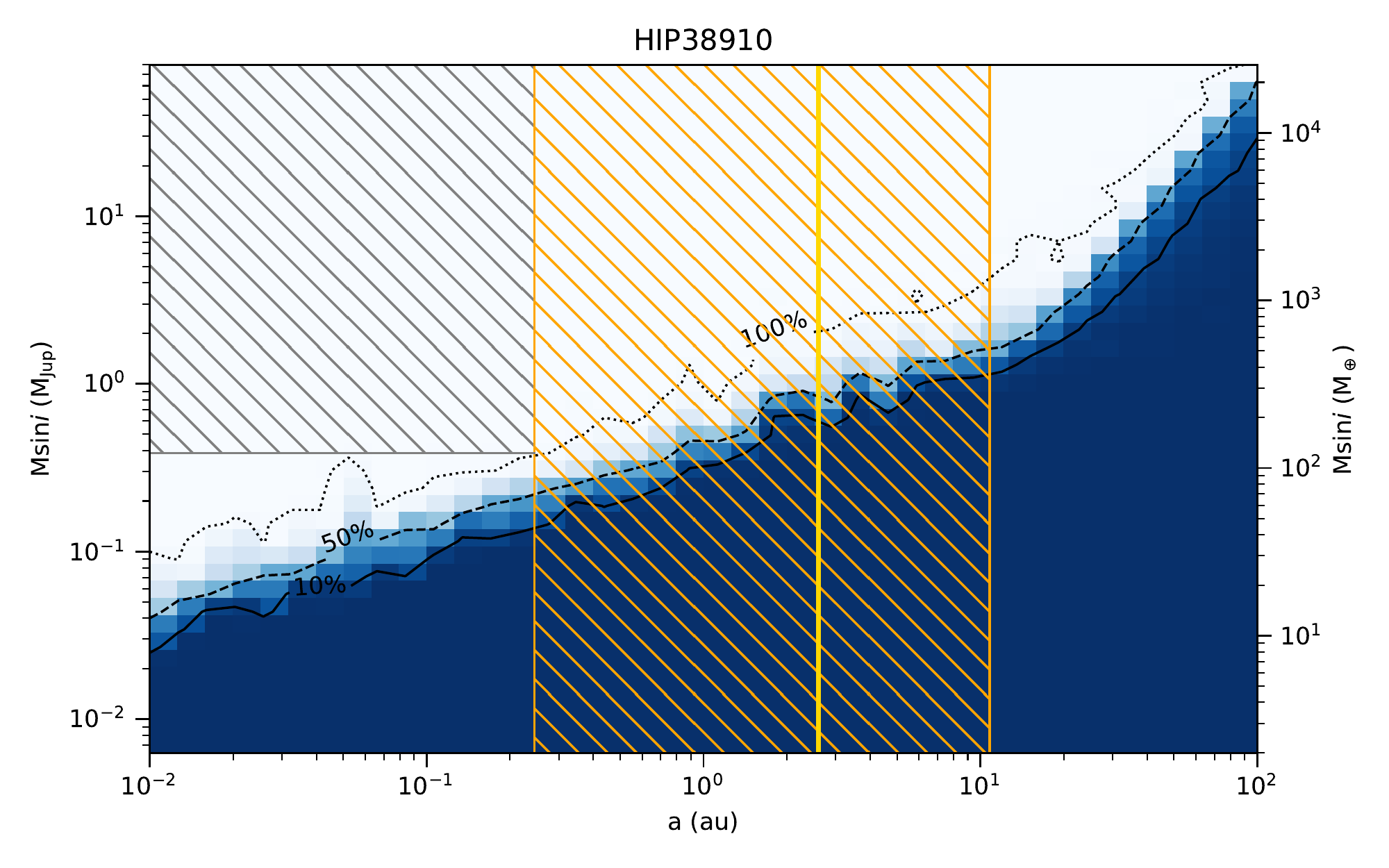}\\
    
    		\includegraphics[width=0.22\linewidth]{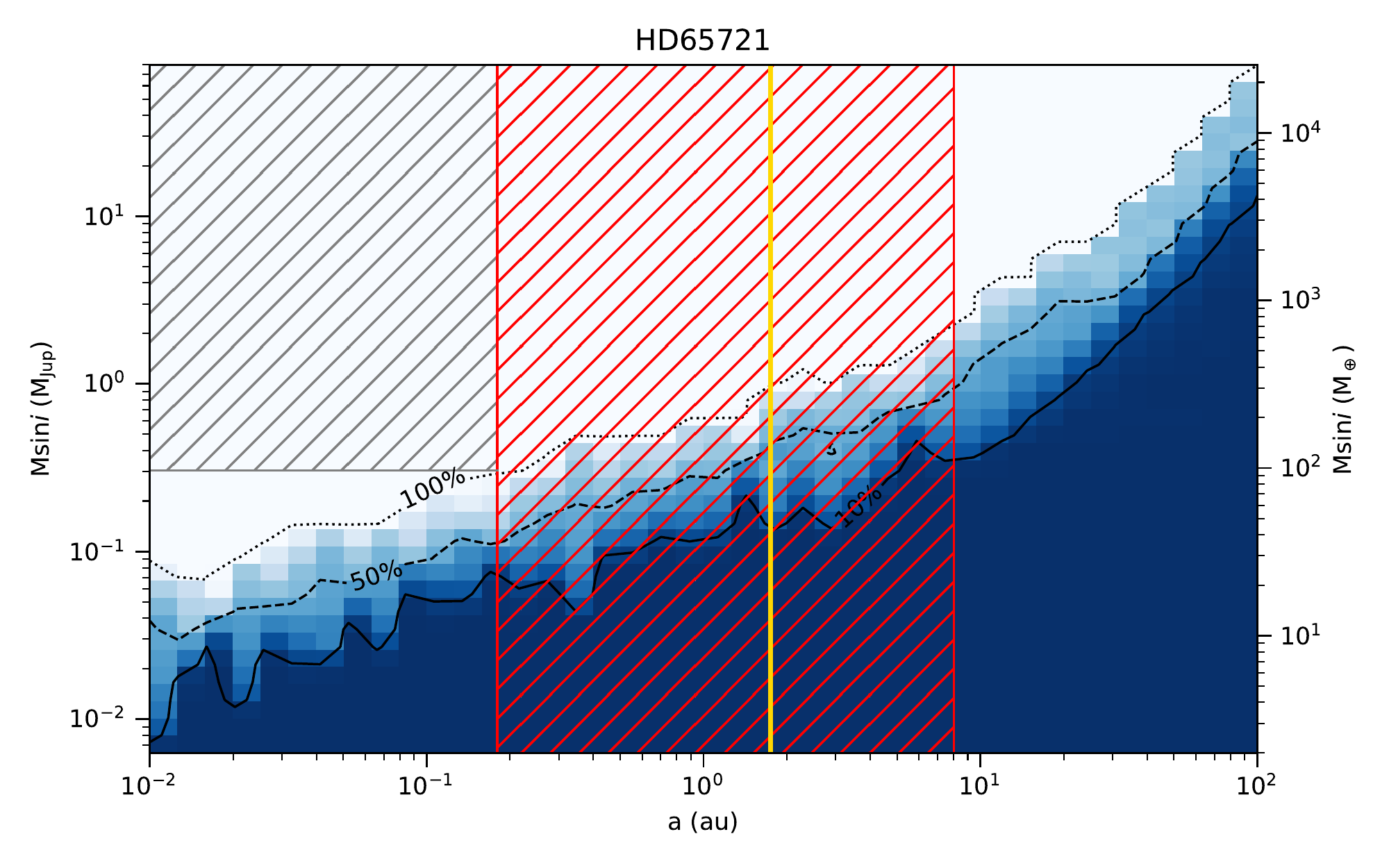}&
    		\includegraphics[width=0.22\linewidth]{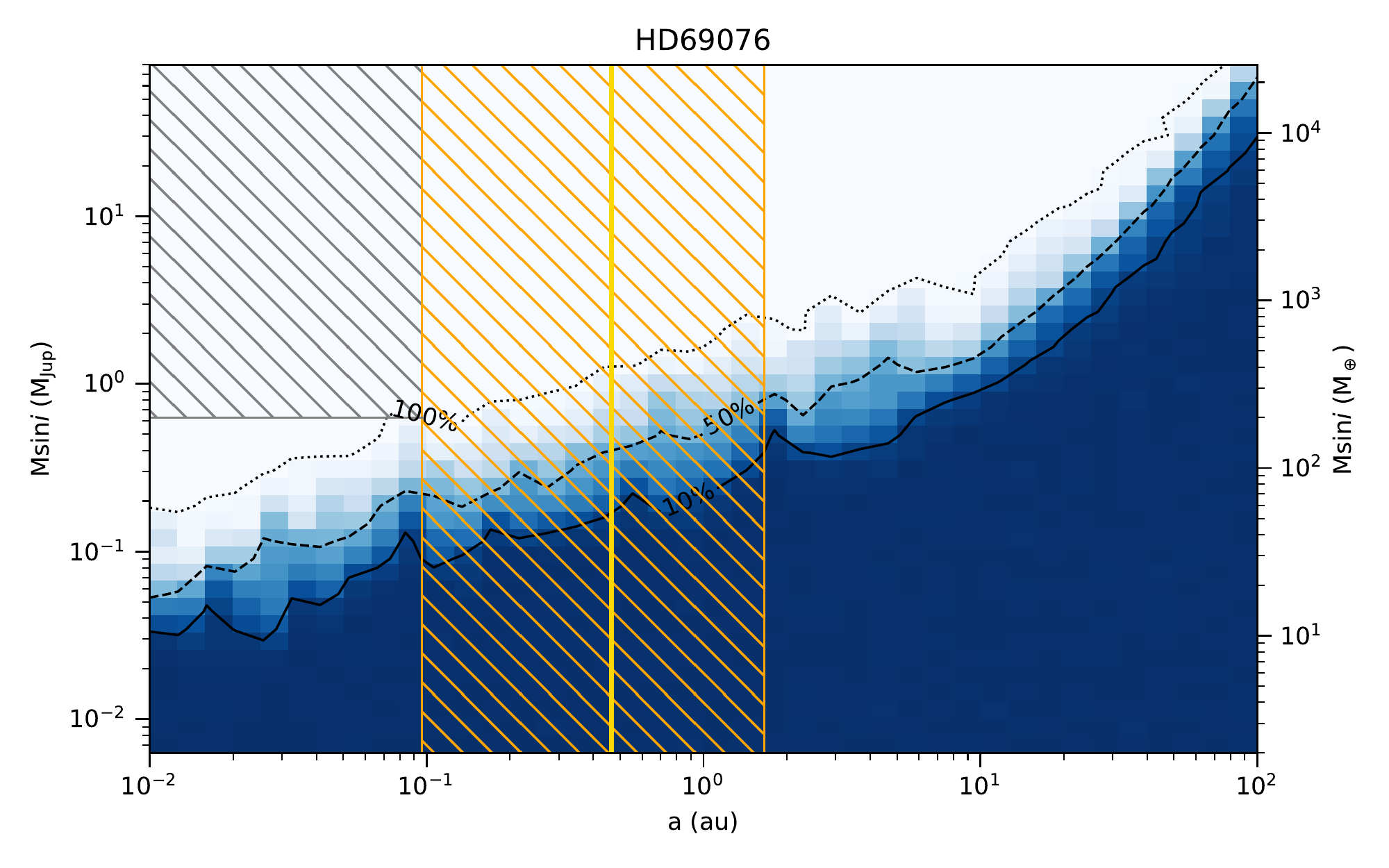}&
    		\includegraphics[width=0.22\linewidth]{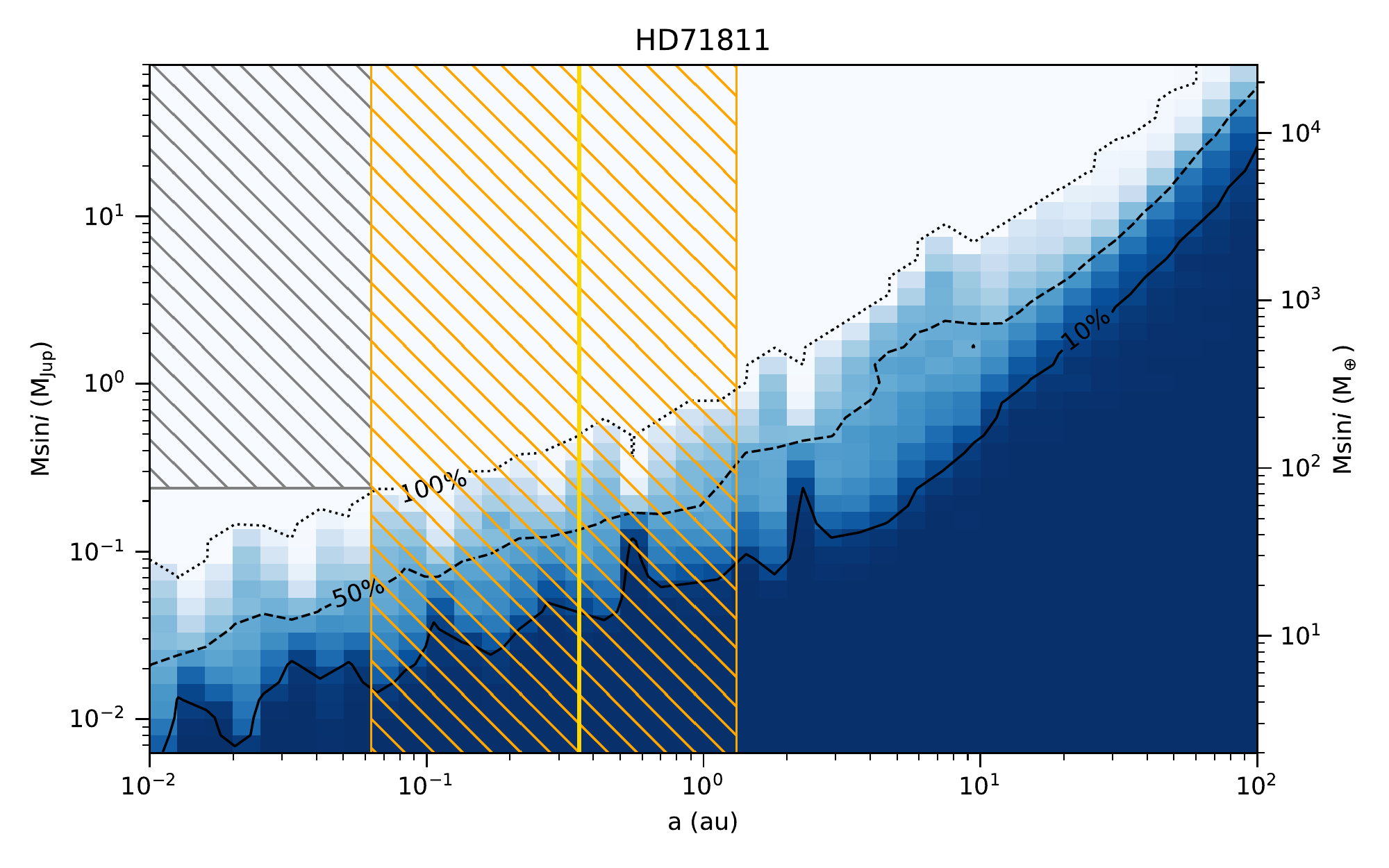}&
    		\includegraphics[width=0.22\linewidth]{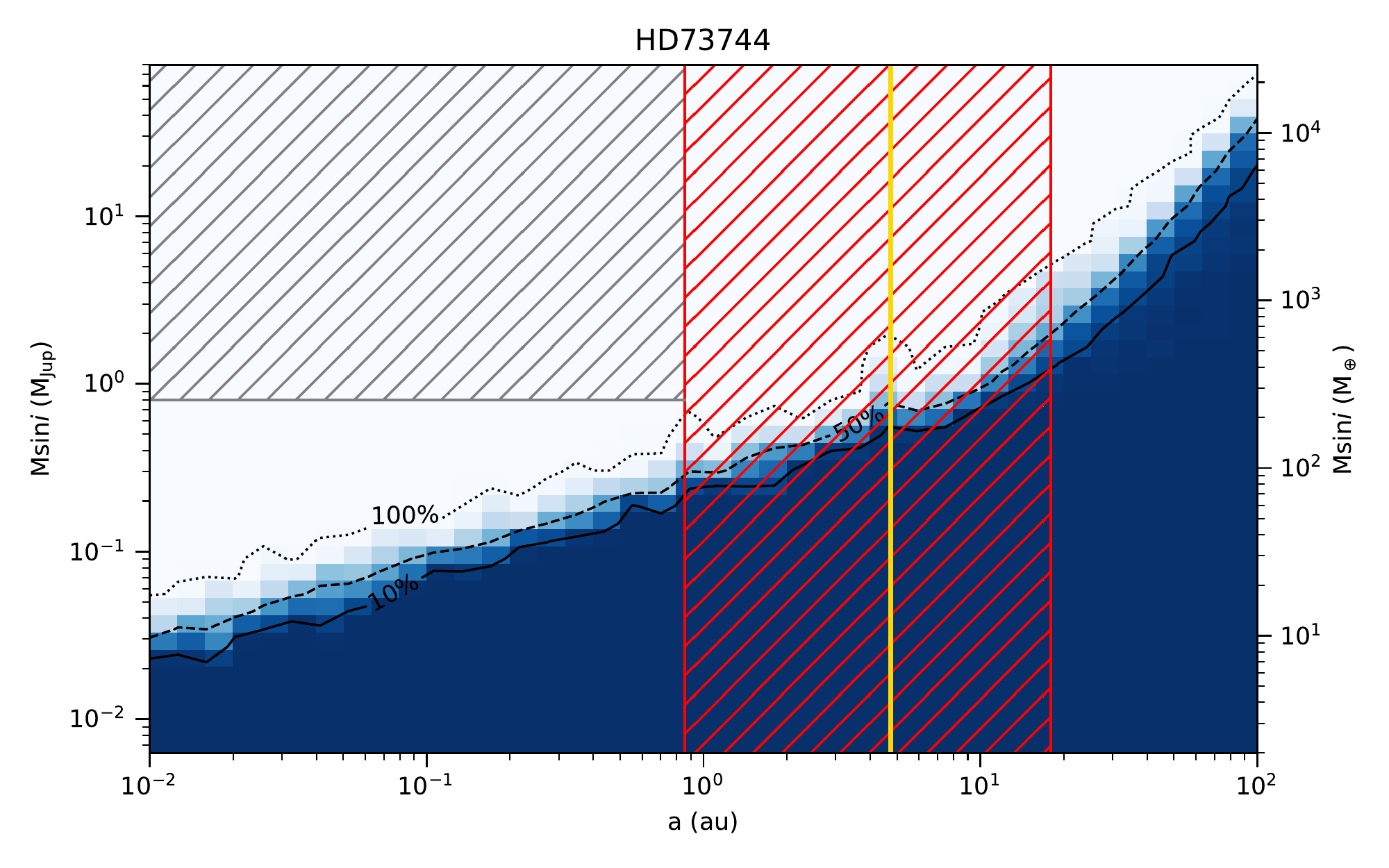}\\
    
    		\includegraphics[width=0.22\linewidth]{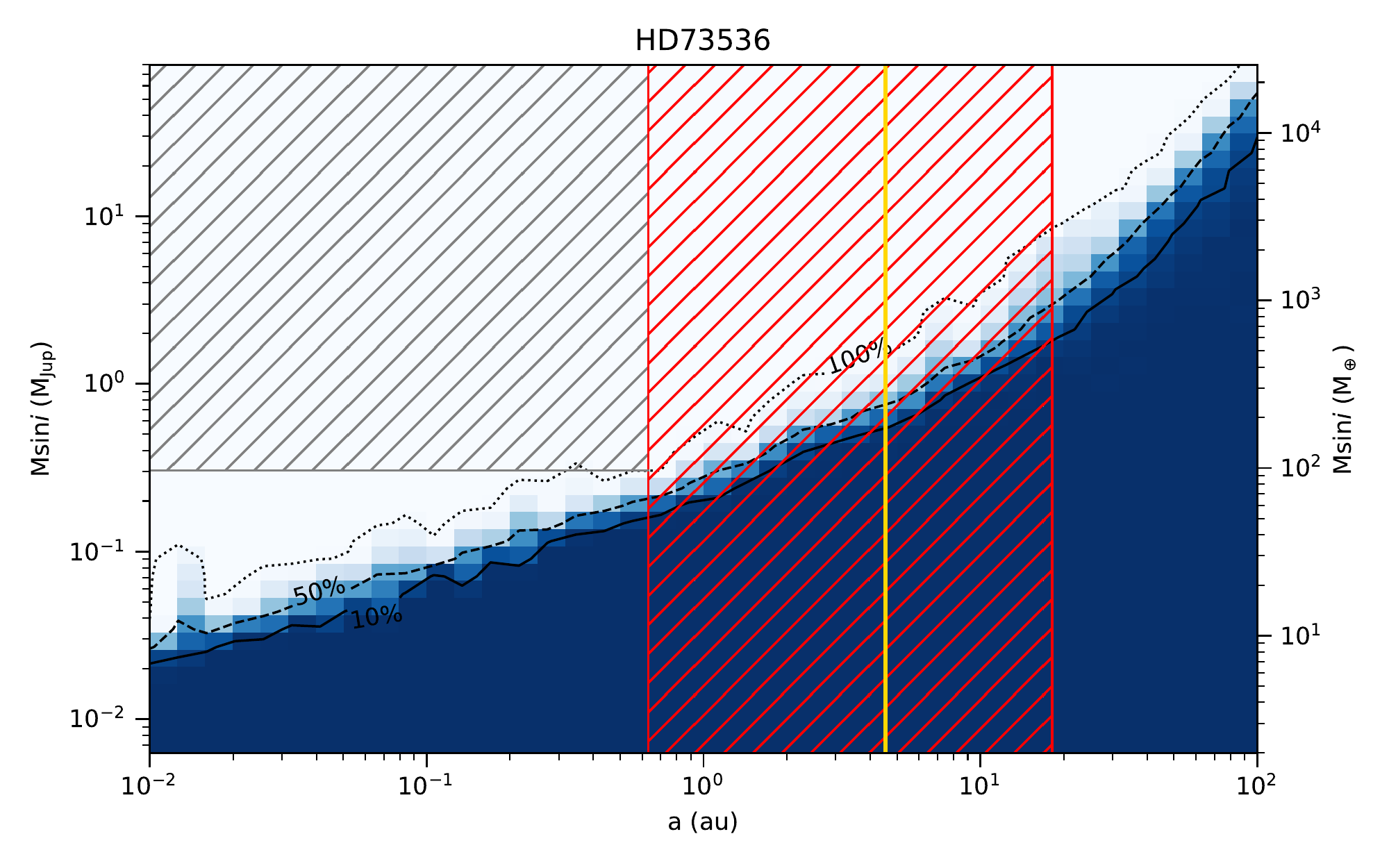}&
    		\includegraphics[width=0.22\linewidth]{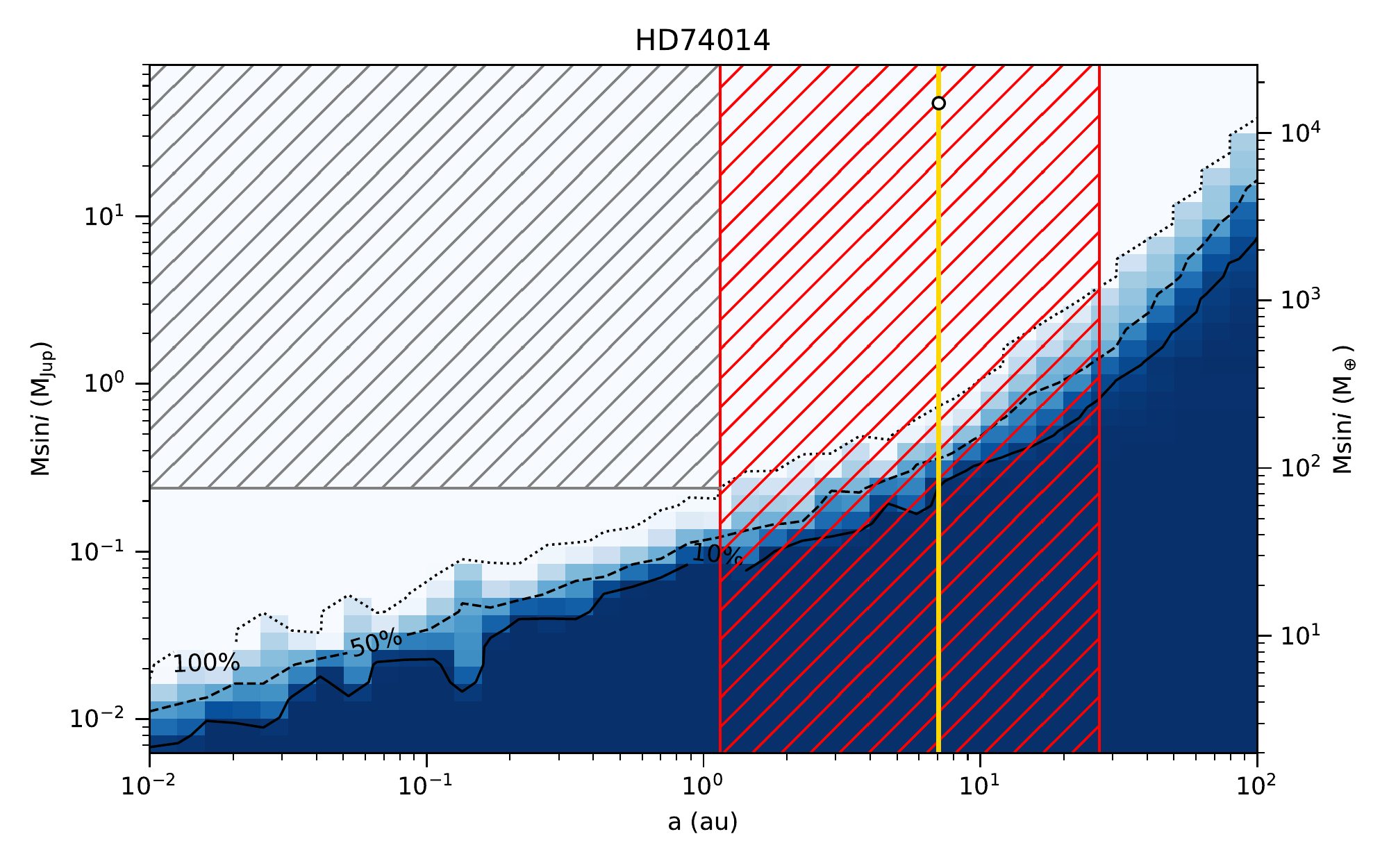}&
    		\includegraphics[width=0.22\linewidth]{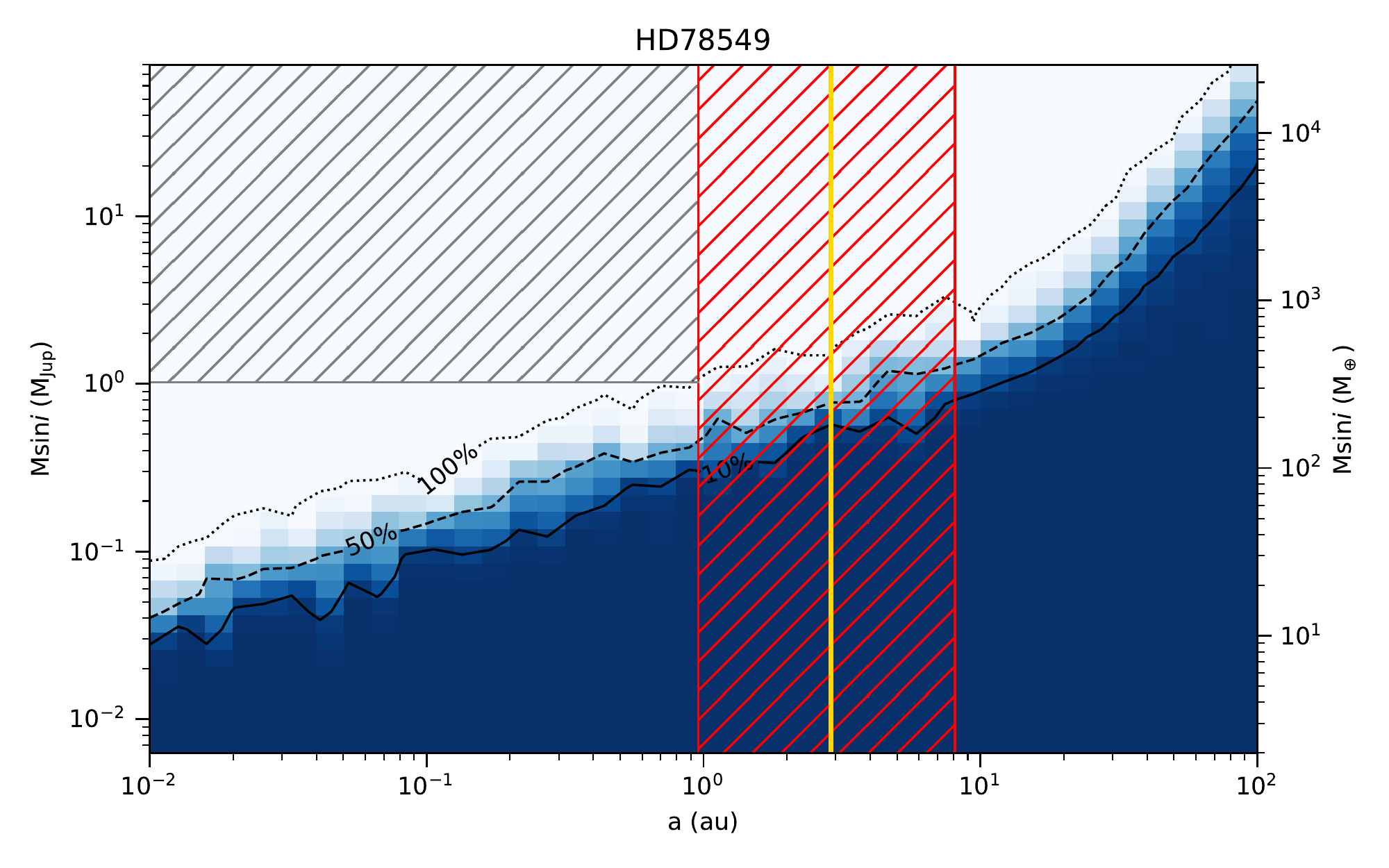}&
    		\includegraphics[width=0.22\linewidth]{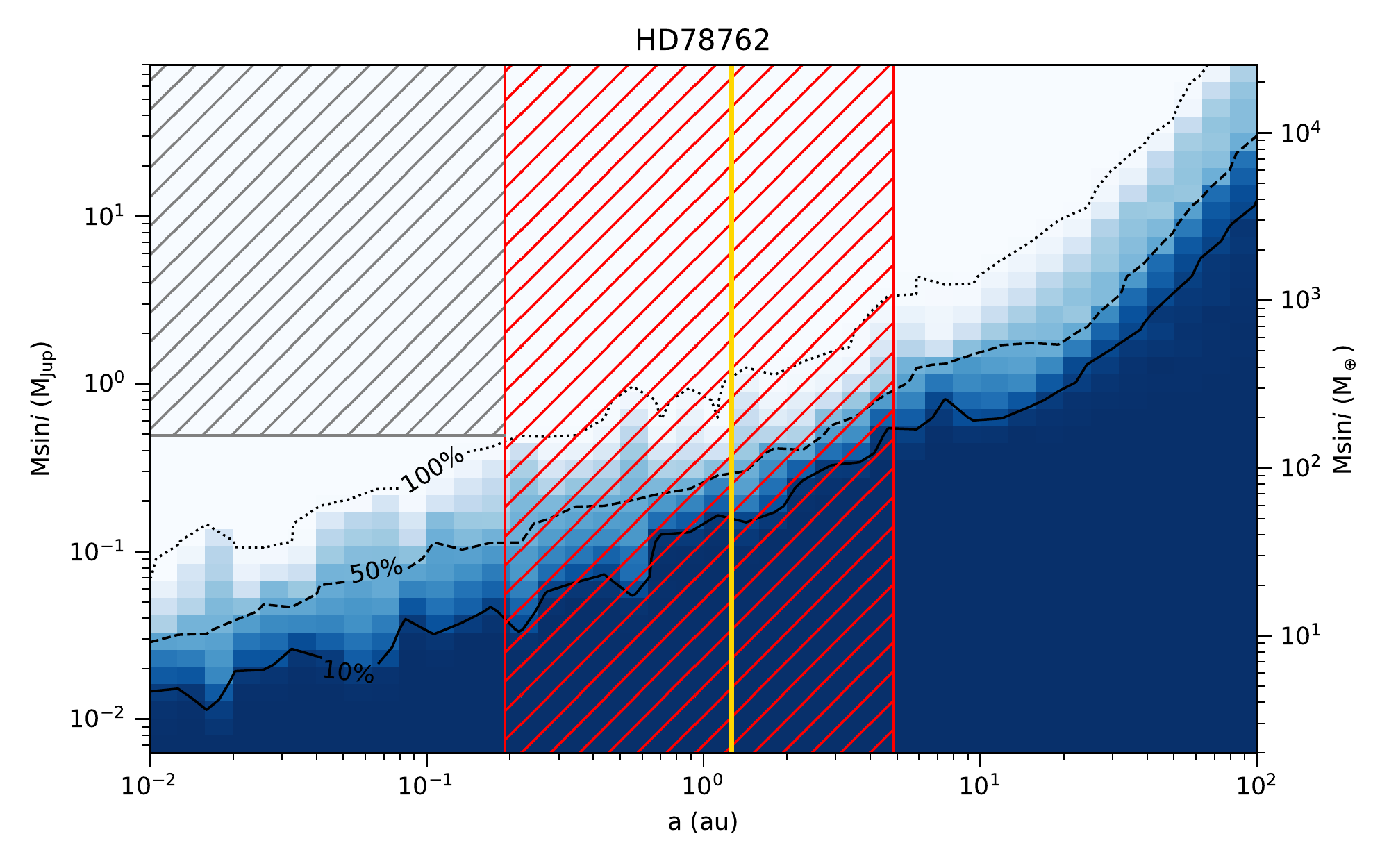}\\
    
    		\includegraphics[width=0.22\linewidth]{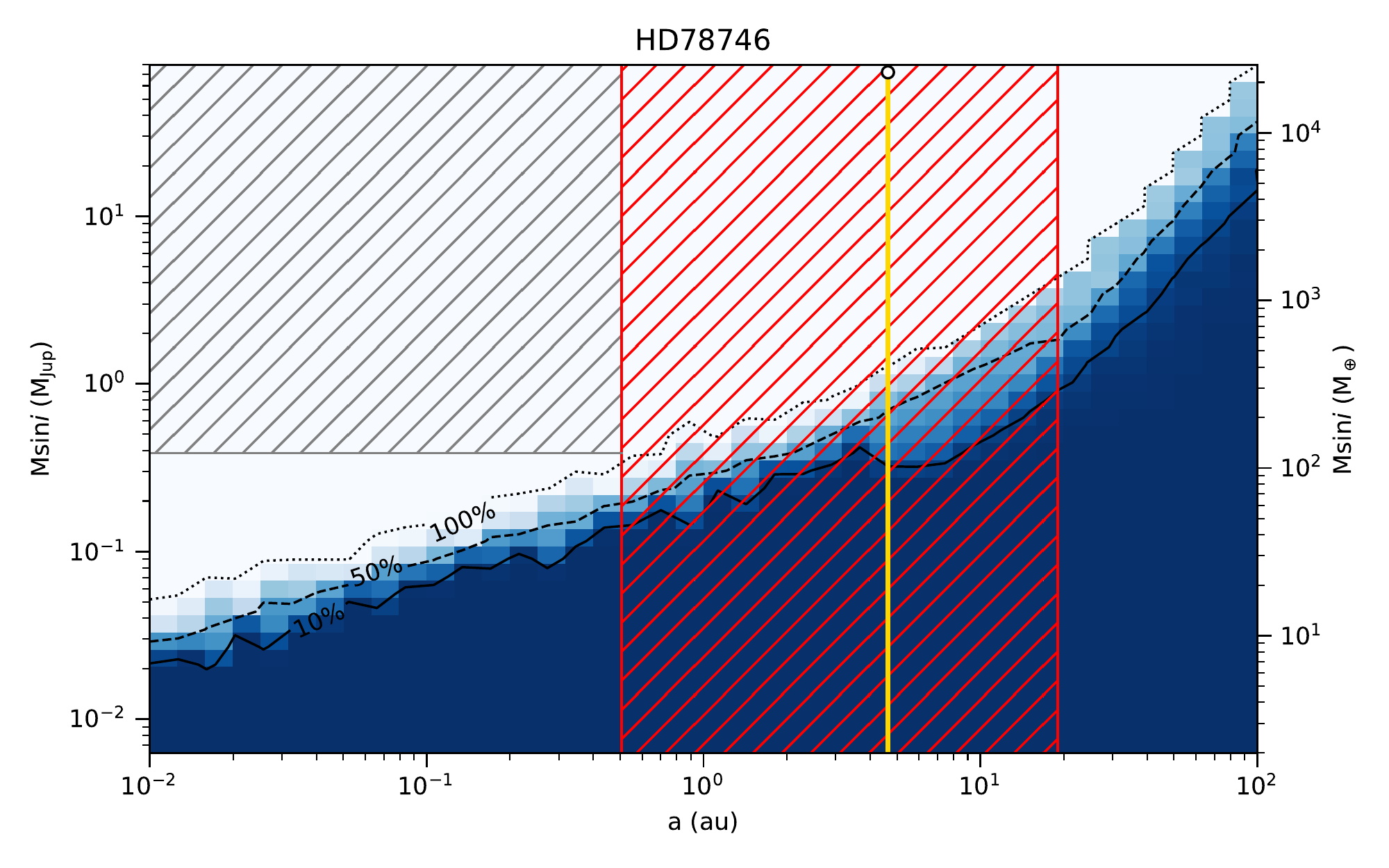}&
    		\includegraphics[width=0.22\linewidth]{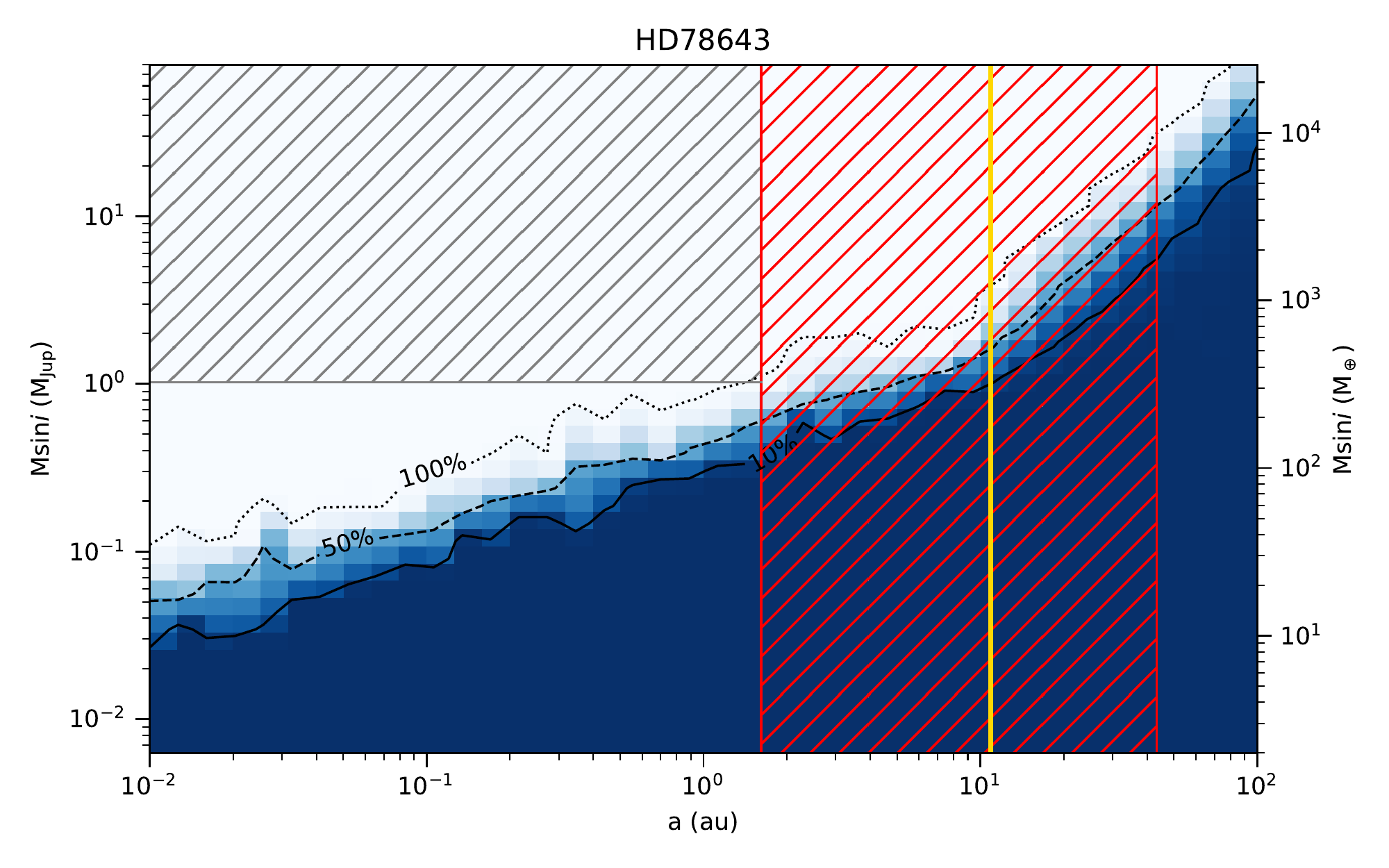}&
    		\includegraphics[width=0.22\linewidth]{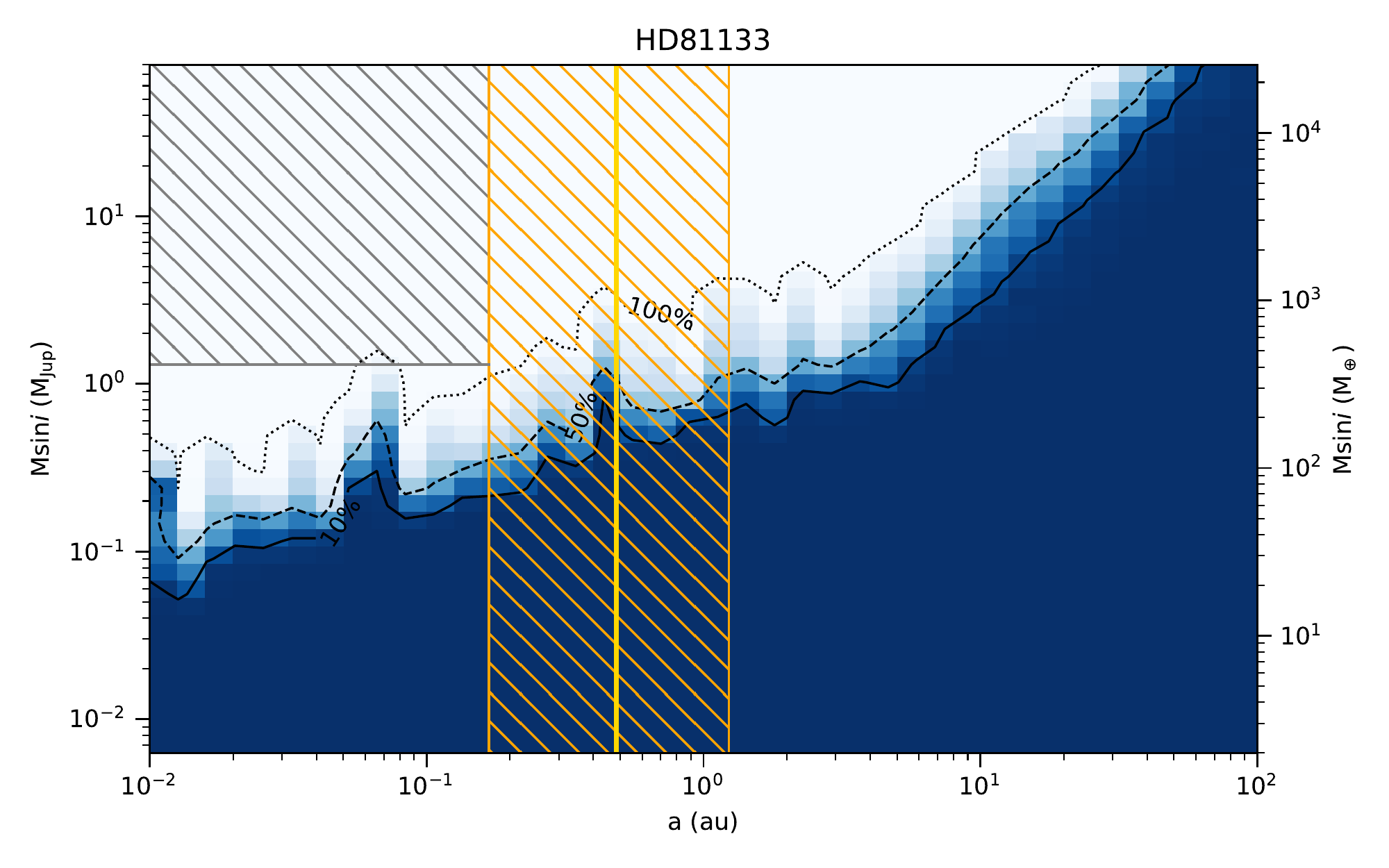}&
    		\includegraphics[width=0.22\linewidth]{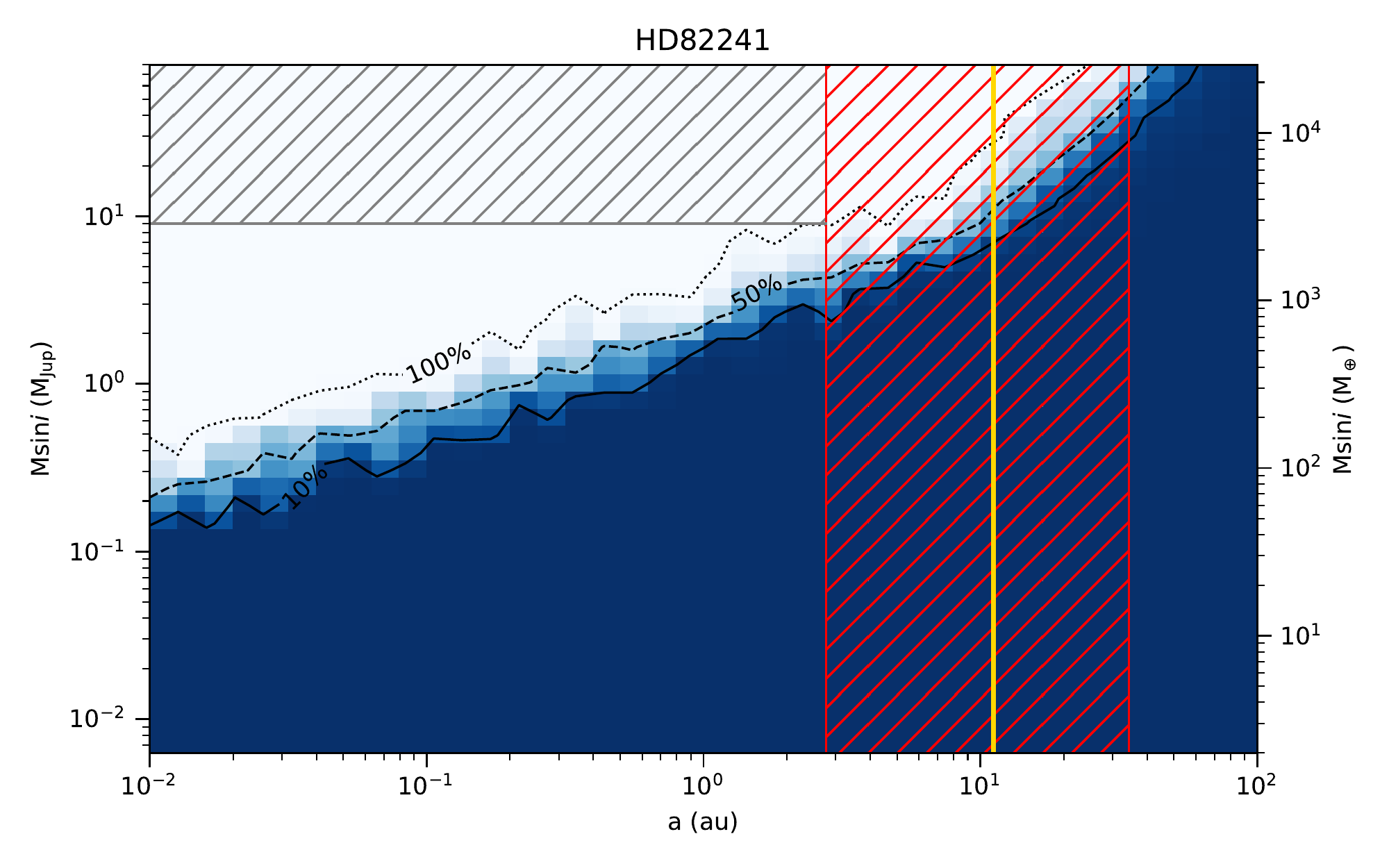}\\
    
    		\includegraphics[width=0.22\linewidth]{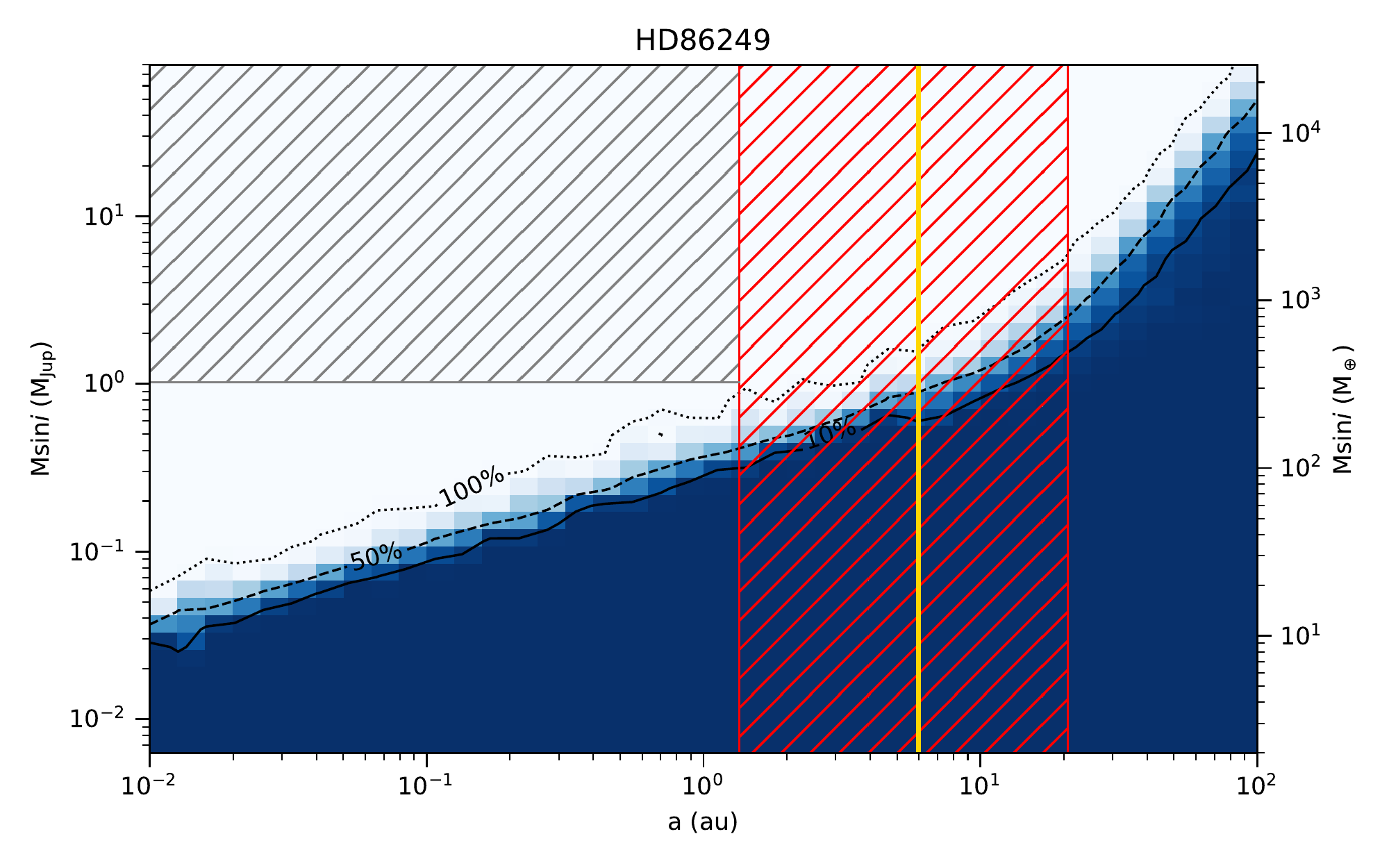}&
    		\includegraphics[width=0.22\linewidth]{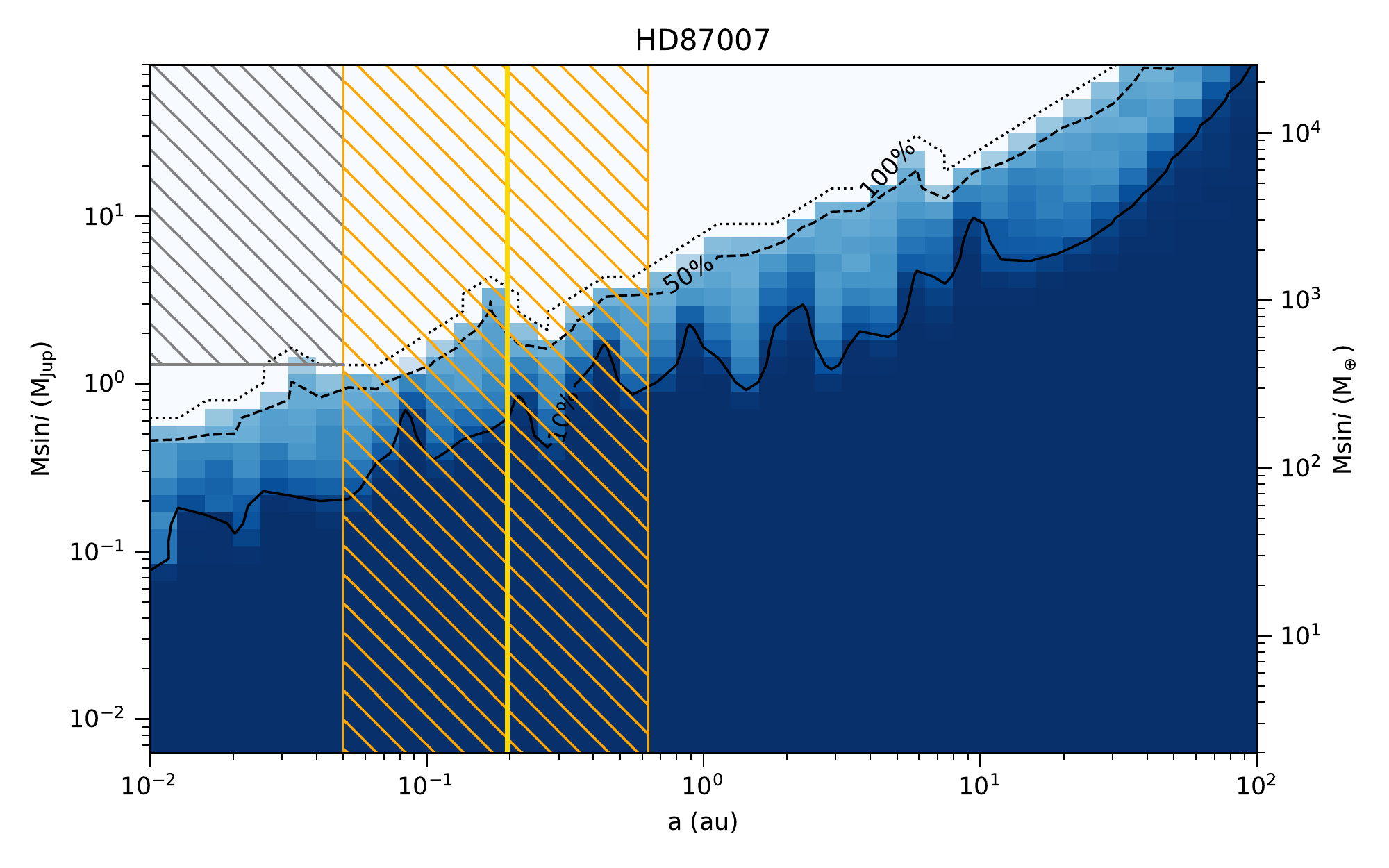}&
    		\includegraphics[width=0.22\linewidth]{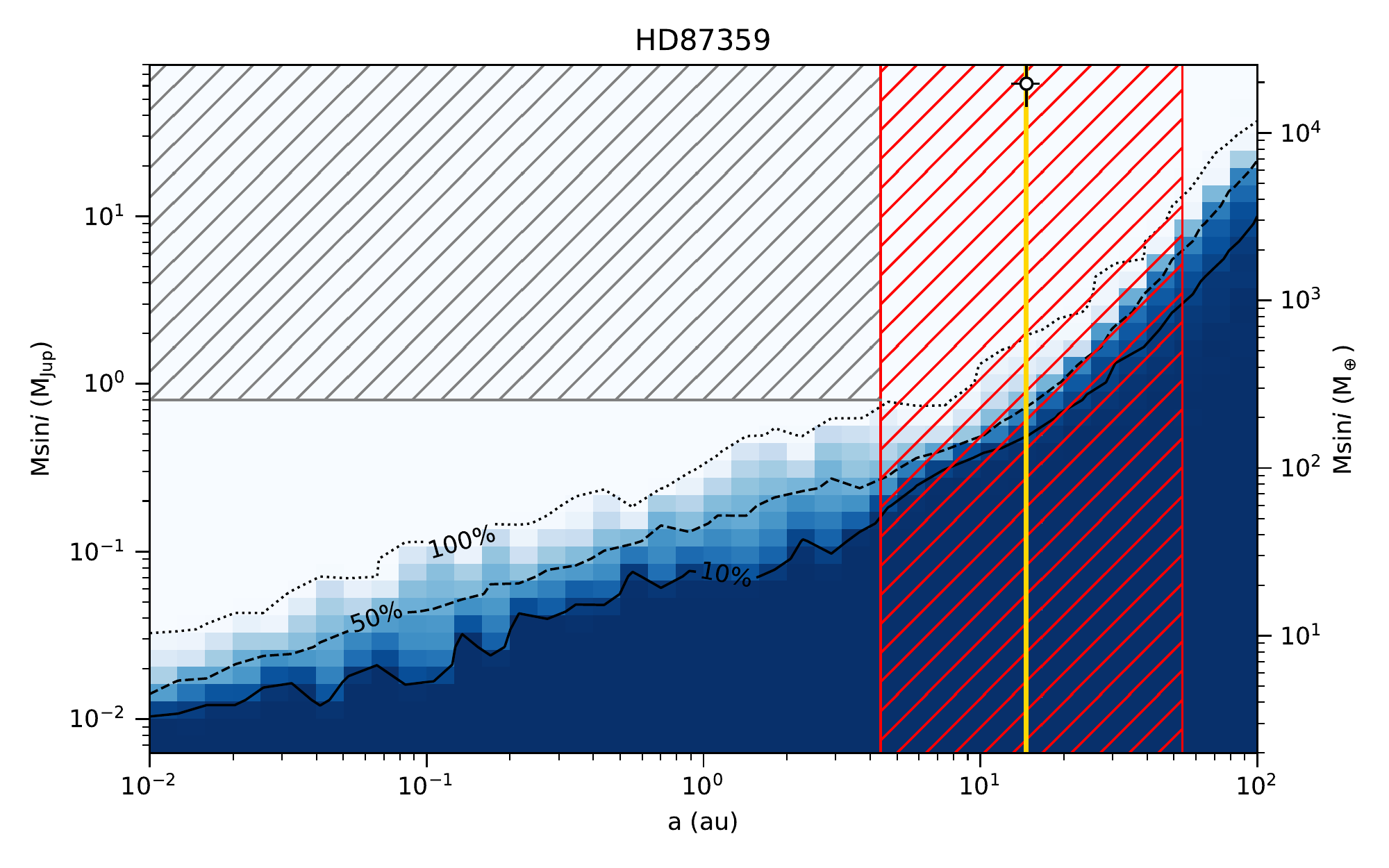}&
    		\includegraphics[width=0.22\linewidth]{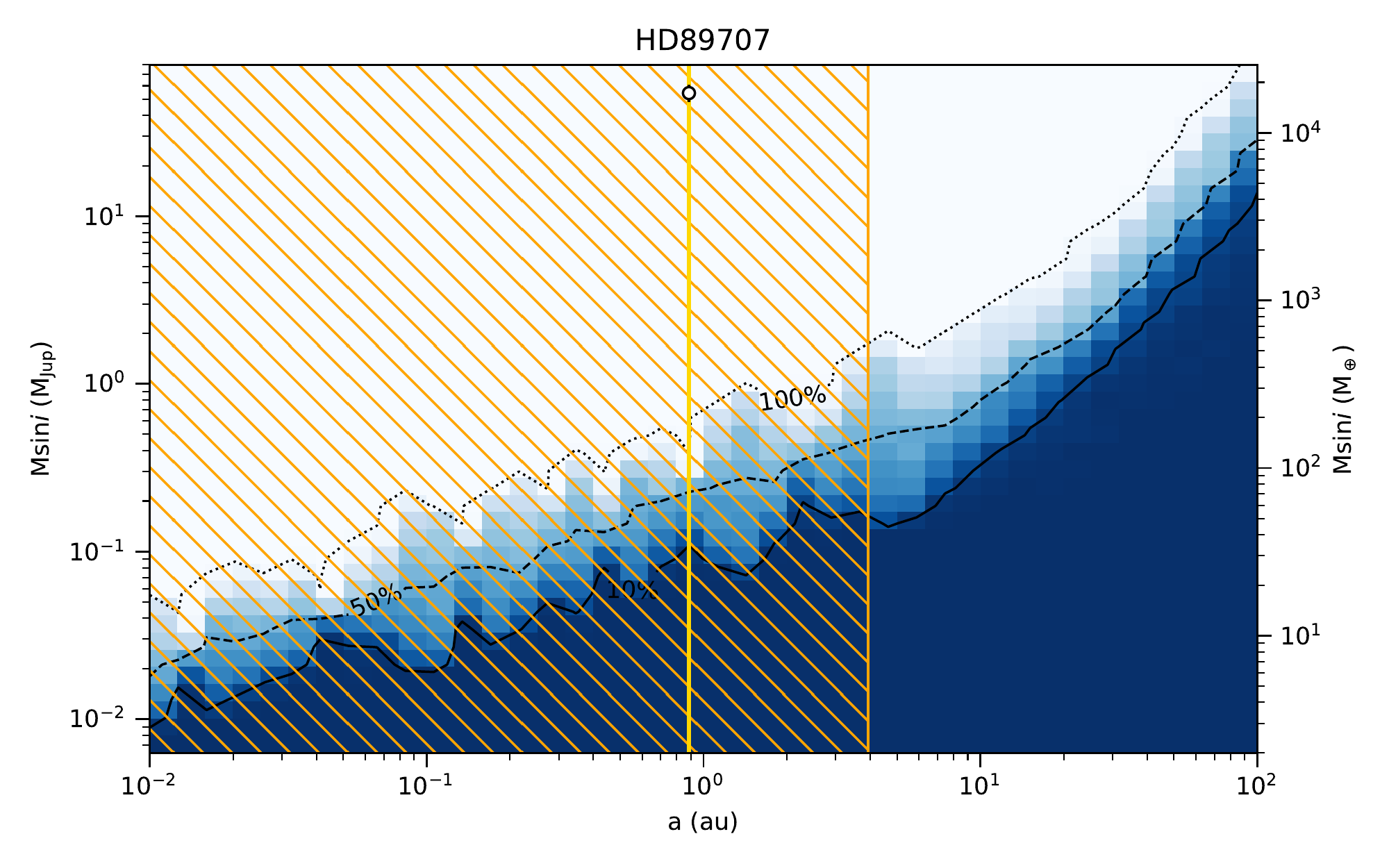}\\
    
    		\includegraphics[width=0.22\linewidth]{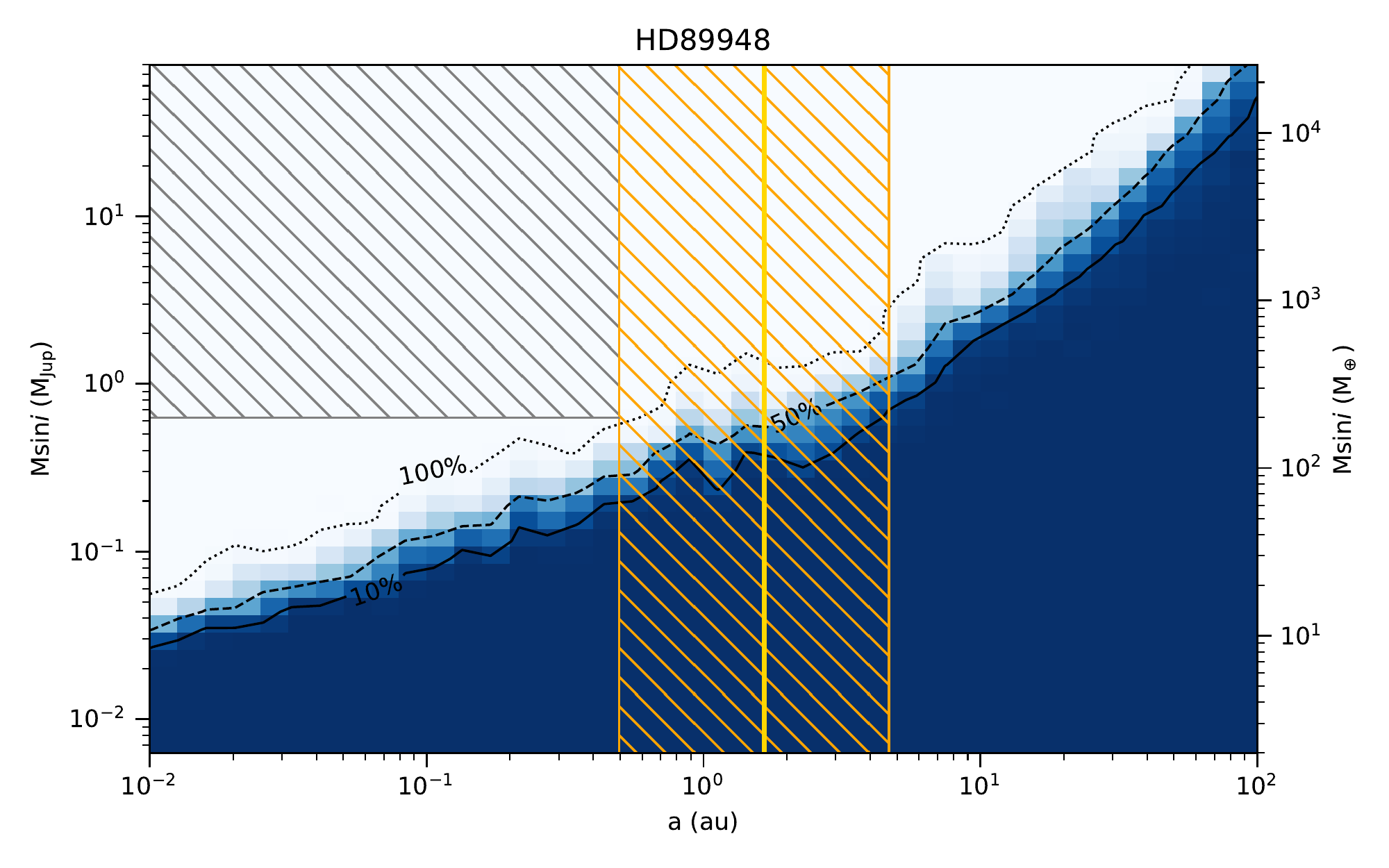}&
    		\includegraphics[width=0.22\linewidth]{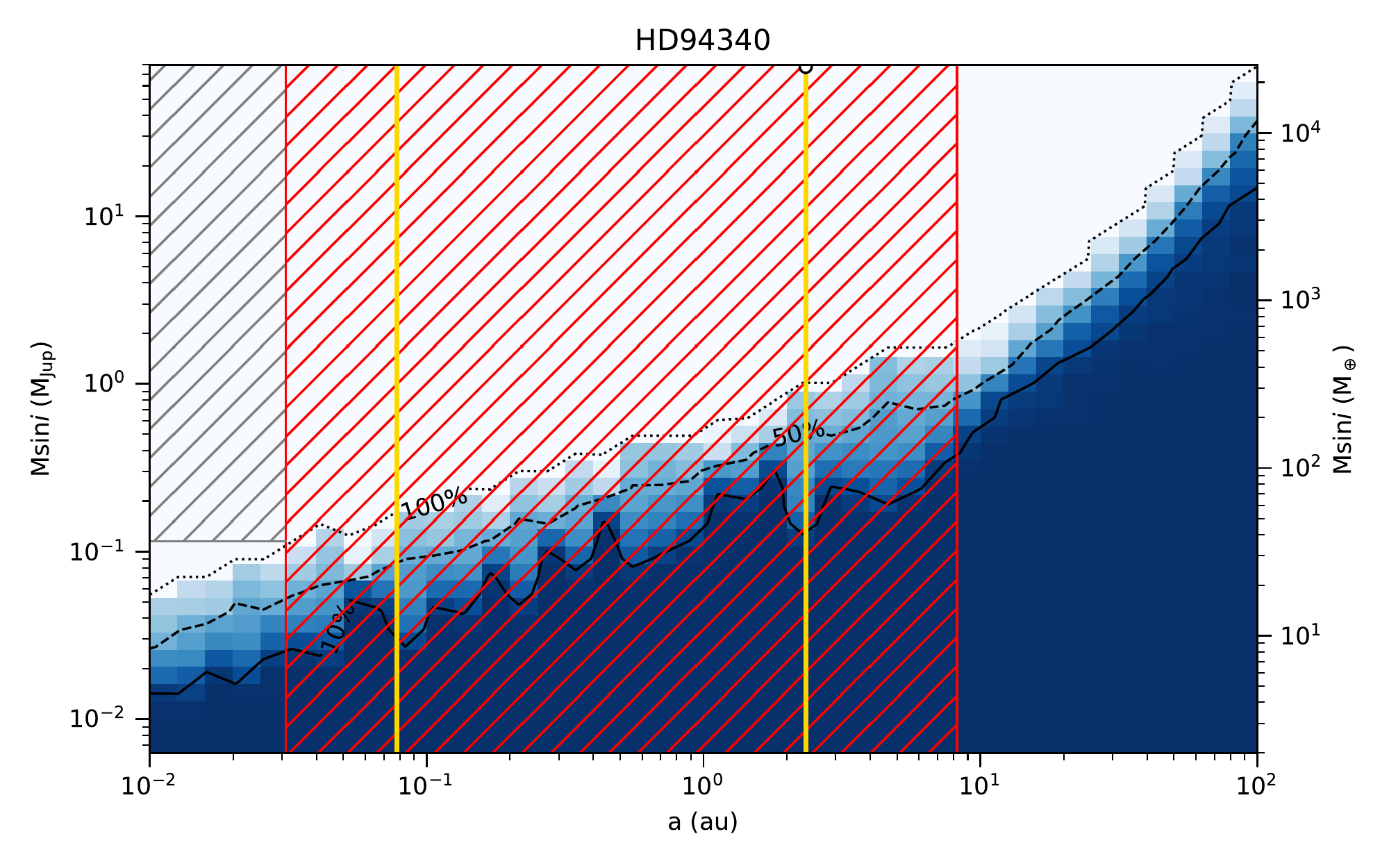}&
    		\includegraphics[width=0.22\linewidth]{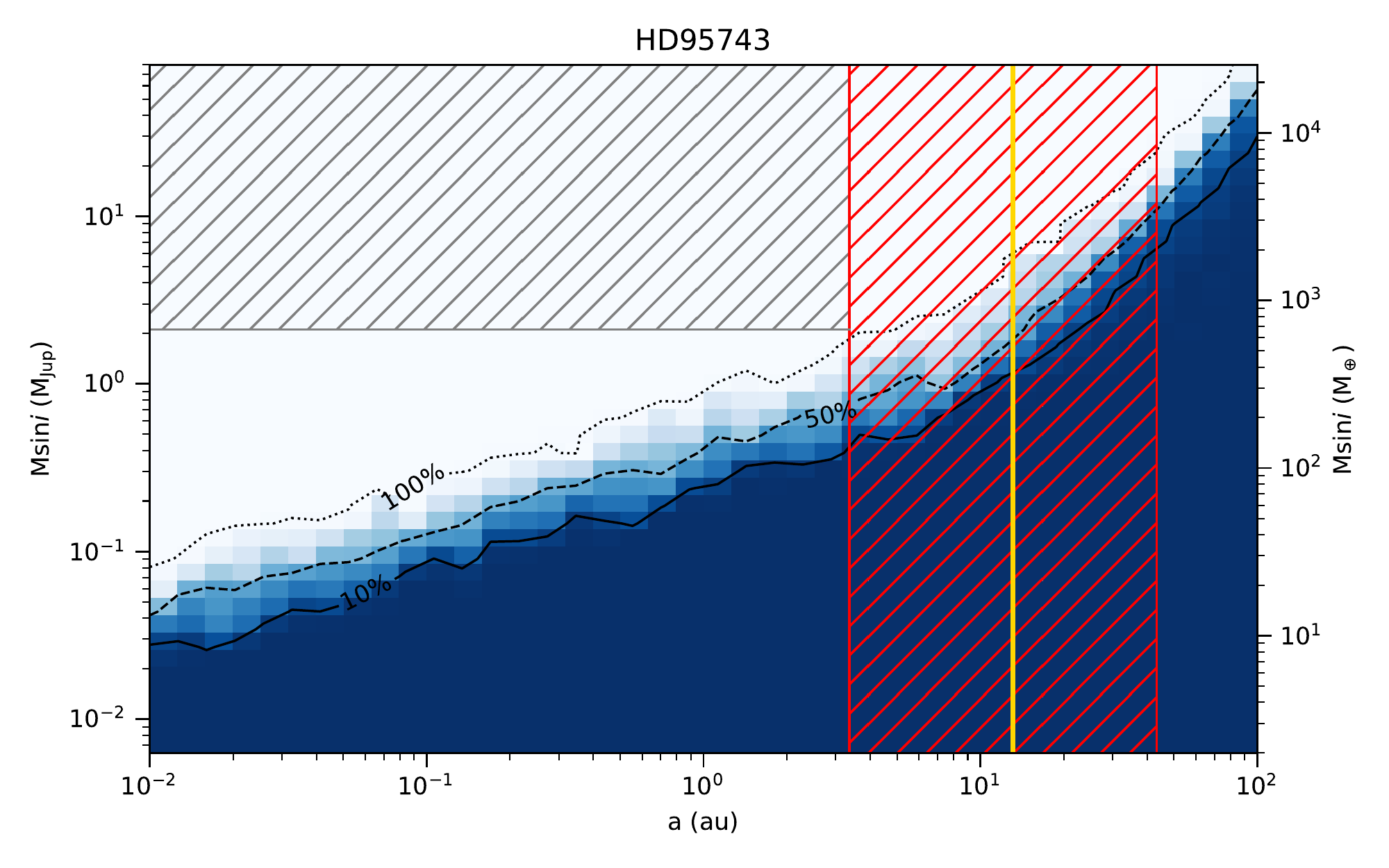}&
    		\includegraphics[width=0.22\linewidth]{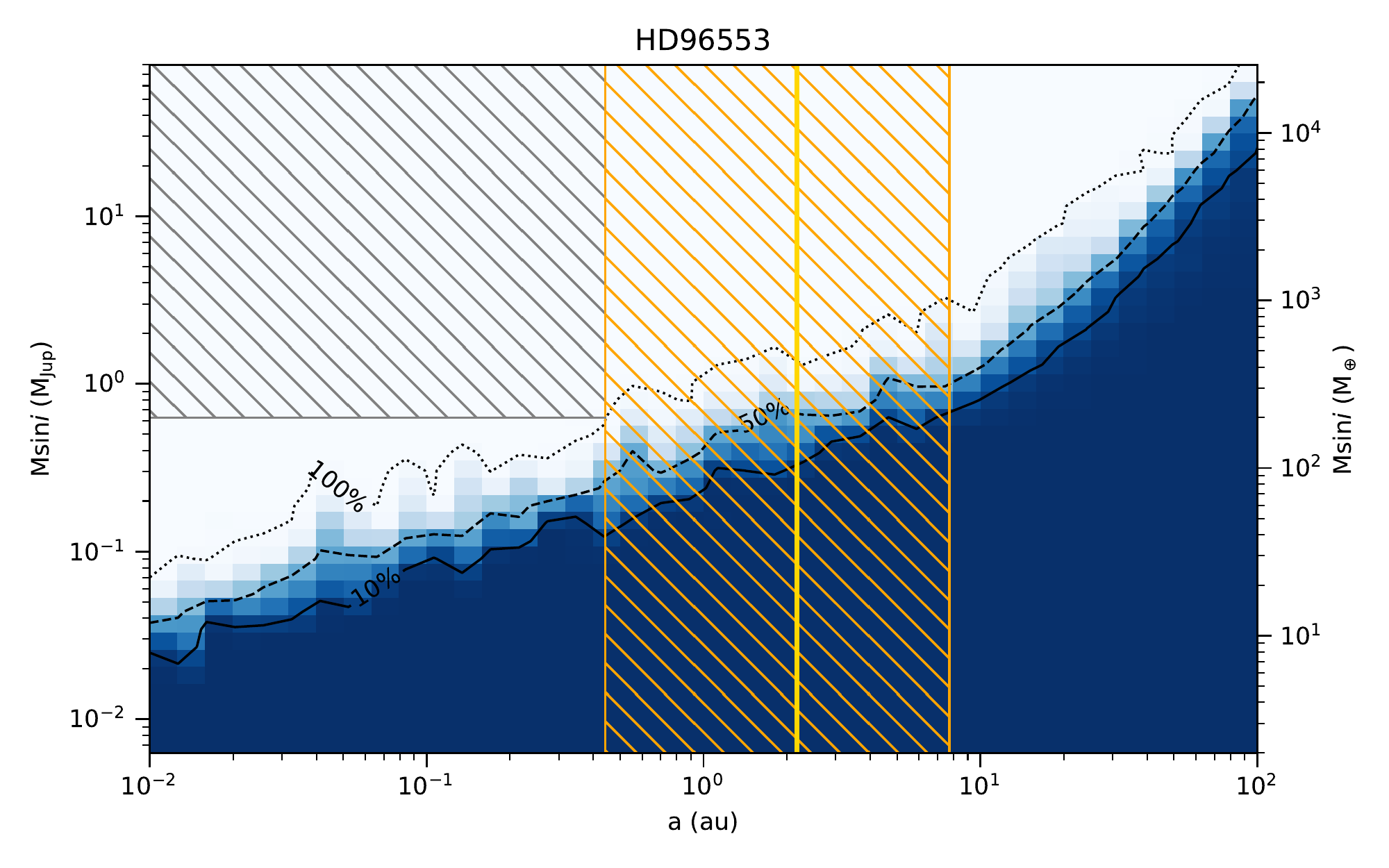}\\
    
    		\includegraphics[width=0.22\linewidth]{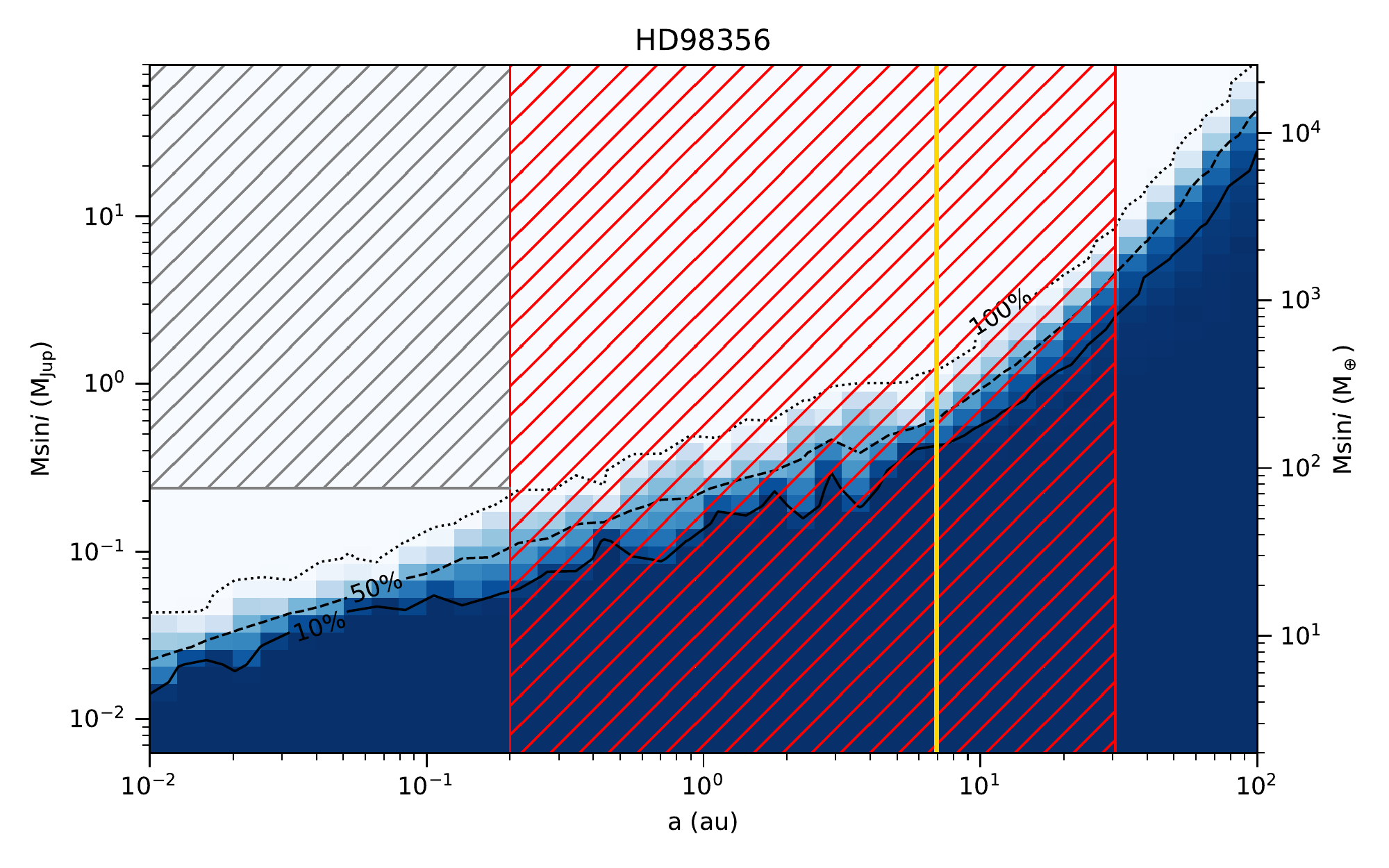}&
    		\includegraphics[width=0.22\linewidth]{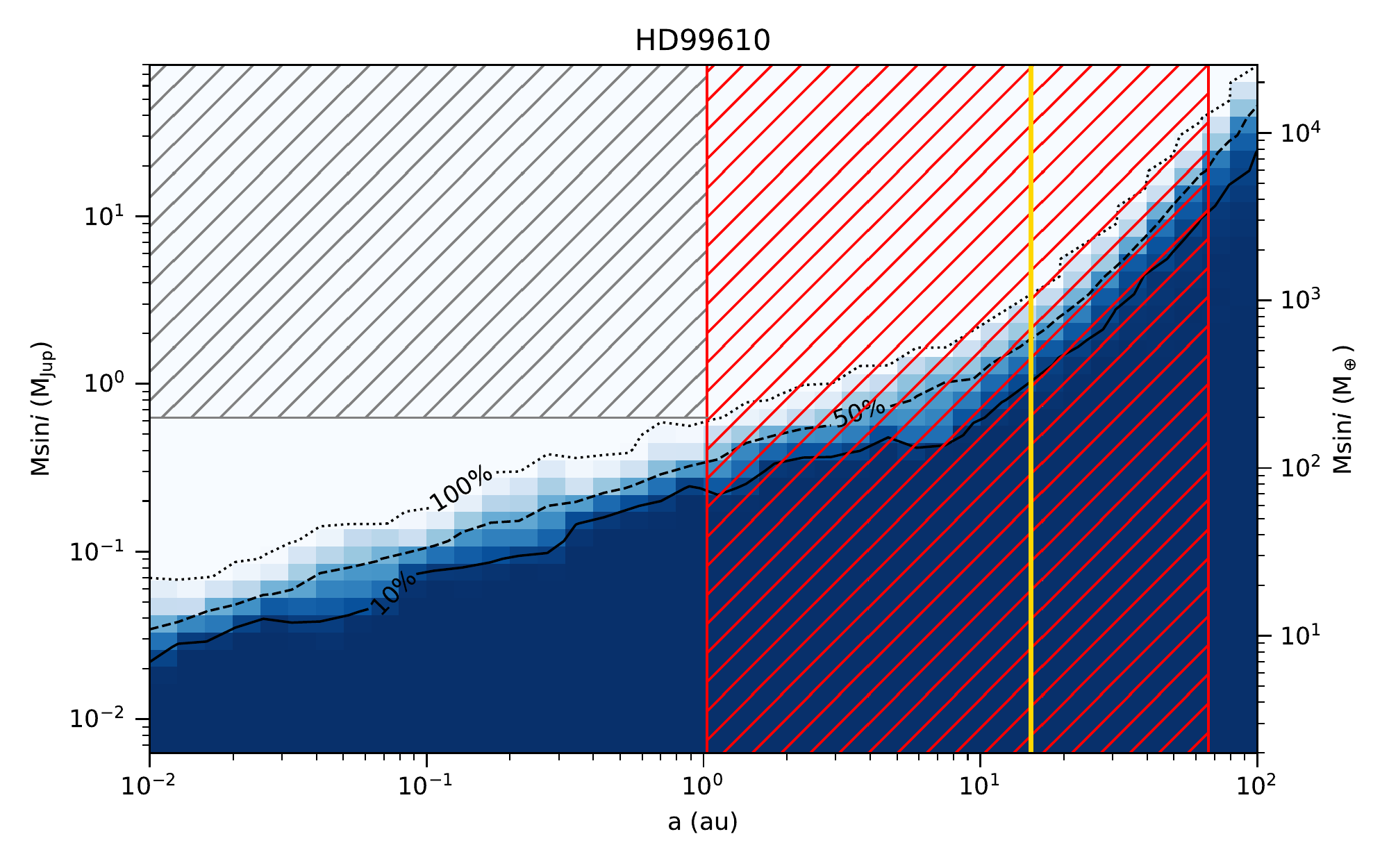}&
    		\includegraphics[width=0.22\linewidth]{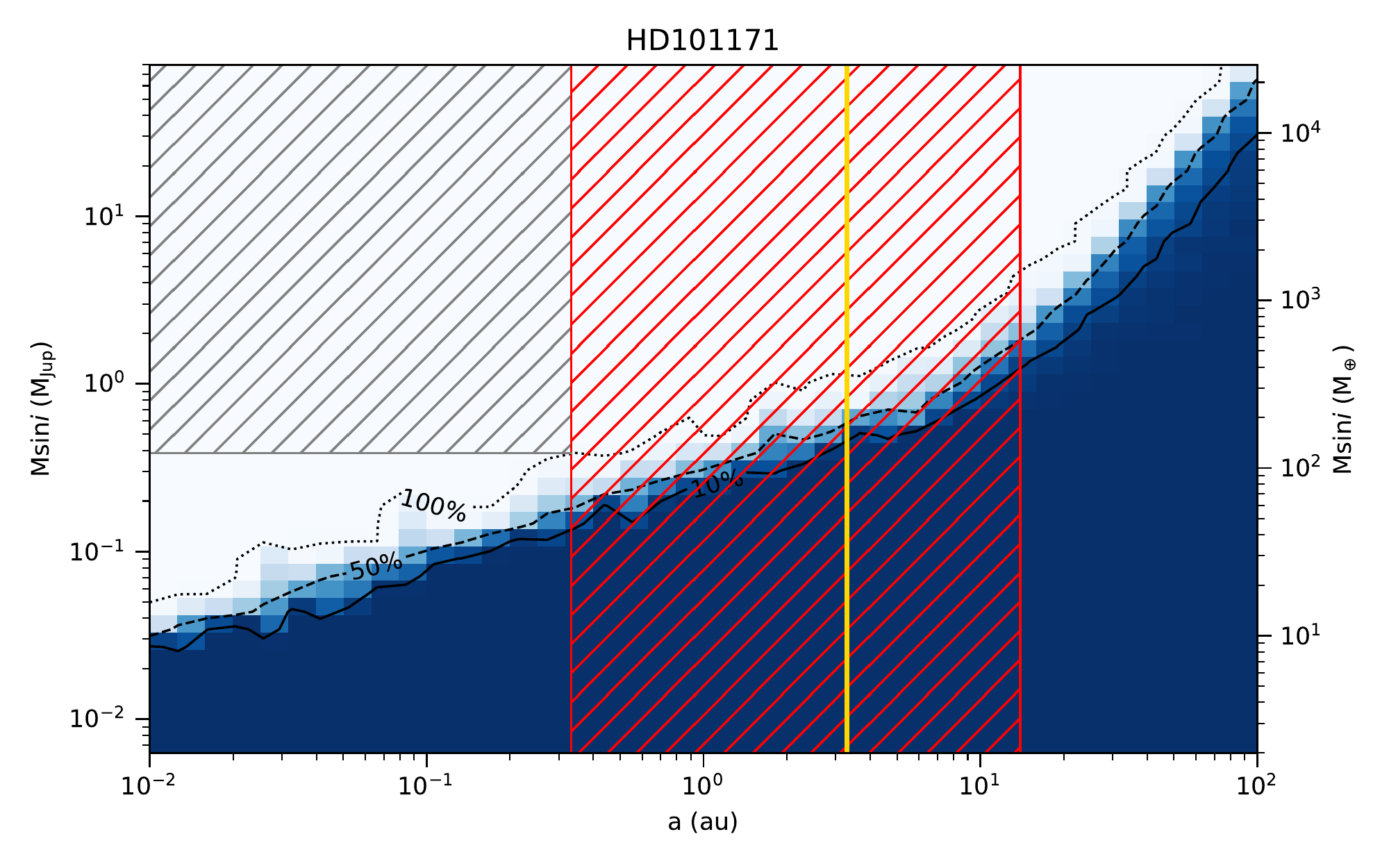}&
    		\includegraphics[width=0.22\linewidth]{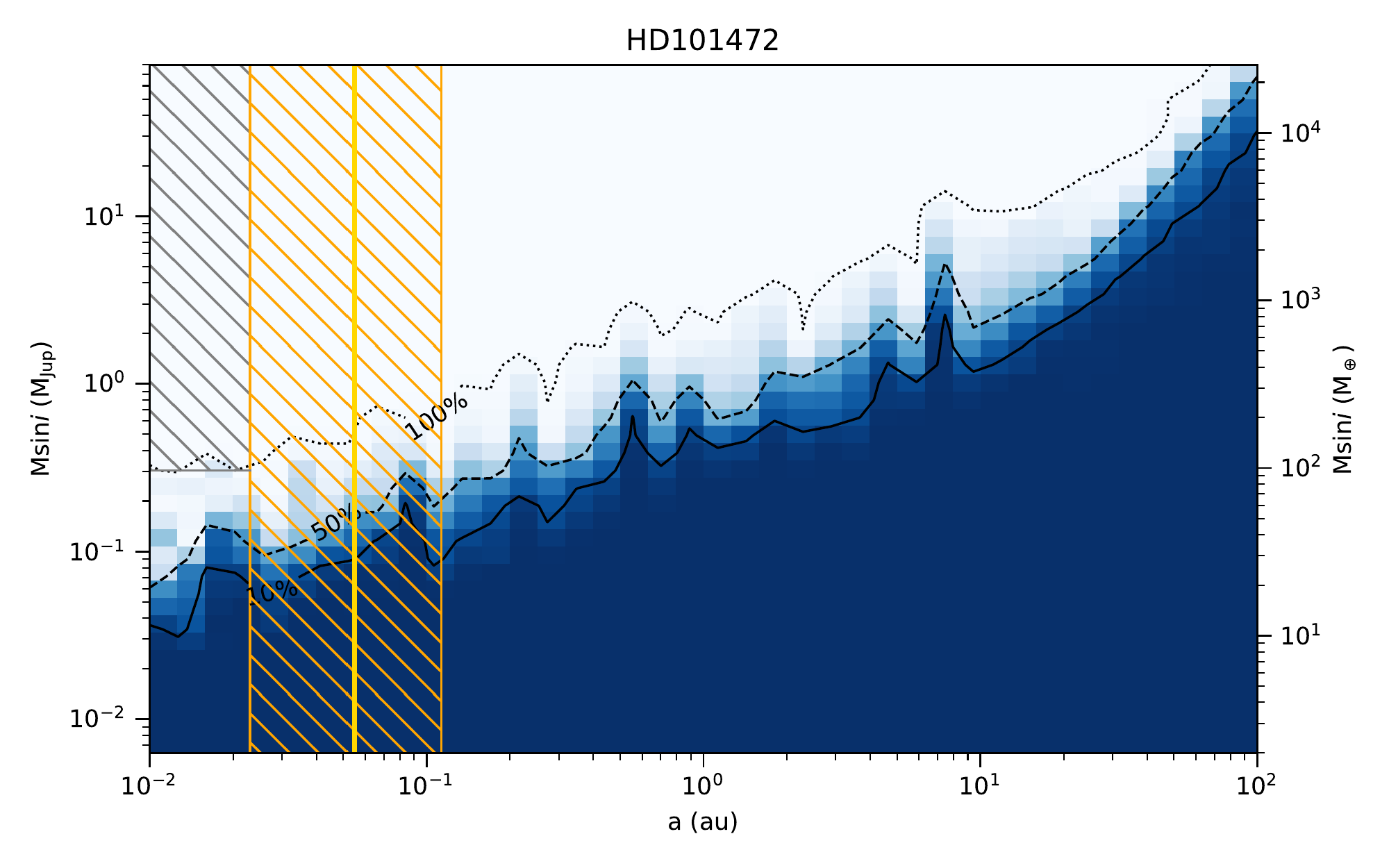}\\
    
    		\includegraphics[width=0.22\linewidth]{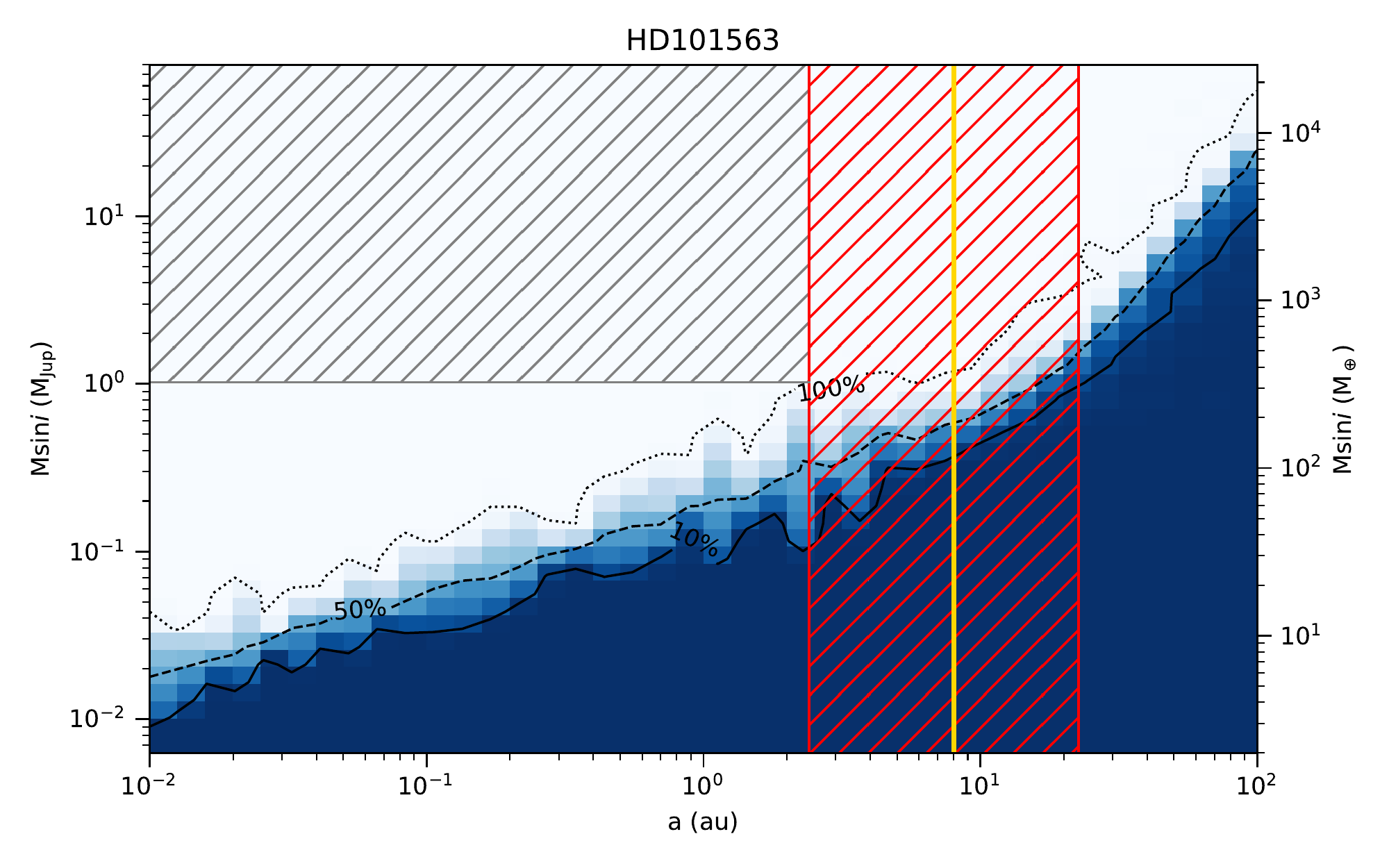}&
    		\includegraphics[width=0.22\linewidth]{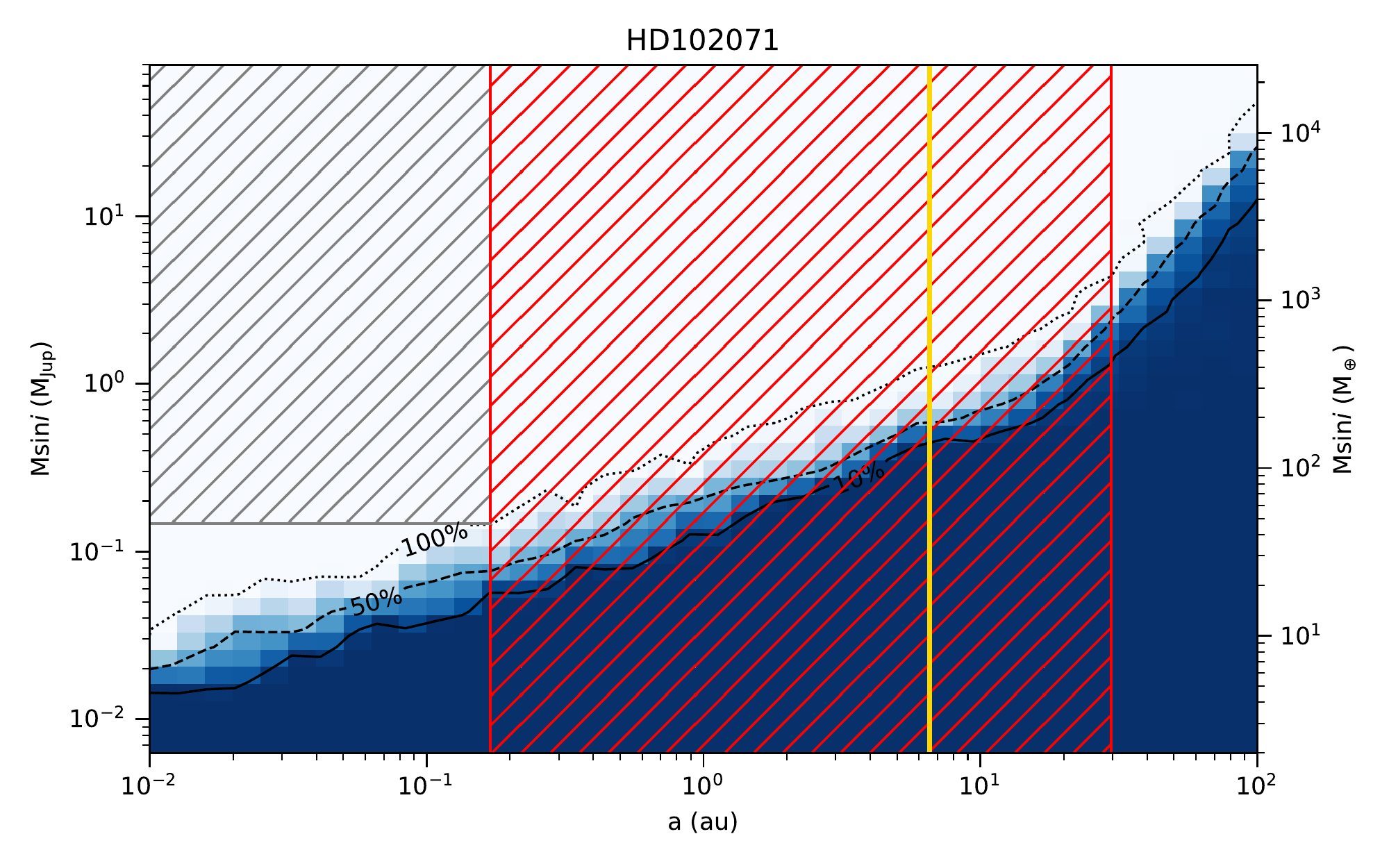}&
    		\includegraphics[width=0.22\linewidth]{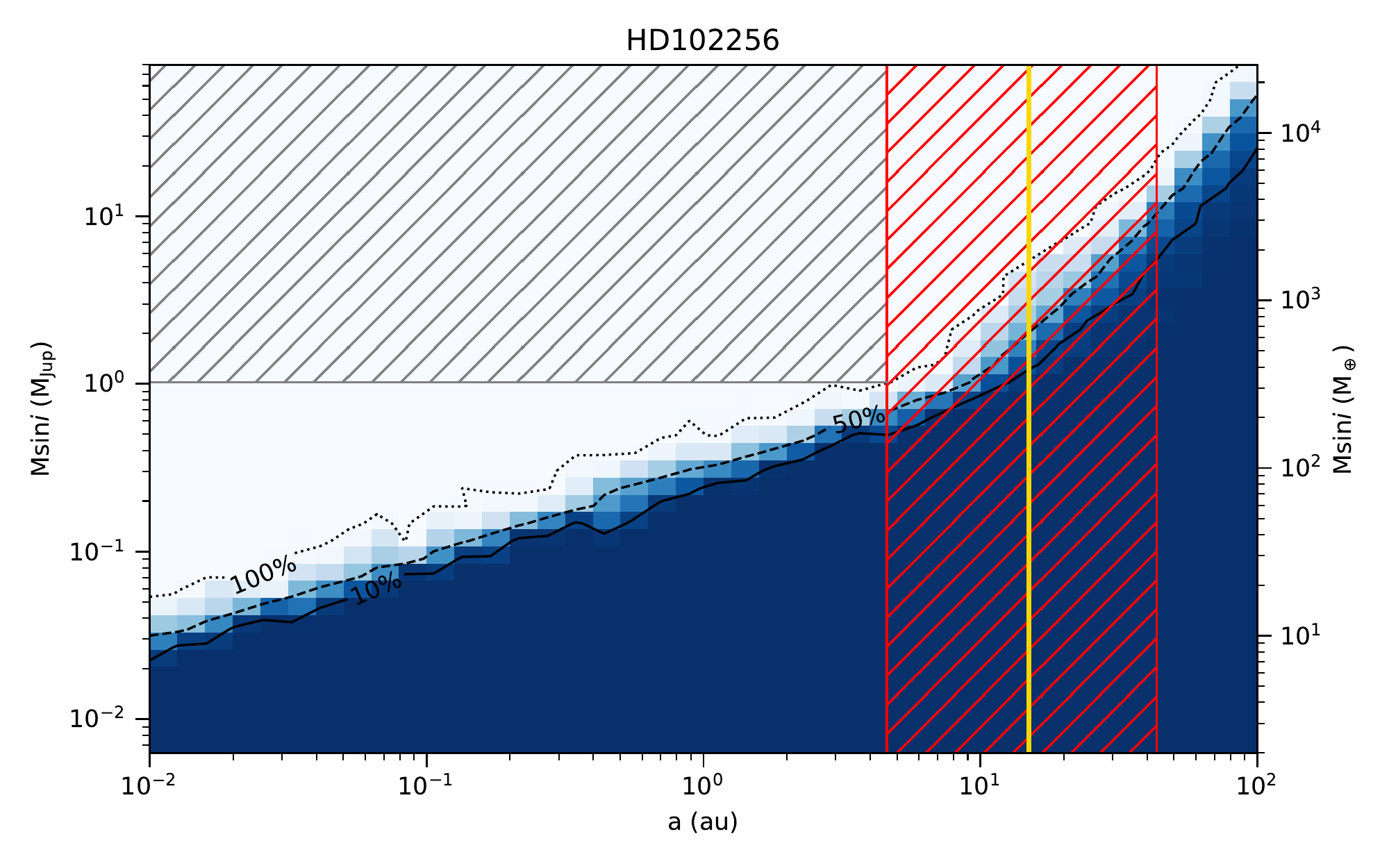}&
    		\includegraphics[width=0.22\linewidth]{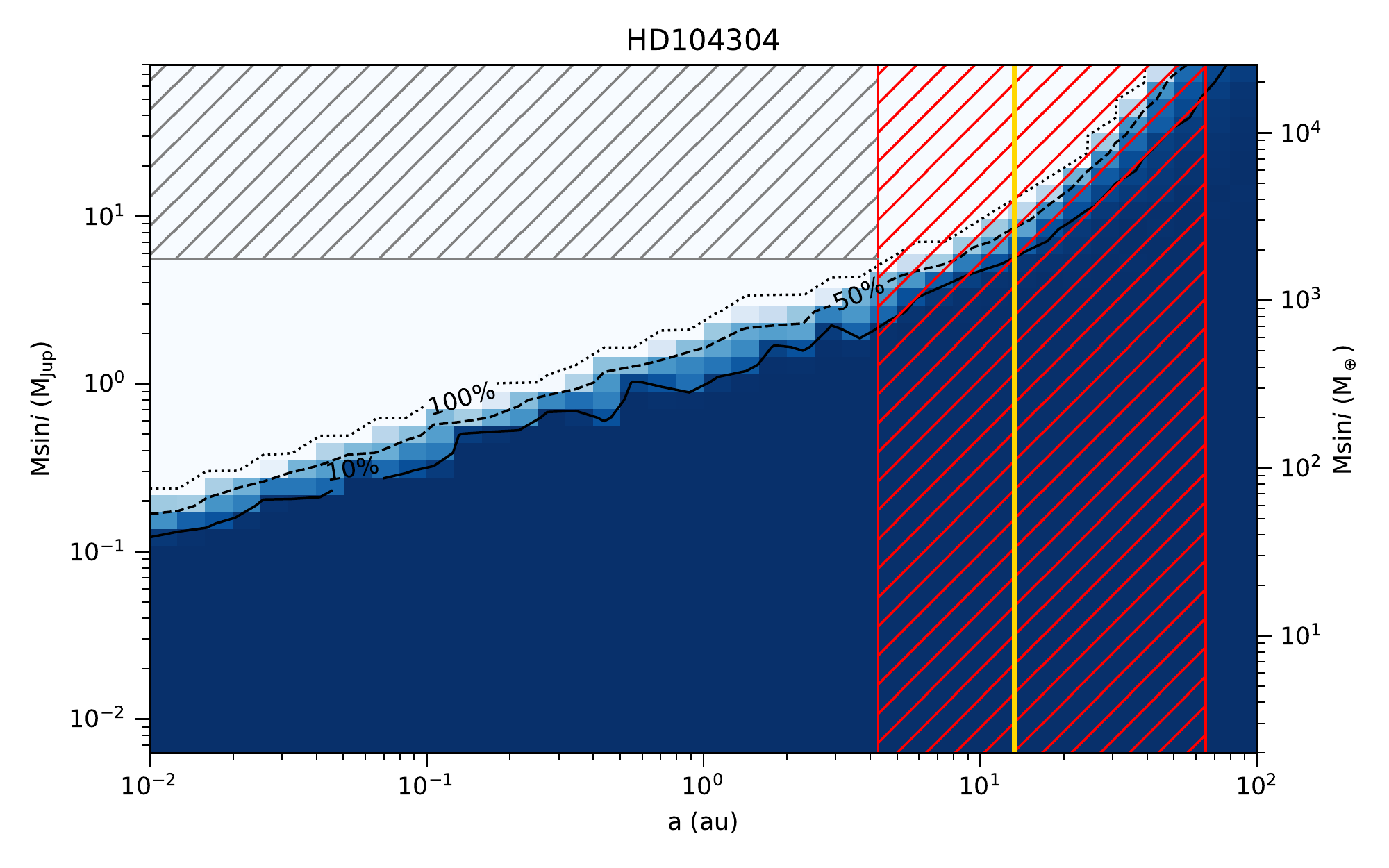}\\
    
    		\includegraphics[width=0.22\linewidth]{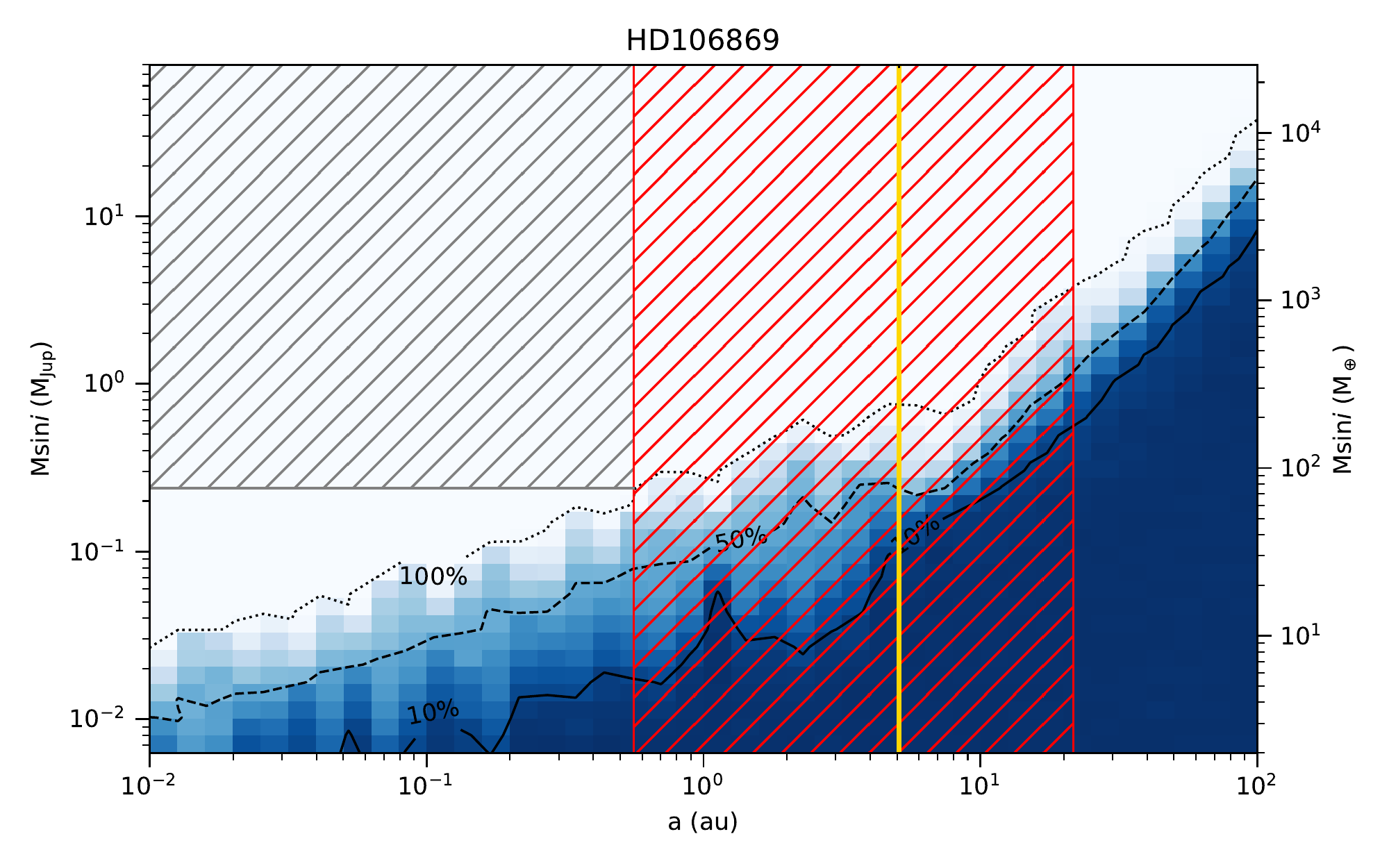}&
    		\includegraphics[width=0.22\linewidth]{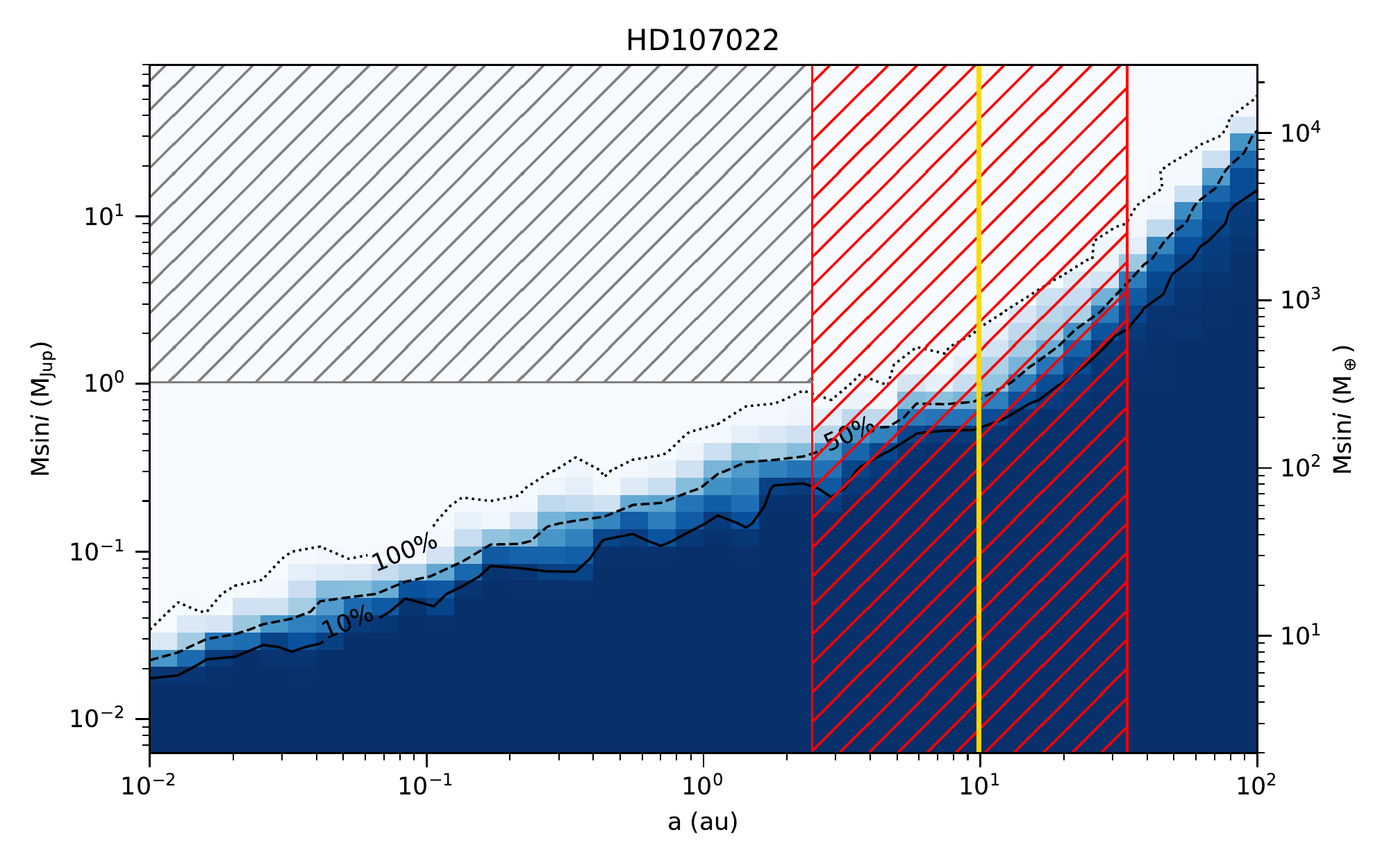}&
    		\includegraphics[width=0.22\linewidth]{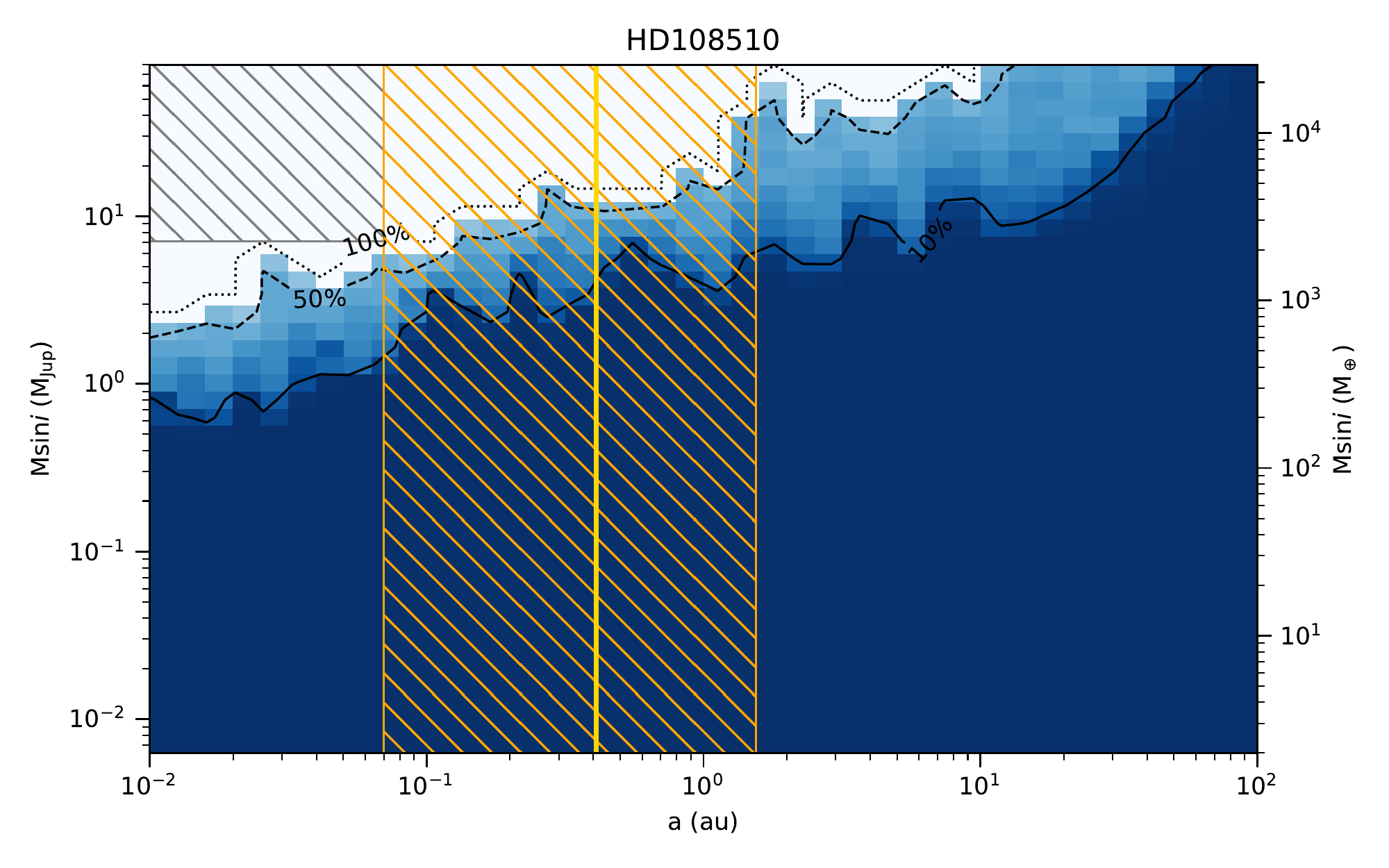}&
    		\includegraphics[width=0.22\linewidth]{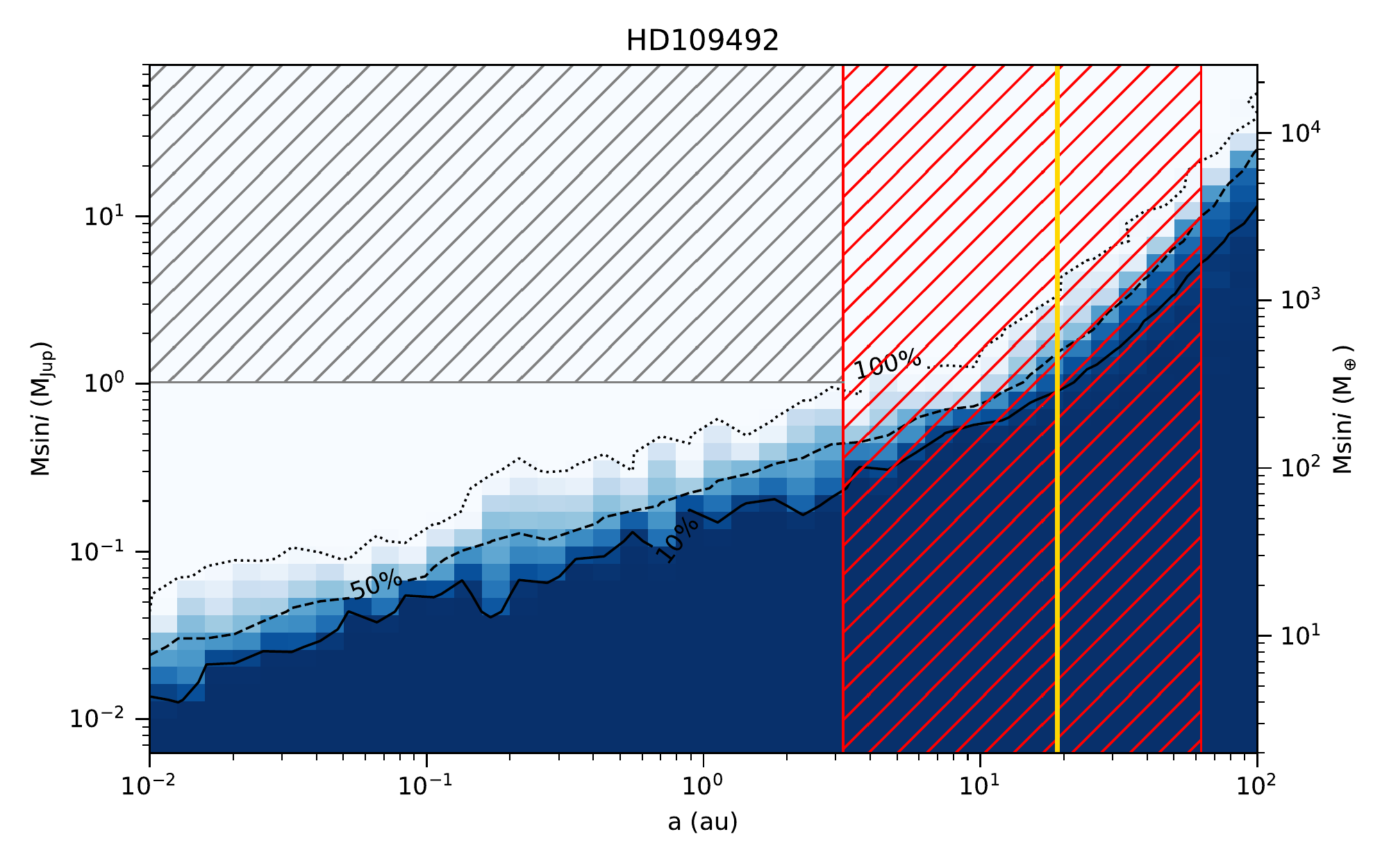}\\
    
    		\includegraphics[width=0.22\linewidth]{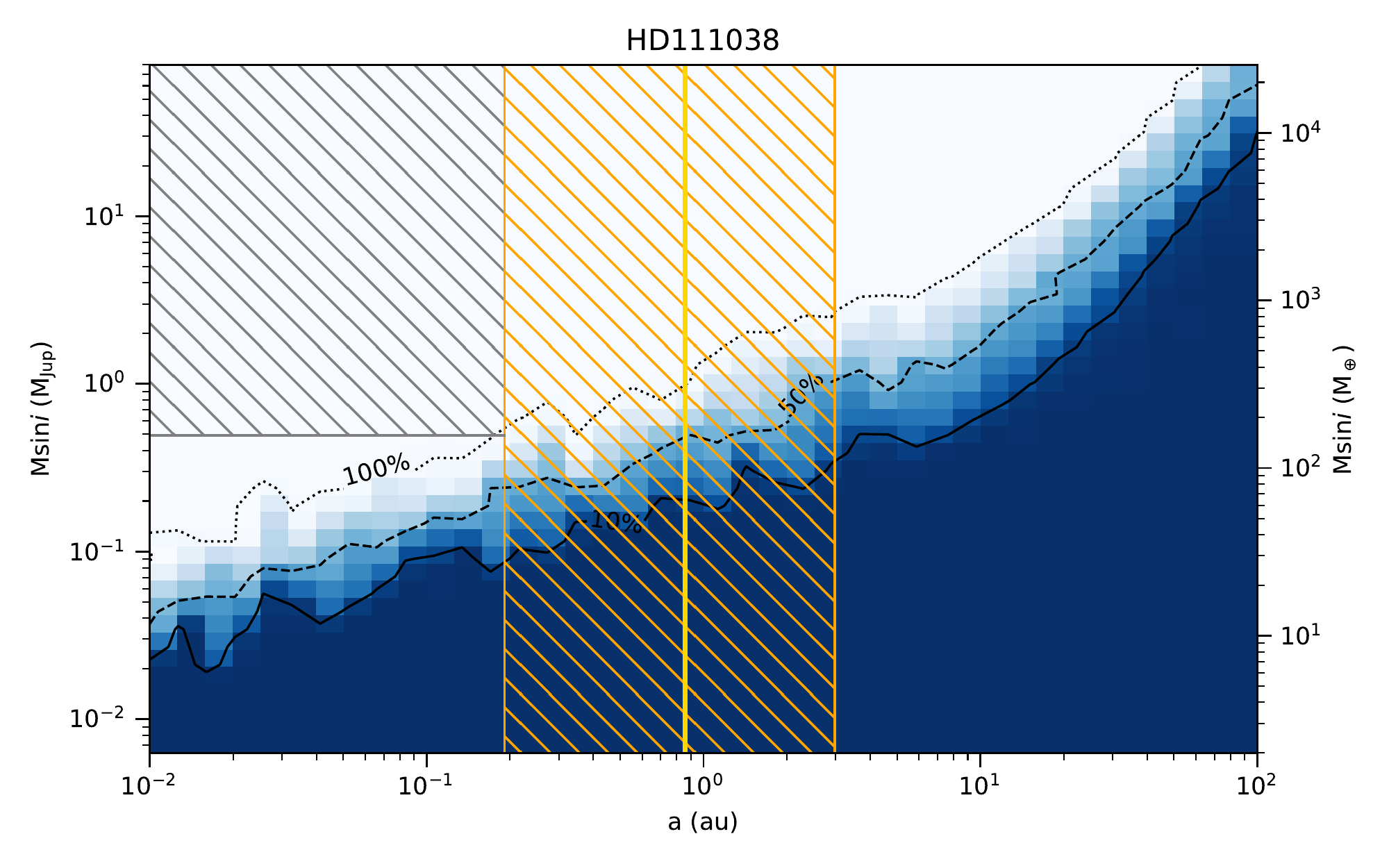}&
    		\includegraphics[width=0.22\linewidth]{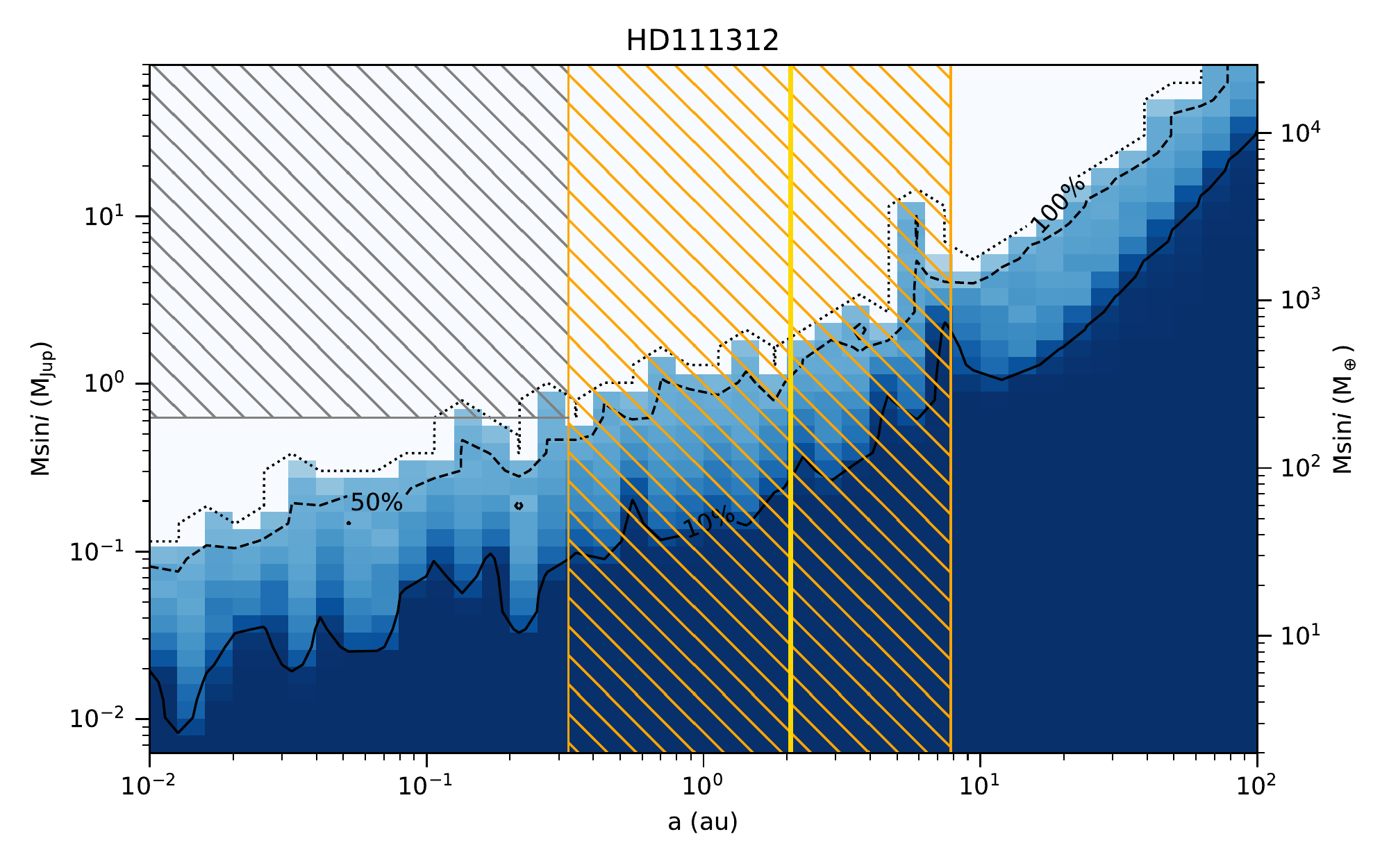}&
    		\includegraphics[width=0.22\linewidth]{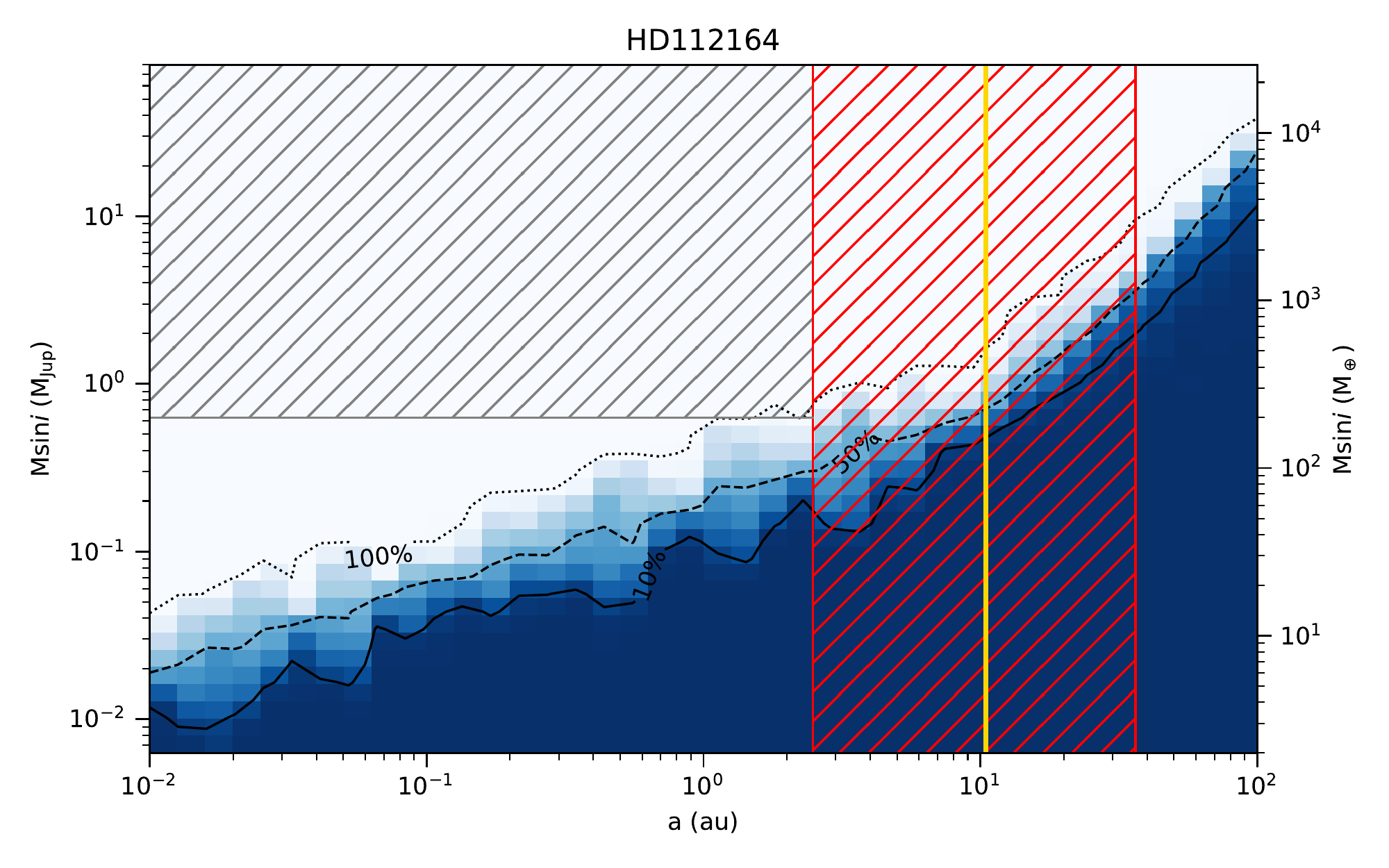}&
    		\includegraphics[width=0.22\linewidth]{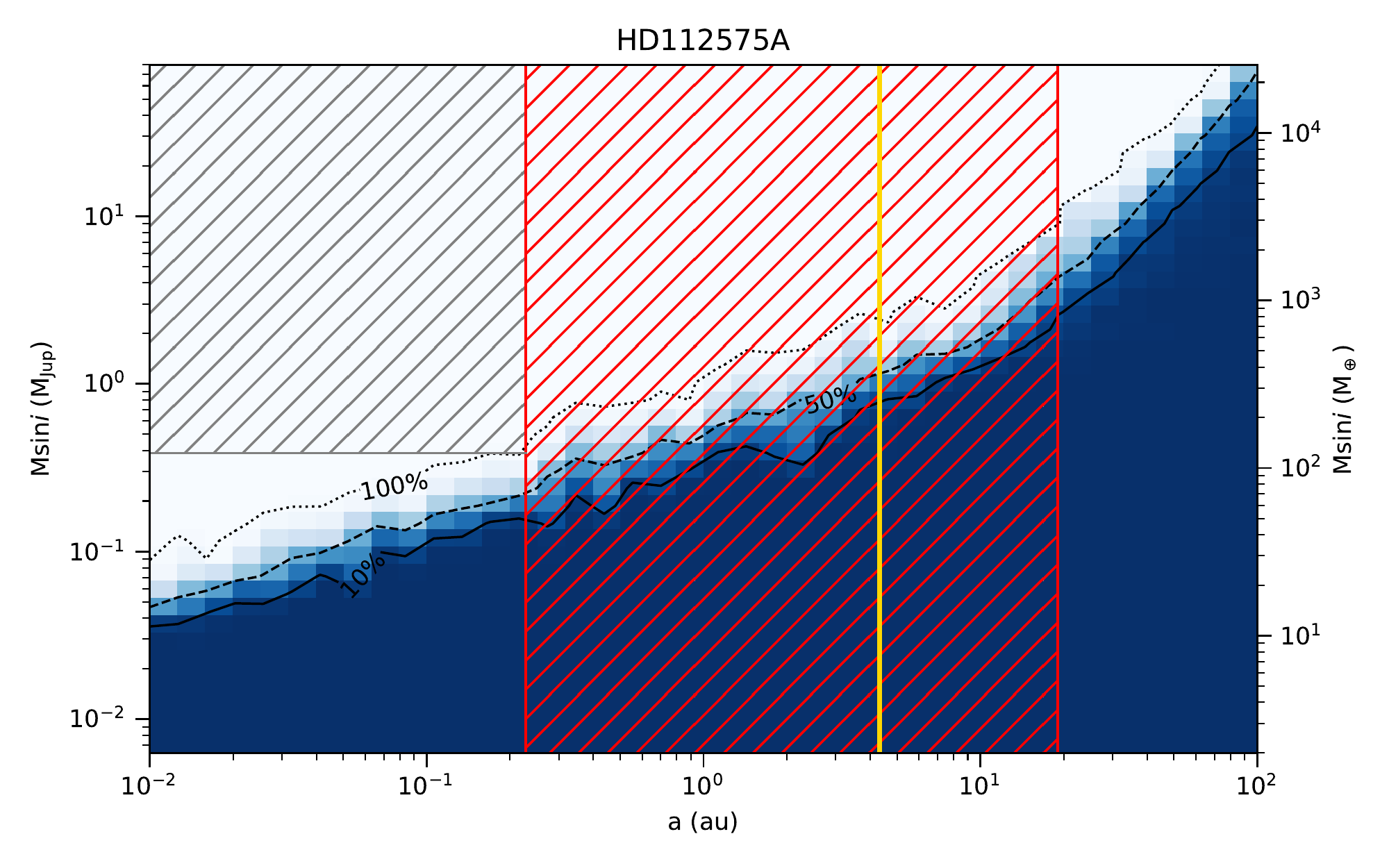}\\
    
    		\includegraphics[width=0.22\linewidth]{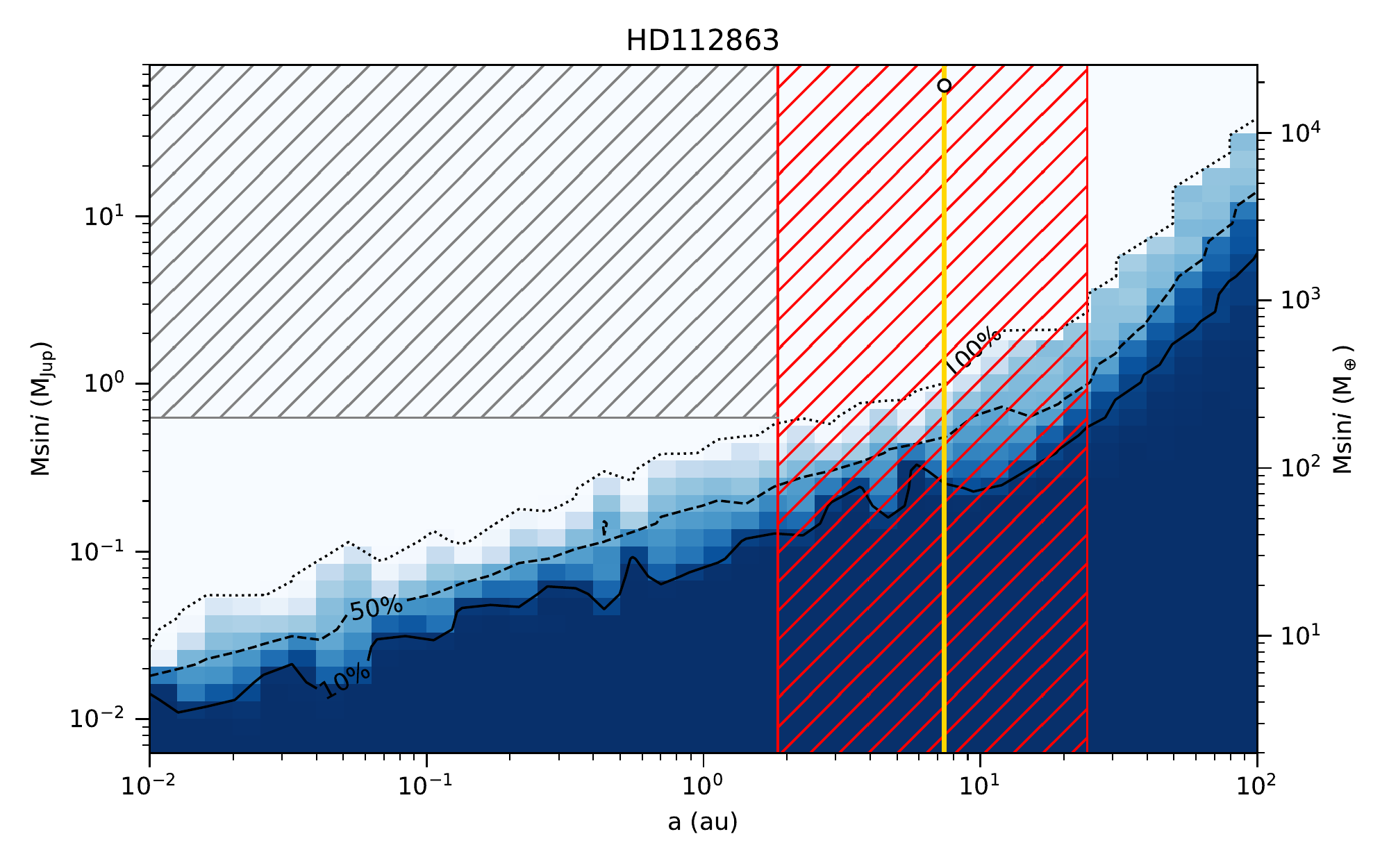}&
    		\includegraphics[width=0.22\linewidth]{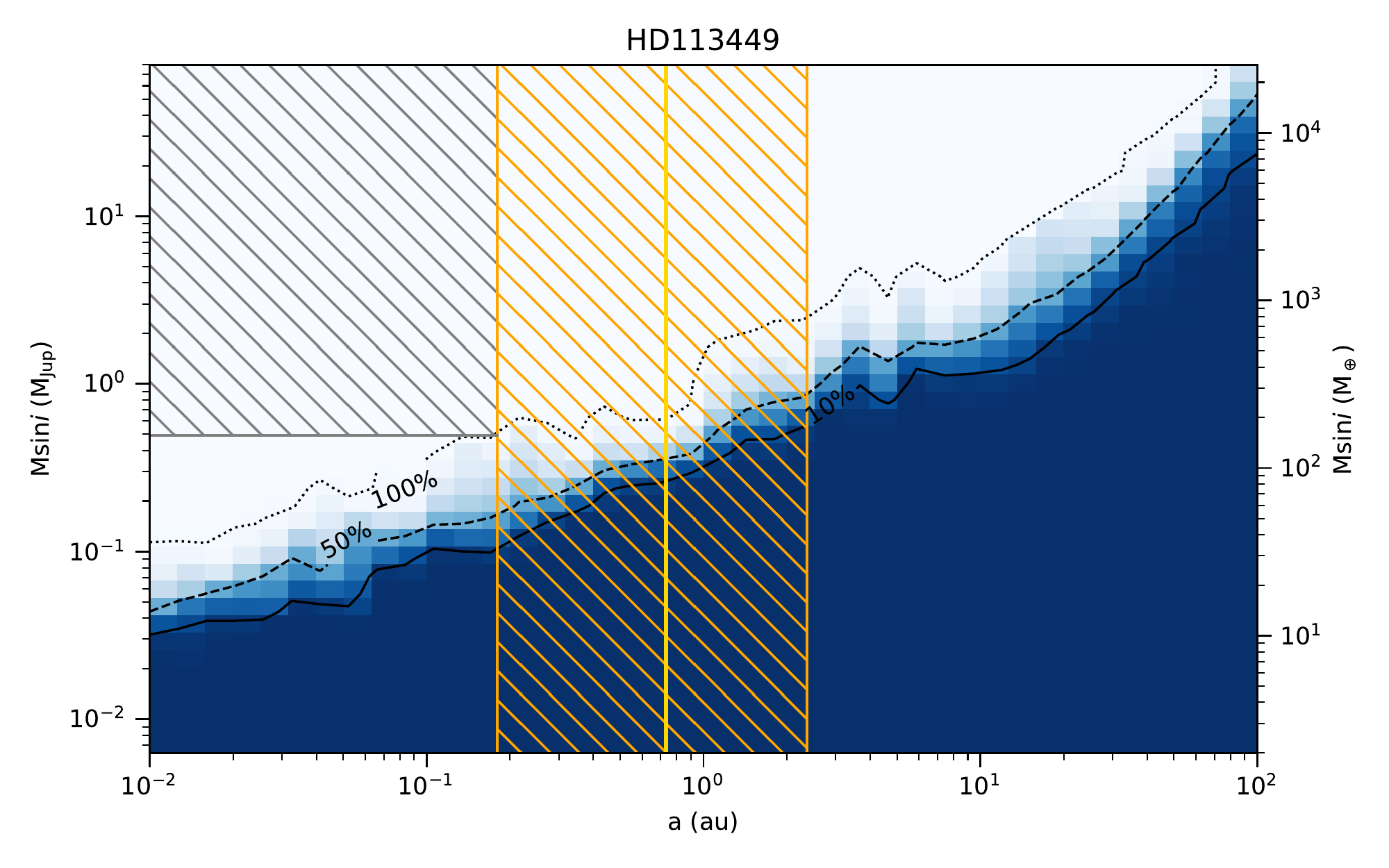}&
    		\includegraphics[width=0.22\linewidth]{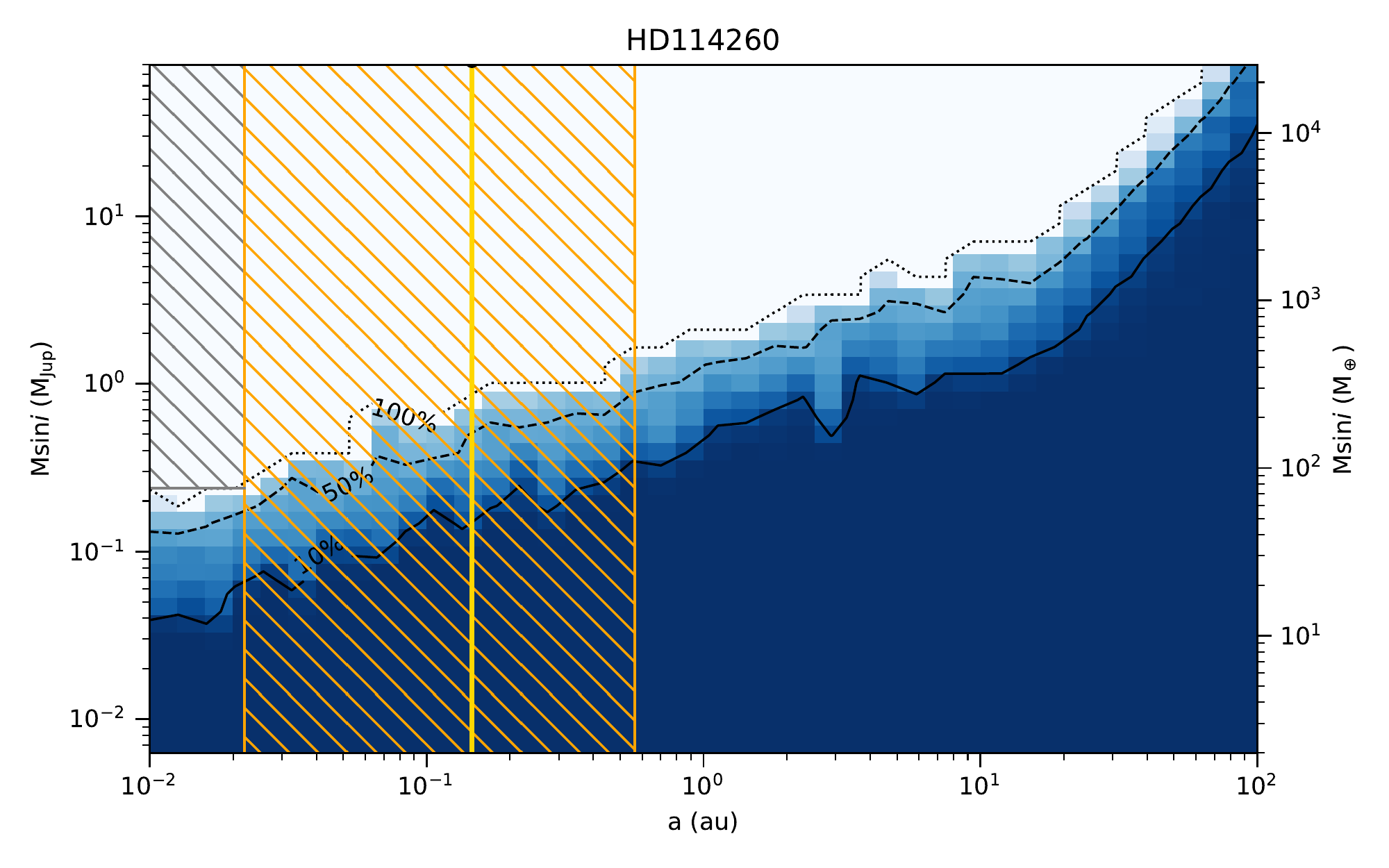}&
    		\includegraphics[width=0.22\linewidth]{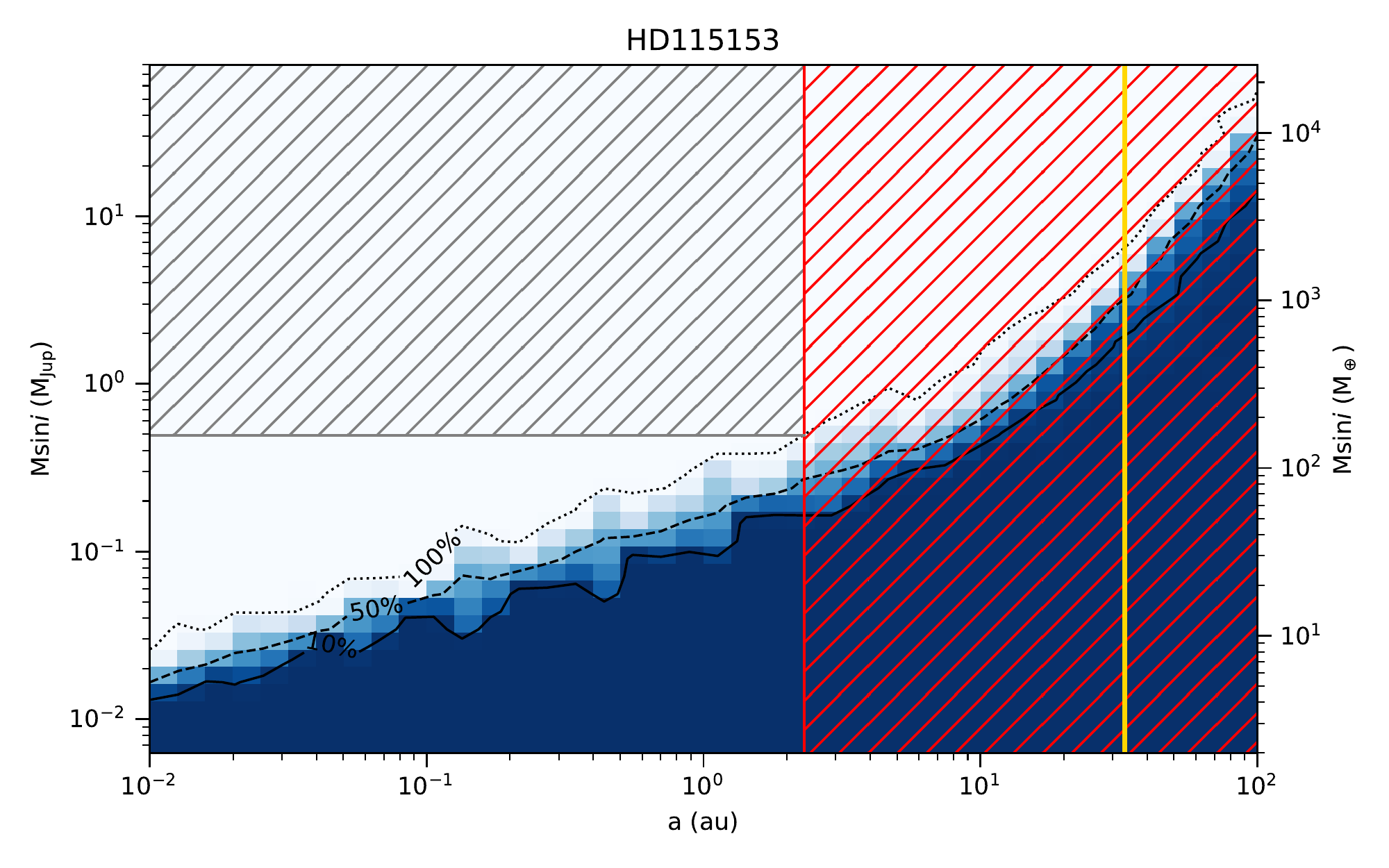}\\
    
    		\includegraphics[width=0.22\linewidth]{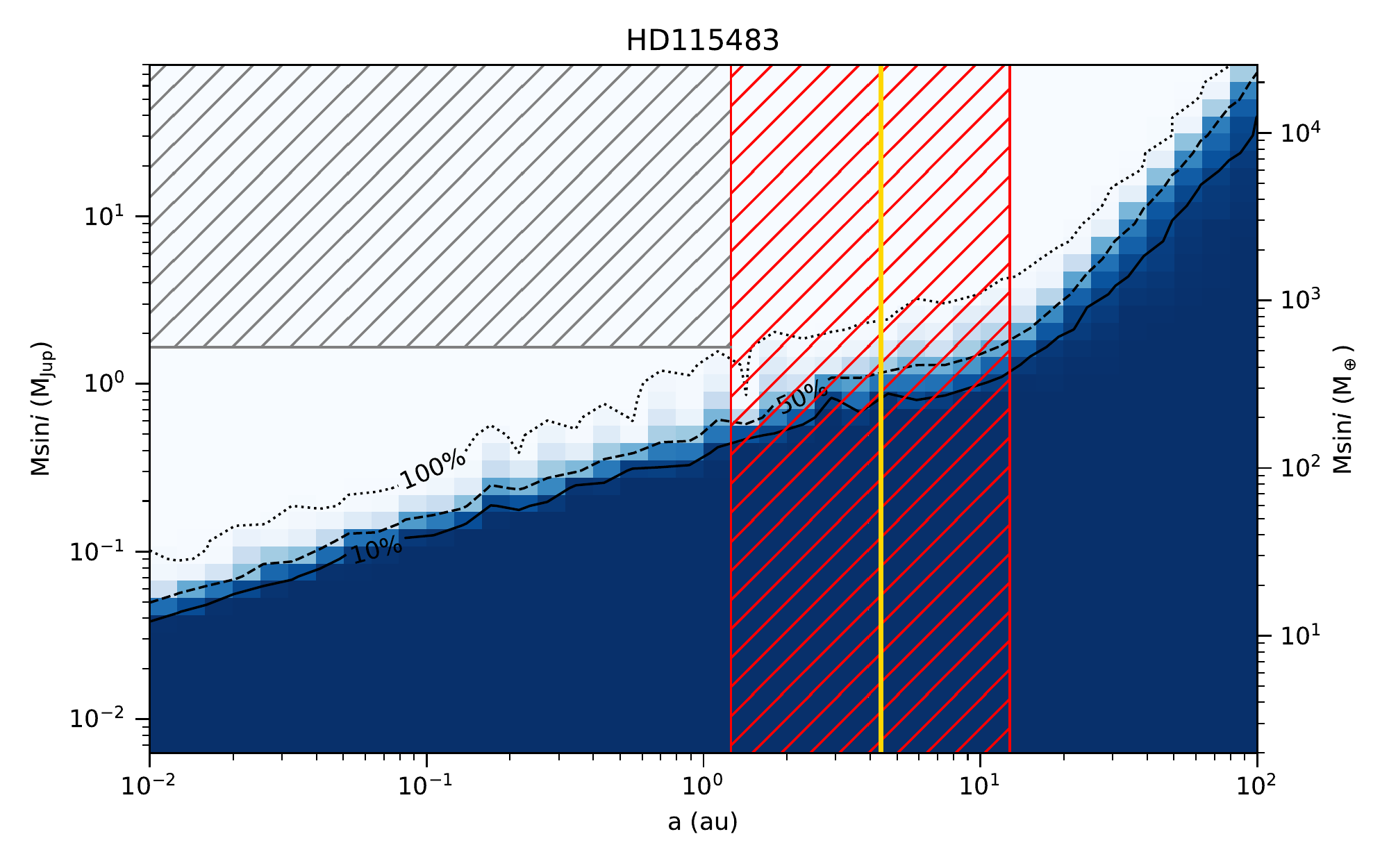}&
    		\includegraphics[width=0.22\linewidth]{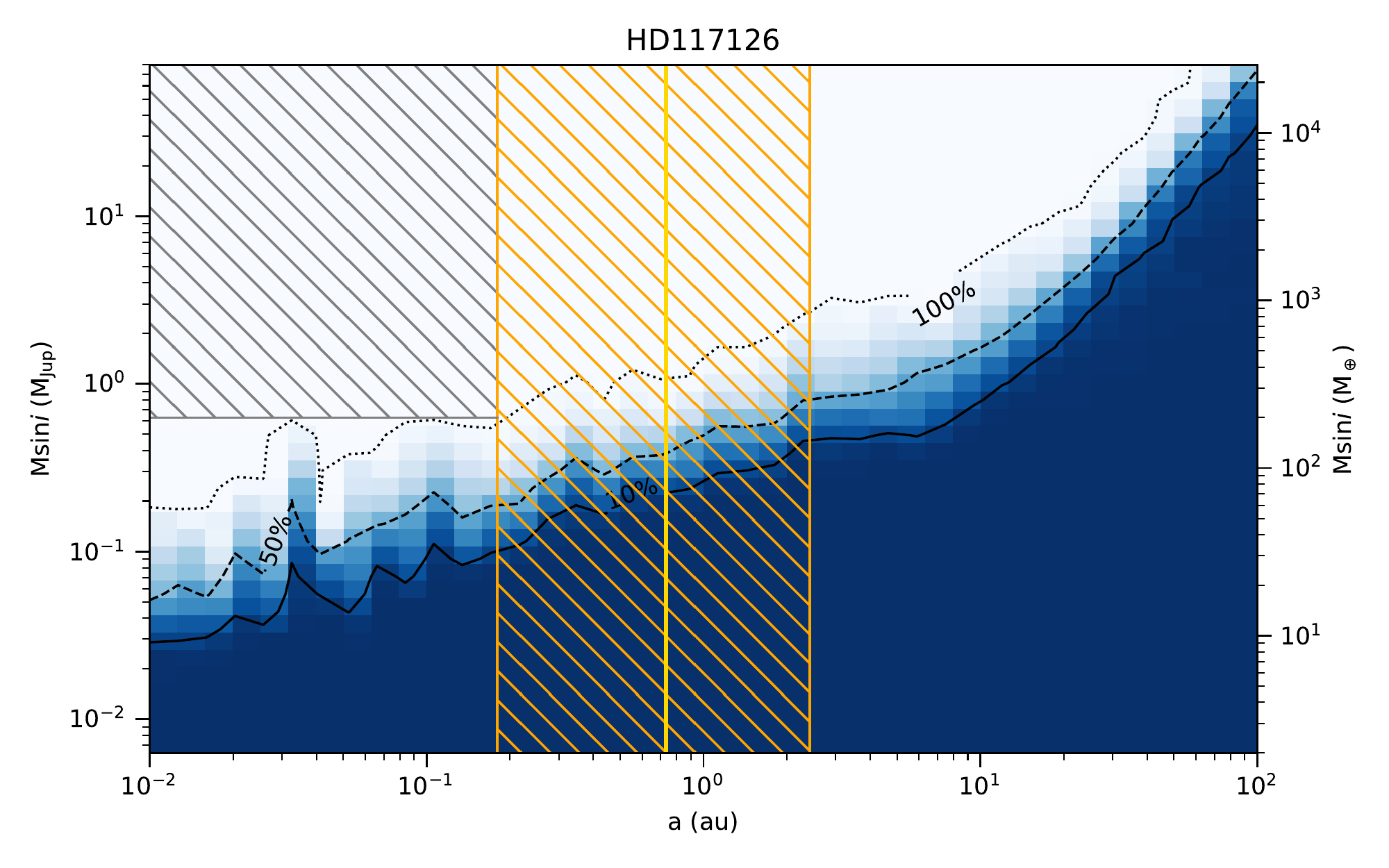}&
    		\includegraphics[width=0.22\linewidth]{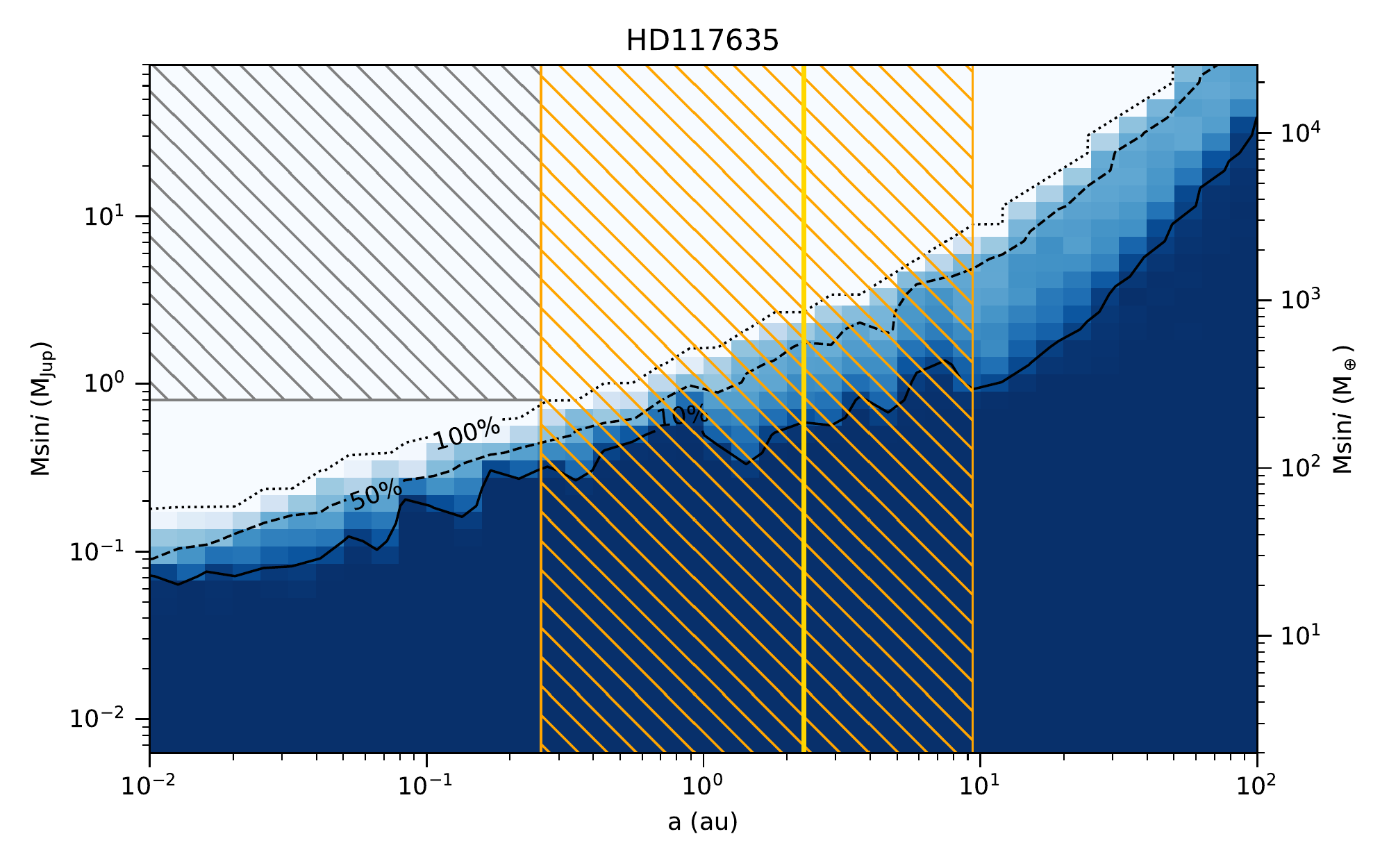}&
    		\includegraphics[width=0.22\linewidth]{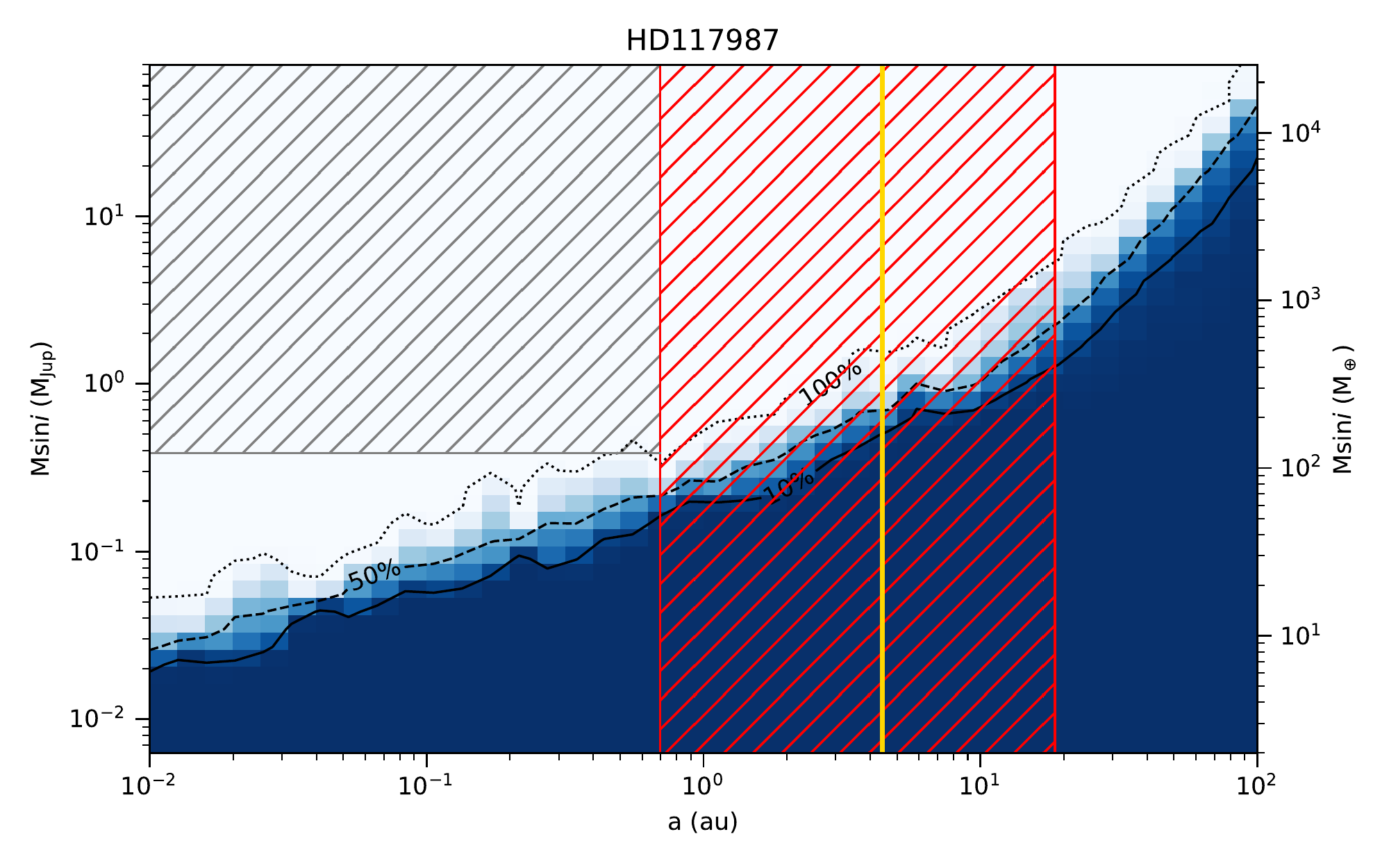}\\
    
    		\includegraphics[width=0.22\linewidth]{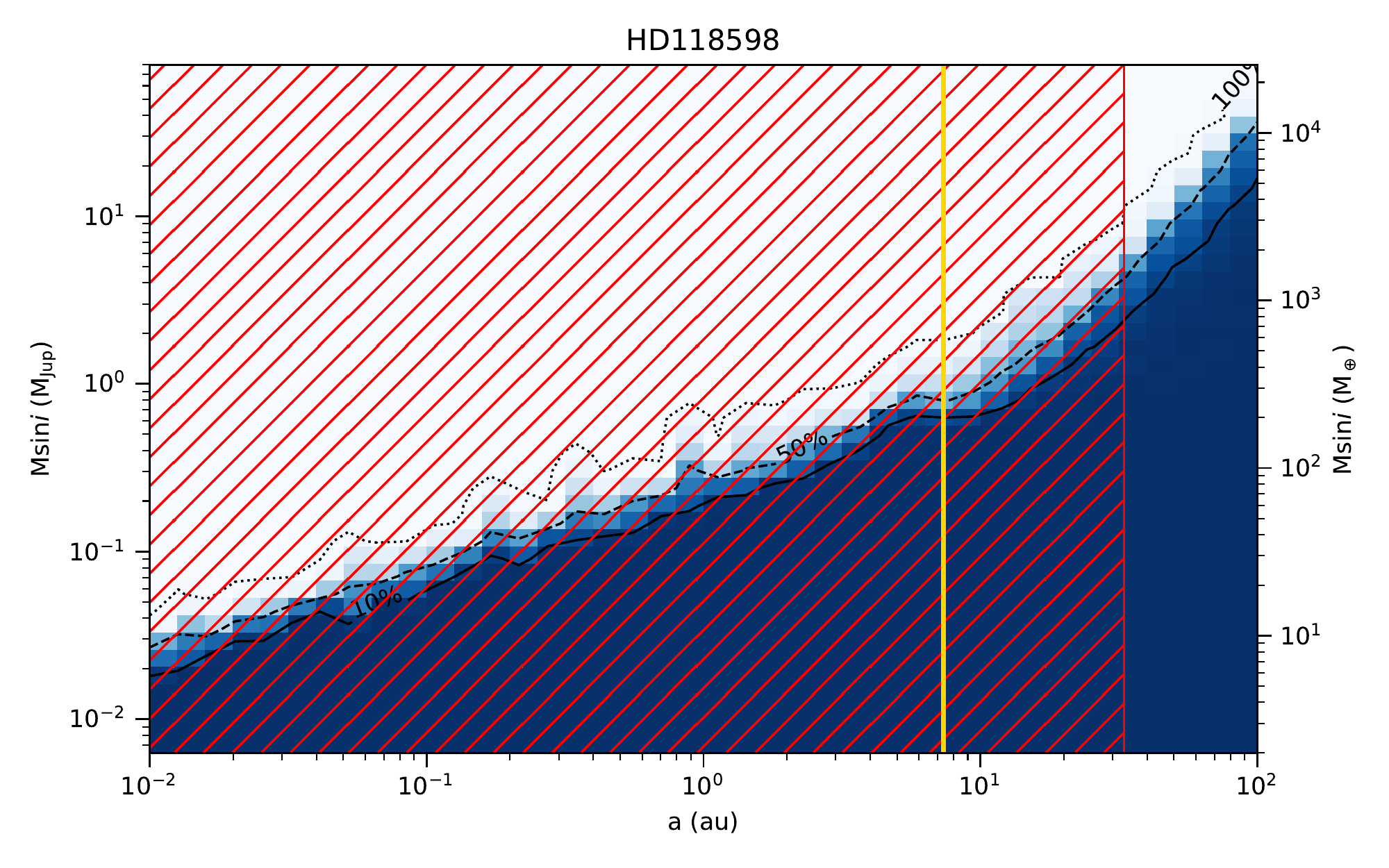}&
    		\includegraphics[width=0.22\linewidth]{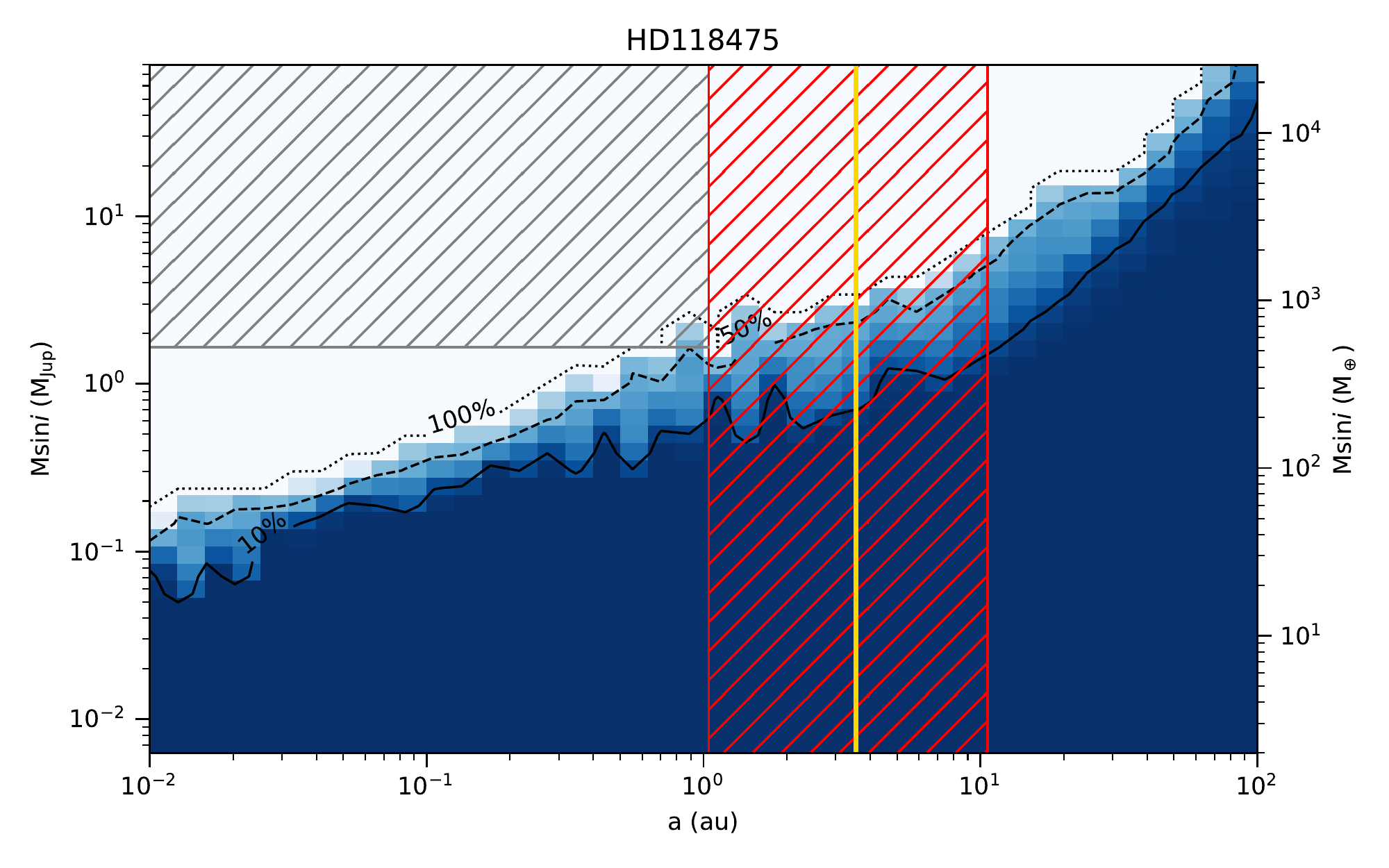}&
    		\includegraphics[width=0.22\linewidth]{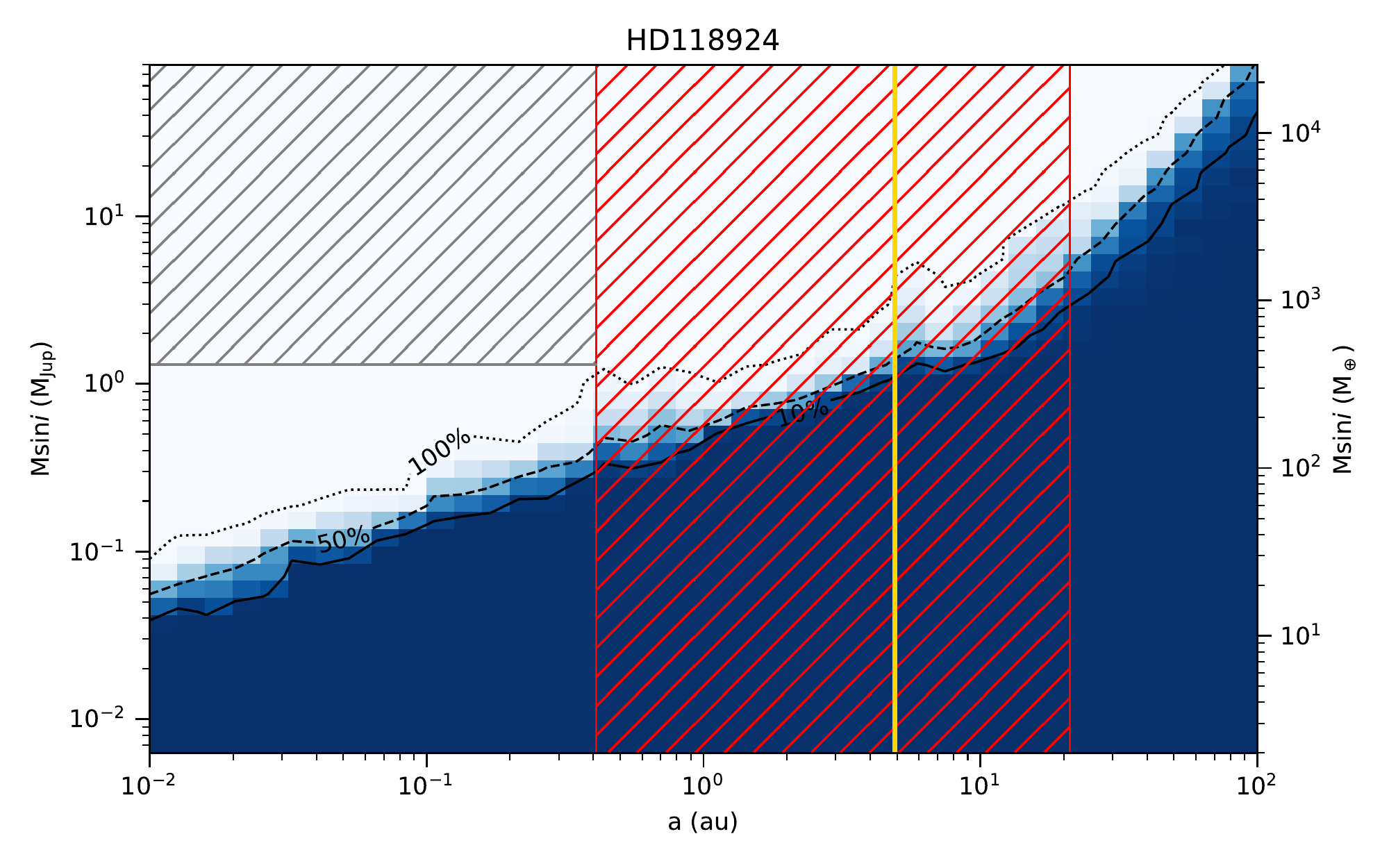}&
    		\includegraphics[width=0.22\linewidth]{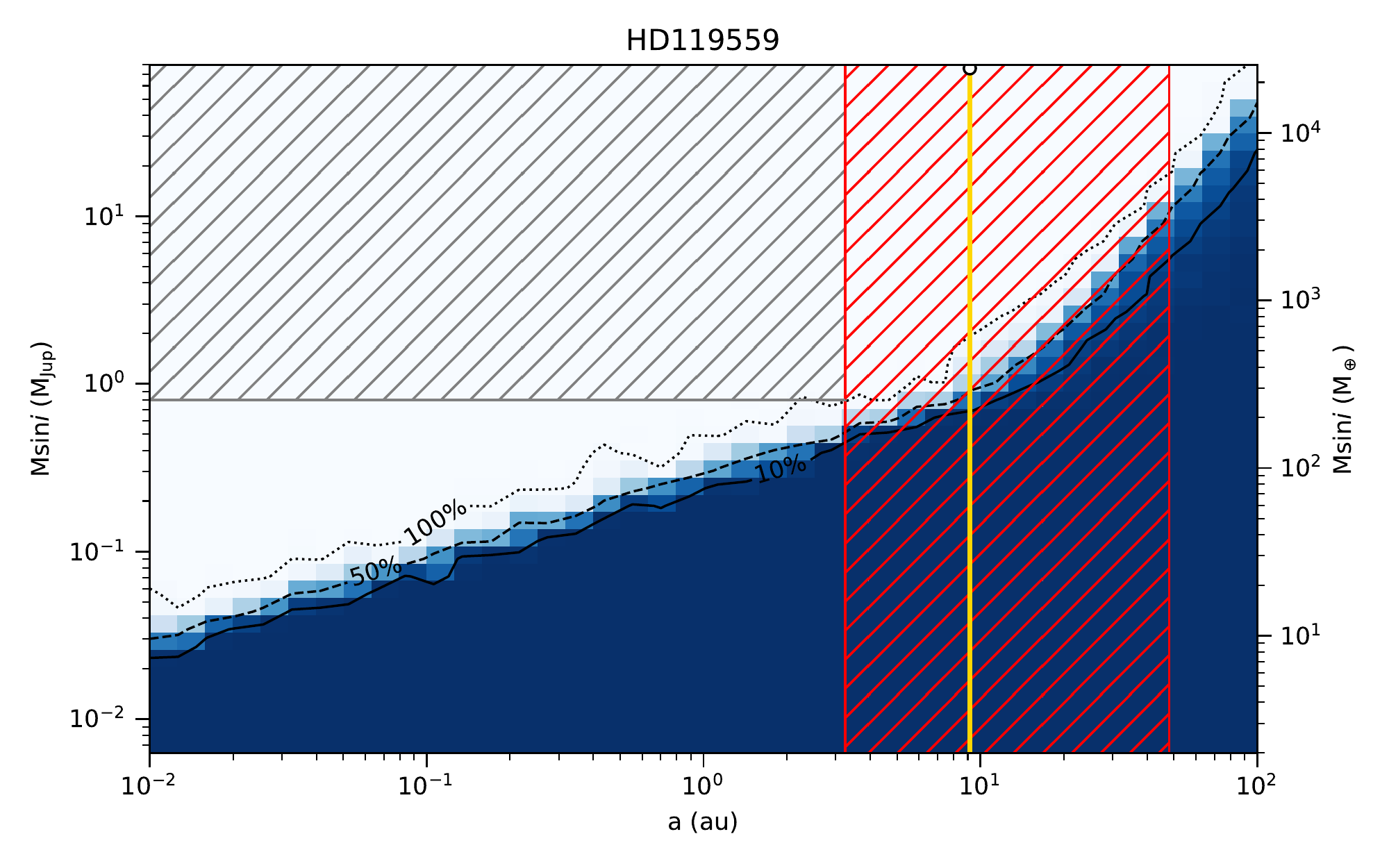}\\
    
    		\includegraphics[width=0.22\linewidth]{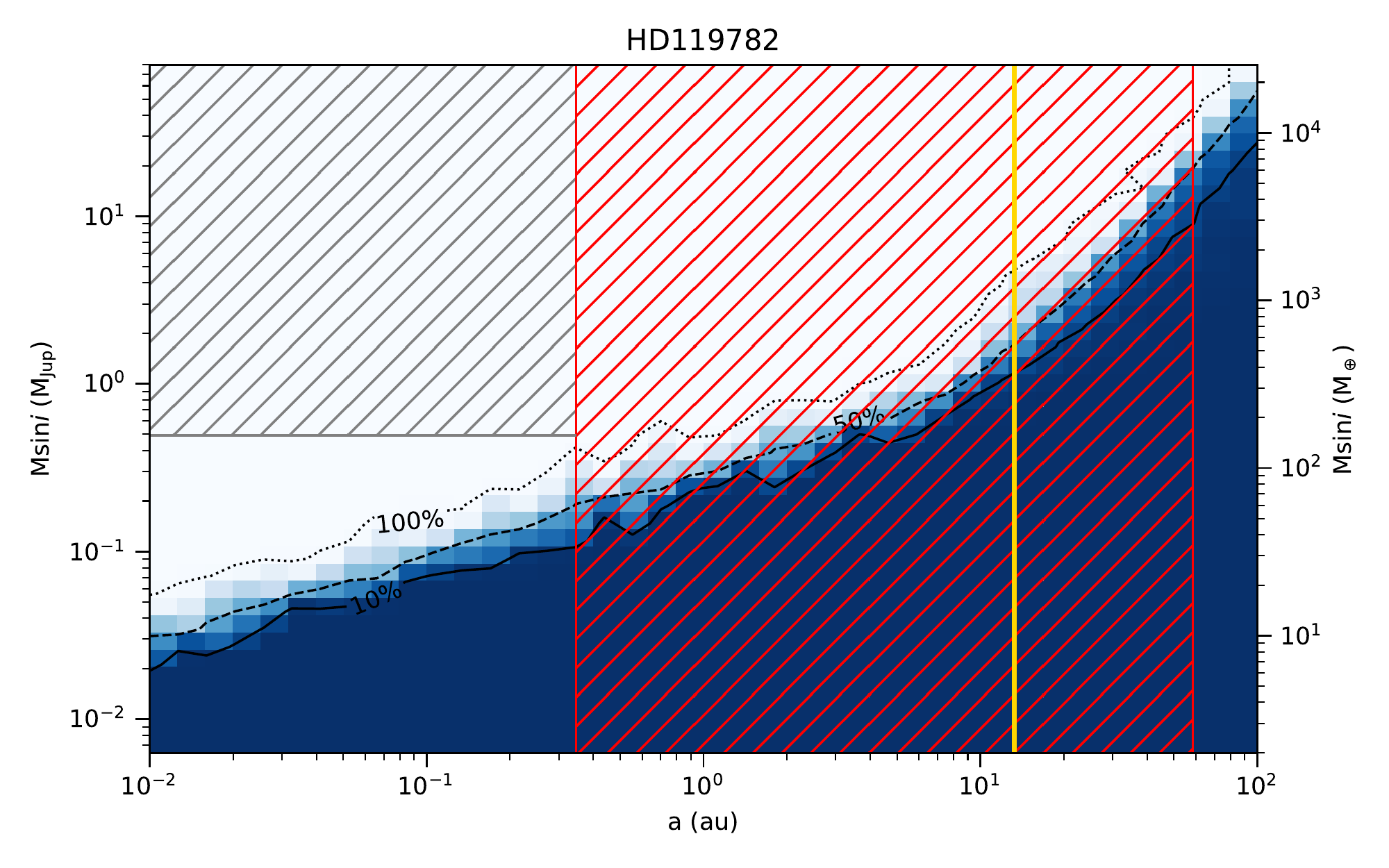}&
    		\includegraphics[width=0.22\linewidth]{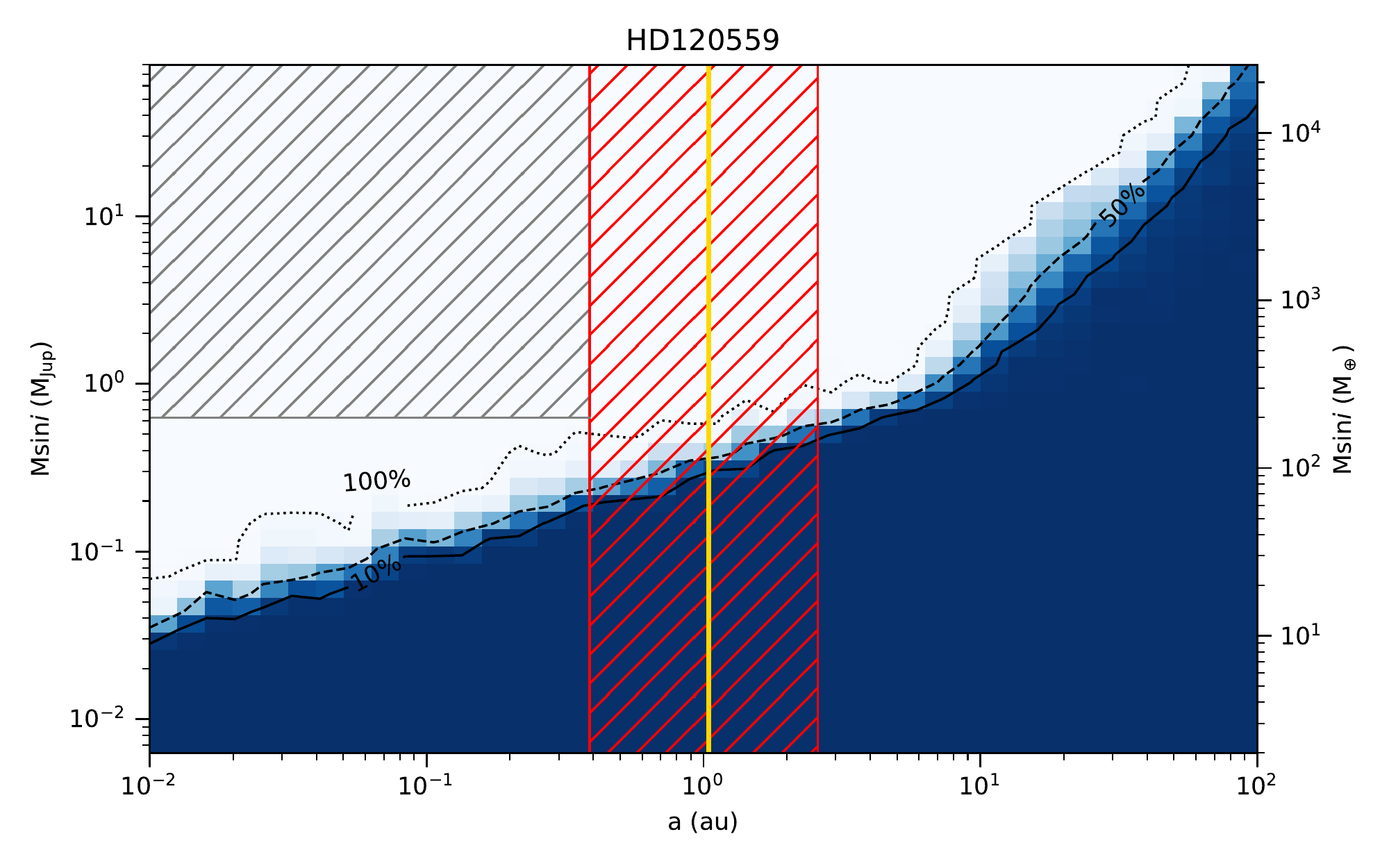}&
    		\includegraphics[width=0.22\linewidth]{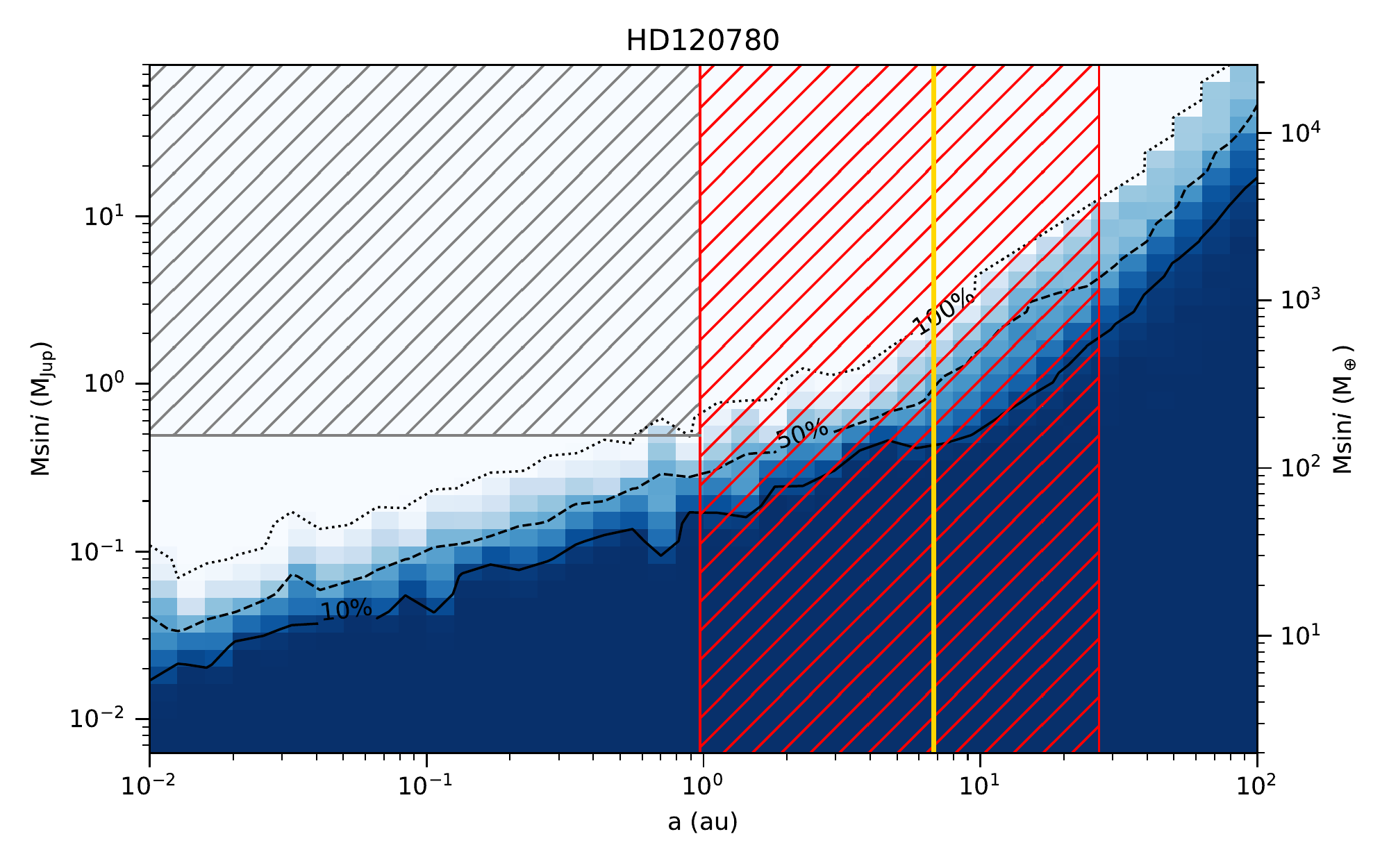}&
    		\includegraphics[width=0.22\linewidth]{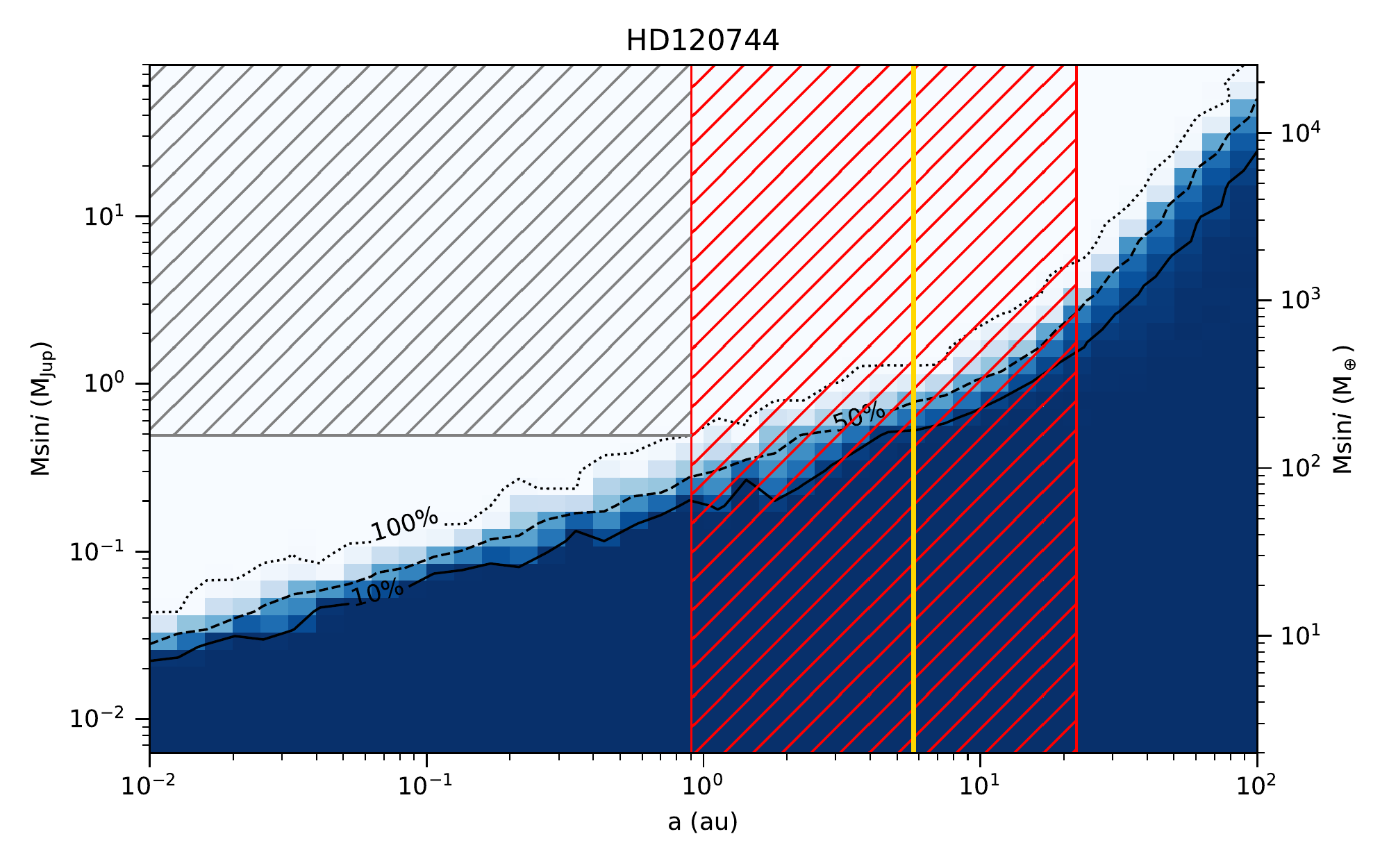}\\
    
    		\includegraphics[width=0.22\linewidth]{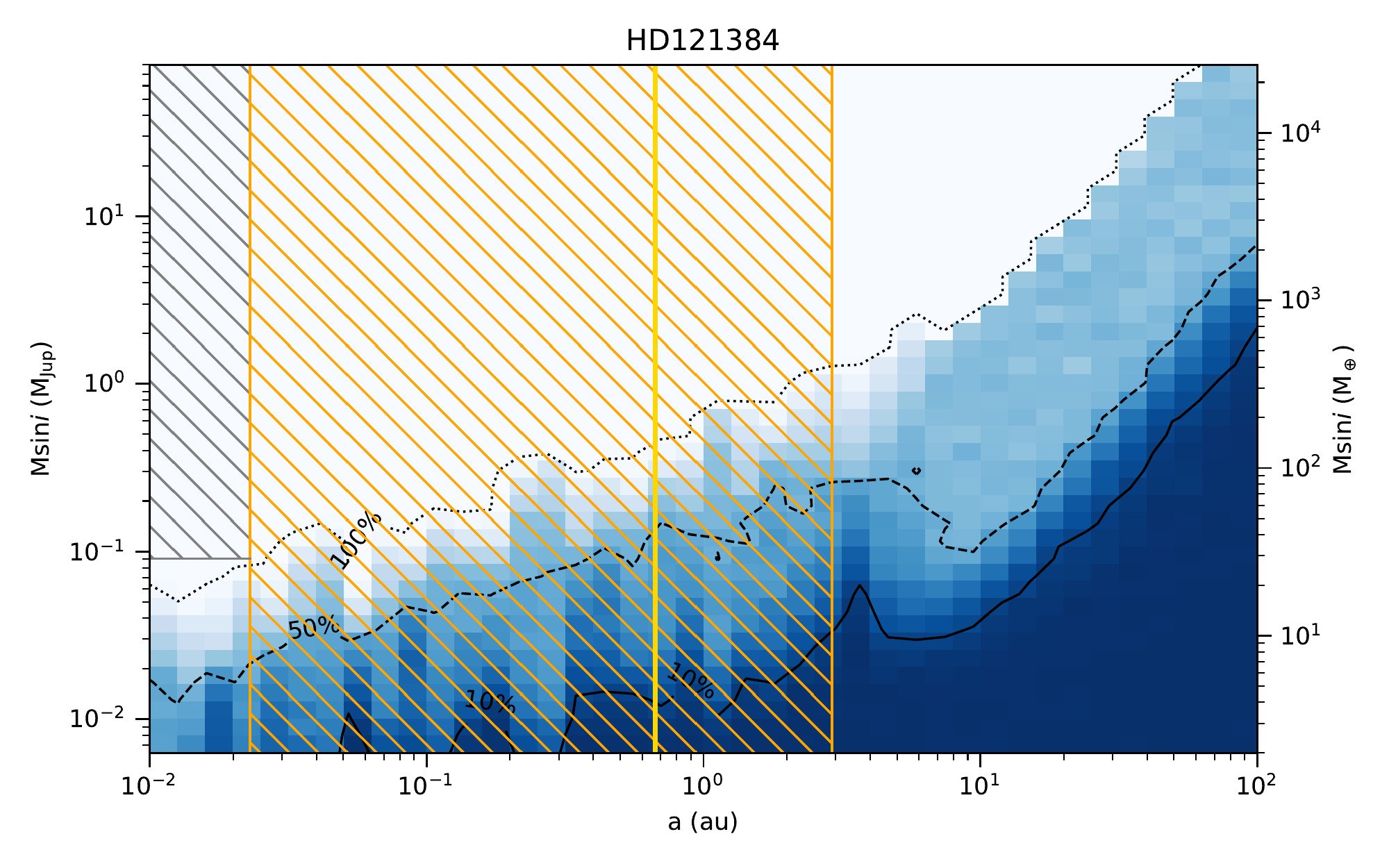}&
    		\includegraphics[width=0.22\linewidth]{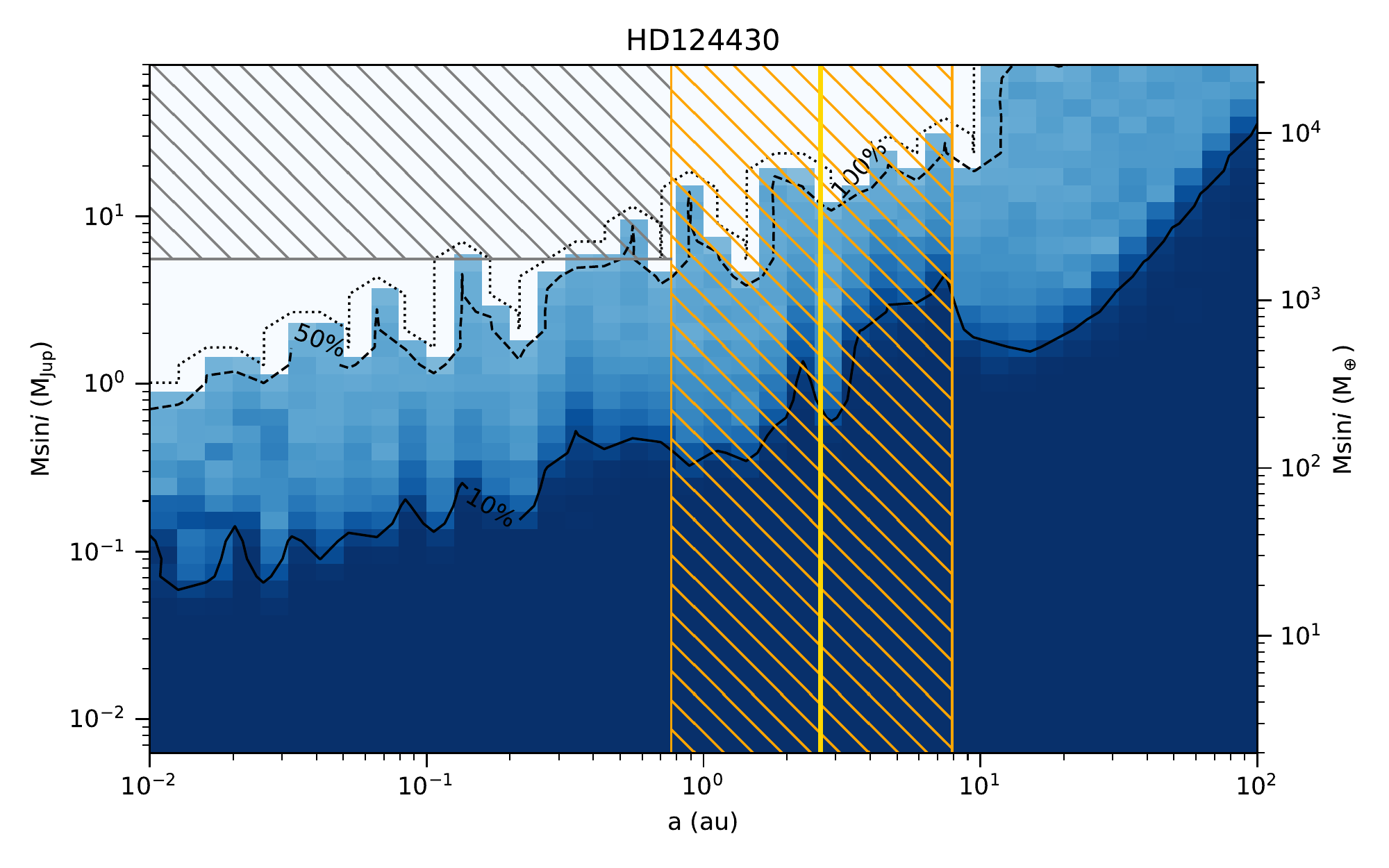}&
    		\includegraphics[width=0.22\linewidth]{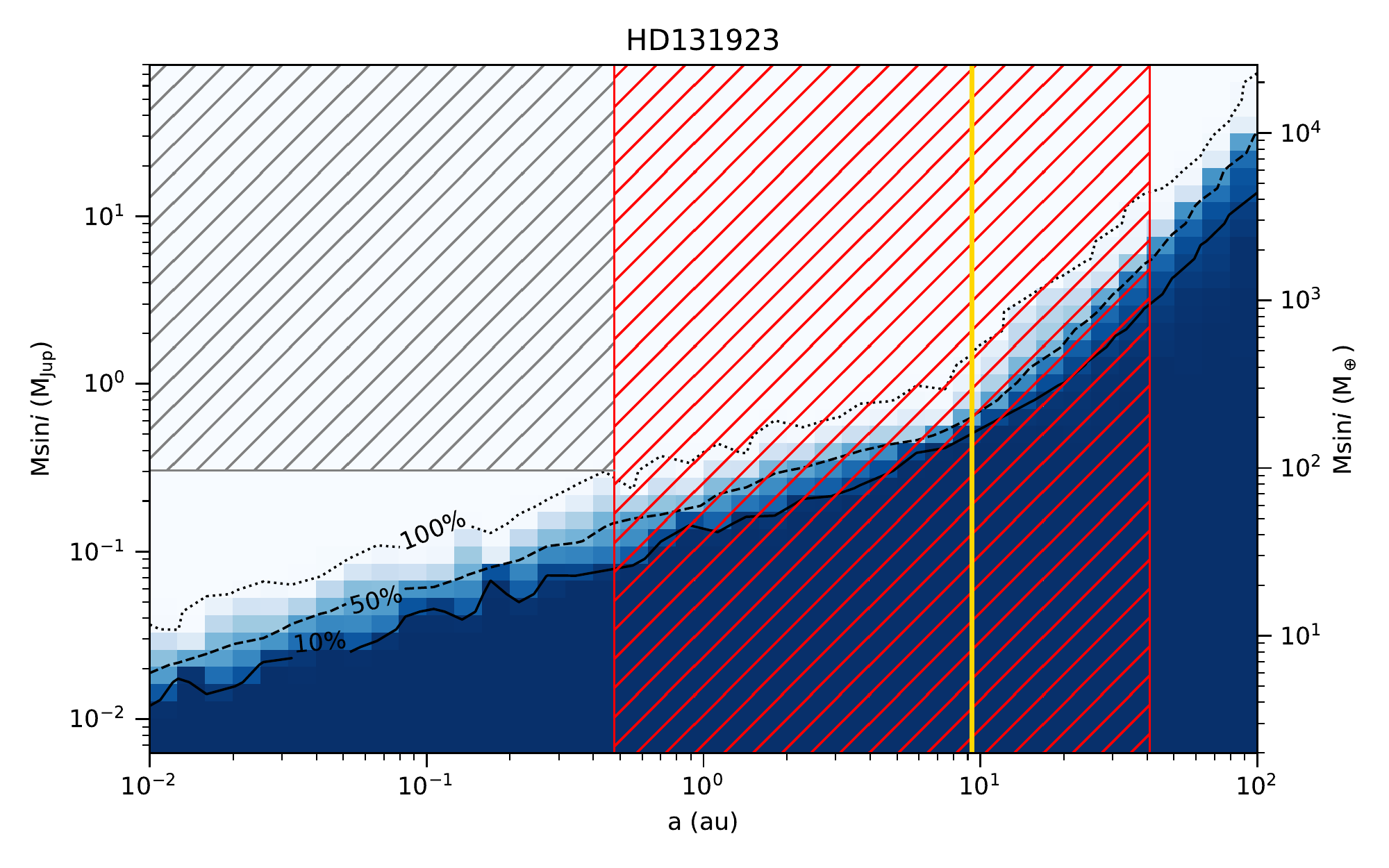}&
    		\includegraphics[width=0.22\linewidth]{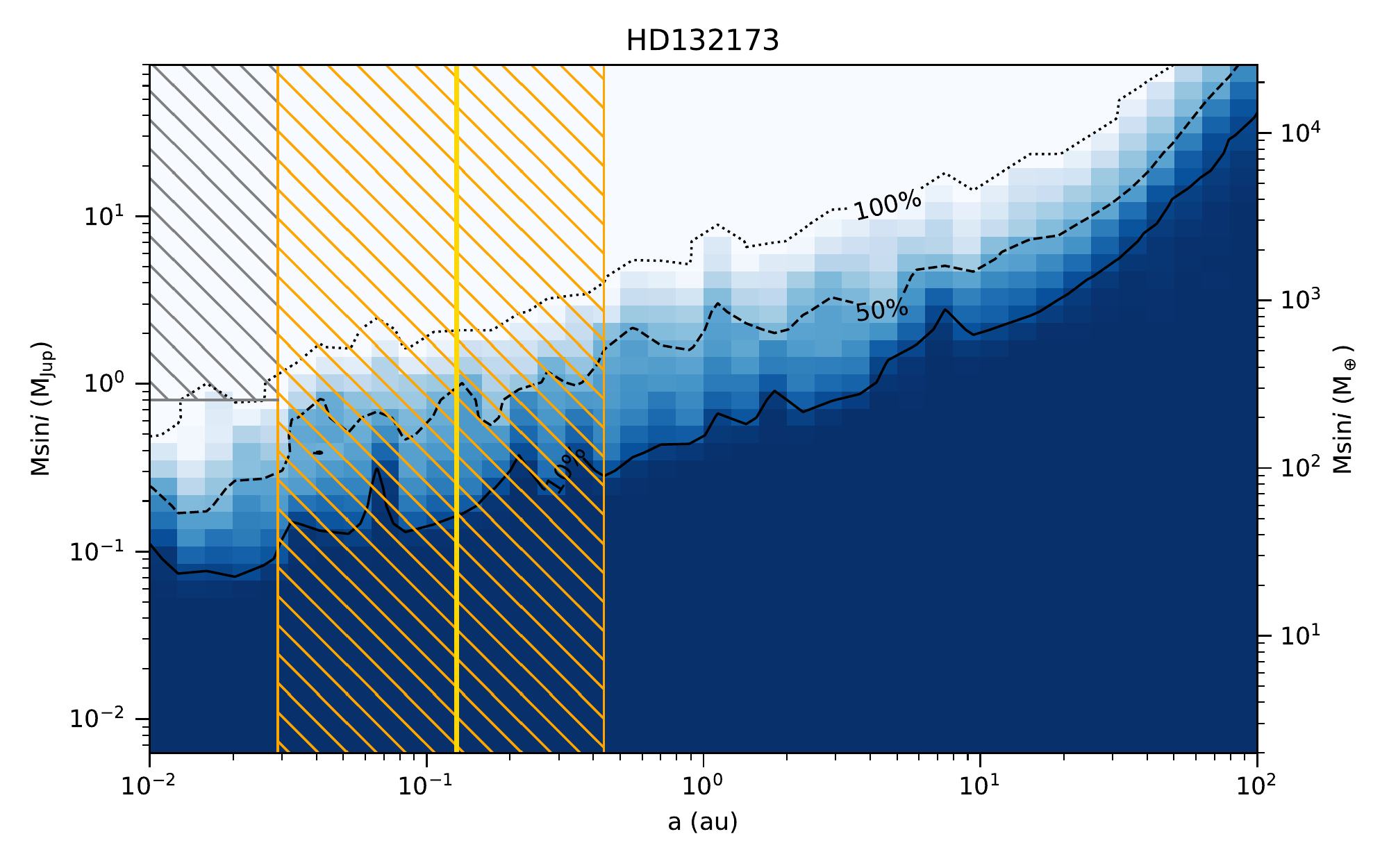}\\
    
    		\includegraphics[width=0.22\linewidth]{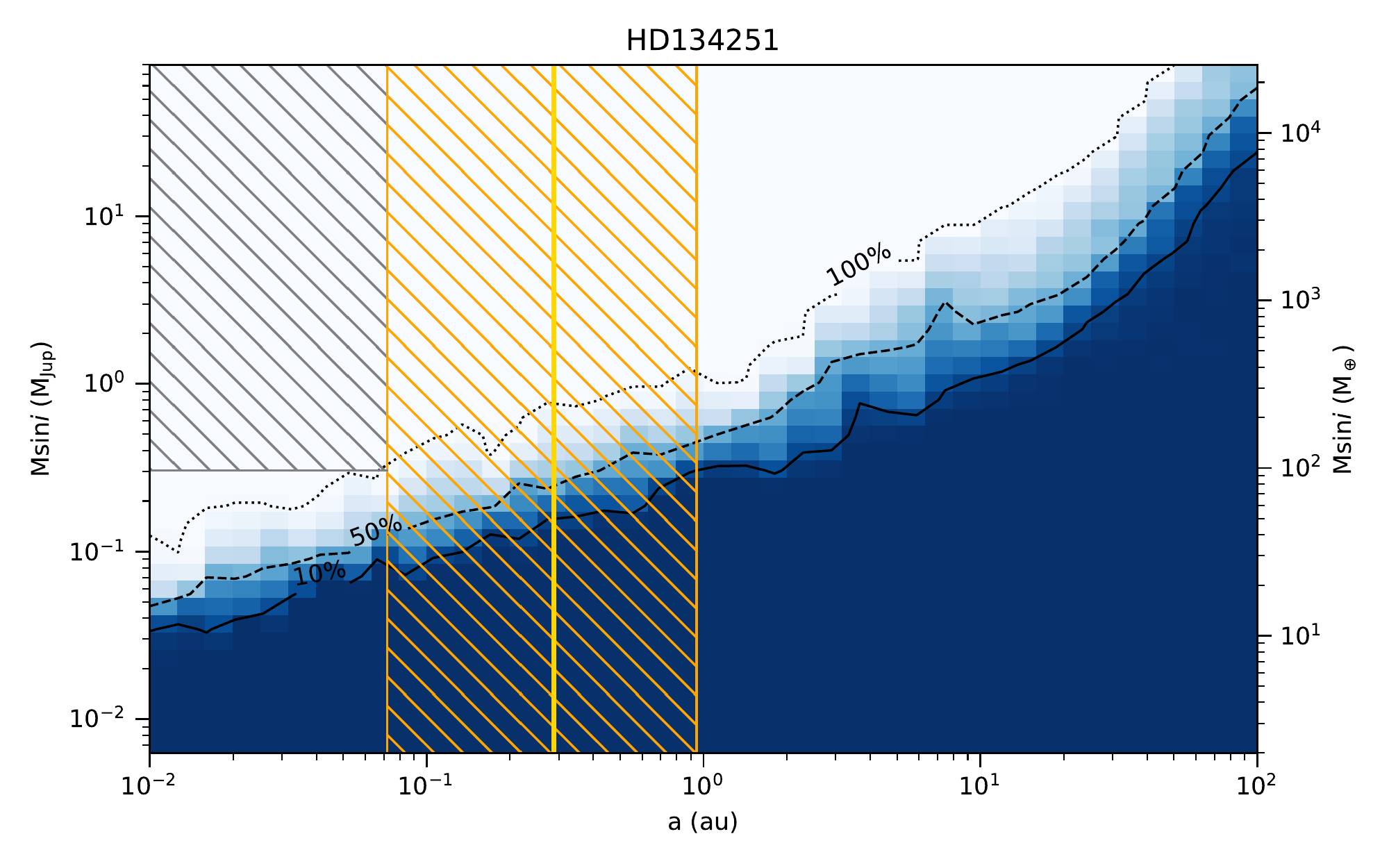}&
    		\includegraphics[width=0.22\linewidth]{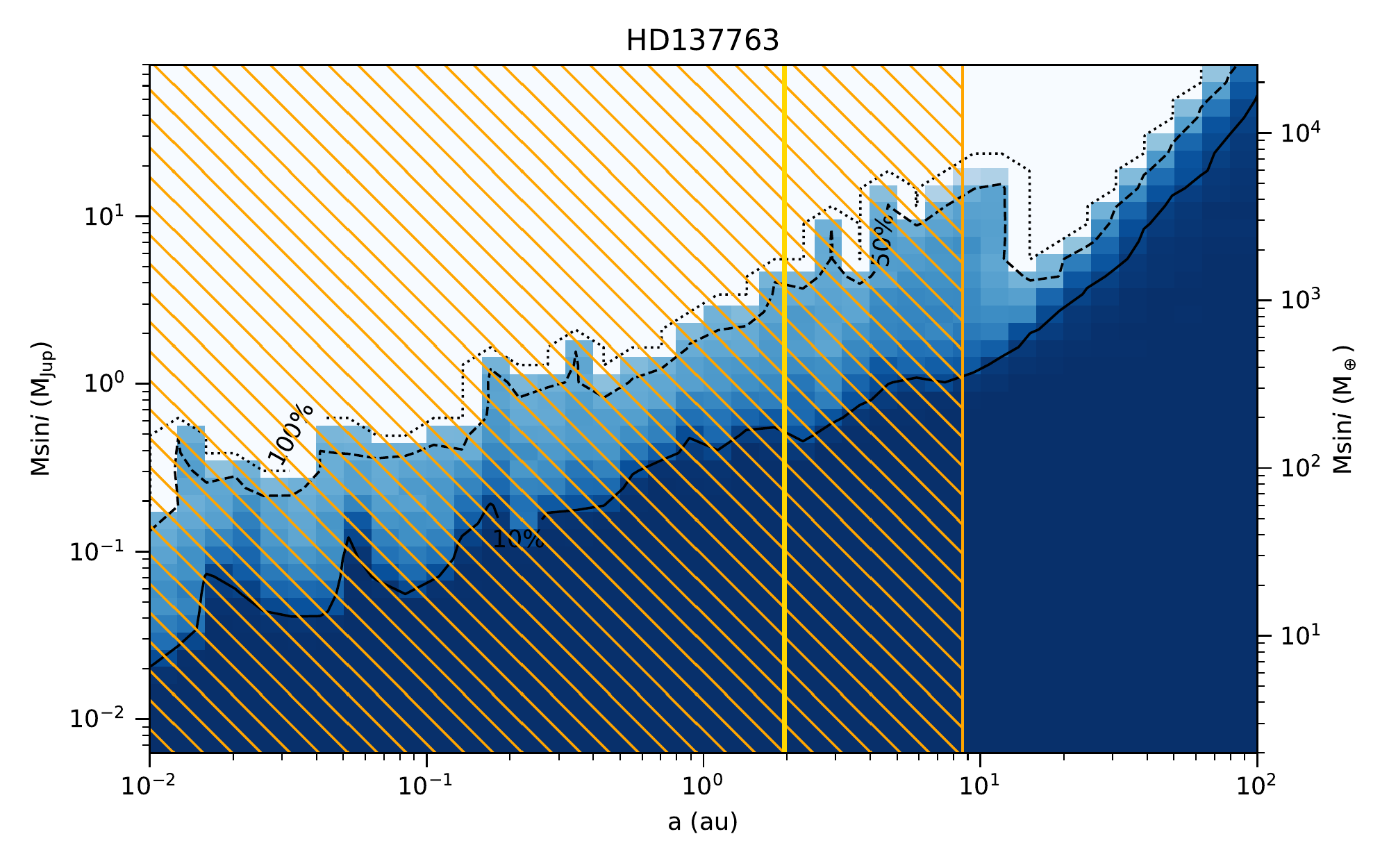}&
    		\includegraphics[width=0.22\linewidth]{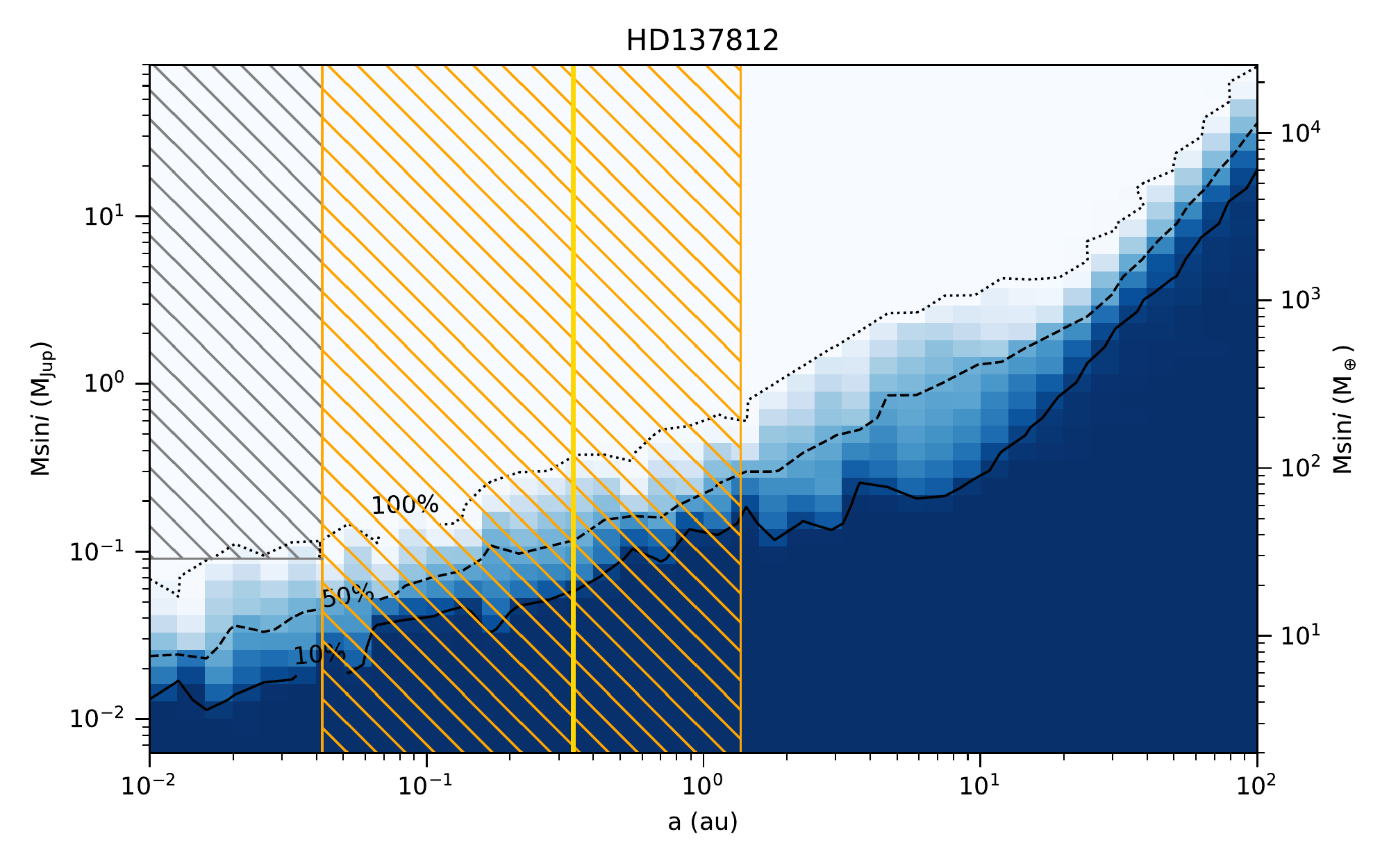}&
    		\includegraphics[width=0.22\linewidth]{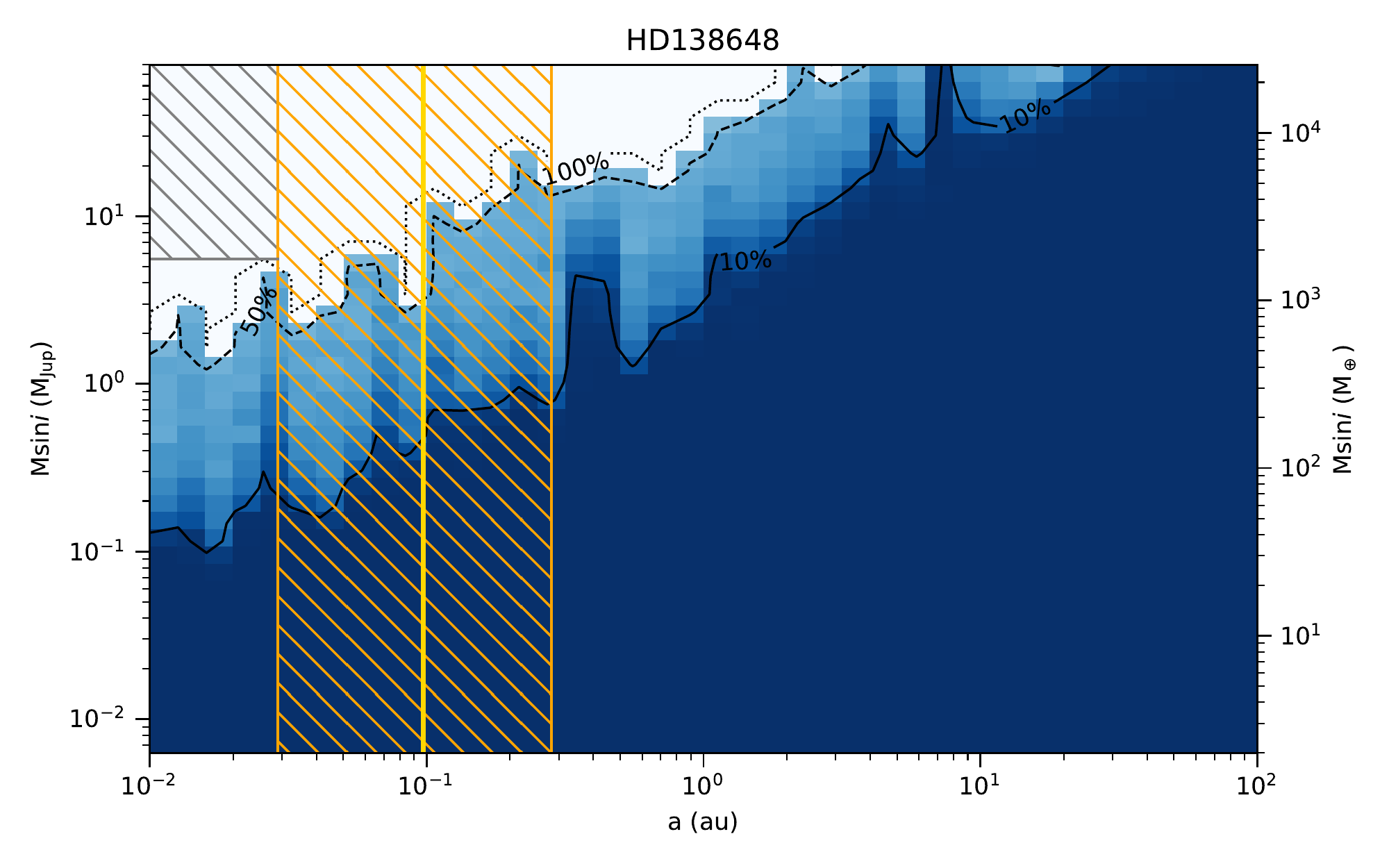}\\
    
    		\includegraphics[width=0.22\linewidth]{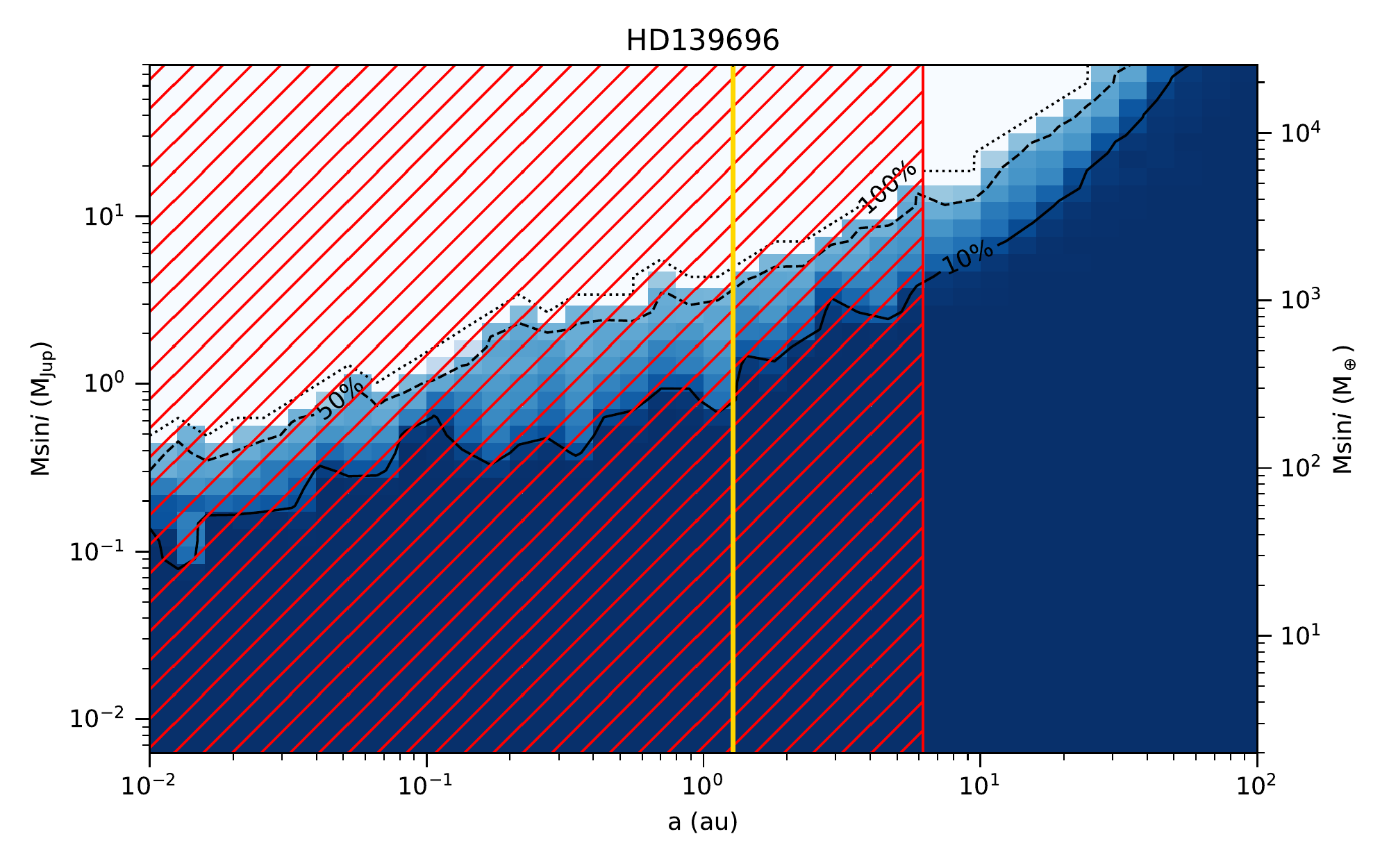}&
    		\includegraphics[width=0.22\linewidth]{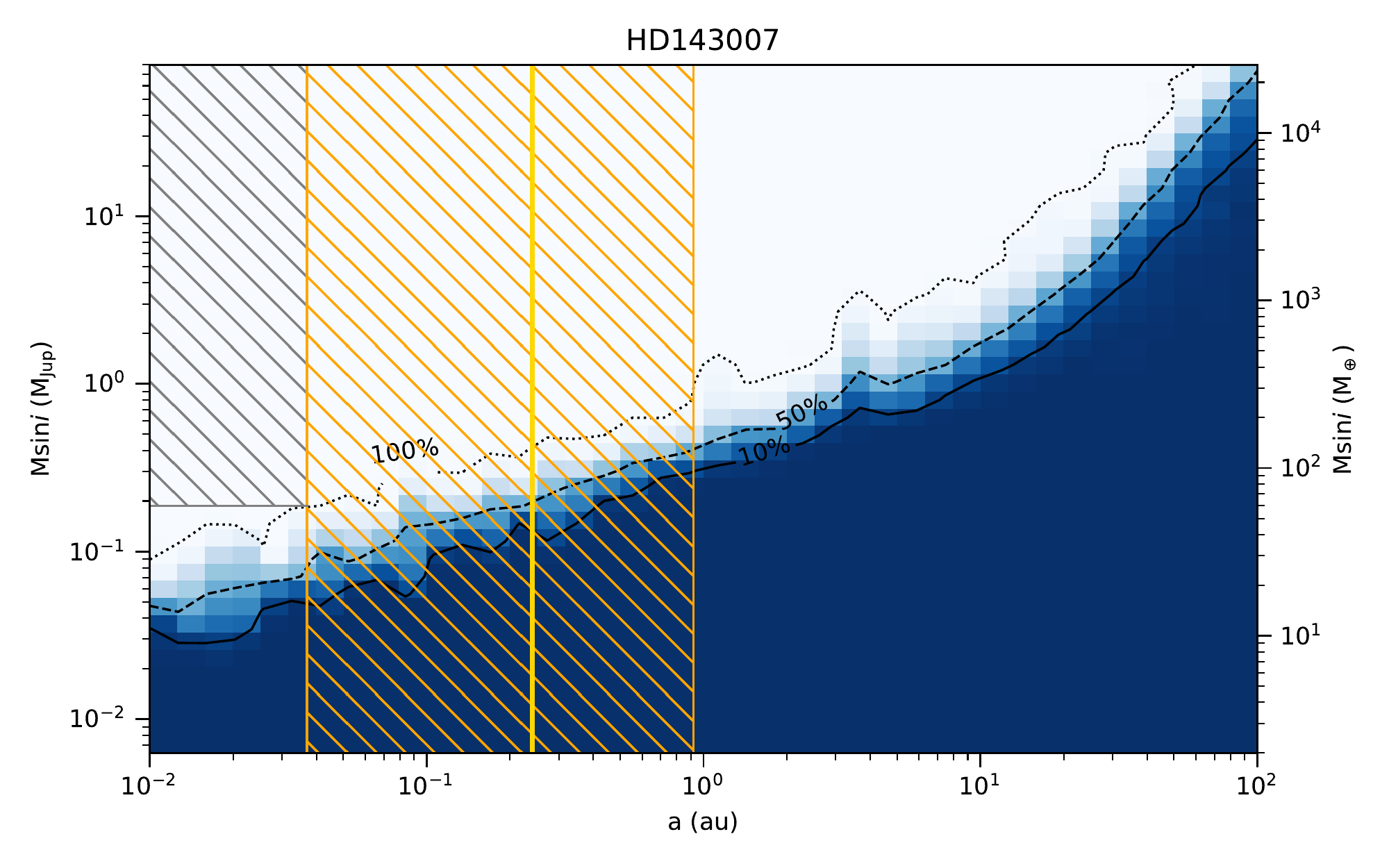}&
    		\includegraphics[width=0.22\linewidth]{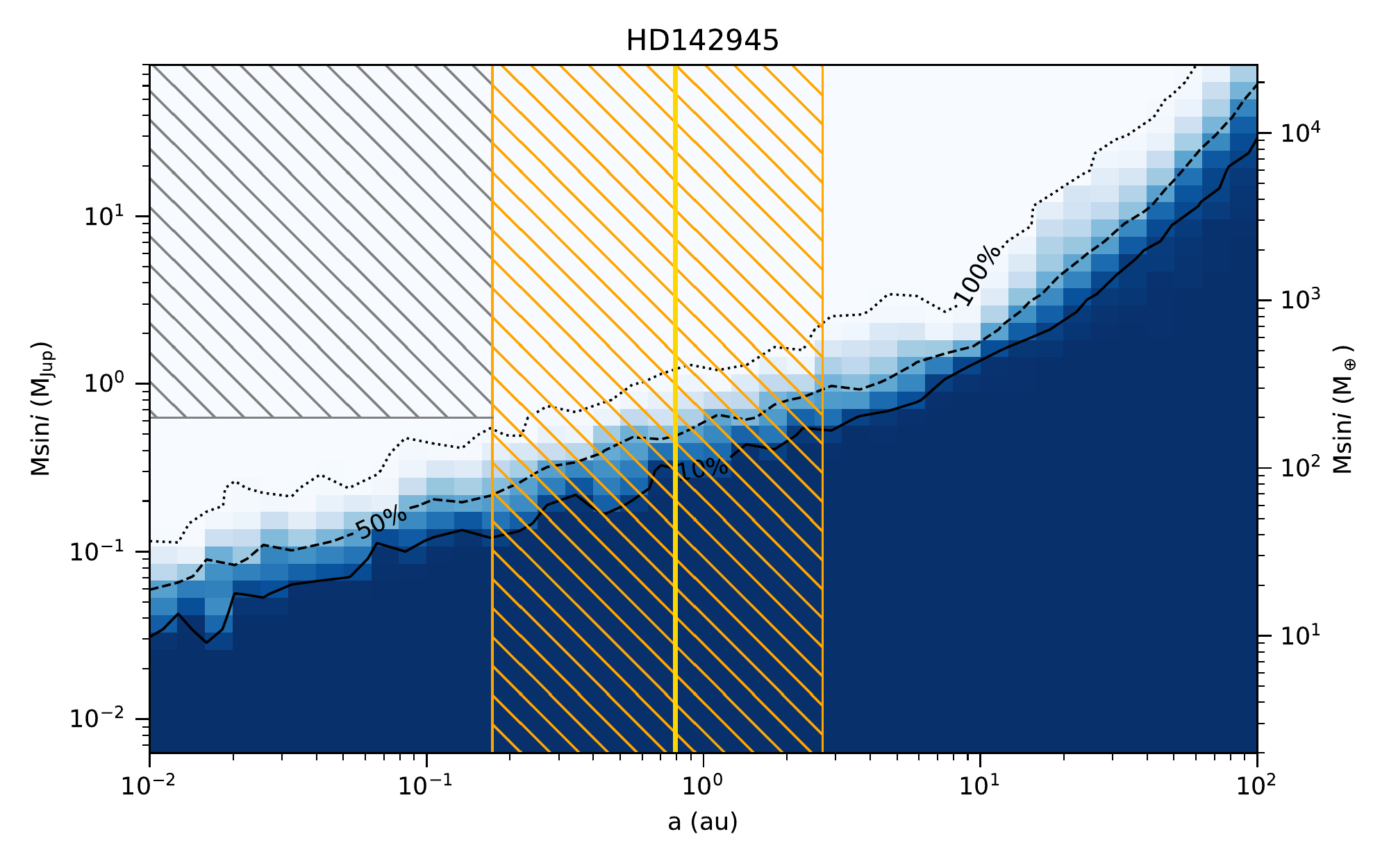}&
    		\includegraphics[width=0.22\linewidth]{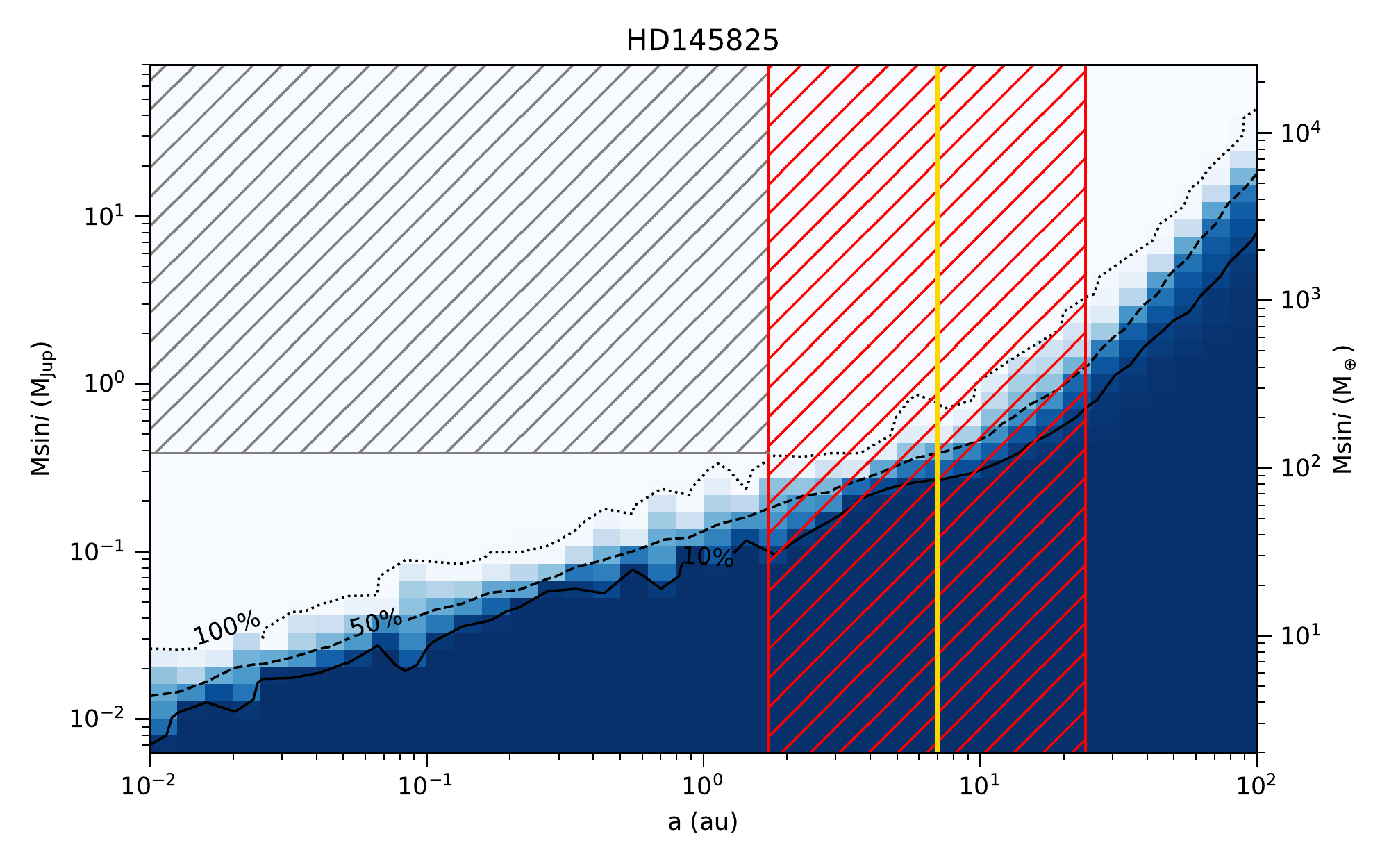}\\
    
    		\includegraphics[width=0.22\linewidth]{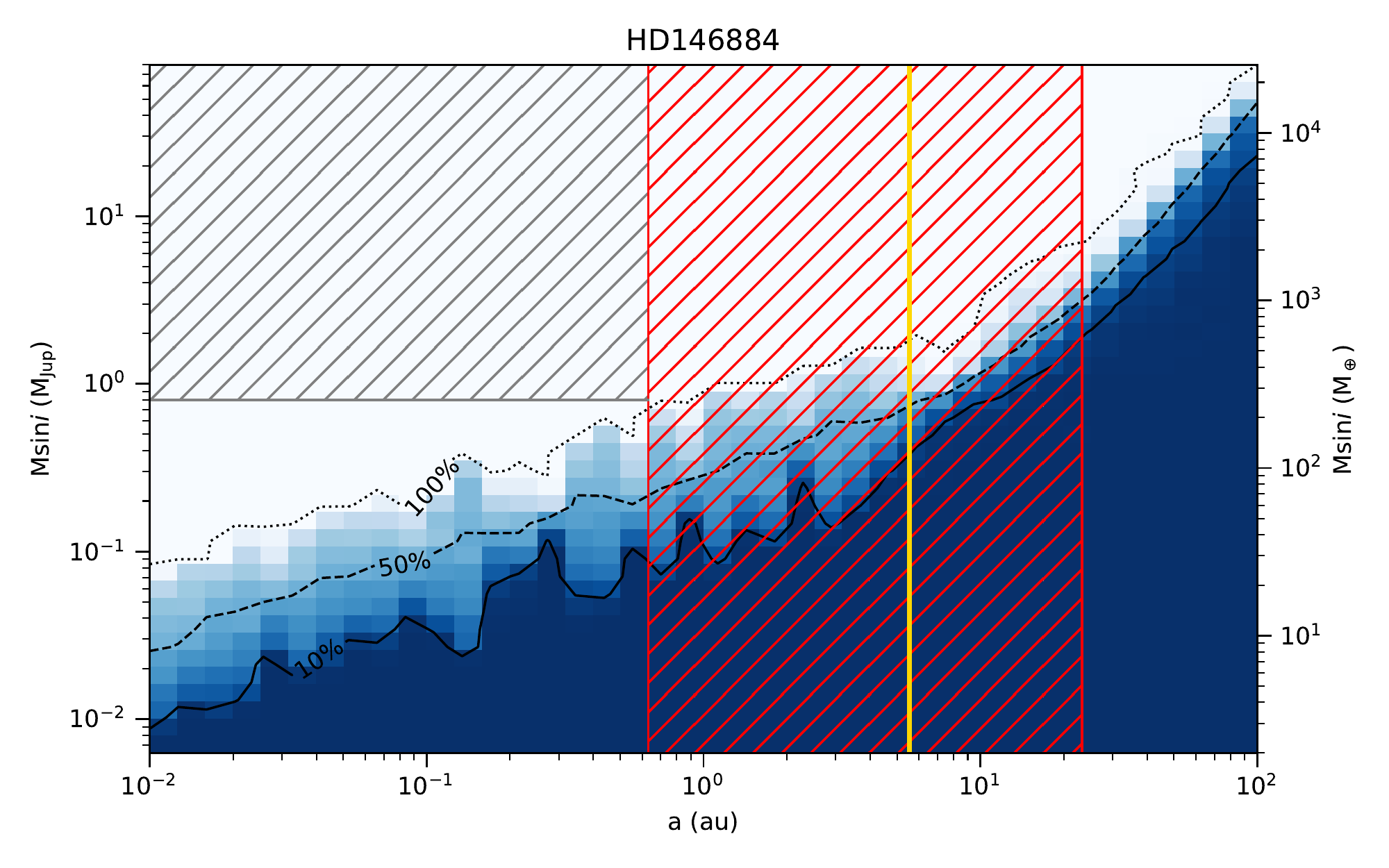}&
    		\includegraphics[width=0.22\linewidth]{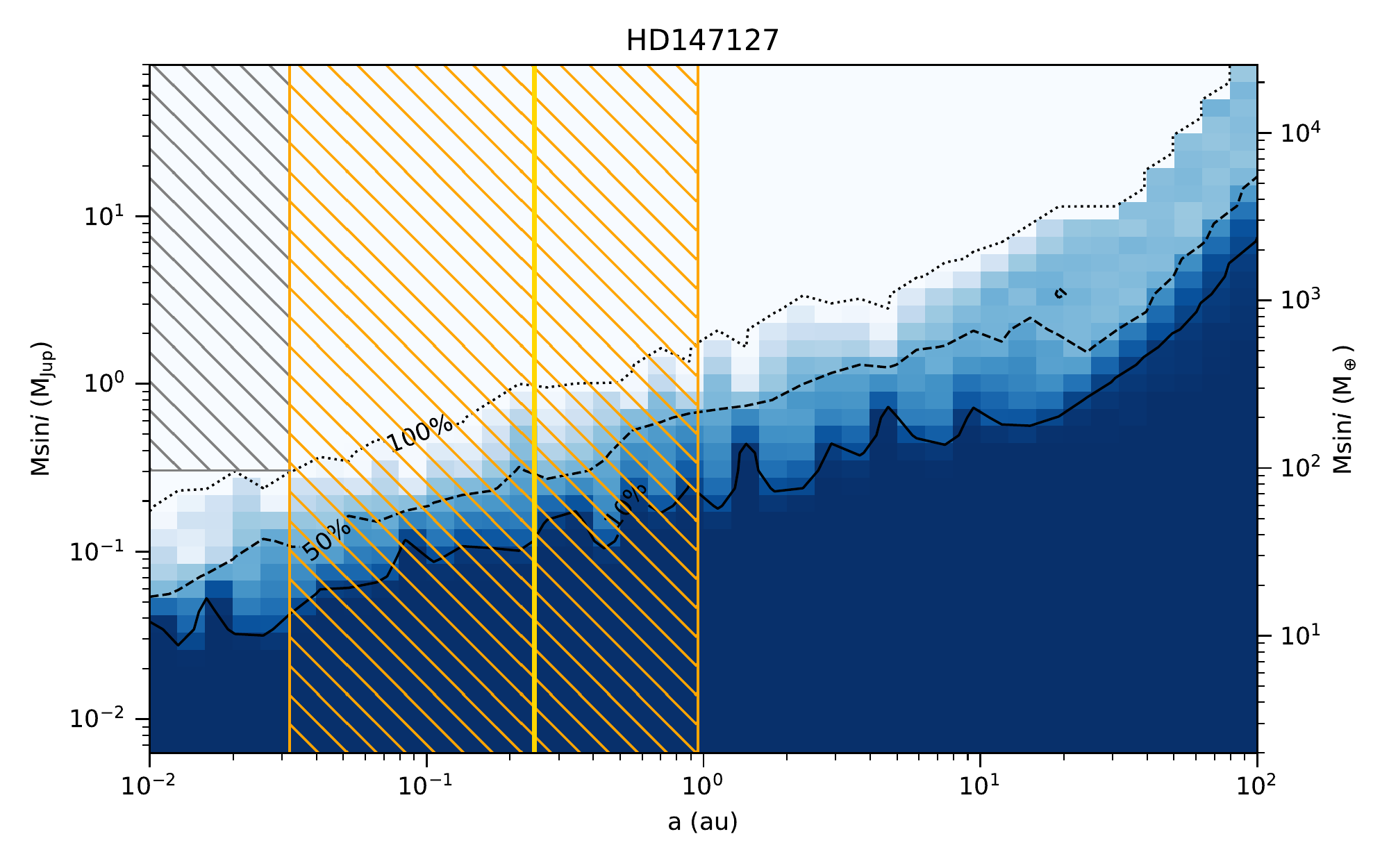}&
    		\includegraphics[width=0.22\linewidth]{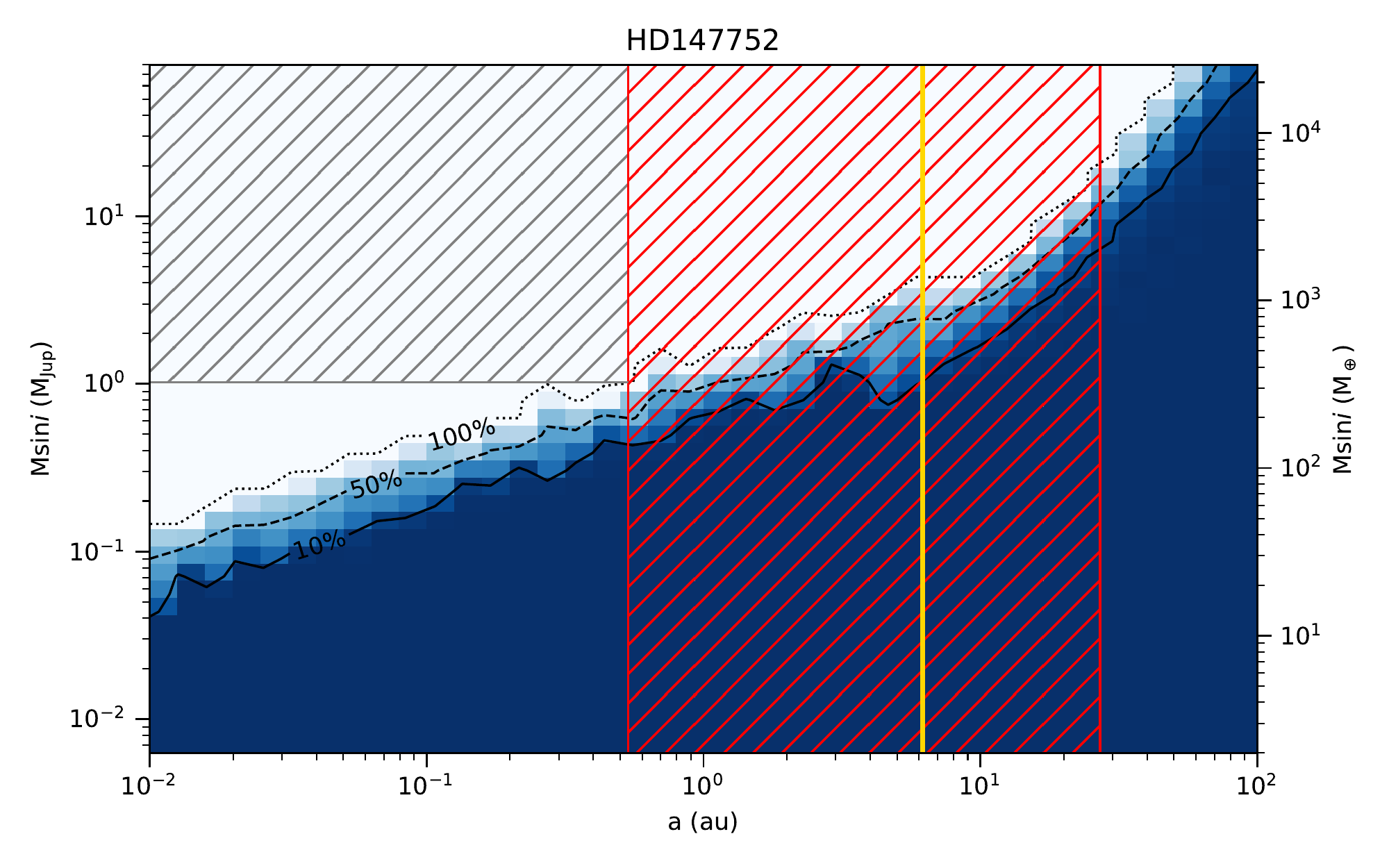}&
    		\includegraphics[width=0.22\linewidth]{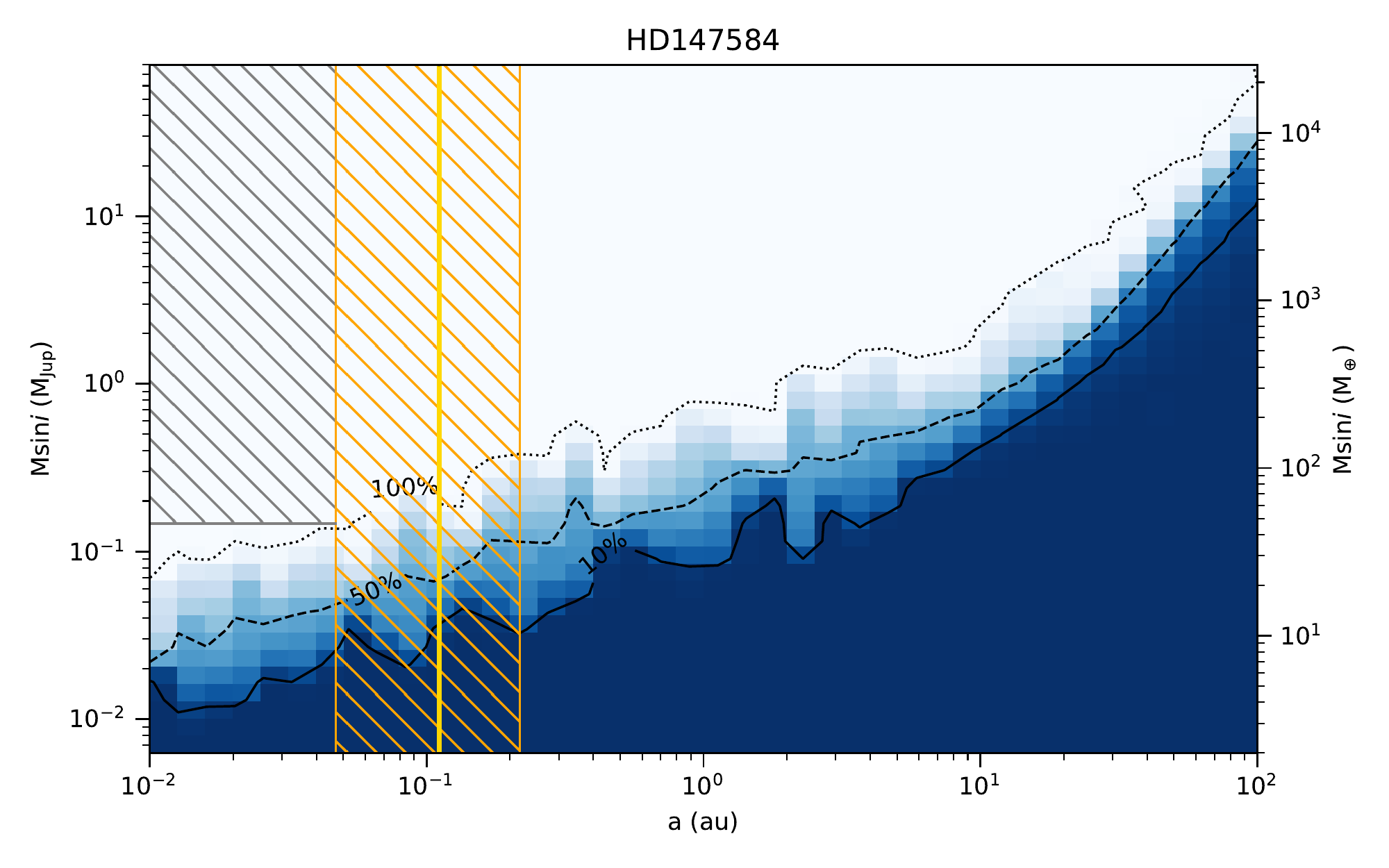}\\
    
    		\includegraphics[width=0.22\linewidth]{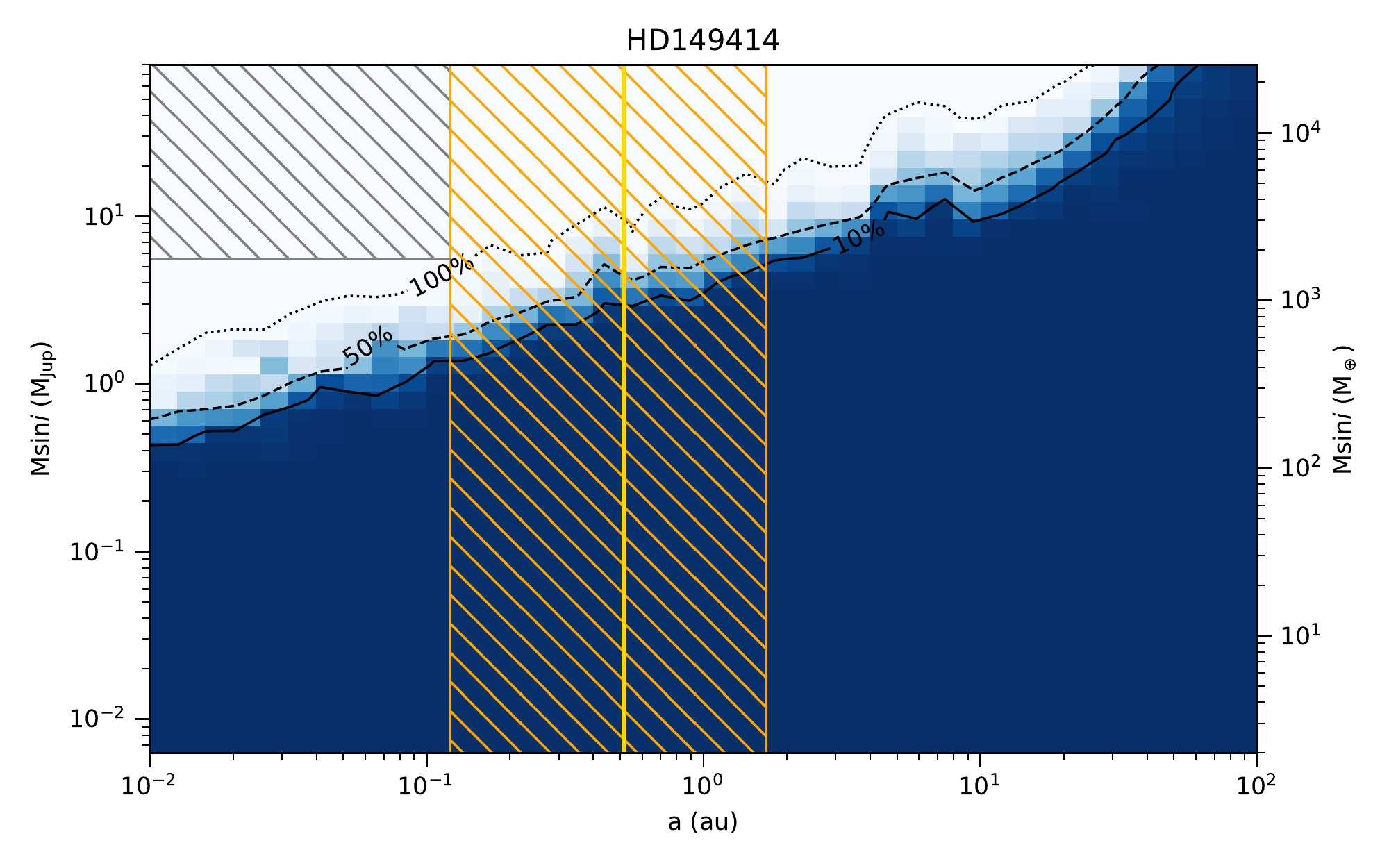}&
    		\includegraphics[width=0.22\linewidth]{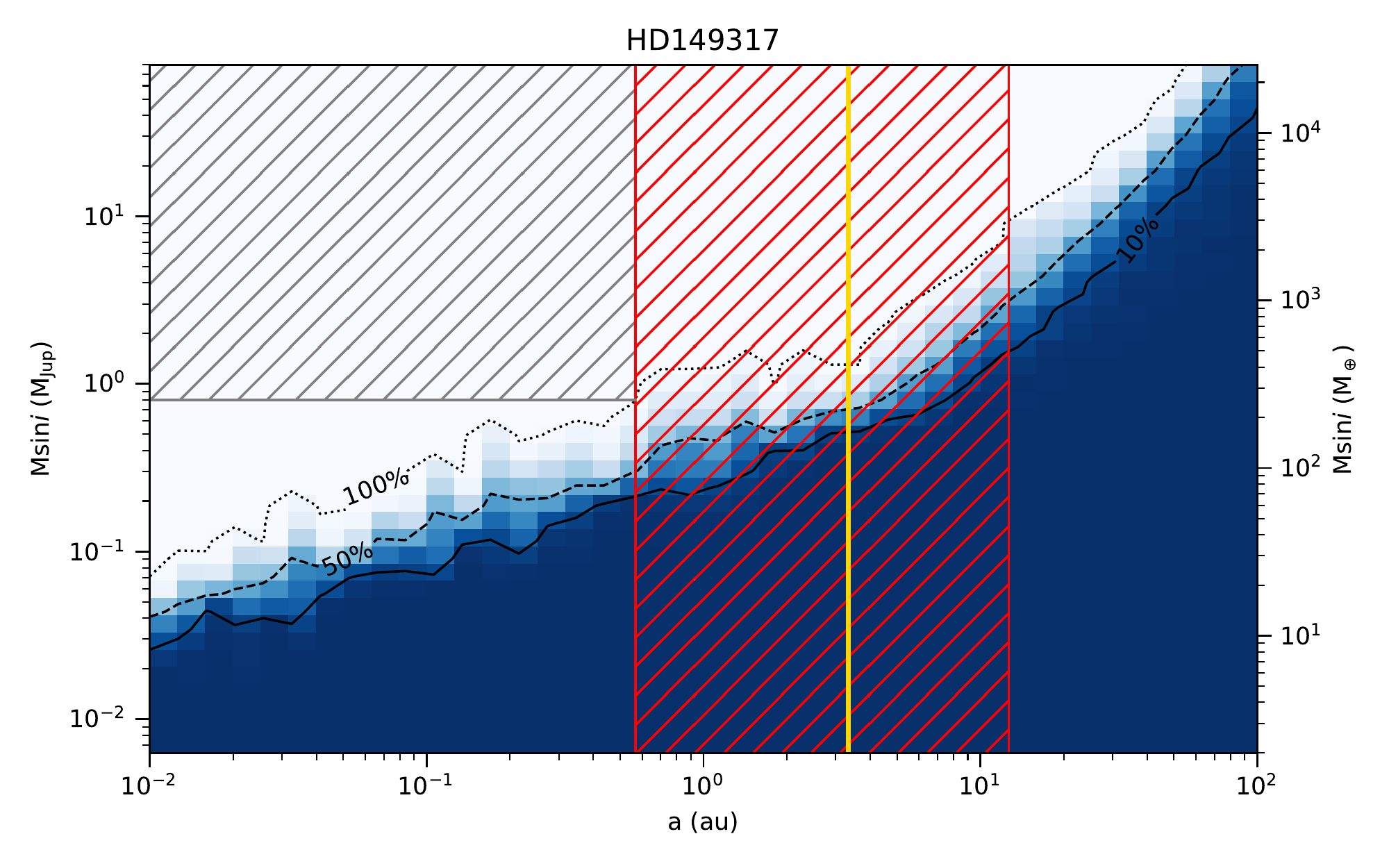}&
    		\includegraphics[width=0.22\linewidth]{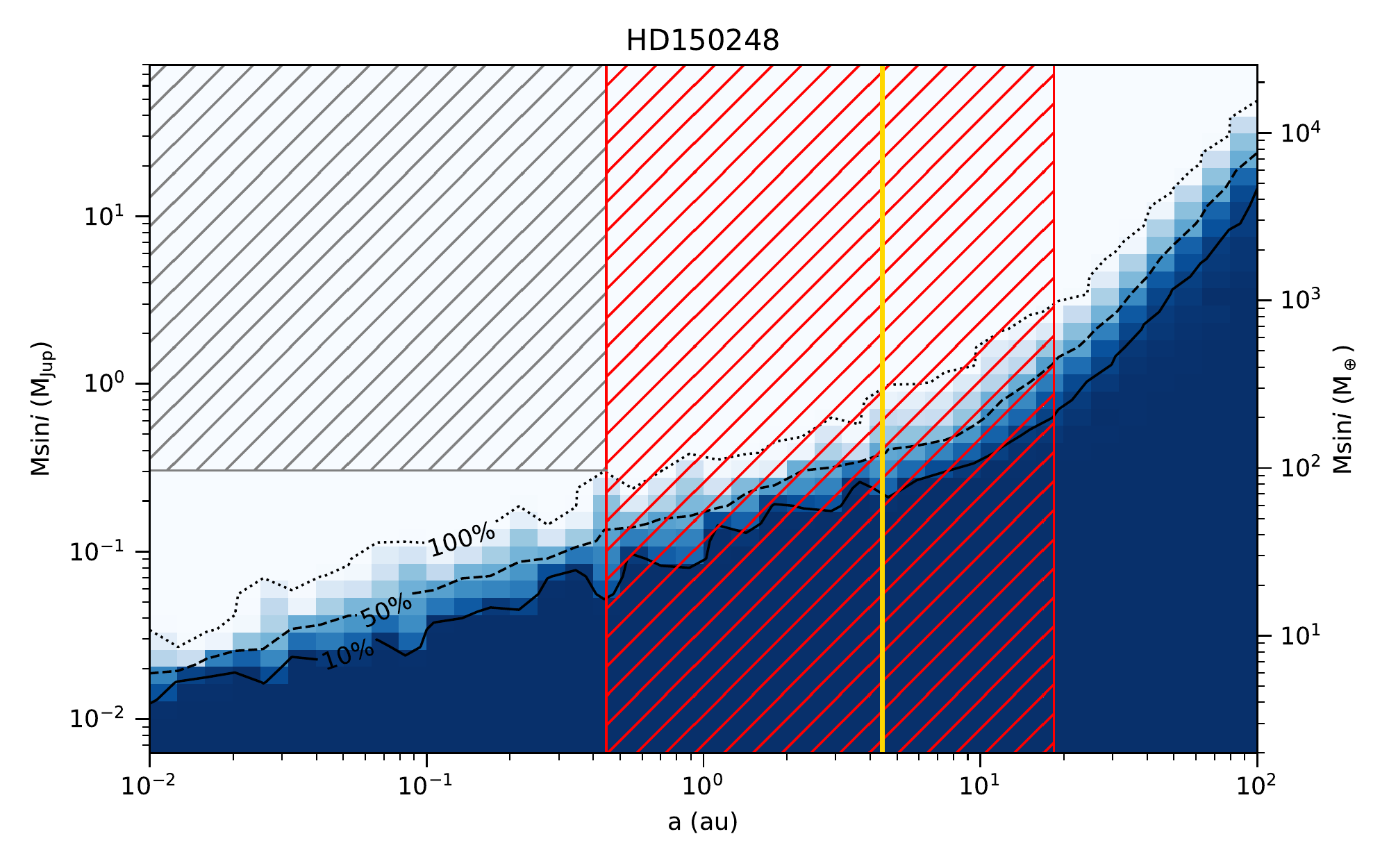}&
    		\includegraphics[width=0.22\linewidth]{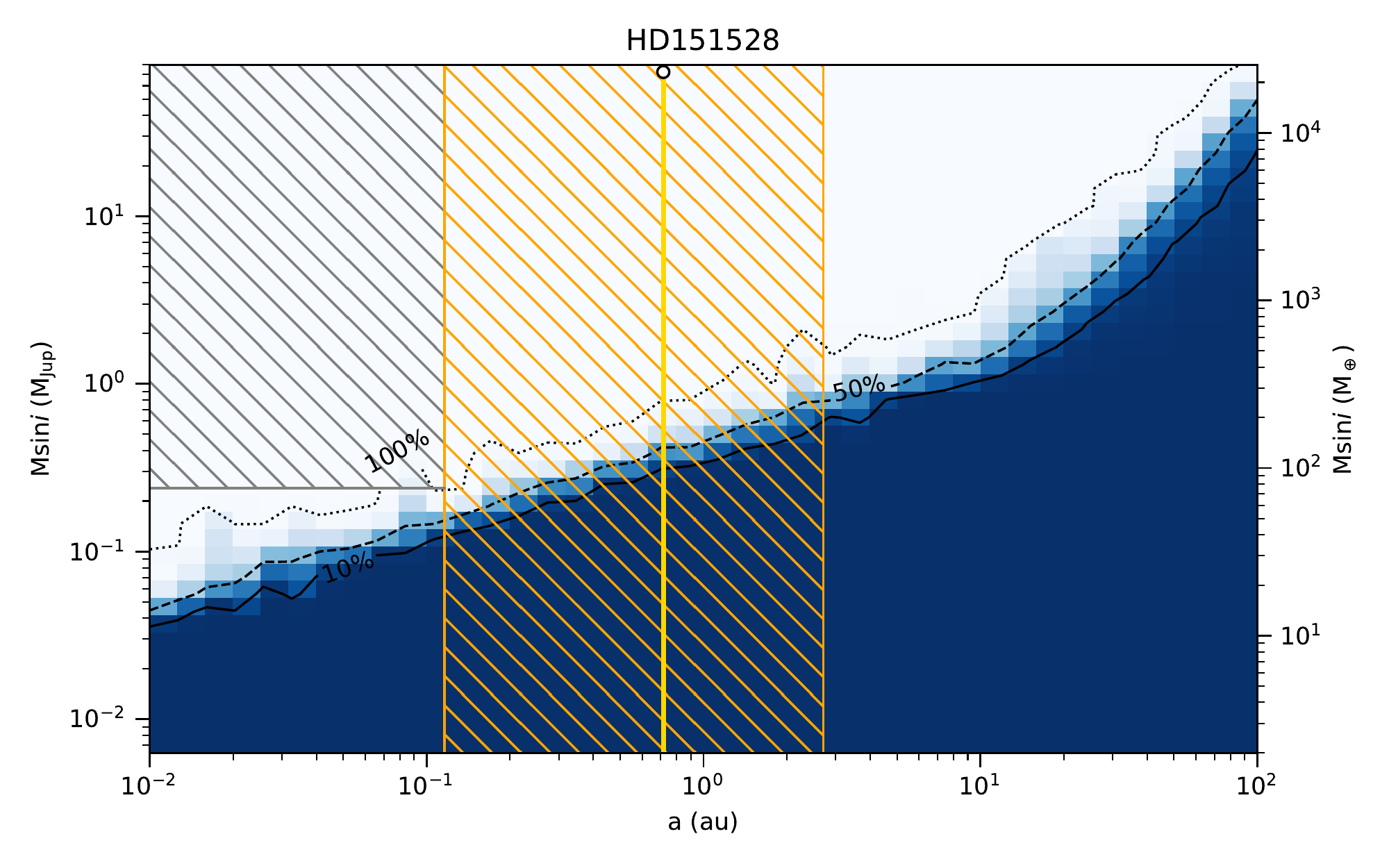}\\
    
    		\includegraphics[width=0.22\linewidth]{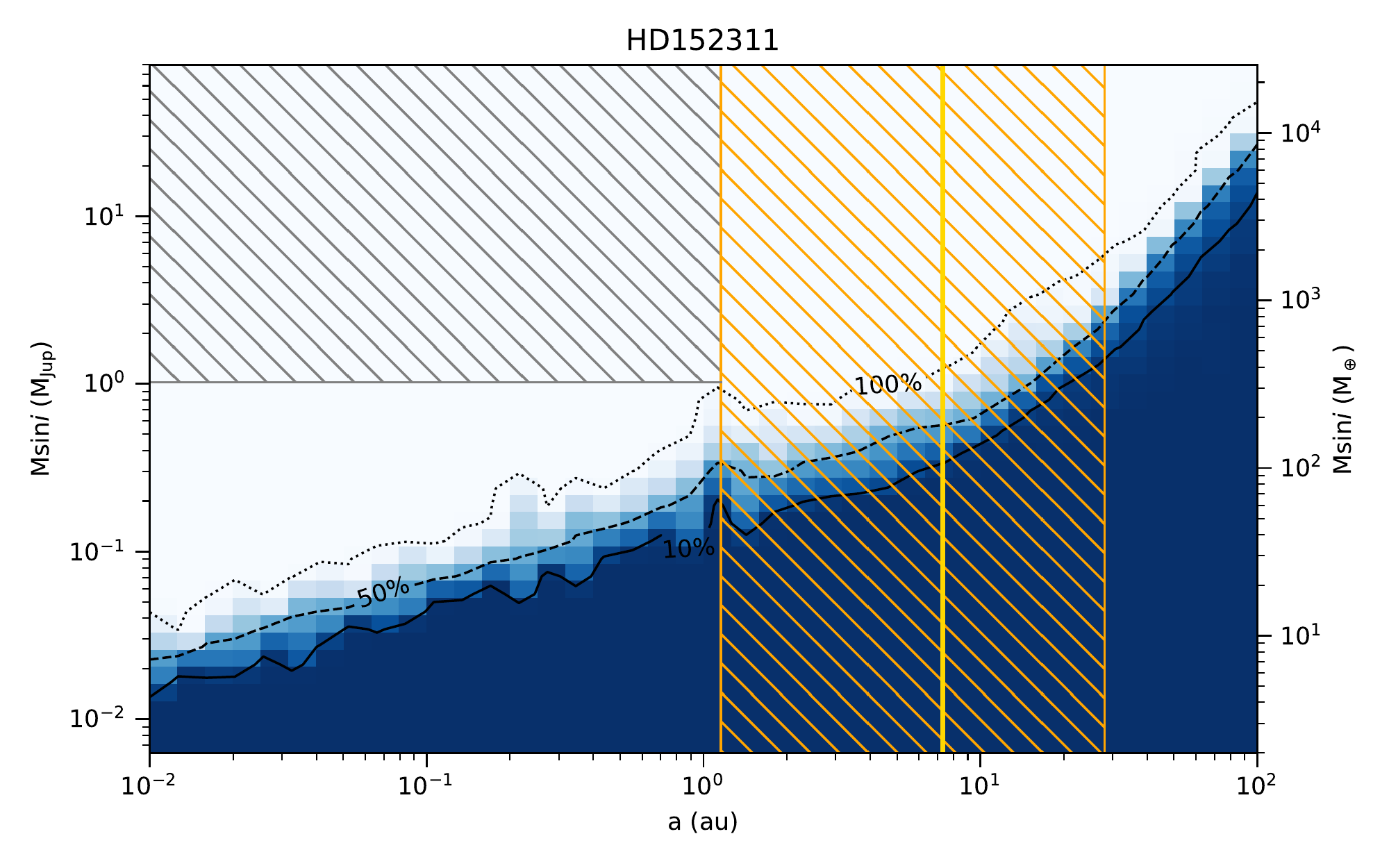}&
    		\includegraphics[width=0.22\linewidth]{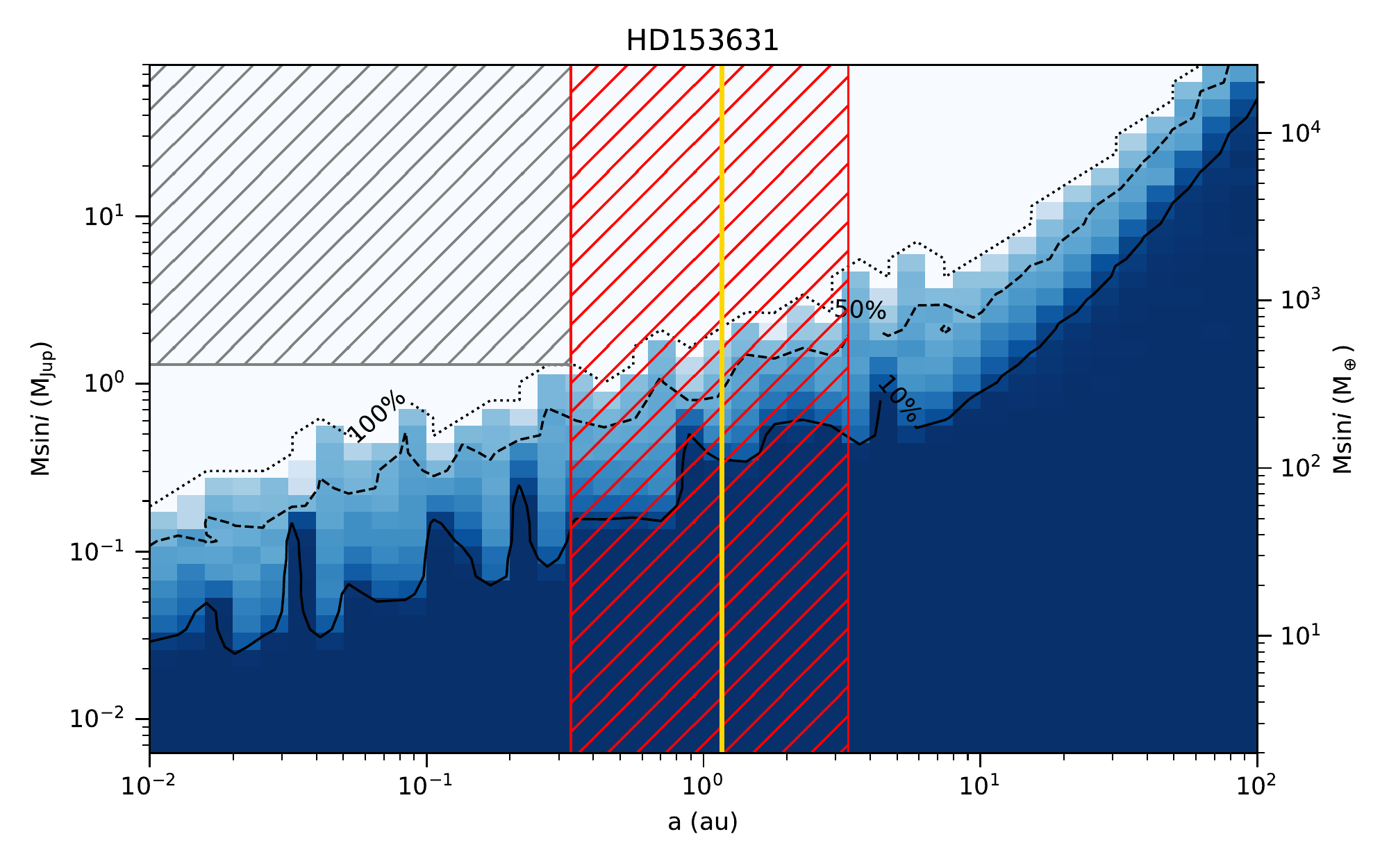}&
    		\includegraphics[width=0.22\linewidth]{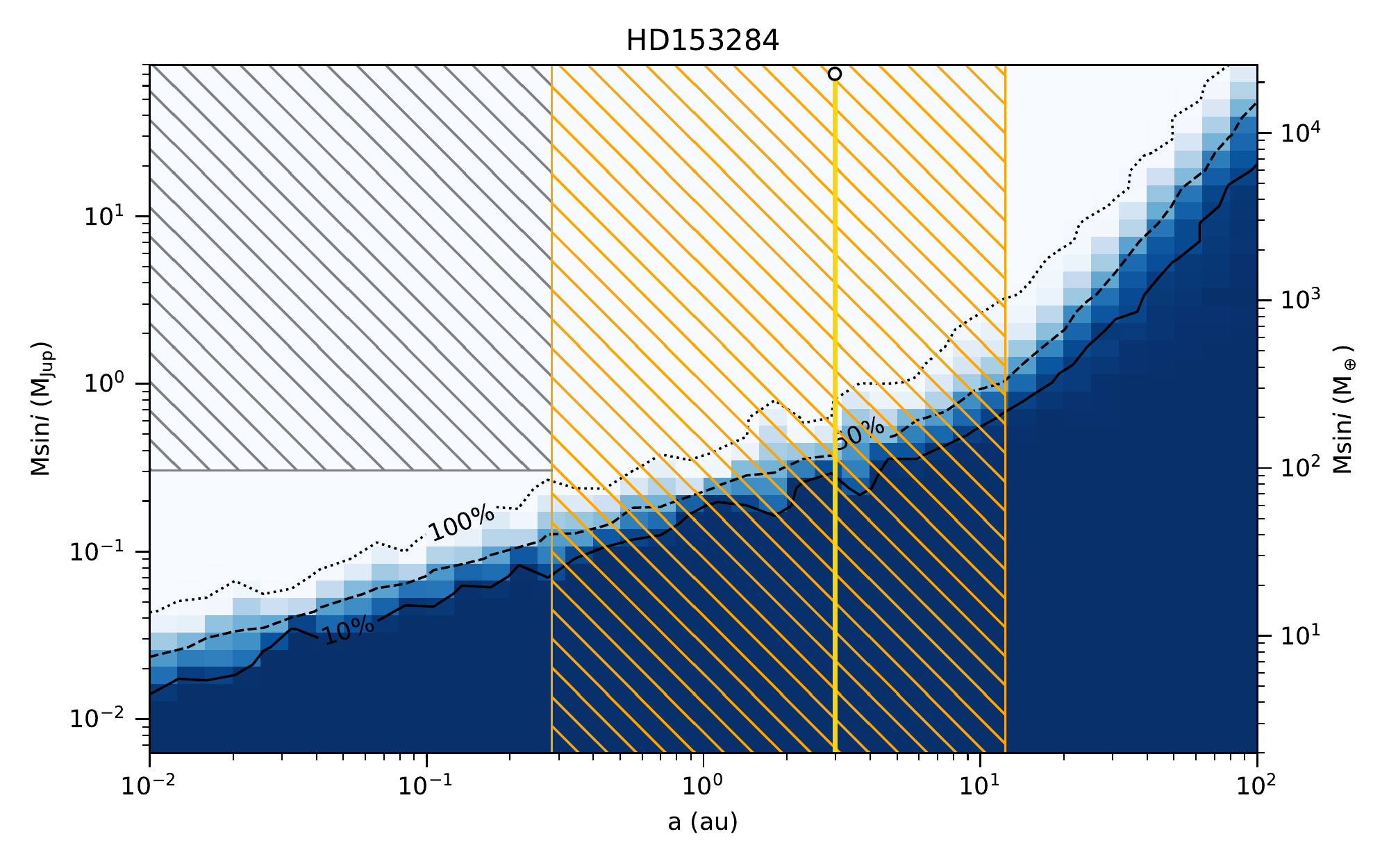}&
    		\includegraphics[width=0.22\linewidth]{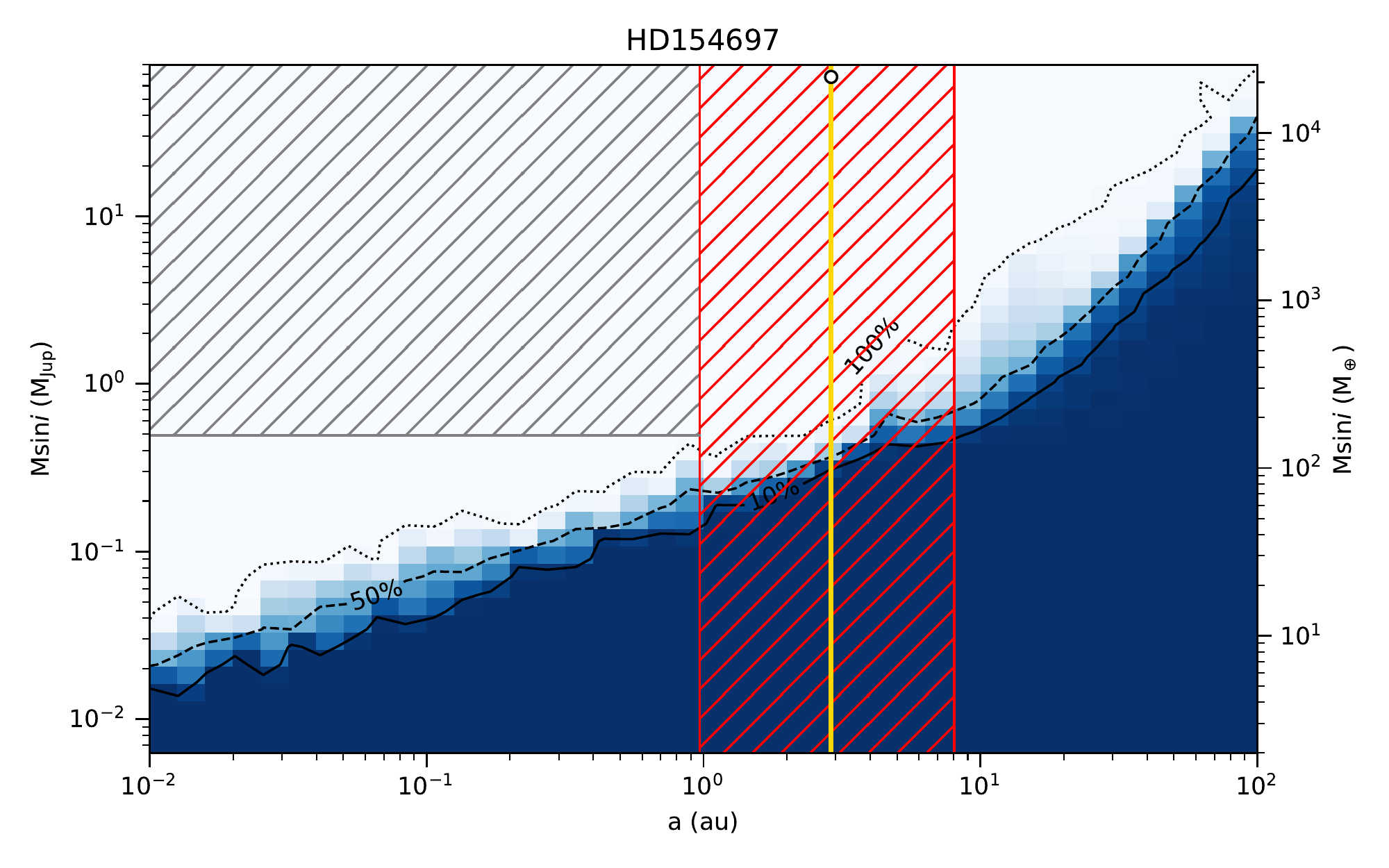}\\
    
    		\includegraphics[width=0.22\linewidth]{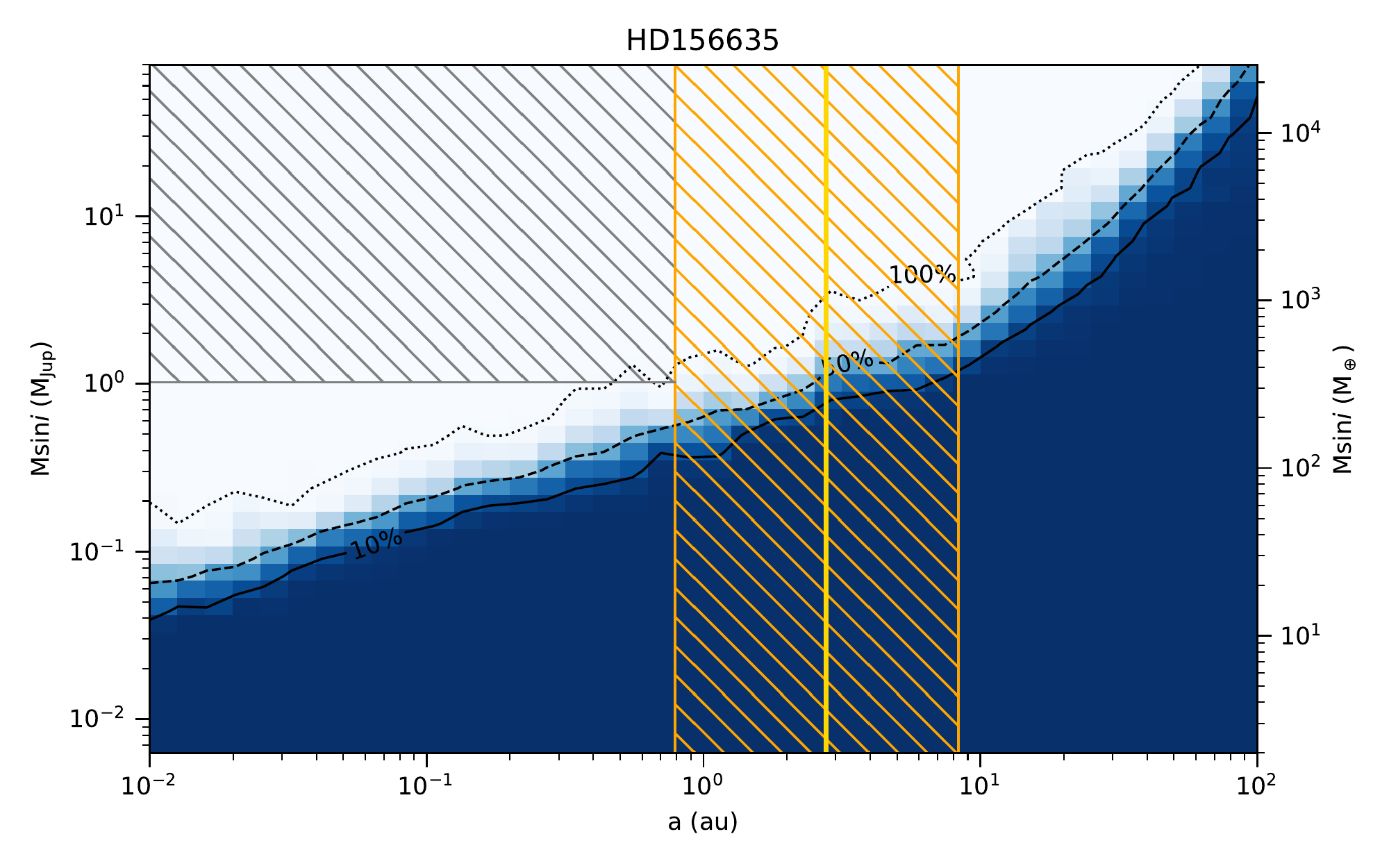}&
    		\includegraphics[width=0.22\linewidth]{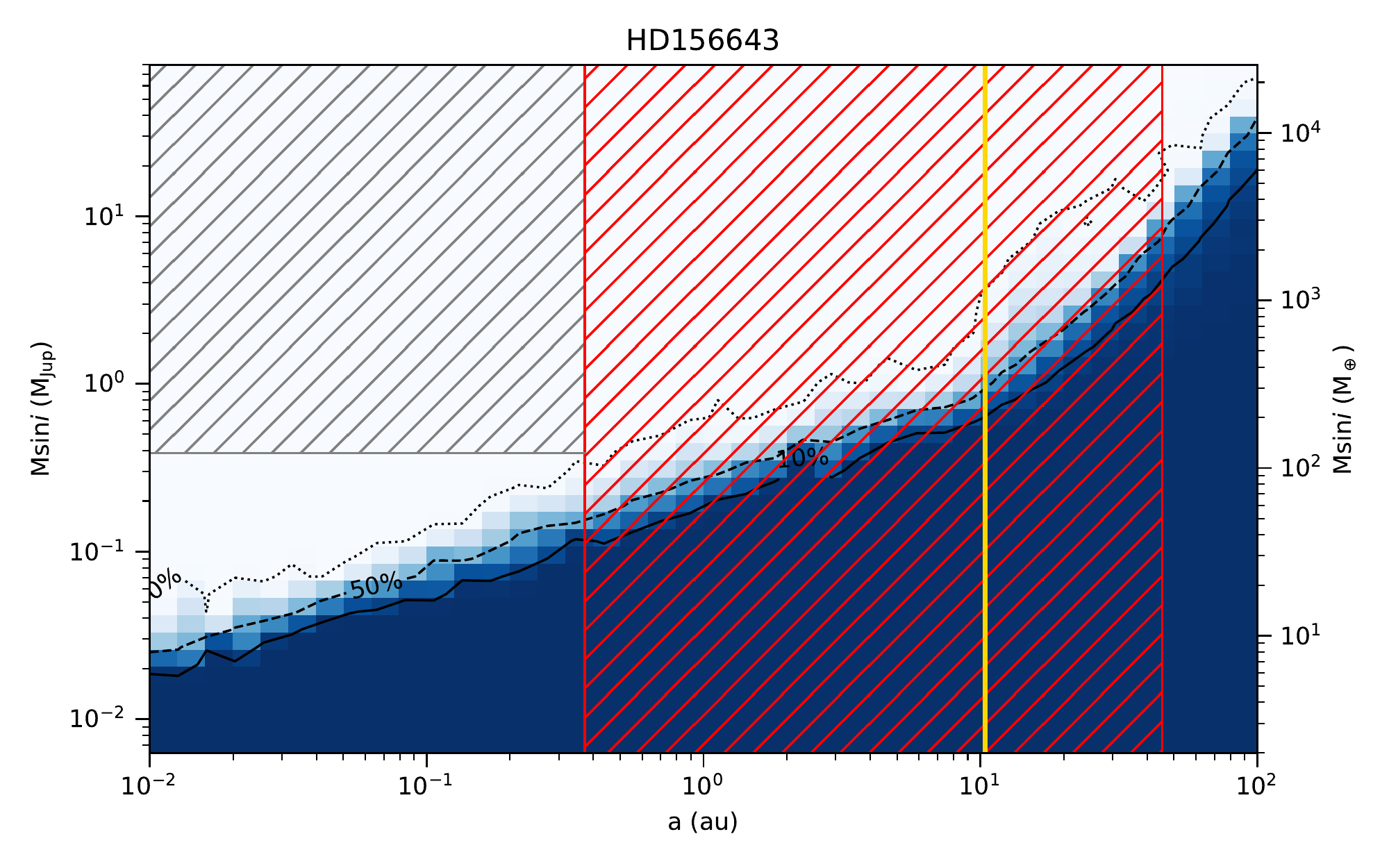}&
    		\includegraphics[width=0.22\linewidth]{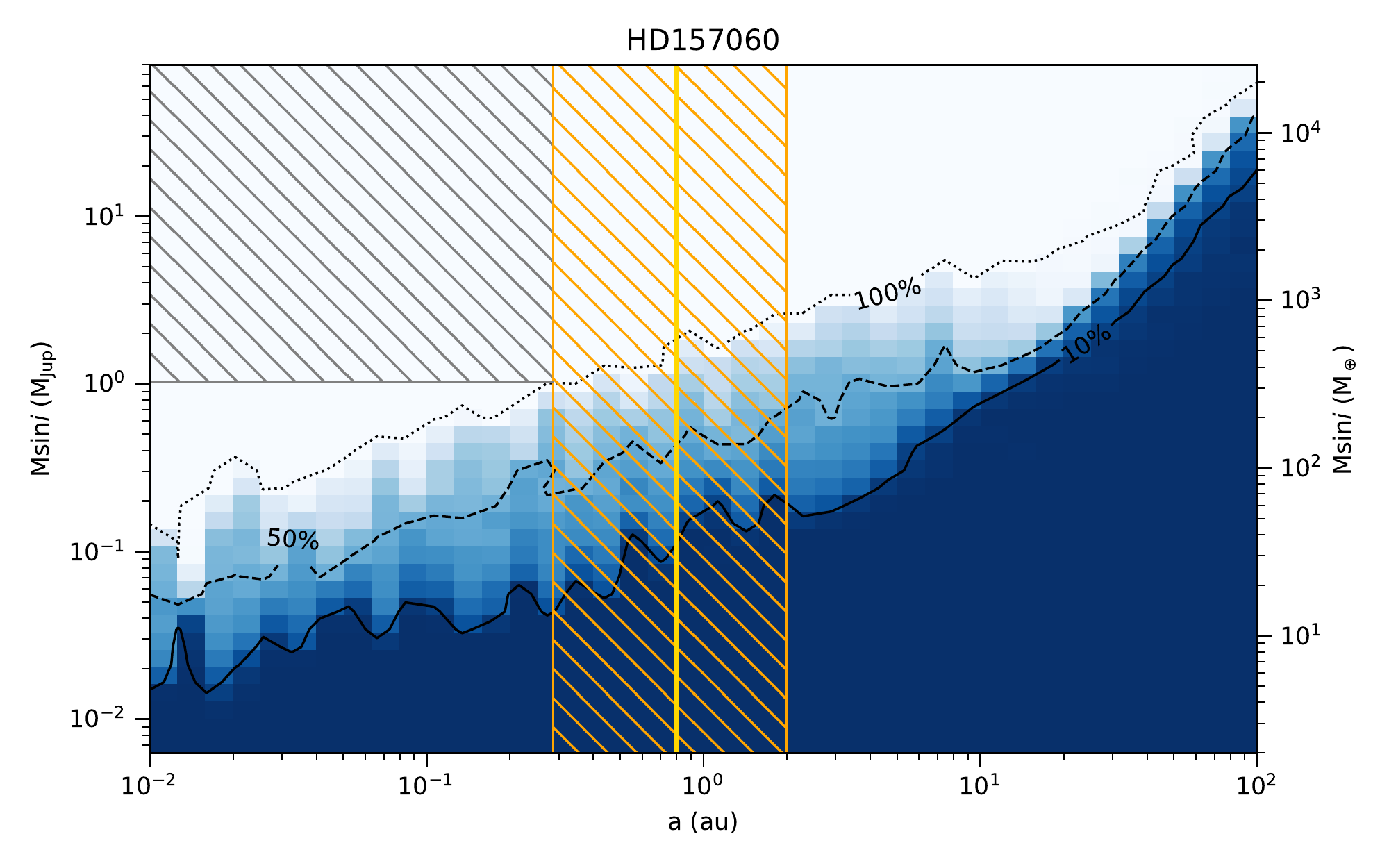}&
    		\includegraphics[width=0.22\linewidth]{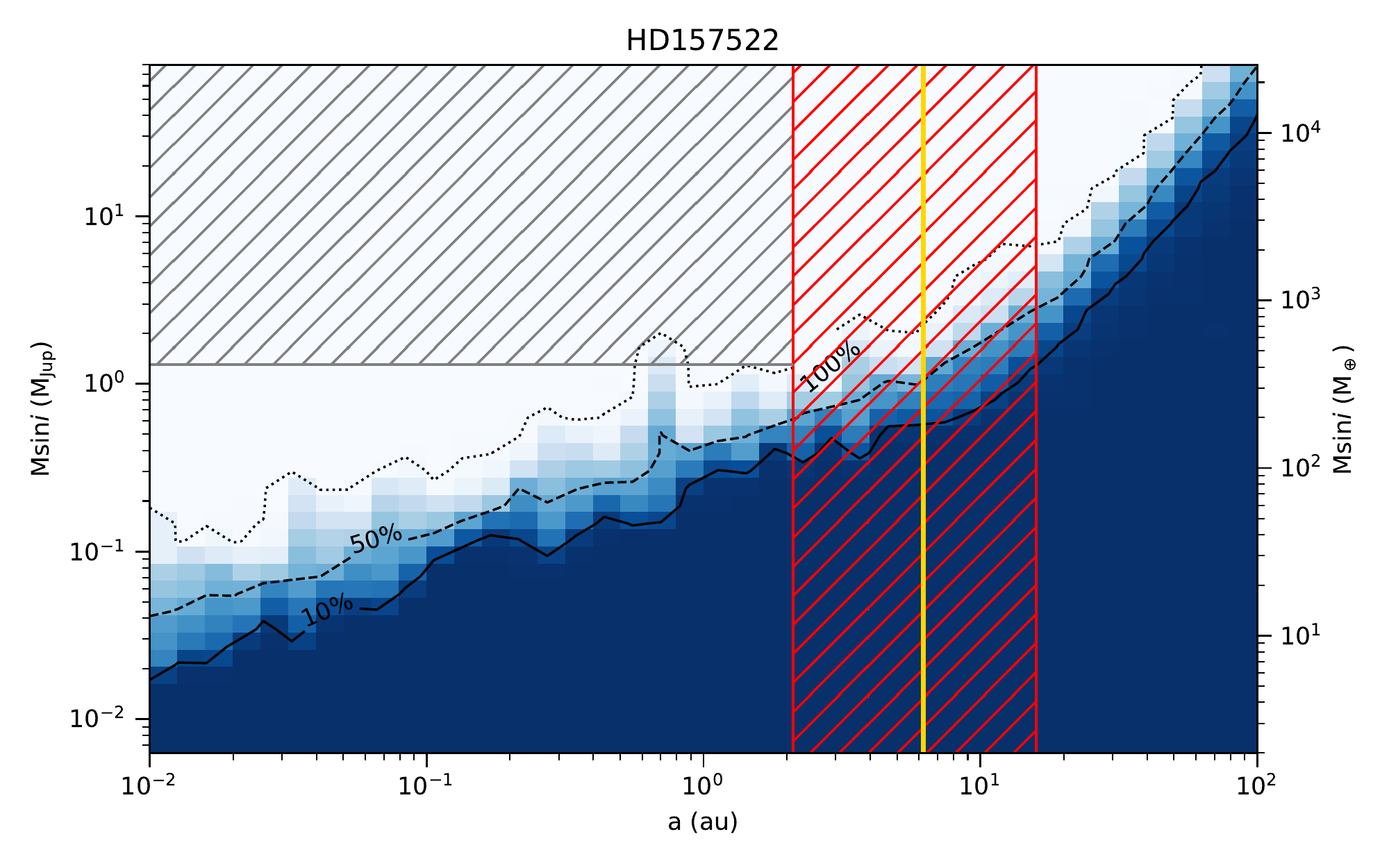}\\
    
    		\includegraphics[width=0.22\linewidth]{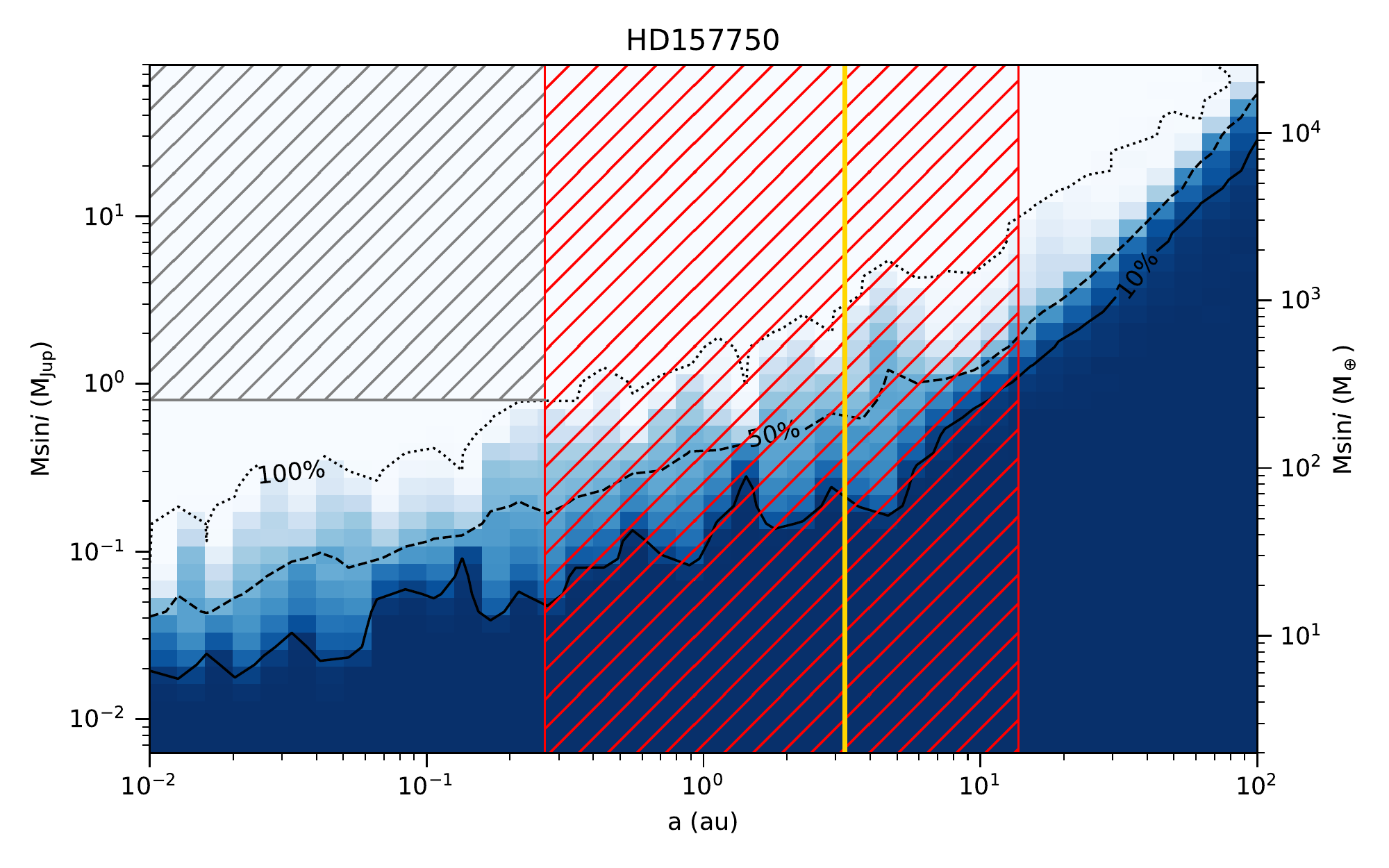}&
    		\includegraphics[width=0.22\linewidth]{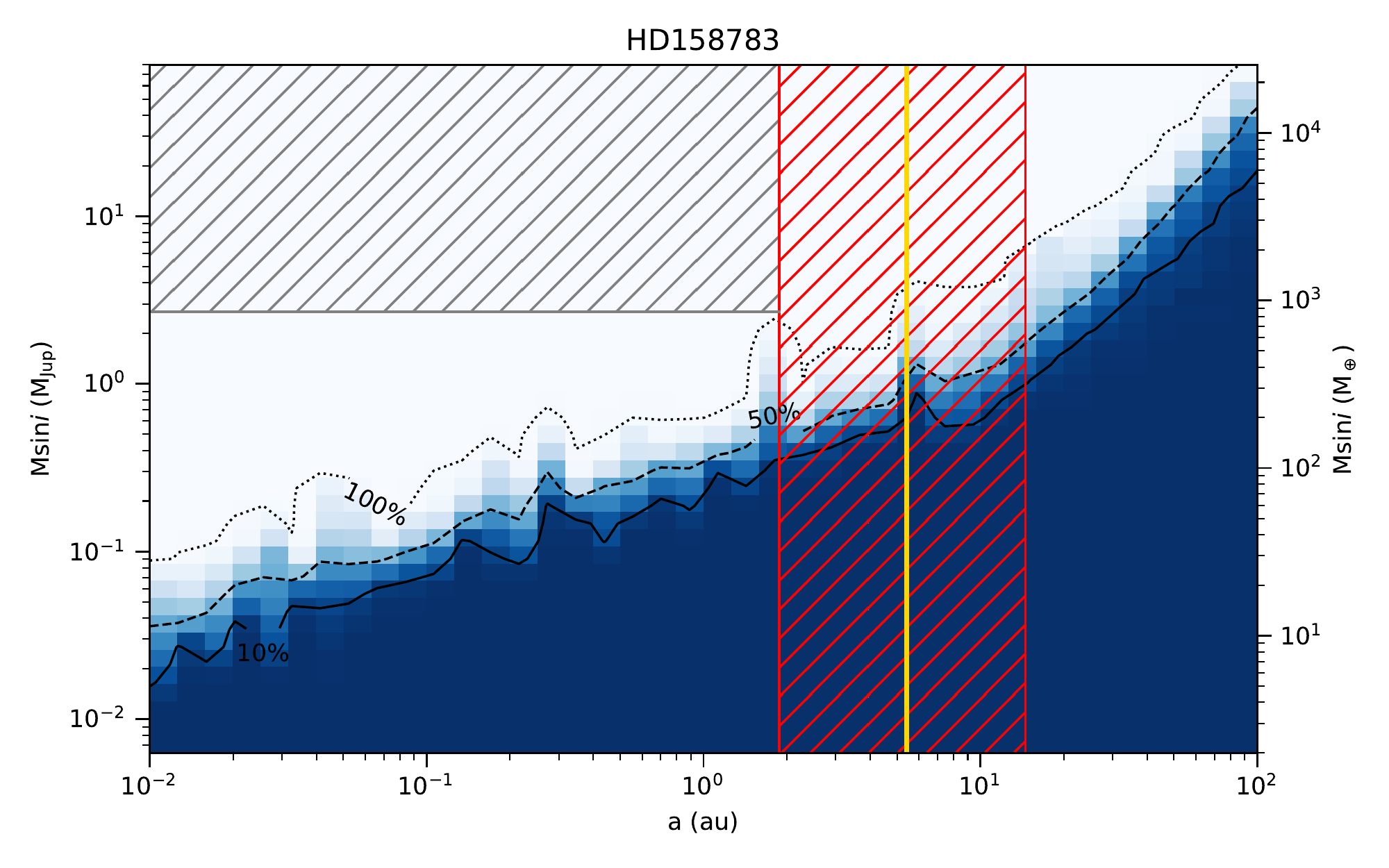}&
    		\includegraphics[width=0.22\linewidth]{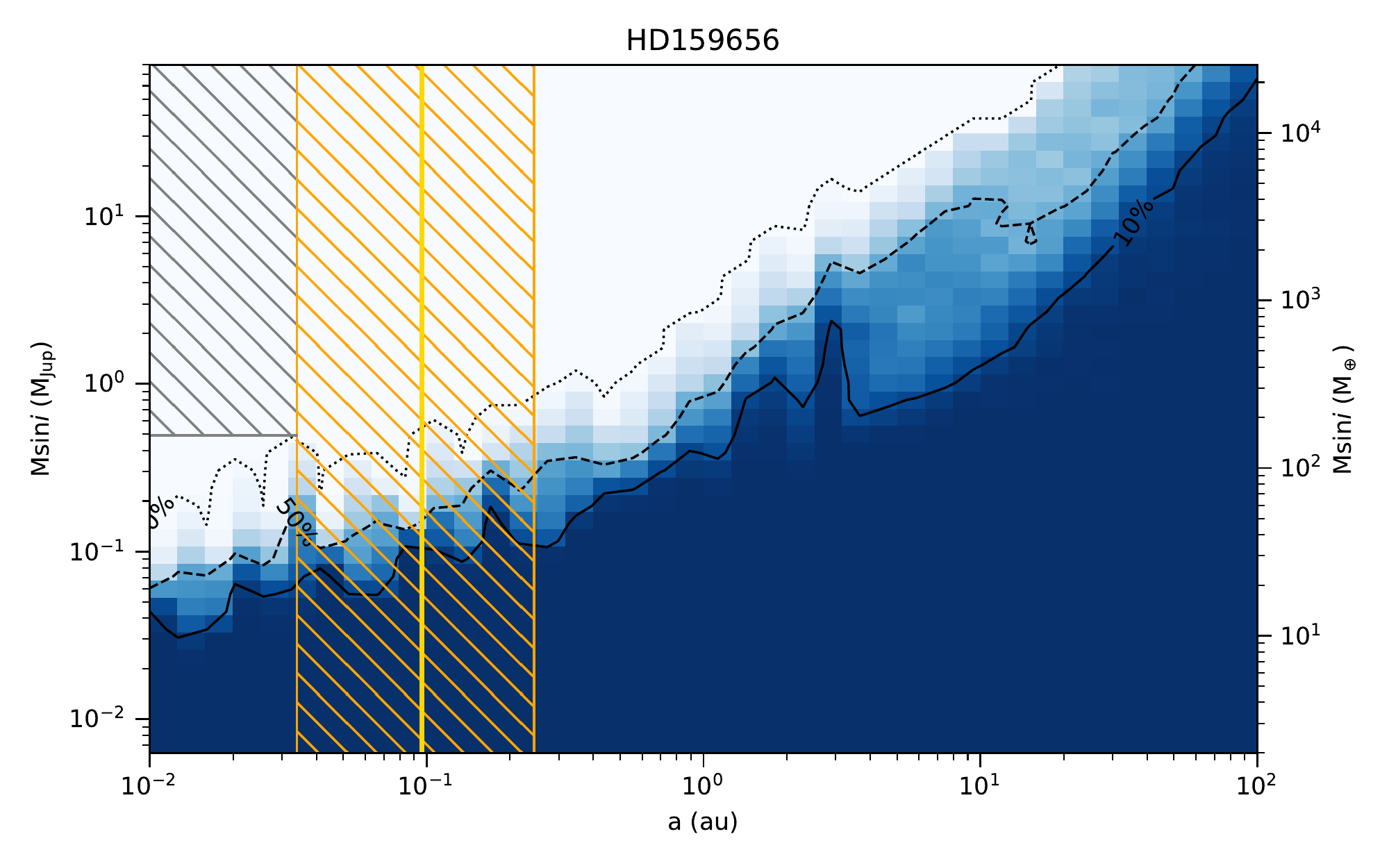}&
    		\includegraphics[width=0.22\linewidth]{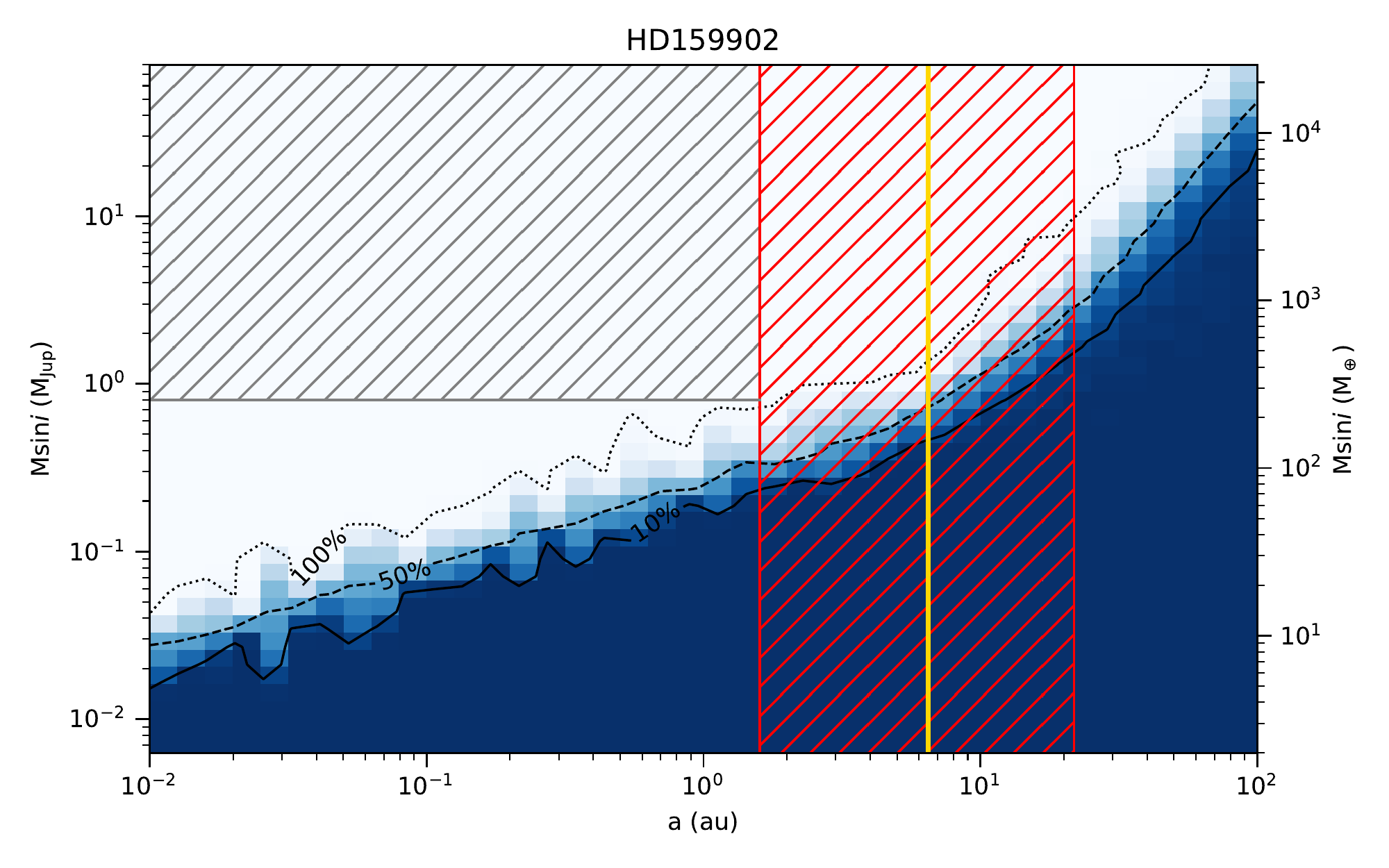}\\
    
    		\includegraphics[width=0.22\linewidth]{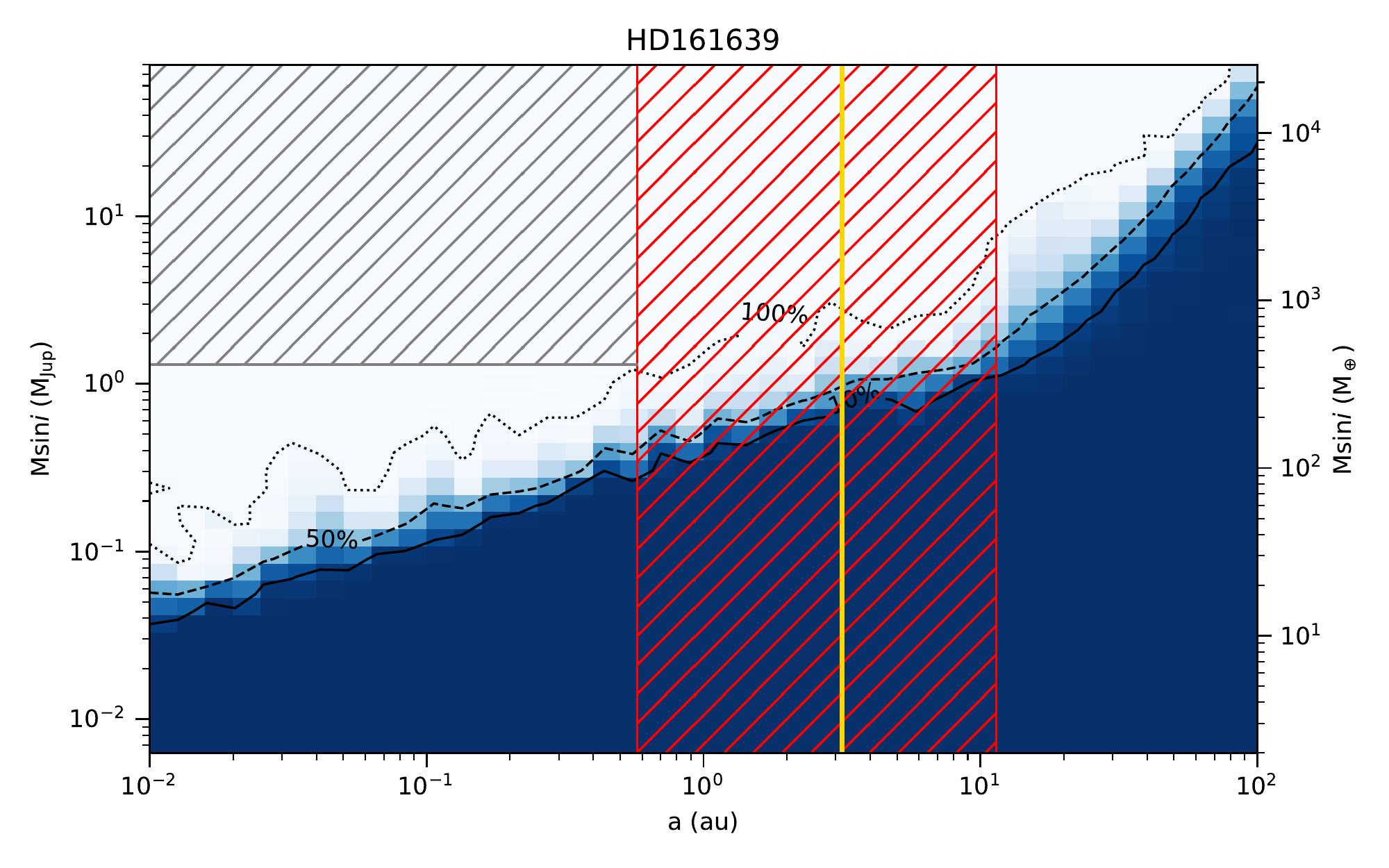}&
    		\includegraphics[width=0.22\linewidth]{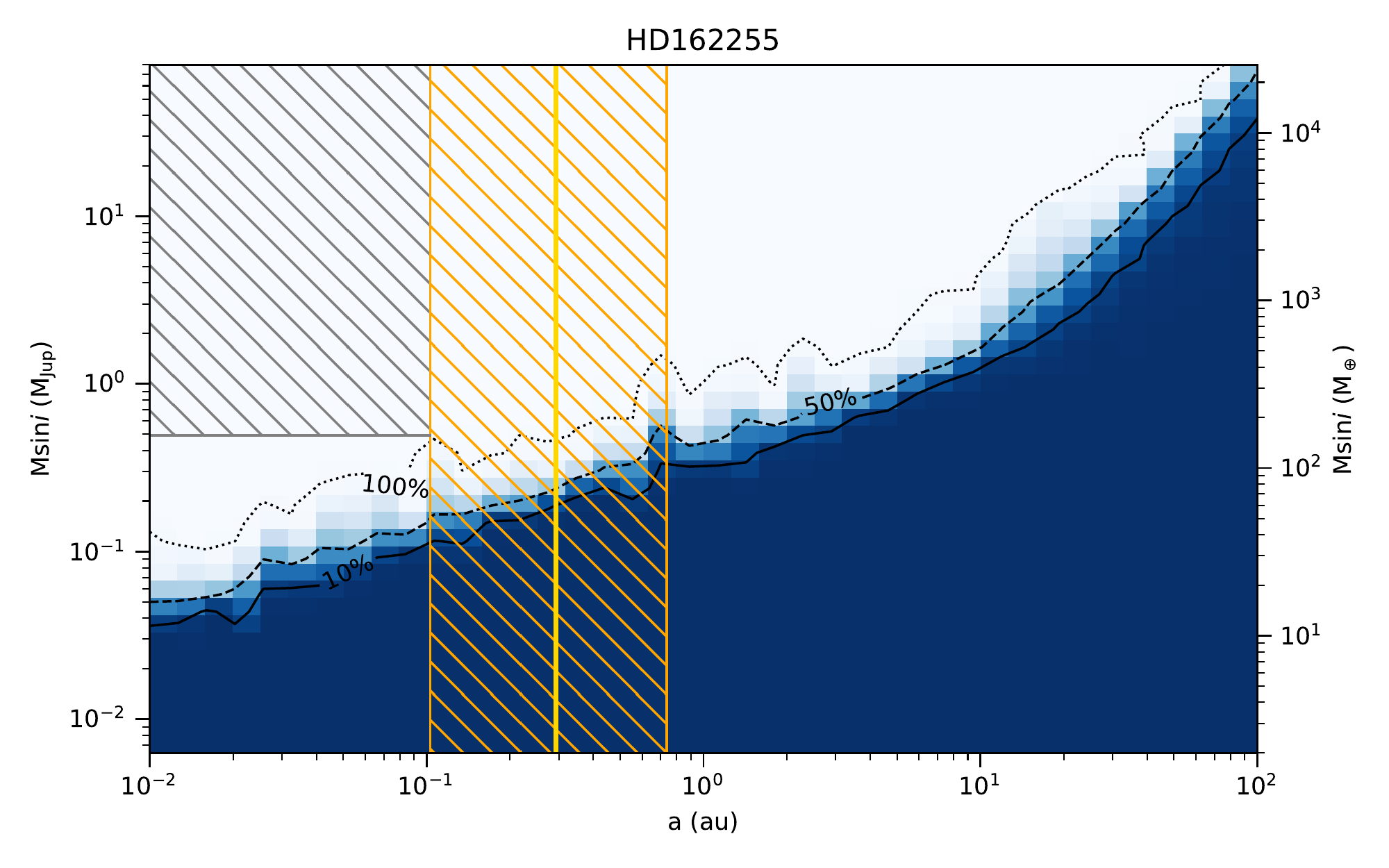}&
    		\includegraphics[width=0.22\linewidth]{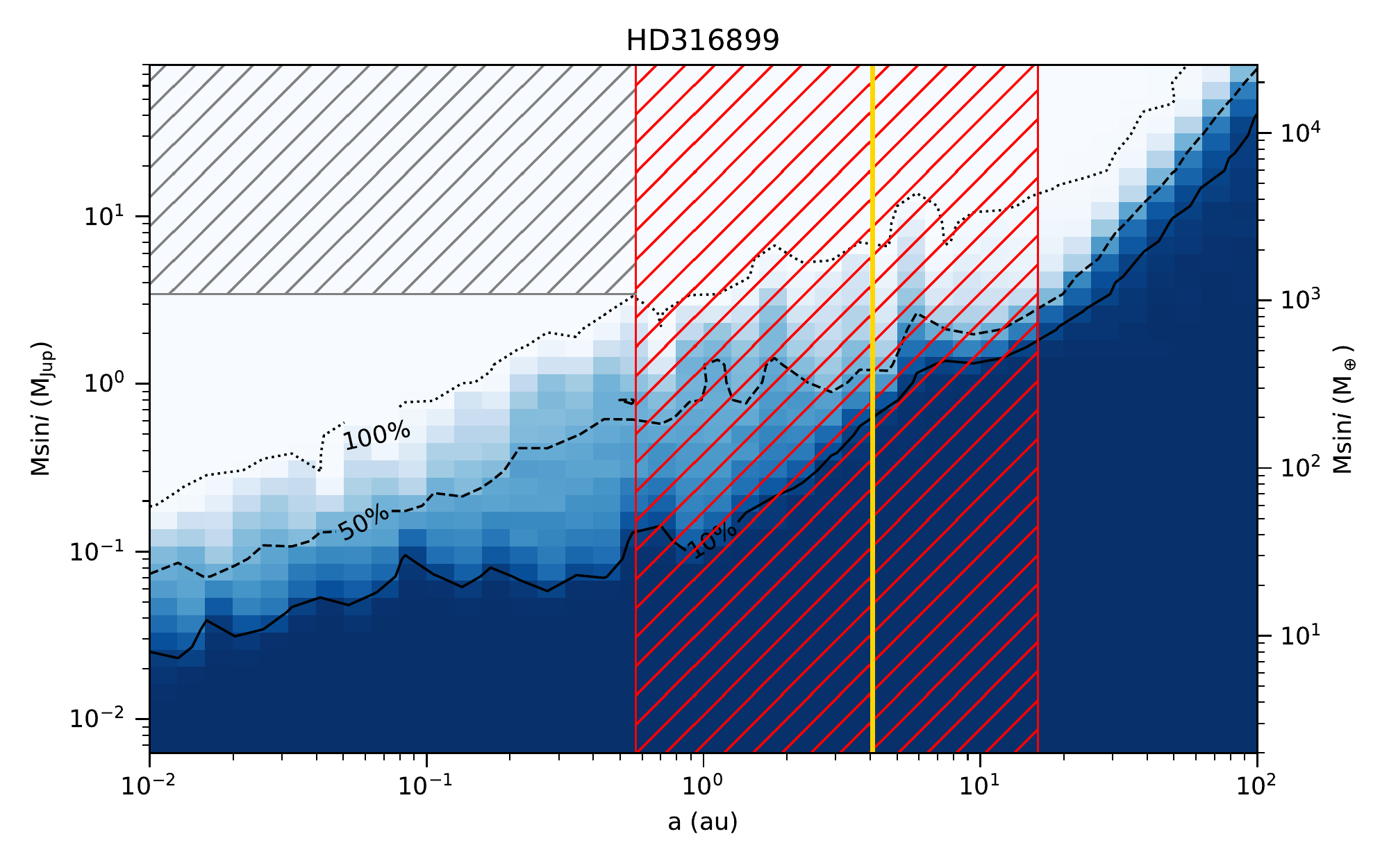}&
    		\includegraphics[width=0.22\linewidth]{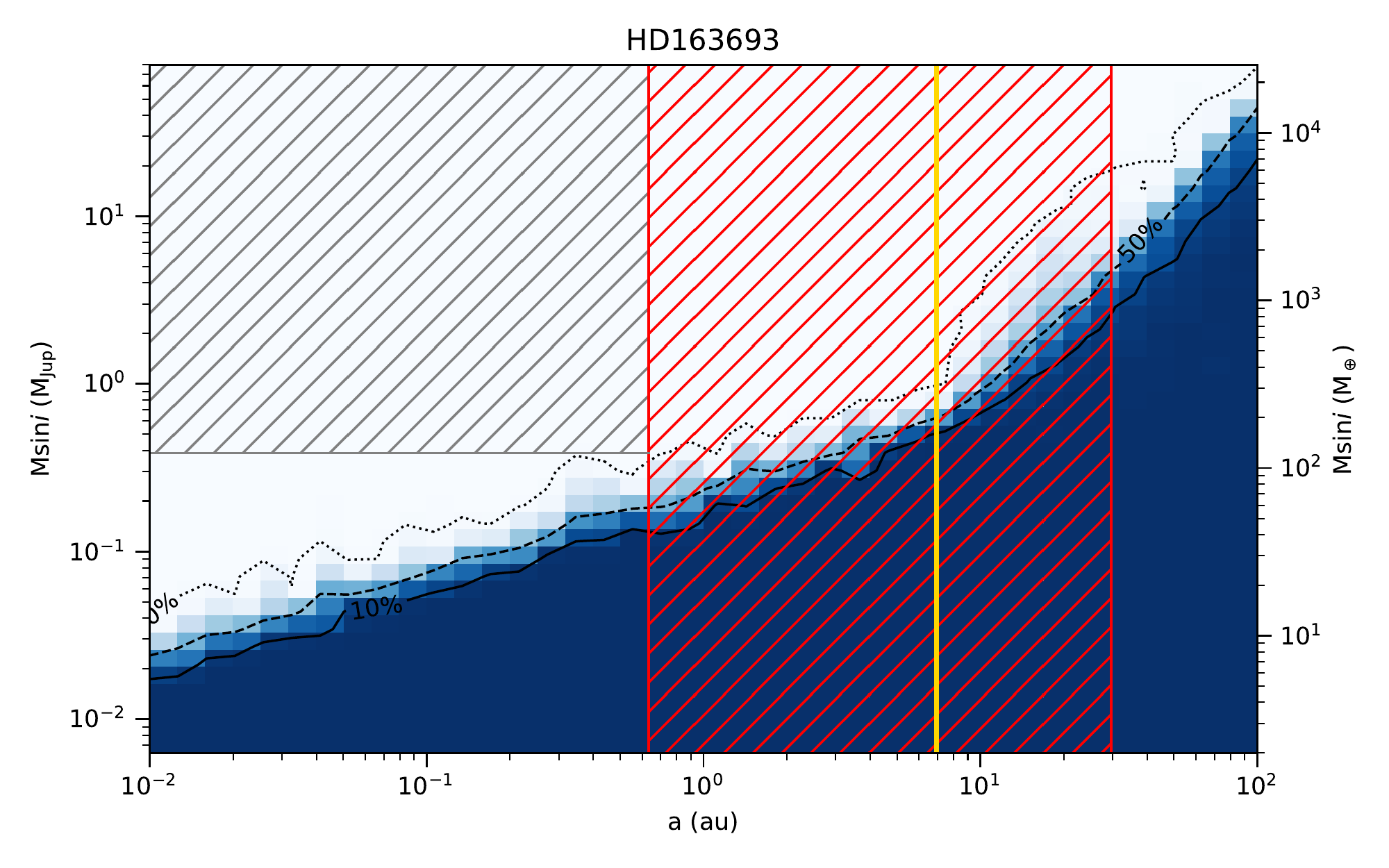}\\
    
    		\includegraphics[width=0.22\linewidth]{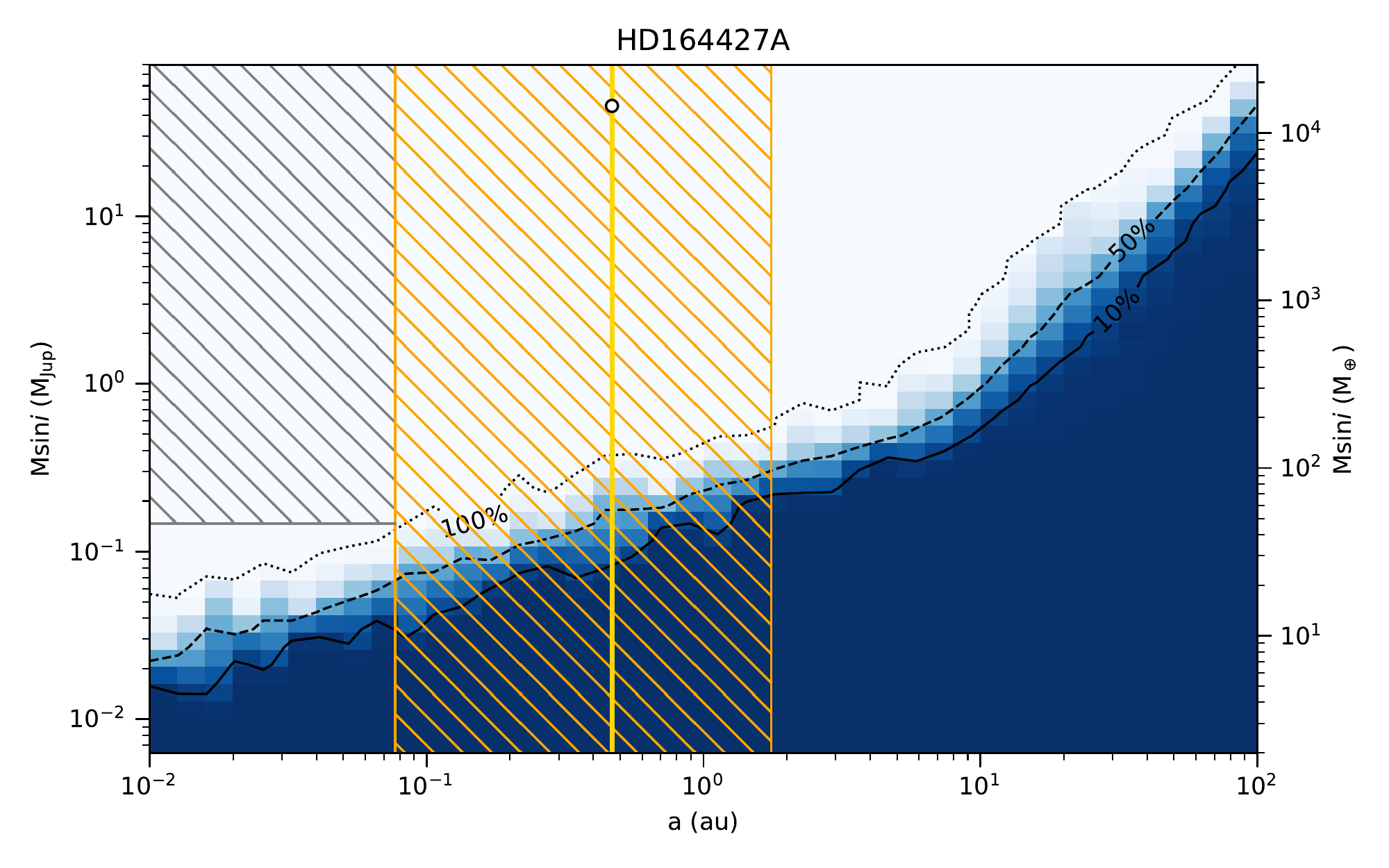}&
    		\includegraphics[width=0.22\linewidth]{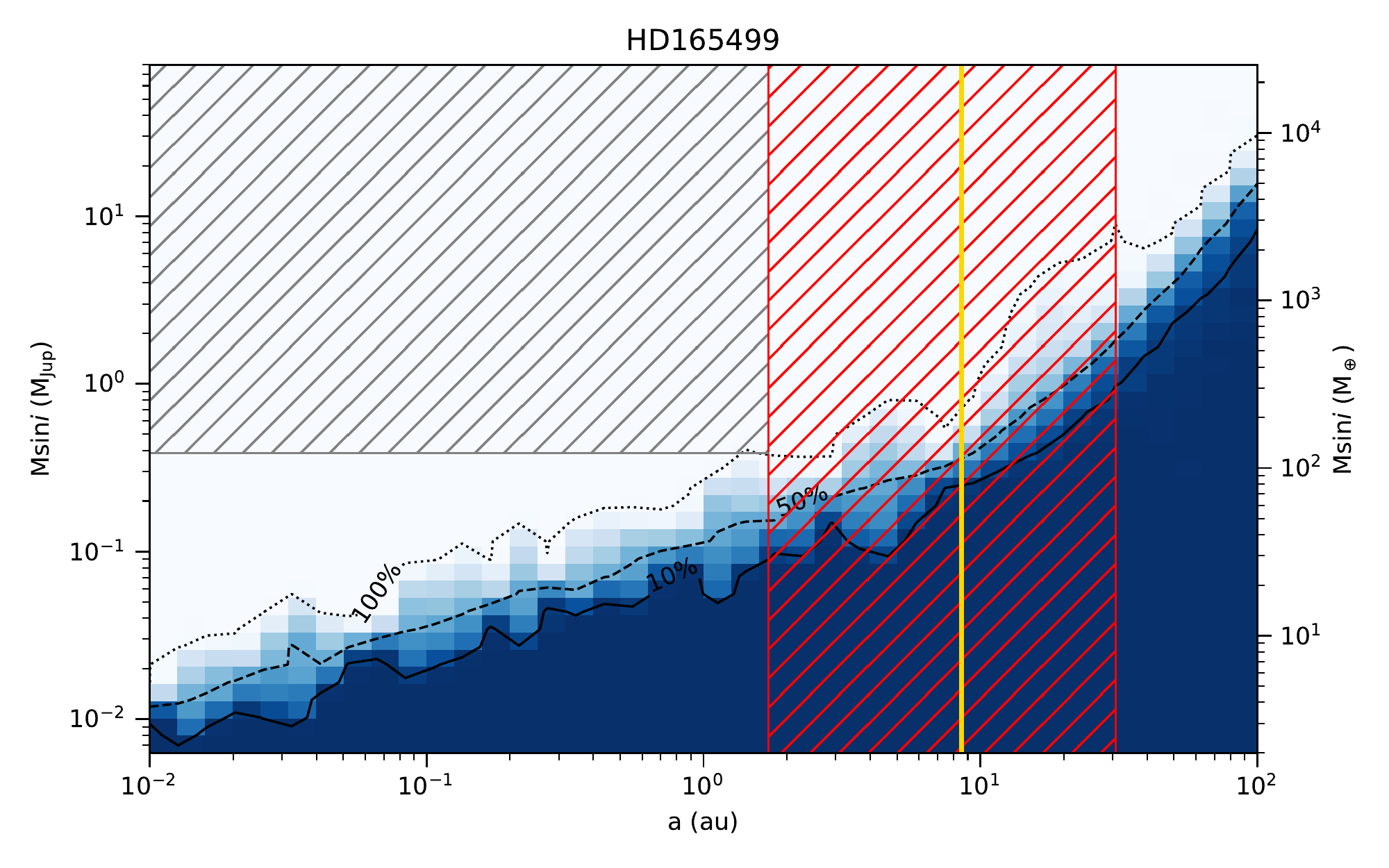}&
    		\includegraphics[width=0.22\linewidth]{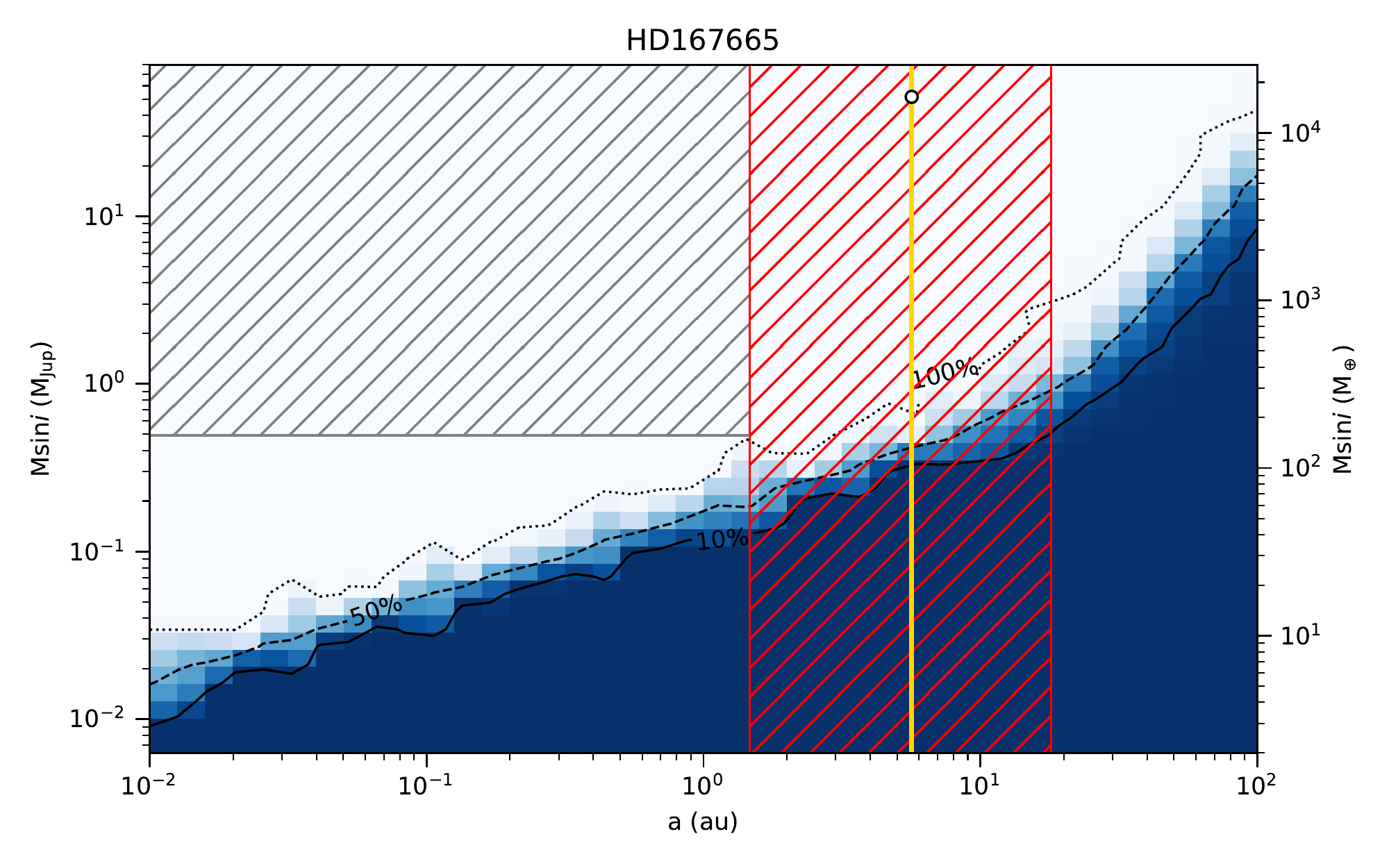}&
    		\includegraphics[width=0.22\linewidth]{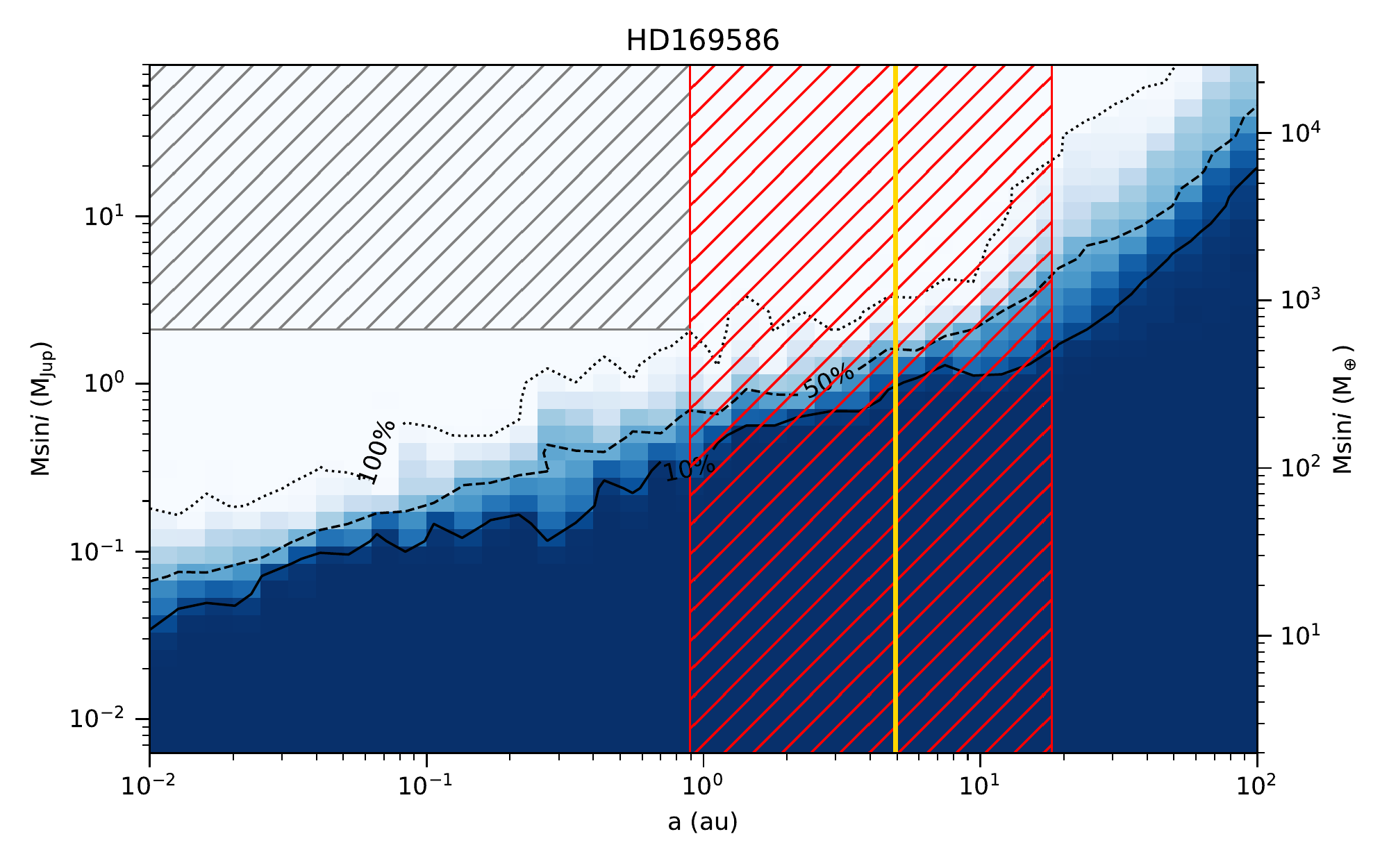}\\
    
    		\includegraphics[width=0.22\linewidth]{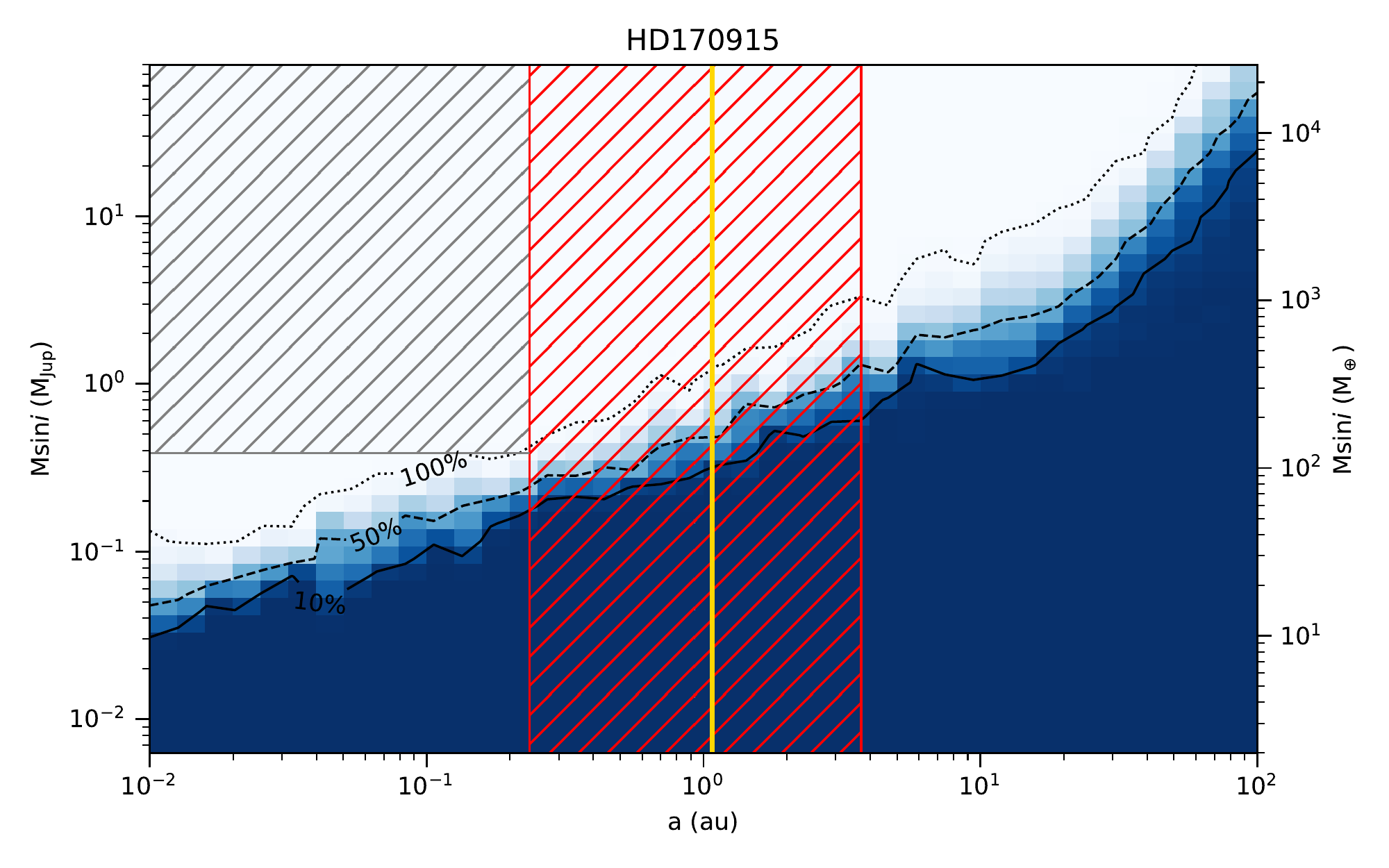}&
    		\includegraphics[width=0.22\linewidth]{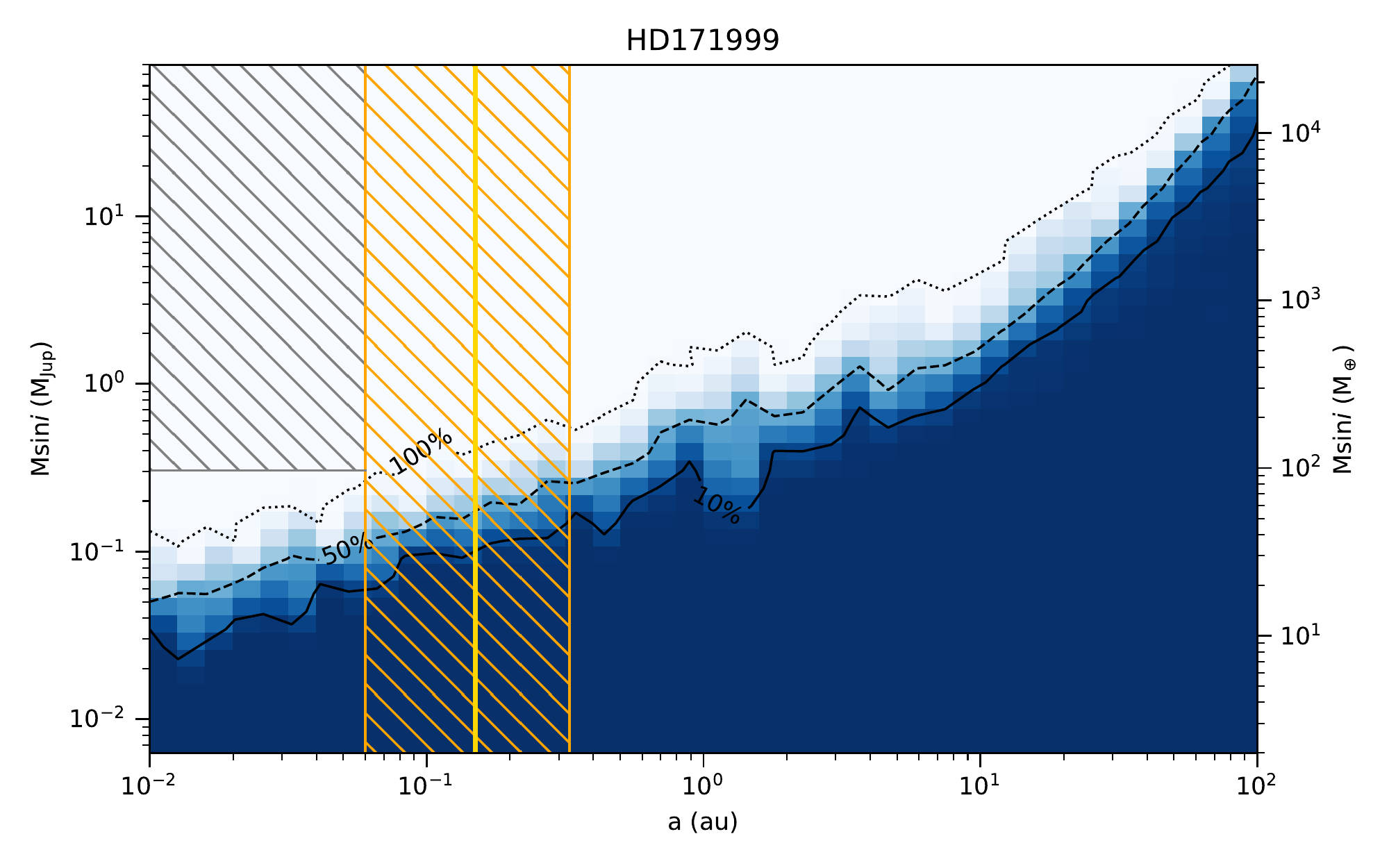}&
    		\includegraphics[width=0.22\linewidth]{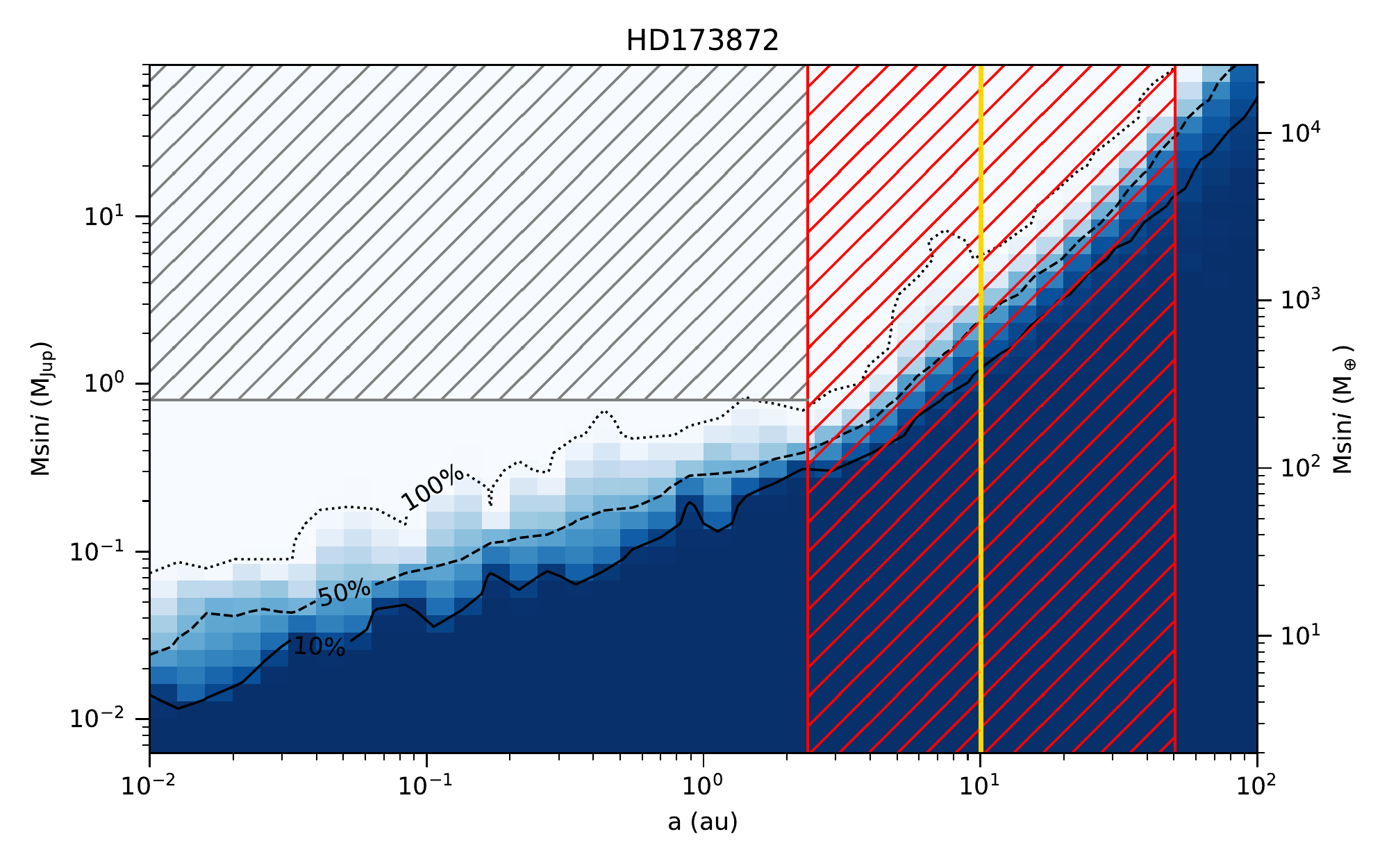}&
    		\includegraphics[width=0.22\linewidth]{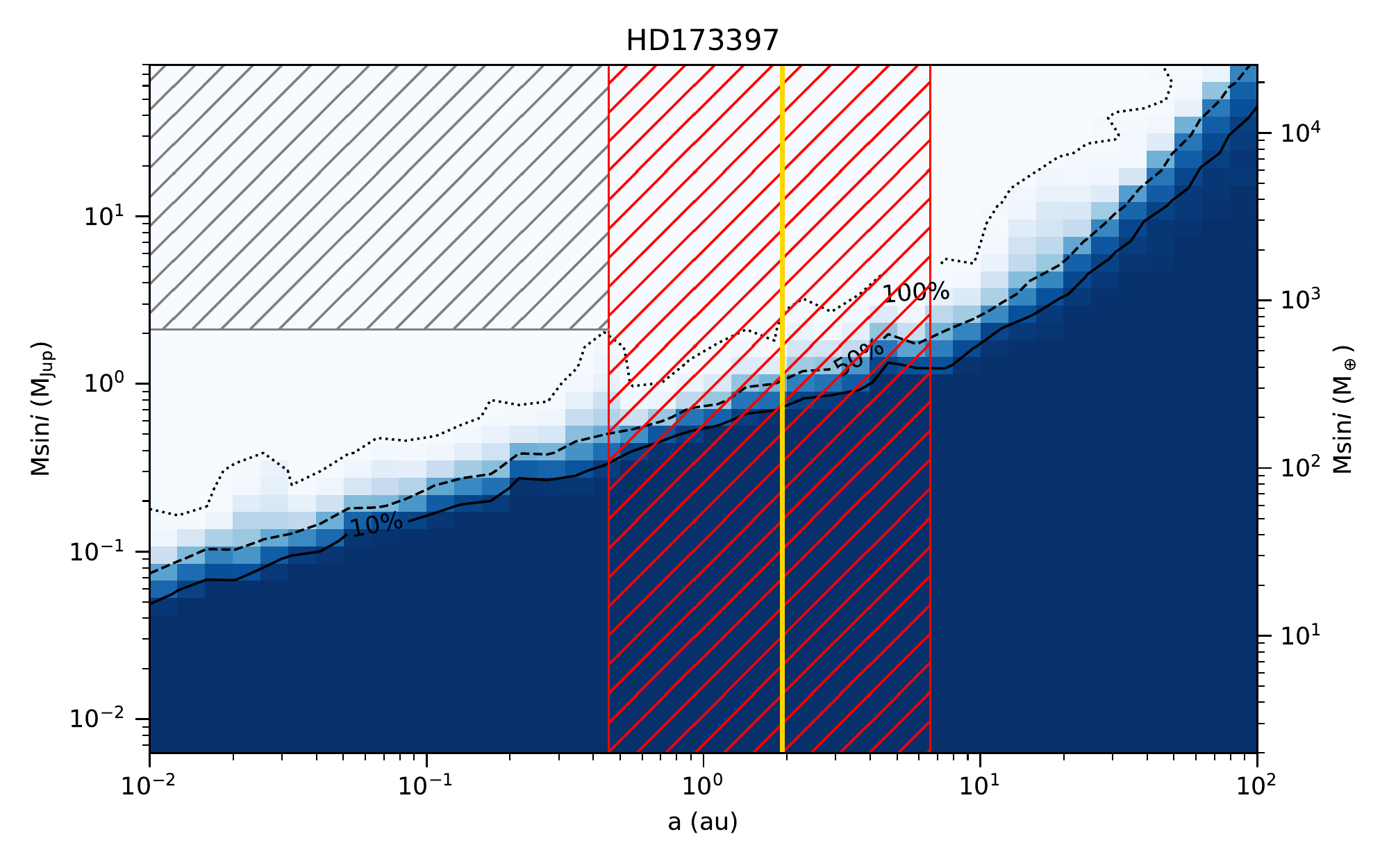}\\
    
    		\includegraphics[width=0.22\linewidth]{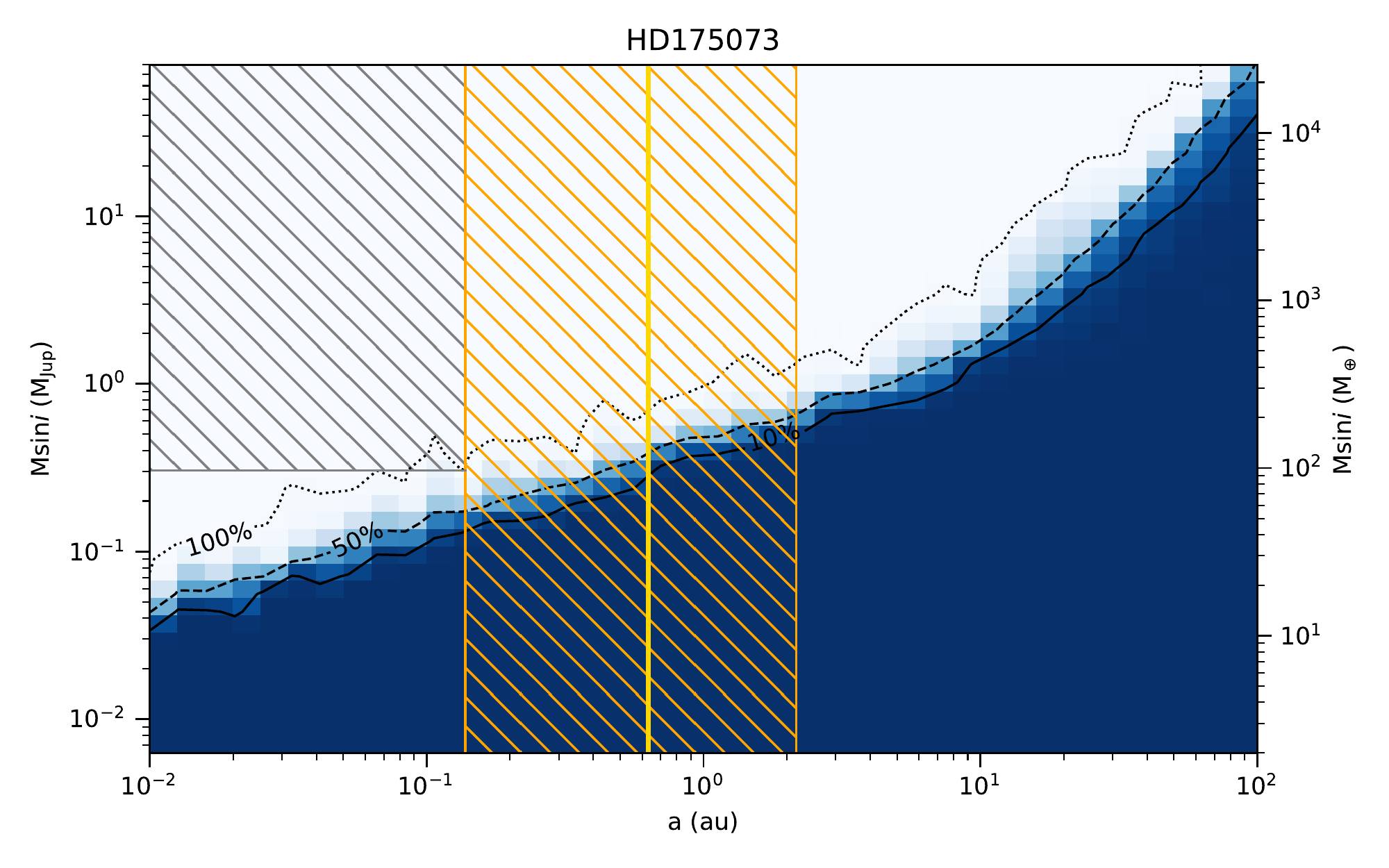}&
    		\includegraphics[width=0.22\linewidth]{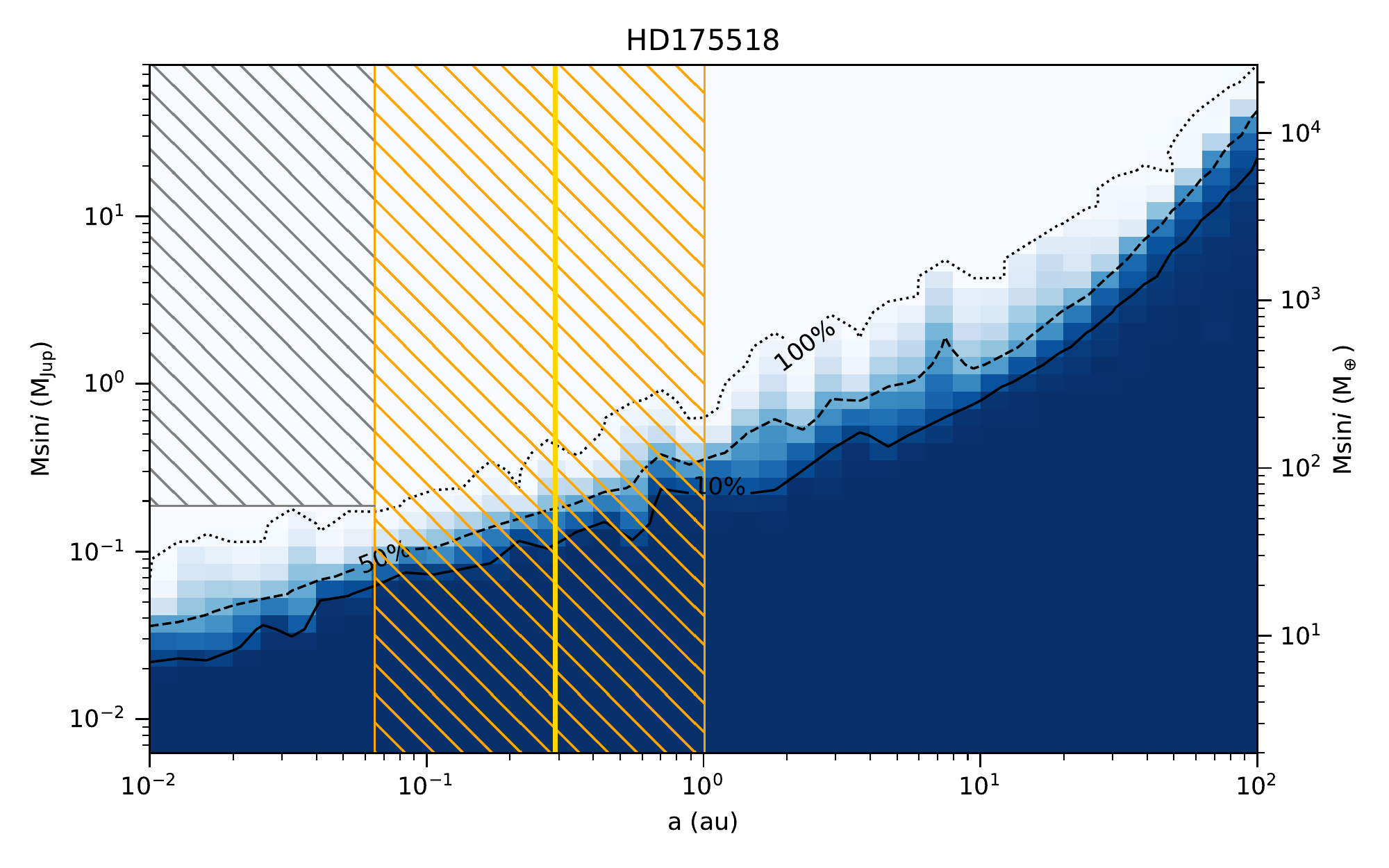}&
    		\includegraphics[width=0.22\linewidth]{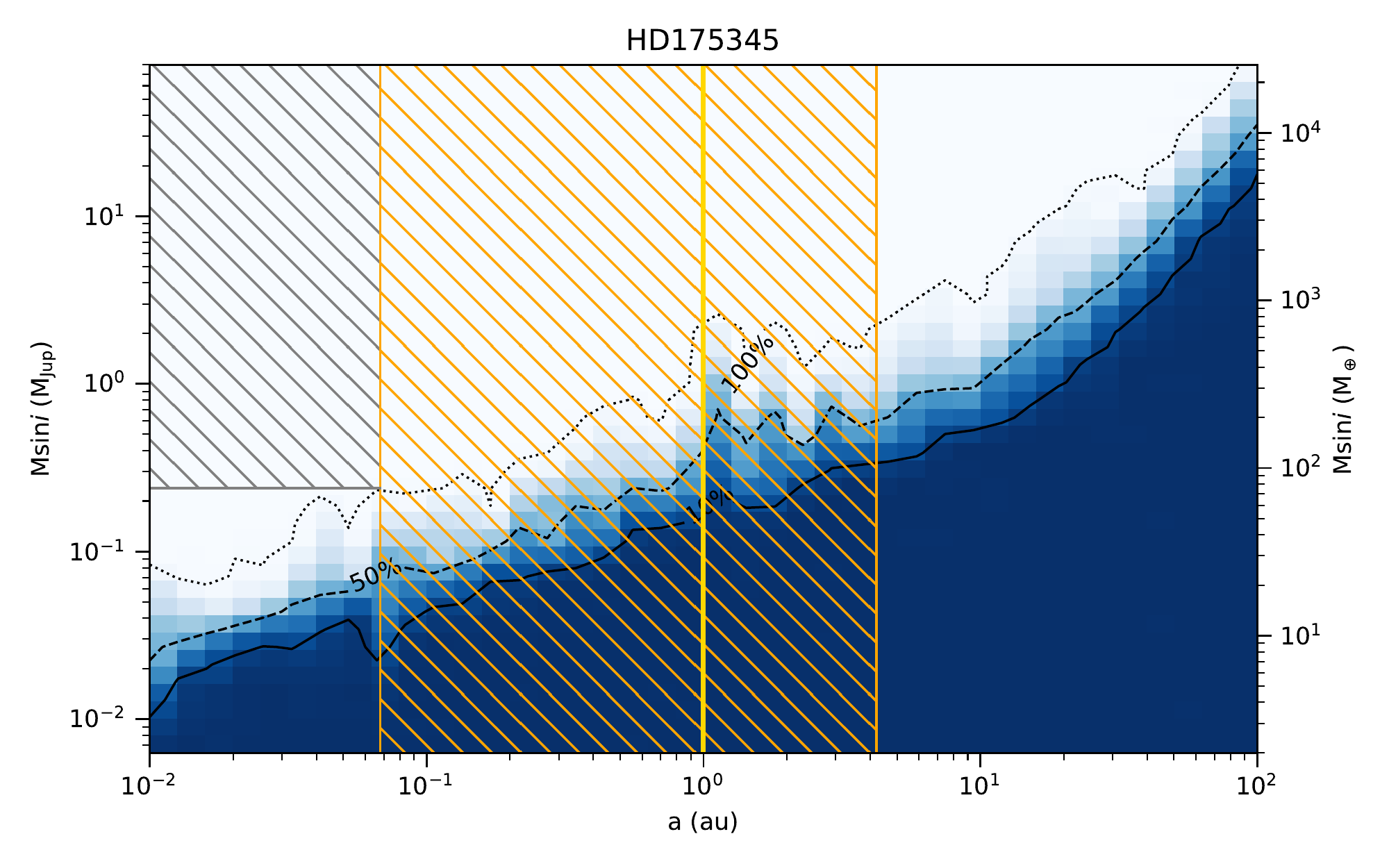}&
    		\includegraphics[width=0.22\linewidth]{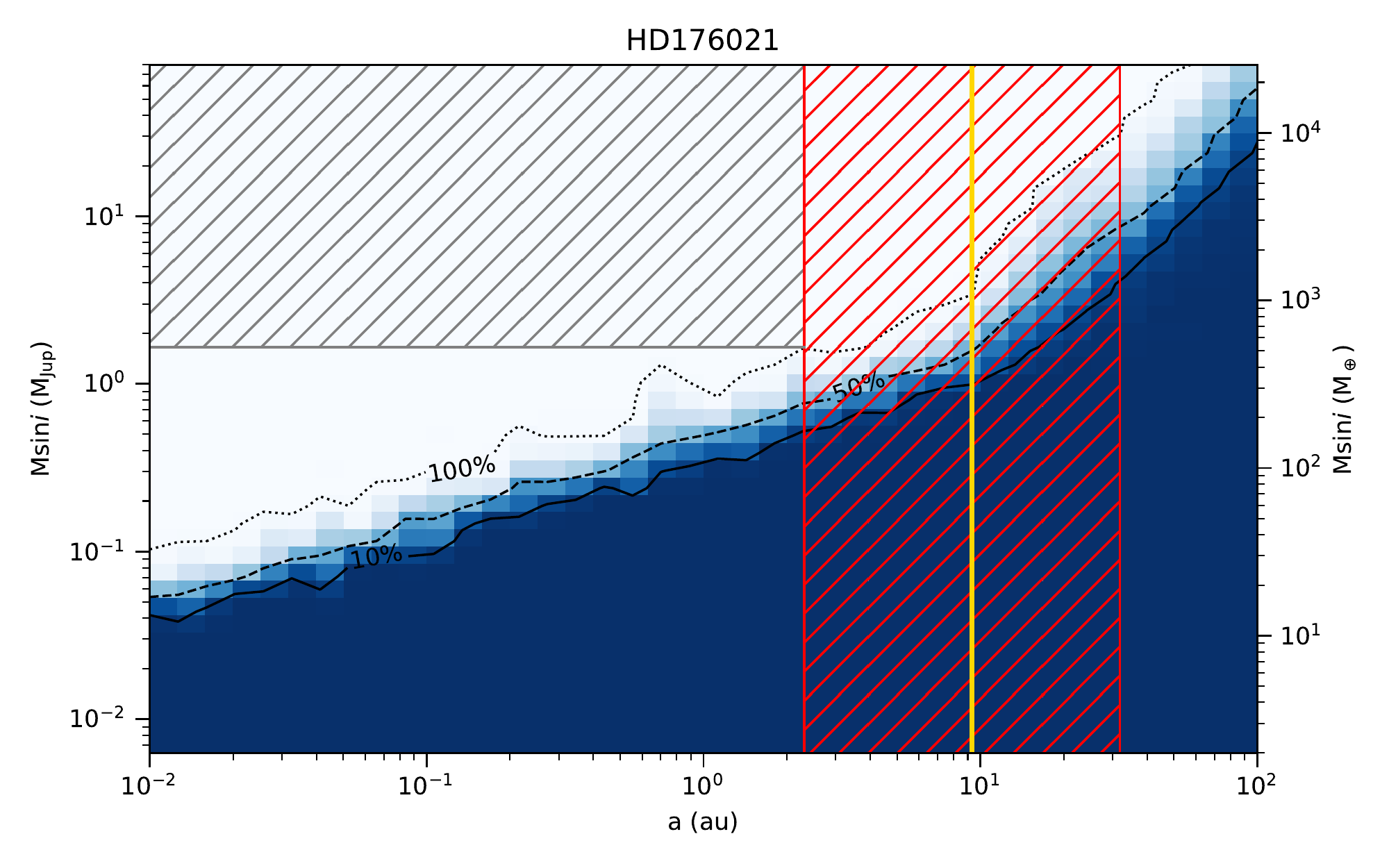}\\
    
    		\includegraphics[width=0.22\linewidth]{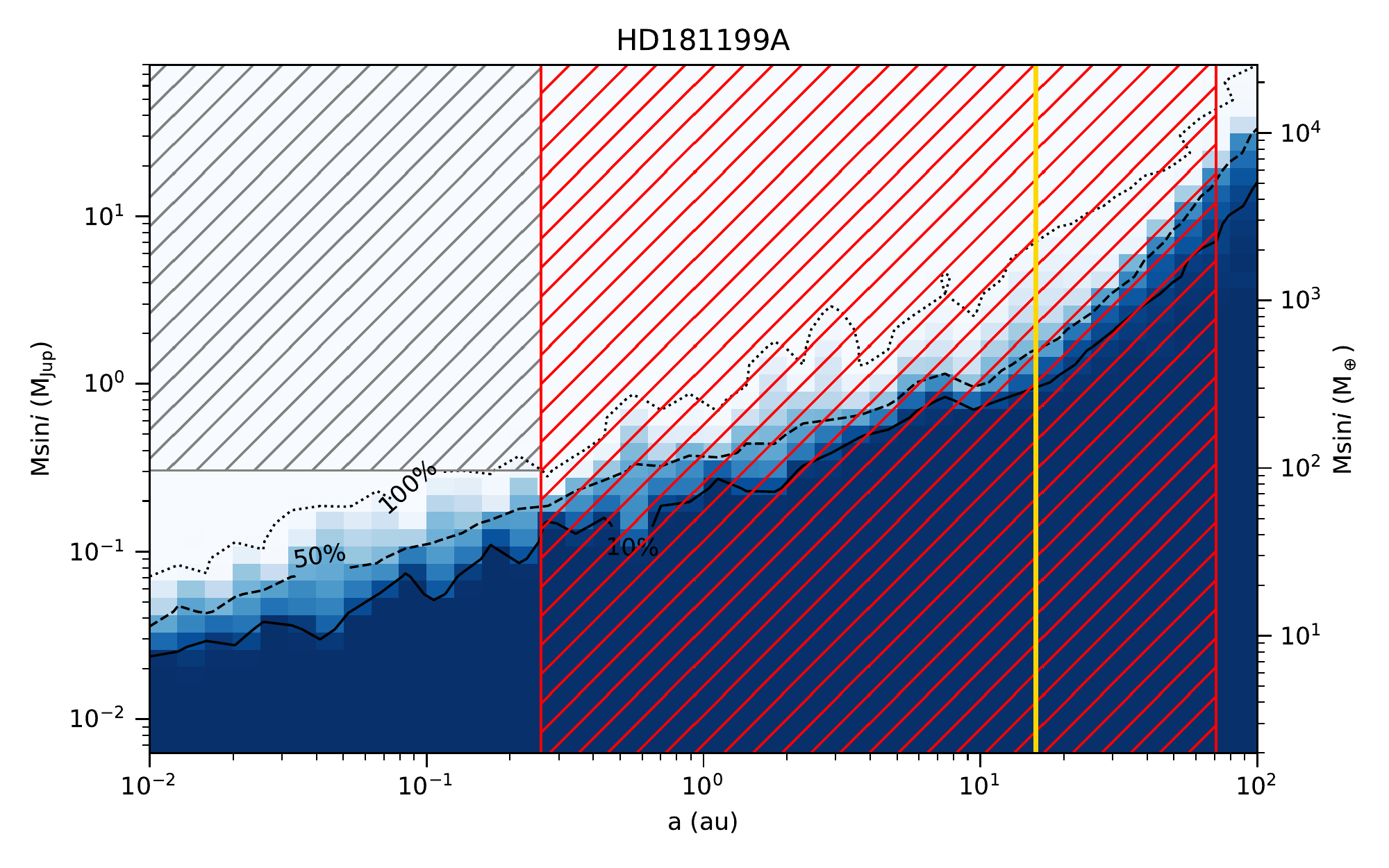}&
    		\includegraphics[width=0.22\linewidth]{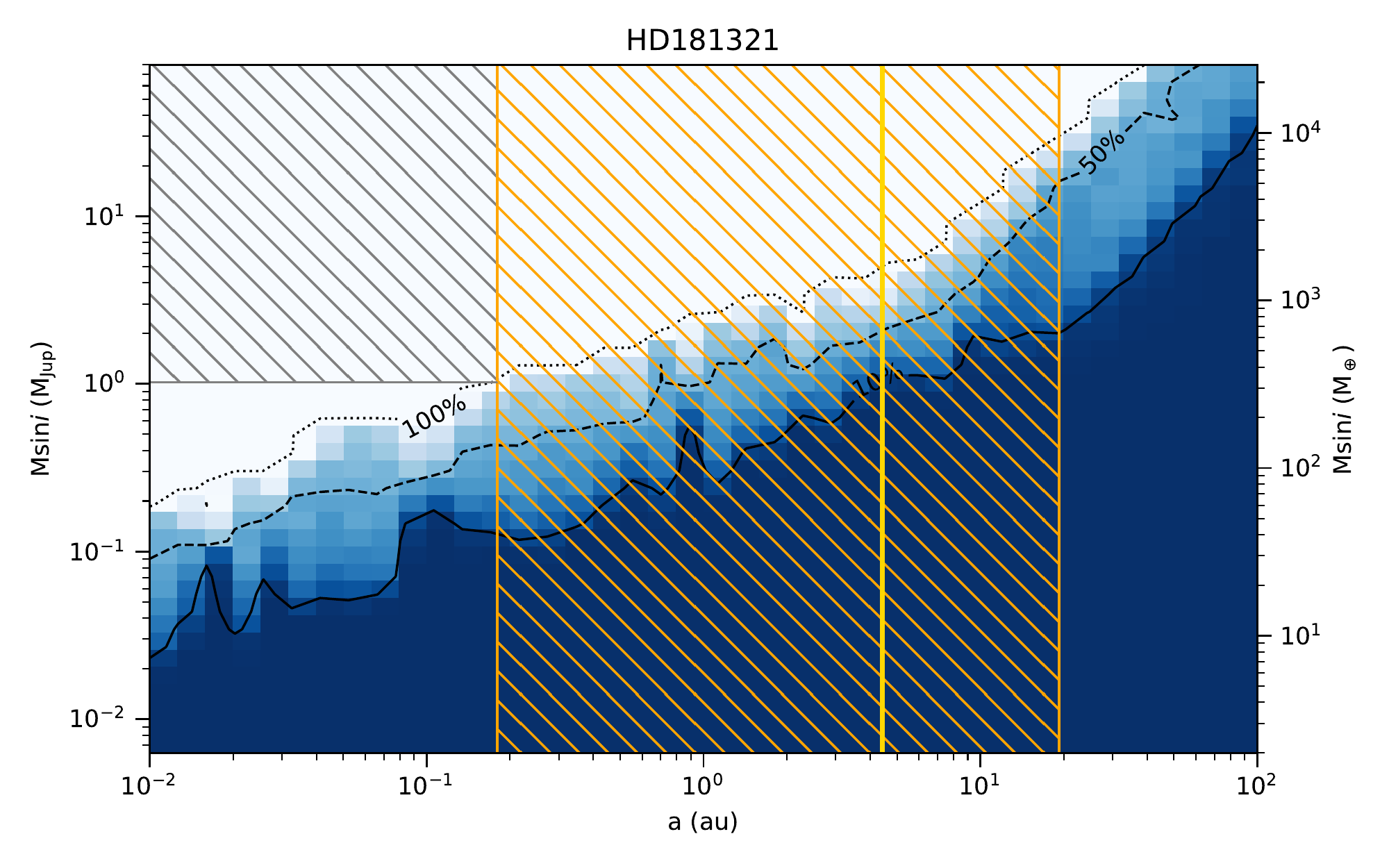}&
    		\includegraphics[width=0.22\linewidth]{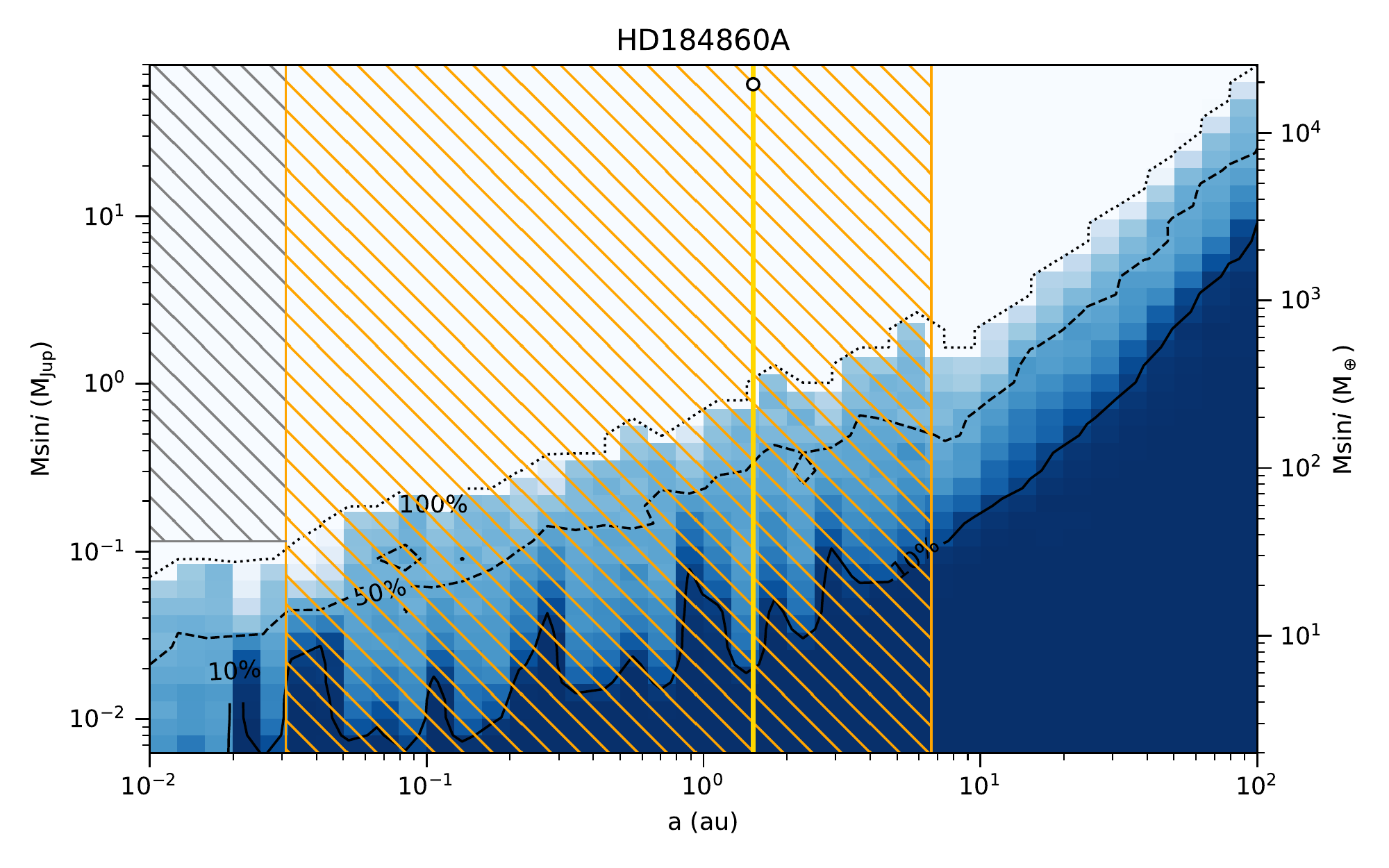}&
    		\includegraphics[width=0.22\linewidth]{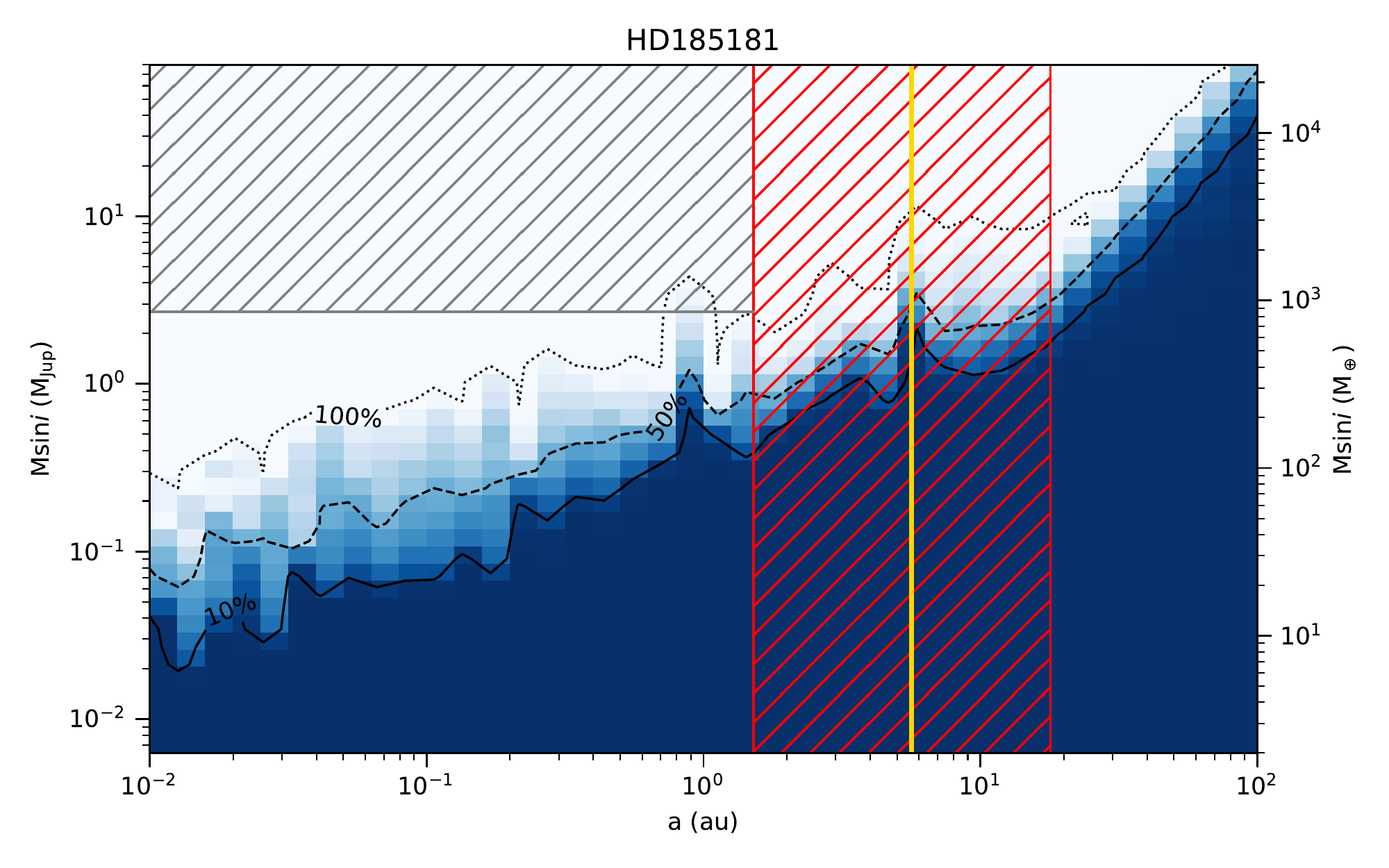}\\
    
    		\includegraphics[width=0.22\linewidth]{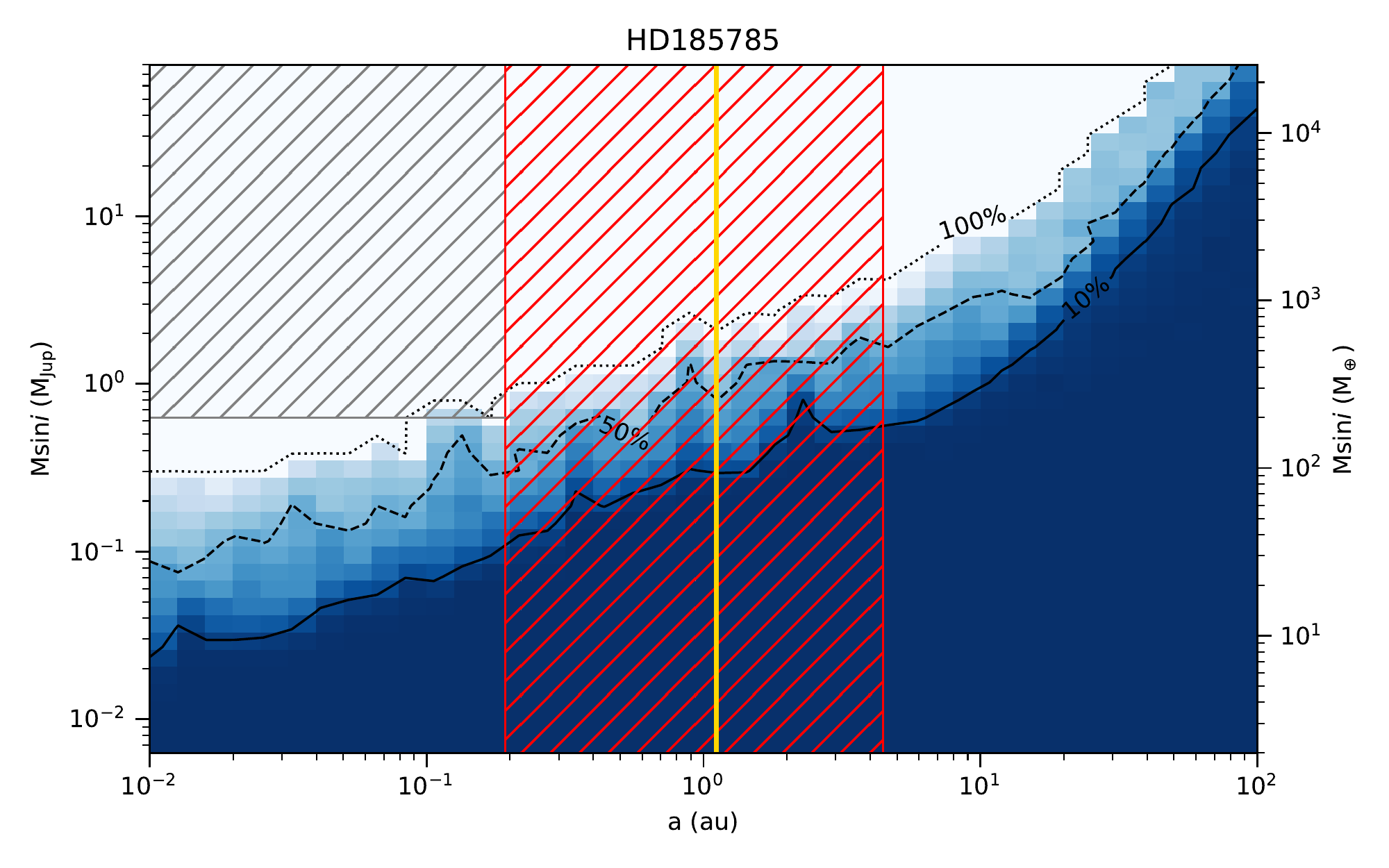}&
    		\includegraphics[width=0.22\linewidth]{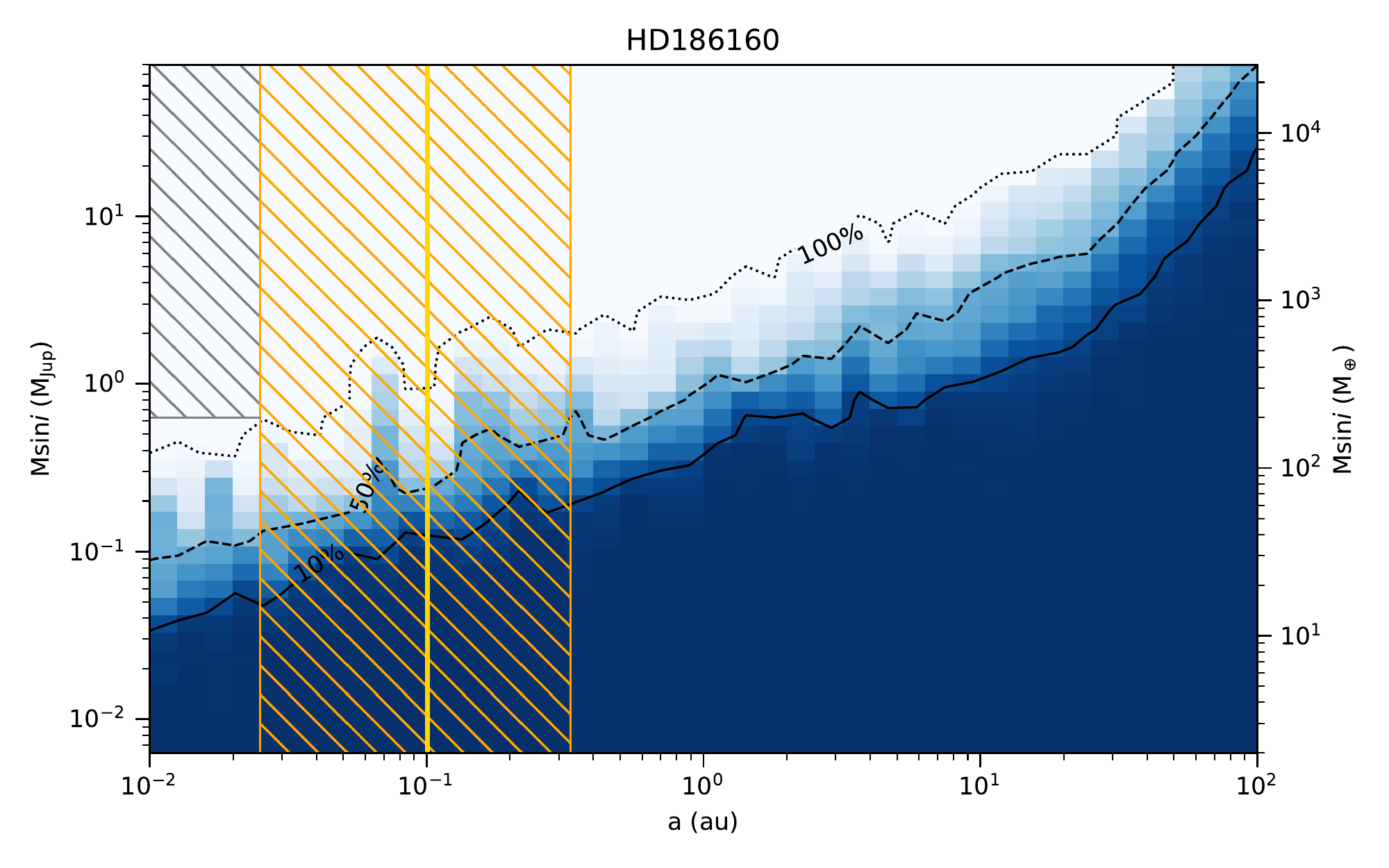}&
    		\includegraphics[width=0.22\linewidth]{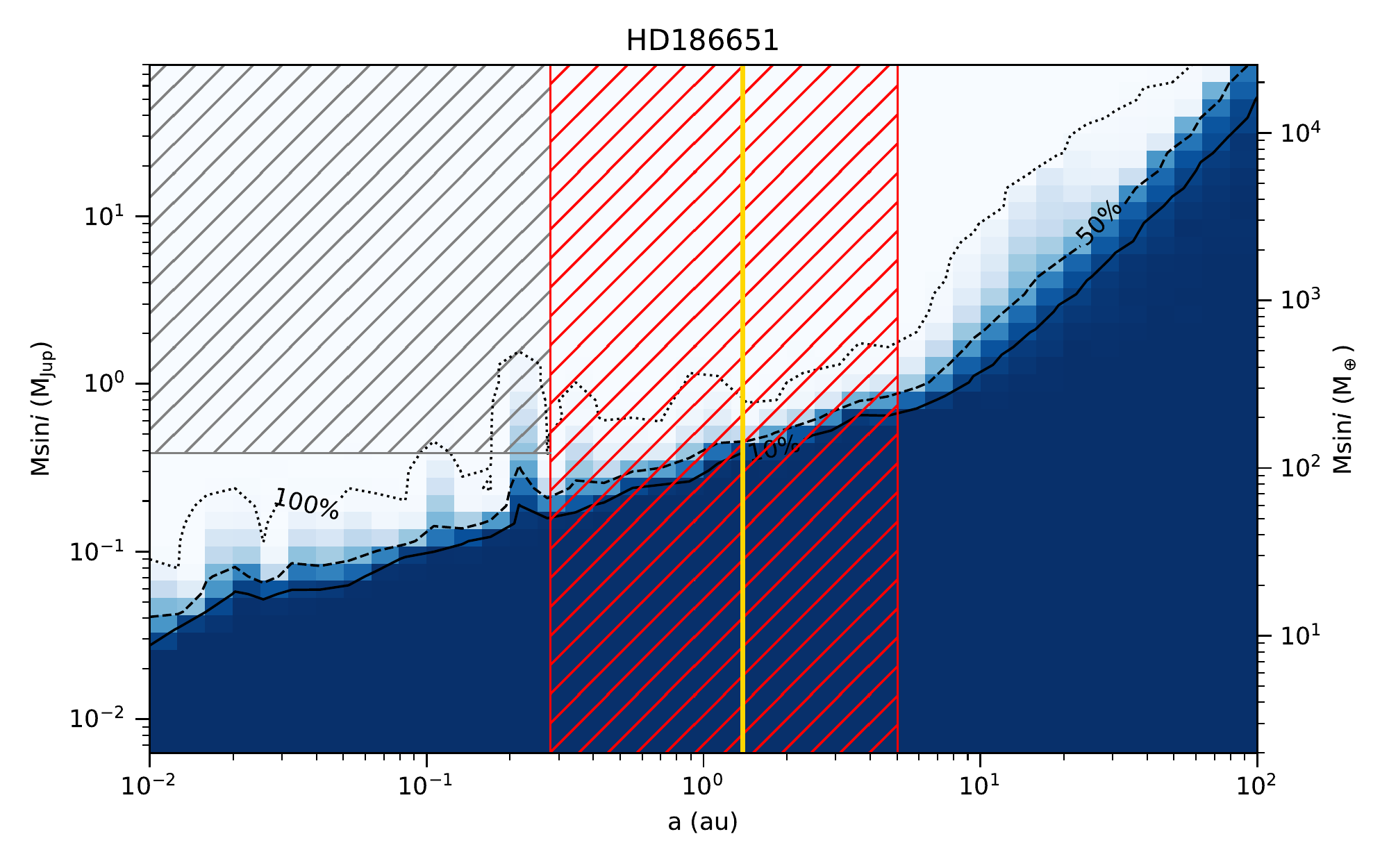}&
    		\includegraphics[width=0.22\linewidth]{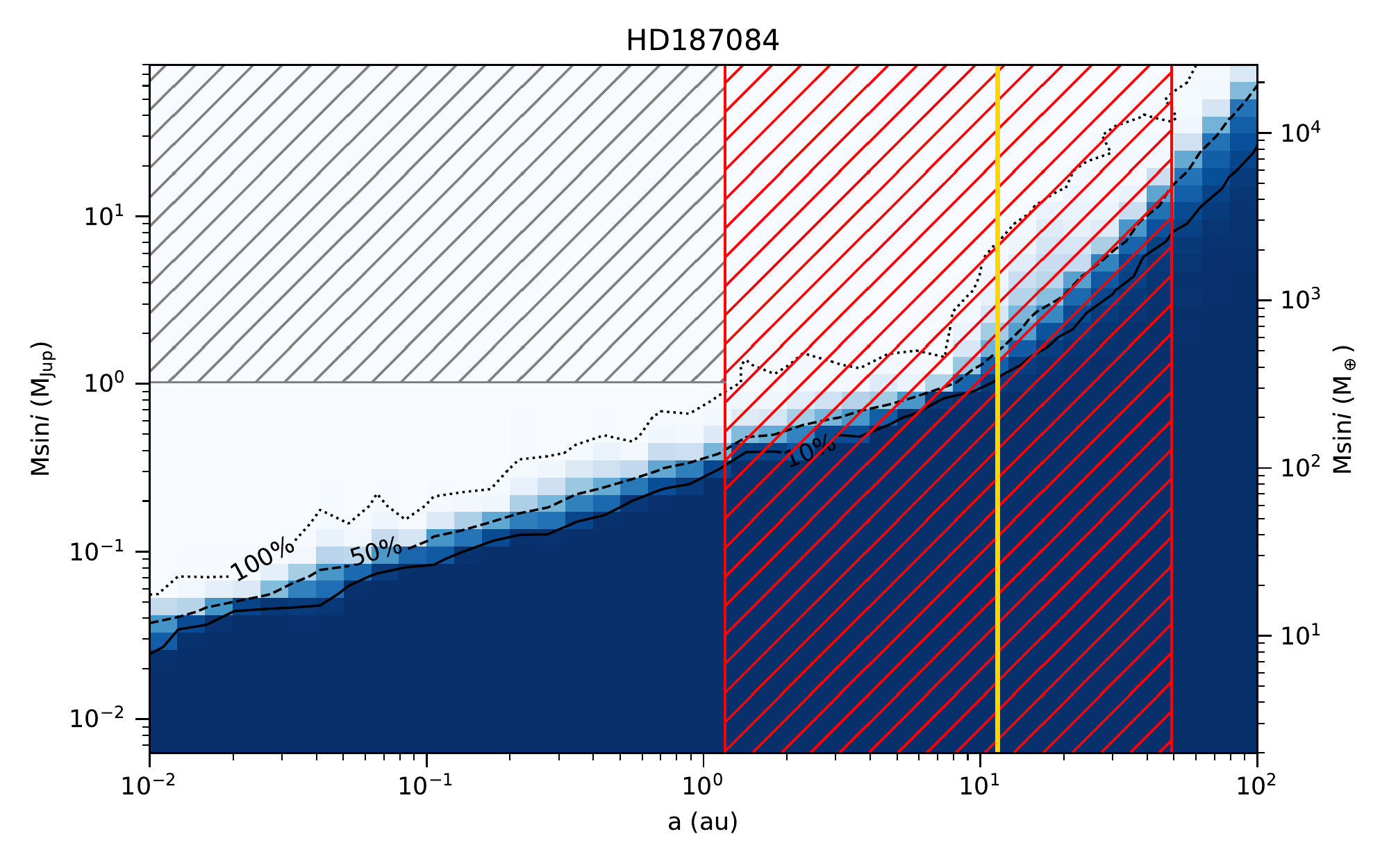}\\
    
    		\includegraphics[width=0.22\linewidth]{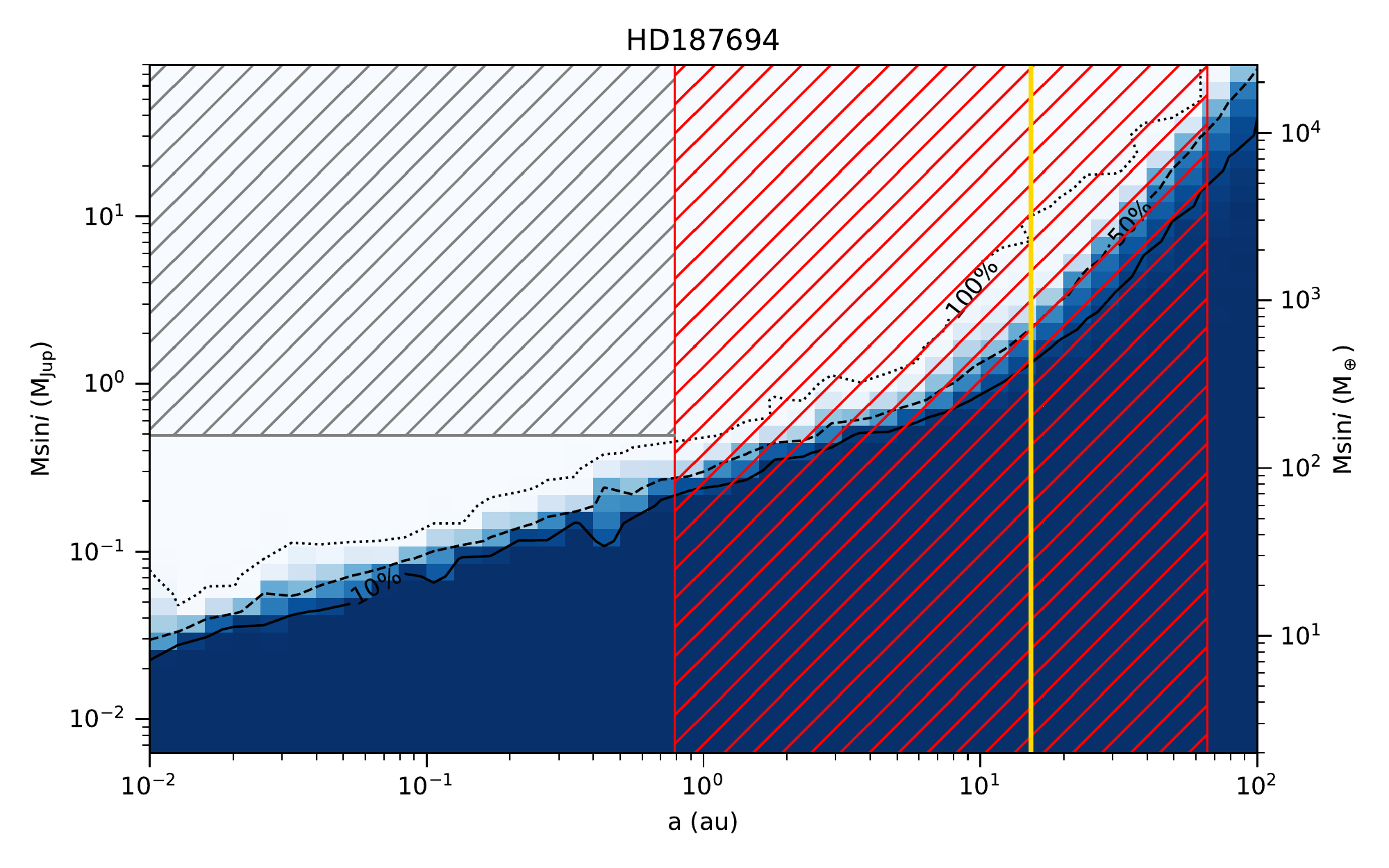}&
    		\includegraphics[width=0.22\linewidth]{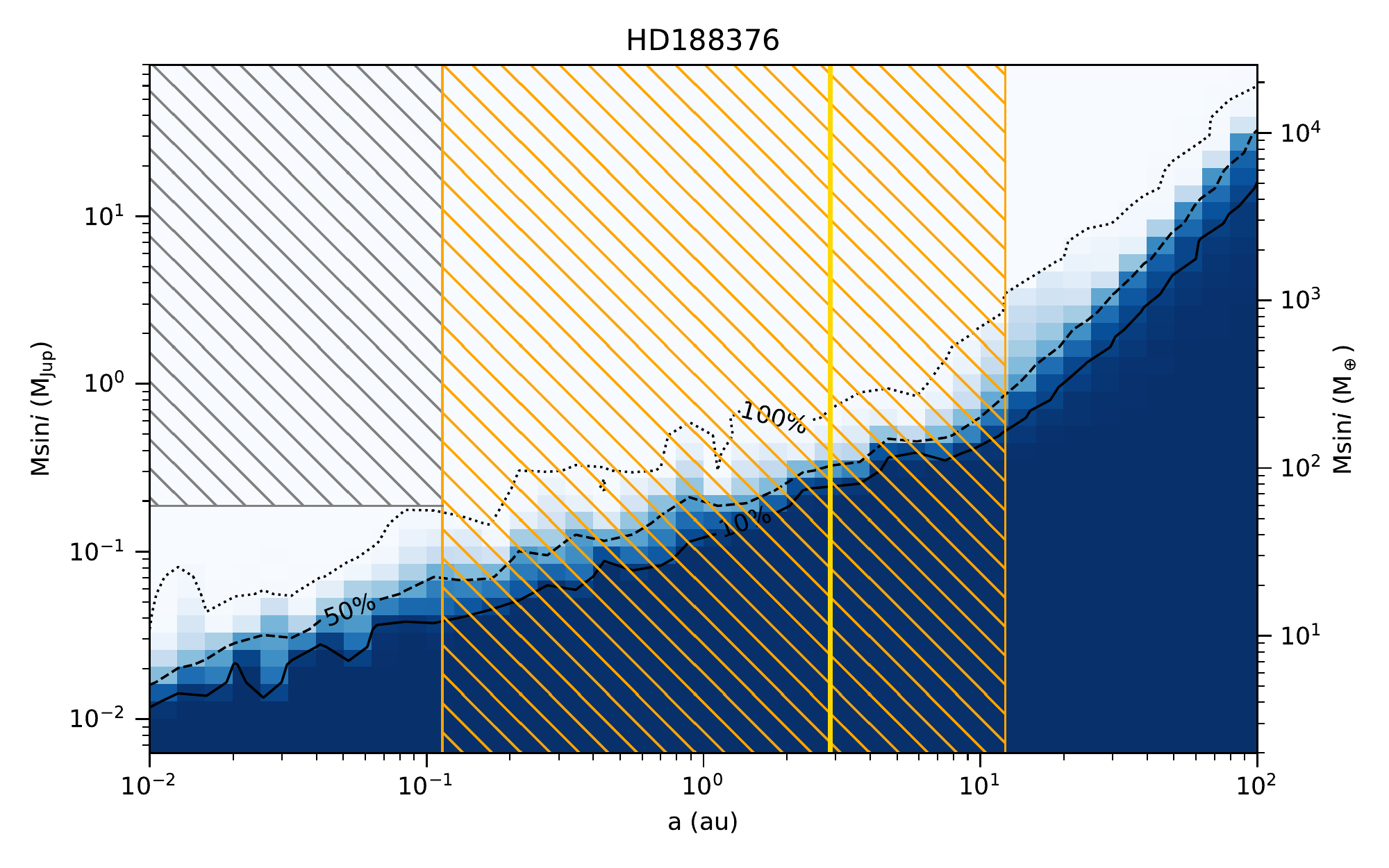}&
    		\includegraphics[width=0.22\linewidth]{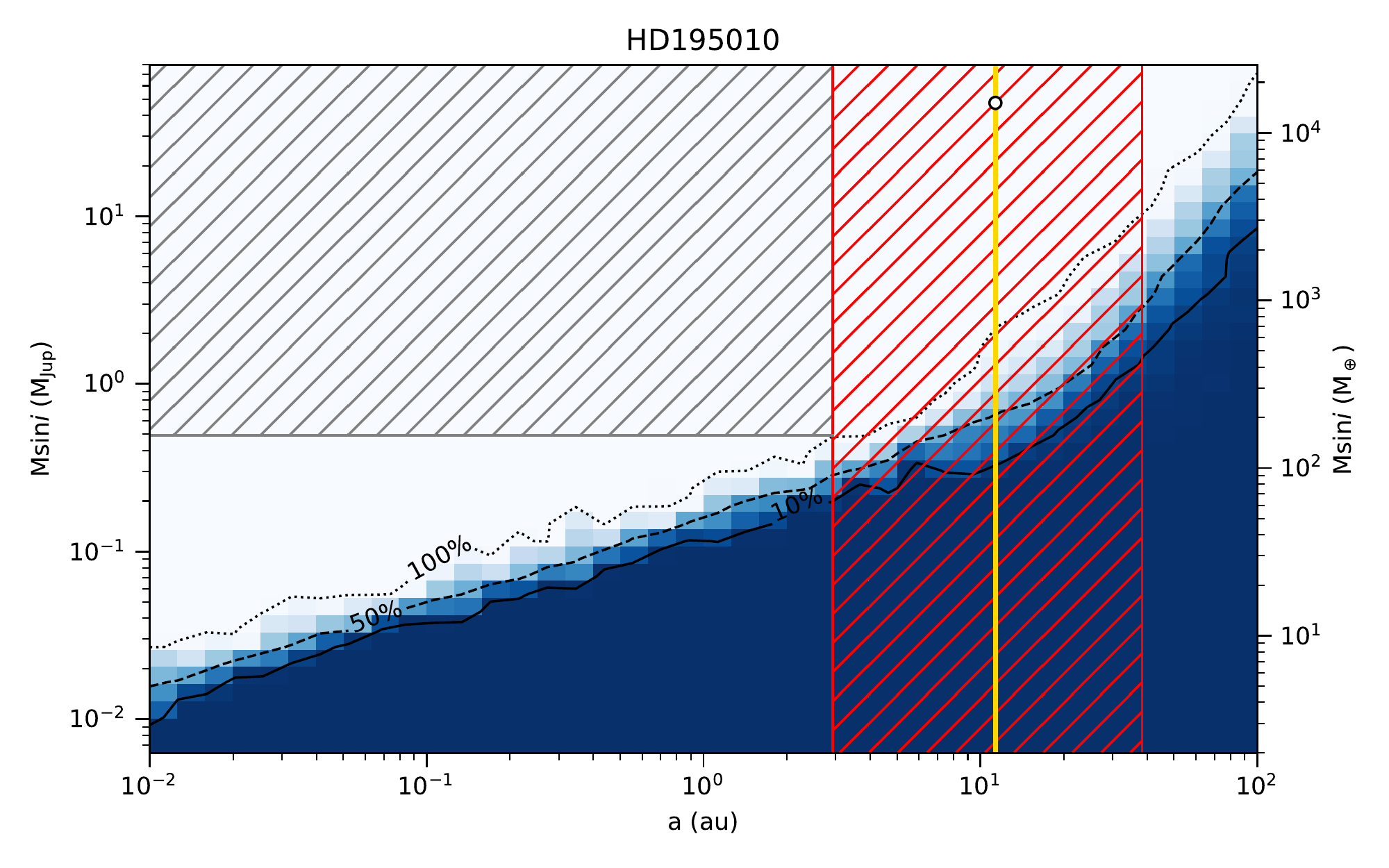}&
    		\includegraphics[width=0.22\linewidth]{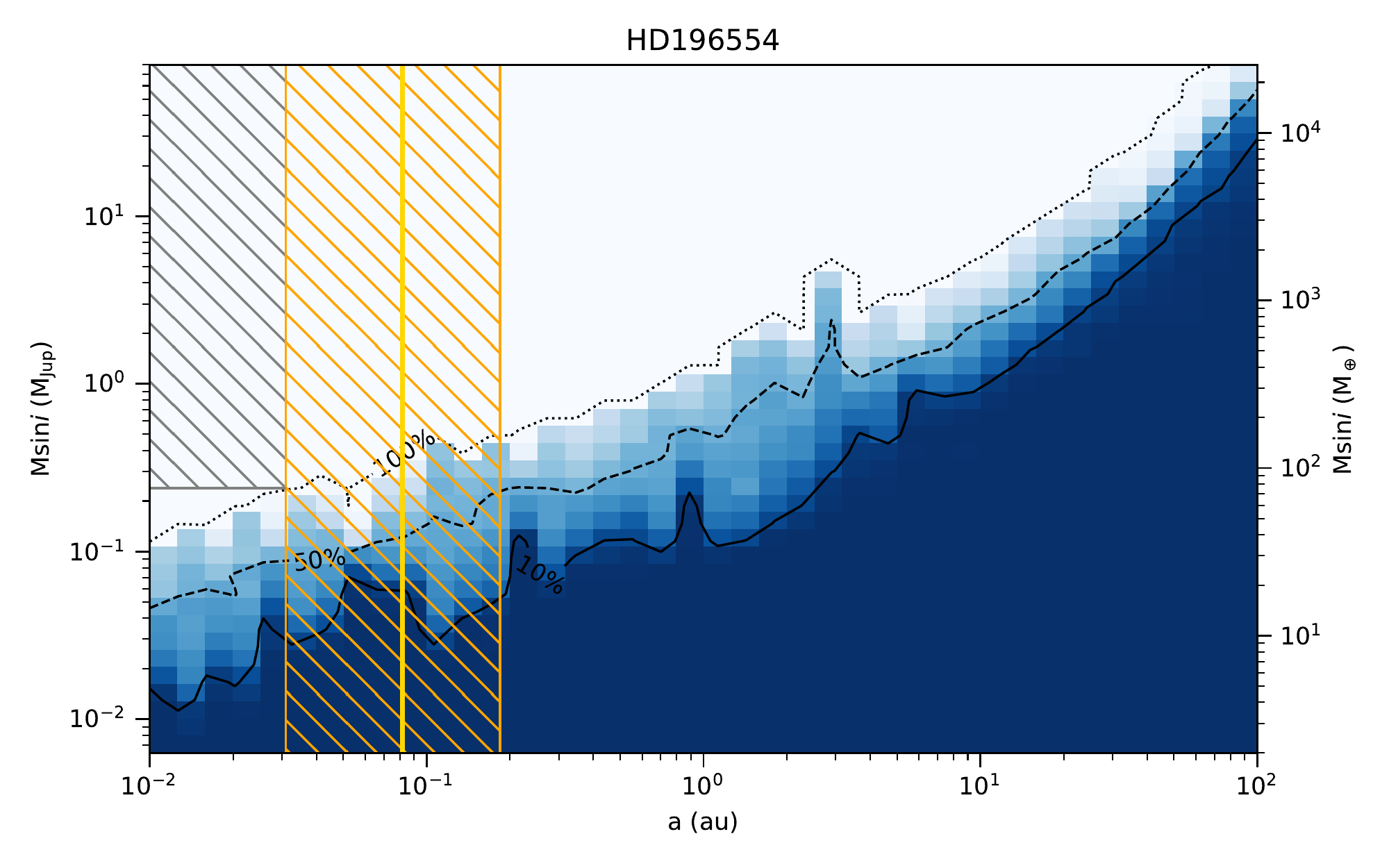}\\
    
    		\includegraphics[width=0.22\linewidth]{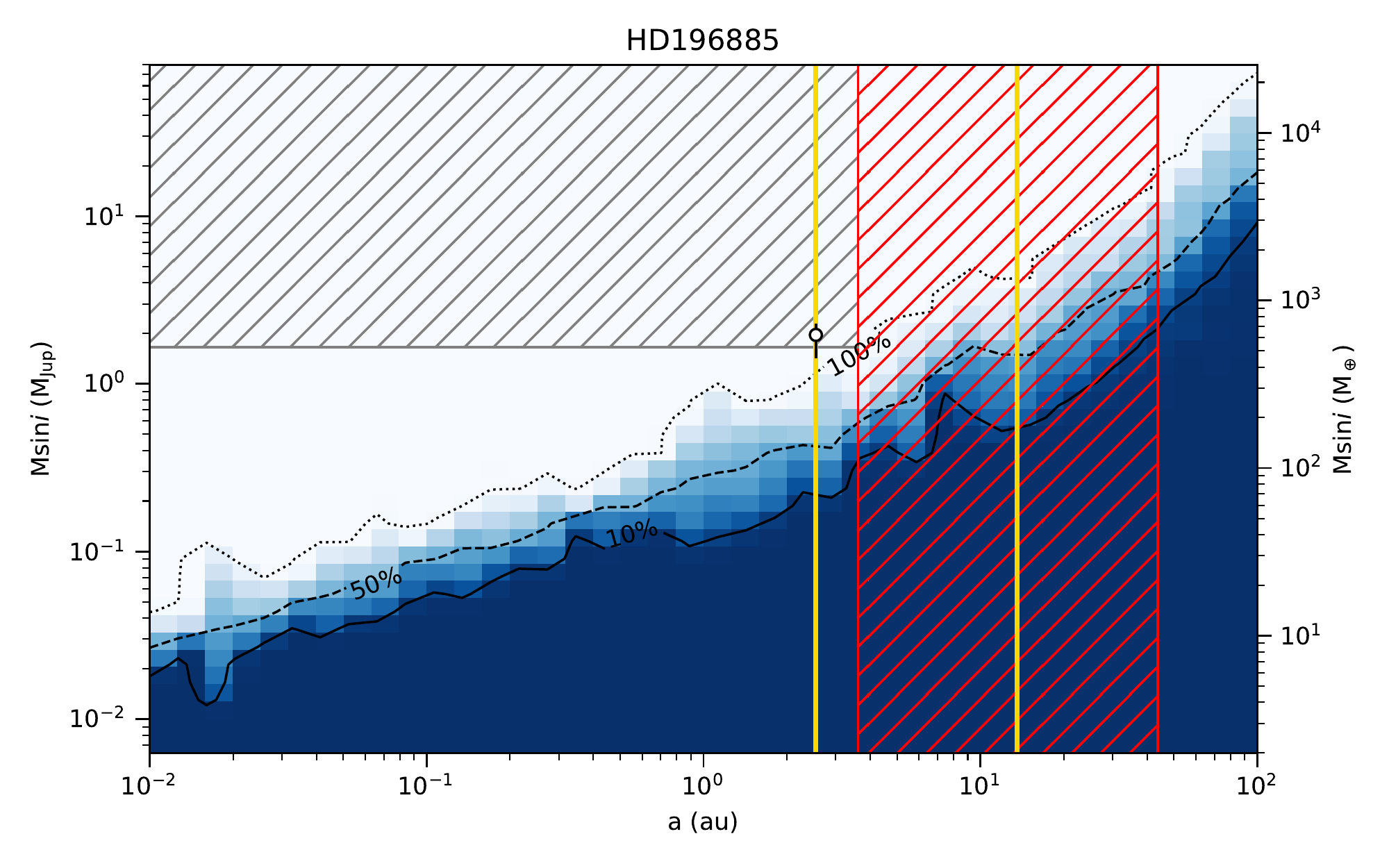}&
    		\includegraphics[width=0.22\linewidth]{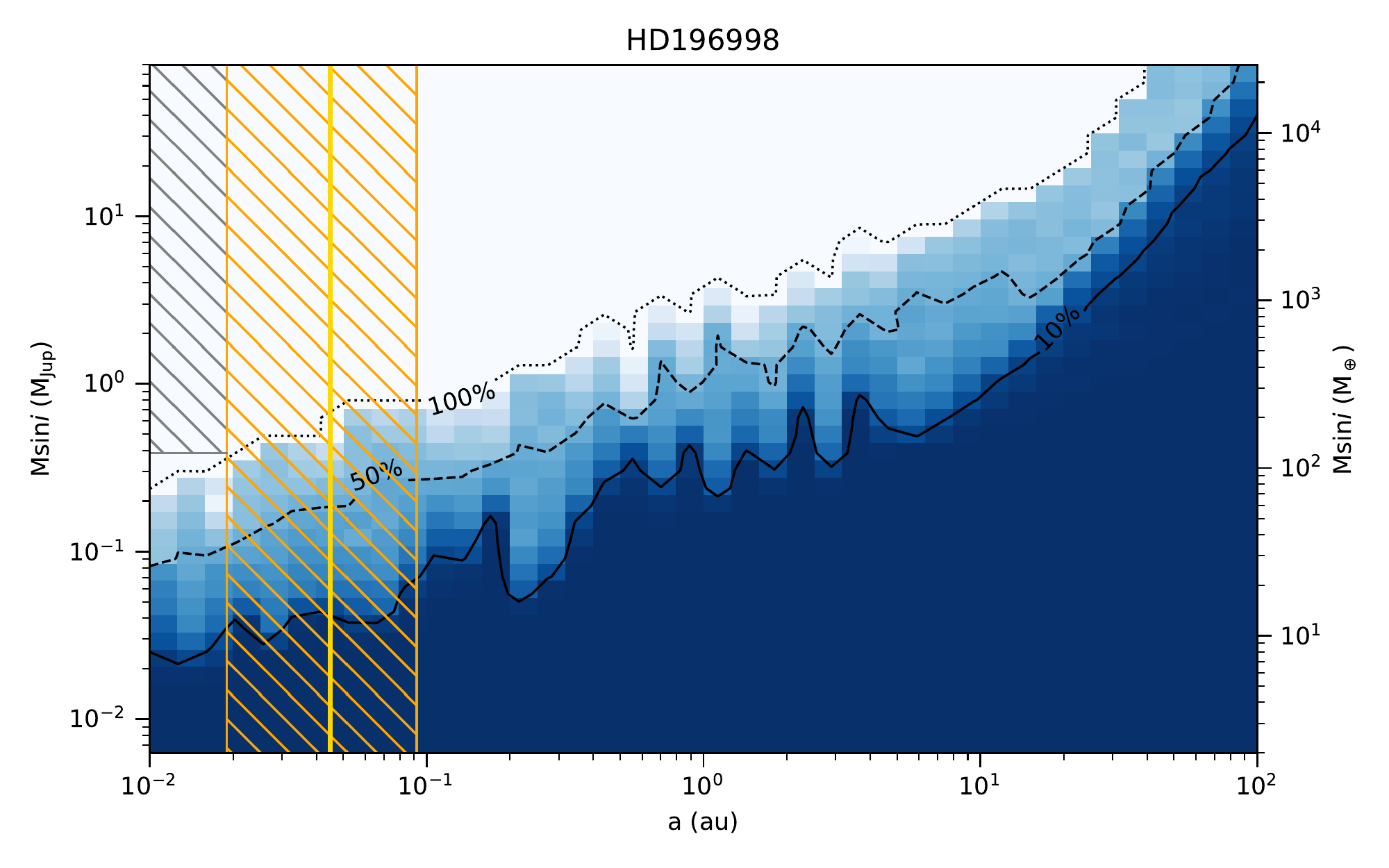}&
    		\includegraphics[width=0.22\linewidth]{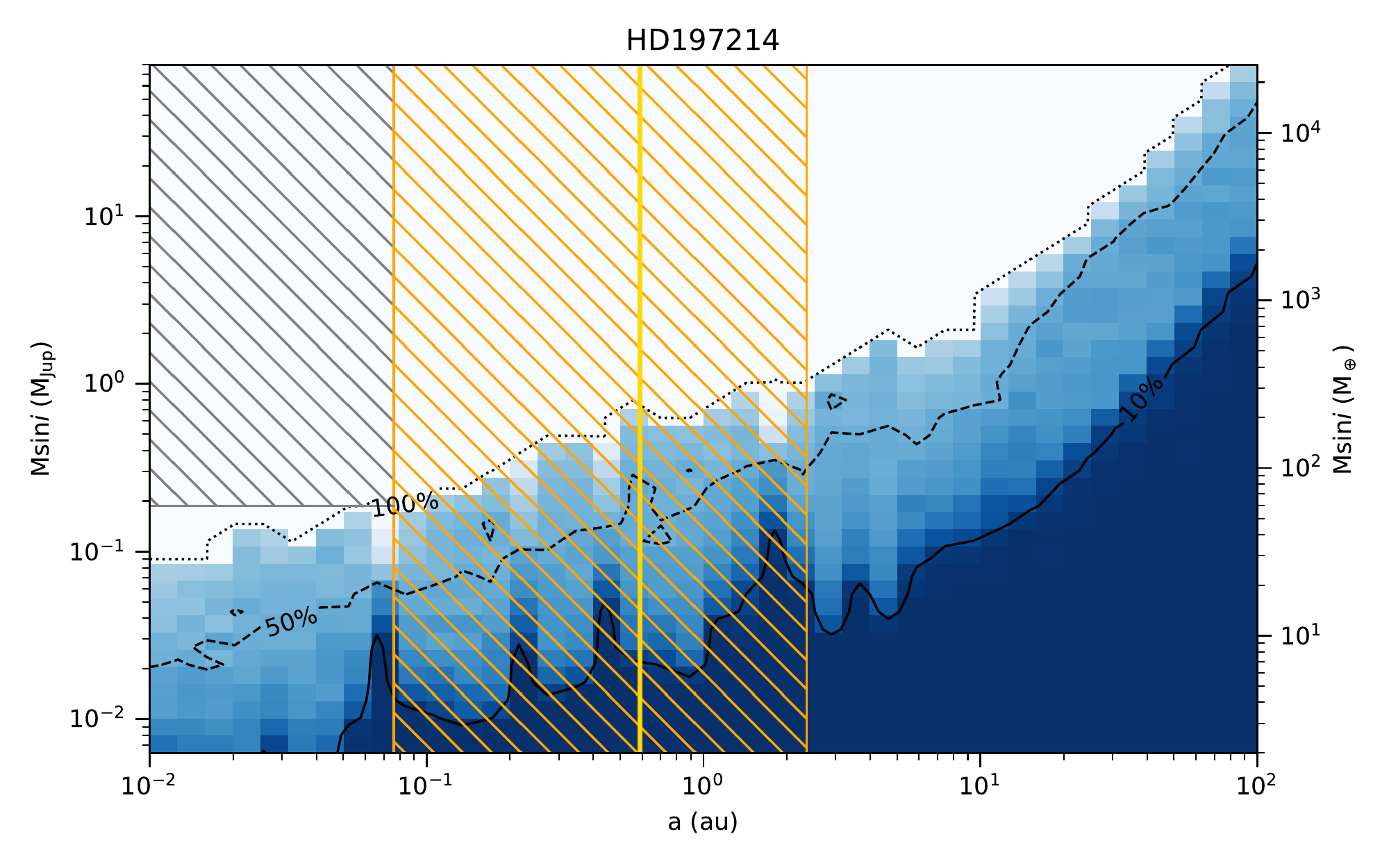}&
    		\includegraphics[width=0.22\linewidth]{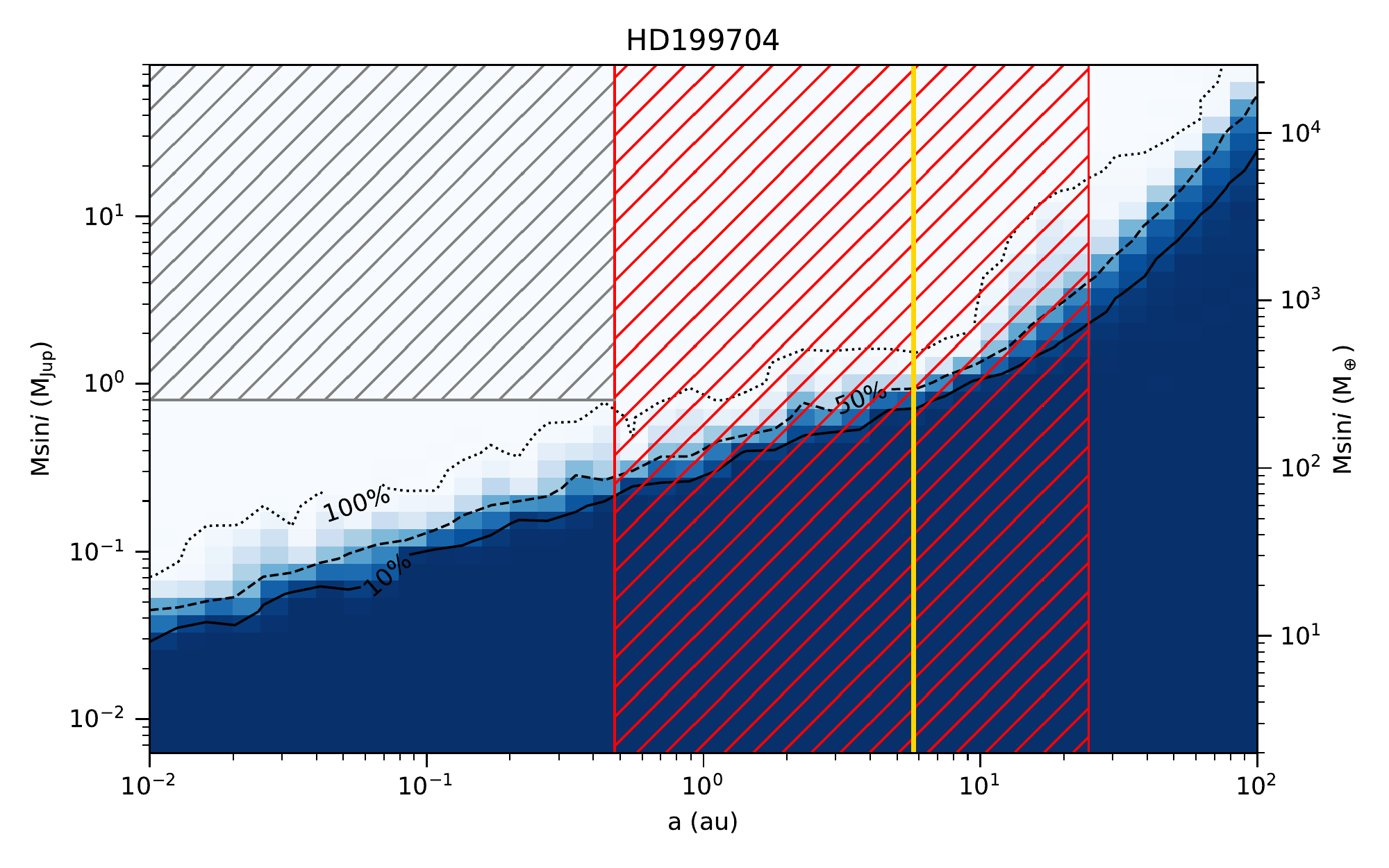}\\
    
    		\includegraphics[width=0.22\linewidth]{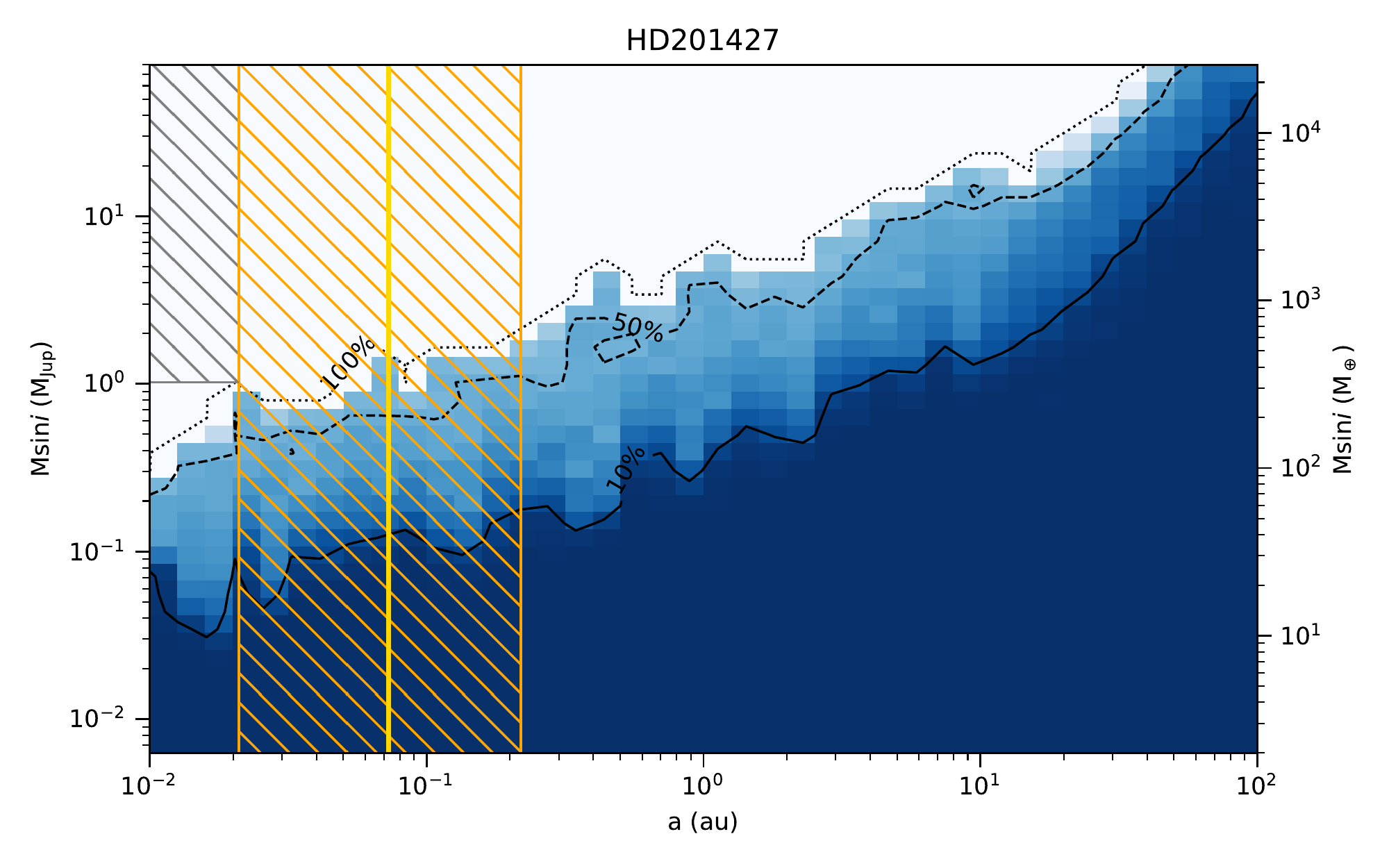}&
    		\includegraphics[width=0.22\linewidth]{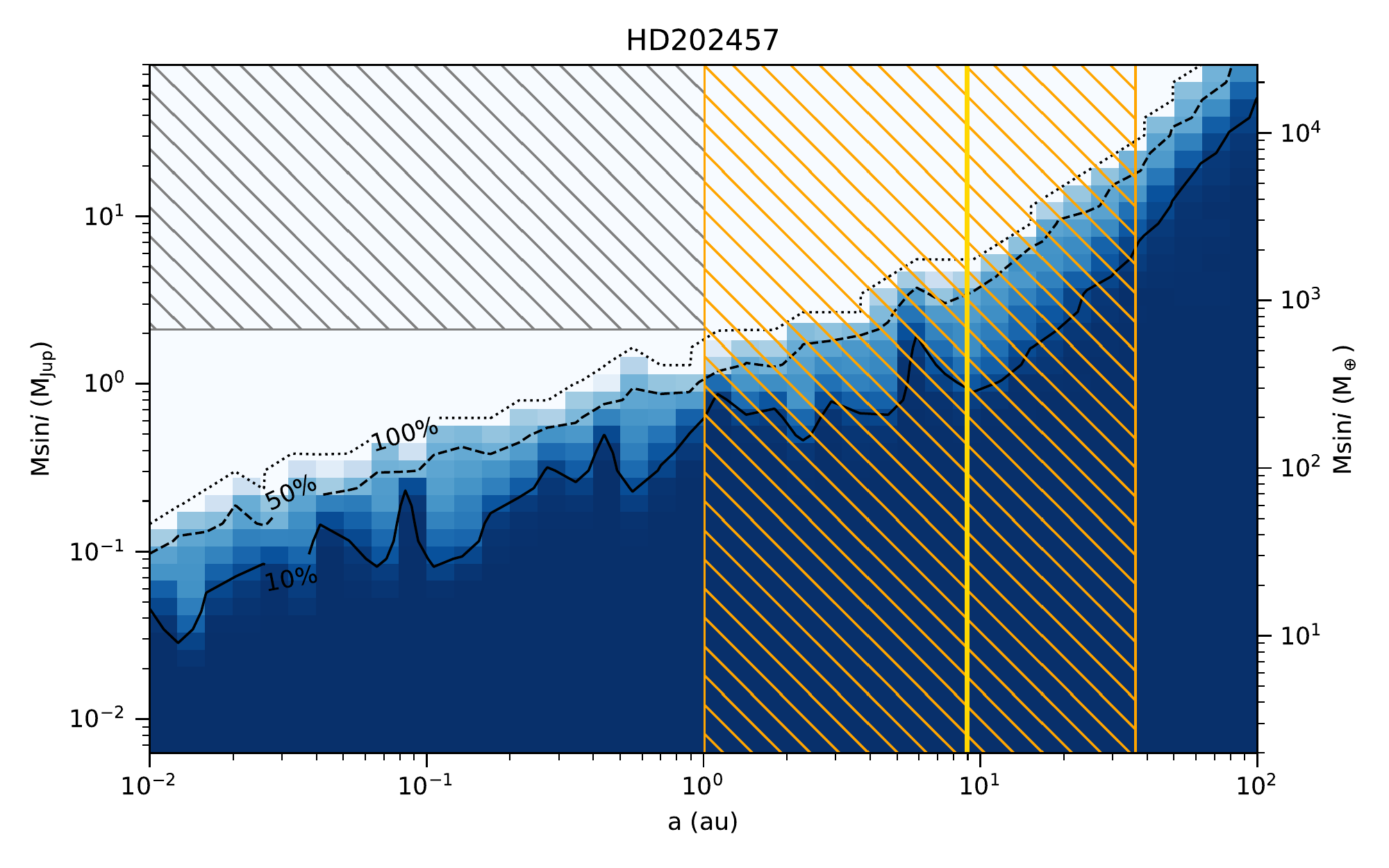}&
    		\includegraphics[width=0.22\linewidth]{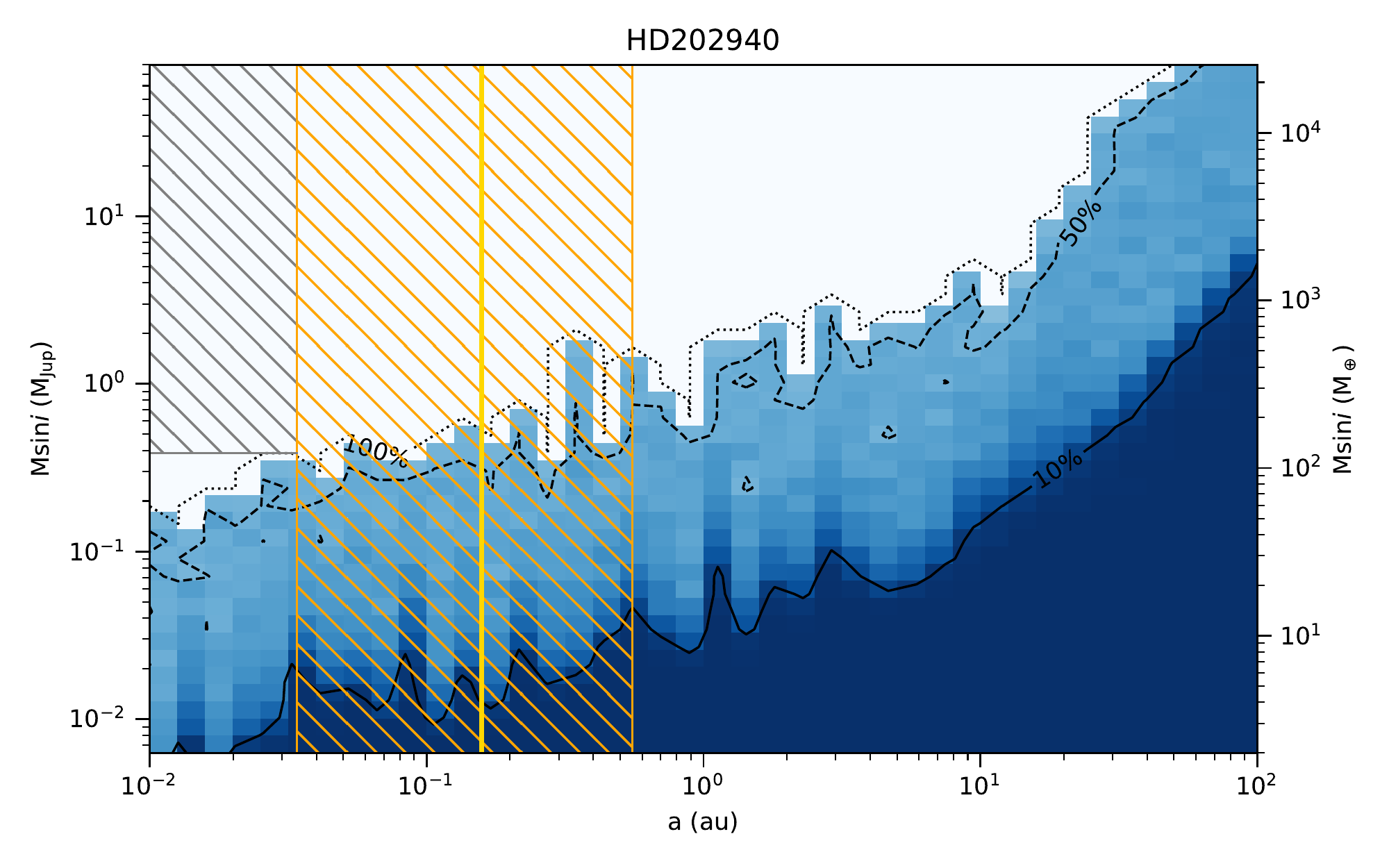}&
    		\includegraphics[width=0.22\linewidth]{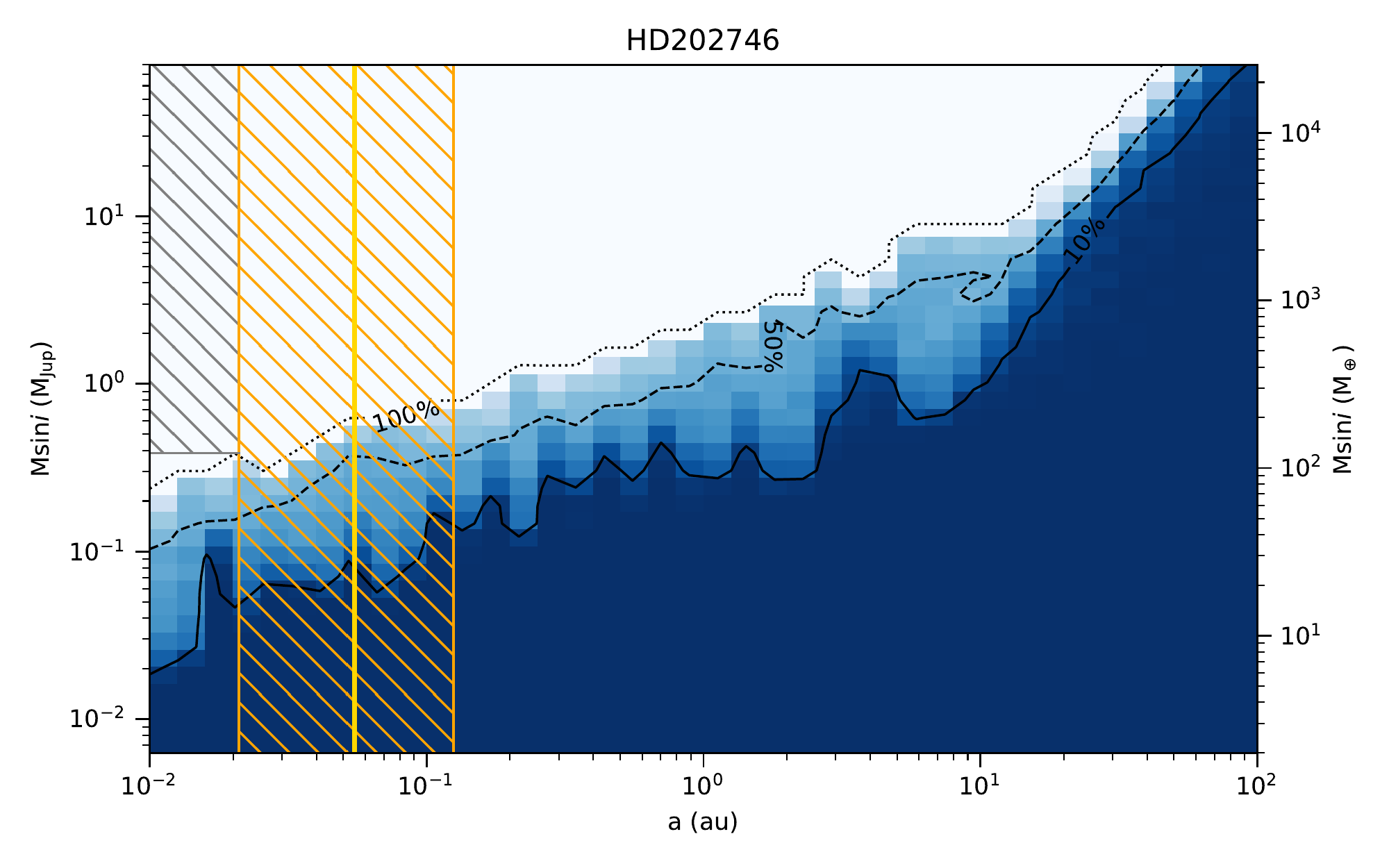}\\
    
    		\includegraphics[width=0.22\linewidth]{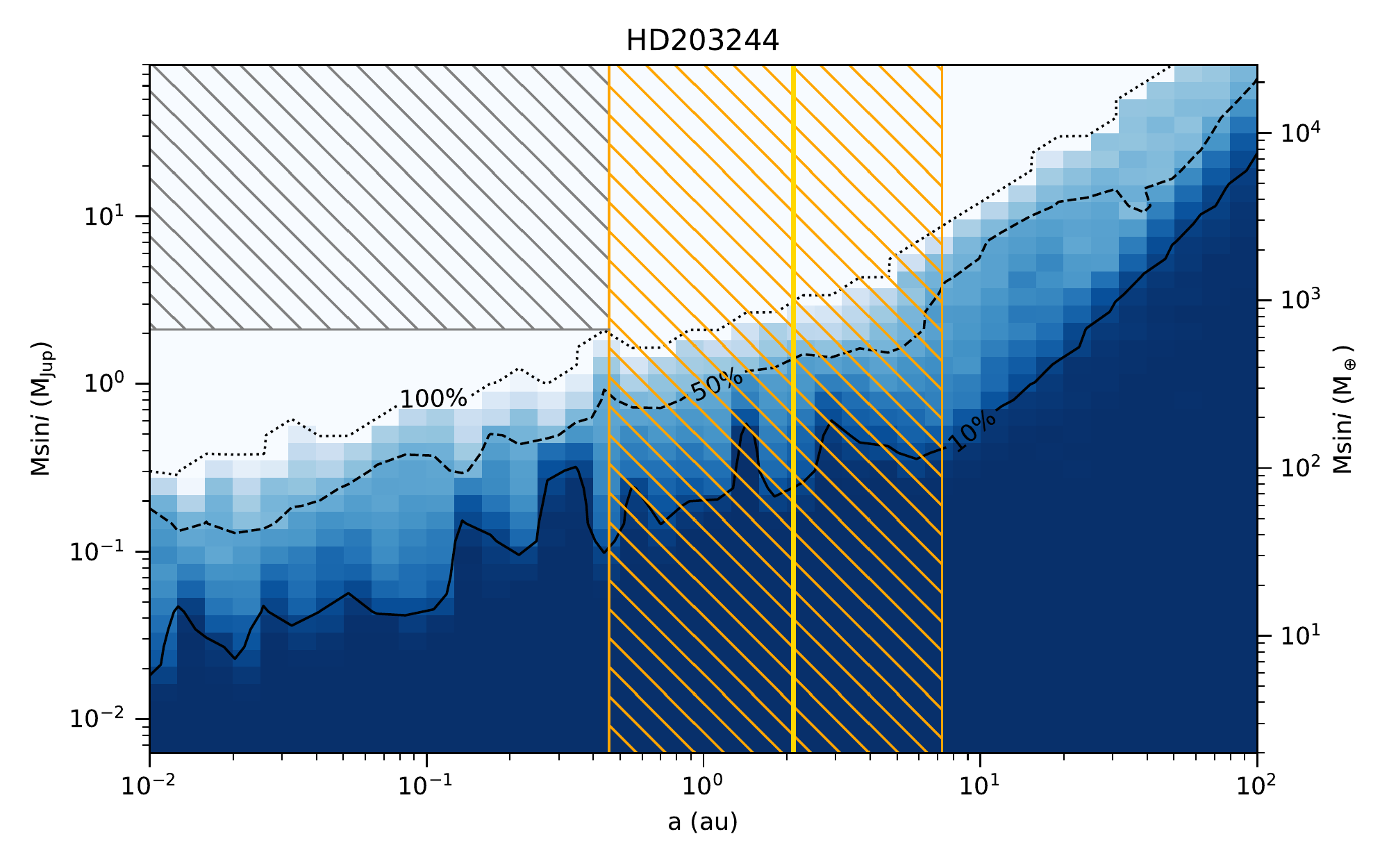}&
    		\includegraphics[width=0.22\linewidth]{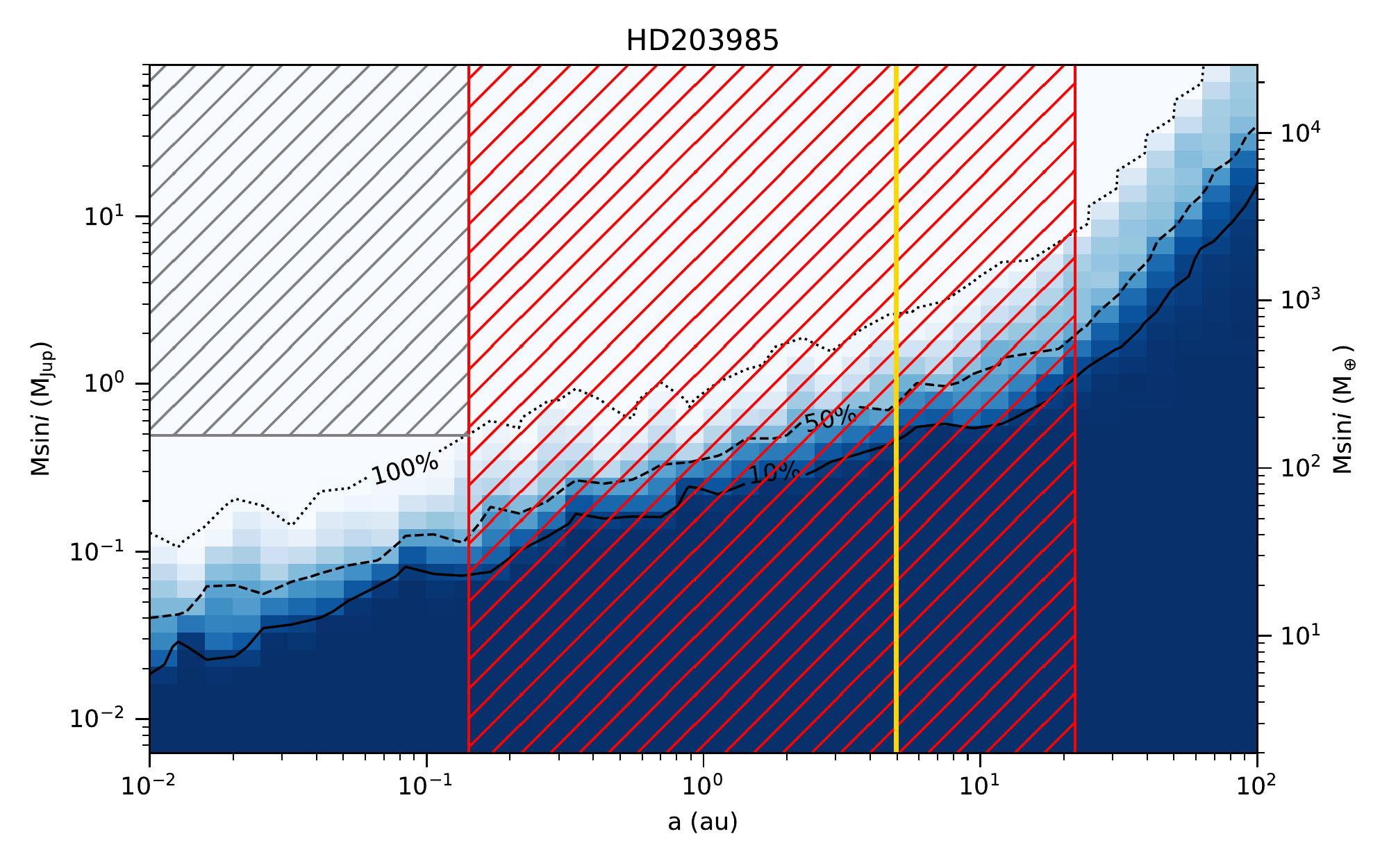}&
    		\includegraphics[width=0.22\linewidth]{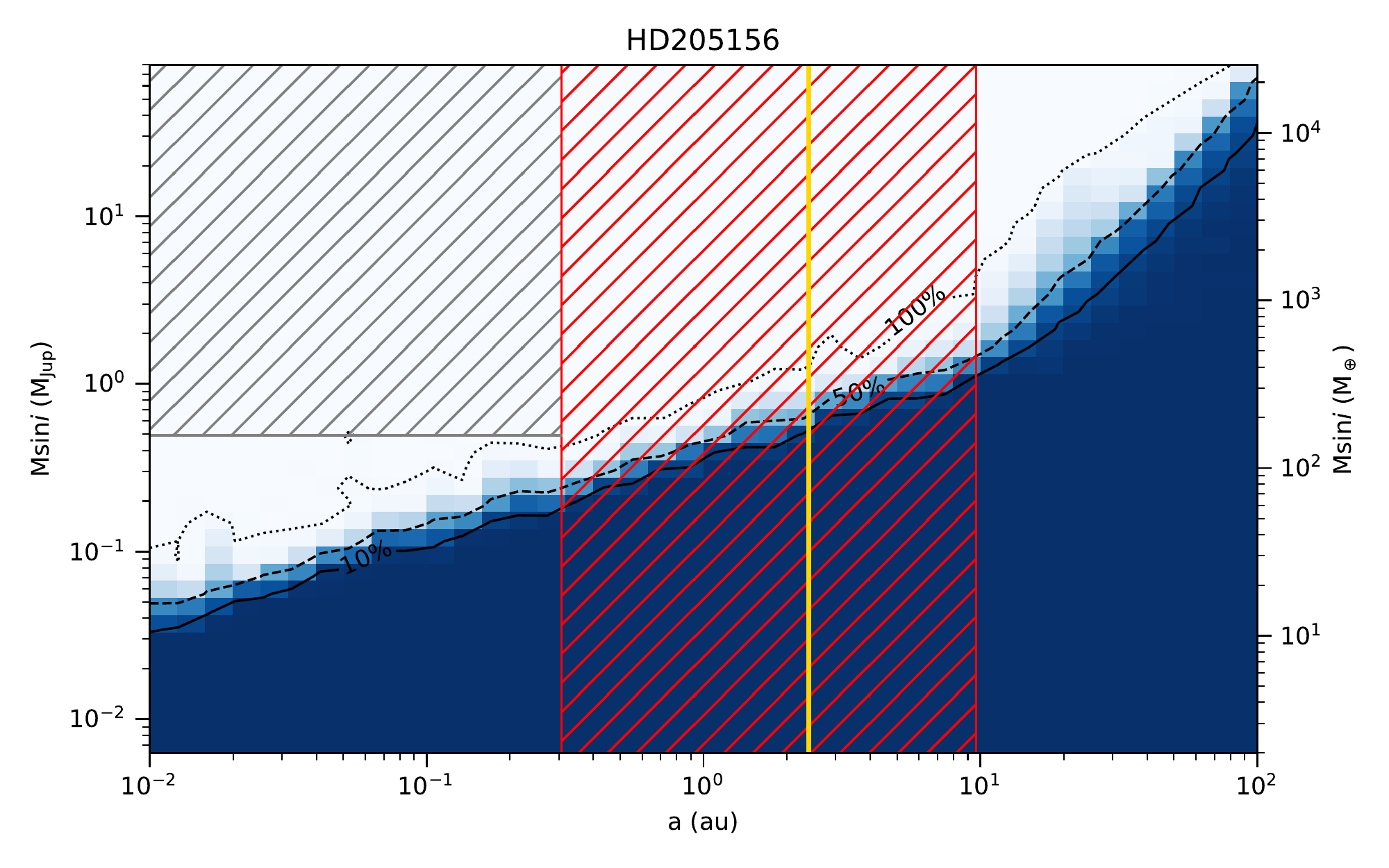}&
    		\includegraphics[width=0.22\linewidth]{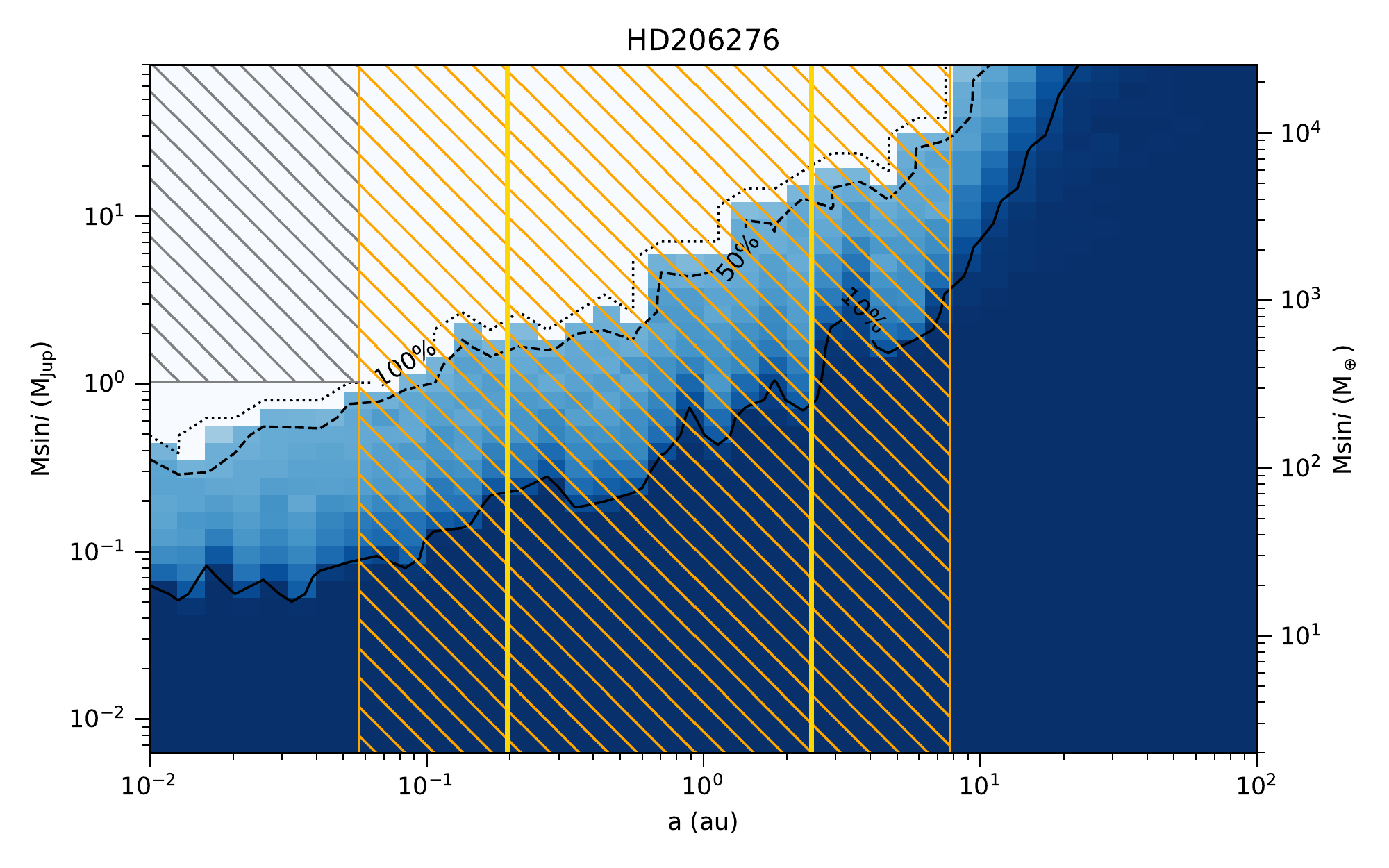}\\
    
    		\includegraphics[width=0.22\linewidth]{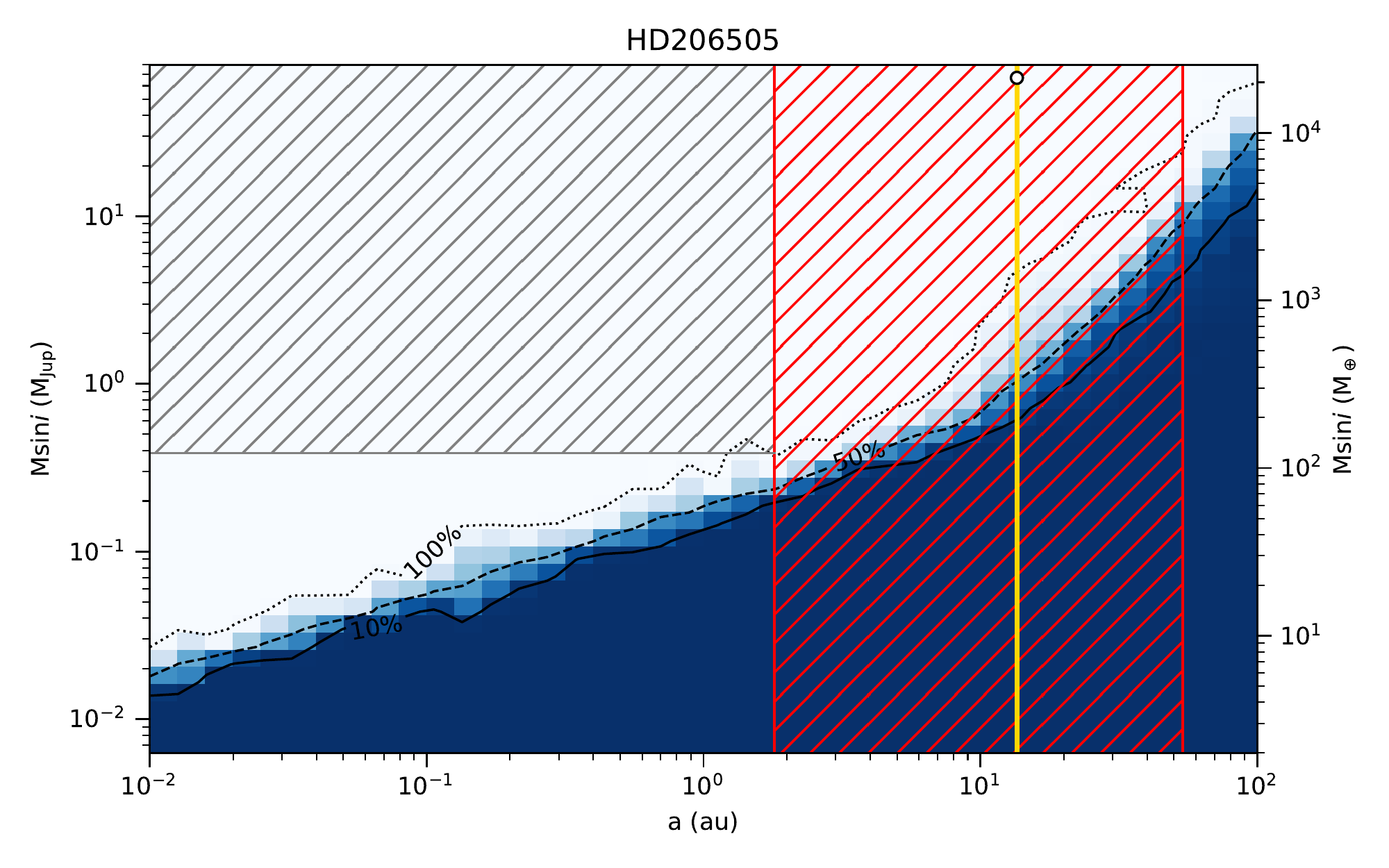}&
    		\includegraphics[width=0.22\linewidth]{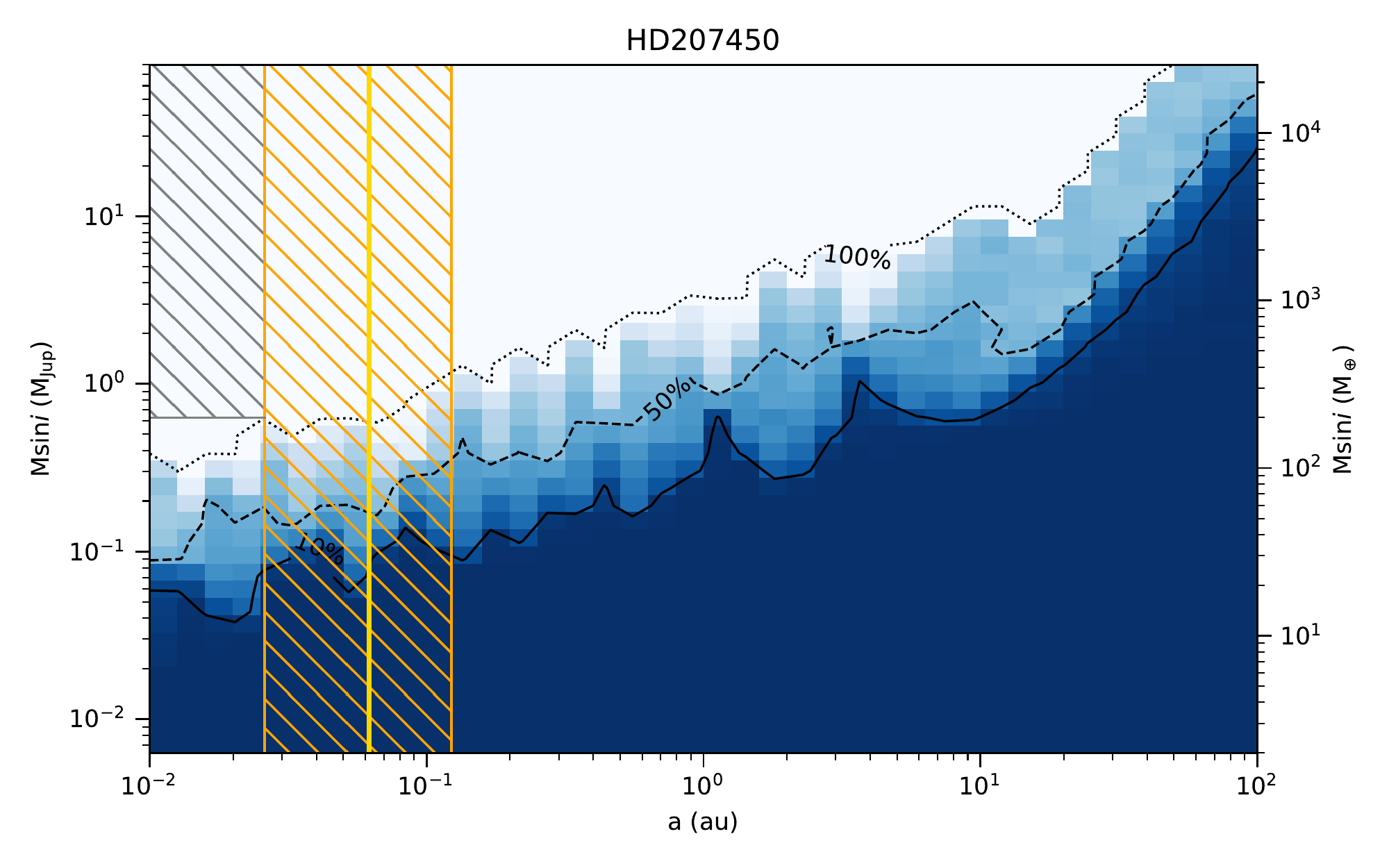}&
    		\includegraphics[width=0.22\linewidth]{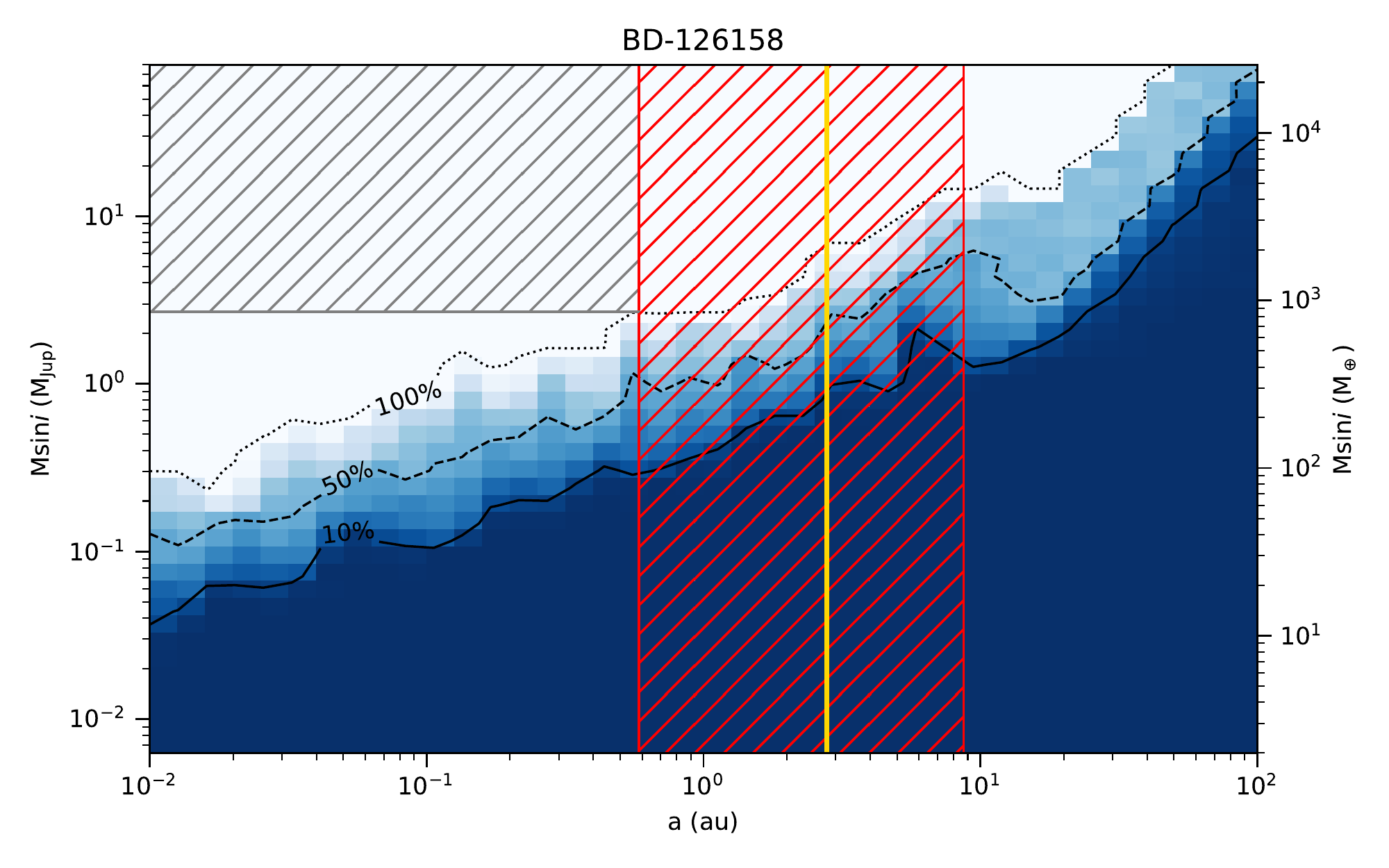}&
    		\includegraphics[width=0.22\linewidth]{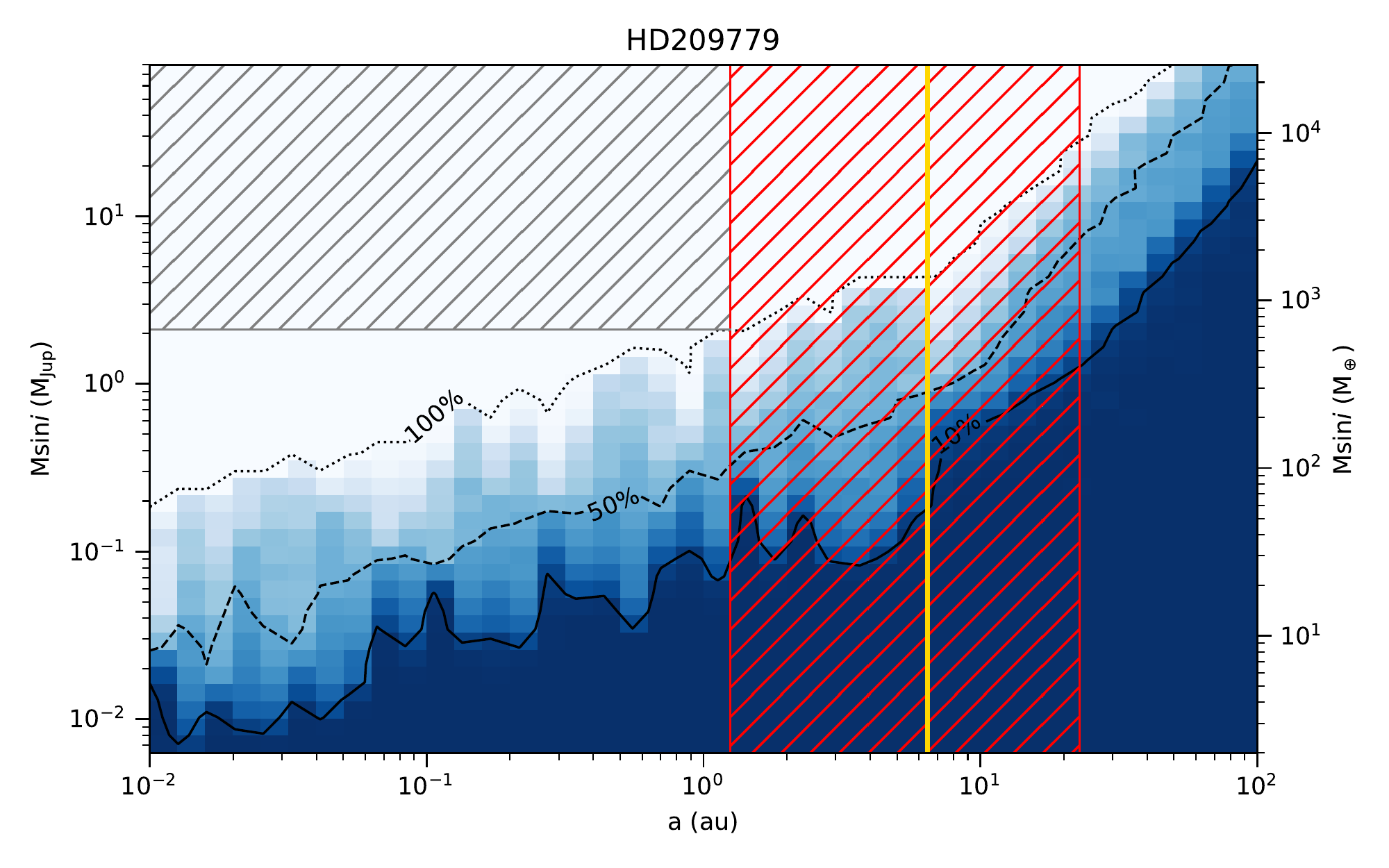}\\
    
    		\includegraphics[width=0.22\linewidth]{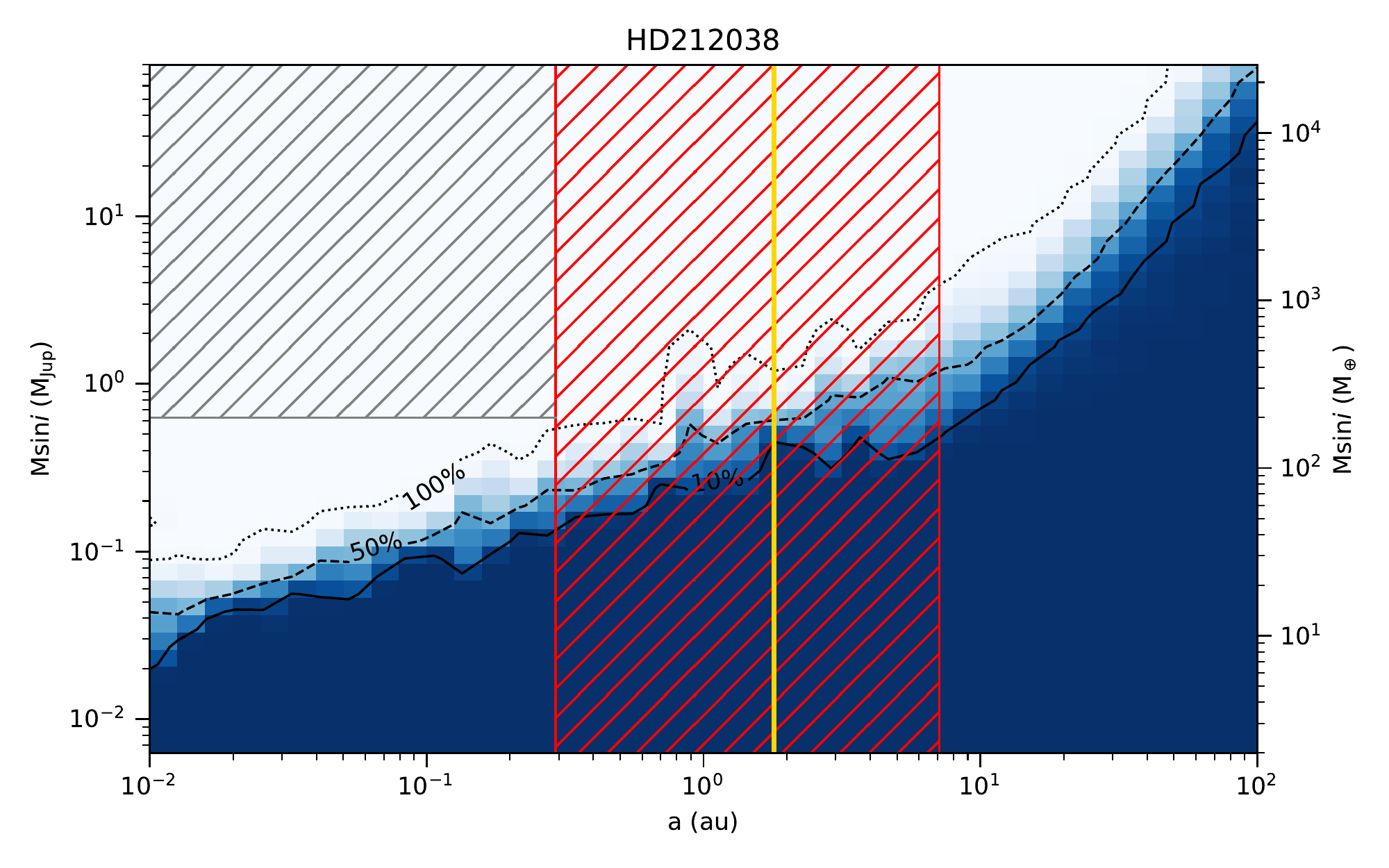}&
    		\includegraphics[width=0.22\linewidth]{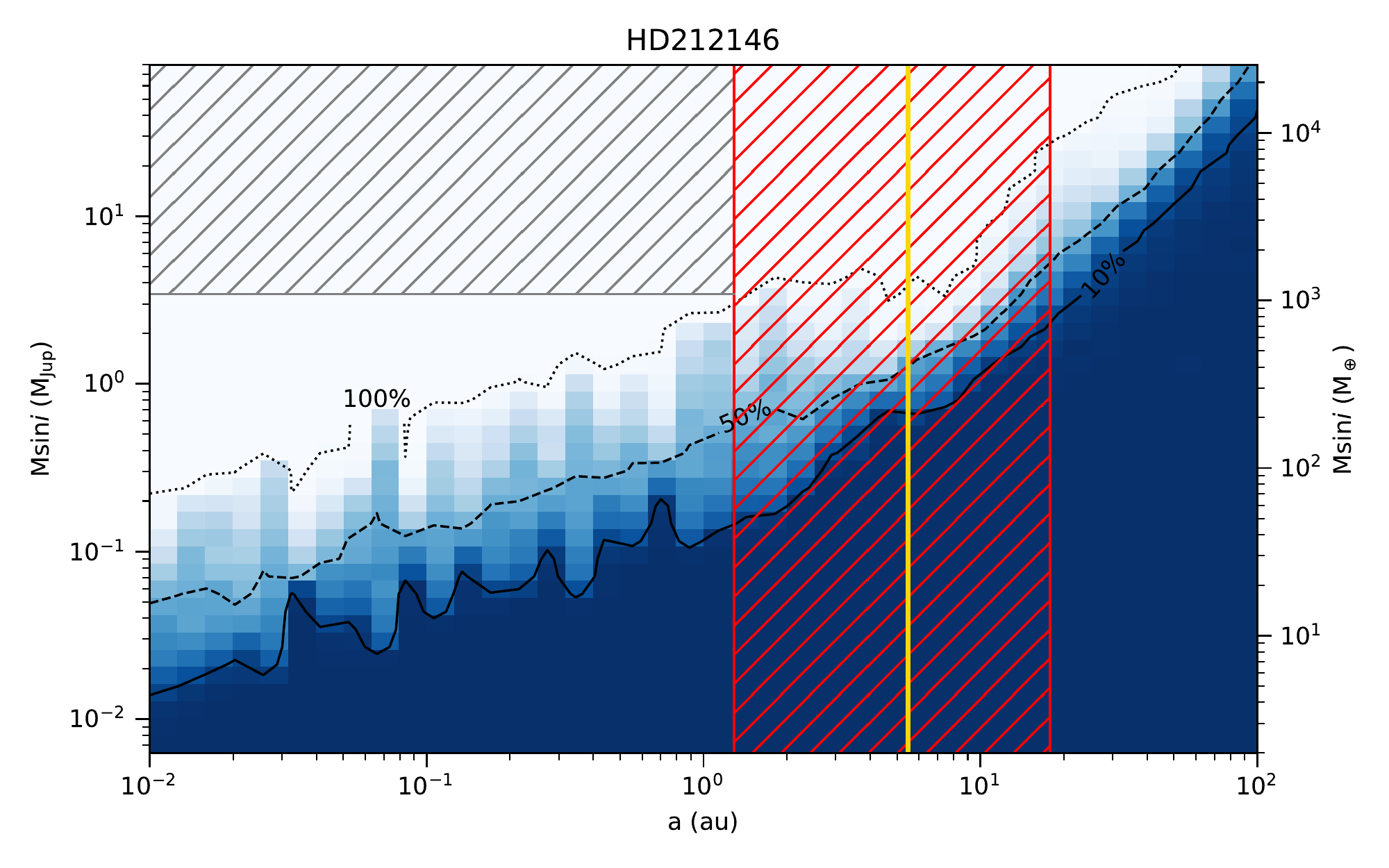}&
    		\includegraphics[width=0.22\linewidth]{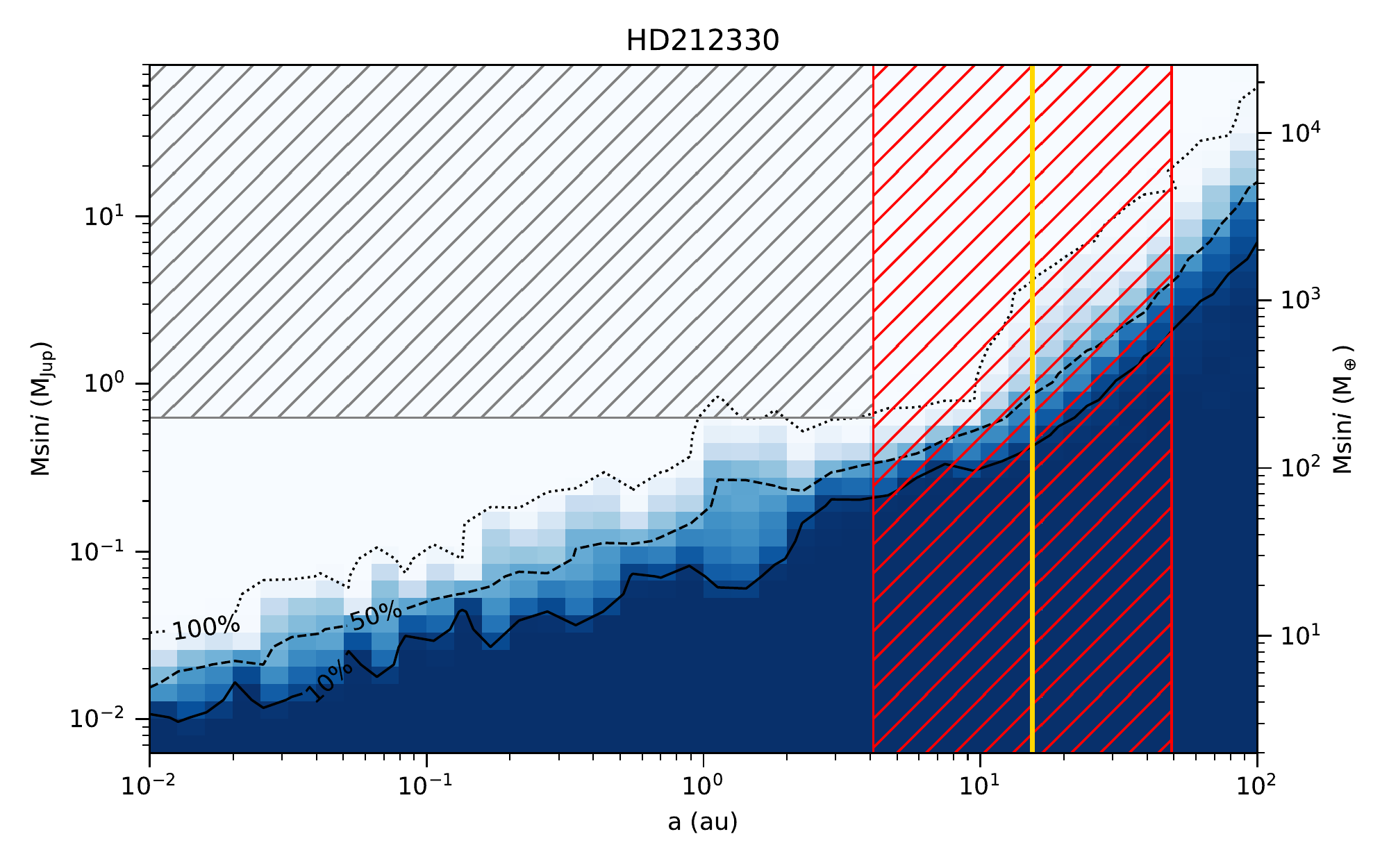}&
    		\includegraphics[width=0.22\linewidth]{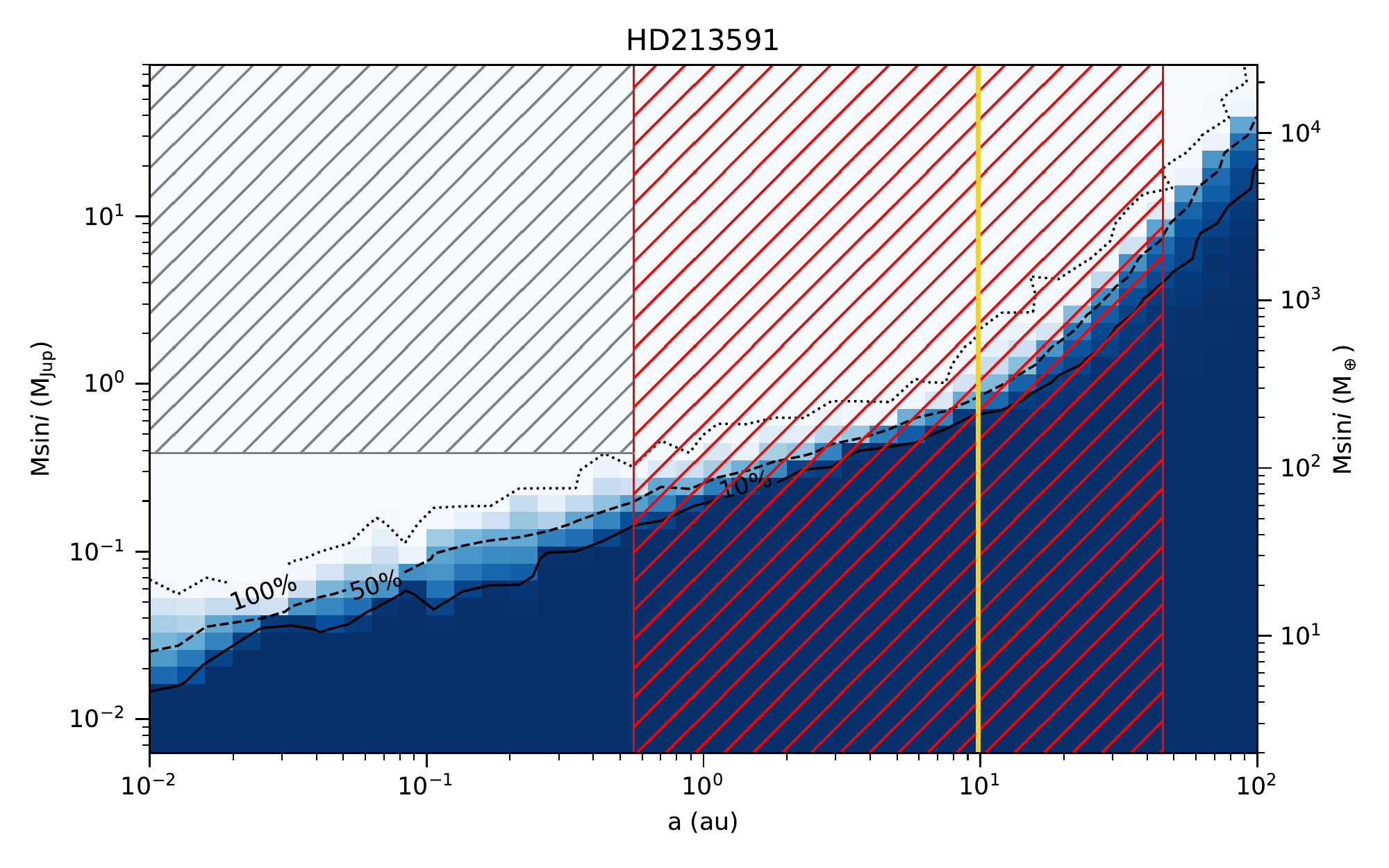}\\
    
    		\includegraphics[width=0.22\linewidth]{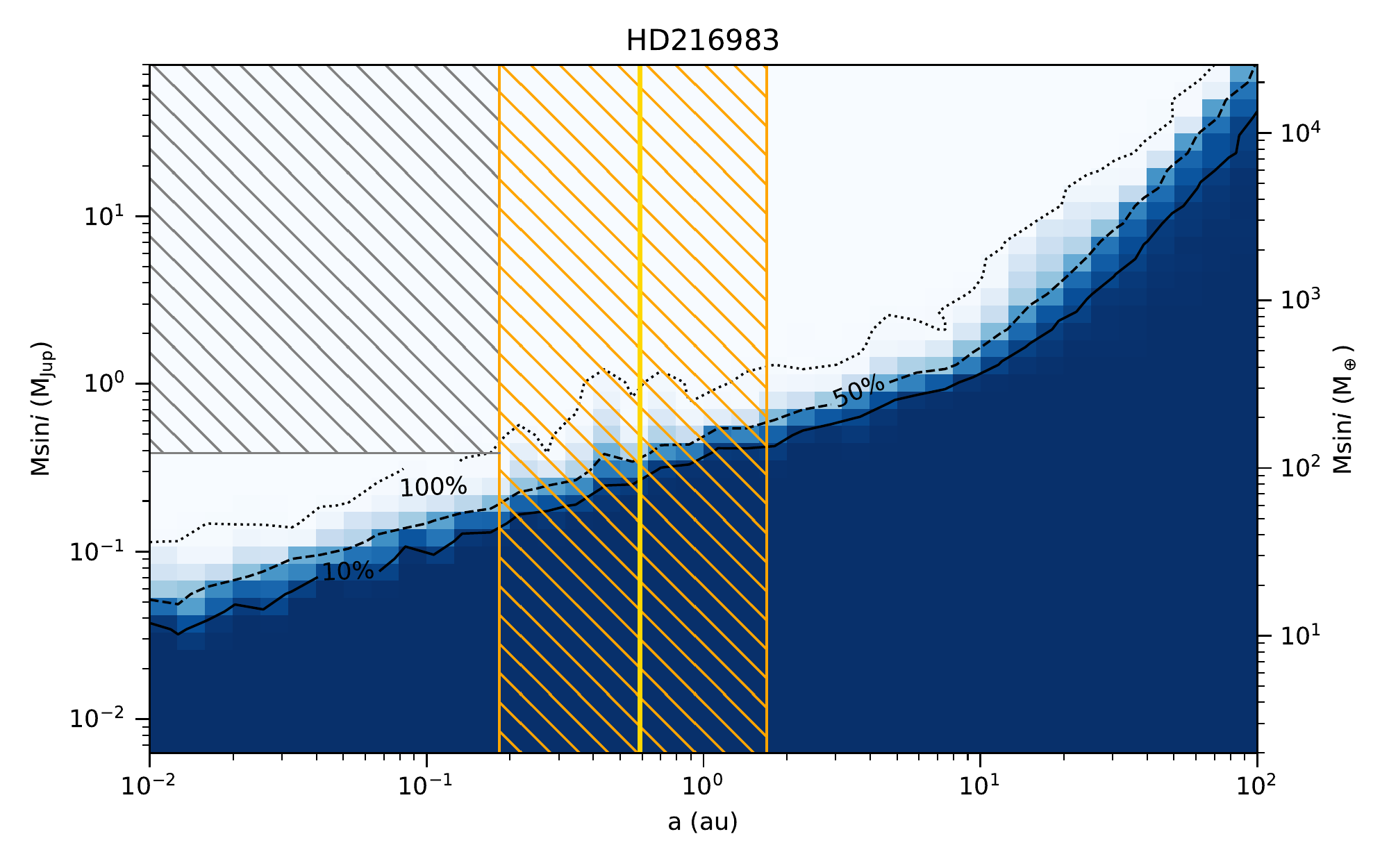}&
    		\includegraphics[width=0.22\linewidth]{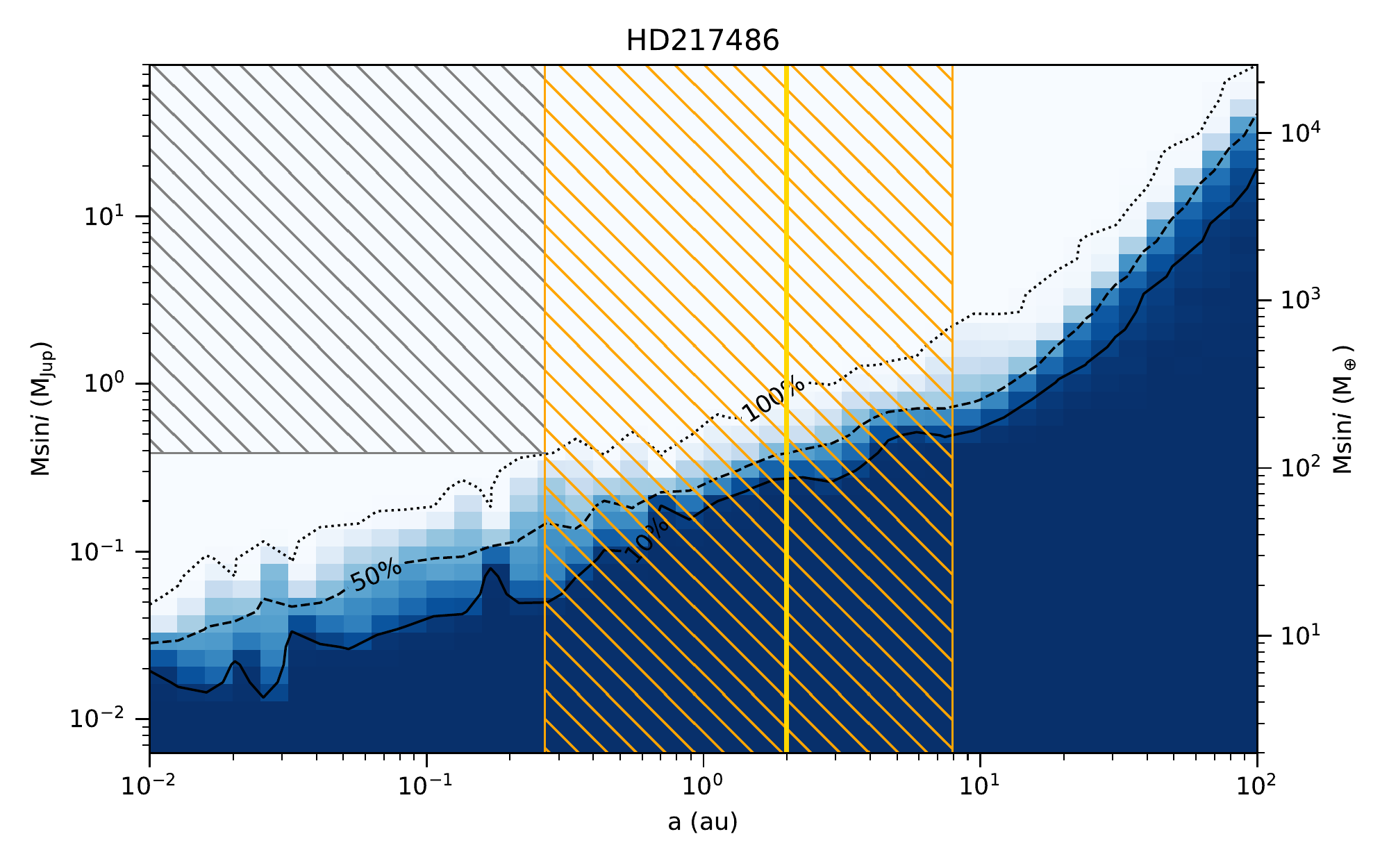}&
    		\includegraphics[width=0.22\linewidth]{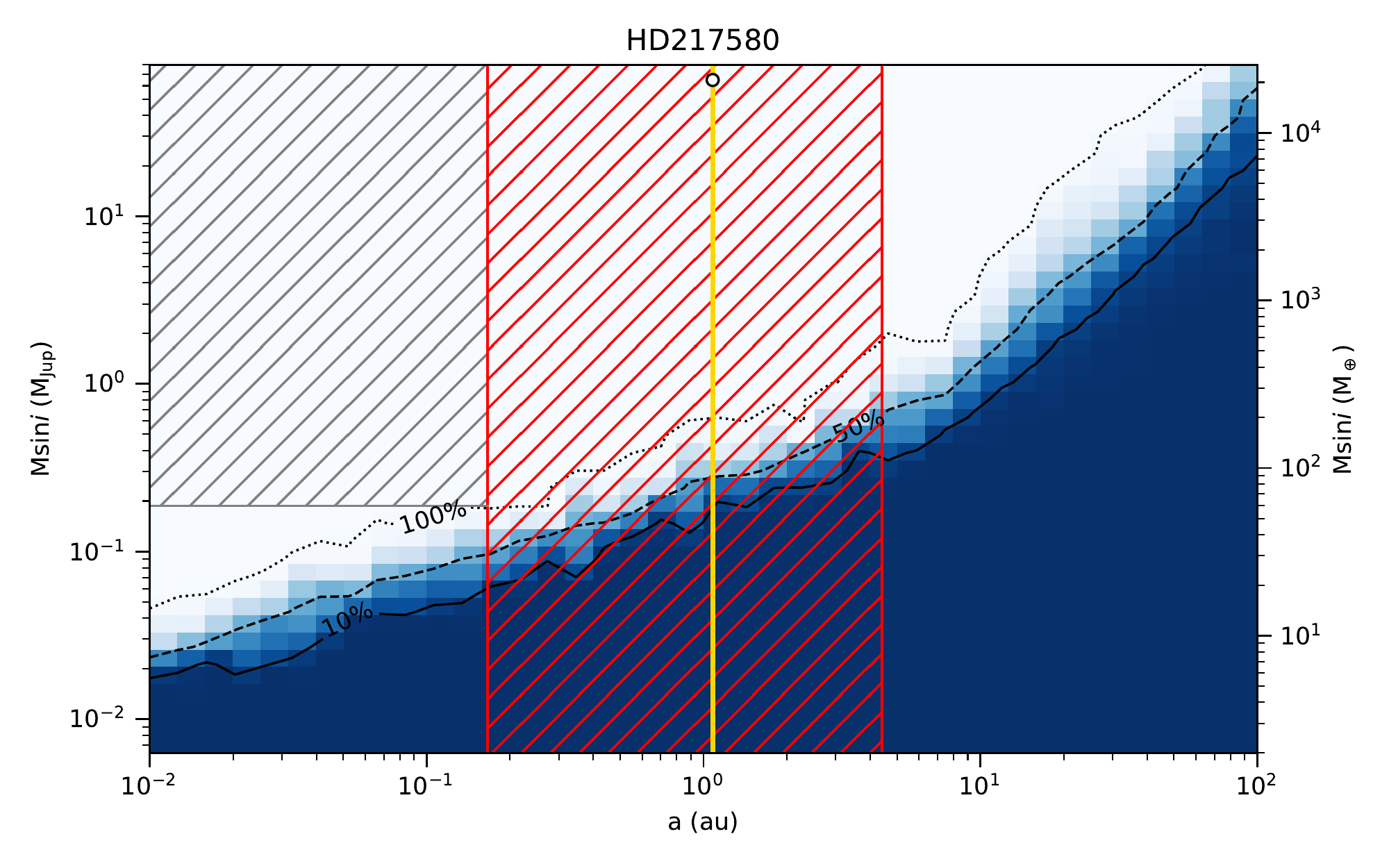}&
    		\includegraphics[width=0.22\linewidth]{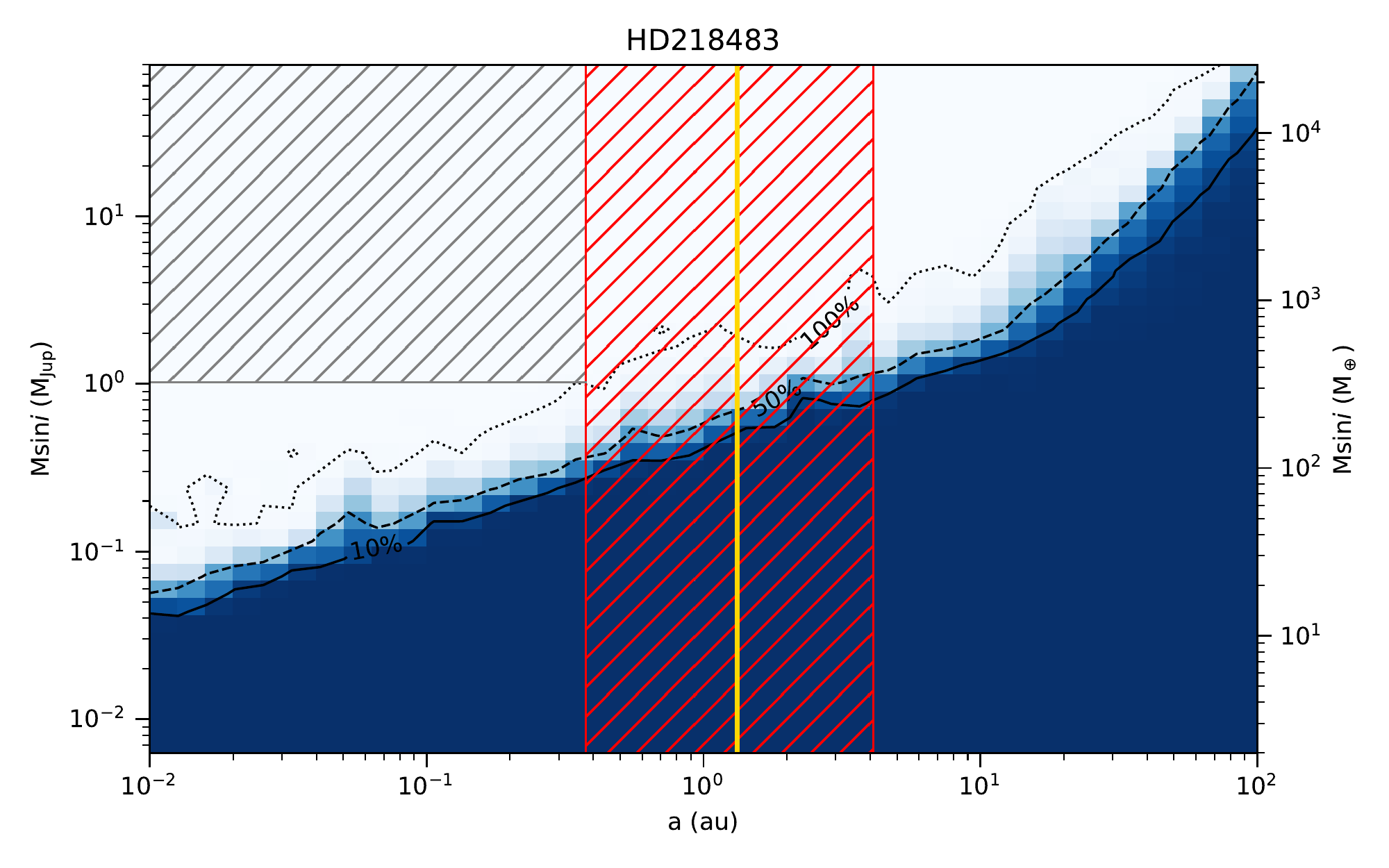}\\
    
    		\includegraphics[width=0.22\linewidth]{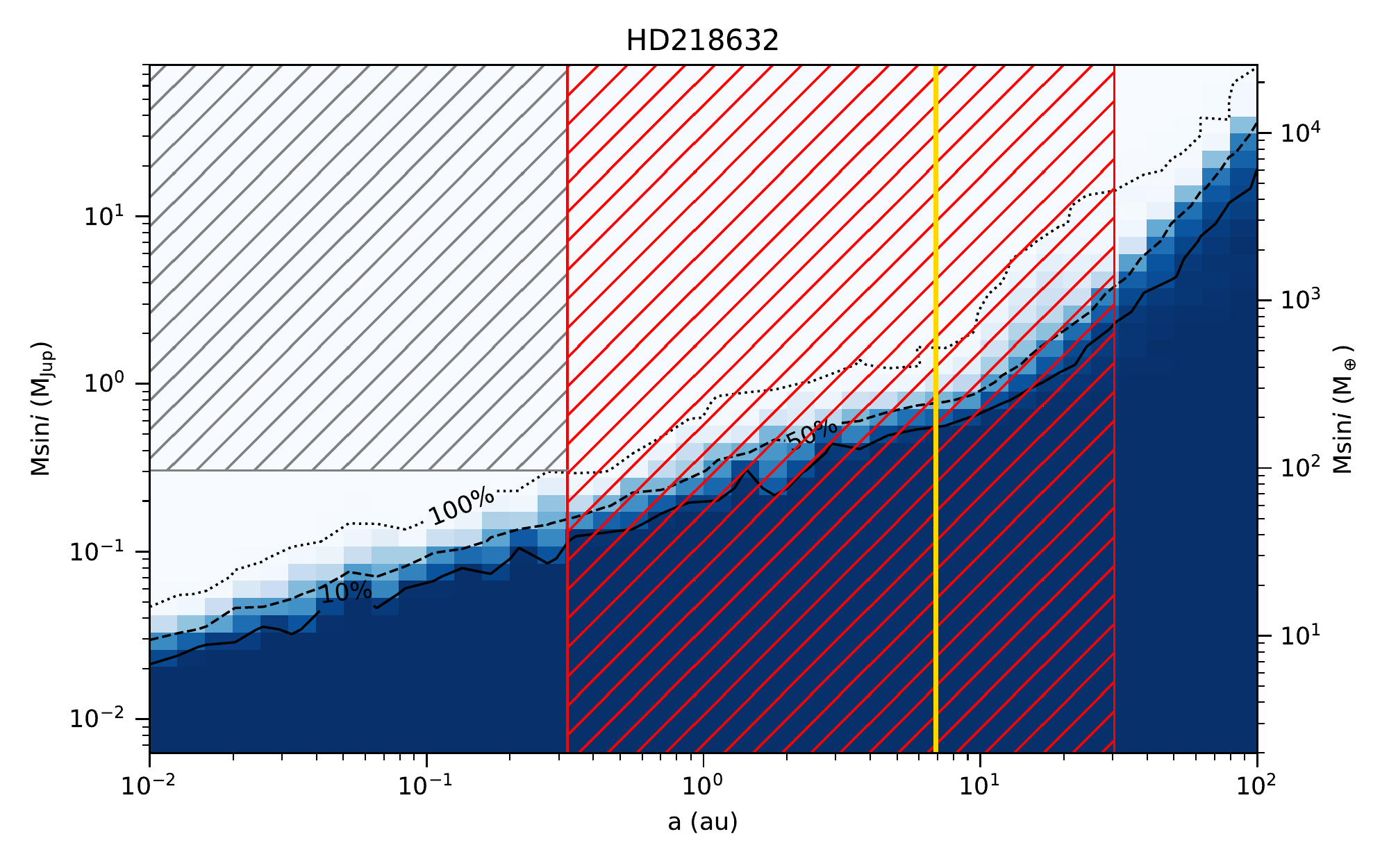}&
    		\includegraphics[width=0.22\linewidth]{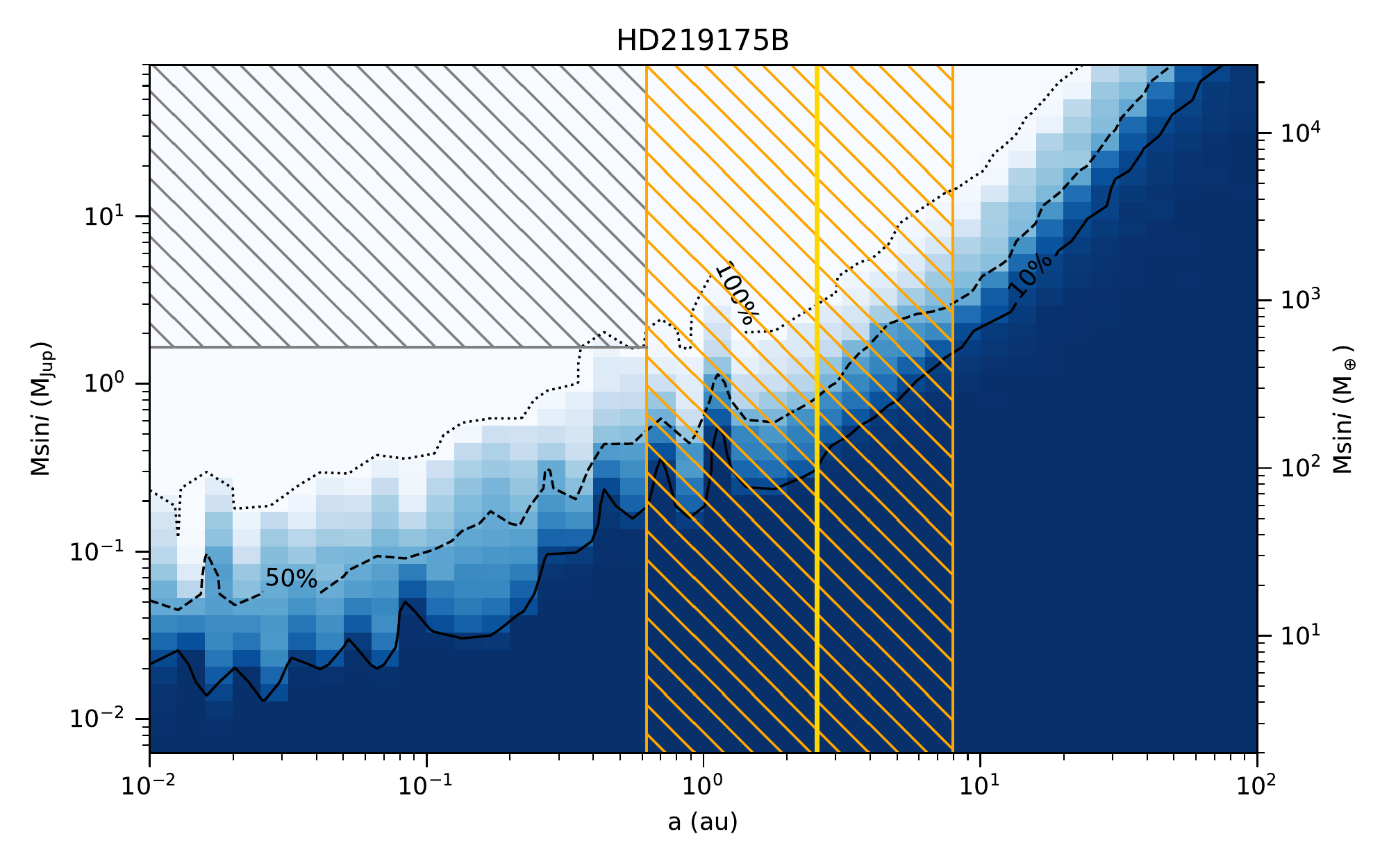}&
    		\includegraphics[width=0.22\linewidth]{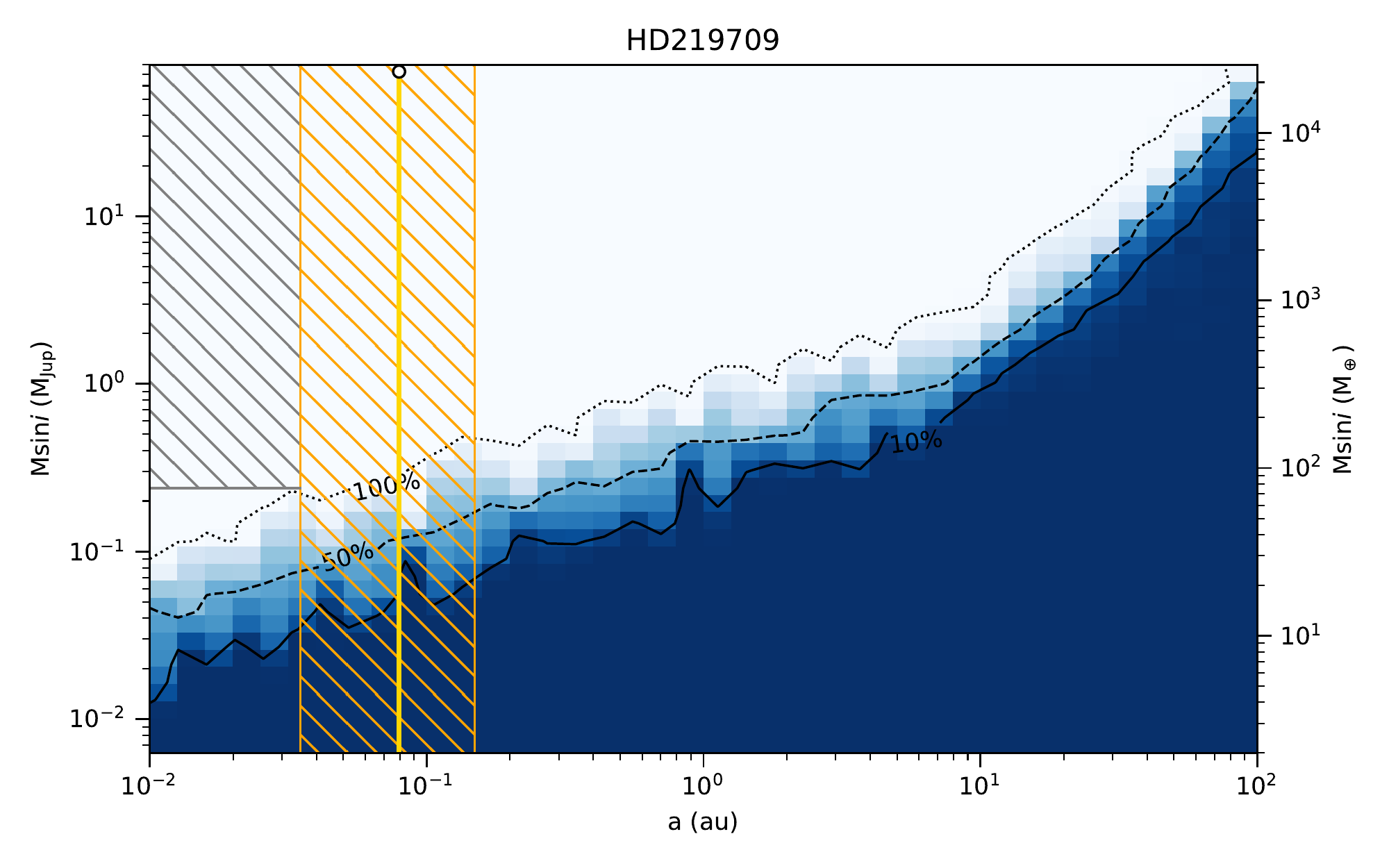}&
    		\includegraphics[width=0.22\linewidth]{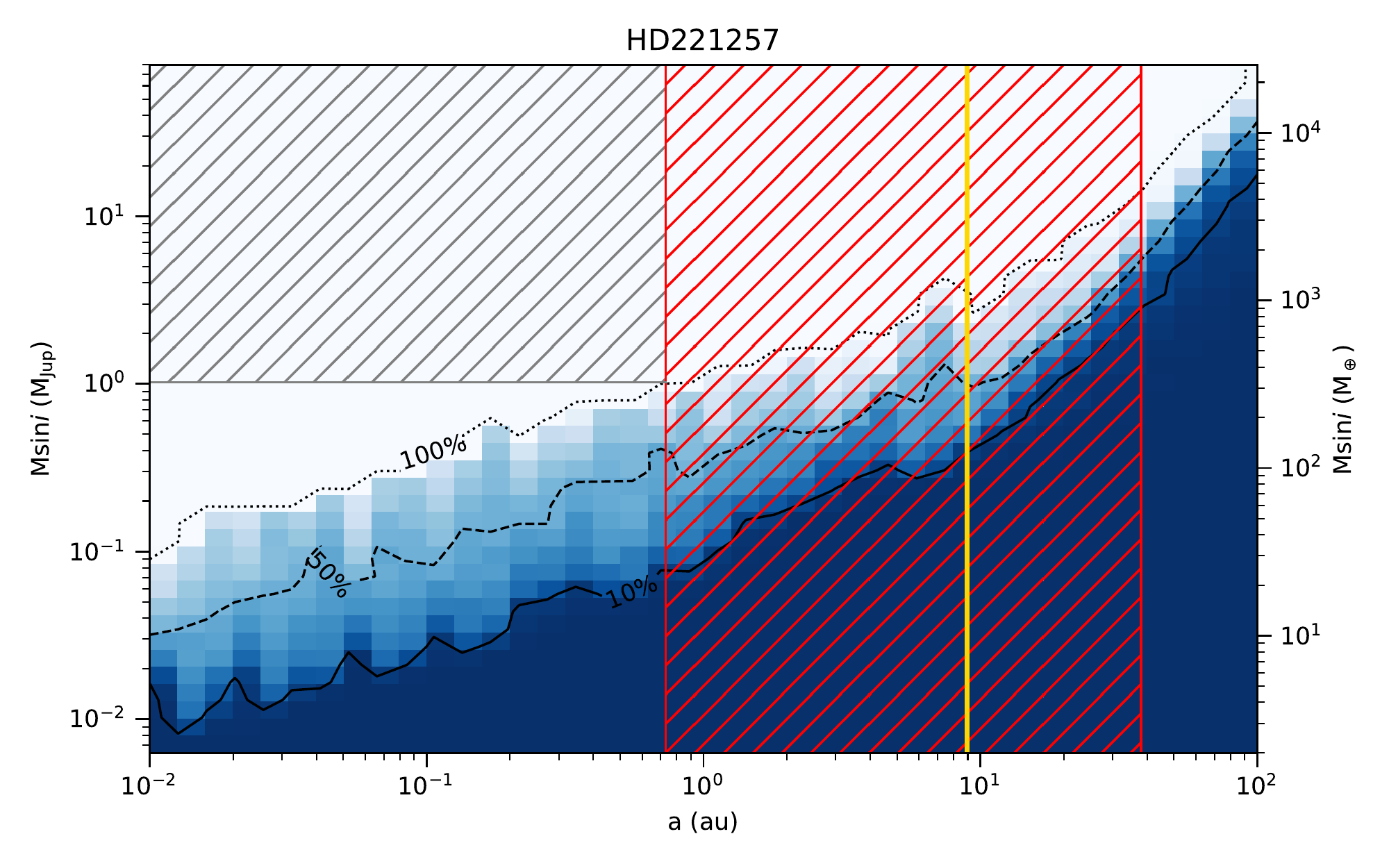}\\
    
    		\includegraphics[width=0.22\linewidth]{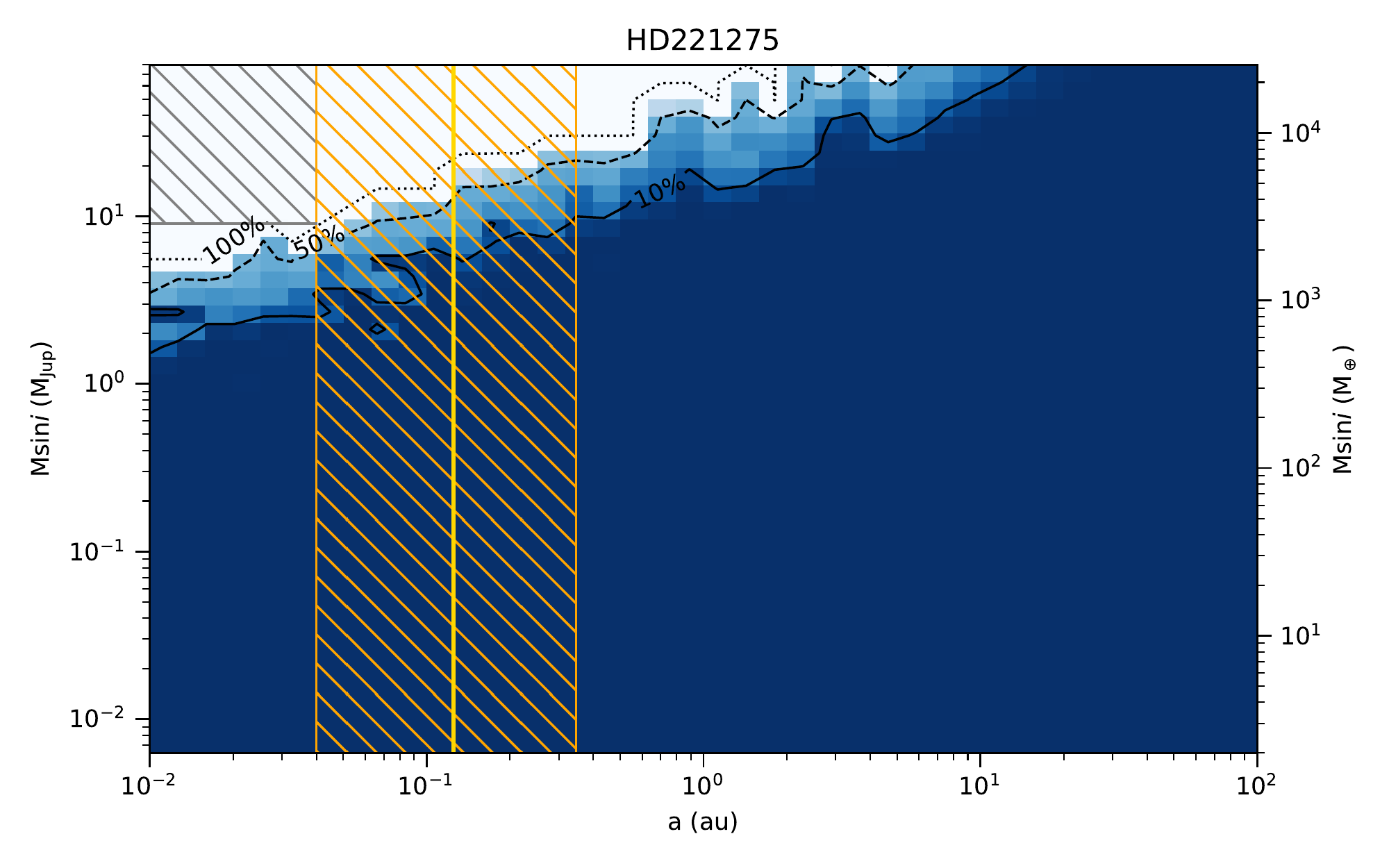}&
    		\includegraphics[width=0.22\linewidth]{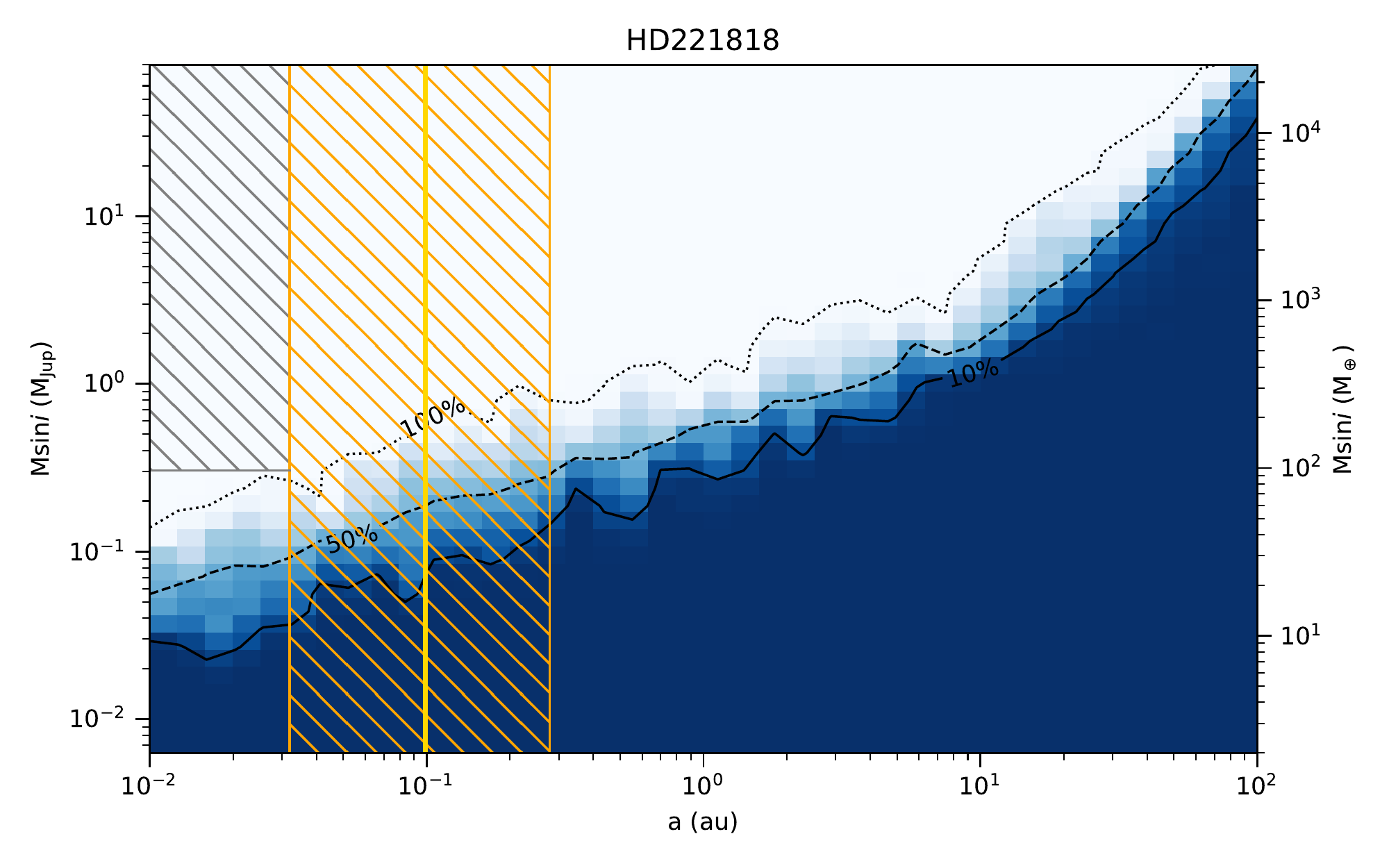}&
    		\includegraphics[width=0.22\linewidth]{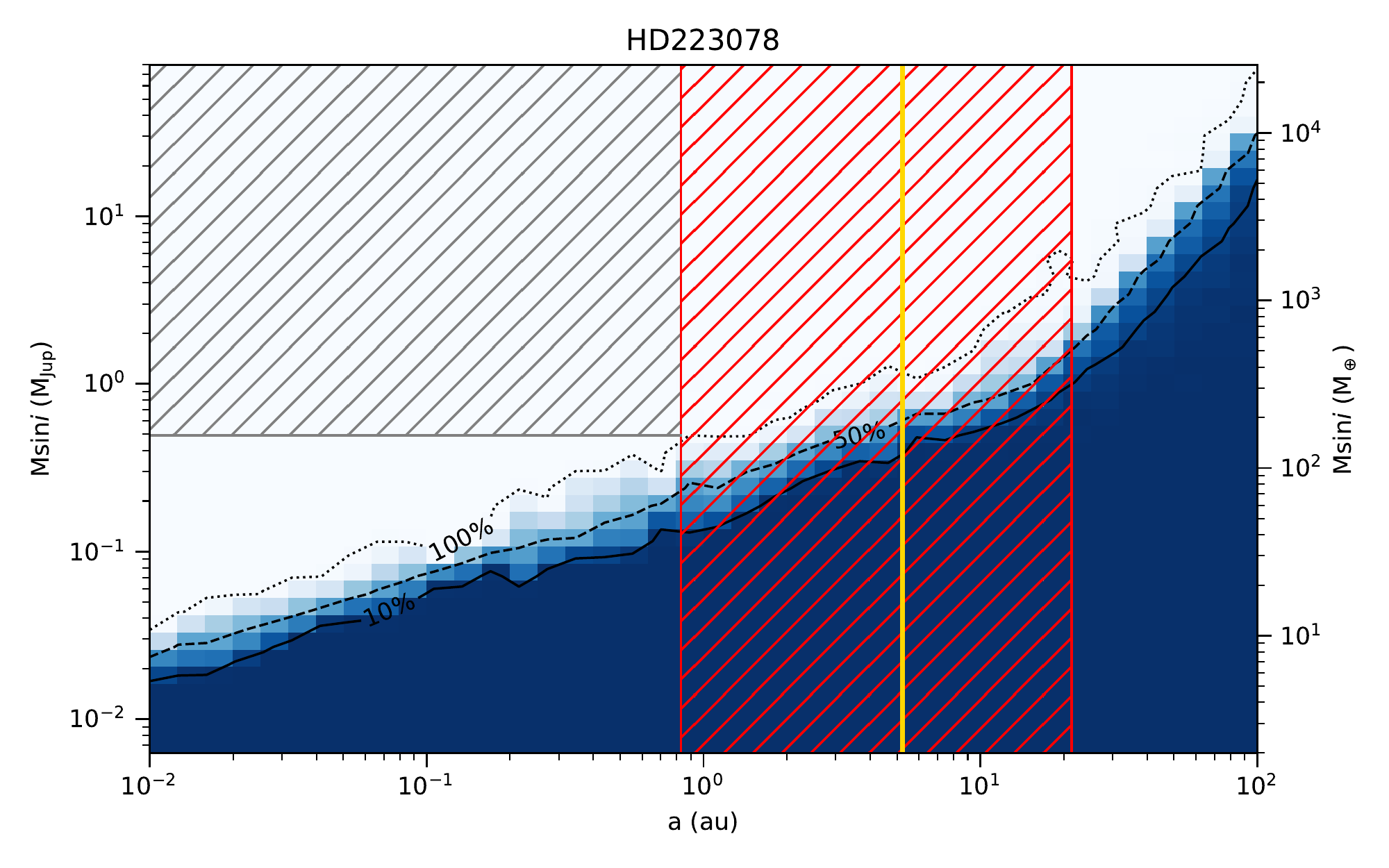}&
    		\includegraphics[width=0.22\linewidth]{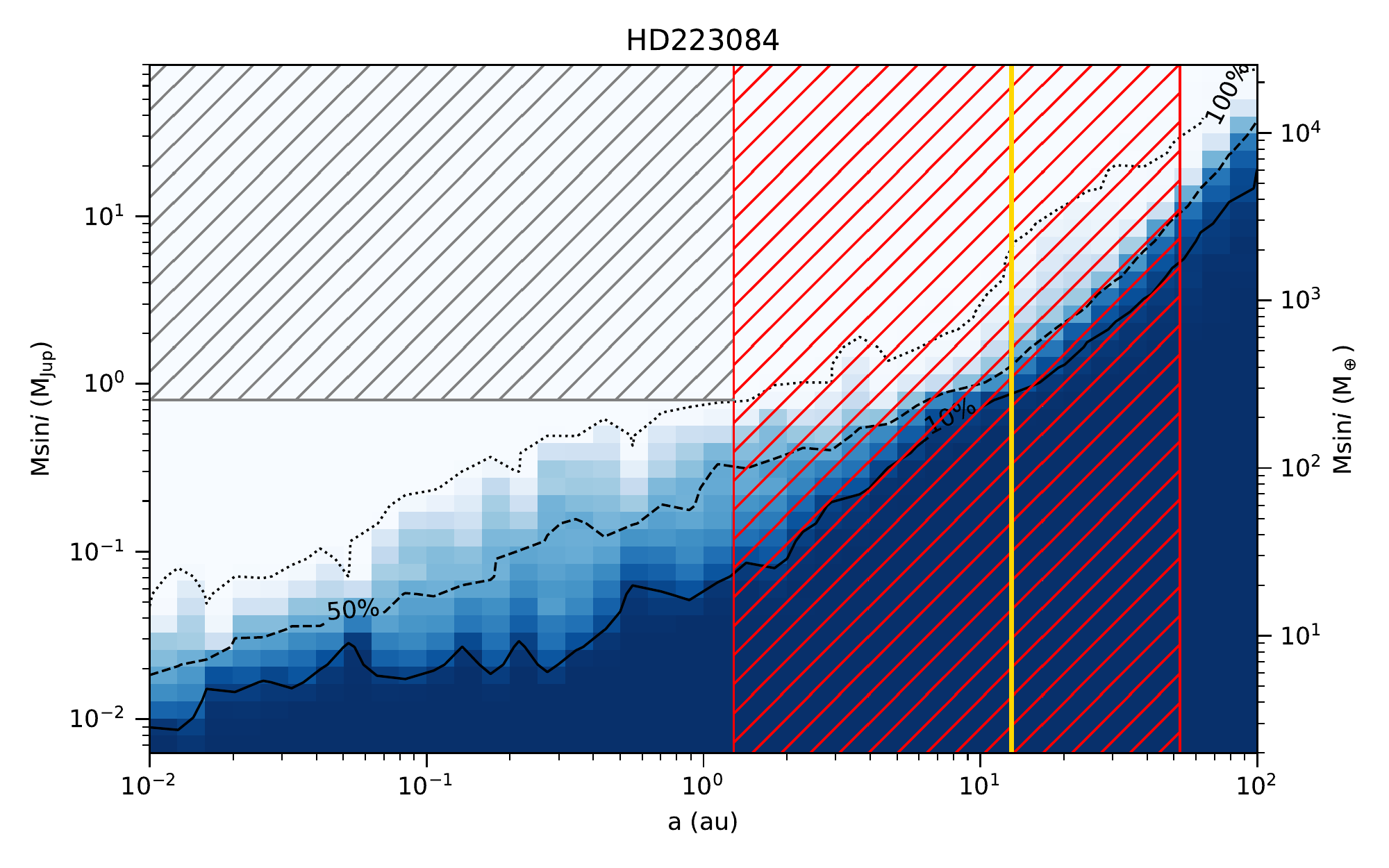}\\
    
    		\includegraphics[width=0.22\linewidth]{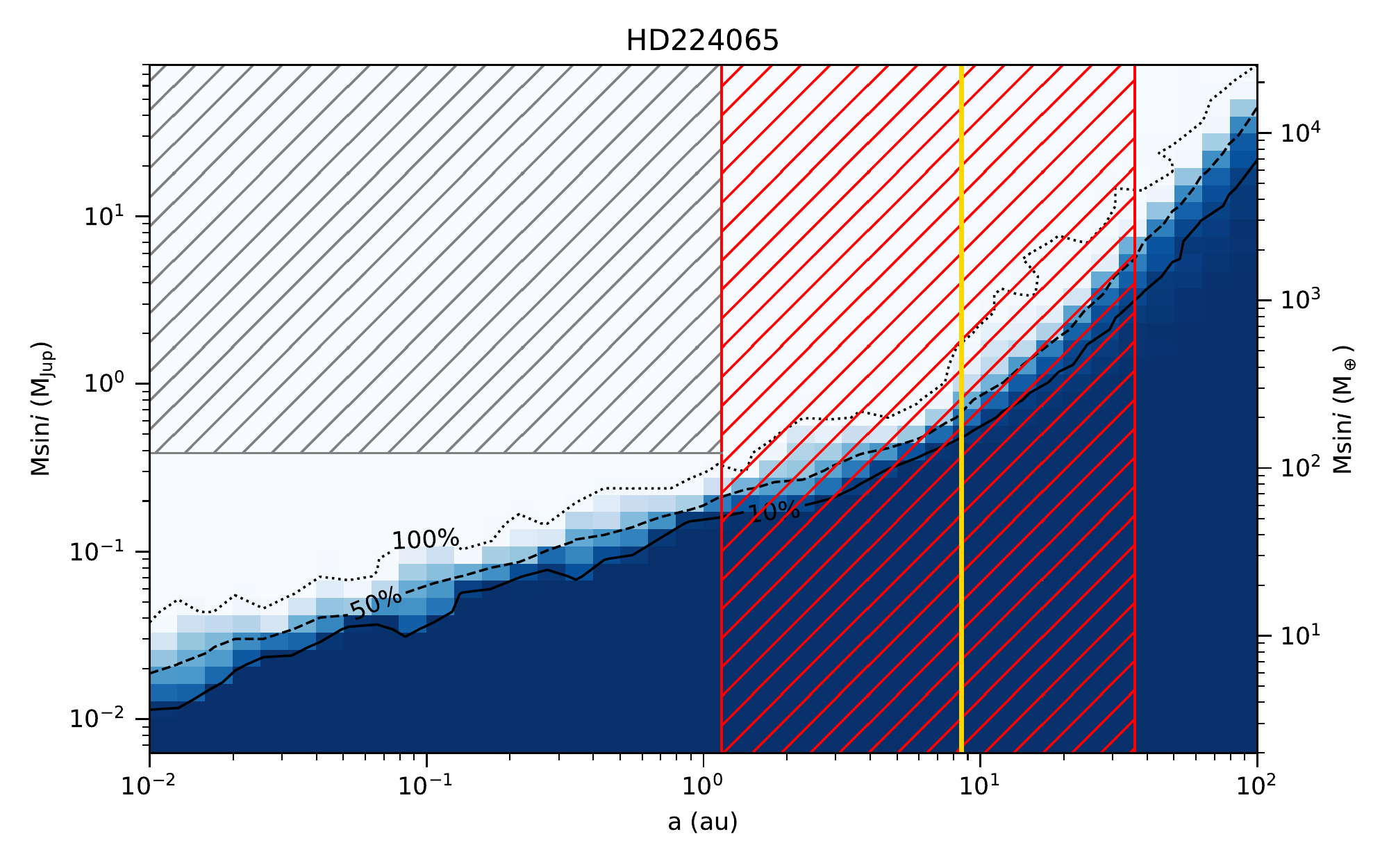}&
    		\includegraphics[width=0.22\linewidth]{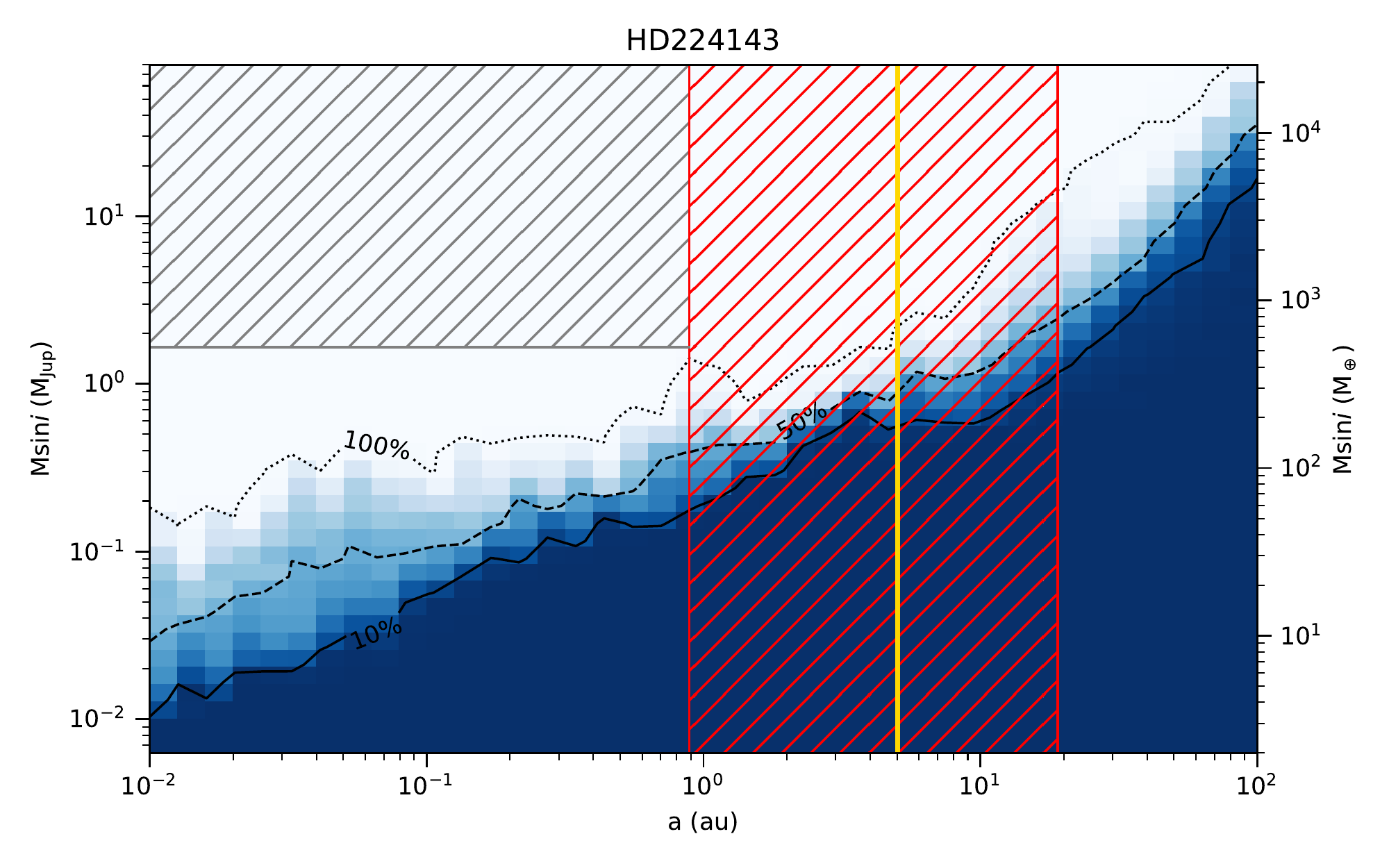}&
    	\end{longtable}
    	\captionof{figure}{Completeness maps of each star in the binary sample, focused on the substellar (2\Mearth$<$\msini$<$80\Mjup) companion regime. Detection frequency contour levels of 10, 50 and 100\% are respectively shown as solid, dashed and dotted curves, while the companions detected around each star are shown as white circles. In each plot the hatched red and orange boxes represent the dynamically unstable semimajor axis range for additional substellar companions computed respectively using the true mass and minimum mass values of the detected companions in the sample as described in Sect.~\ref{subsec:dynamics}, with a vertical yellow line representing the detected companion semimajor axis. The hatched grey boxes show the region of parameter space for which we can exclude the presence of additional circumprimary companions based on the CORAVEL and CORALIE data analysed.}\label{fig:app-detection}
    	}
    \clearpage
    \twocolumn
    
  \end{appendix}

\end{document}